\pdfoutput=1

\documentclass[3p,times]{elsarticle}

\graphicspath{{Merged_Figure/}}
 
\DeclareGraphicsExtensions{.pdf,.png,.jpeg,.eps}

\journal{Eur. Phys. J. C}

\usepackage{amssymb}
\usepackage{graphicx}
\usepackage{amsthm,amsmath,amsfonts}
\usepackage{hyperref}
\usepackage{bm}
\usepackage[figuresright]{rotating}
\usepackage{epstopdf}
\usepackage{xspace}
\usepackage{tabularx}
\usepackage{threeparttable}
\usepackage{lscape}
\usepackage{xcolor}
\usepackage{subfigure}
\biboptions{comma,square,numbers,sort&compress}
\usepackage{natbib}
\usepackage{lineno}
\usepackage{multirow}

\hypersetup{
   pdftitle={},%
    pdfauthor={},%
    pdfsubject={},%
    pdfkeywords={},%
    pdfstartview={},%
    bookmarksopen=true, breaklinks=true, debug=true, %
    colorlinks=true, linkcolor=red, citecolor=blue, urlcolor=blue
  }

\newcommand{\pp}{\ensuremath{\rm pp}\xspace}
\newcommand{\ppbar}{\ensuremath{\rm p\overline{p}}\xspace}
\newcommand{\pa}{{p--A}\xspace}
\newcommand{\pA}{{p--A}\xspace}
\newcommand{\pPb}{{p--Pb}\xspace}
\newcommand{\dAu}{{d--Au}\xspace}
\newcommand{\pb}{{Pb--Pb}\xspace}
\newcommand{\AuAu}{{Au--Au}\xspace}
\newcommand{\CuCu}{{Cu--Cu}\xspace}
\newcommand{\CuAu}{{Cu--Au}\xspace}
\newcommand{\UU}{{U--U}\xspace}
\newcommand{\AAcoll}{{AA}\xspace}
\newcommand{\PbPb}{{Pb--Pb}\xspace}

\newcommand{\snn}{\ensuremath{\sqrt{s_{\mathrm{\scriptscriptstyle NN}}}}\xspace}
\newcommand{\s}{\ensuremath{\sqrt{s}}\xspace}

\newcommand{\jpsi}{\ensuremath{\mathrm{J}/\psi}\xspace}
\newcommand{\chic}{\ensuremath{\chi_c}\xspace}
\newcommand{\chib}{\ensuremath{\chi_b}\xspace}
\newcommand{\psiP}{\ensuremath{\psi\text{(2S)}}\xspace}
\newcommand{\doubleRatioPsi}{\ensuremath{\left.(N_{\psiP}/N_{\jpsi})_{\mathrm{Pb-Pb}}/(N_{\psiP}/N_{\jpsi})_{\mathrm{pp}}\right.\xspace}}
\newcommand{\ups}{\ensuremath{\Upsilon}\xspace}
\newcommand{\upsa}{\ensuremath{\Upsilon\text{(1S)}}\xspace}
\newcommand{\upsb}{\ensuremath{\Upsilon\text{(2S)}}\xspace}
\newcommand{\upsc}{\ensuremath{\Upsilon\text{(3S)}}\xspace}
\newcommand{\upsbc}{\ensuremath{\Upsilon\text{(2S+3S)}}\xspace}
\newcommand{\upsabc}{\ensuremath{\Upsilon\text{(1S+2S+3S)}}\xspace}
\newcommand{\upsn}{\ensuremath{\Upsilon\text{(nS)}}\xspace}

\newcommand{\qqbar}{\ensuremath{{q\overline{q}}}\xspace}
\newcommand{\QQbar}{\ensuremath{{Q\overline{Q}}}\xspace}
\newcommand{\Qcal}{\ensuremath{\Phi\xspace}}
\newcommand{\ccbar}{\ensuremath{{c\overline{c}}}\xspace}
\newcommand{\bbbar}{\ensuremath{{b\overline{b}}}\xspace}

\newcommand{\Dmeson}[1]{\ensuremath{\mathrm{D}^{#1}}\xspace}

\newcommand{\Dzero}{\Dmeson{0}}
\newcommand{\Dplus}{\Dmeson{+}}

\newcommand{\Dstarplus}{\Dmeson{*+}}

\newcommand{\Ds}{\ensuremath{\mathrm{D}^{+}_{s}}\xspace}
\newcommand{\kaon}{\ensuremath{\mathrm{K}}\xspace}
\newcommand{\hfe}{\ensuremath{{\rm HF} \to e^{\pm}}\xspace}
\newcommand{\hfm}{\ensuremath{{\rm HF} \to \mu^{\pm}}\xspace}

\newcommand{\etac}{\ensuremath{\eta_c}\xspace}
\newcommand{\lambdab}{\ensuremath{\Lambda_b}\xspace}

\newcommand{\lambdacplus}{\ensuremath{\Lambda_c^+}\xspace}

\newcommand{\sigmacplus}{\ensuremath{\Sigma_c^+}\xspace}

\newcommand{\lambdabplus}{\ensuremath{\Lambda_b^+}\xspace}

\newcommand{\ee}{\ensuremath{e^+e^-}\xspace}
\newcommand{\mumu}{\ensuremath{\mu^+\mu^-}\xspace}

\newcommand{\electron}{\ensuremath{e}\xspace}

\newcommand{\cquark}{\ensuremath{c}\xspace}
\newcommand{\bquark}{\ensuremath{b}\xspace}
\newcommand{\uquark}{\ensuremath{u}\xspace}
\newcommand{\dquark}{\ensuremath{d}\xspace}
\newcommand{\squark}{\ensuremath{s}\xspace}

\newcommand{\raa}{\ensuremath{R_{\mathrm{AA}}}\xspace}
\newcommand{\rpa}{\ensuremath{R_{\mathrm{pA}}}\xspace}
\newcommand{\spa}{\ensuremath{S_{\mathrm{pA}}}\xspace}
\newcommand{\rppb}{\ensuremath{R_{\mathrm{pPb}}}\xspace}
\newcommand{\rdau}{\ensuremath{R_{\mathrm{dAu}}}\xspace}
\newcommand{\rcp}{\ensuremath{R_{\mathrm{CP}}}\xspace}
\newcommand{\rfb}{\ensuremath{R_{\mathrm{FB}}}\xspace}

\newcommand{\MeV}{\ensuremath{\text{~MeV}}\xspace}
\newcommand{\GeV}{\ensuremath{\text{~GeV}}\xspace}
\newcommand{\TeV}{\ensuremath{\text{~TeV}}\xspace}

\newcommand{\GeVc}{\ensuremath{\text{~GeV}/c}\xspace}

\newcommand{\lumi}{\ensuremath{\mathcal{L}}\xspace}

\newcommand{\mubinv}{\ensuremath{~\mu\text{b}^\text{$-$1}}\xspace}
\newcommand{\nbinv}{\ensuremath{\text{~nb}^\text{$-$1}}\xspace}

\newcommand{\eg}{\emph{e.\,g.}\xspace}%
\newcommand{\ie}{\emph{i.\,e.}\xspace}%
\newcommand{\etal}{\emph{et al.}\xspace}

\newcommand{\dd}{\ensuremath{\mathrm{d}}}

\newcommand{\av}[1]{\ensuremath{\left\langle #1 \right\rangle}}
\newcommand{\bk}{\vec{k}}
\newcommand{\bq}{\vec{q}} 
\newcommand{\sabs}{\ensuremath{\sigma_\mathrm{abs}}\xspace}
\newcommand{\muf}{\ensuremath{\mu_F}\xspace}
\newcommand{\mur}{\ensuremath{\mu_R}\xspace}

\newcommand{\beq}{\begin{equation}}
\newcommand{\eeq}{\end{equation}}
\newcommand{\beqn}{\begin{eqnarray}}
\newcommand{\eeqn}{\end{eqnarray}}

\newcommand{\Npart}{\ensuremath{\mathrm{N_{part}}}\xspace}
\newcommand{\Ncoll}{\ensuremath{\mathrm{N_{coll}}}\xspace}
\newcommand{\taa}{\ensuremath{T_{\mathrm{AA}}}\xspace}
\newcommand{\dEdx}{\ensuremath{{\mathrm{d}}E/{\mathrm{d}}x}\xspace}
\newcommand{\vtwo}{\ensuremath{v_{2}}\xspace}
\newcommand{\pt}{\ensuremath{p_{\mathrm{T}}}\xspace}
\newcommand{\mt}{\ensuremath{m_{\mathrm{T}}}\xspace}
\newcommand{\kt}{\ensuremath{k_{\mathrm{T}}}\xspace}

\newcommand{\iaa}{\ensuremath{I_{\mathrm{AA}}}\xspace}
\newcommand{\ylab}{\ensuremath{y_{\mathrm{lab}}}\xspace}
\newcommand{\ycm}{\ensuremath{y}\xspace}
\newcommand{\xf}{\ensuremath{x_{\mathrm{F}}}\xspace}
\newcommand{\dsdpt}{\ensuremath{{\rm d}\sigma/{\rm d}p_{\mathrm{T}}}\xspace}
\newcommand{\dsdy}{\ensuremath{{\rm d}\sigma/{\rm d}y}\xspace}
\newcommand{\meanpt}{\ensuremath{\langle p_{\mathrm{T}} \rangle}\xspace}

\newcommand{\tf}{\ensuremath{\tau_{\rm f}}\xspace}
\newcommand{\tcross}{\ensuremath{\tau_{\rm cross}}\xspace}

\newcommand{\fig}[1]{\figurename~\ref{#1}}
\newcommand{\tab}[1]{\tablename~\ref{#1}}
\newcommand{\sect}[1]{Section~\ref{#1}}
\newcommand{\eq}[1]{Eq.~(\ref{#1})}
\newcommand{\ci}[1]{Ref.~\cite{#1}}
\newcommand{\cis}[1]{Refs.~\cite{#1}}

\newcommand{\figs}{Figures\xspace}
\newcommand{\tabs}{Tables\xspace}
\newcommand{\sects}{Sections\xspace}
\newcommand{\eqs}{Equations\xspace}

\newcommand{\RunOne}{Run~1\xspace}
\newcommand{\RunTwo}{Run~2\xspace}
\newcommand{\RunThree}{Run~3\xspace}
\newcommand{\LSOne}{LS1\xspace}
\newcommand{\LSTwo}{LS2\xspace}
\newcommand{\LSThree}{LS3\xspace}

\begin{document}

\begin{frontmatter}
%



\title{
Heavy-flavour and quarkonium production in the LHC era: \\from proton-proton to heavy-ion collisions \\
}


\author[1]{A.~Andronic\fnref{ed}}
\author[2,3]{F.~Arleo\fnref{ed}}
\author[4]{R.~Arnaldi\fnref{ed}}
\author[4]{A.~Beraudo}
\author[4]{E.~Bruna}
\author[5]{D.~Caffarri}
\author[6]{Z.~Conesa del Valle\fnref{ed}}
\author[8]{J.G.~Contreras\fnref{ed}}
\author[9]{T.~Dahms\fnref{ed}}
\author[10]{A.~Dainese\fnref{ed}}
\author[11]{M.~Djordjevic}
\author[12]{E.G.~Ferreiro\fnref{ed}}
\author[13]{H.~Fujii}
\author[14]{P.-B.~Gossiaux\fnref{ed}}
\author[2]{R.~Granier~de~Cassagnac}
\author[6]{C.~Hadjidakis\fnref{ed}}
\author[15]{M.~He}
\author[16]{H.~van~Hees}
\author[17]{W.A.~Horowitz}
\author[14,18]{R.~Kolevatov}
\author[19]{B.Z.~Kopeliovich}
\author[6]{J.~P.~Lansberg\fnref{ed}}
\author[20,2Obis]{M.P.~Lombardo}
\author[5]{C.~Louren\c co}
\author[14]{G.~Martinez-Garcia\fnref{ed}}
\author[14,21,6]{L.~Massacrier\corref{cor1}\fnref{ed}}
\ead{massacri@lal.in2p3.fr}
\cortext[cor1]{Corresponding author:}
\author[2]{C.~Mironov}
\author[22,23]{A.~Mischke}
\author[24]{M.~Nahrgang}
\author[2]{M.~Nguyen}
\author[25]{J.~Nystrand\fnref{ed}}
\author[14]{S.~Peign\'e}
\author[26]{S.~Porteboeuf-Houssais\fnref{ed}}
\author[19]{I.K.~Potashnikova}
\author[27]{A.~Rakotozafindrabe}
\author[28]{R.~Rapp}
\author[21]{P.~Robbe\fnref{ed}}
\author[29]{M.~Rosati}
\author[26]{P.~Rosnet\fnref{ed}}
\author[30]{H.~Satz}
\author[31]{R.~Schicker\fnref{ed}}
\author[32]{I.~Schienbein\fnref{ed}}
\author[19]{I.~Schmidt}
\author[4]{E.~Scomparin}
\author[33]{R.~Sharma}
\author[31]{J.~Stachel}
\author[14]{D.~Stocco\fnref{ed}}
\author[34]{M.~Strickland}
\author[35]{R.~Tieulent\fnref{ed}}
\author[8]{B.A.~Trzeciak\fnref{ed}}
\author[37]{J.~Uphoff}
\author[38]{I.~Vitev}
\author[39,39bis]{R.~Vogt}
\author[40,40bis]{K.~Watanabe}
\author[5]{H.~Woehri}
\author[41]{P.~Zhuang}

\fntext[ed]{Section editors}

\address[1]{Research Division and ExtreMe Matter Institute EMMI, GSI Helmholzzentrum f\"ur Schwerionenforschung, Darmstadt, Germany, \\}
\address[2]{Laboratoire Leprince-Ringuet, Ecole Polytechnique, IN2P3-CNRS, Palaiseau, France, \\}
\address[3]{Laboratoire d'Annecy-le-Vieux de Physique Th\'eorique (LAPTh), Universit\'e de Savoie, CNRS, Annecy-le-Vieux, France, \\}
\address[4]{INFN, Sezione di Torino, Torino, Italy, \\}
\address[5]{European Organization for Nuclear Research (CERN), Geneva, Switzerland, \\}
\address[6]{Institut de Physique Nucl\'eaire d'Orsay (IPNO), Universit\'e Paris-Sud, CNRS/IN2P3, Orsay, France, \\}
\address[8]{Faculty of Nuclear Sciences and Physical Engineering, Czech Technical University in Prague, Prague, Czech Republic, \\}
\address[9]{Excellence Cluster Universe, Technische Universit\"at M\"unchen, Munich, Germany, \\}
\address[10]{INFN, Sezione di Padova, Padova, Italy, \\}
\address[11]{Institute of Physics Belgrade, University of Belgrade, Belgrade, Serbia, \\}
\address[12]{Departamento de F\'isica de Part\'iculas and IGFAE, Universidad de Santiago de Compostela, Santiago de Compostela, Spain, \\}
\address[13]{Institute of Physics, University of Tokyo, Tokyo, Japan, \\}
\address[14]{SUBATECH, Ecole des Mines de Nantes, Universit\'e de Nantes, CNRS-IN2P3, Nantes, France, \\}
\address[15]{Department of Applied Physics, Nanjing University of Science and Technology, Nanjing, China, \\}
\address[16]{FIAS and Institute for Theoretical Physics, Frankfurt, Germany, \\}
\address[17]{Department of Physics, University of Cape Town, Cape Town, South Africa, \\}
\address[18]{Department of High Energy Physics, Saint-Petersburg State University Ulyanovskaya 1, Saint-Petersburg, Russia, \\}
\address[19]{Departamento de F\'isica, Universidad T\'ecnica Federico Santa Mar\'ia; and Centro Cient\'ifico-Tecnol\'ogico de Valpara\'iso , Valpara\'iso, Chile, \\ }
\address[20]{INFN, Laboratori Nazionali di Frascati, Frascati, Italy,\\}
\address[20bis]{INFN, Sezione di Pisa, Pisa,Italy, \\}
\address[21]{LAL, Universit\'e Paris-Sud, CNRS/IN2P3, Orsay, France, \\}
\address[22]{Institute for Subatomic Physics, Faculty of Science, Utrecht University, Utrecht, the Netherlands, \\ }
\address[23]{National Institute for Subatomic Physics, Amsterdam, the Netherlands, \\}
\address[24]{Department of Physics, Duke University, Durham, USA\\}
\address[25]{Department of Physics and Technology, University of Bergen, Bergen, Norway, \\}
\address[26]{Laboratoire de Physique Corpusculaire (LPC), Universit\'e Clermont Auvergne, Universit\'e Blaise Pascal, CNRS/IN2P3, Clermont-Ferrand, France, \\}
\address[27]{Comissariat \`a l'Energie Atomique, IRFU, Saclay, France, \\}
\address[28]{Cyclotron Institute and Department of Physics and Astronomy, Texas A\&M University, College Station, USA, \\}
\address[29]{Iowa State University, Ames, USA, \\}
\address[30]{Fakult\"at f\"ur Physik, Universit\"at Bielefeld, Bielefeld, Germany, \\}
\address[31]{Physikalisches Institut, Ruprecht-Karls-Universit\"at Heidelberg, Heidelberg, Germany, \\}
\address[32]{Laboratoire de Physique Subatomique et de Cosmologie, Universit\'e Grenoble-Alpes, CNRS/IN2P3, Grenoble, France, \\}
\address[33]{Department of Theoretical Physics, Tata Institute of Fundamental Research, Mumbai, India, \\}
\address[34]{Department of Physics, Kent State University, Kent, United States, \\}
\address[35]{Universit\'e de Lyon, Universit\'e Lyon 1, CNRS/IN2P3, IPN-Lyon, Villeurbanne, France, \\}
\address[37]{Institut f\"ur Theoretische Physik, Johann Wolfgang Goethe-Universit\"at, Frankfurt am Main, Germany, \\}
\address[38]{Theoretical Division, Los Alamos National Laboratory, Los Alamos, USA, \\}
\address[39]{Physics Division, Lawrence Livermore National Laboratory, Livermore, USA, \\}
\address[39bis]{Physics Department, University of California, Davis, USA, \\}
\address[40]{Institute of Physics, University of Tokyo, Tokyo, Japan, \\}
\address[40bis]{Key Laboratory of Quark and Lepton Physics (MOE) and Institute of Particle Physics, Central China Normal University, Wuhan, China, \\}
\address[41]{Physics department, Tsinghua University and Collaborative Innovation Center of Quantum Matter, Beijing, China. \\ \newpage}


\begin{abstract}
This report reviews the study of open heavy-flavour and quarkonium production in high-energy hadronic collisions, as tools to investigate fundamental aspects of Quantum Chromodynamics, from the proton and nucleus structure at high energy to deconfinement and the properties of the Quark-Gluon Plasma. Emphasis is given to the lessons learnt from LHC \RunOne results, which are reviewed in a global picture with the results from SPS and RHIC at lower energies, as well as to the questions to be addressed in the future. The report covers heavy flavour and quarkonium production in proton-proton, proton-nucleus and nucleus-nucleus collisions. This includes discussion of the effects of hot and cold strongly interacting matter, quarkonium photo-production in nucleus-nucleus collisions and perspectives on the study of heavy flavour and quarkonium with upgrades of existing experiments and new experiments. The report results from the activity of the SaporeGravis network of the I3 Hadron Physics programme of the European Union 7$^{\rm th}$ Framework Programme. 
\end{abstract}

\begin{keyword}
Heavy Flavour \sep
Charm \sep
Beauty \sep
Quarkonium \sep
Hadron Collision \sep
Heavy-Ion Collision \sep
Ultra Peripheral Collision
Quark--Gluon Plasma \sep
LHC \sep
RHIC
\end{keyword}

\end{frontmatter}





\newpage
\tableofcontents
\newpage






\section{Introduction}
\label{sec:introduction}

Heavy-flavour hadrons, containing open or hidden charm and beauty flavour, are among the most important tools
for the study of Quantum Chromodynamics (QCD) in high-energy hadronic collisions, from the 
production mechanisms in proton--proton collisions and their modification in proton--nucleus collisions
to the investigation of the properties of the hot and dense strongly-interacting Quark-Gluon Plasma (QGP)
in nucleus--nucleus collisions.

Heavy-flavour production in pp collisions provides important tests of our understanding of various aspects
of QCD. The heavy-quark mass acts as a long distance cut-off so that the partonic hard scattering process can be
calculated in the framework of perturbative QCD down to low transverse momenta (\pt). When the heavy-quark
pair forms a quarkonium bound state, this process is non-perturbative as it involves long distances and soft
momentum scales. Therefore, the detailed study of heavy-flavour production and the comparison to experimental data
provides an important testing ground for both perturbative and non-perturbative aspects of QCD calculations. 

In nucleus--nucleus collisions, open and hidden heavy-flavour production constitutes a sensitive probe 
of the hot strongly-interacting medium, because hard scattering processes take place in the early stage of the collision
on a time-scale that is in general shorter than the QGP thermalisation time.
Disentangling the medium-induced effects and relating them to its properties requires an accurate study of the so-called cold nuclear matter
(CNM) effects, which modify the production of heavy quarks in nuclear collisions with respect to proton--proton collisions.
CNM effects, which can be measured in proton--nucleus interactions, include: the modification of the effective partonic luminosity in nuclei (which can be described using nuclear-modified parton densities),
due to saturation of the parton kinematics phase space; the multiple scattering of partons in the nucleus before and after the hard scattering;
the absorption or break-up of quarkonium states, and the interaction with other particles produced in the collision (denoted 
as comovers). 
 
The nuclear modification of the parton distribution functions can also be studied, in a very clean environment, using 
quarkonium
photo-production in ultra-peripheral nucleus--nucleus collisions, 
in which a photon from the coherent electromagnetic field of an accelerated nucleus interacts with the coherent gluon field of
the other nucleus or with the gluon field of a single nucleon in the other nucleus.

During their propagation through the QGP produced in high-energy 
nucleus--nucleus collisions, heavy quarks interact with the constituents of this medium
 and lose a part of their momentum, thus being able to reveal some of the QGP properties.
QCD energy loss is expected to occur
via both inelastic (radiative energy loss, via medium-induced gluon radiation) and elastic (collisional energy loss)
processes. Energy loss is expected to depend on the parton colour-charge and mass. Therefore, charm and beauty quarks
provide important tools to investigate the energy loss mechanisms, in addition to the QGP properties.
Furthermore, low-\pt heavy quarks could participate, through their interactions with the medium,
in the collective expansion of the system and possibly reach thermal equilibrium with its constituents.

In nucleus--nucleus collisions, 
quarkonium production is expected to be significantly suppressed 
as a consequence of the 
 colour screening of
the force that binds the \ccbar (\bbbar) state. 
In this
scenario, quarkonium suppression should occur sequentially, according to the
binding energy of each state. 
As a consequence, the in-medium dissociation
probability of these states are expected to provide an estimate of the initial
temperature reached in the collisions. 
At high centre-of-mass energy, a new production mechanism 
could be at work in the case of charmonium:
the
abundance of $c$ and $\overline{c}$ quarks might lead to charmonium
production by (re)combination of these quarks. 
An observation of the recombination of heavy quarks would therefore directly
point to the
existence of a deconfined QGP.

The first run of the Large Hadron Collider (LHC), from 2009 to 2013, has provided a wealth of measurements
in pp collisions with unprecedented centre-of-mass energies $\sqrt s$ from 2.76 to 8~TeV, in p--Pb collisions at $\snn=5.02$~TeV
per nucleon--nucleon interaction, in Pb--Pb collisions at $\snn= 2.76$~TeV, as well as in photon-induced collisions.
In the case of heavy-ion collisions, with respect to the experimental programmes at SPS and RHIC, the LHC programme has 
not only extended by more than one order of magnitude the range of explored collision energies, but it has also largely
enriched the studies of heavy-flavour production, with a multitude of new observables and improved 
precision. Both these aspects were made possible by the energy increase, on the one hand, and by the 
excellent performance of the LHC and the experiments, on the other hand.

This report results from the activity of the SaporeGravis network\footnote{\url{https://twiki.cern.ch/twiki/bin/view/ReteQuarkonii/SaporeGravis}}
of the I3 Hadron Physics programme of the European Union $7^{\rm th}$ FP.
The network was structured in working groups, that are reflected in the structure of this review, and it focused on supporting and strengthening the interactions between the experimental and theoretical communities. 
This goal was, in particular, pursued by organising
two large workshops, in Nantes (France)\footnote{\url{https://indico.cern.ch/event/247609}}
in December 2013 and in Padova (Italy)\footnote{\url{https://indico.cern.ch/event/305164}} in 
December 2014.

The report is structured in eight sections. 
Sections~\ref{pp section}, \ref{Cold nuclear matter effects}, \ref{OHF}, \ref{sec:quarkonia} and \ref{sec:UPC} review, respectively: heavy-flavour and quarkonium production in pp collisions, the cold nuclear matter 
effects on heavy-flavour and quarkonium production in proton--nucleus collisions, the QGP effects on open heavy-flavour 
production in nucleus--nucleus collisions, the QGP effects on quarkonium production in nucleus--nucleus collisions,
and the production of charmonium in  photon-induced collisions. \sect{sec:upgrade}
presents an outlook of future heavy-flavour studies with 
the LHC and RHIC detector upgrades and with new experiments. A short summary concludes the report in \sect{sec:summary}.

\newpage

\section{Heavy flavour and quarkonium production in \pp collisions}
\label{pp section}

\subsection{Production mechanisms of open and hidden heavy-flavour in \pp collisions}
\label{sec:pp:Theory}

\subsubsection{Open-heavy-flavour production}
\label{sec:pp:Theory:OpenHF}

 
Open-heavy-flavour production in hadronic collisions provides important 
tests of our understanding of various aspects of Quantum Chromodynamics (QCD). 
First of all, the heavy-quark mass ($m_Q$) acts as a long distance cut-off so that 
this process can be calculated in the framework of perturbative QCD 
down to low \pt and it is possible to compute the  
total cross section by integrating over \pt. 
Second, the presence of multiple hard scales ($m_Q$, \pt) allows us to study 
the perturbation series in different kinematic regions ($\pt < m_Q$, $\pt\sim m_Q$, $\pt \gg m_Q$). 
Multiple hard scales are also present in other collider processes of high interest 
such as weak boson production, Higgs boson production and many cases of 
physics Beyond the Standard Model. 
Therefore, the detailed study of heavy-flavour production and the comparison to experimental data provides an important  
testing ground for the theoretical ideas that deal with this class of problems. 
 
On the phenomenological side, the (differential) cross section for open-heavy-flavour production is sensitive to 
the gluon and the heavy-quark content in the nucleon, so that LHC data in \pp and \pPb collisions can provide valuable  
constraints on these parton-distribution functions (PDFs) inside the proton and the lead nucleus, respectively. 
In addition, these cross sections in \pp and \pA collisions establish the baseline for the study of heavy-quark production 
in heavy-ion collisions. This aspect is a central point in heavy-ion physics since the suppression of heavy quarks at large \pt is 
an important signal of the QGP (see \sect{OHF}). 
Finally, let us also mention that a solid understanding of open-charm production is needed in cosmic-ray and neutrino astrophysics~\cite{Bhattacharya:2015jpa}. 
In the following, we will focus on \pp collisions and review the different theoretical approaches  
to open-heavy-flavour production. 
 
\paragraph{Fixed-Flavour-Number Scheme}
Conceptionally, the simplest scheme is the Fixed-Flavour-Number Scheme (FFNS)  
where the heavy quark is not an active parton in the proton. 
Relying on a factorisation theorem, the differential cross section for the inclusive production of a heavy quark $Q$ 
can be calculated as follows: 
\begin{equation} 
\dd\sigma^{Q+X}[s,\pt,y, m_Q] \simeq \sum_{i,j} \int_0^1 \dd x_i \int_0^1 dx_j\,  f^A_i(x_i,\mu_F) f^B_j(x_j,\mu_F)   
\dd\tilde \sigma_{ij\to Q+X}[x_i,x_j,s,\pt,y,m_Q,\mu_F,\mu_R]\, ,  
\label{eq:xsQ} 
\end{equation} 
or in short 
\begin{equation} 
\dd\sigma^{Q+X} \simeq \sum_{i,j} f^A_i \otimes f^B_j \otimes  
\dd\tilde \sigma_{ij\to Q+X}\, , 
\label{eq:xsQ2} 
\end{equation} 
where \pt and $y$ are the transverse momentum and the rapidity of the heavy quark and $s$ is the 
square of the hadron centre-of-mass energy. 
The PDFs $f_i^A$ ($f_j^B$) give the number density of the parton of flavour `$i$' (`$j$') inside the hadron `$A$' (`$B$') 
carrying a fraction $x_i$ ($x_j$) of the hadron momentum 
at the factorisation scale \muf. Furthermore, the short-distance cross section $\dd \tilde \sigma$ is the partonic cross section  
from which the so-called mass singularities or collinear singularities associated to the light quarks and the gluon have been removed  
via the mass-factorisation procedure and which therefore also depends on \muf.  
The partonic cross section also depends on the strong coupling constant $\alpha_s$, which is evaluated at the renormalisation scale  
$\mu_R$. 
As a remainder of this procedure, the 
short-distance cross section will depend on logarithms of the ratio of \muf with the hard scale. In order to avoid large 
logarithmic contributions, the factorisation scale \muf should be chosen in the vicinity of the hard scale. Also the renormalisation scale 
$\mu_R$ is determined by the hard scale. 
The tilde is used to indicate that the finite collinear logarithms of the heavy-quark mass present in the partonic cross section 
have not been removed by the mass-factorisation procedure. These logarithms are therefore not resummed to all orders in the FFNS 
but are accounted for in Fixed-Order (FO) perturbation theory.  
The error of the approximation in~\eq{eq:xsQ} is suppressed by an inverse power of the hard scale 
which is set by the mass or the transverse momentum of the heavy quark, \ie it is  
on the order of ${\cal O}((\Lambda/\muf)^p)$ where $\Lambda\sim 200\MeV$ is a typical hadronic scale, and $p=1$ or 2. 
 
In \eq{eq:xsQ}, a sum over all possible sub-processes $i+j \to Q + X$ is understood, where $i,j$ are the  
active partons in the proton: 
$i,j \in \{q,\overline q=(u,\overline u, d, \overline d, s, \overline s), g\}$ for a FFNS with three active flavours (3-FFNS) usable for both 
charm and beauty production, and 
$i,j \in \{q,\overline q=(u,\overline u, d, \overline d, s, \overline s, c, \overline c), g\}$ in the case of four active flavours (4-FFNS)  
often used for beauty production. In the latter case, the charm quark is also an active parton (for $\muf > m_c$) 
and the charm-quark mass is neglected in the hard-scattering cross section $\dd \tilde \sigma$ whereas the beauty 
quark mass $m_b$ is retained. 
At the leading order (LO) in $\alpha_S$, there are only two sub-processes which contribute:  
{\it (i)} $q+\overline q \to Q+ \overline Q$, {\it (ii)} $g + g \to Q+ \overline Q$. 
At the next-to-leading order (NLO), the virtual one-loop corrections to these $2\to 2$ processes have to be included 
in addition to the following $2\to 3$ processes: 
{\it (i)} $q+\overline q \to Q+ \overline Q + g$, {\it (ii)} $g + g \to Q+ \overline Q + g$, {\it (iii)} $g+q\to q + Q+ \overline Q$ and $g+\overline q\to \overline q + Q+ \overline Q$. 
Complete NLO calculations of the integrated/total cross section  
and of one-particle inclusive distributions 
were performed in the late 80's~\cite{Nason:1987xz,Nason:1989zy,Beenakker:1988bq,Beenakker:1990ma}. 
These calculations form also the basis for more differential observables/codes~\cite{Mangano:1991jk}  
(where the phase space of the second heavy quark has not been integrated out) allowing us to study the correlations  
between the heavy quarks -- sometimes referred to as NLO MNR. 
They are also an important ingredient to the other theories discussed below (FONLL, GM-VFNS, POWHEG, MC@NLO).


The typical range of applicability of the FFNS at NLO is roughly $0 \le \pt \lesssim 5 \times m_Q$. 
A representative comparison with data has been made for $\rm B^+$ production in~\cite{Kniehl:2015fla} 
where it is clear that the predictions of the FFNS at NLO using the branching fraction $B(b \to {\rm B}) = 39.8\%$  
starts to overshoot the Tevatron data for $\pt \gtrsim 15\GeV/c$  
even considering the theoretical uncertainties here that has been evaluated by varying the renormalisation and factorisation scales 
by factors of 2 and 1/2 around the default value\footnote{ 
We stress here that this widespread procedure to assess theoretical uncertainties associated with pQCD computations does not provide 
values which should be interpreted as coming from a statistical procedure.  
This is only an estimate of the missing contributions at higher order QCD corrections. 
} $\muf =\mur =m_{\rm T}$ with $m_{\rm T}=\sqrt{m_Q^2+\pt^2}$. 
 
%
Such a kind of discrepancies at increasing \pt can be attributed to the shift of the momentum between the $b$ quark and 
the $B$ meson which can be accounted for by a fragmentation function (FF). 
Indeed, the scope of the FFNS can be extended to slightly larger \pt by convolving the differential cross section 
for the production of the heavy quark $Q$ with a suitable, scale-independent, FF $D_Q^H(z)$ describing the 
transition of the heavy quark with momentum $p_Q$ into the observed heavy-flavoured hadron $H$  
with momentum $p_H=z\,p_Q$ 
(see~\cite{Kniehl:2015fla}): 
\begin{equation} 
\dd\sigma^H = \dd\sigma^Q \otimes D_Q^H(z)\, . 
\label{eq:xsB} 
\end{equation} 
At large transverse momenta, the differential cross section falls off with a power-like behaviour  
$\dd\sigma^Q/\dd\pt \propto 1/\pt^n$ with $n=4,5$ so that the convolution with the fragmentation function (FF) effectively 
corresponds to a multiplication with the fourth or fifth Mellin moment of this FF which lowers the cross section and leads to  
an improved agreement with the data at large \pt. 
It should be noted that this FF is included on a purely phenomenological basis and 
there are ambiguities on how the convolution prescription is implemented ($E_H = z\, E_Q$, $\vec{p}_{\rm T}^H = z\, \vec{p}_{\rm T}^Q$) leading to differences at $\pt \simeq m_Q$. Furthermore, at NLO, a harder FF should be used than at LO 
in order to compensate for the softening effects of the gluon emissions. 
Apart from this, it is generally believed that this scale-independent FF is universal and can be extracted from data, \eg from 
$\ee$ data. 
 
The same conclusions about the range of applicability of the FFNS apply at the LHC  
where the heavy-quark production is dominated by the $gg$-channel (see, \eg Figure~3 in~\cite{Kniehl:2015fla}) over 
the \qqbar one. 
As can be seen, the uncertainty at NLO due to the scale choice is very large (about a factor of two).  
For the case of top pair production, complete NNLO calculations are now available for both 
the total cross section~\cite{Czakon:2013goa} and, most recently, differential distributions~\cite{Czakon:2014xsa}. 
To make progress, it will be crucial to have NNLO predictions for charm and beauty production as well. 
%
 
 


\paragraph{ZM-VFNS} 
For $\pt \gg m_Q$, the logarithms of the heavy-quark mass ($\frac{\alpha_s}{2\pi} \ln (\pt^2/m_Q^2)$) 
become large and will eventually have to be resummed to all orders in the perturbation theory.  
This resummation is realised by absorbing the large logarithmic terms into the PDFs and FFs whose scale-dependence is governed by  
renormalisation group equations, the DGLAP evolution equations. 
This approach requires that the heavy quark is treated as an active parton for factorisation scales 
$\muf \ge \mu_T$ where the transition scale $\mu_T$ is usually (for simplicity) identified with the heavy-quark mass.  
Such a scheme, where the number of active flavours is changed when crossing the transition scales is 
called a Variable-Flavour-Number Scheme (VFNS). 
If, in addition, the heavy-quark mass $m_Q$ is neglected in the calculation of the short-distance cross sections, 
the scheme is called Zero-Mass VFNS (ZM-VFNS). 
The theoretical foundation of this scheme is provided by a well-known factorisation theorem 
and the differential cross section for the production of a heavy-flavoured hadron ($A+B \to H +X$) 
is calculated as follows: 
\begin{equation} 
\dd\sigma^{H+X} \simeq \sum_{i,j,k} \int_0^1 \dd x_i \int_0^1 \dd x_j\, \int_0^1 \dd z\  
f^A_i(x_i,\mu_F) f^B_j(x_j,\mu_F)   
\dd\hat \sigma_{ij\to k+X} D_k^H(z,\mu_F') 
+{\cal O}(m_Q^2/\pt^2) 
\, . 
\label{eq:xsH} 
\end{equation} 
Because the heavy-quark mass is neglected in the short-distance cross sections ($\dd\hat \sigma$), the predictions 
in the ZM-VFNS are expected to be reliable only for very large transverse momenta. 
The sum in \eq{eq:xsH} extends over a large number of sub-processes $i+j\to k+X$ since $a,b,c$ can be gluons, light quarks, 
and heavy quarks. A calculation of all sub-processes at NLO has been performed in the late 80's~\cite{Aversa:1988vb}. 
 
Concerning the FFs into the heavy-flavoured hadron $H={\rm D,B},\Lambda_c,\ldots$, two main approaches are employed in the literature: 
\begin{itemize} 
\item In the  Perturbative-Fragmentation Functions (PFF) approach~\cite{Cacciari:1993mq},  
the FF $D_k^H(z,\mu_F')$ is given by a convolution of a PFF accounting for the fragmentation of the parton $k$ into the heavy quark $Q$, $D_k^Q(z,\mu_F')$, with a scale-independent FF $D_Q^H(z)$ describing the hadronisation of the heavy quark into the hadron $H$: 
\begin{equation} 
D_k^H(z,\mu_F') = D_k^Q(z,\mu_F') \otimes D_Q^H(z)\, . 
\label{eq:pff} 
\end{equation} 
The PFFs resum the final-state collinear logarithms of the heavy-quark mass. 
Their scale-dependence is governed by the DGLAP evolution equations and the boundary conditions for the PFFs 
at the initial scale are calculable in the perturbation theory. 
On the other hand, the scale-independent FF is a non-perturbative object (in the case of heavy-light flavoured hadrons) 
which is assumed to be universal. It is usually determined by a fit to $\ee$ data, although approaches exist in the literature which  
attempt to compute these functions. 
It is reasonable to identify the scale-independent fragmentation function in~\eq{eq:xsB} with the one in~\eq{eq:pff}. 
This function describing the hadronisation process involves long-distance physics and might be modified in the presence of a  
QGP, whereas the PFFs (or the unresummed collinear logarithms $\ln \pt^2/m_Q^2$ in the FFNS) involve only short-distance  
physics and are the same in \pp, \pA, and \AAcoll collisions. 
\item In the Binnewies-Kniehl-Kramer (BKK) approach~\cite{Binnewies:1998xq,Binnewies:1997gz,Binnewies:1998vm},  
the FFs are not split up into a perturbative and a non-perturbative piece. 
Instead, boundary conditions at an initial scale $\mu_F' \simeq m_Q$  
are determined from \ee data for the full non-perturbative FFs, $D_k^H(z,\mu_F')$, 
in complete analogy with the treatment of FFs into light hadrons (pions, kaons).  
These boundary conditions are again evolved to larger scales $\mu_F'$ with the help of the DGLAP equations. 
\end{itemize} 
It is also noteworthy that the BKK FFs ($D(z,\mu_F')$) are directly determined as functions in $z$-space whereas the FFs in the PFF approach are determined in Mellin-N-space where the N-th Mellin moment of a function $f(z)$ ($0<z<1$) is defined as $f(N) = \int_0^1 dz\ z^{N-1} f(z)$. 
\paragraph{GM-VFNS} 
The FFNS and the ZM-VFNS are valid only in restricted and complementary regions of the transverse momentum. 
For this reason, it is crucial to have a unified framework which combines the virtues of the massive FO calculation in the FFNS  
and the massless calculation in the ZM-VFNS. 
The General-Mass VFNS (GM-VFNS)~\cite{Kniehl:2004fy,Kniehl:2005mk}  
is such a framework which is valid in the entire kinematic range from the smallest to the largest transverse momenta ($\pt \ll m_Q, \pt \simeq m_Q, \pt \gg m_Q$). 
It is very similar to the ACOT  
heavy-flavour scheme~\cite{Aivazis:1993kh,Aivazis:1994pi} which has been formulated for structure functions in deep inelastic scattering (DIS). 
Different variants of the ACOT scheme exist like the S-ACOT scheme~\cite{Kramer:2000hn} and the (S)-ACOT$_\chi$ scheme \cite{Tung:2001mv} 
which are used in global analyses of PDFs by the CTEQ Collaboration and 
the ACOT scheme has been extended to higher orders in Refs.\ \cite{Kretzer:1998ju,Stavreva:2012bs,Guzzi:2011ew}. 
The theoretical basis for the ACOT scheme has been laid out in an all-order proof of a factorisation theorem with massive quarks by  
Collins~\cite{Collins:1998rz}. 
While the discussion in~\cite{Collins:1998rz} deals with inclusive DIS, it exemplifies the general principles for the treatment of heavy quarks  
in perturbative QCD (see also~\cite{Thorne:2008xf,Olness:2008px}) which should be applicable to other processes as well. 
Therefore, it is very important to test these ideas also in the context of less inclusive observables. 
First steps in this direction had been undertaken in~\cite{Kretzer:1997pd,Kretzer:1998nt} 
where the ACOT scheme had been applied to inclusive D meson production in DIS. 
The case of hadroproduction in the ACOT scheme had been studied for the first time in~\cite{Olness:1997yc}  
taking into account the contributions from the NLO calculation in the FFNS  
combined with the massless contributions in the ZM-VFNS from all other sub-processes at ${\cal O}(\alpha_s^2)$  
resumming the collinear logarithms associated to the heavy quark at the leading-logarithmic (LL) accuracy. 
In contrast, the GM-VFNS has a NLO+NLL accuracy. It has been worked out for $\gamma\gamma$, \pp, \ppbar, $\ee$, $e$p, and $\gamma$p  
collisions in a series of papers~\cite{Kramer:2001gd,Kramer:2003cw,Kramer:2003jw,Kniehl:2004fy,Kniehl:2005ej,Kneesch:2007ey,Kniehl:2008zza,Kniehl:2009mh,Kniehl:2011bk,Kniehl:2012ti,Kniehl:2015fla} 
and has been successfully compared to experimental data from LEP, HERA, Tevatron and the LHC. 
Furthermore, inclusive lepton spectra from heavy-hadron decays have been studied for \pp collisions at the LHC at 2.76 and 7\TeV 
centre-of-mass energy~\cite{Bolzoni:2012kx} and compared to data from ALICE, ATLAS and CMS.  
In addition, predictions have been obtained for D mesons produced at \s = 7\TeV from B decays~\cite{Bolzoni:2013vya}. 
A number of comparisons with  hadroproduction data are discussed in \sect{sec:pp:Data}.

The cross section for inclusive heavy-flavour hadroproduction in the GM-VFNS is calculated using a factorisation formula similar to the one in \eq{eq:xsH}: 
\begin{equation} 
\dd\sigma^{H+X} \simeq \sum_{i,j,k} \int_0^1 \dd x_i \int_0^1 \dd x_j\, \int_0^1 \dd z\  
f^A_i(x_i,\mu_F) f^B_j(x_j,\mu_F)   
\dd\hat \sigma_{ij\to k+X}[\pt,m_Q] D_k^H(z,\mu_F') 
\, . 
\label{eq:xsH2} 
\end{equation} 
In particular, the same sub-processes as in the ZM-VFNS are taken into account.  
However, the finite heavy-quark-mass terms (powers of $m_Q^2/\pt^2$) are  retained in the short-distance cross sections 
of sub-processes involving heavy quarks. 
More precisely,  
the heavy-quark-mass terms are taken into account in the 
sub-processes $q+\overline q \to Q+ X$, $g + g \to Q+X$, $g+q\to Q+ X$ and $g+\overline q\to Q+X$ which are also present 
in the FFNS.  
However, in the current implementation, they are neglected in the heavy-quark-initiated sub-processes  
($Q+g\to Q+X$, $Q+g\to g+X$, \ldots) as it is done in the S-ACOT scheme~\cite{Kramer:2000hn}. 
The massive hard-scattering cross sections are defined in a way that they approach,  
in the limit $m_Q/\pt \to 0$, the massless hard-scattering cross sections defined in the $\overline{\rm MS}$ scheme. 
Therefore, the GM-VFNS approaches the ZM-VNFS at large $\pt \gg m_Q$. 
It can be shown that the GM-VFNS converges formally to the FFNS at small \pt. 
However, while the S-ACOT scheme works well for the computation of DIS structure functions 
at NLO, this scheme causes problems in the hadroproduction case at low \pt because the  
massless $\bquark$-quark initiated cross sections diverge in the limit $\pt \to 0$.  
This problem can be circumvented by a suitable choice for the factorisation scale 
so that the heavy-quark PDF is switched off sufficiently rapidly and the GM-VFNS approaches 
the FFNS at small $\pt$~\cite{Kniehl:2015fla}.

\paragraph{FONLL} 
Similar to the GM-VFNS, the Fixed-Order plus Next-to-Leading  Logarithms (FONLL) approach~\cite{Cacciari:1998it}  
is a unified framework which is valid in the entire kinematic range ($\pt \ll m_Q, \pt \simeq m_Q, \pt \gg m_Q$). 
This approach has also been applied to DIS structure functions and is used in the global analyses of PDFs 
by the NNPDF Collaboration~\cite{Forte:2002fg,Ball:2011mu}. Predictions for $\cquark$ and $\bquark$ quark production at the LHC with a centre-of-mass energy 
of 7\TeV have been presented in~\cite{Cacciari:2012ny}. 
The FONLL scheme is based on the matching of the massive NLO cross section in the FFNS (=FO) 
with the massless NLO calculation in the ZM-VNFS (=RS) according to the prescription 
\begin{equation} 
\dd\sigma_{\rm FONLL} = \dd\sigma_{\rm FO} + G(m_Q,\pt) \times \left(\dd\sigma_{\rm RS} - \dd\sigma_{\rm FOM0}\right) 
\end{equation} 
where $\dd\sigma_{\rm FOM0}$ is the cross section $\dd\sigma_{\rm FO}$ in the asymptotic limit 
$\pt \gg m_Q$ where the finite power-like mass terms can be neglected and the cross section is dominated 
by the collinear logarithm of the heavy-quark mass. 
 
The condition $\dd\sigma_{\rm FONLL} \to d\sigma_{\rm RS}$ for $\pt \gg m_Q$ implies that 
the matching function $G(m_Q,p_T)$ has to approach unity in this limit. 
Furthermore, in the limit of small transverse momenta $\dd\sigma_{\rm FONLL}$ has to approach the fixed-order 
calculation $\dd\sigma_{\rm FO}$. 
This can be achieved by demanding that $G(m_Q,p_T)\to 0$ for $\pt \to 0$ which effectively suppresses the contribution from the divergent $b$-quark initiated contributions in $\dd \sigma_{RS}$. 
In the FONLL, the interpolating function is chosen to be 
$G(m_Q,\pt) = \pt^2/\left(\pt^2 + a^2 m_Q^2\right)$ 
where the constant is set to $a=5$ on phenomenological grounds. 
In this language the GM-VFNS is given by 
$\dd\sigma_{\rm GM-VFNS} = \dd\sigma_{\rm FO} + \dd\sigma_{\rm RS} - \dd\sigma_{\rm FOM0}$ , 
\ie no interpolating factor is used.  
 
Other differences concern the non-perturbative input. In particular, the FONLL scheme uses fragmentation functions  
in the PFF formalism whereas the GM-VFNS uses fragmentation functions which are determined in the $z$-space 
in the BKK approach. 

\paragraph{Monte Carlo generators} 
The GM-VFNS and FONLL calculations are mostly analytic and provide a precise description of the inclusive production of a 
heavy hadron or its decay products at NLO+NLL accuracy. 
Compared to this, general-purpose Monte-Carlo generators like PYTHIA~\cite{Sjostrand:2007gs} or HERWIG~\cite{Corcella:2000bw} 
allow for a more complete description of the hadronic final state but only work at LO+LL accuracy. 
However, in the past decade, NLO Monte Carlo generators have been developed using the MC@NLO~\cite{Frixione:2003ei} 
and POWHEG~\cite{Frixione:2007nw} methods for a consistent matching of NLO calculations with parton showers. 
They, therefore, have all the strengths of Monte Carlo generators, which allow for a complete modelling of the 
hadronic final state (parton showering, hadronisation, decay, detector response), while, at the same time, the NLO accuracy 
in the hard scattering is kept and the soft/collinear regimes are resummed at the LL accuracy. 
A comparison of POWHEG NLO Monte Carlo predictions for heavy-quark production in \pp collisions  
at the LHC with the ones from the GM-VFNS and FONLL can be found in~\cite{Klasen:2014dba}. 
 

\subsubsection{Quarkonium-production mechanism}
\label{sec:pp:Theory:Onia}

The theoretical study of quarkonium-production processes involves 
both pertubative and non-perturbative aspects of QCD. On one side, the production 
of the heavy-quark pair, \QQbar, which will subsequently form the quarkonium, is expected to be perturbative 
since it involves momentum transfers at least as large as the mass of the considered heavy quark, as 
for open-heavy-flavour production discussed in the previous section. 
On the other side, the evolution of the \QQbar pair into the physical quarkonium state is non-perturbative, over long distances,  
with typical momentum scales such as the momentum of the heavy-quarks in  
the bound-state rest frame, $m_Q v$ and their binding energy $m_Q v^2$, $v$ being the typical velocity of the 
heavy quark or antiquark in the quarkonium rest frame  ($v^2\sim 0.3$ for the charmonium and 0.1 for the bottomonium).
 
In nearly all the models or production mechanisms discussed nowadays, the idea of a factorisation 
between the pair production and its binding is introduced. Different approaches differ essentially 
in the treatment of the hadronisation, although some may also introduce new ingredients in the description  
of the heavy-quark-pair production. In the following, we briefly describe three of them which can be distinguished 
in their treatment of the non-perturbative part: the Colour-Evaporation Model (CEM), the Colour-Singlet Model (CSM), the Colour-Octet  
Mechanism (COM), the latter two being encompassed in an effective theory referred to as Non-Relativistic QCD (NRQCD).

\paragraph{The Colour-Evaporation Model (CEM)} 
 
This approach is in line with the principle of quark-hadron duality~\cite{Fritzsch:1977ay,Halzen:1977rs}. As such, 
the production cross section of quarkonia is expected to be directly connected to that to  produce a \QQbar pair 
in an invariant-mass region where its hadronisation into a quarkonium is possible, that is 
between  the kinematical threshold to produce a quark pair, $2m_Q$, and that  
to create the lightest open-heavy-flavour hadron pair, $2m_{H}$.  
 
The cross section to produce a given quarkonium state is then supposed to be obtained after 
a multiplication by a phenomenological factor $F_{\cal Q}$ related to a process-independent probability that the 
pair eventually hadronises into this state. One assumes that a number of non-perturbative-gluon emissions 
occur once the $Q \overline Q$ pair is produced and that the quantum state of the pair at its hadronisation 
is essentially decorrelated --at least colour-wise-- with that at its production. From the reasonable 
assumption~\cite{Amundson:1995em} that one ninth --one colour-{\it singlet} \QQbar  
configuration out of 9 possible-- of the pairs in the suitable kinematical region  
hadronises in a quarkonium,  a simple statistical counting~\cite{Amundson:1995em} was proposed 
based on the spin $J_{\cal Q}$ of the quarkonium ${\cal Q}$, 
$F_{\cal Q}= {1}/{9} \times {(2 J_{\cal Q} +1)}/{\sum_i (2 J_i +1)}$, 
where the sum over $i$ runs  over all the charmonium states below the open heavy-flavour threshold. It was shown to  
reasonably account for existing \jpsi hadroproduction data of the late 90's and, in fact, is comparable to  
the fit value in~\cite{Bedjidian:2004gd}. 
 
Mathematically, one has 
\begin{equation} 
\sigma^{\rm (N)LO}_{\cal Q}= F_{\cal Q}\int_{2m_Q}^{2m_H}  
\frac{\dd\sigma_{Q\overline Q}^{\rm (N)LO}}{\dd m_{Q\overline Q}}\dd m_{Q\overline Q} 
\label{eq:sigma_CEM} 
\end{equation} 
In the latter formula, a factorisation between the short-distance \QQbar-pair production and its hadronisation is the  
quarkonium state is of course implied although it does not rely on any factorisation proof. In spite of this,  
this model benefits --as some figures will illustrate it in the next section-- from a successful phenomenology but for the absence of predictions for polarisation  
observables and discrepancies in some transverse momentum spectra.   
 
\paragraph{The Colour-Singlet Model (CSM)} 
 
The second simplest model to describe quarkonium production relies on the rather opposite assumption  
that the quantum state of the pair does {\it not} evolve between its production and its hadronisation, neither 
in spin, nor in colour~\cite{Chang:1979nn,Baier:1981uk,Baier:1983va} -- gluon emissions from the heavy-quark are 
suppressed by powers of $\alpha_s(m_Q)$. In principle, they are taken into account in the (p)QCD corrections  
to the hard-scattering part account for the \QQbar-pair production. 
If one further assumes that the quarkonia are non-relativistic bound states 
with a highly peaked wave function in the momentum space, it can be shown that   
partonic cross section for quarkonium production should then be expressed as that for the production of a heavy-quark  
pair with zero relative velocity, $v$, in a colour-singlet state and  
in the same angular-momentum and spin state as that of the to-be produced quarkonium,  
and the square of the Schr\"odinger wave function at the origin in the position space. In the case of hadroproduction,  
which interests us most here,  
one should further account for the parton $i,j$ densities in the  
colliding hadrons, $f_{i,j}(x)$, in order to get the following hadronic cross section  
\begin{equation} 
d\sigma[{\cal Q}+X] 
=\sum_{i,j}\!\int\! \dd x_{i} \,\dd x_{j} \,f_{i}(x_i,\mu_F) \,f_{j}(x_j,\mu_F) \dd\hat{\sigma}_{i+j\rightarrow (Q\overline{Q})+X} (\mu_R,\mu_F) |\psi(0)|^2 
\label{eq:sigma_CSM} 
\end{equation} 
In the case of $P$-waves, $|\psi(0)|^2$ vanishes and, in principle, one should consider its derivative and that of  
the hard scattering. In the CSM, $|\psi(0)|^2$ or $|\psi'(0)|^2$ also appear in decay processes and can be extracted 
from decay-width measurements. The model then becomes  fully predictive but for the usual unknown values of the non-physical 
factorisation and renormalisation scales and of the heavy-quark mass entering the hard part. 
A bit less than ten years ago, appeared the first evaluations of the QCD  
corrections~\cite{Campbell:2007ws,Artoisenet:2007xi,Gong:2008sn,Gong:2008hk,Artoisenet:2008fc} 
to the yields of \jpsi and \ups (also commonly denoted $\cal Q$) in  
hadron  collisions in the CSM. It is now widely accepted~\cite{Lansberg:2008gk,ConesadelValle:2011fw,Brambilla:2010cs}  
that $\alpha^4_s$ and $\alpha^5_s$ corrections to the CSM are significantly larger than the LO  contributions at $\alpha^3_s$ 
at mid and large \pt and that they should systematically be accounted for in any study of their \pt spectrum.  
 
Possibly due to its high predictive power, the CSM has faced several phenomenological issues although 
it accounts reasonably well for the bulk of hadroproduction data from RHIC to LHC energies~\cite{Brodsky:2009cf,Lansberg:2010cn,Feng:2015cba},  
\ee data at $B$ factories~\cite{Ma:2008gq,Gong:2009kp,He:2009uf} and photo-production data at HERA~\cite{Aaron:2010gz}.  
Taking into account NLO --one loop-- corrections and approximate NNLO contributions (dubbed as NNLO$^\star$ in the following)  
has reduced the most patent discrepancies in particular for \pt up to a couple of $m_{\cal Q}$~\cite{Lansberg:2010vq,Lansberg:2011hi,Lansberg:2012ta,Lansberg:2013iya}. 
A full NNLO computation (\ie~at $\alpha^5_s$) is however needed to confirm this trend. 
 
It is however true that the CSM is affected by infra-red divergences in the case of $P$-wave decay at NLO, which were  
earlier regulated by an ad-hoc binding energy~\cite{Barbieri:1976fp}. These can nevertheless  
be rigorously cured~\cite{Bodwin:1992ye} in the more general framework of NRQCD which we discuss now and which 
introduce the concept of colour-octet states. 
 
\paragraph{The Colour-Octet Mechanism (COM) and NRQCD} 
 
Based on the effective theory NRQCD~\cite{Bodwin:1994jh}, one can express in a more rigorous way the hadronisation probability 
of a heavy-quark pair into a quarkonium via long-distance matrix elements (LDMEs). In addition to the usual 
expansion in powers of $\alpha_s$, NRQCD further introduces an expansion in $v$. It is then natural to account 
for the effect of higher-Fock states (in $v$) where the \QQbar pair is in an octet state with a different  
angular-momentum and spin states --the sole consideration of the {\it leading} Fock state (in $v$) amounts to the CSM, which is thus 
{\it a priori} the {\it leading} NRQCD contribution (in $v$). However, this opens the possibility for non-perturbative transitions between 
these coloured states and the physical meson. One of the virtues of this is the consideration of $^3S_1^{[8]}$ states 
in $P$-wave productions, whose contributions cancel the aforementioned divergences in the CSM. The necessity for  
such a cancellation does not however fix the relative importance of these contributions. In this precise case,  
it depends on an non-physical scale $\mu_\Lambda$.  
 
As compared to the \eq{eq:sigma_CSM}, one has to further 
consider additional quantum numbers (angular momentum, spin and colour), generically denoted $n$, involved in the production mechanism: 
\begin{equation} 
\dd\sigma[{\cal Q}+X] 
=\sum_{i,j,n}\!\int\! \dd x_{i} \,\dd x_{j} \,f_{i}(x_i,\mu_F) \,f_{j}(x_j,\mu_F) \dd\hat{\sigma}_{i+j\rightarrow (Q\overline{Q})_{n}+X} (\mu_R,\mu_F,\mu_\Lambda) \langle{\cal O}_{\cal Q}^{n} \rangle. 
\label{eq:sigma_NRQCD} 
\end{equation} 
 
Instead of the Schr\"odinger wave function at the origin squared, the former equation involves  
the aforementioned LDMEs, $\langle{\cal O}_{\cal Q}^{n} \rangle$, which {\it cannot}  
be fixed by decay-width measurements nor lattice studies\footnote{We however note that a first attempt to evaluate colour octet {\it decay} LDMEs was made in~\cite{Bodwin:2005gg}. In principle they can be related by crossing symmetry to the {\it production} LDMEs which are relevant for the present discussion. Quantitative results are however still lacking.}-- but the leading CSM ones of course. Only  
relations based on Heavy-Quark Spin Symmetry (HQSS) can relate some of them. 
 
Three groups (Hamburg~\cite{Butenschoen:2012px}, IHEP~\cite{Gong:2012ug} and PKU~\cite{Chao:2012iv})  
have, in the recent years, carried out a number of NLO studies\footnote{A recent LO study has also been performed including LHC data in  
the  used sample~\cite{Sharma:2012dy}.}  of cross section fits to determine 
the NRQCD LDMEs. A full description  
of the differences between these analyses is beyond the scope of this review, it is however important 
 to stress that they somehow contradict each other in their results as regards the polarisation observables. 
In particular, in the case of the \jpsi, the studies of the Hamburg group, which is the only one to fit low \pt data from  
hadroproduction, electroproduction and \ee collisions at $B$ factories, predict a strong transverse polarised yield at variance with the experimental data. 
 
\paragraph{Theory prospects}

Although NRQCD is 20 years old, there does not exist yet a complete proof of factorisation, in particular, in the case of  
hadroproduction.  
A discussion of the difficulties in establishing NRQCD factorisation can be found in~\cite{Brambilla:2010cs}. 
A first step was achieved in 2005 by the demonstration~\cite{Nayak:2005rt,Nayak:2005rw} that, in the large-\pt region where a  
description in terms of  
fragmentation functions is justified, the infra-red poles at NNLO could be absorbed in the NRQCD LDMEs, 
provided that the NRQCD production operators were modified to include nonabelian phases.  
 
As mentioned above, it seems that the mere expansion of the hard matrix elements in $\alpha_s$ is probably 
not optimal since higher QCD corrections receive contributions which are enhanced by powers 
of $\pt/m_{\cal Q}$. It may therefore be expedient to organise the quarkonium-production cross section in powers of $\pt/m_{\cal Q}$ 
before performing the  $\alpha_s$-expansion of the short distance coefficients for the \QQbar production. This  
 is sometimes referred to as the fragmentation-function approach (see \cite{Kang:2011mg,Ma:2014svb}) which offers new 
perspectives in the theoretical description of quarkonium hadroproduction especially at mid and large \pt. 
Complementary information could also be obtained from similar studies based on Soft Collinear Effective Theory (SCET), see~\cite{Fleming:2012wy}. 

At low \pt, it was recently emphasised in \cite{Feng:2015cba} that one-loop results show 
an intriguing energy dependence which might hint at a break-down of NRQCD factorisation in this kinematical region. 
In any case, as for now, past claims that colour-octet transitions are the dominant source of the low-\pt  
 \jpsi and \ups cannot be confirmed at one loop accuracy. 
Approaches such as the \kt factorisation based on the Lipatov action in the quasi multi Regge kinematics 
(see~\cite{Hagler:2000dd,Yuan:2000qe} for quarkonium studies), the TMD factorisation  
(see~\cite{Boer:2012bt,Ma:2012hh} for recent applications to quarkonium production) or 
the combined use of the CGC formalism and NRQCD \cite{Kang:2013hta,Ma:2014mri}  may therefore bring  
their specific bricks in the building  
of a consistent theory of quarkonium production. Finally, let us mention the relevance of the colour-transfer  
mechanism~\cite{Nayak:2007mb}, beyond NRQCD, in the case of production of a  
quarkonium in the vicinity of another heavy quark.

\subsection{Recent cross section measurements at hadron colliders}
\label{sec:pp:Data}

Due to their short lifetimes (at most a picosecond), the production of open-heavy-flavour particles is studied through their decay products. 
Four main techniques are used: 
\begin{enumerate} 
\item Fully reconstruction of  exclusive decays, such as ${\rm B}^0 \to \jpsi \, {\rm K^0_S}$ or $\Dzero \to {\rm K}^- \, \pi^+$. 
\item Selection of specific (semi-)inclusive decays.  
For beauty production, one looks for a specific particle, for example \jpsi, and imposes it to point to a secondary vertex displaced  
a few hundred~\footnote{ 
For larger \pt or $y$, such a distance can significantly be larger. 
} $\mu$m from the primary vertex.  
Such {\it displaced} or {\it non-prompt} mesons are therefore supposed to come from $b$-decay only. 
\item Detection of leptons from these decays. This can be done  
{\it (i)} by subtracting a cocktail of known/measured sources (photon conversions, Dalitz decays of $\pi^0$ and $\eta$ in the case of electrons, light hadron, Drell-Yan pair, \jpsi,\dots) to the lepton yield.  
Alternatively, the photon conversion and Dalitz decay contribution can be evaluated via an invariant mass analysis of the \ee pairs.  
{\it (ii)} By selecting displaced leptons with a track pointing to a secondary vertex separated by few hundred $\mu$m from the primary vertex. 
\item Reconstruction of $\cquark$- and $\bquark$-jets. Once a jet is reconstructed, a variety of reconstructed objects, such as tracks, vertices and identified leptons, are used to distinguish between jets from light or from heavy flavour. A review of $\bquark$-tagging methods used by the CMS Collaboration can be found in~\cite{Chatrchyan:2012jua}. 
\end{enumerate} 
Different methods are used in different contexts, depending on the detector information available, the trigger strategy, the corresponding statistics  
(hadronic decays are less abundant than leptonic ones), the required precision (only exclusive decay channels  
allow for a full control of the kinematics), the kinematical range ($\bquark$-tagged jets give  
access to very large \pt whereas exclusive-decay channels allow for differential studies down to \pt equals to 0). A fifth method based on the indirect extraction of  the total charm- and beauty-production  
from {\it dileptons} -- as opposed to single leptons-- (see \eg~\cite{Abreu:2000nj}) is not discussed in this review.

Hidden-heavy-flavour, \ie quarkonia, are also analysed through their decay products. 
The triplet $S$-waves are the most studied since they decay up to a few per cent of the time 
in dileptons. This is the case for $\jpsi$, $\psiP$, $\upsa$, $\upsb$, $\upsc$. The triplet $P$-waves,  
such as the $\chic$ and $\chib$, are  
usually reconstructed via their radiative decays into a triplet $S$-wave.  
For other states, such as the singlet $S$-wave, studies are far more complex. The very first inclusive 
hadroproduction study of $\eta_c$ was just carried out this year in the \ppbar  
decay channel by the LHCb Collaboration~\cite{Aaij:2014bga}.

\begin{figure}[!t] 
\begin{center} 
\includegraphics[width=0.4\textwidth]{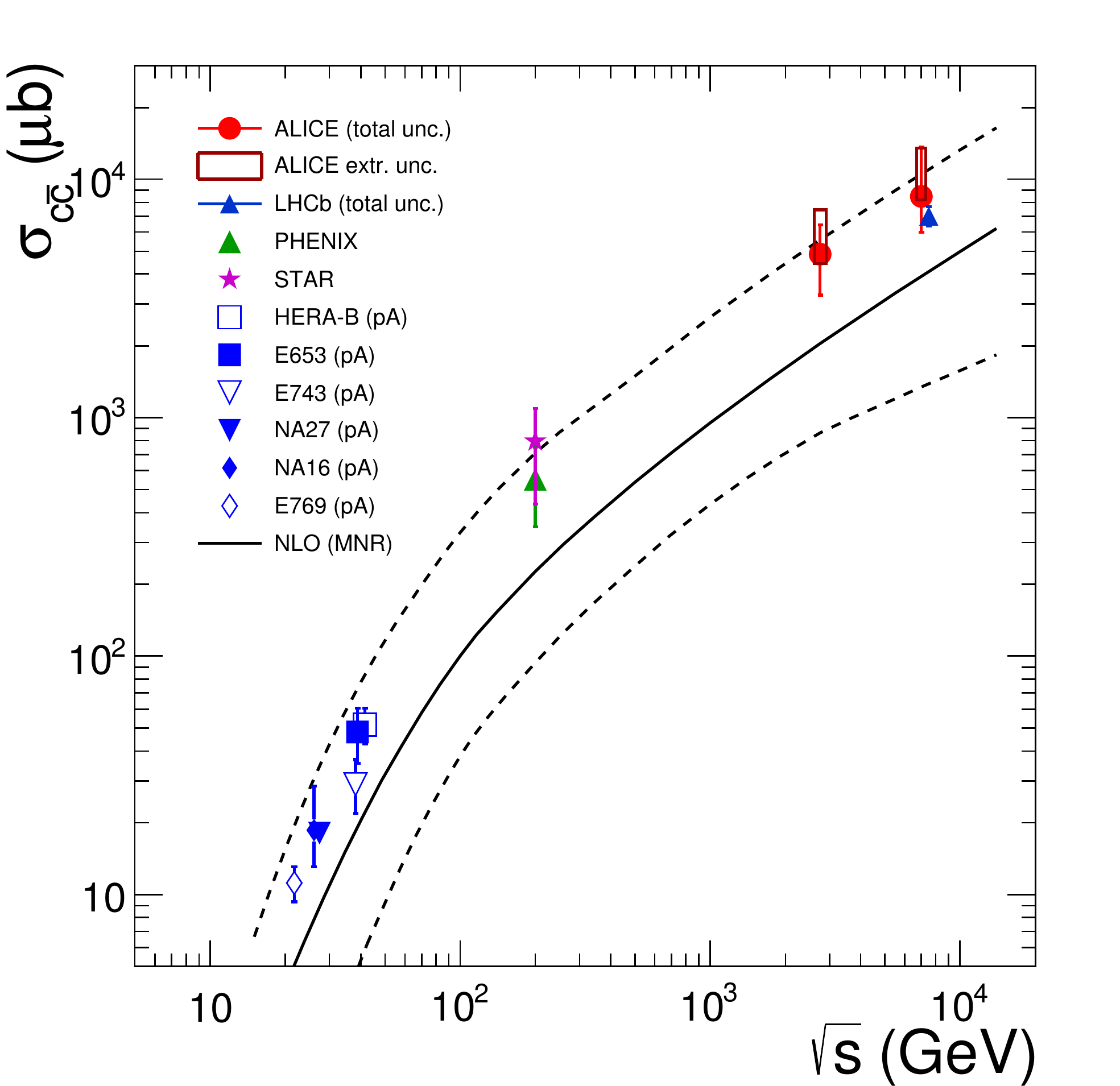} 
\includegraphics[width=0.41\textwidth]{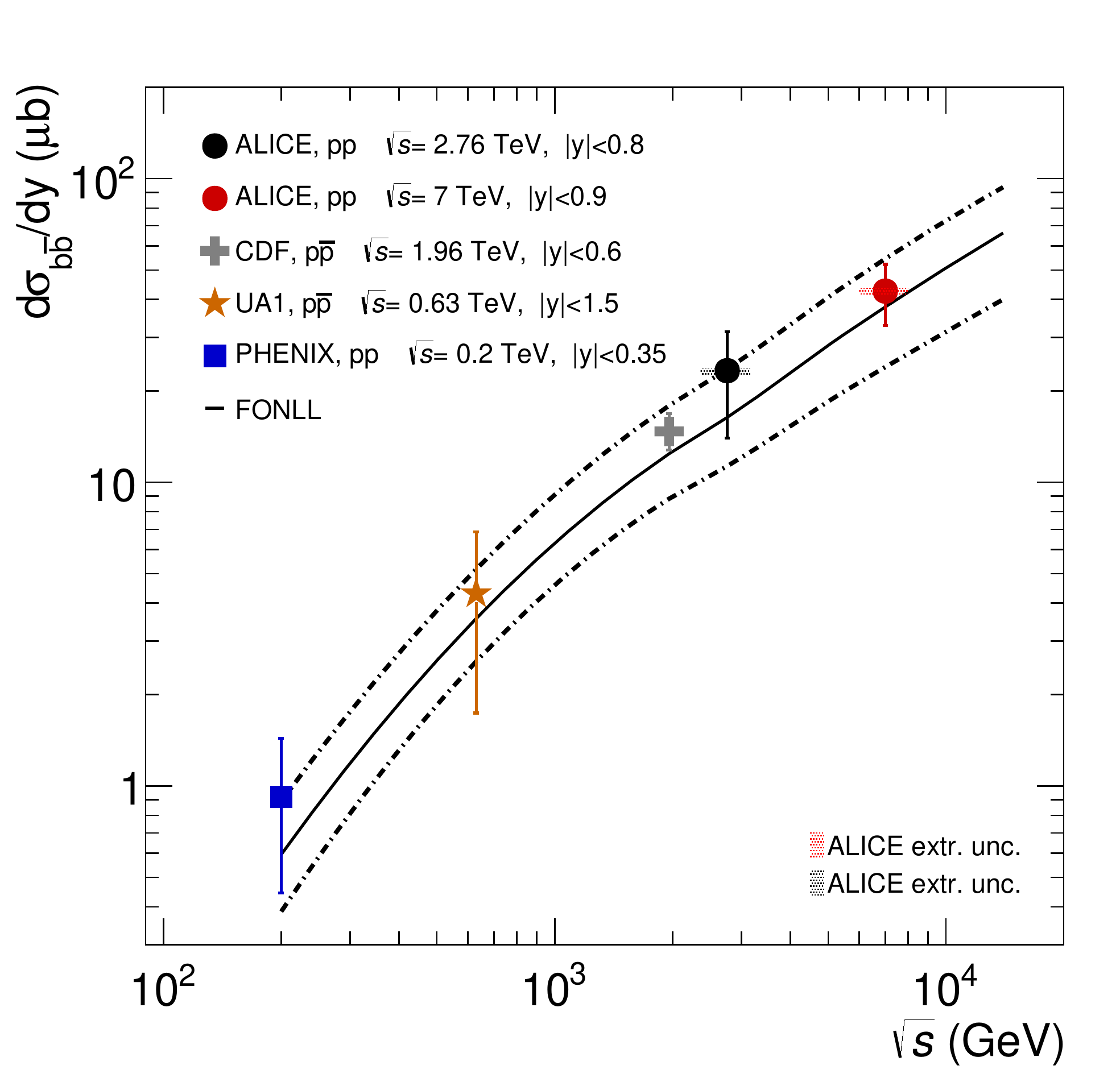} 
\caption[Compilation of total beauty cross section measurements]{ 
\label{fig:pp:CharmAndBottomXsec} 
Left: Total (extrapolated) $\ccbar$ cross section as a function of \s~\cite{Bortoletto:1988kw,Barate:1999bg,Acosta:2003ax,Adamczyk:2012af,Abelev:2012vra,
LHCb:2010lga,Aaij:2013mga}.  
Data in proton--nucleus (\pA) or deuteron--nucleus (d--A) collisions were scaled down assuming no nuclear effect.  
Right: A compilation of the $\bbbar$ differential cross section measurements at mid-rapidity in \pp and \ppbar collisions~\cite{Abelev:2014hla,Abelev:2012gx,Albajar:1990zu,Abe:1995dv,Adare:2009ic}. 
Results are compared to pQCD calculations, NLO MNR~\cite{Mangano:1991jk} and FONLL~\cite{Cacciari:2003uh, Cacciari:2012ny} for $\ccbar$ and $\bbbar$, respectively.  
} 
\end{center} 
\end{figure}

A compilation of the measurements of the \pt-integrated \ccbar and \bbbar cross section,  
$\sigma_{\ccbar}$ and $\sigma_{\bbbar}$, is shown in \fig{fig:pp:CharmAndBottomXsec} from  
SPS to LHC energies. Let us stress that most of the \pt-integrated results and nearly all $y$-integrated ones 
are based on different extrapolations, which significantly depend on theoretical inputs and which are not necessarily identical 
in the presence of nuclear effects.  The results are described within the uncertainties by pQCD calculations,  
NLO MNR~\cite{Mangano:1991jk}  and FONLL~\cite{Cacciari:2003uh, Cacciari:2012ny} for the $\ccbar$ and $\bbbar$,  
respectively. Note that most of the experimental results for  $\sigma_{\ccbar}$, in particular 
at high energies, lie on the upper edge of the NLO MNR uncertainties. 

%
%
%
%
\subsubsection{Leptons from heavy-flavour decays} 
\label{subsubsec:pp:HFLeptons} 
 
\begin{figure}[!ht] 
\begin{center} 
\subfigure[PHENIX, $\hfe$, \s = 200\GeV]{ 
	\label{fig:pp:PHENIXHFe} 
		\includegraphics[width=0.31\columnwidth,height=0.31\columnwidth]{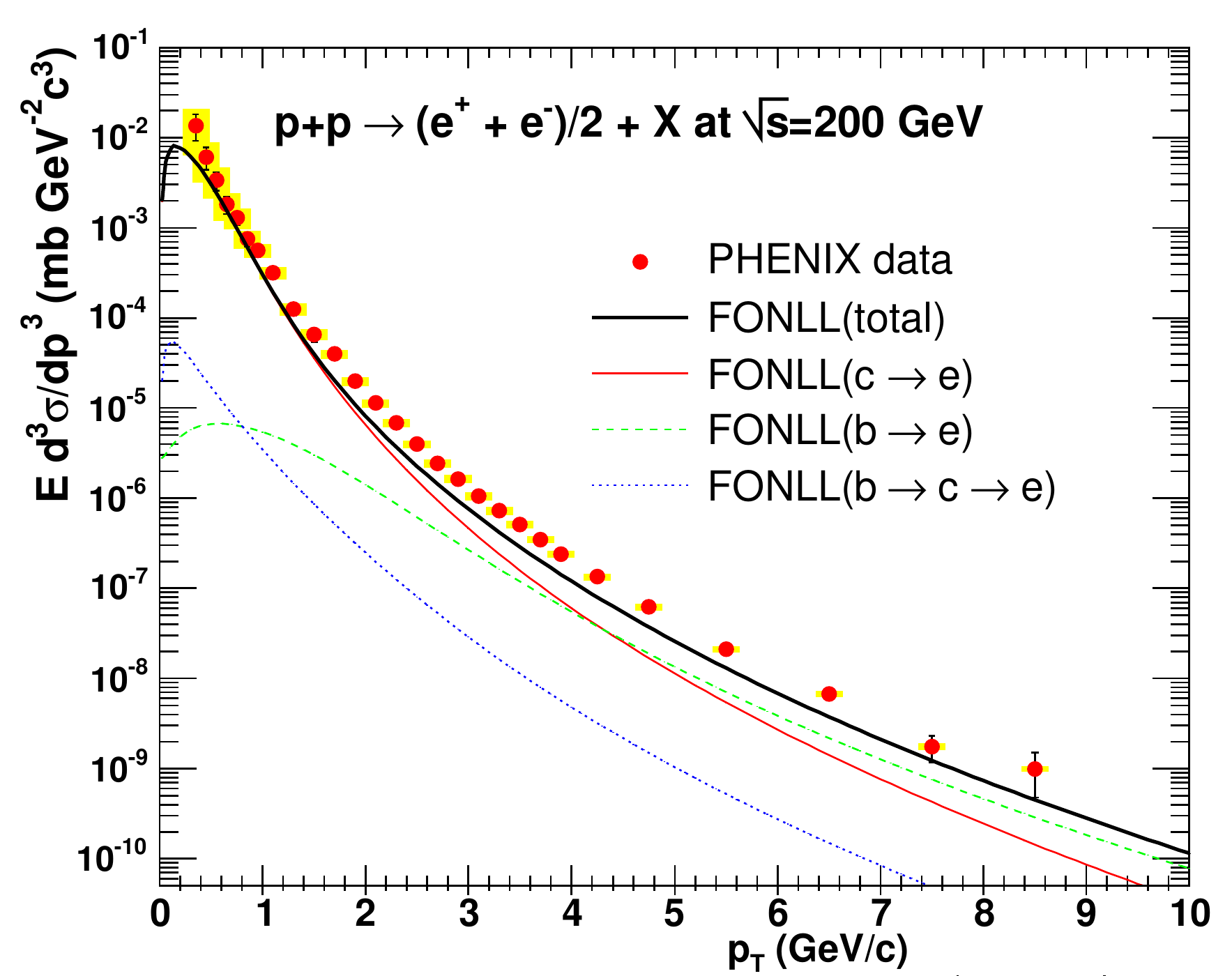} 
	}  
\subfigure[ALICE, $\hfe$, $\s=2.76$\TeV]{ 
	\label{fig:pp:ALICE:HFe} 
	\includegraphics[width=0.31\columnwidth,height=0.31\columnwidth]{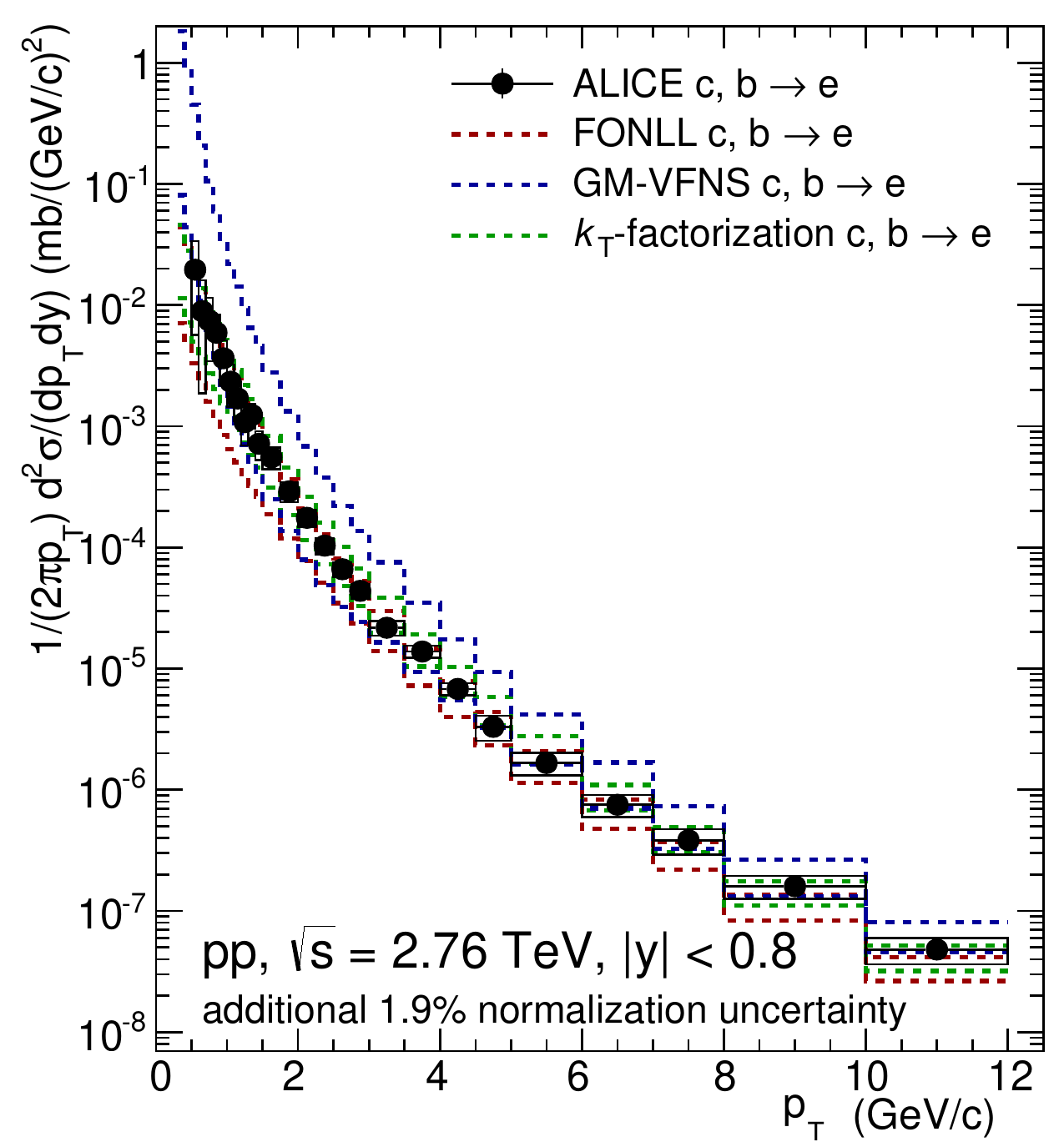} 
	}  
\subfigure[ALICE, $\hfm$, $\s=2.76$\TeV]{ 
	\label{fig:pp:ALICE:HFm} 
	\includegraphics[width=0.31\columnwidth,height=0.31\columnwidth]{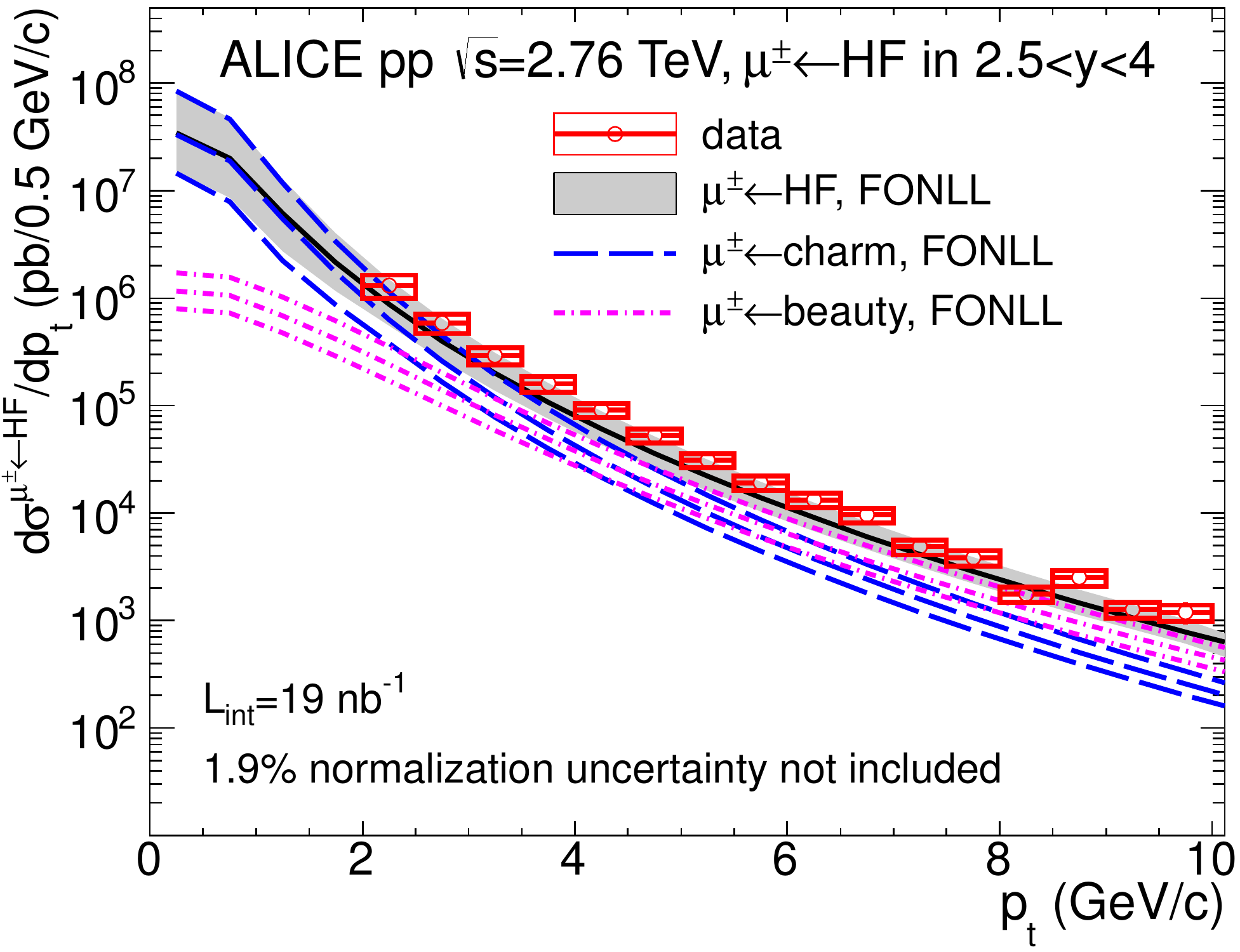} 
	}  
\caption{ 
	\label{fig:pp:HFlepton} 
	$\dsdpt$ for heavy-flavour decay leptons in \pp collisions: (a) electrons at mid-rapidity for \s = 200\GeV from PHENIX~\cite{Adare:2006hc}, (b) electrons at mid-rapidity for \s = 2.76\TeV~\cite{Abelev:2014gla} and (c) muons at forward-rapidity for \s = 2.76\TeV from ALICE~\cite{Abelev:2012pi}.  
	FONLL~\cite{Cacciari:2003uh, Cacciari:2012ny} predictions are also shown; in (a) and (c) the calculations for leptons from charm and beauty decays are shown separately (without theoretical uncertainty bands in (a)).  
	GM-VFNS~\cite{Kniehl:2004fy,Kniehl:2005mk} and \kt-factorisation~\cite{Maciula:2013wg} calculations are also drawn in (b).  
} 
\end{center} 
\end{figure}

The first open-heavy-flavour measurements in heavy-ion collisions were performed by  
exploiting heavy-flavour decay leptons at RHIC by the PHENIX and STAR Collaborations. These were done  
both in \pp and \AAcoll collisions~\cite{Adare:2006hc,Adare:2010ud,Adare:2013xlp,Aggarwal:2010xp,Agakishiev:2011mr}.  
At the LHC, the ATLAS and ALICE Collaborations have also performed such studies in heavy-ion collisions~\cite{Abelev:2012xe,Abelev:2014gla,Abelev:2012pi,Abelev:2012qh,Aad:2011rr}.  
A selection of the \pt-differential production cross sections of heavy-flavour decay leptons in \pp collisions at different rapidities and energies is presented in \fig{fig:pp:HFlepton}.  
The measurements are reported together with calculations of FONLL~\cite{Cacciari:2003uh, Cacciari:2012ny} for  \s = 0.2 and 2.76 \TeV, GM-VFNS~\cite{Kniehl:2004fy,Kniehl:2005mk} and \kt-factorisation~\cite{Maciula:2013wg} at \s = 2.76\TeV.  
The POWHEG predictions~\cite{Klasen:2014dba}, not shown in this figure, show a remarkable agreement with the FONLL ones. 
The differential cross sections of heavy-flavour-decay leptons are well described by pQCD calculations.

In addition, leptons from open charm and beauty production can be separated out via: 
{\it (i)} a cut on the lepton impact parameter, \ie~the distance between the origin of the lepton and the collision primary vertex,  
{\it (ii)} a fit of the lepton impact parameter distribution using templates of the different contributions to the inclusive spectra,  
{\it (iii)} studies of the azimuthal angular correlations between heavy-flavour decay leptons and charged hadrons (see \eg~\cite{Abelev:2012sca,Abelev:2014hla}). These measurements are also described by pQCD calculations. 
 
%

%
%
%
%
\subsubsection{Open charm} 
\label{subsubsec:pp:OpenCharm} 
 
Recently, D-meson production has been studied at RHIC, Tevatron and LHC energies~\cite{Adamczyk:2012af,
Acosta:2003ax,Reisert:2007zza,
ALICE:2011aa,Abelev:2012vra,Abelev:2012tca,
Aaij:2013mga}. 
The measurements were performed by fully reconstructing the hadronic decays of the D mesons, \eg~$\Dzero  \to {\rm K}^-\pi^+$ and charge conjugates.  
D-meson candidates are built up of pairs or triplets of tracks with the appropriate charge sign combination. The analyses exploit the detector particle identification abilities to reduce the combinatorial background, which is important at low \pt.  
For the measurements at Tevatron and LHC, 
the background is also largely reduced by applying topological selections on the kinematics of the secondary decay vertices, typically displaced by few hundred $\mu$m from the interaction vertex.  
The results at RHIC energies report the {\it inclusive} D-meson yields~\cite{Adamczyk:2012af}, \ie those from both $\cquark$ and $\bquark$ quark fragmentation. The former are called {\it prompt}, and the later {\it secondary} D mesons. The measurements at Tevatron and LHC energies report prompt D-meson yields. Prompt yields include both the {\it direct} production and the {\it feed-down} from excited charmed resonances. 
The secondaries contribution to the D-meson yields is evaluated and subtracted by:  
{\it (i)} either scrutinising the D-meson candidates impact parameter distribution, exploiting the larger lifetime of $\bquark$- than $\cquark$-flavoured hadrons~\cite{Acosta:2003ax,Reisert:2007zza,Aaij:2013mga}, which requires large statistics,  
{\it (ii)} or evaluating the beauty hadron contribution using pQCD-based calculations~\cite{ALICE:2011aa,Abelev:2012vra,Abelev:2012tca}, advantageous strategy for smaller data samples but limited by the theoretical uncertainties. 
 
\begin{figure}[!ht] 
\begin{center} 
\subfigure[STAR, inclusive $\Dzero$ and $\Dstarplus$, \s = 200\GeV]{ 
	\label{fig:pp:STAR:D} 
	\includegraphics[height=0.24\columnwidth]{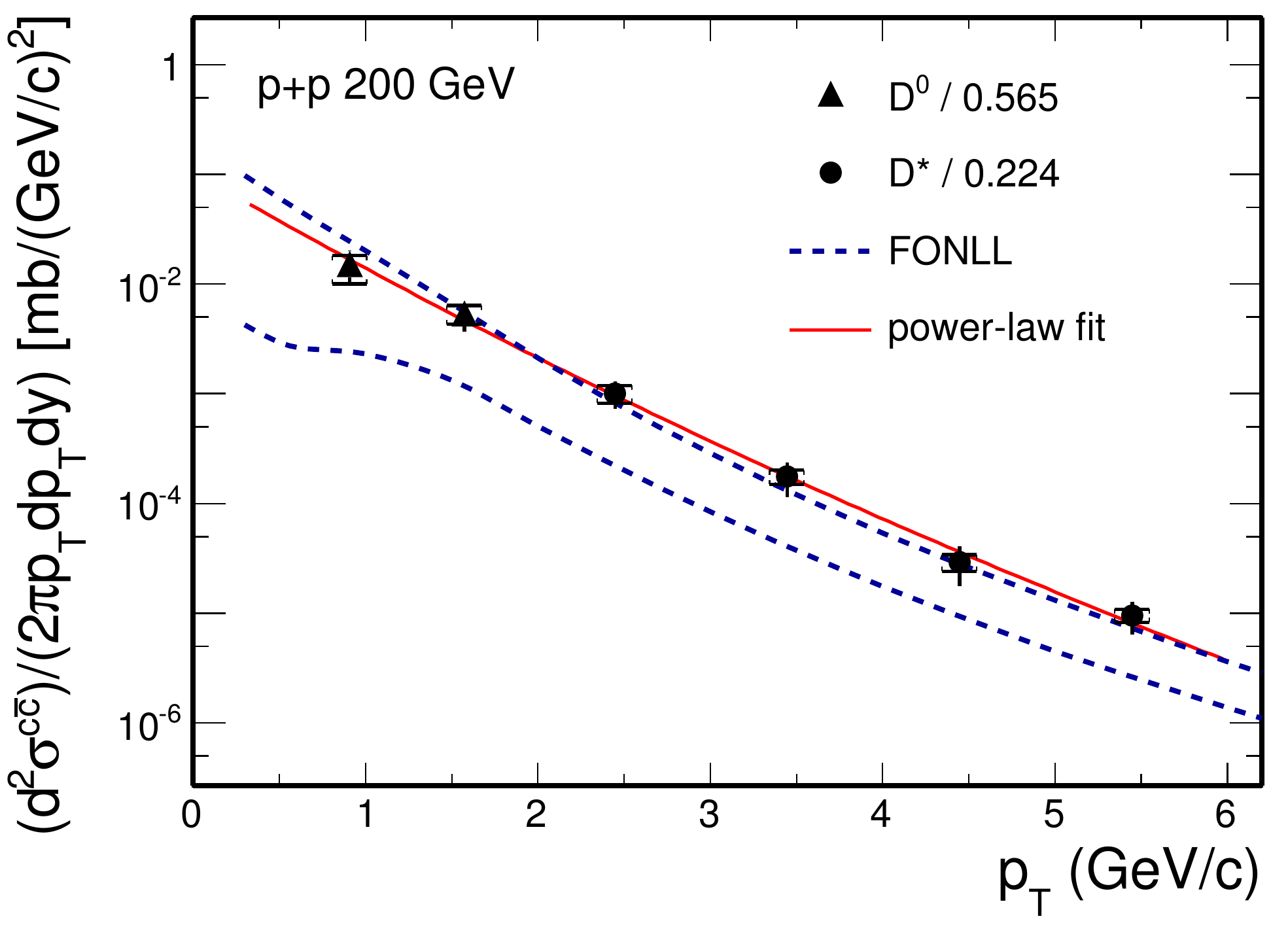} 
	}  
\subfigure[ALICE, prompt $\Dplus$, \s = 7\TeV]{ 
	\label{fig:pp:ALICE:Dplus} 
	\includegraphics[height=0.24\columnwidth]{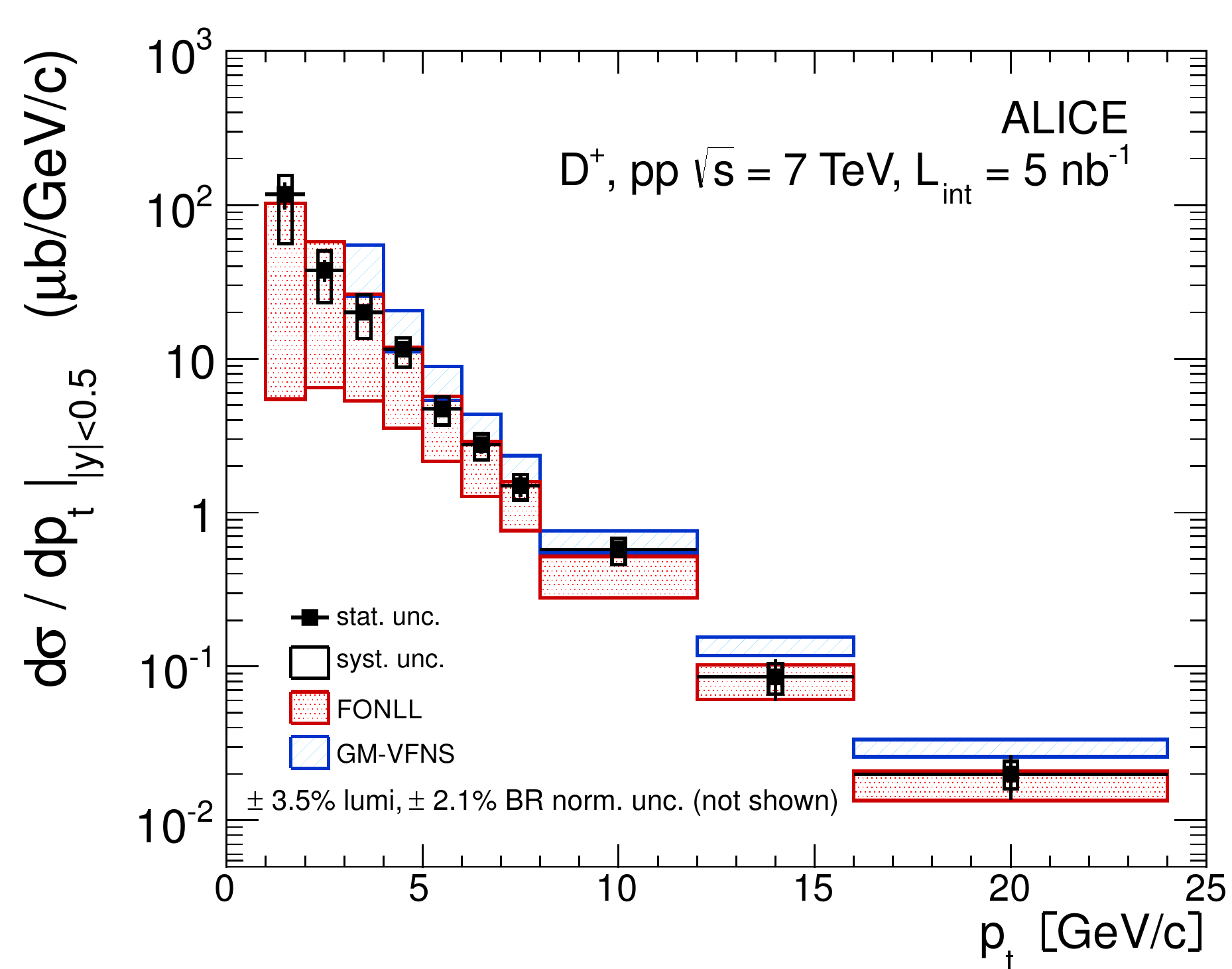} 
	}  
\subfigure[LHCb, prompt $\Ds$, \s = 7\TeV]{ 
	\label{fig:pp:LHCb:Ds} 
	\includegraphics[height=0.24\columnwidth]{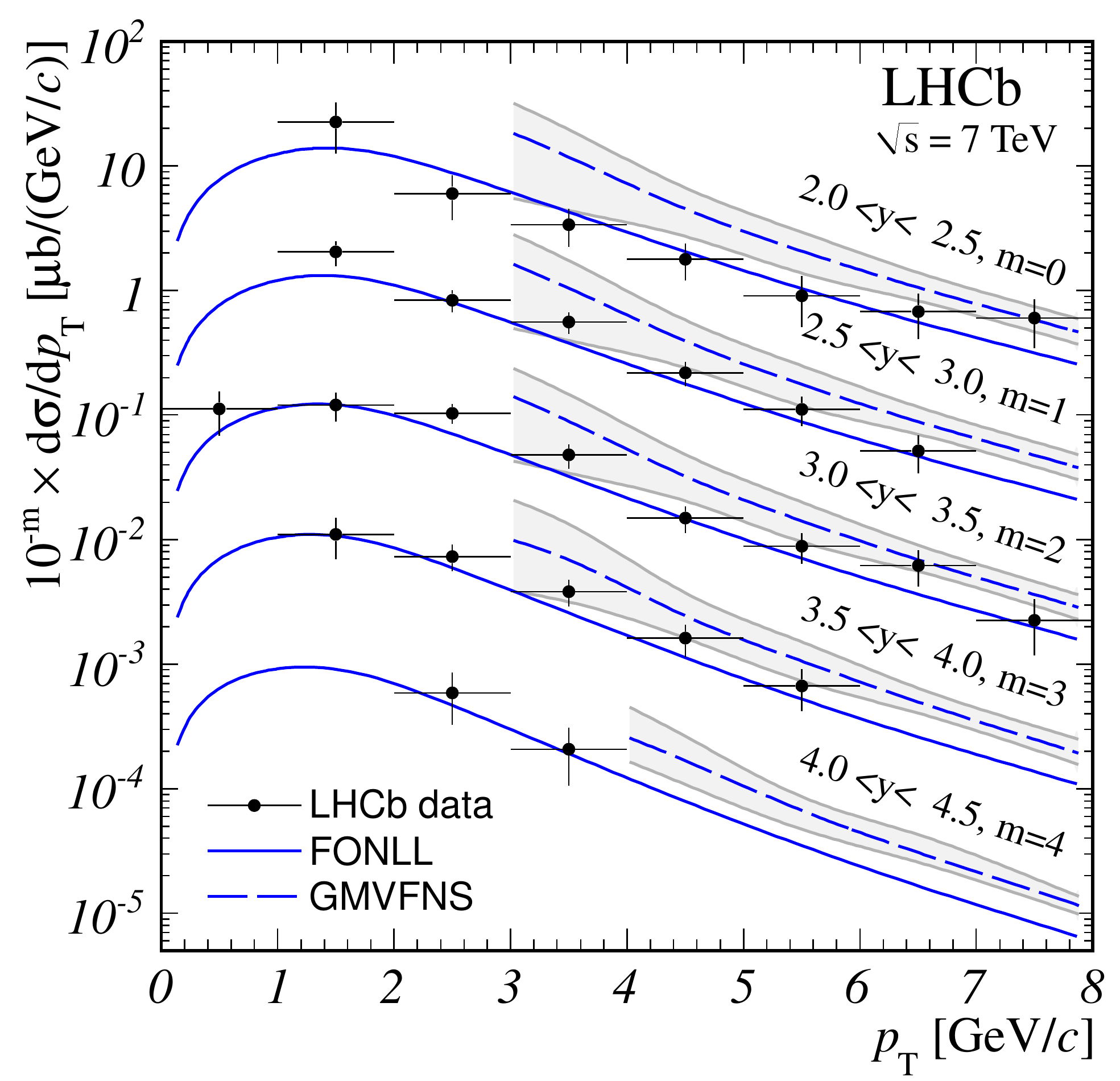} 
	}  
\caption{ 
\label{fig:pp:Dmeson} 
$\dsdpt$ for D meson production  in \pp collisions at different energies.  
(a): $\Dzero$ and $\Dstarplus$ measurements at \s = 200\GeV with the STAR detector~\cite{Adamczyk:2012af}.  
(b): $\Dplus$ data at \s = 7\TeV with the ALICE detector~\cite{ALICE:2011aa}.  
(c): $\Ds$ data at \s = 7\TeV with the LHCb detector~\cite{Aaij:2013mga}, where data from different $y$ ranges are scaled by factors $10^{-m}$, with $m$ shown in the plot.  
The measurements are compared to FONLL~\cite{Cacciari:2003uh, Cacciari:2012ny} and GM-VFNS~\cite{Kniehl:2004fy,Kniehl:2005mk} calculations.  
} 
\end{center} 
\end{figure}

\begin{figure}[!ht] 
\begin{center} 
\subfigure[$\sigmacplus$ and $\lambdacplus$ mass difference]{ 
	\label{fig:pp:LcMassDifference} 
	\includegraphics[height=0.3\columnwidth]{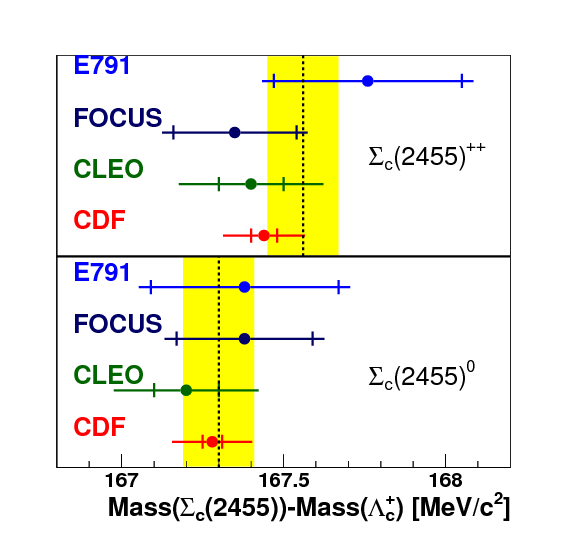} 
	}  
\subfigure[LHCb, prompt $\lambdacplus$, \s = 7\TeV]{ 
	\label{fig:pp:LambdaC} 
	\includegraphics[height=0.3\columnwidth]{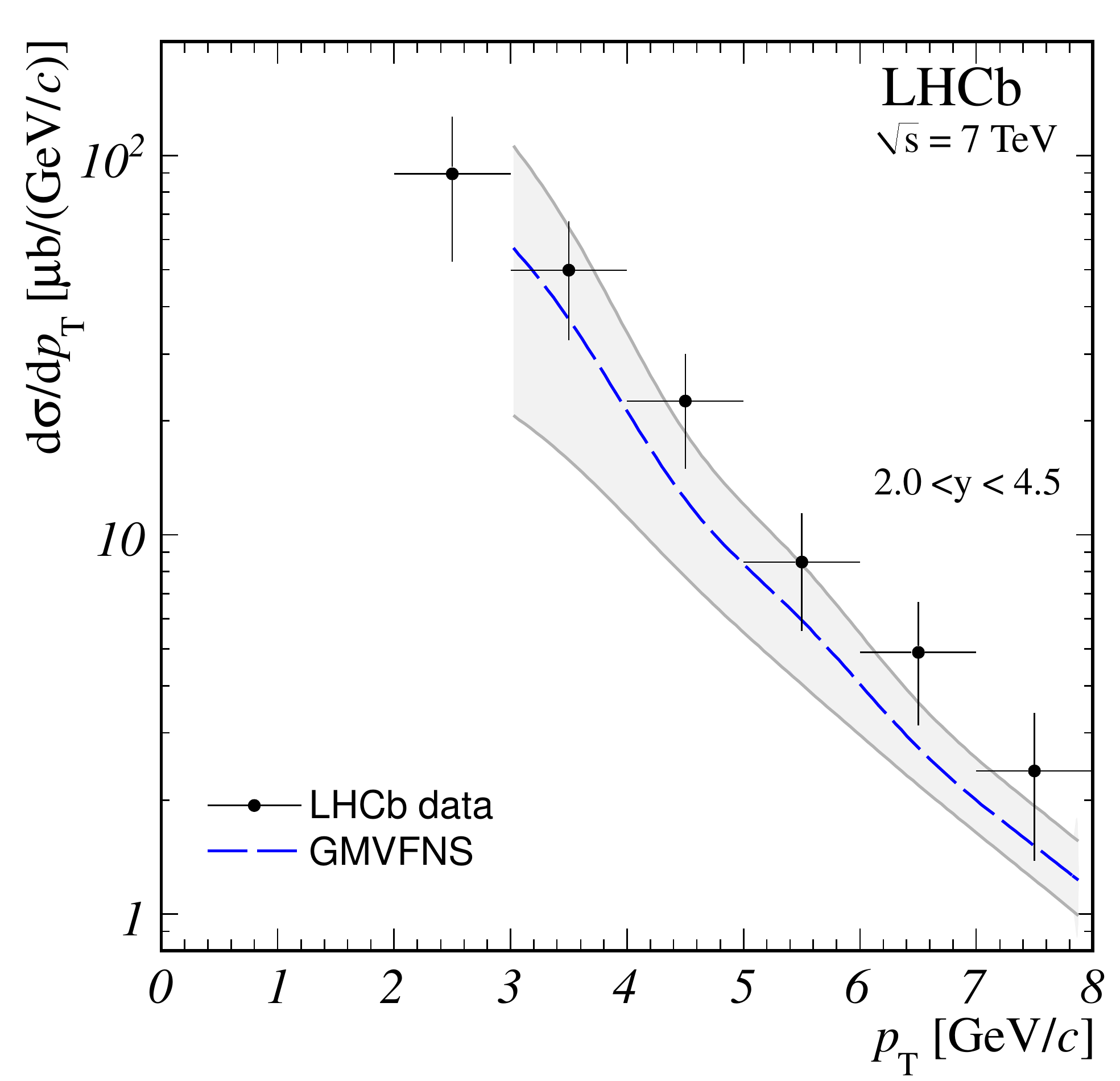} 
	}  
\caption{ 
\label{fig:pp:EtaLambdaC} 
(a): $\lambdacplus$ and $\sigmacplus(2455)$ mass differences as measured by CDF~\cite{Aaltonen:2011sf}, Fermilab E791~\cite{Aitala:1996cy}, FOCUS~\cite{Link:2001ee}, and CLEO~\cite{Artuso:2001us}.  
The vertical dashed line together with the surrounding box symbolizes the world average value and its uncertainty taken from~\ci{Agashe:2014kda}.  
(b) : $\dsdpt$ for $\lambdacplus$ production in \pp collisions at \s = 7\TeV~\cite{Aaij:2013mga} compared to GM-VFNS~\cite{Kniehl:2004fy,Kniehl:2005mk} calculations.  
} 
\end{center} 
\end{figure} 
 
\fig{fig:pp:Dmeson} presents a selection of the D meson measurements compared to pQCD calculations.  
The $\Dzero$, $\Dplus$ and $\Dstarplus$ $\dsdpt$ are reproduced by the theoretical calculations within uncertainties.  
Yet, FONLL~\cite{Cacciari:2003uh, Cacciari:2012ny} and POWHEG~\cite{Frixione:2007nw} predictions tend to underestimate the data, whereas GM-VFNS~\cite{Kniehl:2004fy,Kniehl:2005mk} calculations tend to overestimate the data (see Figure~3 and 4 in~\cite{Klasen:2014dba}).  
At low \pt, where the quark mass effects are important, the FONLL and POWHEG predictions show a better agreement with data. 
At intermediate to high \pt, where the quark mass effects are less important, all the FONLL, POWHEG, GM-VFNS and \kt-factorisation  
calculations agree with data.  
The agreement among the FONLL and POWHEG calculations is better for heavy-flavour decay leptons than for charmed mesons, which seems to be related to the larger influence of the fragmentation model on the latter.  
The $\Ds$ $\pt$-differential cross section is compared to calculations in \fig{fig:pp:LHCb:Ds}.  
The $\Ds$ measurements are also reproduced by FONLL, GM-VFNS and \kt-factorisation predictions, but POWHEG calculations predict a lower production cross section than data.  

Charmed baryon production measurements in hadron colliders are scarce.  
The properties and decay branching ratios of the $\Lambda_c$, $\Sigma_c$ and $\Xi_c$ states have been studied at the charm- and B-factories and fixed target experiments, see \eg~\cite{Briere:2006em,Klempt:2009pi,Butler:2013kdw,Bevan:2014iga}.  
An example are the results by Fermilab E791~\cite{Aitala:1996cy}, FOCUS~\cite{Link:2001ee}, and CLEO~\cite{Artuso:2001us} Collaborations.  
The CDF Collaboration measured charmed baryons in \ppbar collisions at \s = 1.96\TeV, see for example~\cite{Aaltonen:2011sf}. For illustration, a compilation of the $\sigmacplus$ and $\lambdacplus$ mass difference is shown in \fig{fig:pp:LcMassDifference}.  
The LHCb Collaboration measured the \pt and $y$ differential production cross section of $\Lambda_c$ in \pp collisions at \s = 7\TeV~\cite{Aaij:2013mga}.  
\fig{fig:pp:LambdaC} shows the $\pt$-differential cross section compared to GM-VFNS calculations.  
No dedicated FONLL calculation is available for $\Lambda_c$ production due to the lack of knowledge of the fragmentation function. The GM-VFNS predictions include the fragmentation functions resulting from a fit to $\ee$ collider data~\cite{Kneesch:2007ey}, where the prompt and secondary contributions to the measurements were not separated.

%
%
%
%
\subsubsection{Open beauty} 
\label{subsubsec:pp:OpenBeauty}

Open-beauty production is usually measured by looking for $\bquark$-jets or for beauty hadrons  
via their hadronic decays, similarly to D mesons. They have been traditionally studied  
at the $\ee$ B-factories (see \eg~\cite{Butler:2013kdw,Bevan:2014iga}), where, despite the small $\bquark$-quark production cross section,  
the large luminosity allows precise measurements, such as those of the CKM  
matrix. Yet, heavier states like the $\mathrm{B}_s$, $\mathrm{B}_c$ or $\Lambda_c$  
cannot be produced at the B-factories. They are however studied at Tevatron and at the 
 LHC hadron colliders. The higher collision energy increases their production cross section,  
although their reconstruction is more difficult at hadron colliders due to the larger  
combinatorics compared to the $\ee$ environment. It should also be kept in mind that the experiments  
optimised to study the high-$\pt$ probes, like top production, are not as good for low-$\pt$  
measurements, and often require the usage of dedicated triggers.

\begin{figure}[!ht] 
\begin{center} 
\subfigure[ATLAS, $\psiP$, \s = 7\TeV]{ 
	\label{fig:pp:NonPromptOnia:Pis2S} 
	\includegraphics[width=0.31\columnwidth,height=0.31\columnwidth]{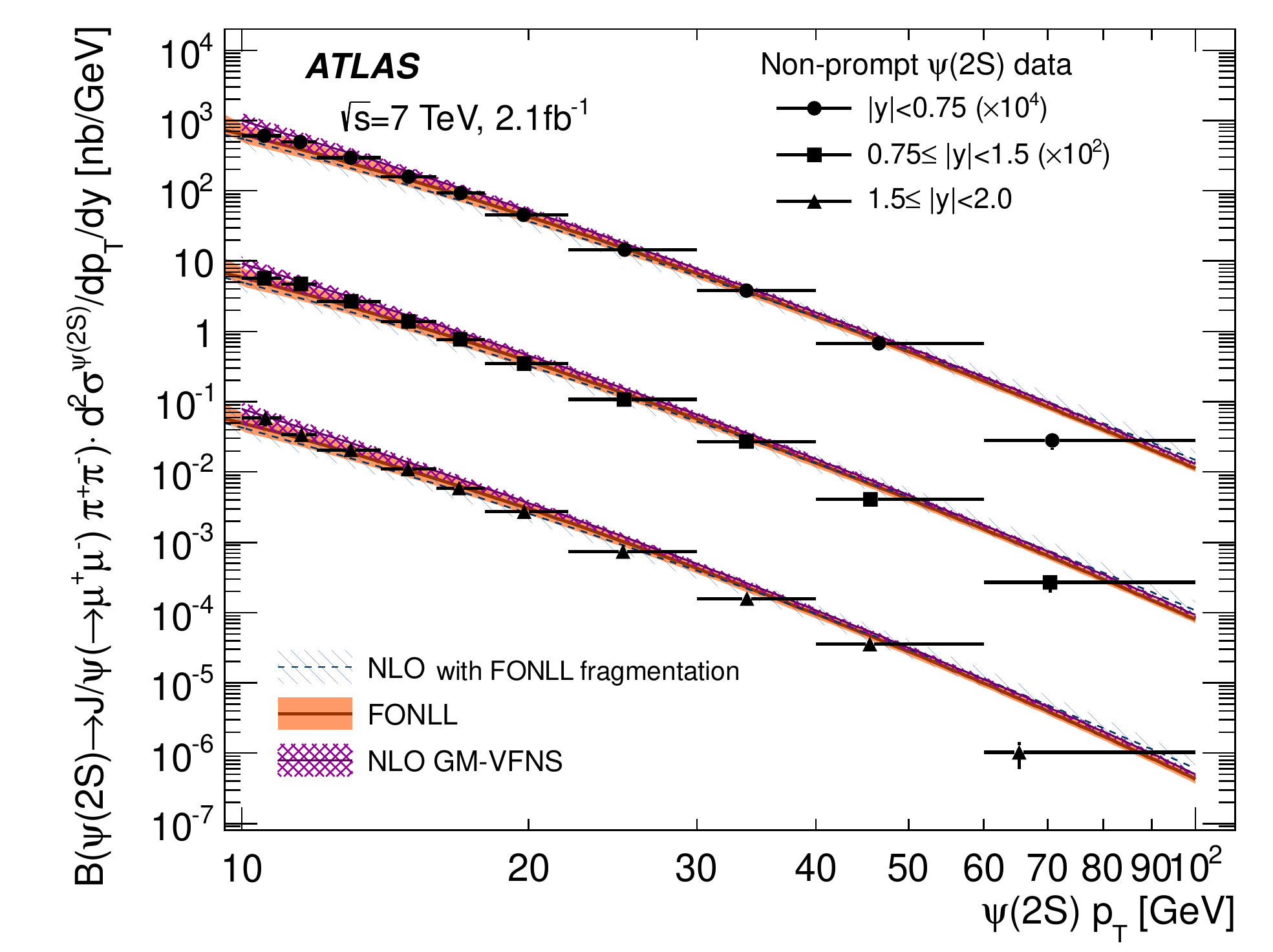} 
	}  
\subfigure[ATLAS, $\chi_{c1}$ and $\chi_{c2}$, \s = 7\TeV]{ 
	\label{fig:pp:NonPromptOnia:Chic1} 
	\includegraphics[width=0.31\columnwidth,height=0.31\columnwidth]{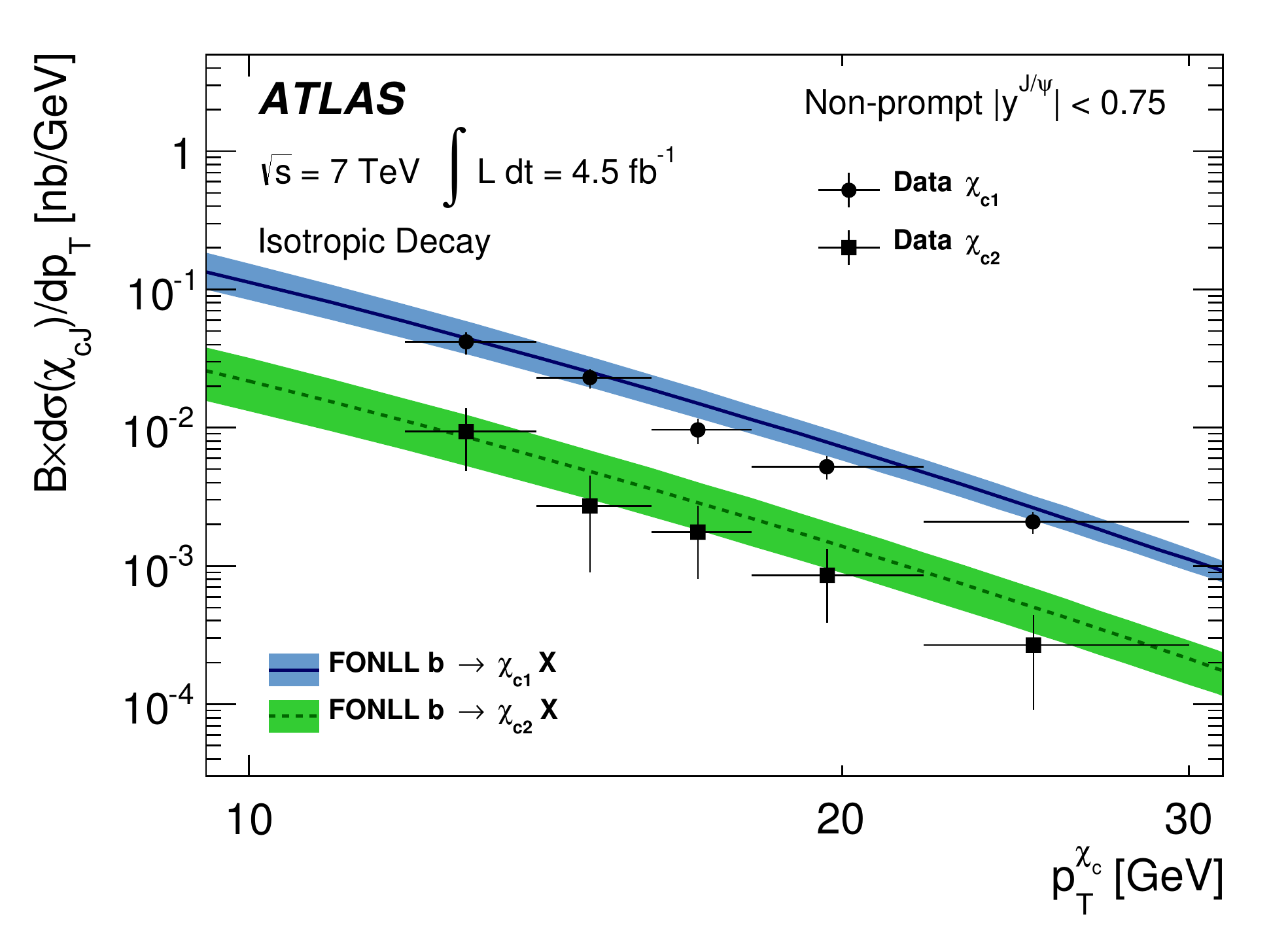} 
	}  
\subfigure[LHCb, $\etac$ (red filled circles) and $\jpsi$ (blue open circles), $2.0 < y < 4.5$, \s = 7\TeV]{ 
	\label{fig:pp:NonPromptOnia:Etac} 
	\includegraphics[width=0.31\columnwidth,height=0.31\columnwidth]{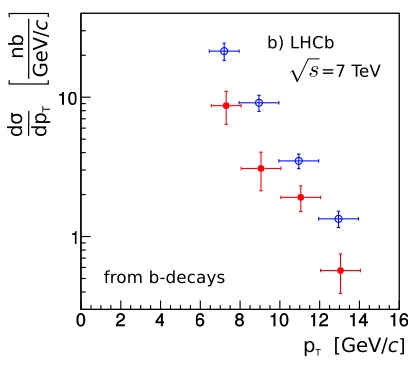} 
	}  
\caption{ 
 \label{fig:pp:NonPromptOnia} 
$\pt$-differential cross sections for non-prompt charmonia --assumed to come from $\bquark$ decays-- for 
(a) $\psiP$~\cite{Aad:2014fpa} ${\rm d}^2\sigma/{\rm d}\pt{\rm d}y$ by ATLAS compared to NLO~\cite{Mangano:1991jk},  
FONLL~\cite{Cacciari:2003uh, Cacciari:2012ny} and GM-VFNS~\cite{Kniehl:2004fy,Kniehl:2005mk} calculations,  
(b) $\chi_{c1}$ and $\chi_{c2}$  $\dsdpt$ by ATLAS~\cite{ATLAS:2014ala} compared to FONLL~\cite{Cacciari:2012ny}, 
(c) $\etac$ and $\jpsi$ $\dsdpt$ by LHCb~\cite{Aaij:2014bga}. 
} 
\end{center} 
\end{figure}

As discussed in the \sect{sec:pp:Theory:OpenHF}, predictions for open-beauty cross sections rely on   
the fragmentation functions derived from fits to $\ee$ data~\cite{Mangano:1997ri,Kniehl:2008zza}. 
A high accuracy on the $\ee$ measurements and on the fragmentation function parametrisations is  
required to calculate the $\bquark$-hadron production cross section at hadron colliders. $\bquark$-jet measurements  
have the advantage to be the least dependent on the $\bquark$-quark fragmentation characteristics.  

In addition, measurements of the B cross section via a displaced charmonium have been performed  
multiple times at Tevatron and at LHC. Charmonia from beauty decays are selected by  
fitting the pseudo-proper decay length distribution of the charmonium candidates,  
$L_{xy} \left( m / \pt \right)_{\jpsi}$. \fig{fig:pp:NonPromptOnia} presents a selection  
of the LHC results: the non-prompt $\pt$-differential cross section of $\jpsi$, $\psiP$, $\etac$,  
$\chi_{c1}$ and $\chi_{c2}$ in \pp collisions at \s = 7\TeV~\cite{Aad:2014fpa,ATLAS:2014ala,Aaij:2014bga}.  
The results at intermediate to low \pt are well reproduced by the FONLL~\cite{Cacciari:2003uh, Cacciari:2012ny}, NLO GM-VFNS~\cite{Kniehl:2004fy,Kniehl:2005mk} and NLO~\cite{Mangano:1991jk} with FONLL  
fragmentation calculations. At high-$\pt$ the predictions tend to overestimate data. 
This could be related to the usage of the $\ee$ fragmentation functions in an unexplored kinematic range.  
\fig{fig:pp:NonPromptOnia:Etac}, which reports the first measurement of non-prompt charmonium in a purely hadronic decay channel 
at hadron colliders, shows a similar transverse-momentum spectrum for non-prompt singlet and triplet $S$-wave charmonia.

 \begin{figure}[!ht]
\begin{center}
\subfigure[CMS, $\pt$ differential]{
	\label{fig:pp:CMS:Bplus:pt}
	\includegraphics[width=0.31\columnwidth,height=0.31\columnwidth]{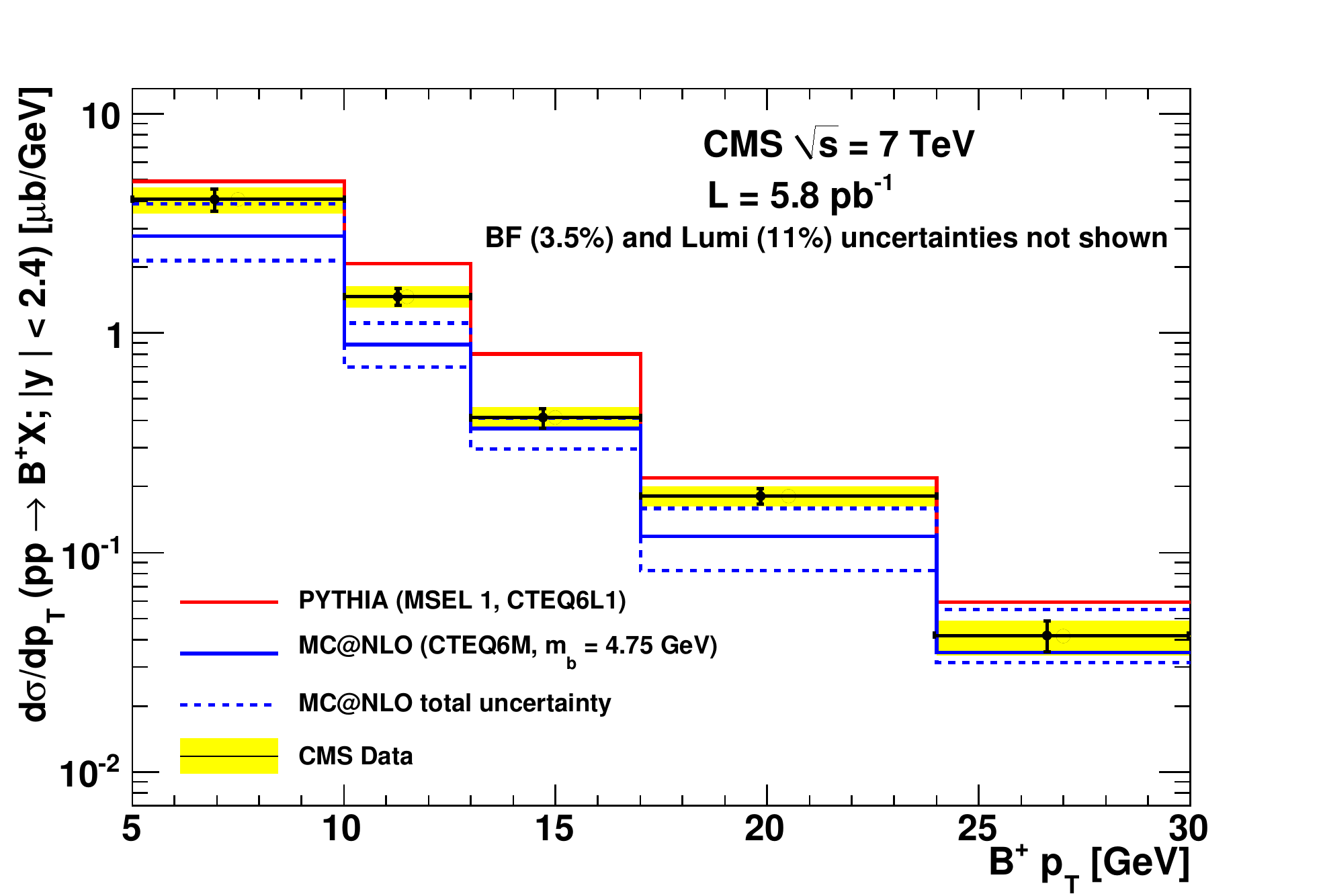}
	} 
\subfigure[ATLAS, CMS, $\pt$ differential]{
     \label{fig:pp:ATLAS:Bplus:pt}
	\includegraphics[width=0.31\columnwidth,height=0.31\columnwidth]{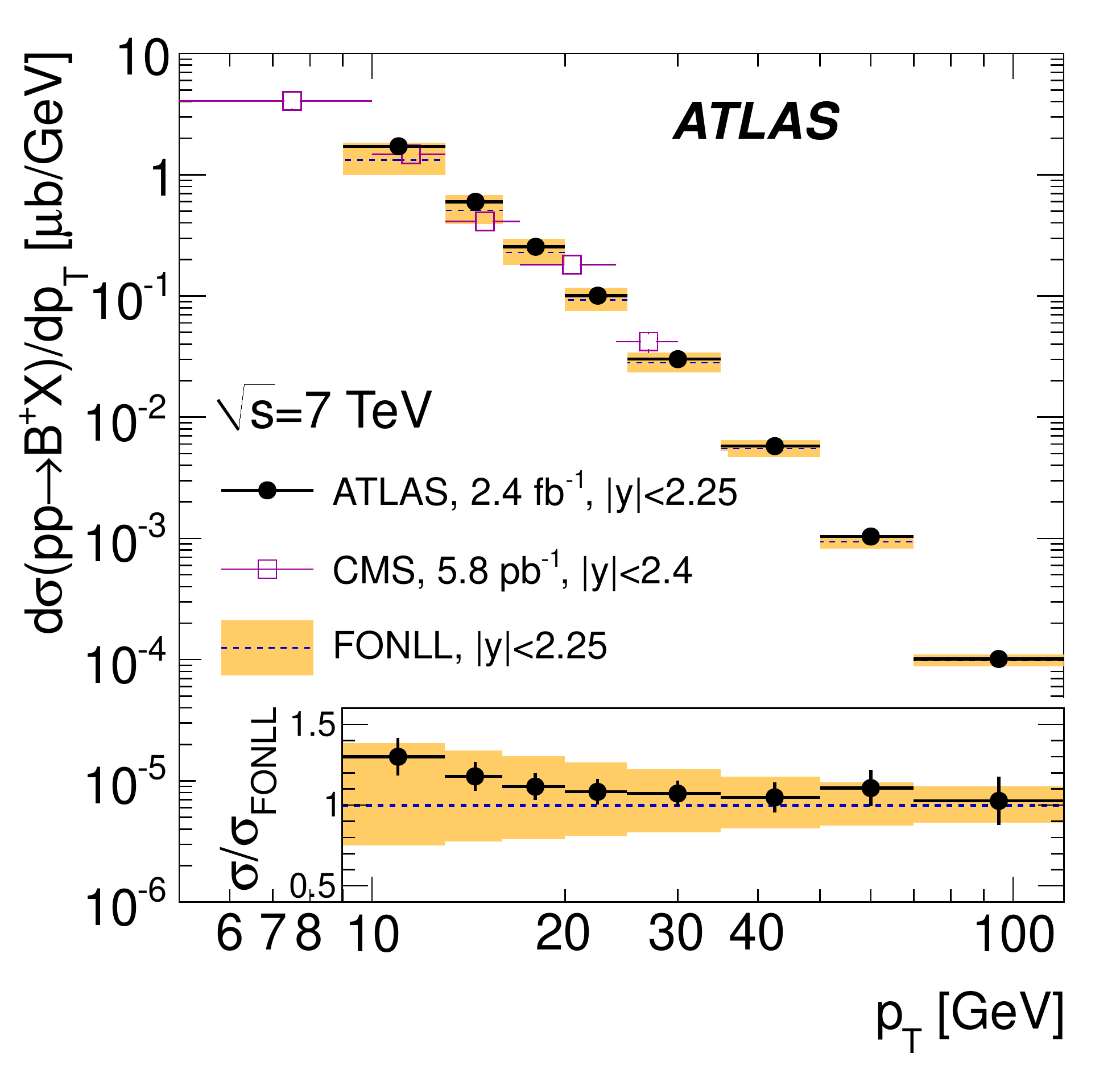}
	} 
\subfigure[ATLAS, $y$ differential]{
	\label{fig:pp:ATLAS:Bplus:y}
	\includegraphics[width=0.31\columnwidth,height=0.31\columnwidth]{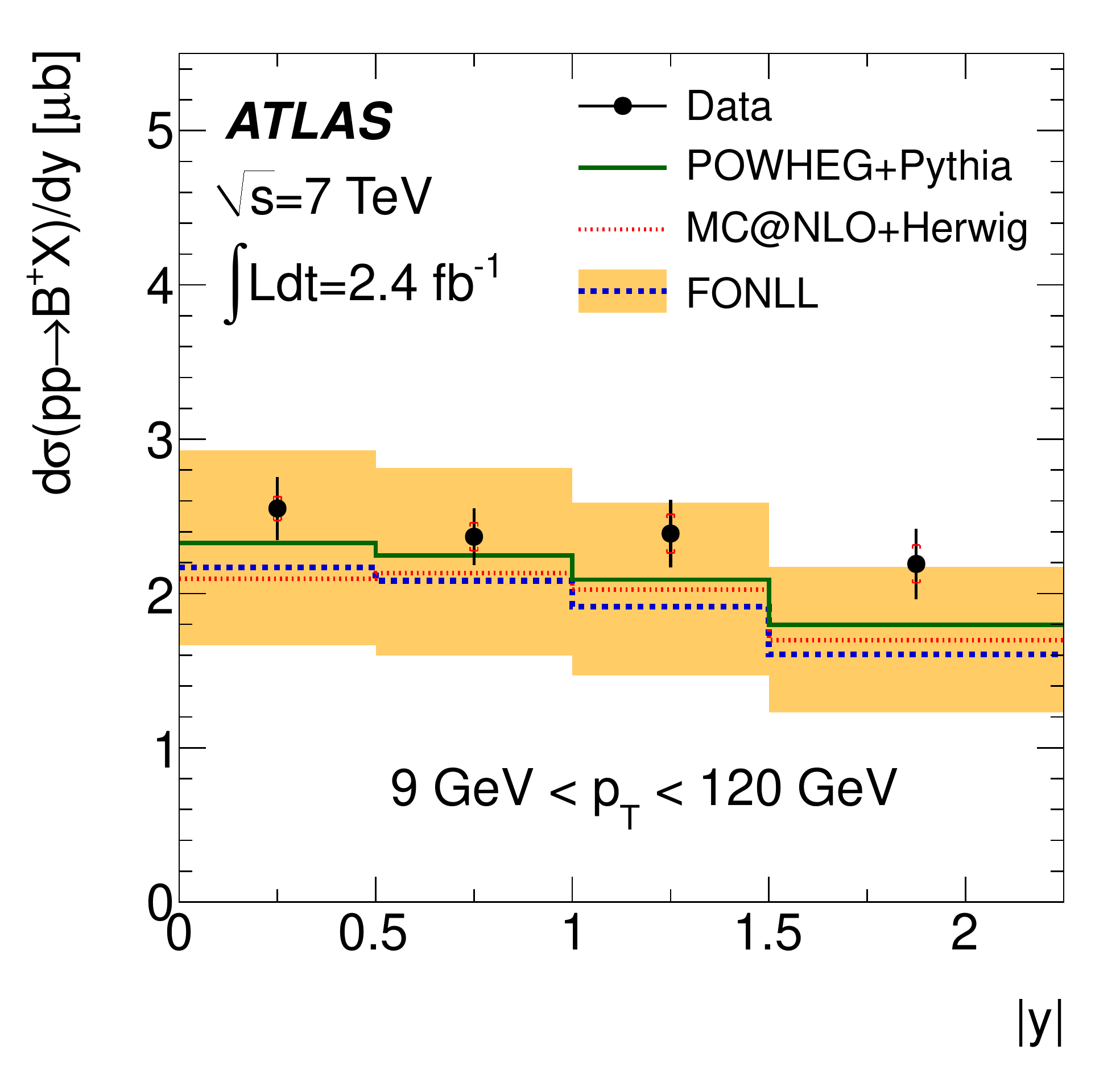}
	} 
\caption{\label{fig:pp:Bplusmeson}
$\mathrm{B}^+$differential cross sections $\dsdpt$ (a), (b) and $\dsdy$ (c) in \pp collisions at \s = 7\TeV compared to theory predictions~\cite{Khachatryan:2011mk,ATLAS:2013cia}. 
In (c) the error bars correspond to the differential cross-section measurement with total uncertainty (lines on the error bars indicate the statistical component).
}
\end{center}
\end{figure}

\begin{figure}[!ht] 
\begin{center} 
\subfigure[CMS, $\lambdabplus$ $\dsdpt$, \s = 7\TeV]{ 
	\label{fig:pp:Lambdab:dsdpt} 
	\includegraphics[height=0.275\columnwidth]{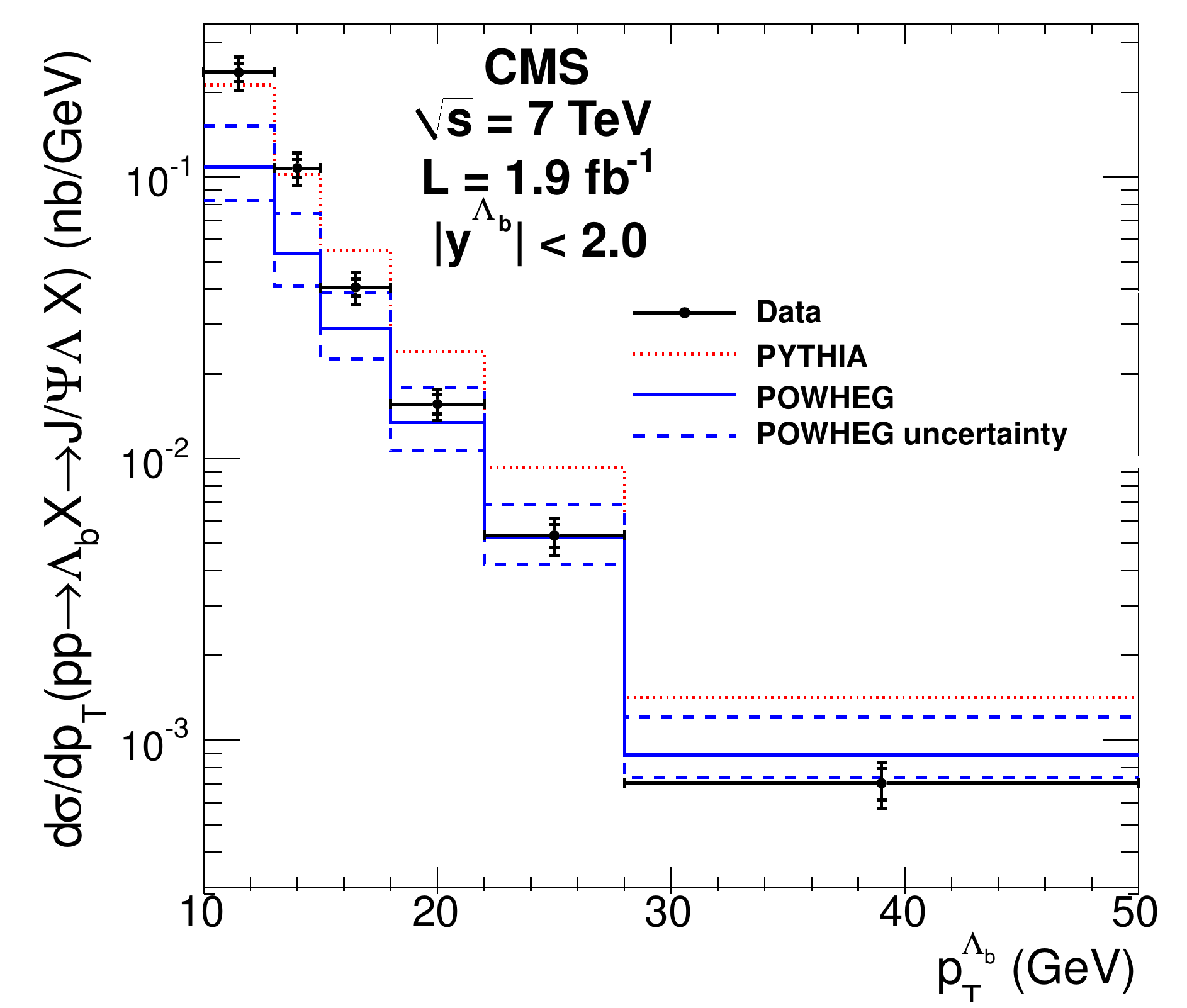} 
	}  
\subfigure[CMS, $\lambdabplus$ $\dsdy$, \s = 7\TeV]{ 
	\label{fig:pp:Lambdab:dsdy} 
	\includegraphics[height=0.275\columnwidth]{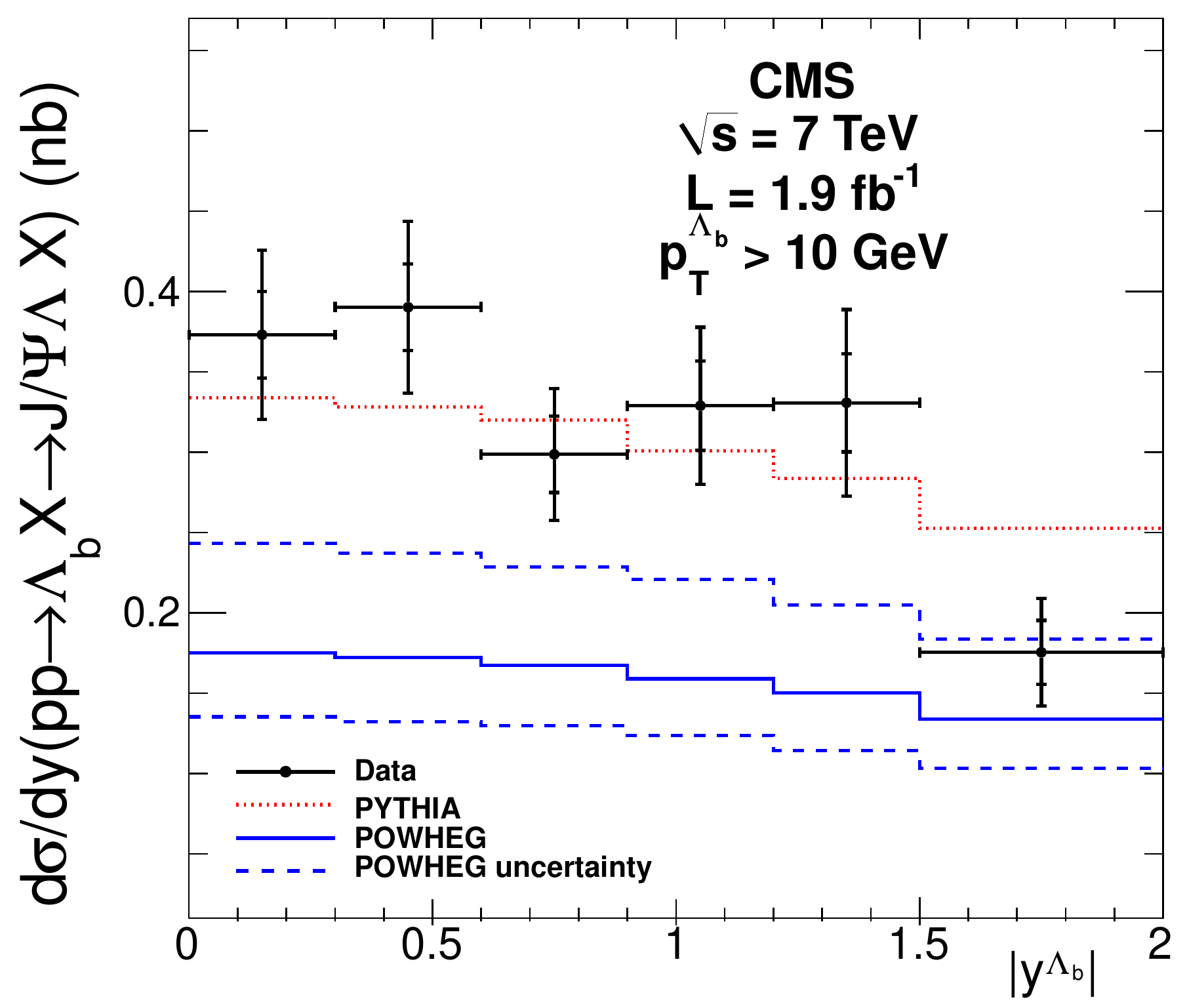} 
	}  
\subfigure[CMS, self-normalised $\dsdpt$, \s = 7\TeV]{ 
	\label{fig:pp:Lambdab:B0Bp} 
	\includegraphics[height=0.275\columnwidth]{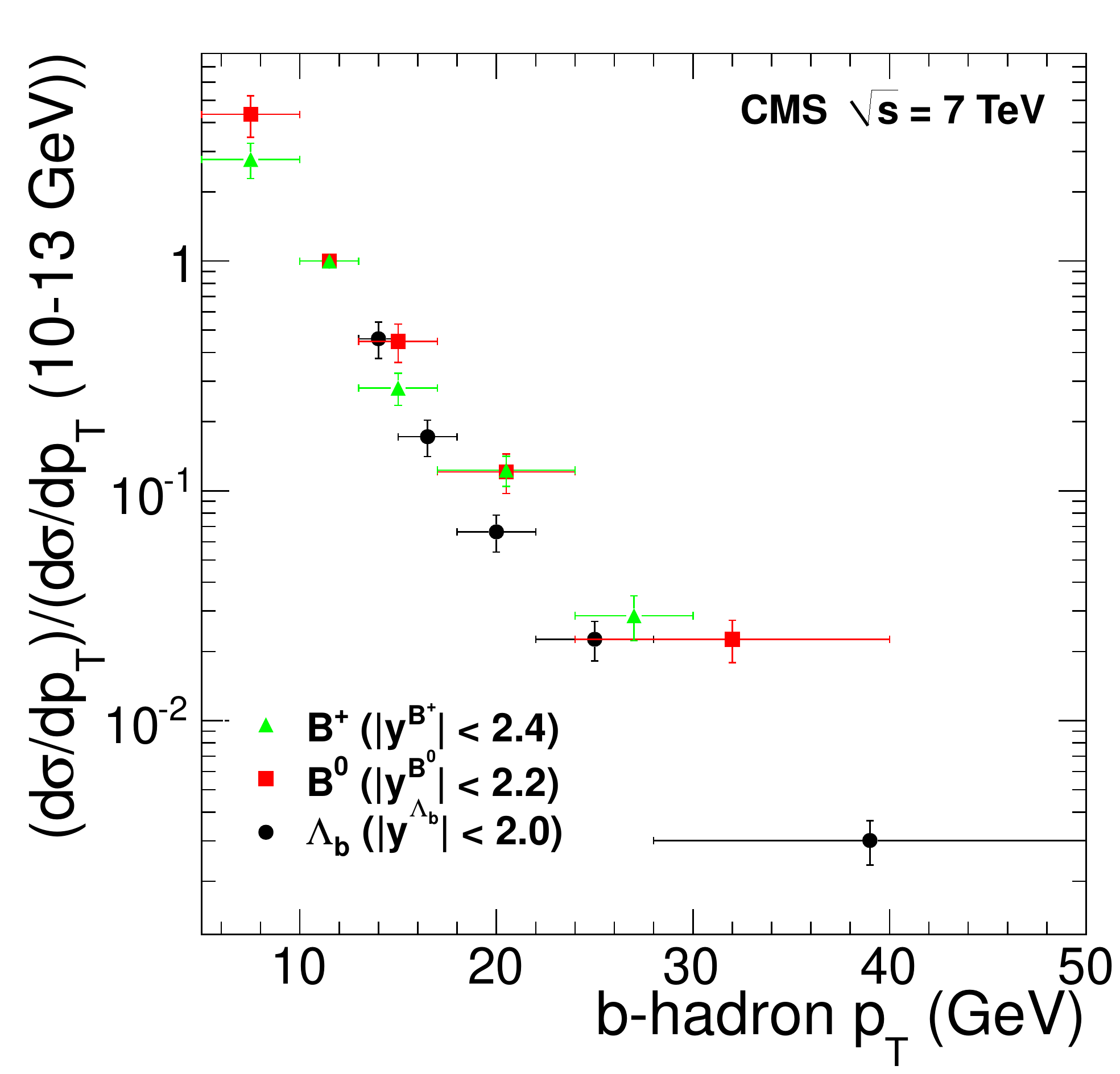} 
	}  
\caption{ 
	\label{fig:pp:Lambdab} 
$\lambdabplus$ studies at the LHC in \pp collisions at  \s = 7\TeV by CMS~\cite{Chatrchyan:2012xg} compared 
to PYTHIA and POWHEG calculations for (a) $\dsdpt$ for $|y|<2.0$ and for (b) $\dsdy$ for $\lambdabplus$ for $\pt > 10$\GeV. (c) Comparison of the (self-normalised) \pt differential cross section for $\mathrm{B}^+$, $\mathrm{B}^0$ and \lambdabplus. 
} 
\end{center} 
\end{figure} 

Studies of open-beauty production have also been performed in exclusive channels at Tevatron and at the LHC, \eg in the case of $\mathrm{B}^{\pm}$, $\mathrm{B}^{0}$ and $\mathrm{B}^0_s$~\cite{Aaltonen:2009tz, Abazov:2013xda, ATLAS:2013cia, Aaij:2011ep, Aaij:2013noa, Aaij:2012jd, Khachatryan:2011mk, Chatrchyan:2011pw, Chatrchyan:2011vh}.  
As example, \fig{fig:pp:Bplusmeson} presents the $\mathrm{B}^+$ $\pt$ and $y$ differential cross section in \pp collisions at \s = 7\TeV compared to theory predictions~\cite{Khachatryan:2011mk,ATLAS:2013cia}.  
PYTHIA (D6T tune), that has LO + LL accuracy, does not provide a good description of the data.  
This could be explained by the choice of $m_b$ and by the fact that for $\pt \simeq m_b$,  
NLO and resummation effects become important, which are, in part, accounted for in FONLL~\cite{Cacciari:2003uh, Cacciari:2012ny} or MC{@}NLO.  
POWHEG and MC{@}NLO calculations are quoted with an uncertainty of the order of 20-40\%, from $m_b$ and the renormalisation and factorisation scales, and describe the data within uncertainties. 
The FONLL prediction provides a good description of the measurements within uncertainties.

\begin{figure}[!ht] 
\begin{center} 
\subfigure[CMS, $\bquark$-jets, \s = 7\TeV]{ 
	\label{fig:pp:CMS:Bjet} 
	\includegraphics[height=0.3\columnwidth]{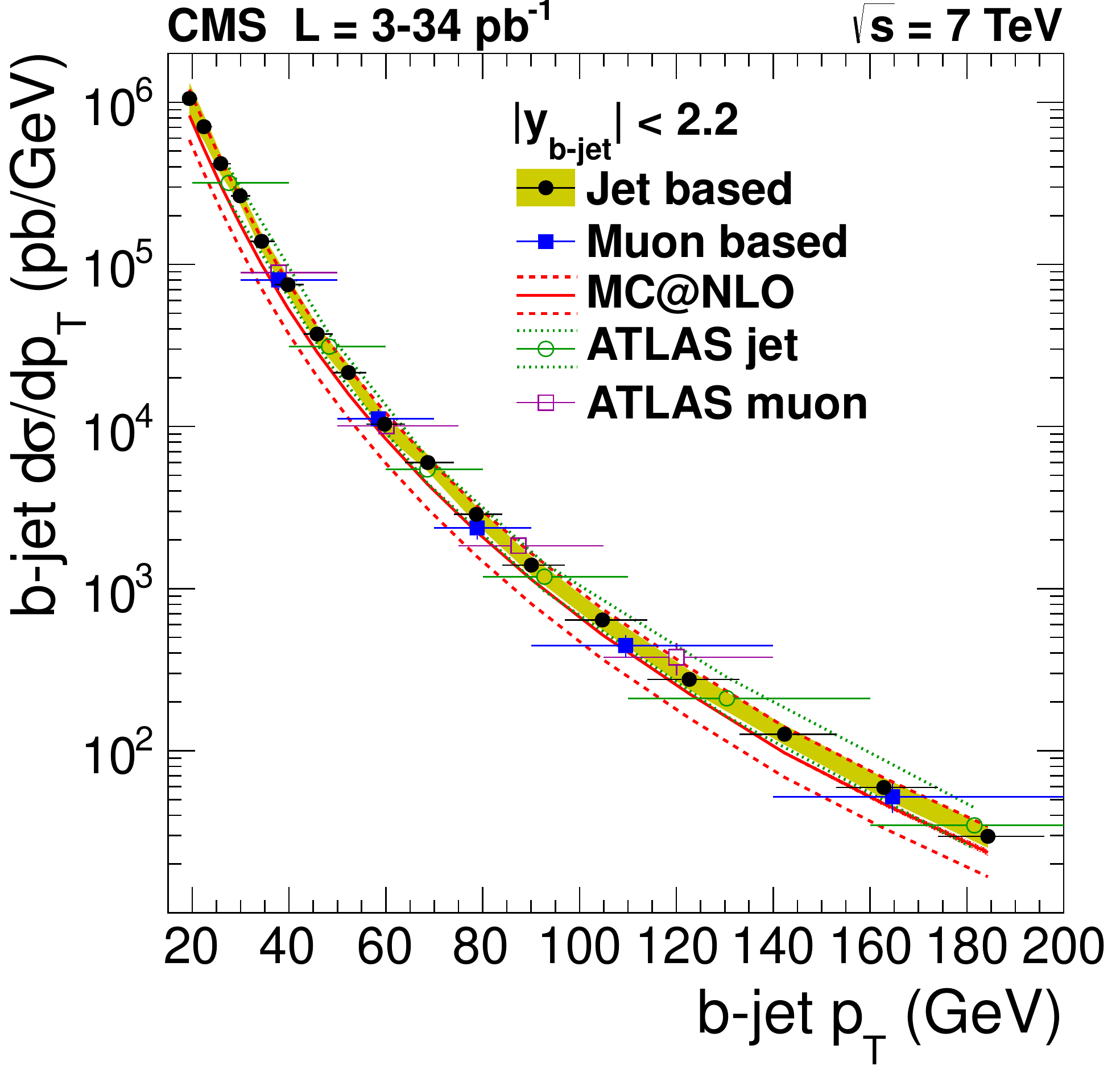} 
	}  
\subfigure[ATLAS, $\bquark$-jet, \s = 7\TeV]{ 
	\label{fig:pp:ATLAS:Bjet} 
	\includegraphics[height=0.3\columnwidth]{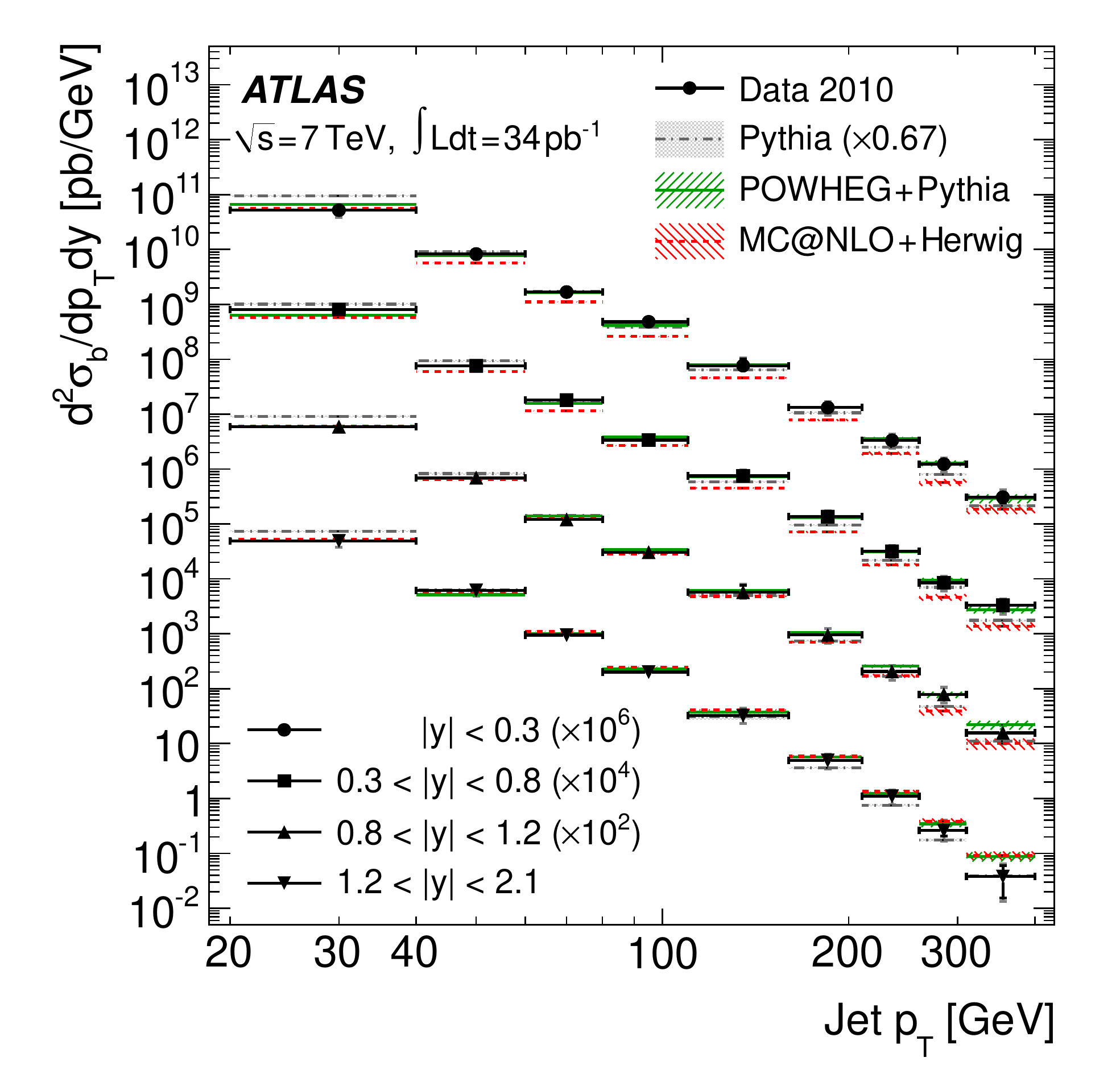} 
	}  
\caption{ 
\label{fig:pp:Bjet} 
$\bquark$-jet cross section as a function of \pt in \pp collisions at \s = 7\TeV: (a) $\dsdpt$ from the lifetime-based and muon-based analyses  
by CMS~\cite{Chatrchyan:2012dk} and ATLAS~\cite{ATLAS:2011ac} compared to the MC{@}NLO calculation, and 
(b) ${\rm d}^2\sigma/{\rm d}\pt {\rm d}y$ by ATLAS from the lifetime-based analysis~\cite{ATLAS:2011ac} compared to the predictions of PYTHIA, POWHEG (matched to PYTHIA) and MC{@}NLO (matched to HERWIG)~\cite{Sjostrand:2006za,Alioli:2010xd,Frixione:2002ik,Frixione:2003ei,Corcella:2000bw}.  
} 
\end{center} 
\end{figure}

Measurements of the beauty and charm baryon production in \ppbar collisions at \s = 1.96\TeV are summarised in~\ci{Aaltonen:2014wfa}.  
In particular, the doubly strange $\bquark$-baryon $\Omega_b^-$ and measurement of the $\Xi_b^-$ and $\Omega_b^-$ properties can be found in~\cis{Abazov:2008qm,Aaltonen:2009ny}.  
Such measurements have also been performed at the LHC in \pp collisions at \s = 7\TeV and 8\TeV. For example, the observation of the $\Xi_b$ baryon states was reported in~\cis{Chatrchyan:2012ni,Aaij:2014yka}.  
The measured mass and width of the $\Xi_b$ baryon states is consistent with theoretical expectations~\cite{Ebert:2005xj,Liu:2007fg,Jenkins:2007dm,Karliner:2008sv,Lewis:2008fu,Karliner:2008in,Zhang:2008pm,Klempt:2009pi,Detmold:2012ge}.  
The $\lambdabplus$ $\pt$ and $y$ differential production cross section in \pp collisions at \s = 7\TeV by CMS~\cite{Chatrchyan:2012xg} is reported in \fig{fig:pp:Lambdab}. The $\lambdab$ $\dsdpt$ and $\dsdy$ are not reproduced by neither PYTHIA (Z2 tune) nor POWHEG calculations: PYTHIA expects a harder \pt-distribution and flatter $y$ distribution than data, while POWHEG underestimates its production cross section, particularly at low \pt, see \fig{fig:pp:Lambdab:dsdpt} and~\fig{fig:pp:Lambdab:dsdy}.  
The measured $\lambdab$ $\pt$-spectrum at mid-rapidity seems to fall more steeply than the $\mathrm{B}^0$ and $\mathrm{B}^+$ ones, see \fig{fig:pp:Lambdab:B0Bp}, and falls also faster than predicted by PYTHIA and POWHEG.  
As discussed for the non-prompt charmonium measurements, this could be influenced by the lack of data to extract the fragmentation functions in this kinematic region.  

The fragmentation of the $\bquark$ quark is relatively hard compared to that of lighter flavours, with the $\bquark$-hadron taking about 70\% of the parton momentum on average at the Z-pole~\cite{DELPHI:2011aa}.   
Identification of jets from beauty quark fragmentation or ``\bquark-tagging'' can be achieved by direct reconstruction of displaced vertices, although the efficiency for doing so is limited.  Higher efficiency can be achieved by looking for large impact parameter tracks inside jets, or by a combination of the two methods, which are collectively known as lifetime tagging.  Leptons inside jets can also be used for $\bquark$-tagging, but, due to the branching fraction, are usually only used as a calibration of the lifetime methods.   
At the LHC, both ATLAS and CMS have performed measurements of the $\bquark$-jet cross section~\cite{ATLAS:2011ac, Chatrchyan:2012dk}. Theoretical comparisons can be made to models which calculate fully exclusive final states, which can be achieved by matching NLO calculations to parton showers~\cite{Nason:2004rx}.  
\fig{fig:pp:Bjet} shows the $\bquark$-jet cross section measurement by ATLAS and CMS in \pp collisions at \s = 7\TeV. The measurements are shown as a function of \pt and in several bins of rapidity. Calculations from POWHEG~\cite{Alioli:2010xd} (matched to PYTHIA~\cite{Sjostrand:2006za}) and MC{@}NLO~\cite{Frixione:2002ik,Frixione:2003ei} (matched to HERWIG~\cite{Corcella:2000bw}), are found to reproduce the data. Measurements from both lifetime- and lepton-based tagging methods are shown.

%
%
%
%
\subsubsection{Prompt charmonium} 
\label{subsubsec:exp_onium_prompt} 
 
In this section, we show and discuss a selection of experimental measurements of prompt 
charmonium production at RHIC and LHC energies. We thus focus here on the production channels which do not involve beauty decays; these were discussed in the \sect{subsubsec:pp:OpenBeauty}.

\begin{figure}[!h] 
\begin{center} 
\subfigure[]{ 
        \label{fig:pp:LHCb:Jpsi} 
        \includegraphics[width=0.4\columnwidth]{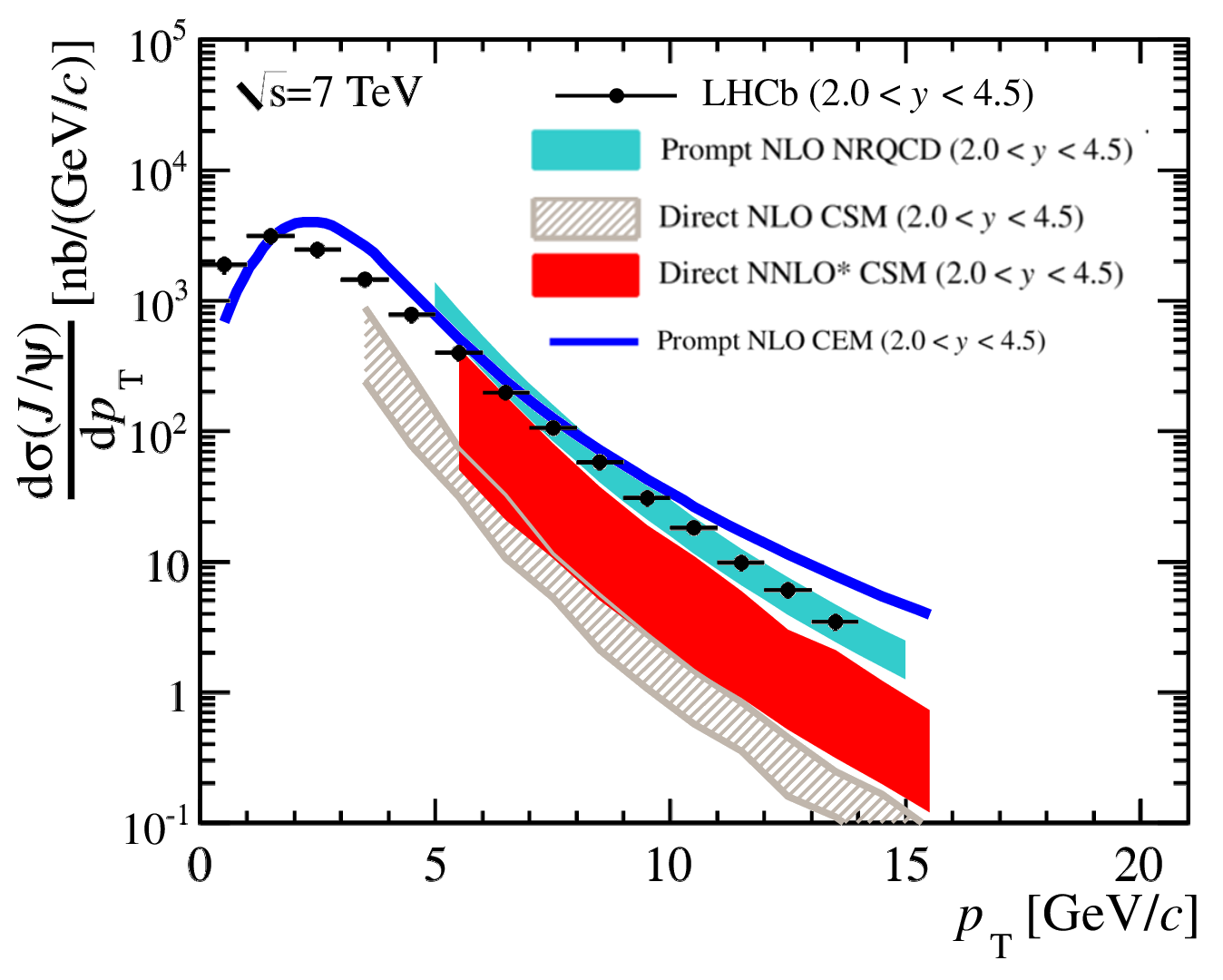} 
}\quad 
\subfigure[]{ 
        \label{fig:pp:ALICE:Jpsi} 
       \includegraphics[width=0.343\columnwidth]{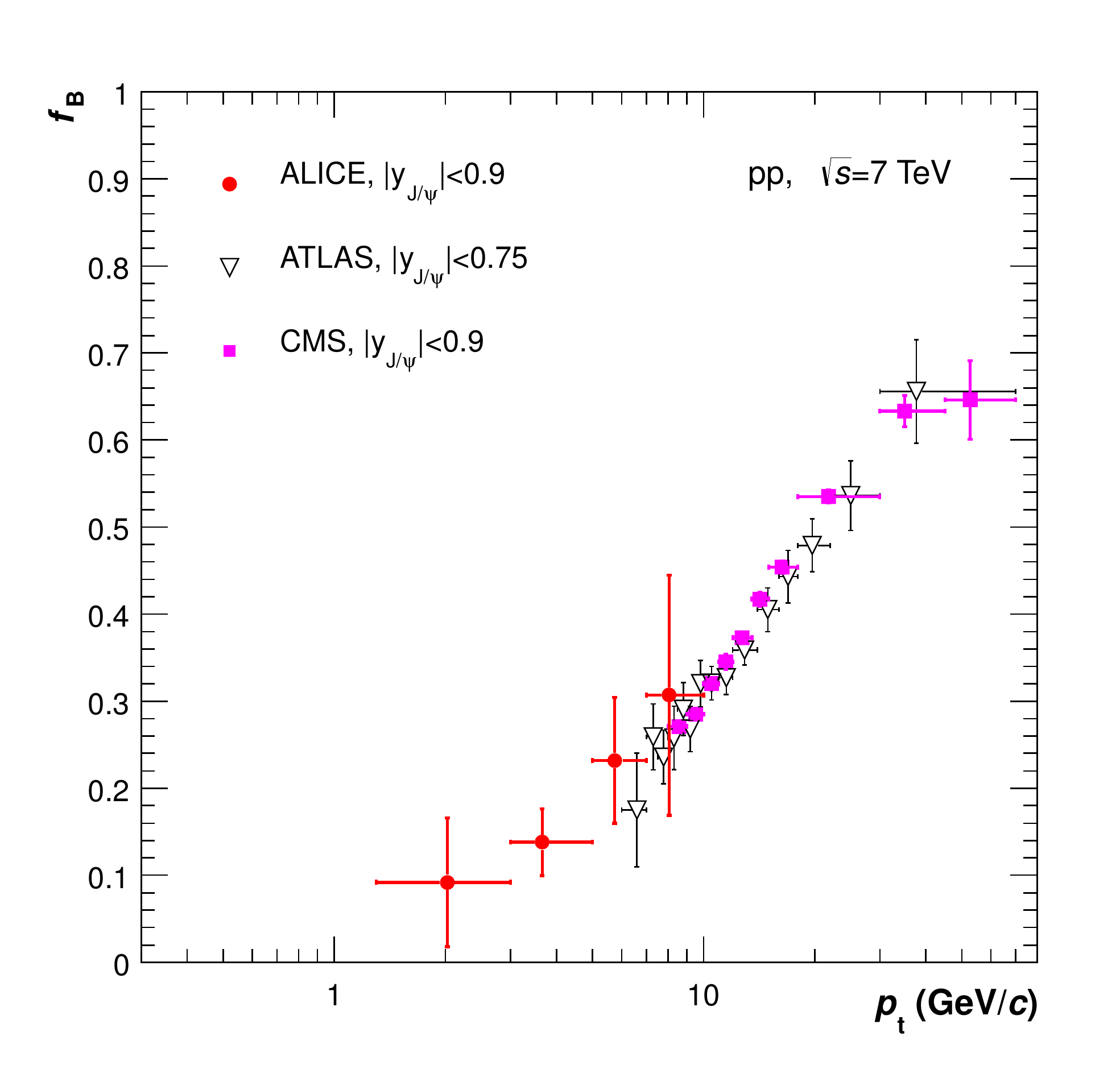} 
        } 
\caption{ 
        \label{fig:pp:prompt_dimuon} 
(a) Prompt $\jpsi$ yield as measured by LHCb~\cite{Aaij:2011jh} at \s = 7\TeV compared to different theory predictions referred to as ``prompt NLO NRQCD''\cite{Ma:2010yw}, ''DirectNLO CS''\cite{Campbell:2007ws,Artoisenet:2007xi}, ``Direct NNLO$^\star$ CS'' \cite{Artoisenet:2008fc,Lansberg:2008gk} and  ``Prompt NLO CEM'' \cite{Frawley:2008kk}. (b) Fraction of $\jpsi$ from B as measured by ALICE\cite{Abelev:2012gx}, ATLAS~\cite{Aad:2011sp} and CMS~\cite{Chatrchyan:2011kc} at \s = 7\TeV in the central rapidity region. 
} 
\end{center} 
\end{figure}

Historically, promptly produced $\jpsi$ and $\psiP$ have always been studied in the dilepton channels. 
Except for the PHENIX, STAR and ALICE experiments, the recent studies in fact only 
consider dimuons which offer a better signal-over-background ratio and a purer triggering. There are many recent 
experimental studies. In \fig{fig:pp:prompt_dimuon}, we show only two of these. First we  show $\dsdpt$ for 
prompt $\jpsi$ at \s = 7\GeV as measured by LHCb compared to a few predictions for the prompt yield from the CEM 
and from NRQCD at NLO\footnote{Let us stress that the NRQCD band in \fig{fig:pp:LHCb:Jpsi} is not drawn for \pt lower 
than 5 GeV because such a NLO NRQCD fit overshoots the data in this region and since data at low \pt are in fact not used in this fit. For a complete discussion of NLO CSM/NRQCD results for the \pt-integrated yields, see \cite{Feng:2015cba}. As regards the CEM curves, an uncertainty band should also be drawn (see for instance~\cite{Nelson:2012bc}).} as well as the direct yield\footnote{ 
The expected difference between prompt and direct 
is discussed later on.} compared to a NNLO$^\star$ CS evaluation. 
Our point here is to emphasise the precision of the data and to illustrate that at low and mid \pt ~--which is the region 
where heavy-ion studies are carried out-- none of the models can simply be ruled out owing to their theoretical 
uncertainties (heavy-quark mass, scales, non-perturbative parameters, unknown QCD and relativistic corrections, ...). 
Second, we show the fraction of \jpsi from $\bquark$ decay for $y$ close to 0 at \s = 7\TeV as 
function of \pt as measured by ALICE~\cite{Abelev:2012gx}, ATLAS~\cite{Aad:2011sp} and 
CMS~\cite{Chatrchyan:2011kc}. At low \pt, the difference between the inclusive and prompt yield should not 
exceed 10\% -- from the determination of the $\sigma_{\bbbar}$, it is expected to be a few percent at RHIC 
energies~\cite{Adare:2009ic}. It however steadily grows with \pt. At the 
highest \pt reached at the LHC, the majority of the inclusive \jpsi is from $\bquark$ decays. At $\pt \simeq$ 10\GeV, 
which could be reached in future quarkonium measurements in \pb collisions, it is already 3 times higher 
than at low \pt: 1 \jpsi out of 3 comes from $\bquark$ decays.

\begin{figure}[!htbp] 
\begin{center} 
\subfigure[Theory and ATLAS, $\psiP$, \s=7\TeV]{ 
        \label{fig:pp:ATLAS:Psi2S} 
        \includegraphics[height=0.225\columnwidth]{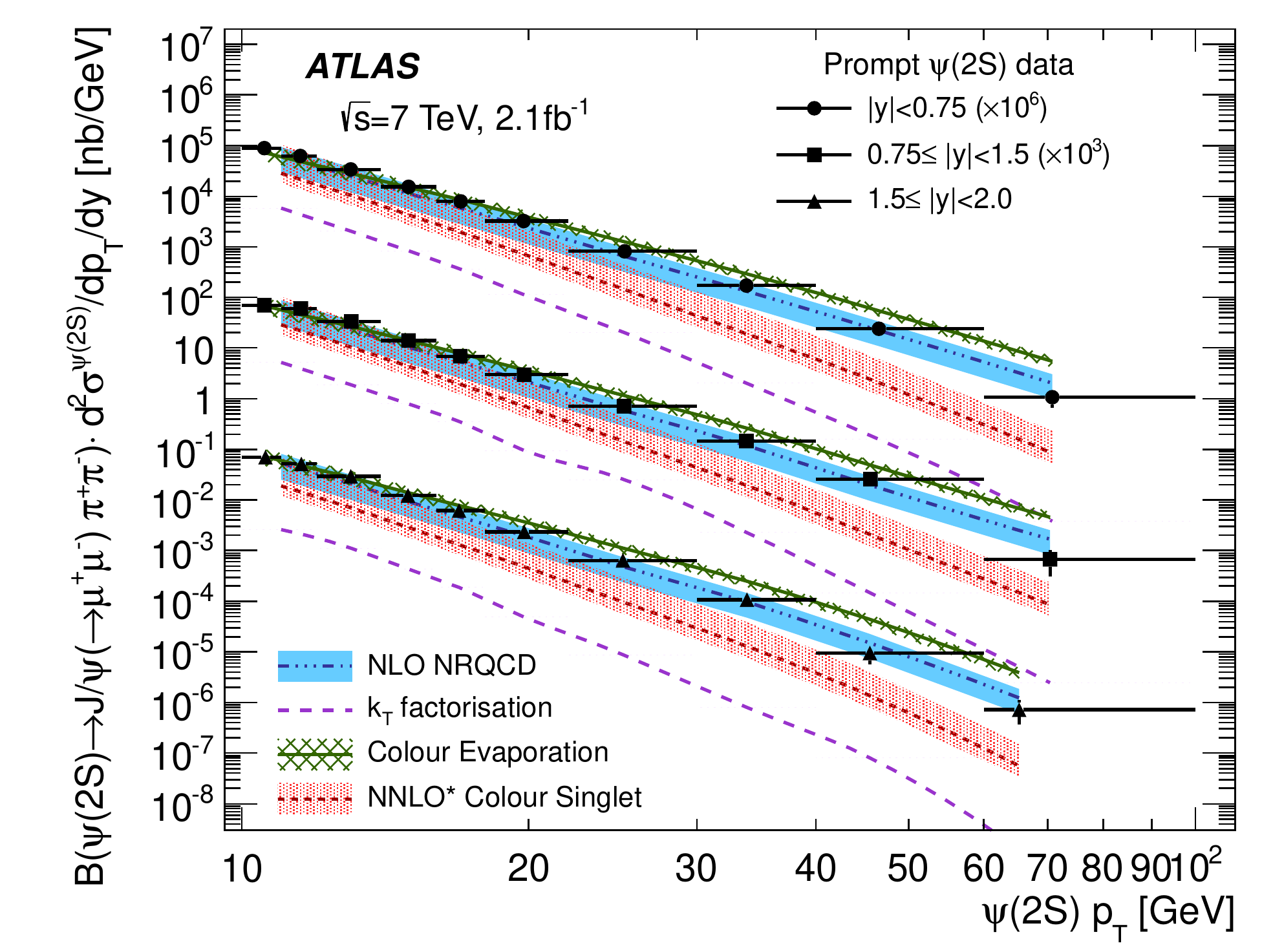} 
        } 
\subfigure[Prompt X(3872) compilation and theory]{ 
        \label{fig:pp:X3872} 
        \includegraphics[height=0.225\columnwidth]{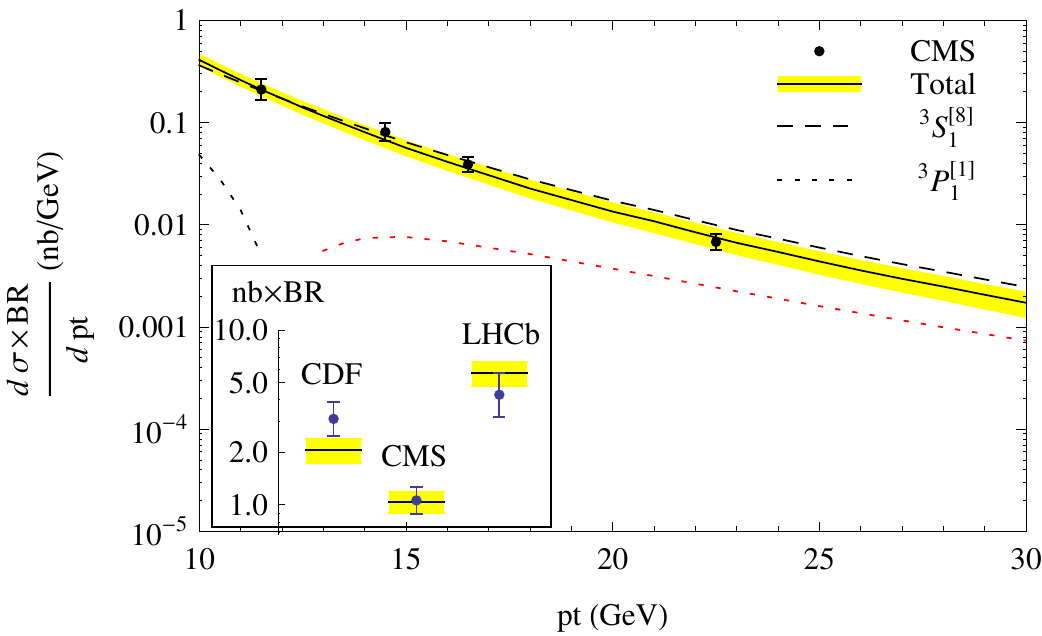} 
        } 
\subfigure[Theory and LHCb, $\etac$, \s=8\TeV]{ 
        \label{fig:pp:ppbardecay} 
        \includegraphics[height=0.225\columnwidth]{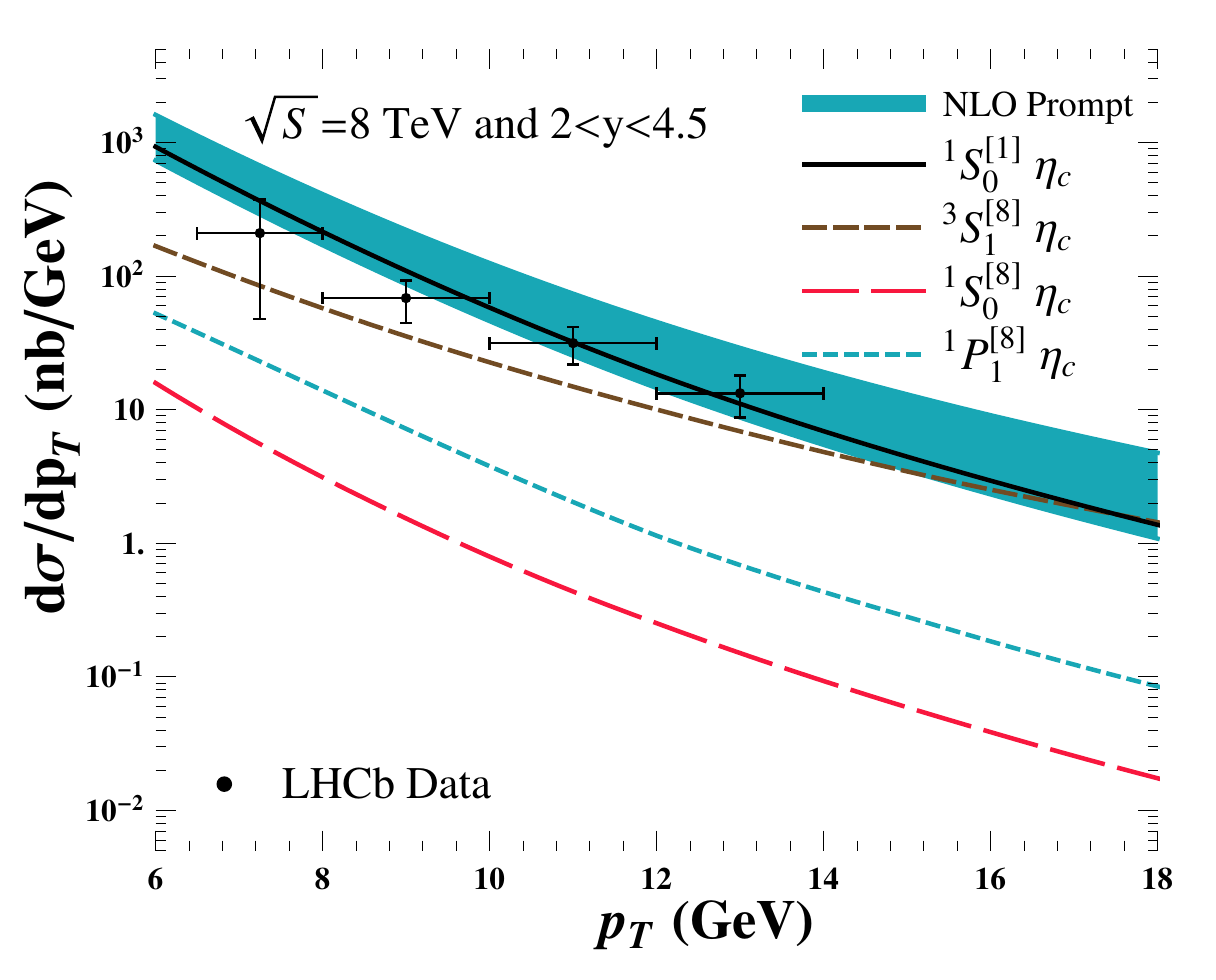} 
        } 
\caption{ 
        \label{fig:pp:OniaRare} 
        (a): ATLAS $\psiP$ differential cross section~\cite{Aad:2014fpa} compared to different theoretical curves. 
        (b): prompt X(3872) production cross section measured by the CDF~\cite{Acosta:2003zx,Bauer:2004bc}, CMS~\cite{Chatrchyan:2013cld}, and LHCb~\cite{Aaij:2011sn} Collaborations compared with NLO NRQCD allowing the CS contribution 
to differ from that from HQSS~\cite{Butenschoen:2013pxa}. 
        (c): Prompt-$\etac$ transverse-momentum cross section in \pp collisions at \s = 8\TeV measured by LHCb~\cite{Aaij:2014bga} 
compared to the CS contribution following HQSS and fitted CO contributions at NLO~\cite{Han:2014jya}. 
} 
\end{center} 
\end{figure} 
 
For excited states, there is an interesting alternative to the sole dilepton channel, 
namely $\jpsi+\pi\pi$. This is particularly relevant since more than 50\% of 
the $\psiP$ decay in this channel. The decay chain $\psiP \to \jpsi+\pi\pi \to \mumu +\pi\pi$ 
is four times more likely than  $\psiP \to  \mumu$. The final state 
$\jpsi+\pi\pi$ is also the one via which the $X(3872)$ was first seen at \pp 
colliders~\cite{Acosta:2003zx,Abazov:2004kp}. ATLAS 
released~\cite{Aad:2014fpa} the most precise study to date of $\psiP$ production up to 
\pt of 70\GeV at \s = 7\TeV, precisely in this channel. The measured differential cross section is 
shown for three rapidity intervals in~\fig{fig:pp:ATLAS:Psi2S} with four theoretical predictions. Along 
the same lines, the CDF, CMS and LHCb Collaborations measured 
the prompt $X(3872)$ yields at different values of \pt (see \fig{fig:pp:X3872}). In the NRQCD framework, these 
measurements tend to contradict~\cite{Butenschoen:2013pxa} a possible assignment of the $X(3872)$ as a 
radially excited $P$-wave state above the open-charm threshold. Such a statement should, however, be 
considered with care owing the recurrent issues in understanding prompt quarkonium production. 
In addition, LHCb determined the $X(3872)$ quantum numbers to be $J^{PC}=1^{++}$, excluding explanation of the $X(3872)$ as a conventional $\eta_{c2}(1^1 D_2)$ state~\cite{Aaij:2013zoa}. 
A brief survey of the new charmonium states above he $D\bar{D}$ threshold and their interpretation can be found in Ref.~\cite{Agashe:2014kda}.
 
Ultimately the best channel to look at all $n=1$ charmonium yields at once is that of 
baryon-antibaryon decay. Indeed, all $n=1$ charmonia can decay in this channel with a similar branching ratio, 
which is small, \ie~on the order of $10^{-3}$. LHCb is a pioneer in such a study with 
the first measurement of $\jpsi$ into \ppbar, made along that of the $\eta_c$. The latter case is the first measurement of the inclusive production of the charmonium ground state. 
It indubitably opens a new era in the study of quarkonia at colliders. The resulting cross section is 
shown in~\fig{fig:pp:ppbardecay} and was shown to bring about constraints~\cite{Butenschoen:2014dra,Han:2014jya,Zhang:2014ybe} on the existing  global fits of NRQCD LDMEs by virtue 
of heavy-quark spin symmetry (HQSS) which is an essential property of NRQCD. As for now, it seems that the CS 
contributions to $\eta_c$ are large --if not dominant-- in the region covered by the LHCb data and the different CO have to cancel each others  not to overshoot the measured yield.

The canonical channel used to study $\chi_{c1,2}$ production at hadron colliders corresponds to the studies involving $P$ waves decaying into $\jpsi$ and a photon. Very recently the measurement of $\chi_{c0}$ relative yield was performed by LHCb 
\cite{Aaij:2013dja} despite the very small branching ratio $\chi_{c0} \to \jpsi +\gamma$ of the order of one percent, that is 30 (20) times smaller 
than that of $\chi_{c1}$ ($\chi_{c2}$). LHCb found out that $\sigma(\chi_{c0})/\sigma(\chi_{c2})$ is compatible with unity for $\pt > $4\GeVc, in striking contradiction with statistical counting, 1/5. 
 
Currently, the experimental studies are focusing on the ratio of the $\chi_{cJ}$ yields 
which are expected to be less sensitive to the photon acceptance determination. They 
bring about constraints on production mechanism but much less than the absolute 
cross section measurements which can also be converted into the fraction of $\jpsi$ 
from $\chi_{cJ}$. This was the first measurement of this 
fraction at the Tevatron by CDF in 1997~\cite{Abe:1997yz} which confirmed that our 
understanding of quarkonium production at colliders was incorrect (for reviews 
see \eg~\cite{Kramer:2001hh,Lansberg:2006dh}). It showed that the $\jpsi$   
yield at Tevatron energies was mostly from {\it direct} $\jpsi$ and {\it not} from 
$\chi_{cJ}$ decays. The latter fraction was found out to be at most 30\%. Similar 
information are also fundamental to use charmonia as probes of QGP, especially for 
the interpretation of their possible sequential suppression. It is also very important 
to understand the evolution of such a fraction as function of $\s$, $y$ and $\pt$. 
 
\begin{figure}[!t] 
\begin{center} 

\subfigure[Typical source of prompt \jpsi and low and mid \pt]{ 
        \label{fig:pp:FeedDownFraction-psi} 
\includegraphics[width=0.25\columnwidth]{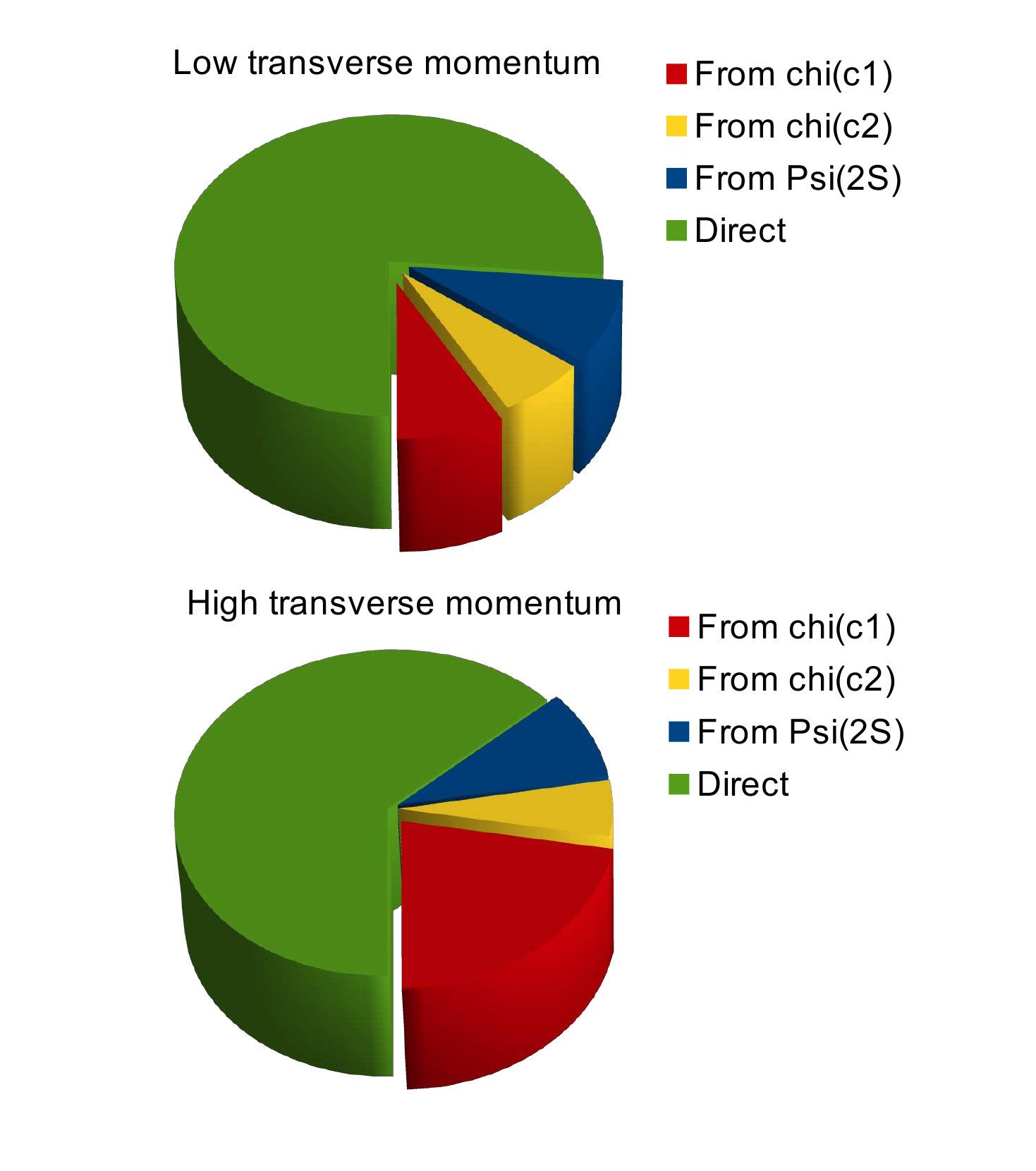}
        } 
\subfigure[$\chi_{c2}/\chi_{c1}$ compilation, \s = 7\TeV]{ 
        \label{fig:pp:ChiCRatio} 
        \includegraphics[width=0.35\columnwidth]{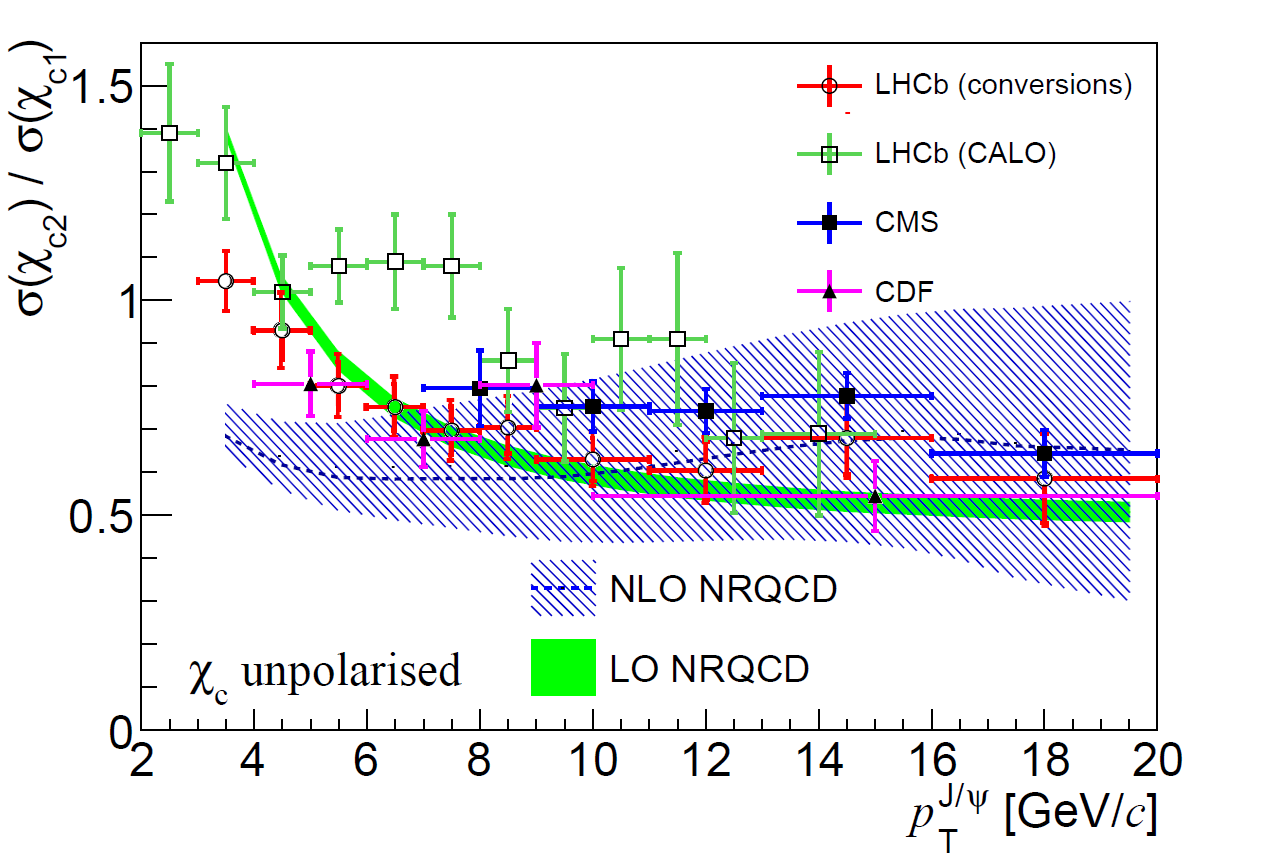} 
        } 
\subfigure[ATLAS, $\chi_{c1}$ prompt cross section, \s = 7\TeV]{ 
        \label{fig:pp:PromptOnia:Chic1} 
       \includegraphics[width=0.35\columnwidth]{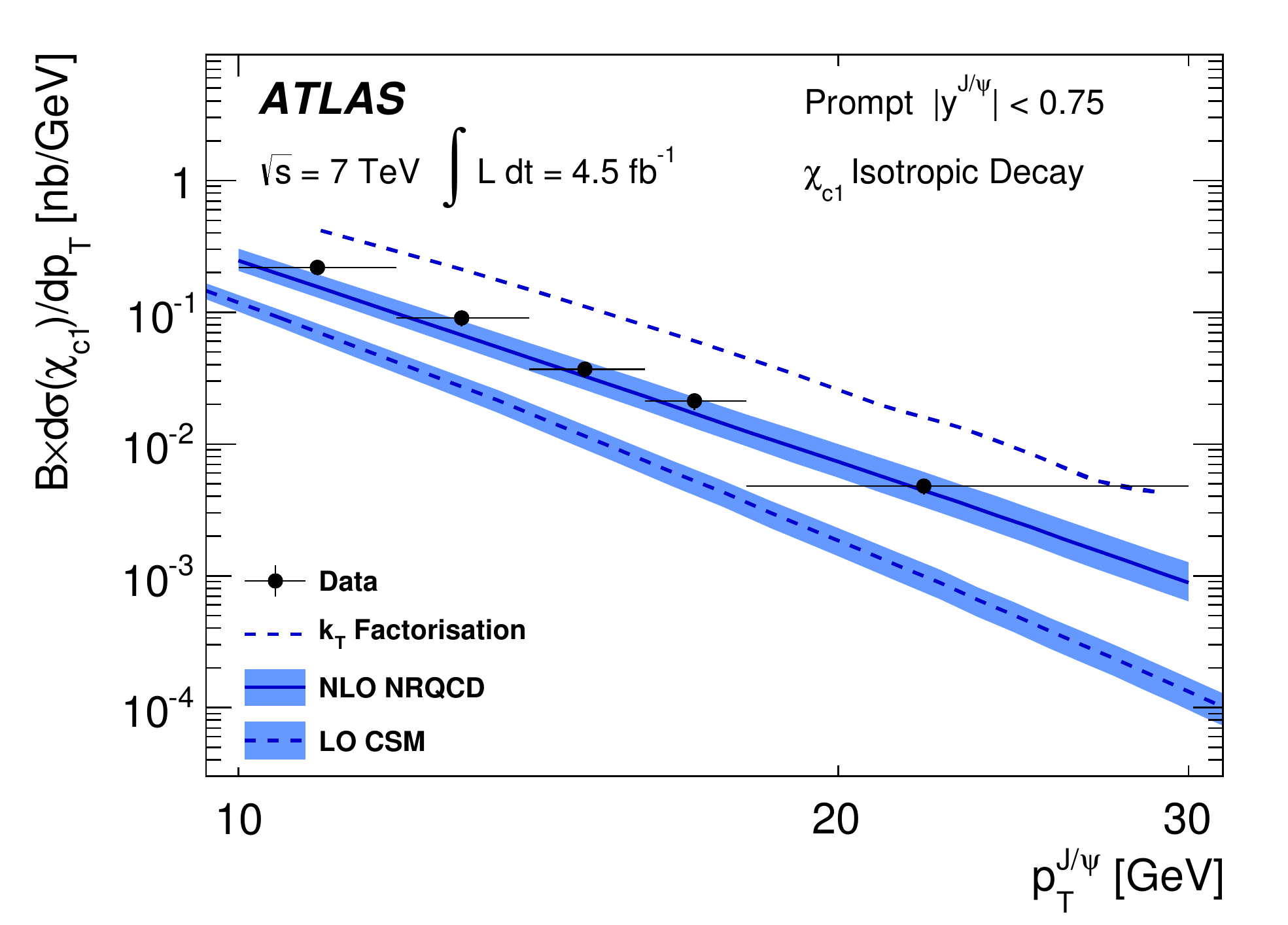} 
        } 
 
\caption{ 
        \label{fig:pp:ChiC} 
        (a): Typical source of prompt \jpsi and low and mid \pt. 
        (b): Ratio of $\chi_{{\rm c}1}$ to $\chi_{{\rm c}2}$ as measured by the LHCb experiment in \pp~collisions at \s = 7\TeV compared to results from other experiments~\cite{Aaij:2013dja,Chatrchyan:2012ub,Abulencia:2007bra} and NRQCD calculations~\cite{Ma:2010vd,Likhoded:2013aya}. 
        (c): $\chi_{{\rm c}1}$ $\dsdpt$~\cite{ATLAS:2014ala} as compared to LO CSM\protect\footnotemark , NLO NRQCD~\cite{Ma:2010vd,Shao:2012iz,Ma:2010yw} and \kt factorization~\cite{Baranov:2011ib,Baranov:2010zz}. 
} 
\end{center} 
\end{figure}

\fig{fig:pp:FeedDownFraction-psi} shows the typical size of the feed-down fraction of the 
$\chi_c$ and $\psiP$ into \jpsi at low and mid \pt, which are different. One 
should therefore expect differences in these fraction between \pt-integrated yields and yields 
measured at $\pt = 10\GeV/c$ and above. \fig{fig:pp:ChiCRatio} shows the ratio of the  $\chi_{c2}$ 
over  $\chi_{c1}$ yields as measured\footnote{ 
The present ratio depends on the polarisation of 
the $\chi_c$ since it induces different acceptance correction.} at the LHC by LHCb, CMS and at 
the Tevatron by CDF. On the experimental side, the usage of the conversion method to detect the photon becomes an advantage. LHCb is able to carry out measurements down to \pt as small as 2\GeVc, where the ratio seems to strongly increase. 
This increase is in line with the Landau-Yang theorem according to which $\chi_{c1}$ production from 
collinear and on-shell gluons at LO is forbidden. Such an increase appears in the LO NRQCD band, 
less in the NLO NRQCD one. At larger \pt, such a measurement helps to fix the value of the NRQCD 
LDMEs (see the pioneering study of Ma et al.~\cite{Ma:2010vd}). As we just discussed, once the 
photon reconstruction efficiencies and acceptance are known, one can derive the $\chi_c$ feed-down 
fractions which are of paramount importance to interpret inclusive \jpsi results. One can 
of course also derive absolute cross section measurements which are interesting to understand 
 the production mechanism of the $P$-wave quarkonia {\it per se}; these may not be 
the same as that of $S$-wave quarkonia. \fig{fig:pp:PromptOnia:Chic1} shows the \pt dependence of the 
yield of the $\chi_{c1}$ measured by ATLAS (under the hypothesis of an isotropic decay), which is 
compared to predictions from the LO CSM\footnote{ 
For the $P$ wave case, the distinction between 
color singlet and color octet transition is not as clear that for the $S$ wave. In particular the 
separation between CS and CO contribution depends on the NRQCD factorisation scale $\mu_\Lambda$.}, NLO 
NRQCD and \kt factorisation. The NLO NRQCD predictions, whose parameters have been fitted to 
reproduce the Tevatron measurement, is in good agreement with the data. 
Similar cross sections have been measured  for the $\chi_{c2}$.

\footnotetext{ 
as in encoded in {\sc ChiGen}: \url{https://superchic.hepforge.org/chigen.html}.}

%
%
%
%
 
\subsubsection{Bottomonium}

The study of bottomonium production at LHC energies offers some advantages. First, 
there is no beauty feed-down. Second, owing to their larger 
masses, their decay products --usually leptons-- are more energetic and more easily 
detectable (detector acceptance, trigger bandwidth, ...). Third, the existence 
of three sets of bottomonia with their principal quantum number $n=1,2,3$ below the open-beauty 
threshold offers a wider variety of states that can be detected in the dilepton decay channel 
 -- this however introduces a complicated feed-down pattern which 
we discuss later on. Fourth, at such high energies, their production rates with respect to those of charmonia 
 are not necessarily much lower. It was for instance noticed~\cite{Gong:2012ah} 
that, for their production in association with a Z boson, the cross sections are similar. 
\begin{figure}[!ht] 
\begin{center} 
\subfigure[LO CSM compared to the extrapolated direct CMS and LHCb \upsa ${\rm d}\sigma/{\rm d}y$, \s = 7\TeV]{ 
        \label{fig:pp:CMSLHCb:Ups} 
       \includegraphics[height=0.3\columnwidth]{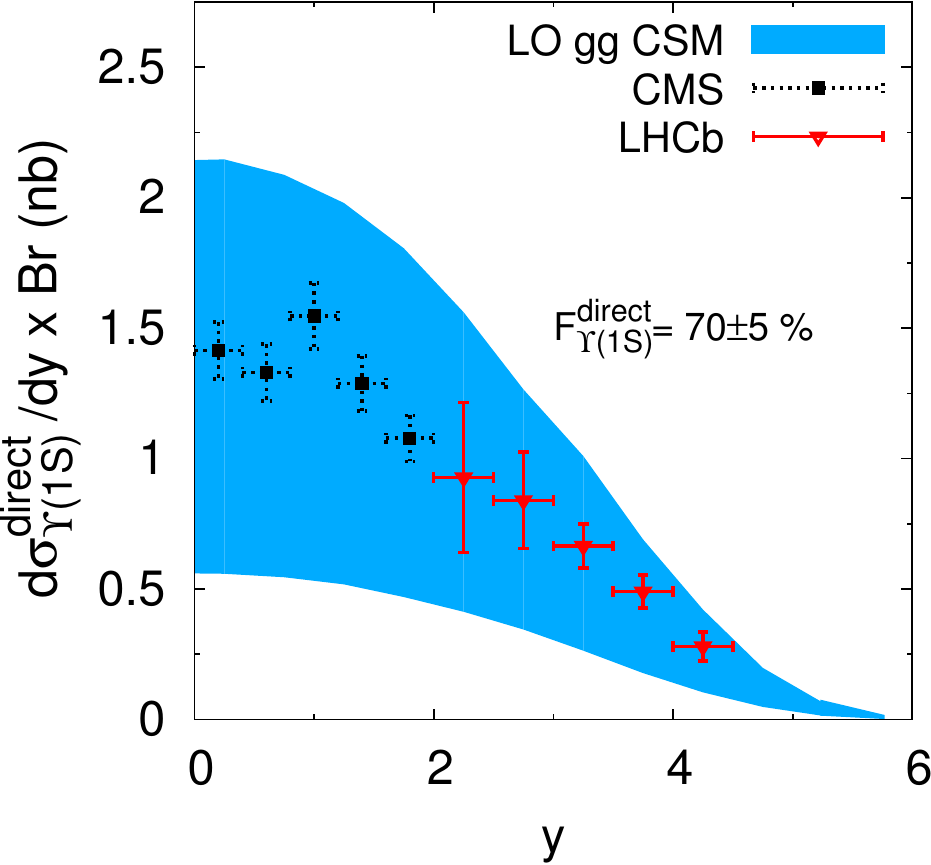} 
        } 
\subfigure[Theory and CMS, \upsa, \s = 7\TeV]{ 
        \label{fig:pp:CMS:Ups1} 
        \includegraphics[height=0.3\columnwidth]{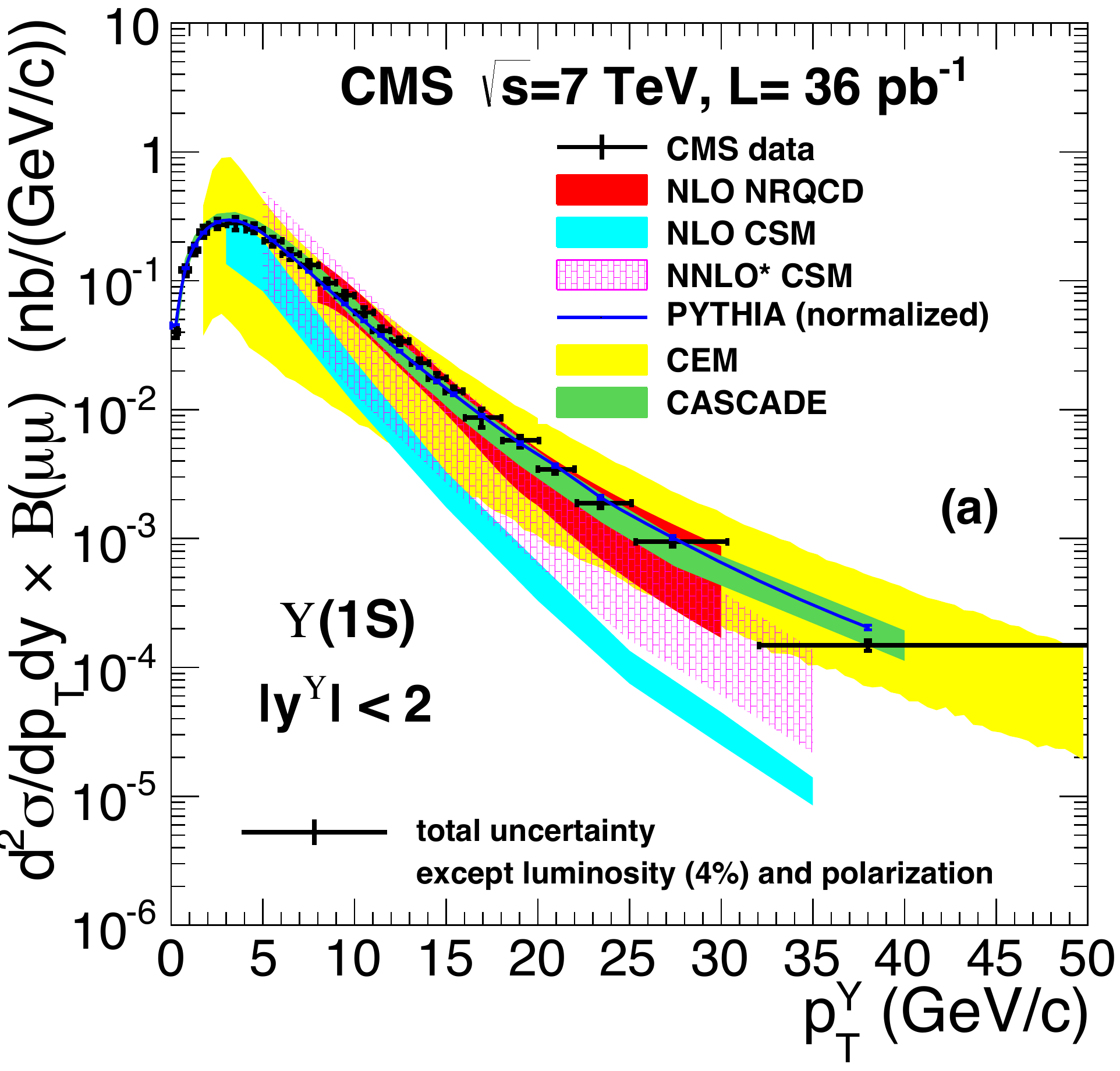} 
        } 
\subfigure[ATLAS and CMS, ratio of the \ups states, \s = 7\TeV]{ 
        \label{fig:pp:CMS:Ups} 
        \includegraphics[height=0.3\columnwidth]{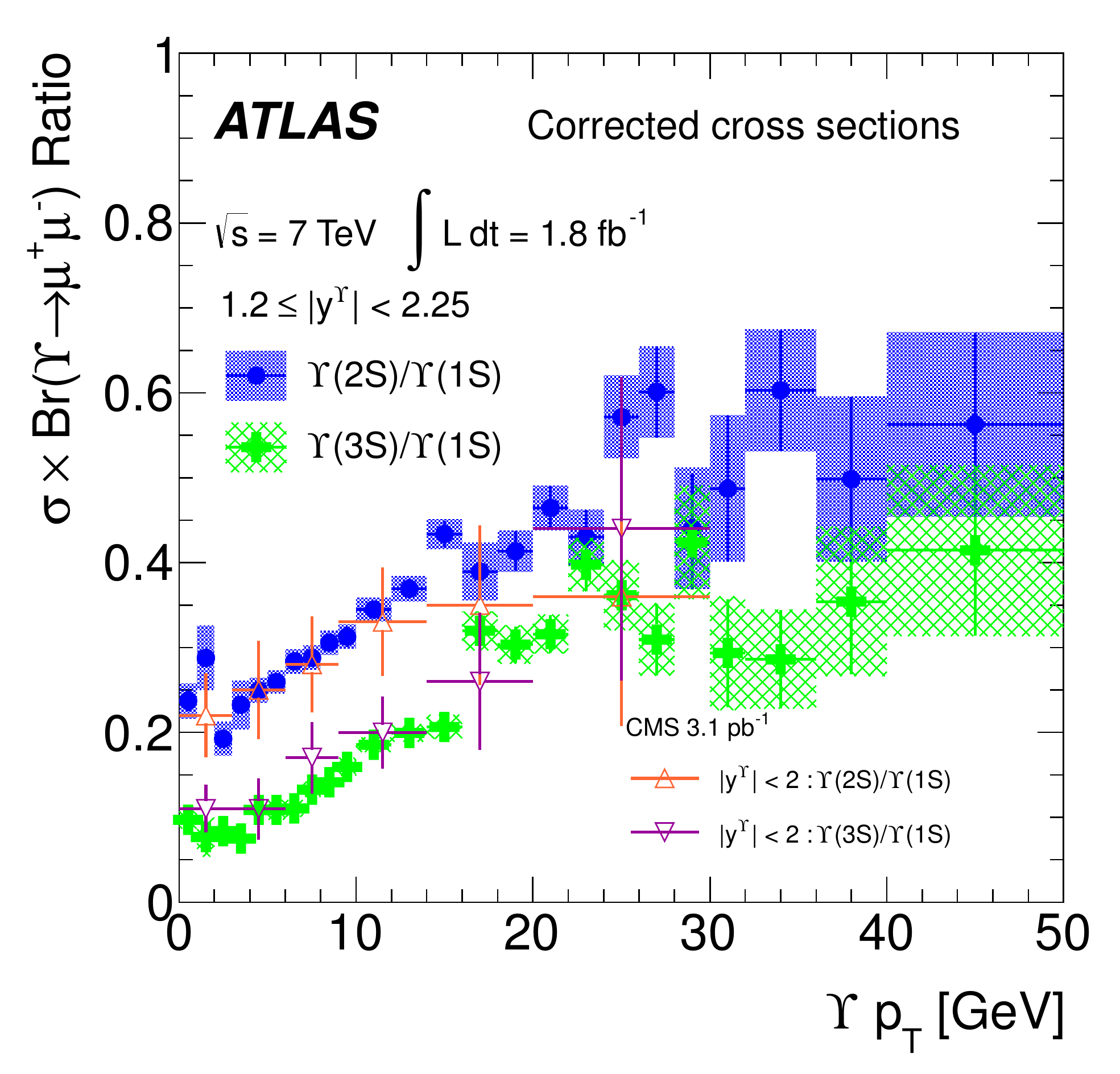} 
        } 
\caption{ 
        \label{fig:pp:Ups} 
        (a): \upsa rapidity differential cross section as measured by ATLAS, CMS and LHCb~\cite{Aad:2012dlq,Chatrchyan:2013yna}. 
        (b) Transverse momentum dependence of the \upsa states as measured by CMS~\cite{Chatrchyan:2013yna}. 
        (c) Transverse momentum dependence of the \ups states ratio as measured by ATLAS~\cite{Aad:2012dlq}. 
} 
\end{center} 
\end{figure}

\fig{fig:pp:CMSLHCb:Ups} shows the rapidity dependence of the \upsa yield from two complementary measurements, 
one at forward rapidities by LHCb and the other at central rapidities by CMS (multiplied by the expected fraction of direct \upsa as discussed below). 
These data are in line with the CS expectations; at least, they do not show an evident need for CO contributions, nor they exclude their presence. 
As for the charmonia, the understanding of their production mechanism for mid and high \pt is a challenge. 
\fig{fig:pp:CMS:Ups1} shows a typical comparison with five theory bands.  
In general, LHC data are much more precise than theory. 
 It is not clear that pushing the measurement to higher \pt would provide striking evidences 
in favour of one or another mechanism -- associated-production channels, which we discuss in~\sect{sec:pp:NewObs}, are probably 
more promising. \fig{fig:pp:CMS:Ups} shows ratios of different $S$-wave bottomonium yields. These are clearly not constant as one might anticipate following the idea of the CEM.  Simple mass effects through feed-down decays can induce an increase of these ratios ~\cite{Lansberg:2012ta,Abelev:2014qha}, but these are likely not sufficient 
to explain the observed trend if all the {\it direct} yields have the same \pt dependence. The $\chi_b$ feed-down, which we discuss in the following, 
can also affect these ratios.

\begin{figure}[!t] 
\begin{center} 
\subfigure[Ratio of $\chib$ states, \s = 8\TeV]{ 
        \label{fig:pp:ChibRatio} 
        \includegraphics[height=0.25\columnwidth]{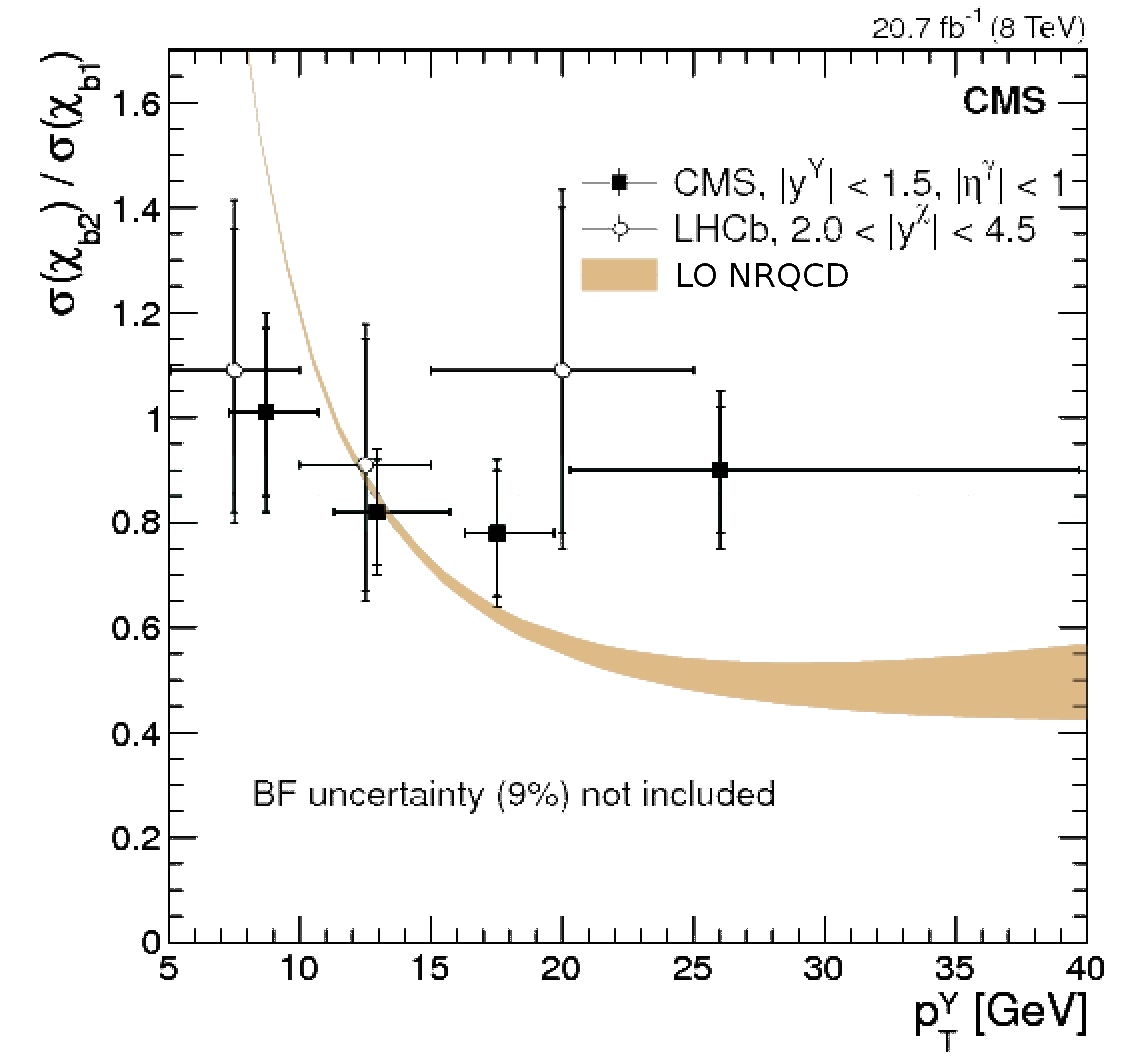} 
        } 
\subfigure[Fraction  of feed-down of $\chib(1,2,3P)$ to \ups, \s = 7 and 8\TeV]{ 
        \label{fig:pp:ChibtoUps1S} 
         \includegraphics[height=0.25\columnwidth]{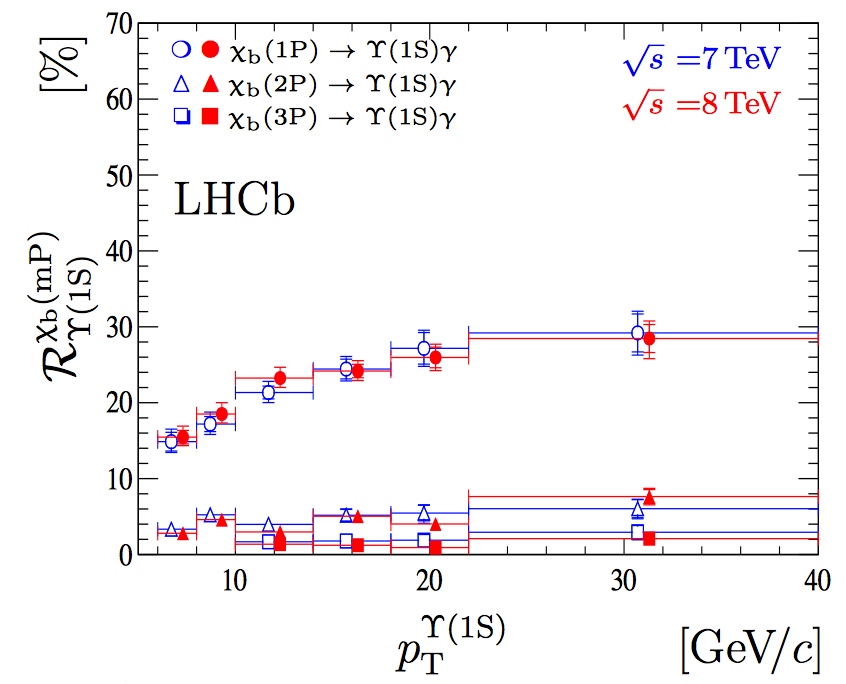} 
        } 
\subfigure[Fraction  of feed-down of $\chib(3P)$ to \upsc, \s = 7 and 8\TeV]{ 
        \label{fig:pp:ChibtoUps3S} 
       \includegraphics[height=0.25\columnwidth]{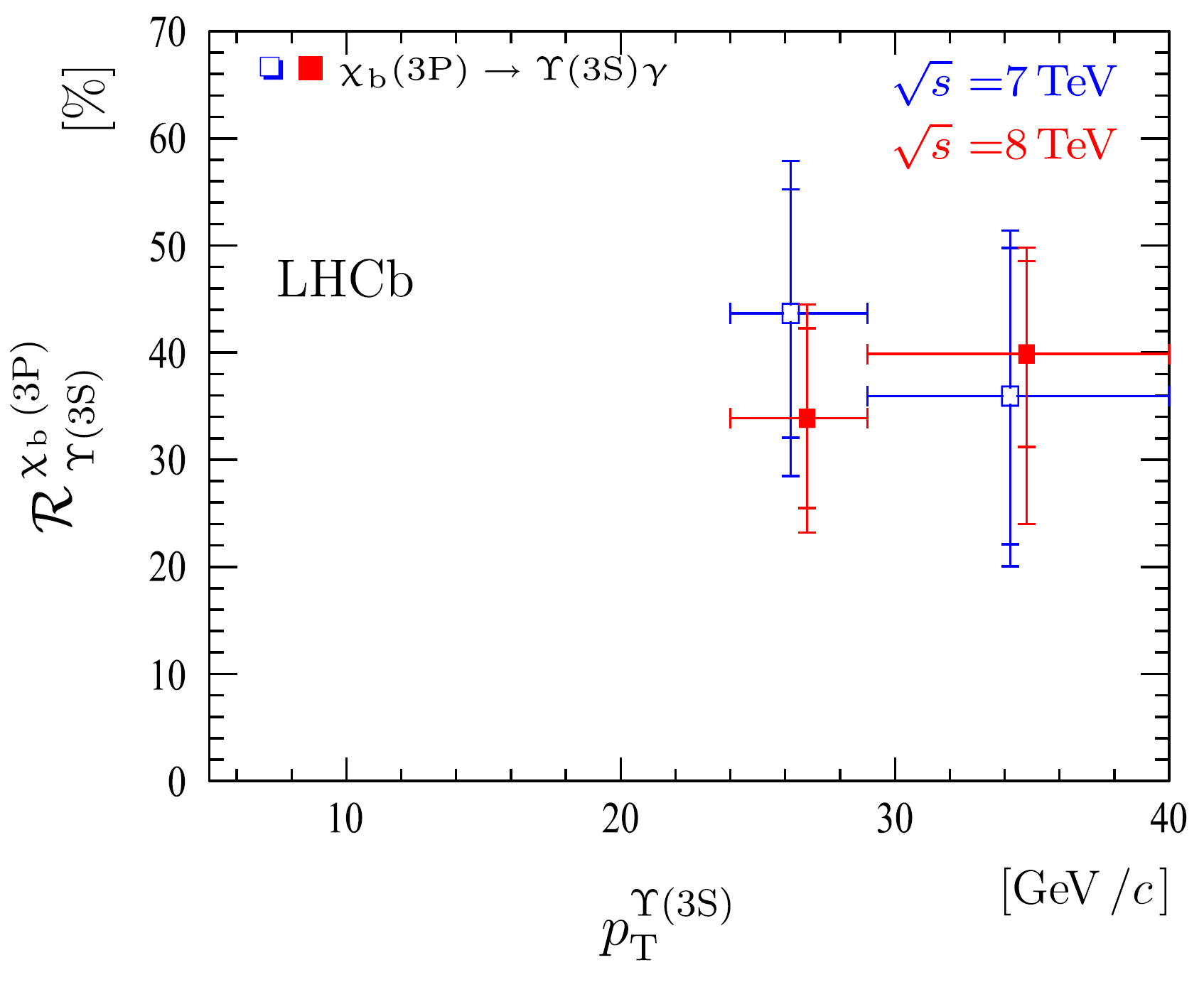} 
        } 
\caption{ 
        \label{fig:pp:Chib} 
(a) Ratio of the production cross section of $\chi_{b2}$ and $\chi_{b1}$ in \pp collisions at \s = 8\TeV~\cite{Khachatryan:2014ofa,Aaij:2014hla}. 
(b) and (c) : Fractions of $\chib$ to \upsa as function of \ups $\pt$~\cite{Aaij:2014caa}. For better visualization the data points are slightly displaced from the bin centres. The inner error bars represent statistical uncertainties, while the outer error bars indicate statistical and systematic uncertainties added in quadrature. 
} 
\end{center} 
\end{figure}

Since the discovery of the $\chi_b(3P)$ by ATLAS~\cite{Aad:2011ih}, we know that the three $n=1,2,3$ families 
likely completely lie under the open-beauty threshold. 
This means, for instance, that we should not only care about $mS\to nS$ and $nP\to nS+\gamma$ transitions but also of  $mP\to nS+\gamma$ ones. 
Obviously, the $n=1$ family is the better known of the three. \fig{fig:pp:ChibRatio} shows the ratio of the production cross section of $\chi_{b2}(1P)$ over that of $\chi_{b1}(1P)$ measured by CMS and LHCb. Although the experimental 
uncertainties are significant, one does not observe the same trend as the LO NRQCD, \ie\ an increase at low \pt\ due to the Landau-Yang theorem. 
Besides, the ratio is close to unity which also seems to be in contradiction to the simple spin-state counting. 
 
\begin{figure}[!htbp] 
\begin{center} 
\subfigure[\upsa]{ 
        \label{fig:pp:FeedDownFraction-upsi1s} 
        \includegraphics[height=0.31\columnwidth]{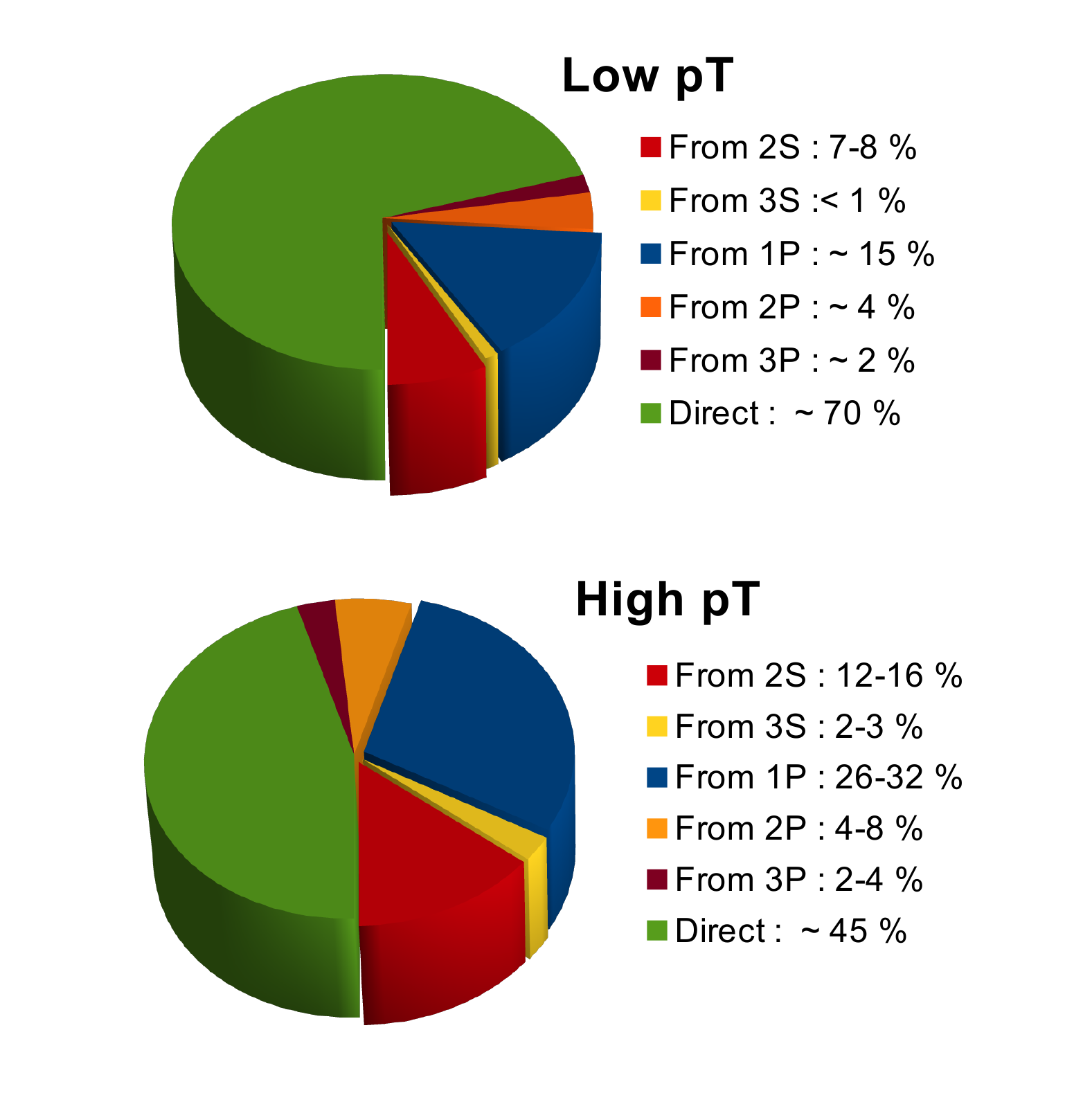} 
        } 
\subfigure[\upsb]{ 
        \label{fig:pp:FeedDownFraction-upsi2s} 
         \includegraphics[height=0.31\columnwidth]{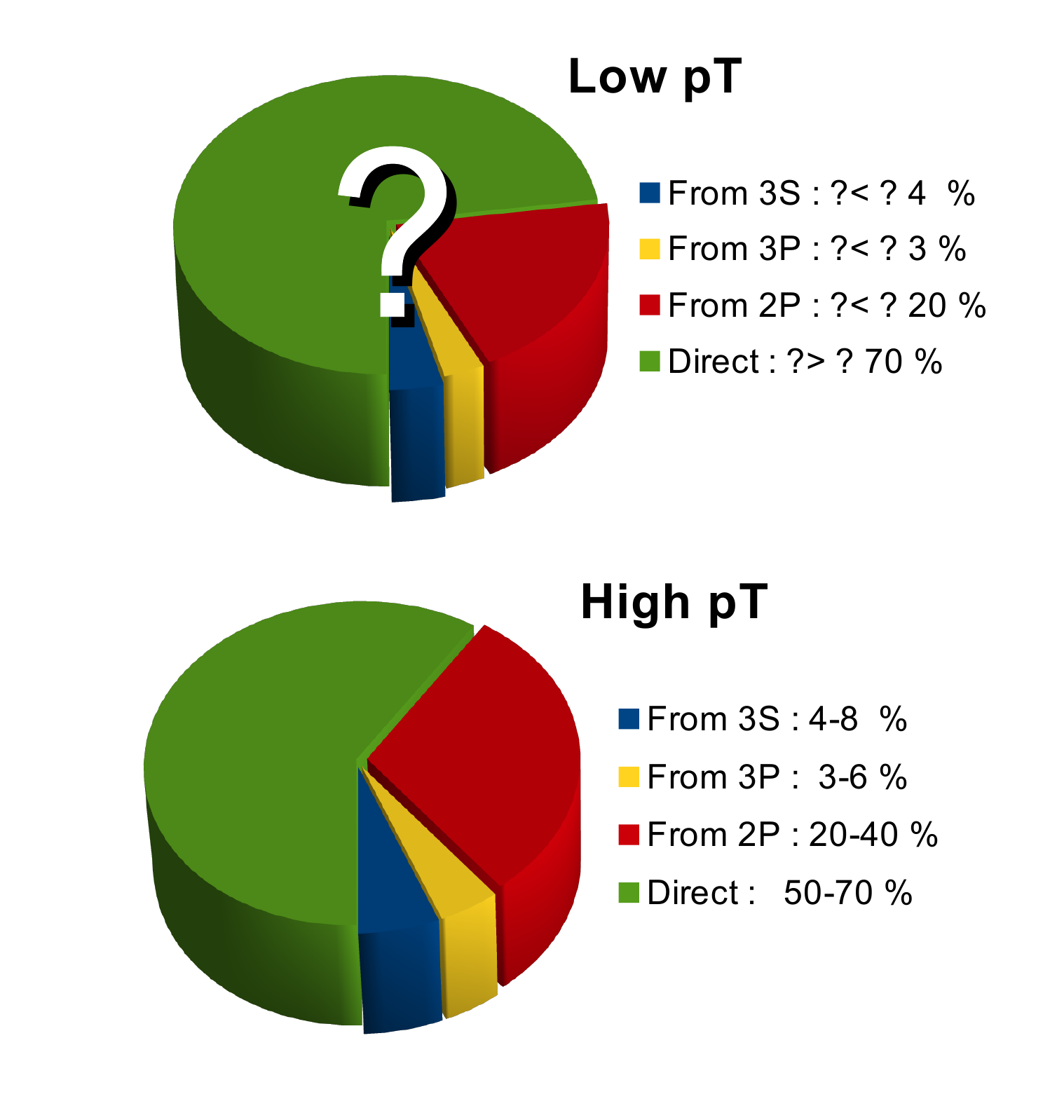} 
        } 
\subfigure[\upsc]{ 
        \label{fig:pp:FeedDownFraction-upsi3s} 
       \includegraphics[height=0.31\columnwidth]{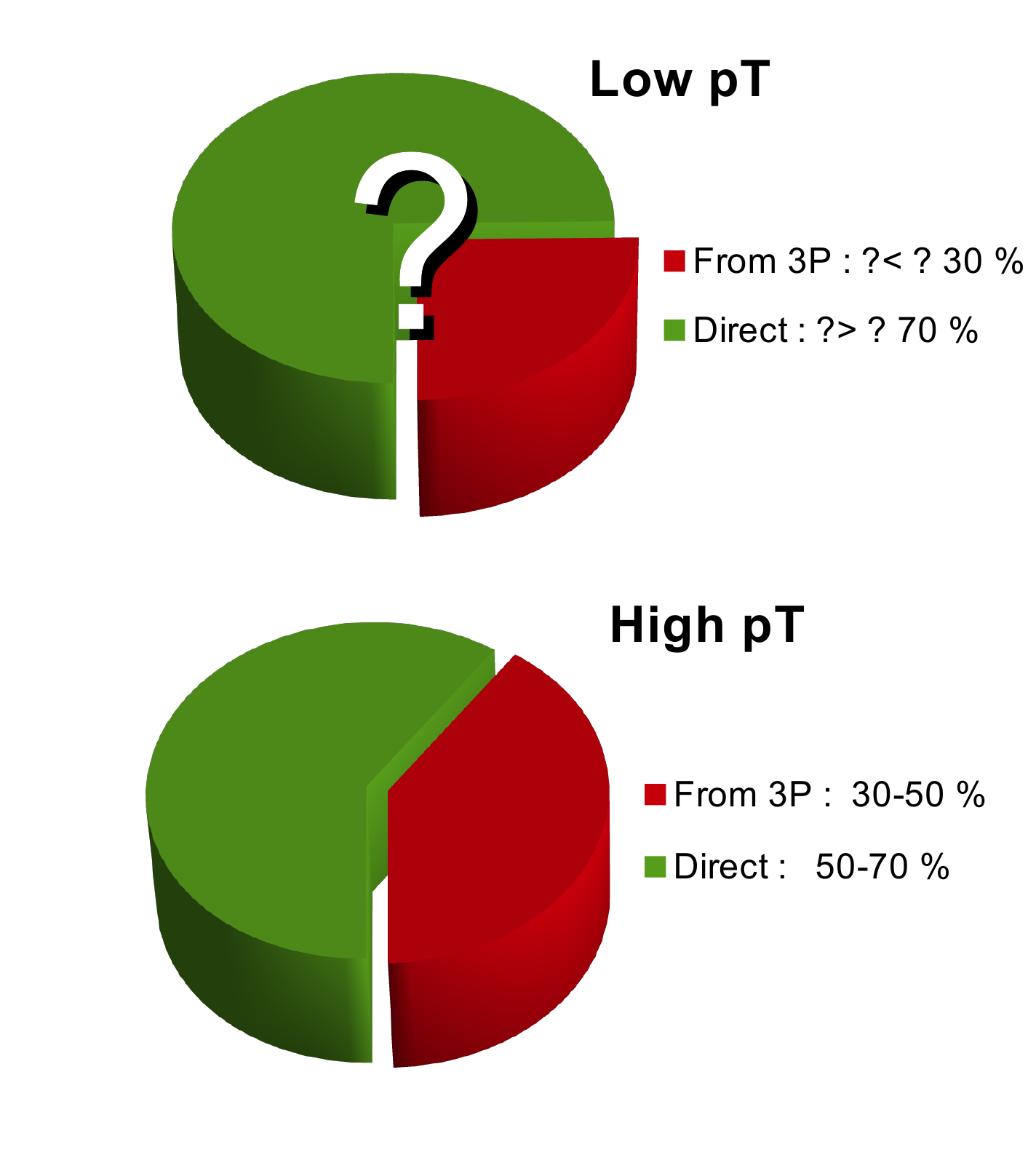} 
        } 
\caption{ 
        \label{fig:pp:FeedDownFraction-upsi} 
Typical sources of $\Upsilon(nS)$ at low and high \pt. These numbers are mostly derived from 
LHC measurements~\cite{Aaij:2014caa,Aad:2012dlq,Aad:2011xv, Chatrchyan:2013yna,Abelev:2014qha,Khachatryan:2010zg,Aaij:2013yaa,Aaij:2012se,LHCb:2012aa} assuming  an absence of a significant rapidity dependence. 
} 
\end{center} 
\end{figure}

Recently, LHCb performed a thorough analysis~\cite{Aaij:2014caa} of all the possible $mP\to nS+\gamma$ 
transitions in the bottomonium system. 
These new measurements along with the precise measurements of \upsb and \upsc \pt-differential cross section show that the feed-down structure is quite different than that commonly accepted 
ten years ago based on the CDF measurement~\cite{Affolder:1999wm}. The latter, made for 
$\pt >$ 8\GeVc~\cite{Affolder:1999wm}, suggested that the  $\chi(nP)\to \Upsilon(1S) + \gamma$ feed-down 
could be as large as 40\% (without excluding values of the order of 25\%) and that only 50\% of the 
\upsa were direct. Based on the LHC results, one should rather say that, at low \pt, where heavy-ion measurements are 
mostly carried out, 70\%  of the \upsa are direct; the second largest source is from $\chi_b(1P)$ -- approximately 
two thirds from $\chi_{b1}(1P)$ and one third from $\chi_{b2}(1P)$ \cite{Khachatryan:2014ofa,Aaij:2014hla}. At larger \pt (above 20\GeVc, 
say), the current picture is similar to the old one, \ie~ less than half of the \upsa are direct and each of 
the feed-down is nearly doubled. For the \upsb, there is no  $\chi_{b}(2P)\to \Upsilon(2S) + \gamma$ measurement at \pt lower than 20\GeVc. 
Above, it is measured to be about 30\% with an uncertainty of 10\%. The feed-down from $\chi_{b}(3P)$ is slightly lower than from \upsc. Taken together they may account for 10 to 15\% of the \upsb yield. For the \upsc, the only existing 
measurement~\cite{Aaij:2014caa} is at large \pt and also shows (see \fig{fig:pp:ChibtoUps3S}) a feed-down fraction of 40\% with a 
significant uncertainty (up to 15\%). 
The situation is schematically summarised on \fig{fig:pp:FeedDownFraction-upsi}.

\begin{figure}[!t] 
\begin{center} 
\subfigure[$\pt$ dependence]{ 
        \label{fig:pp:LHCbBc:pt} 
       \includegraphics[height=0.35\columnwidth]{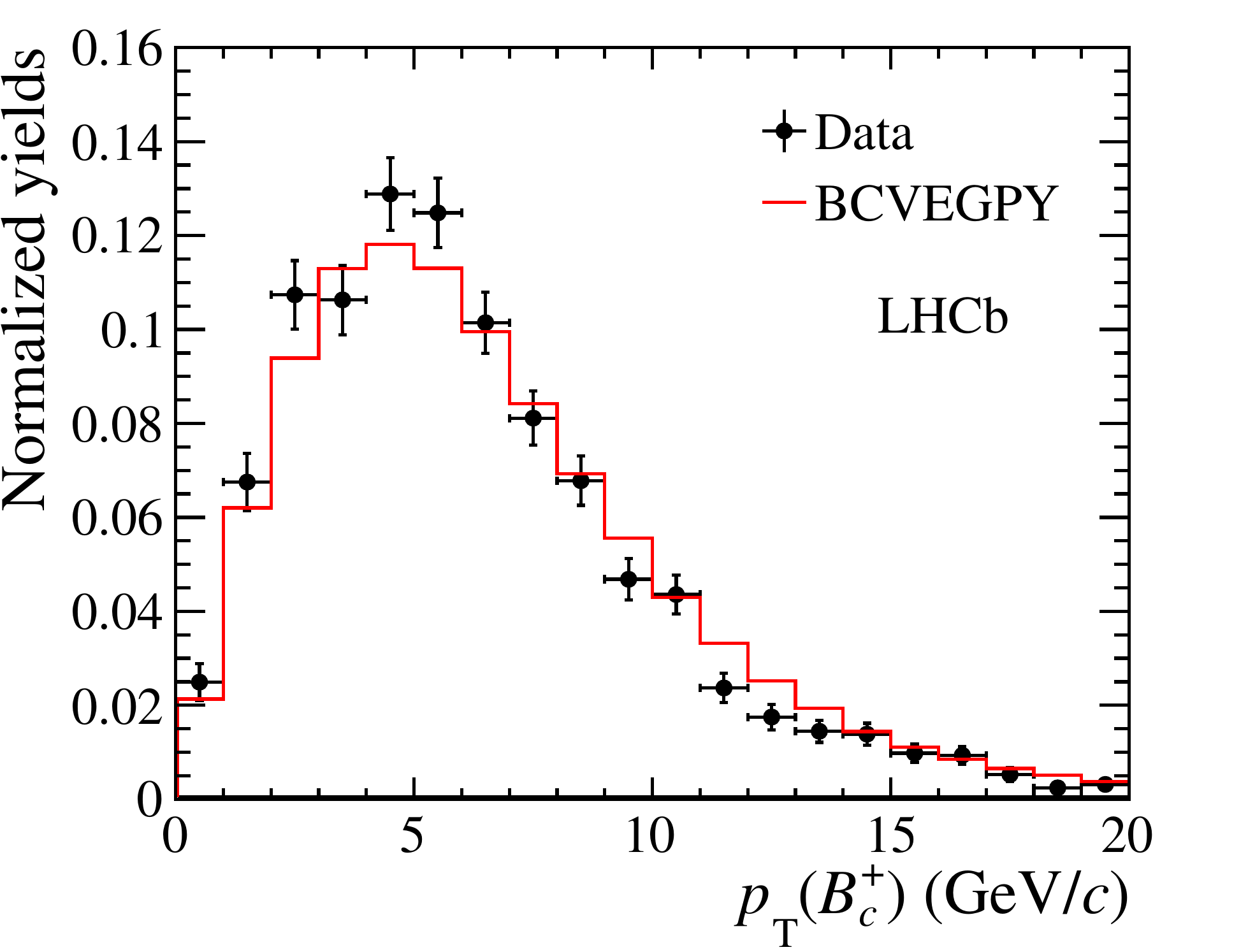} 
        } 
        \subfigure[$y$ dependence]{ 
        \label{fig:pp:LHCbBc:y} 
       \includegraphics[height=0.35\columnwidth]{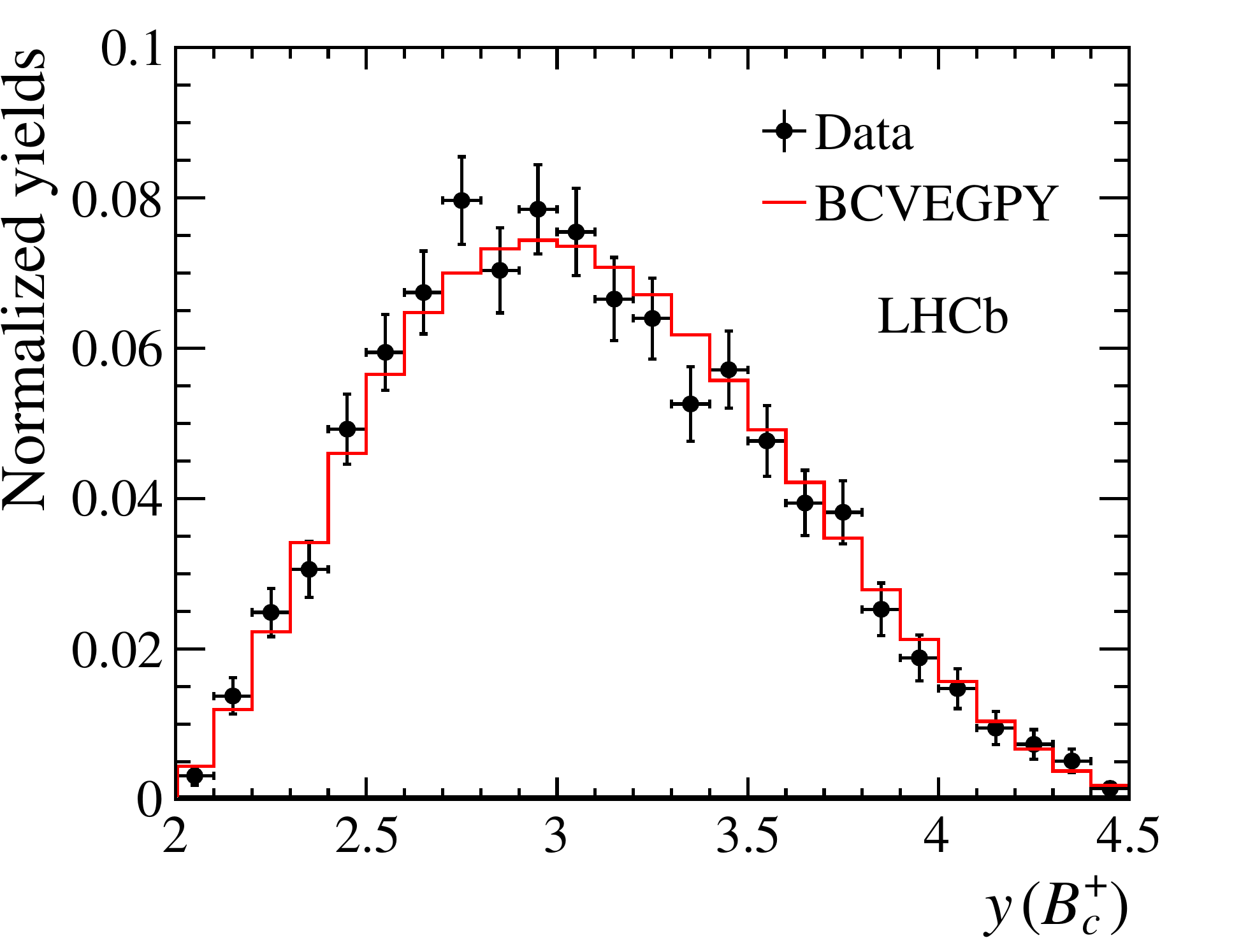} 
        } 
\caption{ 
        \label{fig:pp:LHCbBc} 
       ${\rm B}_c^+$ meson production in \pp collisions at \s = 8\TeV as measured by the LHCb  
Collaboration in its ${\rm B}_c^+ \rightarrow \jpsi \ \pi^+$ decay~\cite{Aaij:2014ija} within  
$0 < \pt < 20$\GeVc and $2.0 < y < 4.5$.        The solid histogram is a theory evaluation based on the 
 complete order-$\alpha_s^4$ calculation --as opposed to fragmentation-function-based computations--, 
implemented in the ${\rm B}_c$ generator BCVEGPY~\cite{Chang:2003cq,Chang:2005hq}.  
} 
\end{center} 
\end{figure}

%
%
%
%
\subsubsection{${\rm B}_c$ and multiple-charm baryons}

After a discovery phase during which the measurement of the mass and the lifetime 
of the ${\rm B}_c$ was the priority, the first measurement of the \pt and $y$ spectra of  
promptly produced ${\rm B}_c^+$ was carried out by the LHCb Collaboration~\cite{Aaij:2014ija}. Unfortunately, as for now, the branching 
${\rm B}_c^+ \to \jpsi \ \pi^+$ is not yet known. This precludes the extraction of $\sigma_{{\rm pp}\to {\rm B}_c^+ +X}$ and the comparison with the existing  
theoretical predictions~\cite{Chang:1992jb,Chang:1994aw,Chang:1996jt,Chang:2005bf,Kolodziej:1995nv,Berezhnoy:1994ba,Berezhnoy:1996an,Baranov:1997wy}. 
Aside from this normalisation issue, the  \pt and $y$ spectra are well reproduced by the theory (see a 
comparison in \fig{fig:pp:LHCbBc} with BCVEGPY~\cite{Chang:2003cq,Chang:2005hq}, which is based on NRQCD  
where the CS contribution is dominant).

%
%
%
%
 
 
Searches for doubly-charmed baryons are being carried out (see \eg~\cite{Aaij:2013voa}) on the existing data sample collected in \pp~collisions at 7 and 8\TeV. As for now, no analysis could confirm the signals seen by the fixed-target experiment SELEX at Fermilab~\cite{Mattson:2002vu,Ocherashvili:2004hi}.

\subsection{Quarkonium polarization studies}
\label{sec:pp:Polarization}

Measurements of quarkonium polarisation can shed more light on the long-standing puzzle  
of the quarkonium hadroproduction. Various models of the quarkonium production, described  
in the previous \sect{sec:pp:Theory:Onia}, are in reasonable agreement with the  
cross section measurements but they usually fail to describe the measured polarisation. 
 
We have collected in this section all results of polarisation measurements performed  
by different experiments at different colliding energies and in different kinematic regions. 
The results for \jpsi and \psiP can be found in \tab{tab:jpsiPolarization} and \tab{tab:psiPolarization} for \pp and \pA collisions. Since there is no known mechanism that  
would change quarkonium polarisation from proton-proton to proton-nucleus collisions, results from \pA  
collisions are also shown in this section. Tables~\ref{tab:ups1Polarization},~\ref{tab:ups2Polarization},~\ref{tab:ups3Polarization}  
gather the results for, respectively, the \upsa, \upsb and \upsc in \pp collisions.  
 
Polarisation of a vector quarkonium state is analysed experimentally via the angular distribution of the leptons  
from the quarkonium dilepton decay, that is parametrised by: 
\begin{equation} 
\frac{\dd^{2}N}{\dd(\cos\theta)\dd\phi} \propto 1+\lambda_\theta \cos^2\theta + \\ \lambda_\phi \sin^2\theta \cos2\phi + \lambda_{\theta\phi}\sin2\theta \cos\phi \,, 
\label{eq:angularDistribution} 
\end{equation} 
$\theta$ is the polar angle between the positive lepton in the quarkonium rest frame and the chosen  
polarisation axis and $\phi$ angle is the corresponding azimuthal angle defined with respect to the  
plane of colliding hadrons. The angular decay coefficients, $\lambda_\theta$, $\lambda_\phi$ and  
$\lambda_{\theta\phi}$, are the polarisation parameters. In the case of an unpolarised yield, one would have  
($\lambda_{\theta}, \lambda_{\phi}, \lambda_{\theta \phi}) = (0,0,0)$ for an isotropic decay angular distribution, whereas 
 $(1,0,0)$ and $(-1,0,0)$ refers to fully transverse and fully longitudinal polarisation, respectively.   
 
It is however very important to bear in mind that the angular distribution of \eq{eq:angularDistribution} is frame dependent 
as the polarisation parameters. All experimental analyses have been carried in a few specific reference frames, 
essentially defined by their polarisation axis\footnote{ 
See~\cite{Faccioli:2010kd} for the definition of the corresponding axes.}, namely: the helicity ($HX$) frame , the Collins-Soper ($CS$)~\cite{Collins:1977iv} frame, the Gottfried-Jackson ($GJ$)~\cite{Gottfried:1964nx} frame as well as the perpendicular helicity ($PX$)~\cite{Braaten:2008mz} frame.  
 
In spite of the frame dependence of $\lambda_{\theta}, \lambda_{\phi}, \lambda_{\theta \phi}$, there exist some combinations which  
are frame invariant~\cite{Faccioli:2010kd,Palestini:2010xu}. An obvious one is the yield, another one is   
$\tilde{\lambda} = (\lambda_{\theta} + 3 \lambda_{\phi}) / (1 - \lambda_{\phi})$~\cite{Faccioli:2010kd}.  
As such, it can be used as a good cross-check  
between measurements done in different reference frames. Different methods have been used  to extract  
the polarisation parameter(s) from the angular dependence of the yields. In the following, we divide them into two groups: 
{\it (i)} $1-D$ technique: fitting $\cos\theta$ distribution with the angular distribution, \eq{eq:angularDistribution}, averaged over the azimuthal $\phi$ angle, and fitting the $\phi$ distribution, \eq{eq:angularDistribution}, averaged over the polar $\theta$ angle {\it (ii)} $2-D$ technique: fitting a two-dimensional $\cos\theta$ {\it vs} $\phi$ distribution with the full angular distribution, \eq{eq:angularDistribution}. 
 
Beyond the differences in the methods employed to extract these parameters, one should also take into consideration that some 
samples are cleaner than others,  
physics-wise\footnote{ 
Irrespective of the experimental techniques used to extract it, a sample of {\it inclusive} low \pt $\psiP$ at energies around 100\GeV is essentially purely direct.}. Indeed, as we discussed in the previous section, a given quarkonium yield 
can come from different sources, some of which are not of specific interests for data-theory comparisons. The most obvious 
one is the non-prompt charmonium yield, which is expected to be the result of quite different mechanism that the prompt yield. 
Nowadays, the majority of the studies are carried out on a prompt sample thanks to a precise vertexing of the events. 
Yet, a further complication also comes from feed-down from the excited states in which case vertexing is of no help.  
As for now, no attempt of removing it 
from \eg prompt \jpsi and inclusive \upsa samples has been made owing its intrinsic complication. We have therefore found it 
important to specify what kind of feed-down could be expected in the analysed sample.

In view of this, the Tables~\ref{tab:jpsiPolarization}--\ref{tab:ups3Polarization} contain, in addition to the information on the collision systems and 
the kinematical coverages, information on the fit technique and a short reminder of the expected feed-down. For each measurement, we also briefly summarise 
the observed trend. The vast majority of the experimental data do not show a significant quarkonium polarisation, neither polar nor azimuthal anisotropy.  
Yet, values as large as $\pm 0.3$ are often not excluded either -- given the experimental uncertainties. Despite these, a simultaneous  
description of both measured quarkonium cross sections and polarisations is still challenging for theoretical models of quarkonium hadroproduction.  
 
\begin{figure}[!t] 
 \begin{center} 
   \subfigure[]{ 
	\includegraphics[angle=0,width=0.48\textwidth]{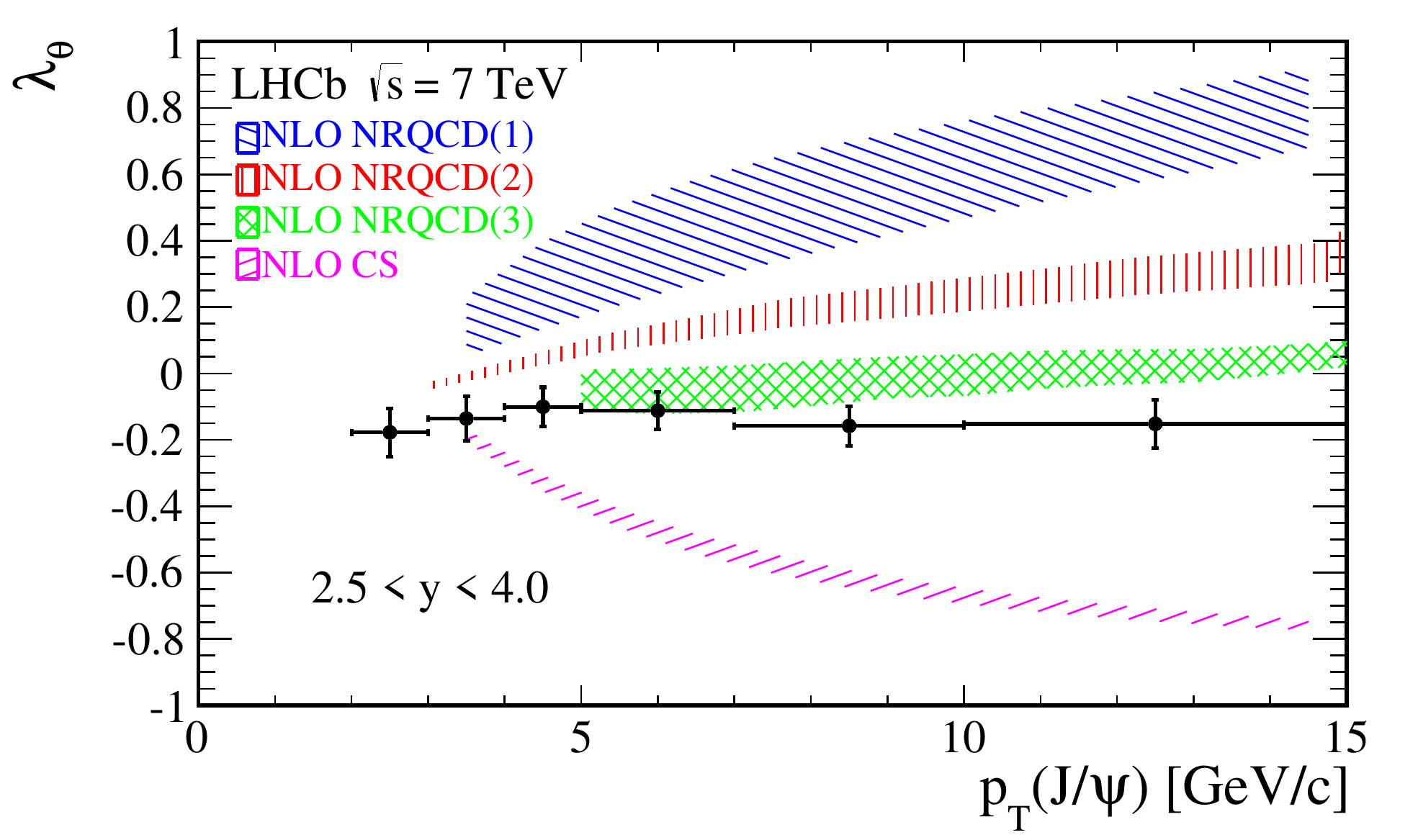}   
	} 
  \subfigure[] { 
   	\includegraphics[angle=0,width=0.48\textwidth]{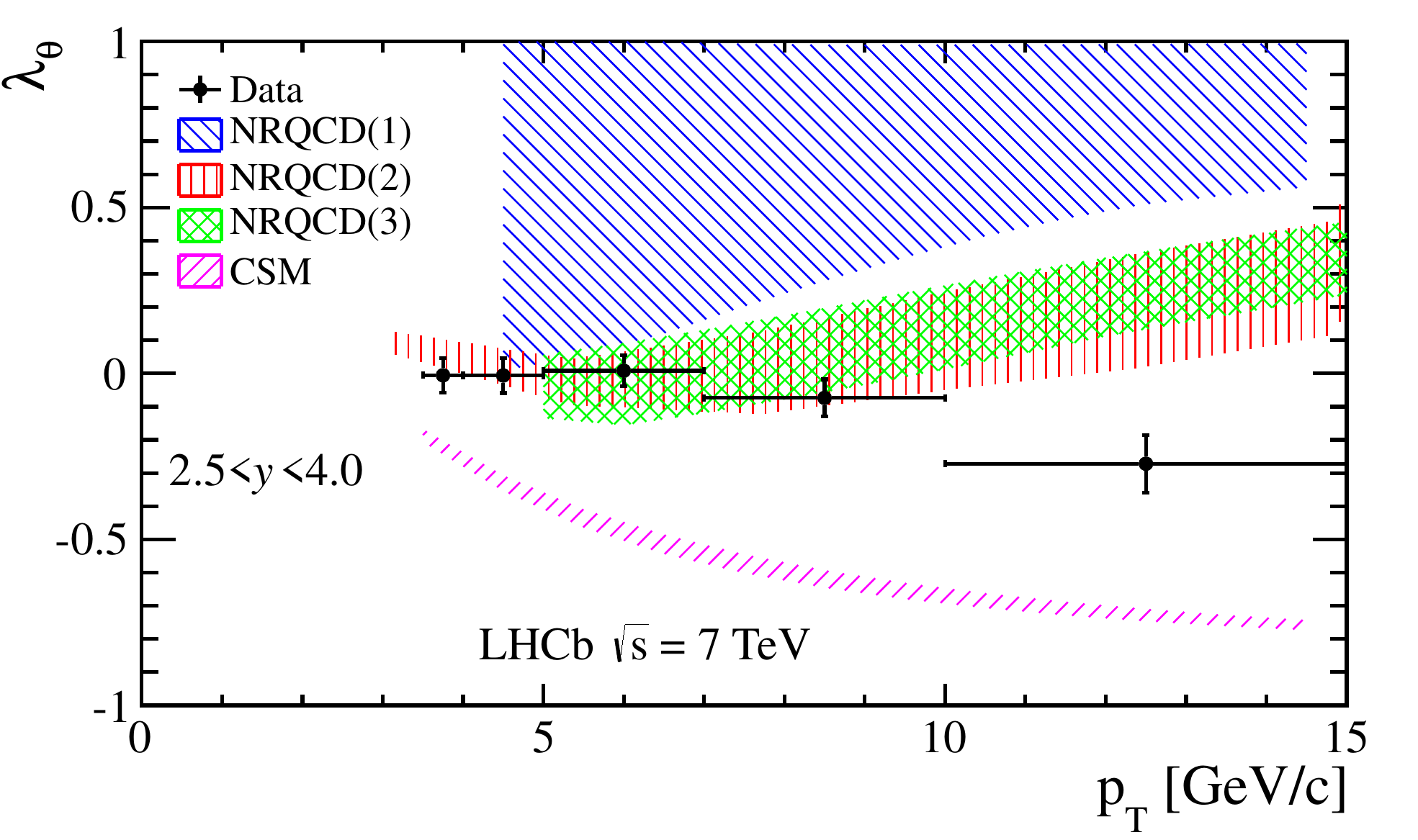} 
	} 
 \end{center} 
 \caption{Polarisation parameter $\lambda_{\theta}$ for prompt \jpsi~\cite{Aaij:2013nlm} (a) and \psiP~\cite{Aaij:2014qea} (b) from LHCb compared to different model predictions: direct NLO CSM~\cite{Butenschoen:2012px} and three NLO NRQCD calculations~\cite{Butenschoen:2012px,Gong:2012ug,Chao:2012iv}, at $2.5<y<4.0$ in the helicity frame.} 
 \label{quarkoniapolarization_LHCb}  
\end{figure}

As example, we show in~\fig{quarkoniapolarization_LHCb} the \pt-dependence of $\lambda_{\theta}$ for prompt \jpsi~\cite{Aaij:2013nlm} (left panel) and \psiP~\cite{Aaij:2014qea} (right panel) measured by LHCb at $2.5<y<4.0$ in the helicity frame compared with a few theoretical predictions. 
NLO NRQCD calculations~\cite{Butenschoen:2012px,Gong:2012ug,Chao:2012iv} show mostly positive or zero values of $\lambda_{\theta}$ with a trend towards  
the transverse polarisation with increasing \pt, and a magnitude of the $\lambda_{\theta}$ depending on the specific calculation and the kinematical region.  
On the other hand, NLO CSM models~\cite{Gong:2008sn,Lansberg:2010vq} tend to predict an unpolarised yield at low \pt and an increasingly  
longitudinal yield ($\lambda_{\theta} < 0$) for increasing \pt. None of these predictions correctly describes the measured \jpsi and \psiP $\lambda_{\theta}$  
parameters and their \pt trends. The NLO NRQCD fits of the PKU group~\cite{Shao:2014yta,Han:2014jya} however open the possibility for an  
unpolarised {\it direct yield} but at the cost of 
not describing the world existing data in $e$p and \ee collisions and data in \pp collisions for $\pt \leq 5$\GeVc.

\begin{figure}[!htb] 
 \begin{center} 
  \subfigure[] 
  {\includegraphics[angle=0,width=0.7\textwidth]{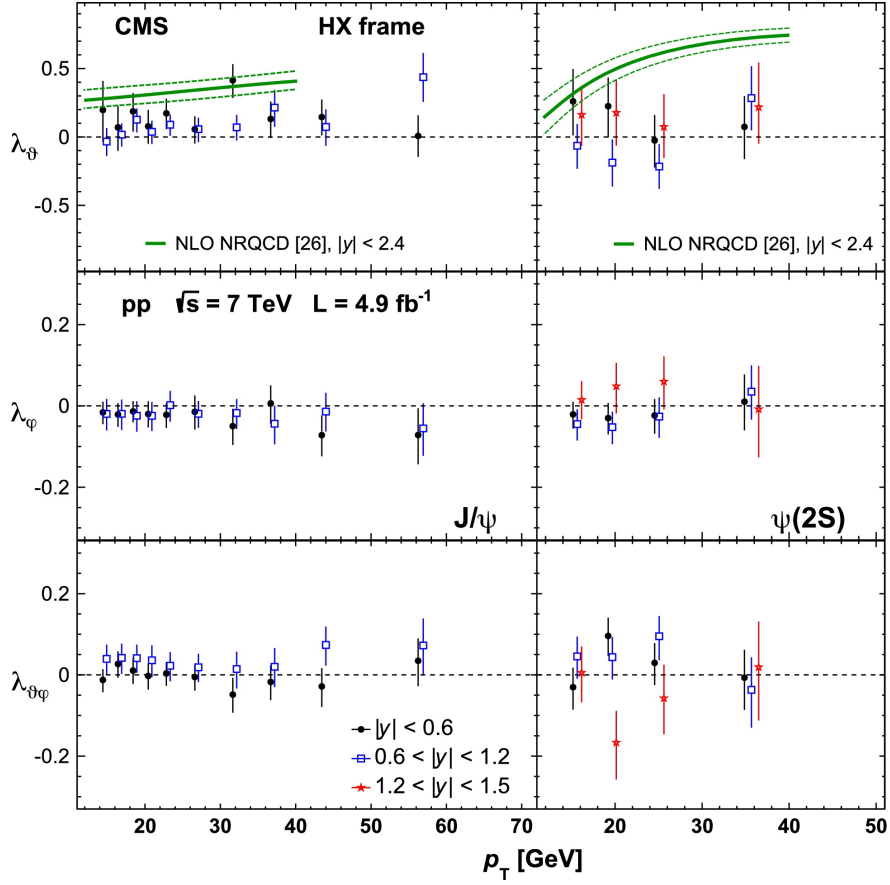}\label{psiPolarization_CMS} } 
\subfigure[] 
{\includegraphics[angle=0,width=0.7\textwidth]{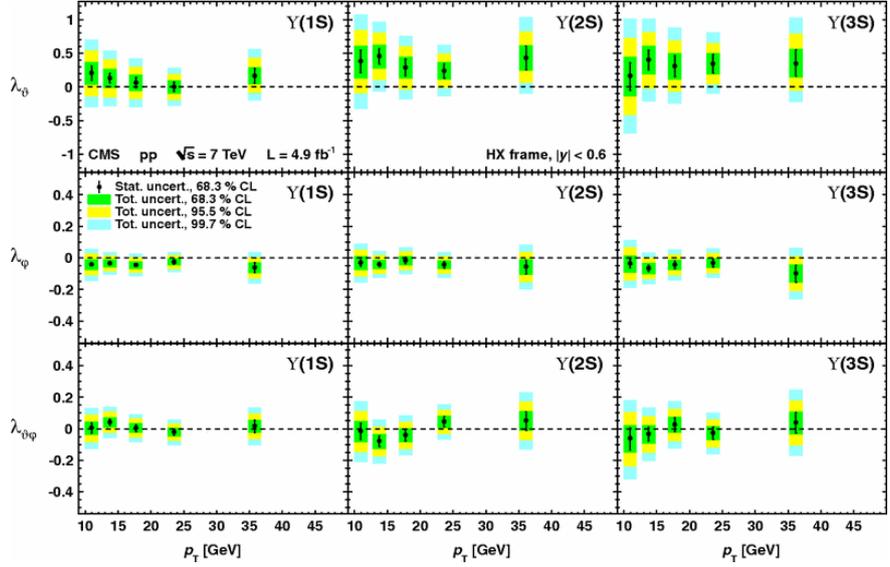}\label{upsilonPolarization_CMS} } 
 \end{center} 
 \caption{ 
 (a) Polarisation parameters, $\lambda_{\theta}$, $\lambda_{\phi}$ and $\lambda_{\theta \phi}$, as a function of \pt measured in the $HX$ frame of  prompt \jpsi, \psiP~\cite{Chatrchyan:2013cla}. Upper panels show also NLO NRQCD calculations~\cite{Gong:2012ug} of $\lambda_{\theta}$ for prompt \jpsi and \psiP for $\vert y \vert <$ 2.4.  
 (b) Polarisation parameters, $\lambda_{\theta}$, $\lambda_{\phi}$ and $\lambda_{\theta \phi}$, as a function of \pt measured in the $HX$ frame of \upsa, \upsb, \upsc~\cite{Chatrchyan:2012woa} for $\vert y \vert < 0.6$. 
} 
  
\end{figure}

In order to illustrate the recent progresses in these delicate studies, let us stress that LHC experiments  
have performed measurements of the three polarisation parameters as well as in different reference frames.  
This has not always been the case before by lack of statistics and of motivation since it is difficult  
to predict theoretically azimuthal effects, \eg $\lambda_{\theta \phi}$. 
\fig{psiPolarization_CMS} and~\ref{upsilonPolarization_CMS}  
show CMS measurements of $\lambda_{\theta}$, $\lambda_{\phi}$ and $\lambda_{\theta \phi}$, in the $HX$ frame for  
\jpsi, \psiP~\cite{Chatrchyan:2013cla} and \upsa, \upsb, \upsc~\cite{Chatrchyan:2012woa} on the left and  
right panel, respectively. CMS has also conducted polarisation measurements in the $CS$ and $PX$ frames,  
in addition to the $HX$ frame and they could cross-check their analysis by obtaining  
the consistency in $\tilde{\lambda}$ in these three frames for different \pt and $y$.  
As for most of the previous measurements, no evidence of a large transverse or longitudinal quarkonium  
polarisation is observed in any reference frame, and in the whole measured kinematic range.  
 
To conclude, let us also mention the importance of measuring the polarisation of $P$-wave states in order to refine our test of \eg NRQCD~\cite{Shao:2012fs}.  
This can be done either directly via the measurement of the angular dependence of the emitted photon or indirectly via that of the polarisation  
of the $S$-wave (\jpsi or \ups) in which they decay~\cite{Faccioli:2011be}. Such studies are very important to constrain  
experimentally the effect of the feed-downs on the polarisation of the available samples. Let us also stress that such a measurement 
in heavy-ion collisions (along the line of the first study in In--In collisions~\cite{Arnaldi:2009ph})  
may also be used as a tool to study a possible sequential suppression of the quarkonia~\cite{Faccioli:2012kp}.

\begin{landscape} 
 
\begin{table}[hbt!] 
\caption{World existing data for \jpsi polarisation in \pp and \pA collisions.} 
\label{tab:jpsiPolarization} 
\begin{threeparttable} 
\small 
\begin{tabularx}{22cm}{ p{1.cm} p{1cm} p{1.6cm} p{1.2cm} p{1.2cm} p{2.8cm} p{0.6cm}  p{1.8cm} X } 
\hline 
\s [GeV] & Colliding system &   Experiment   &   \ycm range  &  \pt range [GeV/$c$]   &    Feed-down   &    Fit     &  Measured parameter(s) & Observed trend  \\ 
\hline 
200  & \pp  &  PHENIX \cite{Adare:2009js} & $\vert y \vert <$ 0.35 & 0 -- 5 & $B$ : $2\div 15\%$ \cite{Adamczyk:2012ey} ~  \chic : $23\div 41\%$ \cite{Adare:2011vq} ~ \psiP: $5\div 20\%$ \cite{Adare:2011vq} & $1-D$ & $\lambda_{\theta}$ {\it vs} \pt in HX \&  GJ& $\lambda_{\theta}$ values from slightly positive (consistent with 0) to negative as \pt increases \\ 
\hline 
200  & \pp  &  STAR \cite{Adamczyk:2013vjy} & $\vert y \vert <$ 1 & 2 -- 6 & $B$ : $2\div 15\%$ \cite{Adamczyk:2012ey} ~  \chic :  $23\div 41\%$ \cite{Adare:2011vq} ~ \psiP: $5\div 20\%$ \cite{Adare:2011vq} & $1-D$ & $\lambda_{\theta}$ {\it vs} \pt in HX & $\lambda_{\theta}$ values from slightly positive (consistent with 0) to negative as \pt increases \\ 
\hline 
1800  & \ppbar  &  CDF \cite{Affolder:2000nn} & $\vert y \vert <$ 0.6 & 4 -- 20  &  \chic : $25\div 35\%$ \cite{Abe:1997yz} ~ \psiP: $10\div 25\%$ \cite{Abe:1997jz} &  $1-D$ & $\lambda_{\theta}$ {\it vs} \pt in HX & Small positive $\lambda_{\theta}$ at smaller \pt then for \pt $>$ 12 \GeV trend towards negative values \\ 
\hline 
1960  & \ppbar  &  CDF \cite{Abulencia:2007us} & $\vert y \vert <$ 0.6 & 5 -- 30 & \chic : $25\div 35\%$ \cite{Abe:1997yz} ~ \psiP: $10\div 25\%$ \cite{Abe:1997jz} &  $1-D$ & $\lambda_{\theta}$ {\it vs} \pt   in HX  & $\lambda_{\theta}$ values from 0 to negative as \pt increases \\ 
\hline 
7000  & \pp  &  ALICE \cite{Abelev:2011md} & 2.5 -- 4.0 & 2 -- 8  &  $B$ : $10\div 30\%$ \cite{Abelev:2012gx} ~  \chic : $15\div 30\%$ \cite{LHCb:2012af} ~ \psiP: $8\div 20\%$ \cite{Abelev:2014qha}  & $1-D$ & $\lambda_{\theta}$, $\lambda_{\phi}$ {\it vs} \pt in HX \& CS& $\lambda_{\theta}$ and $\lambda_{\phi}$ consistent with 0, with a possible hint for a longitudinal polarisation at low \pt in the $HX$ frame \\  
\hline 
7000  & \pp &  LHCb \cite{Aaij:2013nlm} & 2.0 -- 4.5 &  2 -- 15 &  \chic : $15\div 30\%$ \cite{LHCb:2012af} ~ \psiP: $8\div 25\%$ \cite{Aaij:2012ag} & $2-D$ &  $\lambda_{\theta}$, $\lambda_{\phi}$, $\lambda_{\theta \phi}$ {\it vs} \pt and $y$ in HX \& CS& $\lambda_{\phi}$ and $\lambda_{\theta \phi}$ consistent with 0 in the $HX$ frame and $\lambda_{\theta}$ ($\lambda_{\theta} = $ -0.145 $\pm$ 0.027) shows small longitudinal polarisation; $\tilde{\lambda}$ in agreement in the $HX$ and $CS$ frames  \\  
\hline 
7000  & \pp &  CMS \cite{Chatrchyan:2013cla} &  $\vert y \vert <$ 1.2  & 14 -- 70  &  \chic : $25\div 35\%$ ~ \psiP: $15\div 20\%$ \cite{Chatrchyan:2011kc} & $2-D$ & $\lambda_{\theta}$, $\lambda_{\phi}$, $\lambda_{\theta \phi}$, $\tilde{\lambda}$ {\it vs} \pt and in 2 $\vert y \vert$ bins in HX,CS,PX &  In the 3 frames, no evidence of  
large $|\lambda_{\theta}|$ anywhere; $\tilde{\lambda}$ in a good agreement in all reference frames \\  
\hline 
17.2  & \pA  &  NA60 \cite{Arnaldi:2009ph} & 0.28 -- 0.78 &  --  & $B$ : $2\div 8\%$ ~  \chic :$25\div 40\%$ \cite{Abt:2002vq} ~ \psiP: $7\div 10\%$ \cite{Alexopoulos:1995dt} & $1-D$  & $\lambda_{\theta}$, $\lambda_{\phi}$ {\it vs} \pt in HX & $\lambda _{\theta}$ and $\lambda _{\phi}$ consistent with 0; slight increase of the $\lambda _{\theta}$ value with increasing \pt, no \pt dependence $\lambda _{\phi}$ \\ 
\hline 
27.4  & \pA &  NA60 \cite{Arnaldi:2009ph} & -0.17 -- 0.33 &  --  & $B$ : $2\div 8\%$ ~  \chic : $25\div 40\%$ \cite{Abt:2002vq} ~ \psiP: $7\div 10\%$ \cite{Alexopoulos:1995dt} & $1-D$  & $\lambda_{\theta}$, $\lambda_{\phi}$ {\it vs} \pt in HX & $\lambda_{\theta}$ and $\lambda _{\phi}$ consistent with 0, no \pt dependence observed \\  
\hline 
31.5  & $p\mathrm{-Be}$      &  E672/E706 \cite{Gribushin:1999ha} & $x_{F}$: 0.00 -- 0.60 & 0 -- 10 & $B$ : $2\div 8\%$ ~  \chic : $25\div 40\%$ \cite{Abt:2002vq} ~ \psiP: $7\div 10\%$ \cite{Alexopoulos:1995dt} & $1-D$  & $\lambda_{\theta}$ in GJ & $\lambda_{\theta} = $ 0.01 $\pm$ 0.12 $\pm$ 0.09, consistent with no polarisation \\  
\hline 
38.8  & $p\mathrm{-Be}$      &  E672/E706 \cite{Gribushin:1999ha} & $x_{F}$: 0.00 -- 0.60 & 0 -- 10 & $B$ : $2\div 8\%$ ~  \chic : $25\div 40\%$ \cite{Abt:2002vq} ~ \psiP: $7\div 10\%$ \cite{Alexopoulos:1995dt} & $1-D$  & $\lambda_{\theta}$ in GJ & $\lambda_{\theta} = $ -0.11 $\pm$ 0.12 $\pm$ 0.09, consistent with no polarisation  \\ 
\hline 
38.8  & $p\mathrm{-Si}$      &  E771 \cite{Alexopoulos:1997yd} & $x_{F}$: -0.05 -- 0.25 & 0 -- 3.5  & $B$ : $2\div 8\%$ ~  \chic : $25\div 40\%$ \cite{Abt:2002vq} ~ \psiP: $7\div 10\%$ \cite{Alexopoulos:1995dt} & $1-D$  & $\lambda_{\theta}$ in GJ & $\lambda_{\theta} = $ -0.09 $\pm$ 0.12, consistent with no polarisation \\  
\hline 
38.8  & $p\mathrm{-Cu}$      &  E866/NuSea \cite{Chang:2003rz} & $x_{F}$: 0.25 -- 0.9 & 0 -- 4 & \chic : $25\div 40\%$ \cite{Abt:2002vq} ~ \psiP: $7\div 10\%$ \cite{Alexopoulos:1995dt} & $1-D$  & $\lambda_{\theta}$ {\it vs} \pt and $x_{F}$ in CS & $\lambda_{\theta}$ values from small positive to negative with increasing $x_{F}$, no significant \pt dependence observed \\ 
\hline 
41.6  & $p\mathrm{-C,W}$  &  HERA-B \cite{Abt:2009nu} & $x_{F}$: -0.34 -- 0.14 & 0 -- 5.4 &  \chic : $25\div 40\%$ \cite{Abt:2002vq} ~ \psiP: $5\div 15\%$ \cite{Abt:2006va} & $1-D$  &  $\lambda_{\theta}$, $\lambda_{\phi}$, $\lambda_{\theta \phi}$ {\it vs} \pt and $x_{F}$ in HX, CS, GJ & $\lambda _{\theta}$ and $\lambda _{\phi}$ $<0$, $\lambda _{\theta}$ ($\lambda _{\phi}$) decrease (increase) with increasing \pt; no strong $x_{F}$ dependence; for $\pt > 1$\GeVc, $\lambda_{\theta,\phi}$ depends on the frame : 
 $\lambda^{\rm CS}_{\theta}>\lambda^{\rm HX}_{\theta}\simeq 0$ and  $\lambda^{\rm HX} _{\phi} \neq 0$  \\ 
\hline 
\end{tabularx} 
\end{threeparttable} 
\end{table}

\begin{table}[hbt!] 
\caption{World existing data for \psiP polarisation in \pp.} 
\label{tab:psiPolarization} 
\begin{threeparttable} 
\small 
\begin{tabularx}{22cm}{p{1.cm} p{1cm} p{1.6cm} p{1.2cm} p{1.2cm} p{2.8cm} p{0.6cm} p{1.8cm} X } 
\hline 
\s [GeV]   & Colliding system &   Experiment   &   \ycm range  &  \pt range [GeV/$c$]   &    Feed-down   &    Fit     &  Measured parameter(s) & Observed trend  \\ 
\hline 
1800  & \ppbar     &    CDF \cite{Affolder:2000nn}  &  $\vert y \vert <$ 0.6  & 5.5 -- 20 & none &  $1-D$  &   $\lambda_{\theta}$ {\it vs} $p_{T}$ in HX & $\lambda_{\theta}$ consistent with 0  \\ 
\hline 
1960  & \ppbar      &  CDF \cite{Abulencia:2007us} & $\vert y \vert <$ 0.6 & 5 -- 30 & none & $1-D$  & $\lambda_{\theta}$ {\it vs} $p_{T}$ in HX & $\lambda_{\theta}$ values {\it vs} \pt go from slightly positive to small negative  \\ 
\hline 
7000   & \pp    &    LHCb \cite{Aaij:2014qea}  &  2.0 -- 4.5 & 3.5 -- 15  &  none   &  $2-D$ &  $\lambda_{\theta}$, $\lambda_{\phi}$, $\lambda_{\theta \phi}$, $\tilde{\lambda}$ {\it vs} \pt and 3 $\vert y \vert$ bins in HX, CS & no significant polarisation found, with an indication of small longitudinal polarisation - $\tilde{\lambda}$ is negative with no strong \pt and $y$ dependence  \\ 
\hline 
7000   & \pp    &    CMS \cite{Chatrchyan:2013cla} & $\vert y \vert <$ 1.5  & 14 -- 50  &   none  & $2-D$  & $\lambda_{\theta}$, $\lambda_{\phi}$, $\lambda_{\theta \phi}$, $\tilde{\lambda}$ {\it vs} \pt and in 3 $\vert y \vert$ bins in HX, CS, PX & non of the 3 reference frames show evidence of large transverse or longitudinal polarisation, in the whole measured kinematic range; $\tilde{\lambda}$ in a good agreement in all reference frames \\ 
\hline 
\end{tabularx} 
\end{threeparttable} 
\end{table}

\begin{table}[hbt!] 
\caption{World existing data for \upsa polarisation in \pp.} 
\label{tab:ups1Polarization} 
\begin{threeparttable} 
\small 
\begin{tabularx}{22cm}{ p{1.cm} p{1cm} p{1.6cm} p{1.2cm} p{1.2cm} p{2.8cm} p{0.6cm} p{1.8cm} X } 
\hline 
\s [GeV]   & Colliding system &   Experiment   &   \ycm range  &  \pt range [GeV/$c$]   &    Feed-down   &    Fit     &  Measured parameter(s) & Observed trend  \\ 
\hline 
1800  & \ppbar & CDF \cite{Acosta:2001gv}  &  $\vert y \vert <$ 0.4  & 0 -- 20  & \upsb : $6\div 18\%$ \cite{Affolder:1999wm} ~  \upsc : $0.4\div 1.4\%$ \cite{Affolder:1999wm} ~ $\chi_{b}$ : $30\div 45\%$ \cite{Affolder:1999wm} & $1-D$ &   $\lambda_{\theta}$ {\it vs} \pt in HX & $\lambda_{\theta}$ consistent with 0, no significant \pt dependence  \\ 
\hline 
1960  & \ppbar & CDF \cite{CDF:2011ag}  &  $\vert y \vert <$ 0.6  & 0 -- 40  & \upsb : $6\div 18\%$ \cite{Affolder:1999wm} ~  \upsc : $0.4\div 1.4\%$ \cite{Affolder:1999wm} ~ $\chi_{b}$ : $30\div 45\%$ \cite{Affolder:1999wm}  & $2-D$ &   $\lambda_{\theta}$, $\lambda_{\phi}$, $\lambda_{\theta \phi}$, $\tilde{\lambda}$ {\it vs} \pt in HX, CS & the angular distribution found to be nearly isotropic \\ 
\hline 
1960  & \ppbar & D0 \cite{Abazov:2008aa}  &  $\vert y \vert <$ 0.4  & 0 -- 20  & \upsb : $6\div 18\%$ \cite{Affolder:1999wm} ~  \upsc : $0.4\div 1.4\%$ \cite{Affolder:1999wm} ~ $\chi_{b}$ : $30\div 45\%$ \cite{Affolder:1999wm} & $1-D$ &   $\lambda_{\theta}$ {\it vs} \pt in HX & significant negative $\lambda_{\theta}$ at low \pt with decreasing magnitude as \pt increases  \\ 
\hline 
7000  & \pp & CMS \cite{Chatrchyan:2012woa} & $\vert y \vert <$ 1.2  & 10 -- 50  &  \upsb : $10\div 15\%$ ~  \upsc : $0.5\div 3\%$  \cite{Chatrchyan:2013yna} ~ $\chi_{b}$ : $25\div 40\%$ \cite{Aaij:2014caa} & $2-D$  & $\lambda_{\theta}$, $\lambda_{\phi}$, $\lambda_{\theta \phi}$, $\tilde{\lambda}$ {\it vs} \pt and in 2 $\vert y \vert$ bins in HX, CS, PX & no evidence of large transverse of longitudinal polarisation in the whole kinematic range in 3 reference frames  \\ 
\hline 
\end{tabularx} 
\end{threeparttable} 
\end{table}

\begin{table}[hbt!] 
\caption{World existing data for \upsb polarisation in \pp.} 
\label{tab:ups2Polarization} 
\begin{threeparttable} 
\small 
\begin{tabularx}{22cm}{p{1.cm} p{1cm} p{1.6cm} p{1.2cm} p{1.2cm} p{2.8cm} p{0.6cm} p{1.8cm} X } 
\hline 
\s [GeV]   & Colliding system &   Experiment   &   \ycm range  &  \pt range [GeV/$c$]   &    Feed-down   &    Fit     &  Measured parameter(s) & Observed trend  \\ 
\hline 
1960  & \ppbar & D0 \cite{Abazov:2008aa}  &  $\vert y \vert <$ 0.4  & 0 -- 20  & \upsc :  $2.5\div 5\%$ \cite{Chatrchyan:2013yna} ~ $\chi_{b}$ : $25\div 35\%$ \cite{Aaij:2014caa}  & $1-D$ &   $\lambda_{\theta}$ {\it vs} \pt in HX & $\lambda_{\theta}$ consistent with zero at low \pt, trend towards strong transverse polarisation at $\pt>5$\GeVc  \\ 
\hline 
1960  & \ppbar & CDF \cite{CDF:2011ag}  &  $\vert y \vert <$ 0.6  & 0 -- 40  & \upsc :  $2.5\div 5\%$ \cite{Chatrchyan:2013yna} ~ $\chi_{b}$ : $25\div 35\%$ \cite{Aaij:2014caa} & $2-D$ &   $\lambda_{\theta}$, $\lambda_{\phi}$, $\lambda_{\theta \phi}$, $\tilde{\lambda}$ {\it vs} \pt in HX, CS & the angular distribution found to be nearly isotropic    \\ 
\hline 
7000  & \pp & CMS \cite{Chatrchyan:2012woa} & $\vert y \vert <$ 1.2  & 10 -- 50  &  \upsc : $2.5\div 5\%$ \cite{Chatrchyan:2013yna} ~ $\chi_{b}$ : $25\div 35\%$ \cite{Aaij:2014caa} & $2-D$  & $\lambda_{\theta}$, $\lambda_{\phi}$, $\lambda_{\theta \phi}$, $\tilde{\lambda}$ {\it vs} \pt and in 2 $\vert y \vert$ bins in HX, CS, PX & no evidence of large transverse of longitudinal polarisation in whole kinematic range in 3 reference frames  \\ 
\hline 
\end{tabularx} 
\end{threeparttable} 
\end{table} 
 
\begin{table}[hbt!] 
\caption{World existing data for \upsc polarisation in \pp.} 
\label{tab:ups3Polarization} 
\begin{threeparttable} 
\small 
\begin{tabularx}{22cm}{p{1.cm} p{1cm} p{1.6cm} p{1.2cm} p{1.2cm} p{2.8cm} p{0.6cm} p{1.8cm} X } 
\hline 
\s [GeV]   & Colliding system &   Experiment   &   \ycm range  &  \pt range [GeV/$c$]   &    Feed-down   &    Fit     &  Measured parameter(s) & Observed trend  \\ 
\hline 
1960  & \ppbar & CDF \cite{CDF:2011ag}  &  $\vert y \vert <$ 0.6  & 0 -- 40  & $\chi_{b}$ : $30\div 50\%$ \cite{Aaij:2014caa}   & $2-D$ &   $\lambda_{\theta}$, $\lambda_{\phi}$, $\lambda_{\theta \phi}$, $\tilde{\lambda}$ {\it vs} \pt in HX, CS & the angular distribution found to be nearly isotropic   \\ 
\hline 
7000  & \pp & CMS \cite{Chatrchyan:2012woa} & $\vert y \vert <$ 1.2  & 10 -- 50  & $\chi_{b}$ : $30\div 50\%$ \cite{Aaij:2014caa} & $2-D$  & $\lambda_{\theta}$, $\lambda_{\phi}$, $\lambda_{\theta \phi}$, $\tilde{\lambda}$ {\it vs} \pt and in 2 $\vert y \vert$ bins in HX, CS, PX & no evidence of large transverse of longitudinal polarisation in whole kinematic range in 3 reference frames  \\ 
\hline 
\end{tabularx} 
\end{threeparttable} 
\end{table} 
 
\end{landscape}

\subsection{New observables}
\label{sec:pp:NewObs}

Thanks to the large heavy-flavour samples available at hadron colliders, studies of the production of open or hidden heavy-flavour production in association with another particle (light- or heavy-hadrons, quarkonium, or vector boson) are possible. The cross section of these processes is heavily sensitive to the particle production mechanisms and can help distinguishing between them.  
In addition, these final states can also results from multiple parton-parton interactions (or double-parton scatterings, DPS), where several hard parton-parton interactions occur in the same event, without any correlation between them~\cite{Alner:1985wj, Wang:1991us, Sjostrand:1987su, Bartalini:2010su}.  
Analogously, heavy-flavour production dependence with the underlying event multiplicity brings information into their production mechanisms.  
A complete understanding of heavy-flavour production in hadronic collisions is mandatory to interpret heavy-flavour measurements in \pA and \AAcoll collisions, and disentangle cold (see \sect{Cold nuclear matter effects}) and hot (see \sects \ref{OHF} and~\ref{sec:quarkonia}) nuclear matter effects at play.

\subsubsection{Production as a function of multiplicity} 
\label{sec:pp:HadCorrelations}

The correlation of open or hidden heavy-flavour yields with charged particles produced in hadronic collisions can provide insight into their production mechanism and into the interplay between hard and soft mechanisms in particle production.  
In high energy hadronic collisions, multiple parton-parton interactions may also affect heavy-flavour production~\cite{Maciula:2012rs,Porteboeuf:2010dw}, in competition to a large amount of QCD-radiation associated to hard processes.  
In addition to those initial-state effects, heavy-flavour production could suffer from final-state effects due to the high multiplicity environment produced in high energy \pp collisions~\cite{Werner:2010ny, Lang:2013ex}.

\begin{figure}[!b] 
	\centering 
	\includegraphics[width=0.5\columnwidth]{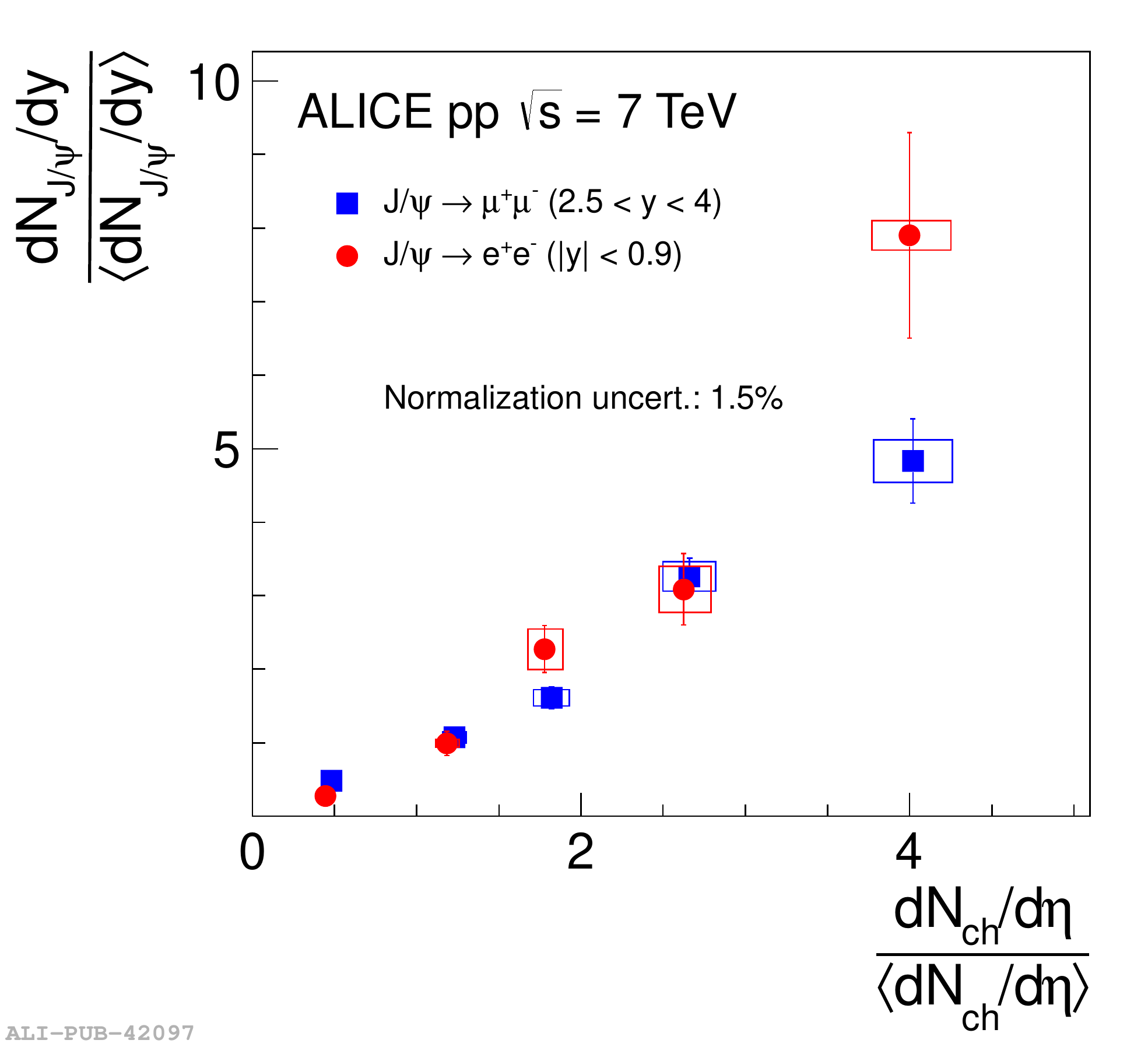} 
	\caption{ 
	\label{fig:ALICE_JPsi_mult} 
	\jpsi yield as a function of the charged-particle density at mid-rapidity in \pp collisions at \s = 7\TeV~\cite{Abelev:2012rz}.  
	Both the yields at forward- ($\jpsi \to \mumu$, $2.5 < y < 4$) and at mid-rapidity ($\jpsi \to \ee$, $|y| < $ 0.9) are shown. 
	} 
\end{figure}

At the LHC, \jpsi yields were measured as a function of charged-particle density at mid-rapidity by the ALICE Collaboration in \pp collisions at \s = 7\TeV~\cite{Abelev:2012rz}. \fig{fig:ALICE_JPsi_mult} shows the \jpsi yields at forward rapidity, studied via the dimuon decay channel at $2.5 < y < 4$, and at mid-rapidity, analysed in its dielectron decay channel at $|y| < 0.9$. The results at mid- and forward-rapidity are compatible within the measurement uncertainties, indicating similar correlations over three units of rapidity and up to four times the average charged-particle multiplicity. The relative \jpsi yield increases with the relative charged-particle multiplicity. This increase can be interpreted in terms of the hadronic activity accompanying \jpsi production, as well as multiple parton-parton interactions, or in the percolation scenario~\cite{Ferreiro:2012fb}.  

\begin{figure}[!t] 
	\centering 
	\includegraphics[width=0.47\columnwidth]{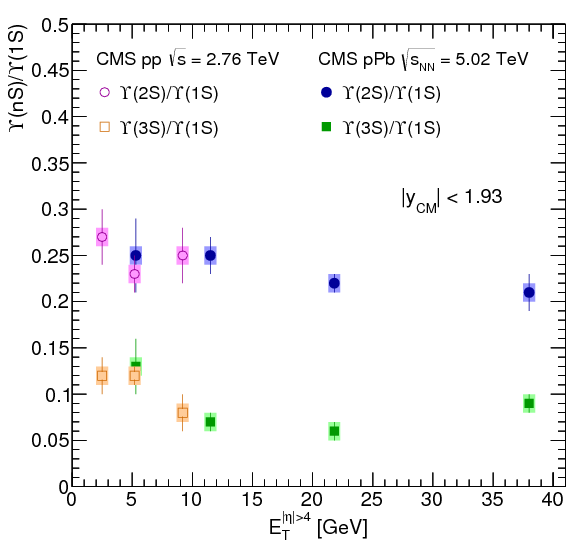} 
	\includegraphics[width=0.47\columnwidth]{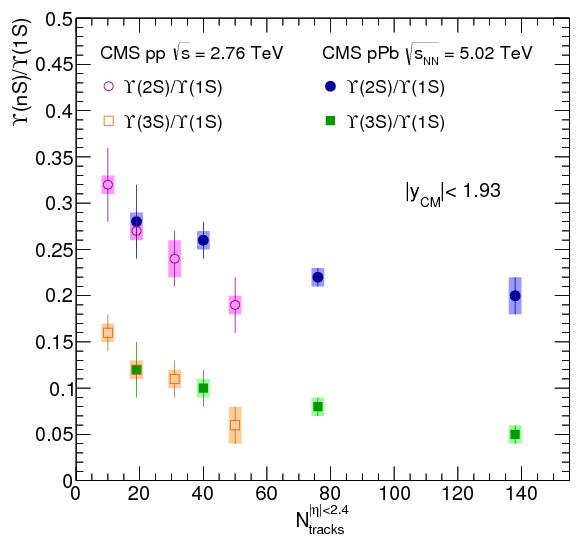} 
	\caption{ 
		\label{fig:CMS_ups_mult} 
	Cross section ratio of $\upsb/\upsa$ and $\upsc/\upsa$ for $|y| < 1.93$ as a function of the transverse energy ($E_{\rm T}$) measured in $4.0 < |\eta| < 5.2$ (left) and the number of charged tracks ($N_{\rm tracks}$) measured in $|\eta|<2.4$ (right), in \pp collisions at \s = 2.76\TeV (open symbols) and \pPb collisions at \snn=5.02\TeV~(filled symbols)~\cite{Chatrchyan:2013nza}. 
	} 
\end{figure}

A similar study of the \ups yields was performed by the CMS Collaboration in \pp collisions at \s = 2.76\TeV~\cite{Chatrchyan:2013nza}.  
The self-normalized cross sections of $\upsa/\langle \upsa \rangle$, $\upsb/\langle \upsb \rangle$ and $\upsc/\langle \upsc \rangle$ at mid-rapidity are found to increase with the charged-particle multiplicity.  
To unveil possible variations of the different \ups states, the ratio of the $\upsb$ and $\upsc$ yields with respect to the $\upsa$ yield is shown in \fig{fig:CMS_ups_mult}. The left figure presents the production cross section ratio as a function of the transverse energy ($E_{\rm T}$) measured in $4.0 < |\eta| < 5.2$, whereas the right figure shows the values with respect to the number of charged tracks ($N_{\rm tracks}$) measured in $|\eta| < 2.4$.  
The excited-to-ground-states cross section ratios seem independent of the event activity when they are evaluated as a function of the forward-rapidity $E_{\rm T}$. These ratios seem to decrease with respect to the mid-rapidity $N_{\rm tracks}$, behaviour that can not be confirmed nor ruled out within the uncertainties.  
The \upsa is produced on average with two extra charged tracks than excited states. Feed-down contribution can not solely explain the observed trend. If \ups states were originated from the same initial partons, the mass difference between the ground and the excited states could generate extra particles produced with \upsa.  
%
%

\begin{figure}[!htbp] 
	\centering 
	\includegraphics[width=0.47\columnwidth]{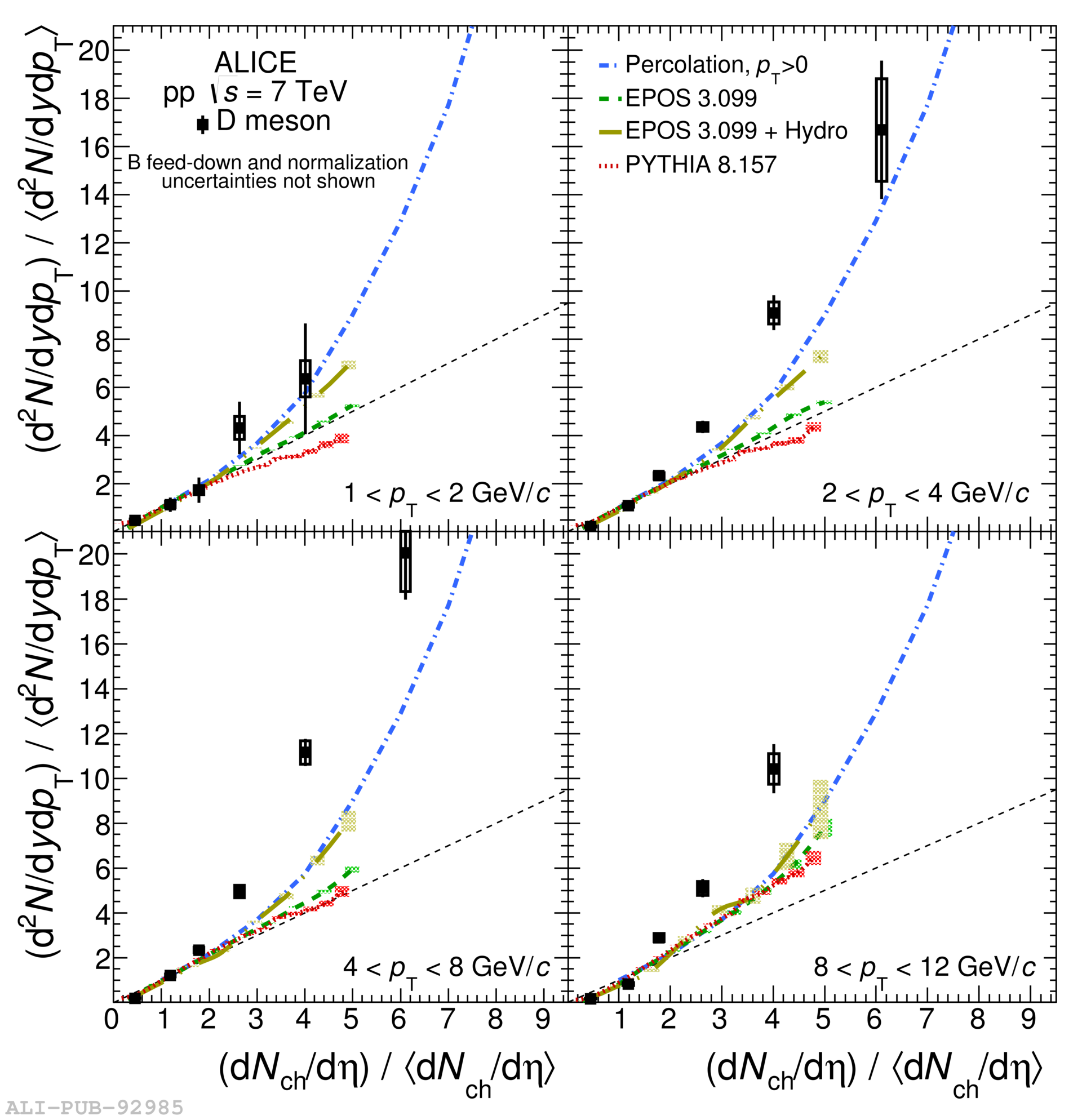} 
	\includegraphics[width=0.47\columnwidth]{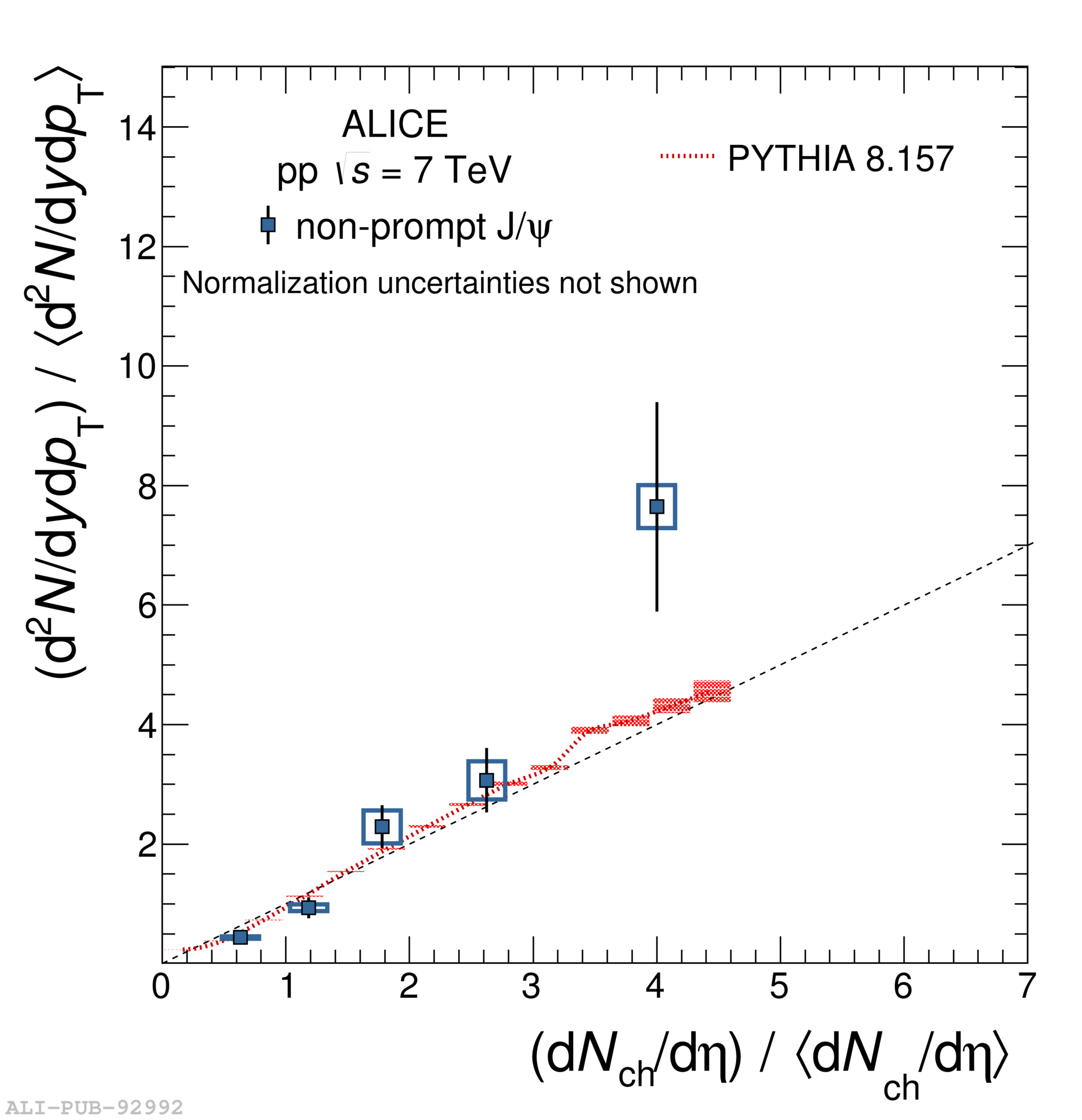} 
	\caption{ 
		\label{fig:ALICE_D_mult} 
	D-meson production (left) and non-prompt \jpsi (right) as a function of charged-particle multiplicity in \pp collisions at \s = 7\TeV ~\cite{Adam:2015ota} compared to PYTHIA~8 ~\cite{Sjostrand:2007gs,Sjostrand:2006za}, EPOS~3 ~\cite{Drescher:2000ha,Werner:2013tya} and the percolation scenario ~\cite{Ferreiro:2012fb, Ferreiro:2015gea}. 
	} 
\end{figure}

The measurement of open heavy-flavour production (via D mesons and non-prompt \jpsi) as a function of charged-particle multiplicity at mid-rapidity in \pp collisions at \s = 7\TeV was recently carried out by the ALICE Collaboration ~\cite{Adam:2015ota}. \fig{fig:ALICE_D_mult} (right) presents the results  
for D mesons in four \pt bins compared to the percolation scenario ~\cite{Ferreiro:2012fb, Ferreiro:2015gea}, EPOS~3 with or without hydro ~\cite{Drescher:2000ha,Werner:2013tya} and PYTHIA~8  simulations ~\cite{Sjostrand:2007gs,Sjostrand:2006za}. D-meson per-event yields are independent of \pt within the  
measurement uncertainties  ($1 < \pt < 12\GeVc$) and increase with multiplicity faster than linearly at high multiplicities. \fig{fig:ALICE_D_mult} (left) shows non-prompt \jpsi yields compared to PYTHIA~8 simulations. D-meson and non-prompt \jpsi yields present a similar increase with charged-particle multiplicity. The heavy-flavour relative yield enhancement as a function of the charged-particle multiplicity is qualitatively described by the percolation model, EPOS 3 and PYTHIA~8 for D mesons and PYTHIA~8 for non-prompt \jpsi.  However, the PYTHIA~8 event generator seems to under-estimate the increase of heavy flavour yields with the charged-particle multiplicity at high multiplicities. Open (D and non-prompt \jpsi) and hidden (inclusive \jpsi) heavy-flavour yields present a similar increase with the charged-particle multiplicity at mid-rapidity. This similarity suggests that the enhancement is likely related to heavy-quark production mechanisms and is not significantly influenced by hadronisation. It could be described by the hadronic  activity associated to heavy-flavour production, multiple parton-parton interactions, or the percolation scenario ~\cite{Maciula:2012rs,Werner:2010ny,Ferreiro:2015gea}. 

Hidden and open heavy-flavour production measurements as a function of the event activity were initiated during the LHC \RunOne leading to unexpected results with impact on our understanding of the production mechanisms and the interpretation of \pPb and \PbPb results. \RunTwo data, with the increased centre-of-mass energy of 13\TeV in \pp collisions and larger luminosities, will allow to reach higher multiplicities and to perform $\pt$-differential studies of hidden and open heavy-flavour hadron production.

\subsubsection{Associated production}  
\label{sec:pp:AngleCorrelations} 
\label{sec:pp:associated} 
 
%
%
%

Heavy-flavour azimuthal correlations in hadronic collisions allow for studies of heavy-quark fragmentation and jet structure at different collision energies, which help to constrain Monte Carlo models, and to understand the different production processes for heavy flavour.  
Heavy quarks can originate from flavour creation, flavour excitation, and parton shower or fragmentation processes of a gluon or a light (anti-)quark including gluon splitting~\cite{Field:2002da}.  
These three different sources of the heavy-flavour production are expected to lead to different correlations between heavy quark and anti-quark, and so a measurement of the opening angle in azimuth ($\Delta \phi$) of two heavy-flavour particles gives an access to different underlying production sub-processes. Azimuthal correlations arising from the flavour creation populate mostly the away-side ($\Delta \phi \approx \pi$), while the near-side region ($\Delta \phi \approx 0$) is sensitive to the presence of the flavour excitation and gluon splitting~\cite{Field:2002da}. Since D--D and B--B correlation measurements are statistically demanding one can also look at angular correlations between heavy-flavour particles with charged hadrons (\eg D--$h$) and correlations between electrons from heavy-flavour decays with charged (\eg $e_{HF}$--$h$) or heavy-flavour hadrons (\eg $e_{HF}$--D).

\begin{figure}[!t] 
 \begin{center} 
  \includegraphics[angle=0,width=0.45\textwidth]{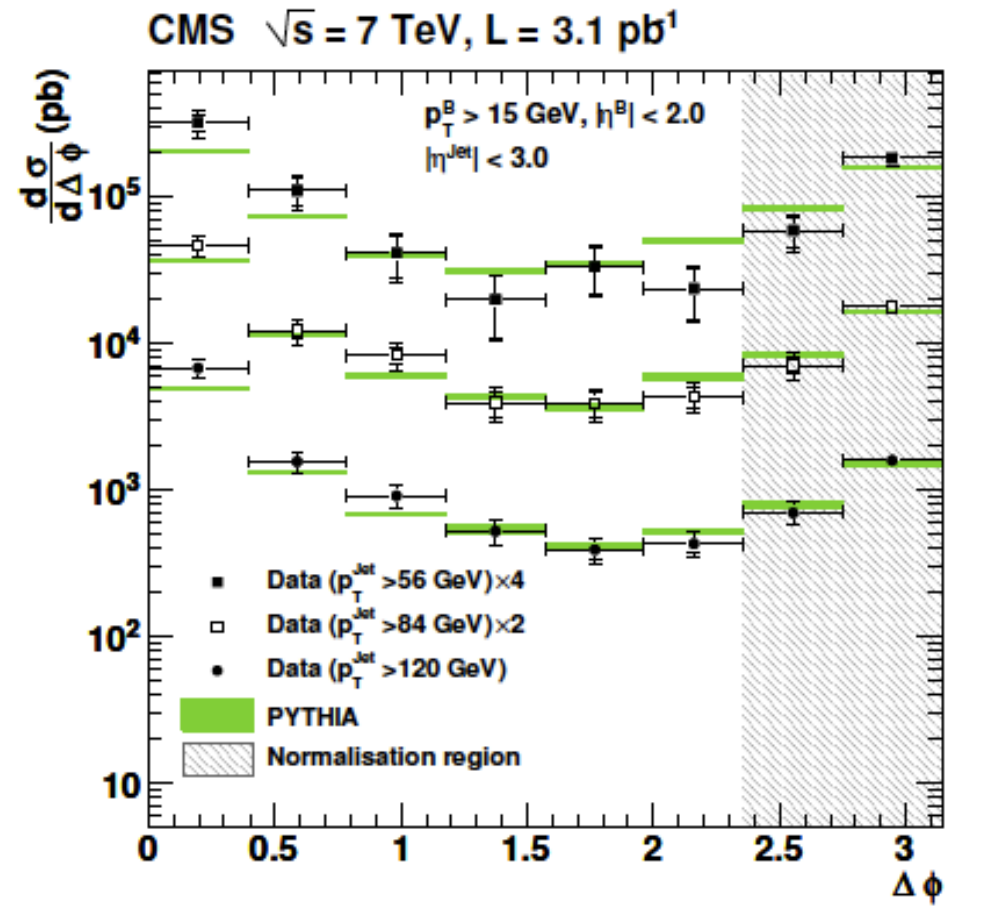} 
 \end{center} 
 \caption{ 
  \label{HFcorrppLHC}  
  Azimuthal correlation of B--$\overline{\rm B}$ mesons measured by CMS in different ranges of the leading jet \pt\ and compared to PYTHIA~\cite{Khachatryan:2011wq}.  
 } 
\end{figure}

Studies of heavy-flavour angular correlations in hadronic collisions were carried out at Tevatron with D--D correlations~\cite{Reisert:2007zza} in \ppbar collisions at \s = 1.96\TeV and at RHIC with $e$--$\mu$ correlations in \pp collisions at \s = 0.2\TeV~\cite{Adare:2013xlp}, where electrons and muons come from heavy-flavour decays and have a large $\eta$ gap. Results on heavy-flavour correlation measurements were also reported by the LHC experiments~\cite{Aaij:2012dz,ATLAS:2011ac,Khachatryan:2011wq} with \pp collisions at \s = 7\TeV, as shown in \fig{HFcorrppLHC}.  
The LHC measurements of the azimuthal correlations between charm (beauty) and anti-charm (anti-beauty) hadrons~(see \eg~\cite{Aaij:2012dz,Khachatryan:2011wq}) show an enhancement at small $\vert \Delta \phi \vert$, not reproduced by PYTHIA, pointing to the importance of the near production (via the gluon splitting mechanism) in addition to the back-to-back production (mostly via flavour creation).  
At RHIC, the comparison of the $e$--$\mu$ azimuthal correlations~\cite{Adare:2013xlp} with PYTHIA suggests that 32\% of $e$--$\mu$ pairs are from the gluon fusion, which agrees with the charm production expectation~\cite{Brambilla:2010cs}. These $e$--$\mu$ correlations show a peak at $\Delta \phi = \pi$ dominated by LO gluon process while the observed continuum is from higher-order contributions, like flavour excitation and gluon splitting.  

\begin{figure}[!t] 
 \begin{center} 
 \subfigure[]{\label{fig:pp:eHcorrppALICE}  
  \includegraphics[angle=0,width=0.4\textwidth]{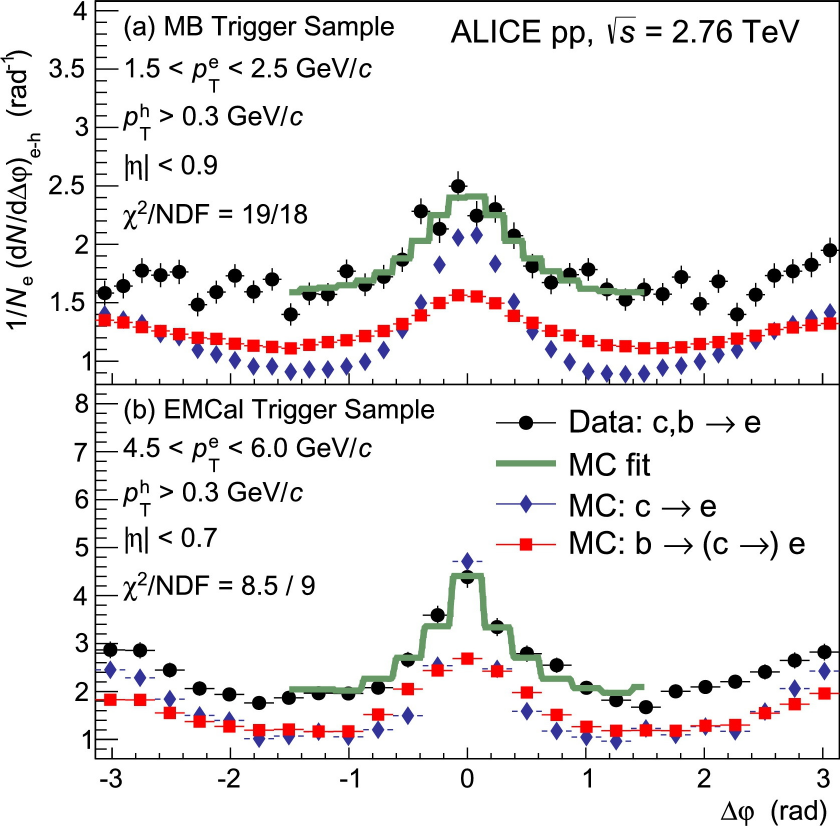}} 
  \subfigure[]{\label{fig:pp:eHcorrppSTAR}  
   \includegraphics[angle=0,width=0.49\textwidth]{eHcorrSTAR_bcontribution.pdf}} 
 \end{center} 
 \caption{ \label{fig:pp:eHcorrpp}  
 (a) : angular correlations between heavy-flavour decay electrons and charged hadrons measured by ALICE in \pp collisions at $\s = 2.76$\TeV, compared to PYTHIA~\cite{Abelev:2014hla}.  
 (b) : Relative beauty contribution to the heavy-flavour electron yield measured by STAR in \pp collisions at $\s = 0.2$\TeV, compared to FONLL calculations~\cite{Aggarwal:2010xp}. 
 } 
\end{figure}

In addition to providing information on the heavy-flavour production mechanisms, the azimuthal correlations of heavy-flavour hadrons with light particles allow to extract the relative contribution of charm and beauty hadron decays to the heavy-hadron yields.  
Due to the different decay kinematics, the azimuthal distribution of the particles produced from B-hadron decays presents a wider distribution at $\Delta \phi \approx 0$ than the one for D decays.  
%
The $e_{HF}$--$h$ angular correlations were measured at mid-rapidity in \pp collisions at $\s = 200$\GeV~\cite{Aggarwal:2010xp,Adare:2009ic,Adare:2010ud} and at $\s = 2.76$\TeV~\cite{Abelev:2014hla}.  
\fig{fig:pp:eHcorrppALICE} presents the azimuthal correlation of $e_{HF}$--h at the LHC.  
PYTHIA calculations of the D and B decay contributions are also shown.  
The contribution of beauty decays to the heavy-flavour electron yield increases with $\pt$ and is described by FONLL pQCD calculations, both at $\s=200$\GeV and at $\s = 2.76$\TeV (see \fig{fig:pp:eHcorrpp}~\cite{Aggarwal:2010xp,Abelev:2014hla}). The beauty contribution to heavy-flavour electron yields becomes as important as the charm one at $\pt \sim 5$\GeVc.  
The results of $e_{HF}$--$\Dzero$ angular correlations at $\s = 200$\GeV are consistent with the $e_{HF}$--h ones~\cite{Aggarwal:2010xp}, see \fig{fig:pp:eHcorrppSTAR}.   
%
%
At the LHC, the preliminary results of D--h angular correlations in \pp collisions at $\s = 7$\TeV are described by various recent PYTHIA tunes~\cite{Bjelogrlic:2014kia}.  
Analogously, the azimuthal correlations of $\jpsi$ with charged hadrons ($\jpsi$--h) can be used to estimate beauty contribution to the inclusive $\jpsi$ production~\cite{Abelev:2009qaa,Adamczyk:2012ey}. The near-side $\jpsi$--h azimuthal correlations originate mostly from non-prompt $\jpsi$ coming from B mesons decays, ${\rm B} \rightarrow \jpsi + X$. 

Recent experimental analyses of associated heavy-flavour production include the measurements of: 
\begin{enumerate} 
\item double $\jpsi$ production at LHCb~\cite{Aaij:2011yc}, D0~\cite{Abazov:2014qba} and CMS~\cite{Khachatryan:2014iia},  
\item open charm hadron plus a $\jpsi$ or another open charm hadron at LHCb~\cite{Aaij:2012dz}, 
\item open charm meson or jet plus a Z boson at LHCb~\cite{Aaij:2014hea} and D0~\cite{Abazov:2013hya}, 
\item open charm hadron plus a W boson at CMS~\cite{Chatrchyan:2013uja}, 
\item $\jpsi$ and W production at ATLAS~\cite{Aad:2014rua}, 
\item $\jpsi$ and Z production at ATLAS~\cite{Aad:2014kba}, 
\item open beauty hadron or jet plus a Z boson at CDF~\cite{Aaltonen:2008mt} and D0~\cite{Abazov:2013uza}, and at ATLAS~\cite{Aad:2011jn}, CMS~\cite{Chatrchyan:2013zja} and LHCb~\cite{Aaij:2014gta}, 
\item the search of production of $\Upsilon(1S)$ associated with W or Z production at CDF~\cite{Aaltonen:2014rda}, 
\item the search of the exclusive decay of $H^0$ into $\jpsi+\gamma$ and  $\Upsilon+\gamma$~\cite{Aad:2015sda}. 
\end{enumerate} 
 
\begin{figure}[!t] 
\begin{center} 
\includegraphics[width=0.35\textwidth]{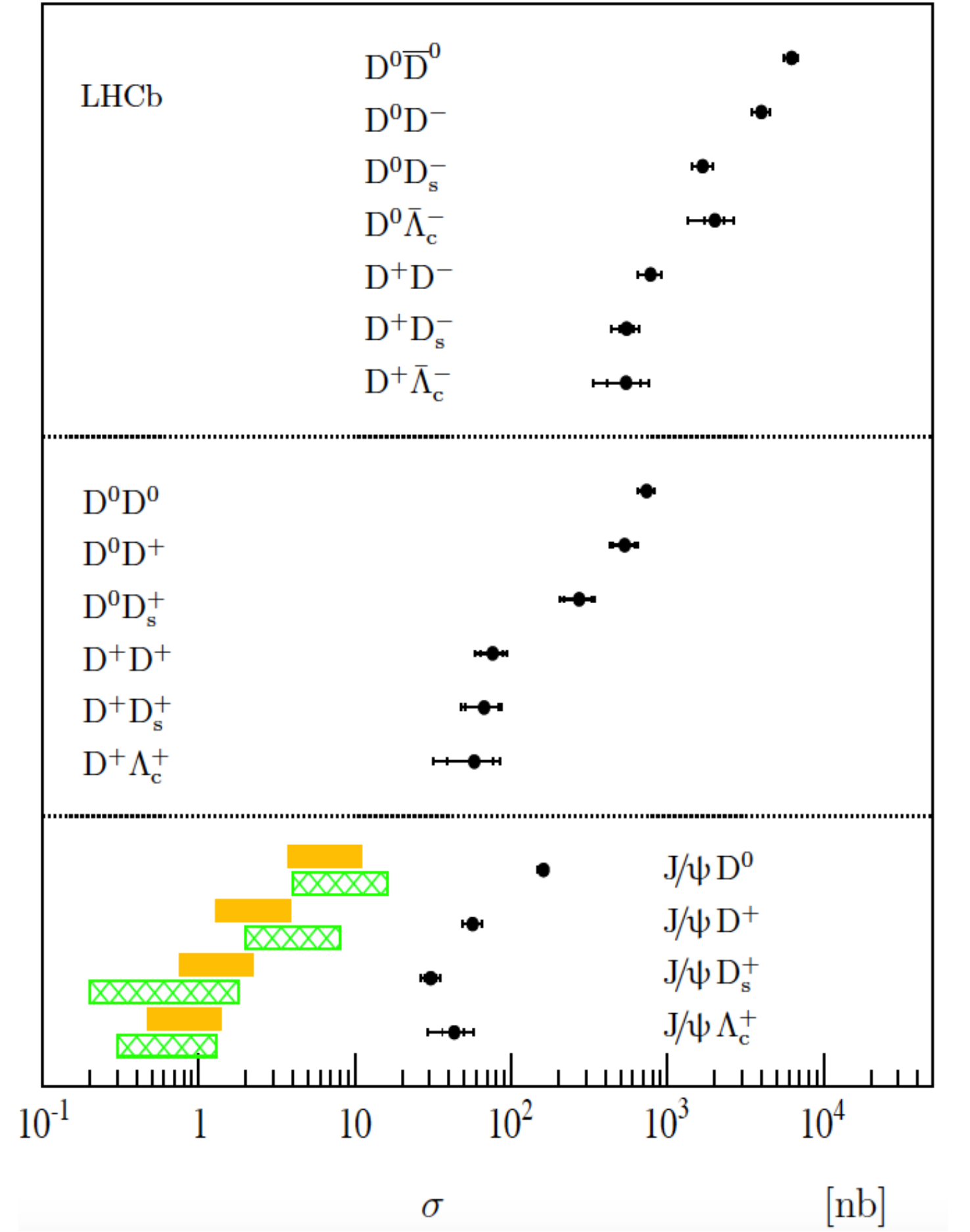} 
\includegraphics[width=0.35\textwidth]{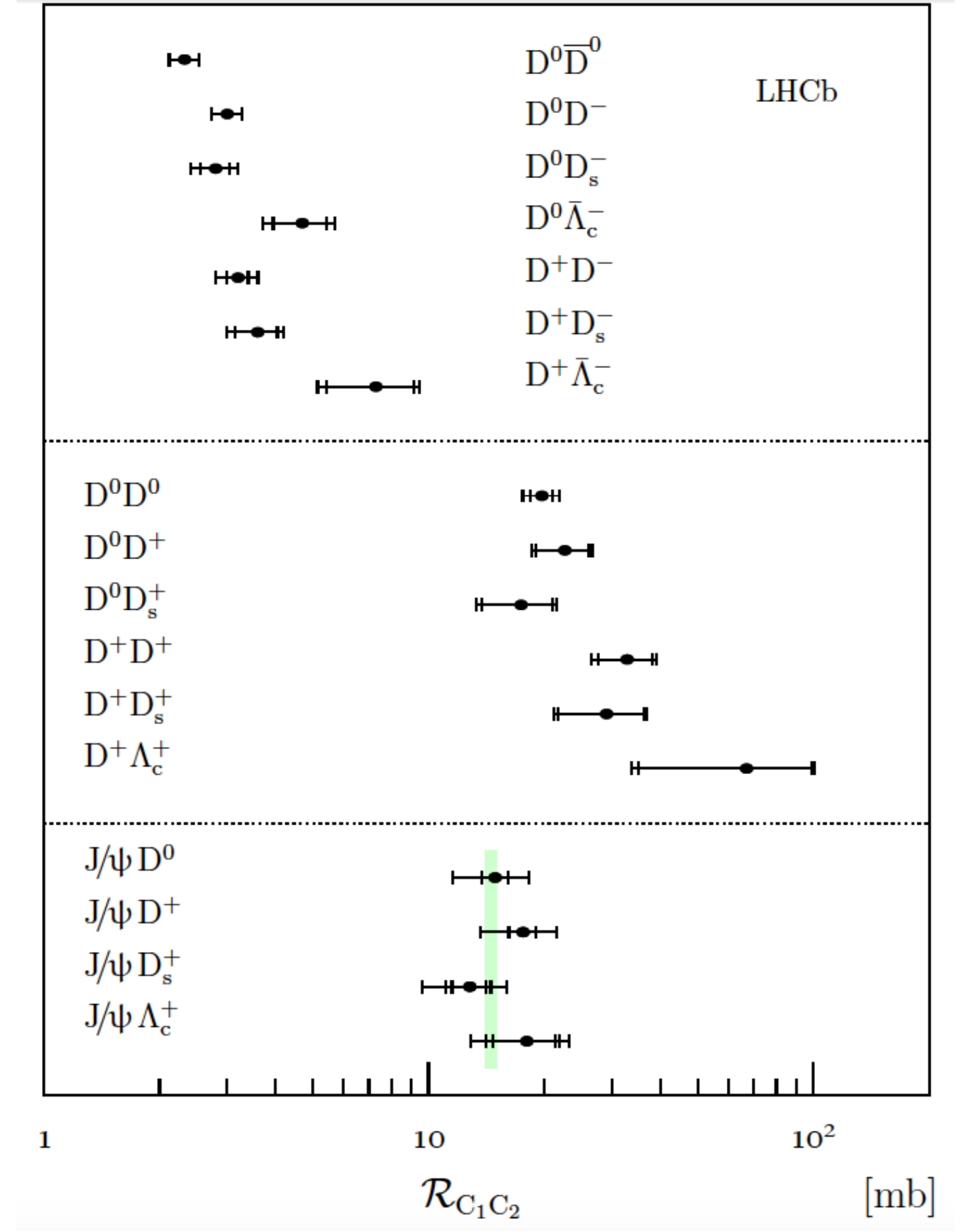} 
\caption[Double charm cross sections]{ 
\label{fig:pp:DoubleCharm} 
Left: Cross sections for double open charm hadron production (top) and open charm hadron plus $\jpsi$ meson (bottom) in \pp collisions at $\s=7$~TeV. 
Right: Measurement of $R_{C_1 \, C_2}$ for double open charm hadron production (top) and open charm hadron plus $\jpsi$ meson (bottom)~\cite{Aaij:2012dz}.}  
\end{center} 
\end{figure}

The measurements of $\jpsi$ plus open charm hadron and of double open charm hadron cross sections are summarised in \fig{fig:pp:DoubleCharm}~(left). The measurements of the production associated with a $\jpsi$ are compared to two computations of the cross sections shown as green hatched areas~\cite{Berezhnoy:1998aa} and yellow shaded areas~\cite{Lansberg:2008gk}. These are calculations of charm production in the hard scattering process of the collision, and underestimate by one order of magnitude the measured cross sections. This suggests that a large contribution to double charm production arises from double-parton scatterings (DPS) where both scatterings involve charm production. Therefore, in addition to providing useful information on the quarkonium-production mechanisms, associate-quarkonium-production observables can also be a rich source of information to understand the physics underlying DPS. 
 
This is also supported by the measurement of the ratio of the double and inclusive production cross sections, defined as $R_{C_1 \, C_2}=\alpha \left( \sigma_{C_1} \sigma_{C_2} / \sigma_{C_1C_2} \right)$, where $\alpha=1/4$ when $C_1$ and $C_2$ are charge conjugates of each other, and $\alpha=1/2$ otherwise. This quantity, which would be equal to $\sigma_{\rm eff}$ in case of a pure DPS yield, was evaluated by LHCb for the different aforementioned observed systems. These are plotted in \fig{fig:pp:DoubleCharm}~(right) and are compared, in the case of $\jpsi+$\,charm, to the results obtained from multi-jet events at the Tevatron, displayed by a green shaded area in the figure.  They point at values close to 15 mb. 
 
The cross section measured by LHCb in the region $2<y<4.5$ and $0<\pt<10$\GeVc is~\cite{Aaij:2011yc} 
\begin{equation} 
\sigma_{{\rm pp} \to \jpsi \, \jpsi + X} = 5.1 \pm 1.0({\rm stat.}) \pm 1.1({\rm syst.})\,{\rm nb}, 
\end{equation} 
and was found to be in agreement with various theoretical models (\eg dominated \cite{Kom:2011bd,Baranov:2011ch,Berezhnoy:2012xq,Baranov:2012re} or not~\cite{Kartvelishvili:1984ur,Humpert:1983yj,Vogt:1995tf,Li:2009ug,Qiao:2009kg,Ko:2010xy,Berezhnoy:2011xy,Li:2013csa,Lansberg:2013qka,Sun:2014gca} by DPS contributions). At this stage,  
the experimental and theoretical uncertainties both on the yield and the invariant mass spectrum are certainly too large to draw any firm conclusion, as recently discussed in~\cite{Lansberg:2013qka,Lansberg:2014swa}. 
 
However, double $\jpsi$ production has recently been studied by D0~\cite{Abazov:2014qba} and CMS~\cite{Khachatryan:2014iia}  
respectively at large rapidity separations and large transverse momenta. As for now, the D0~\cite{Abazov:2014qba} study is  
the only one which really separated out the double- and single-parton-scattering contributions by using the yield dependence on  
the (pseudo)rapidity difference between the J$/\psi$ pair, $\Delta y$, an analysis which was first proposed in Ref.~\cite{Kom:2011bd}.  
The DPS rapidity-separation spectrum is much broader and it dominates at large $\Delta y$. D0 has obtained that, 
in the region where DPS should dominate, the extracted value of $\sigma_{\rm eff}$ is on the order of 5 mb, that is significantly smaller  
than the values obtained with multi-jet events and $\jpsi+$\,charm as just discussed.  At small rapidity separations, the usual 
single-parton-scattering (SPS) contribution is found to be dominant and the yield is  well accounted for by the CSM at  
NLO~\cite{Lansberg:2013qka,Sun:2014gca,Lansberg:2014swa}. CO contributions are only expected to matter at very large transverse momenta, in particular 
at large values of the smaller \pt of both \pt of each \jpsi. 
 
Such a small value of $\sigma_{\rm eff}$ (meaning a large DPS yield) has been shown to be supported by the  
CMS measurement~\cite{Khachatryan:2014iia} at 7 TeV which overshoots by orders of magnitude the  
NLO SPS predictions at large transverse momenta. Indeed, adding the DPS yield obtained  
with $\sigma_{\rm eff}=5$~mb solves~\cite{Lansberg:2014swa} this apparent discrepancy first discussed 
in~\cite{Sun:2014gca}.

\begin{figure}[!t] 
\begin{center} 
\includegraphics[width=0.5\textwidth]{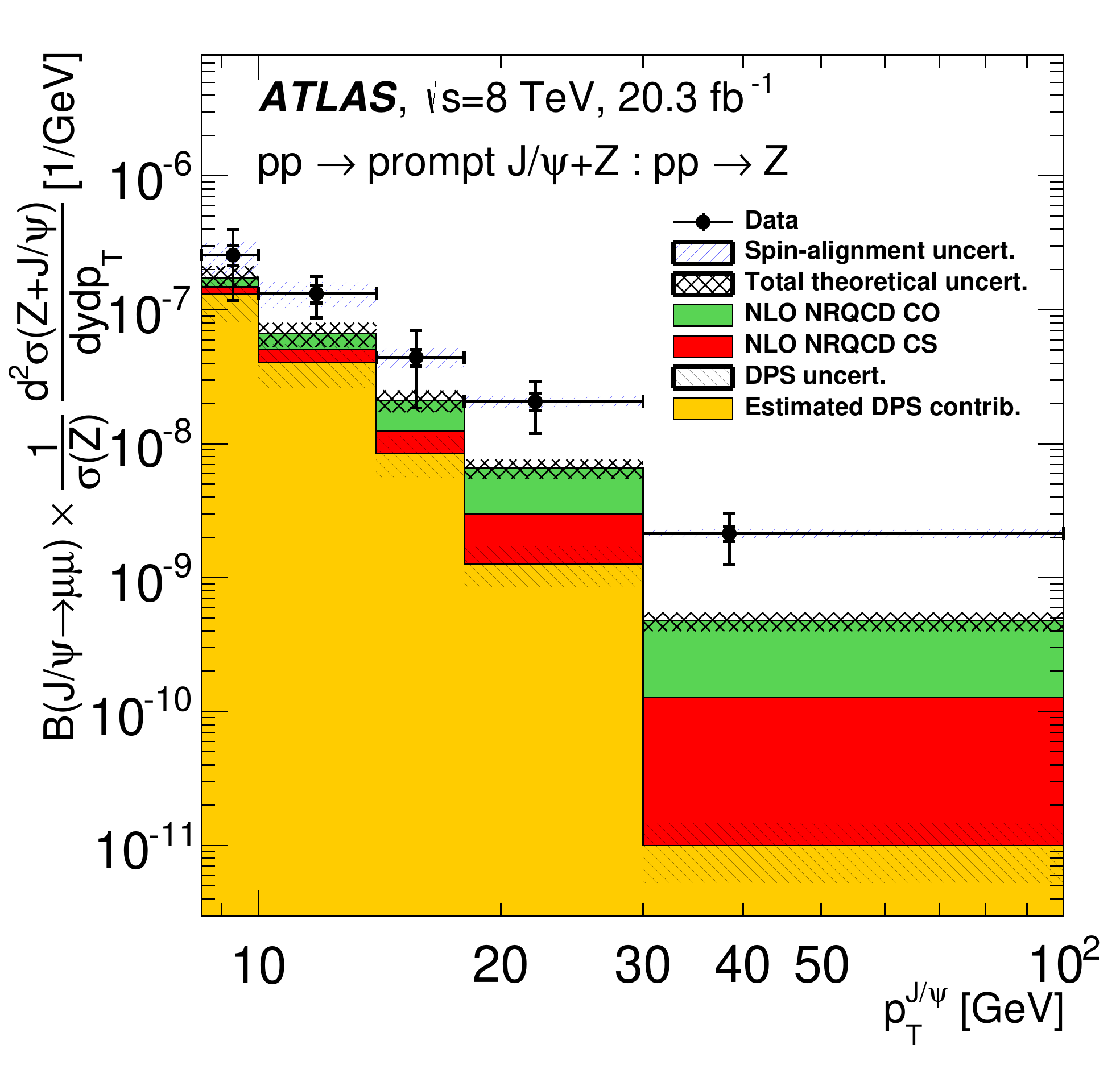} 
\caption[Double charm cross sections]{ 
\label{fig:pp:AtlasZJpsi} 
Production cross section of $\jpsi$ mesons in association with a Z boson (normalised to that of a Z boson) as a function of the $\jpsi$ $\pt$ in \pp collisions at $\s=8$~TeV~\cite{Aad:2014kba} 
compared to CO and CS theoretical predictions~\cite{Gong:2012ah,Mao:2011kf}.}  
\end{center} 
\end{figure}

Finally, measurements of vector bosons, W and Z, associated with a heavy quark or with a $\jpsi$ could also give access to the PDF as well as to DPS studies, in addition to providing complementary information on quarkonium production. As for now, both ATLAS measurements involving a \jpsi and a vector boson~\cite{Aad:2014rua,Aad:2014kba} are difficult to interpret. It seems that the observed yields are systematically higher than the expectations from the DPS and SPS yields as shown for $\jpsi+$Z in \fig{fig:pp:AtlasZJpsi}. 
 
To summarize, the study of associated production of heavy quarks and heavy quarkonia has really taken off with the advent of the LHC and the analysis of the complete data sample taken at the Tevatron. There is no doubt that forthcoming studies will provide much more new information --and probably also puzzles-- on the production of these particles. It is also probable that some of these observables at LHC energies are dominated by DPS contributions and, in such a case, specific nuclear dependences should be observed in proton-nucleus and nucleus-nucleus collisions (see \eg \cite{Strikman:2001gz,d'Enterria:2013ck}).

\subsection{Summary and outlook}
\label{sec:pp:Summary}

The LHC \RunOne provided  a complete set of cross section and polarisation measurements in the charm and beauty sector in \pp collisions at \s = 2.76\TeV,  7\TeV and 8\TeV, that can be summarised as follows: 
\begin {itemize} 
\item Heavy-flavour decay lepton \pt- and \ycm-differential production cross sections are well described by pQCD calculations. 
\item D meson \pt-differential cross sections are well described by pQCD calculations within uncertainties. FONLL and POWHEG central calculations tend to underestimate the data, whereas GM-VFNS tends to overestimate it. The \lambdacplus \pt-differential cross section was measured up to 8 \GeVc and is well described by GM-VFNS. 
\item The \pt-differential cross section of charmonia from beauty decays (non-prompt \jpsi, \psiP, \etac, \ensuremath{\chi_{c1}}\xspace and \ensuremath{\chi_{c2}}\xspace )  at low to intermediate \pt is well described by pQCD calculations. At high \pt the predictions tend to overestimate the data. \pt and \ycm-differential cross section measurements were performed for exclusive decays: $\mathrm{B}^{\pm}$, $\mathrm{B}^{0}$ and $\mathrm{B}^0_s$. $b$-jet cross section measurements are well described by pQCD calculations taking into account matching between NLO calculations and parton showers. 
\item  The ${\rm B}_c^+$ \pt and \ycm-differential cross section was for the first time measured at the LHC and it is well reproduced by theory. 
\item Prompt \jpsi and \psiP differential cross sections were measured, none of the tested models can be ruled out due to large theoretical uncertainties.   
\item \upsa differential cross section description remains a challenge at mid and high \pt, LHC data being more precise than theory.  
\item Quarkonium polarisation studies were performed in various reference frames for \jpsi, \psiP and \ups. At present, none of the models can describe all observed features. 
\end{itemize} 
 
In summary, open charm and beauty differential cross sections are globally well described by pQCD, although the theoretical uncertainties are quite large at low $\pt$, especially in the case of charm production. 
On the other hand, quarkonium production mechanisms remain a puzzle, especially if one aims at describing the \pt- and \ycm-differential cross section and polarisation in the same framework, or predict low and high \pt quarkonium production. The comparison of data with model calculations is still limited by the theoretical uncertainties.  
 
In addition to the \pt- and \ycm-differential production cross sections, the LHC \RunOne has allowed first measurements of heavy-flavour production versus charged-particle multiplicity, azimuthal angular correlations to charged-particles or heavy-flavour hadrons, and of associated heavy-flavour production, giving more insight into the production mechanisms. Those measurements can be summarised as follows: 
\begin {itemize} 
\item Inclusive \jpsi (at central and forward rapidity), prompt D meson and non-prompt \jpsi (at central rapidity) yields were measured at \s =  7\TeV versus charged-particle multiplicity. Heavy-flavour yields increase as a function of charged-particle multiplicity at mid-rapidity; D meson results present a faster-than-linear increase at the highest multiplicities.  
Possible interpretations of these results are the contribution of multiple-parton interactions or the event activity accompanying heavy-flavour hadrons.  
The increase of the prompt D meson yields is qualitatively reproduced by an hydrodynamic calculation with the EPOS event generator and the percolation scenario.   
The \ups measurement at \s =  2.76\TeV also presents an increase with charged-particle multiplicity but the decrease of the fraction of the \upsn to the \upsa state is at present not understood. 
\item Measurements of the azimuthal correlations between charm (beauty) and anti-charm (anti-beauty) point to the importance of the near production via the gluon splitting mechanism in addition to the back-to-back production. 
\item \jpsi plus open charm and double open charm hadron production cross section measurements suggest a non-negligible contribution of double-parton scatterings to double charm production. 
Measurements of vector boson production in association with a \jpsi provide further constrains to model calculations.  
\end{itemize} 
 
The LHC \RunTwo will provide more precise and more differential cross section measurements at the centre-of-mass energy $\sqrt s = 13$\TeV. This will provide strong constraints to the theoretical calculations and further understanding on the production mechanisms.

\newpage


\newcommand{\fa}[1]{{\bf FA : #1}}

\section{Cold nuclear matter effects on heavy flavour and quarkonium production in p-A collisions}
\label{Cold nuclear matter effects}

Characterizing the hot and dense medium produced in heavy-ion (\AAcoll) collisions 
requires a quantitative understanding of the effects induced by the presence of nuclei in the initial-state, the so-called cold nuclear matter (CNM) effects.
These effects
can be studied in proton-nucleus (\pA) or deuteron-nucleus (d--A) collisions\footnote{In the following we will use the generic symbol \pA to denote both \pA and d--A collisions.}.

A way to quantify CNM effects is to measure the nuclear modification factor $\rpa^{\cal C}$ of hard processes, defined as the ratio of their production yield $N_{\rm pA}^{\cal C}$ in \pA\ collisions (in a given centrality class ${\cal C}$) and their \pp\ production cross section $\sigma_{\rm pp}$  at the same energy, scaled by the average nuclear overlap function ${\langle T_{\rm pA} \rangle}_{\cal C}$ (obtained with the Glauber model\cite{Miller:2007ri}),
\begin{eqnarray}
\rpa^{\cal C} = \frac{N_{\rm pA}^{\cal C}}{{\langle T_{\rm pA} \rangle}_{\cal C}\ \sigma_{\rm pp}} \, .
\end{eqnarray}
In ``minimum-bias"  \pA\ collisions (\ie without a selection on centrality), $\rpa^{\cal C}$ reduces to
\begin{eqnarray}
\rpa = \frac{\sigma_{\rm pA}}{A\ \sigma_{\rm pp}} \, .
\end{eqnarray}
The nuclear dependence of a centrality-integrated hard cross section \pA  is sometimes parametrised by $\alpha$ defined as
\begin{eqnarray}
\sigma_{\rm pA}=\sigma_{\rm pp}\,A^\alpha ,
\end{eqnarray}
where $A$ is the mass number. In the absence of CNM effects, the \pA\ production is expected to be proportional to $A$, leading to $\rpa=1$ and  $\alpha=1$.

This section starts (\sect{CNM_Intro}) with a brief introduction to the physics of CNM effects on heavy flavour and with a compilation of available \pA data. 
Next, the different theoretical approaches are discussed in \sect{CNM_theory}, before a review of recent RHIC and LHC experimental results in \sect{CNM_ExpData}.
Afterwards, the extrapolation of CNM effects from \pA to \AAcoll collisions is discussed in section~\sect{CNM_pAtoAA}, from both the theoretical and the experimental points of view.
Finally, \sect{CNM_status} includes a summary and a discussion of short-term perspectives.

\subsection{Heavy flavour in {\rm p--A} collisions}
\label{CNM_Intro}

 
 
Open and hidden heavy flavour production constitutes a sensitive probe of medium effects because heavy quarks are produced in hard processes in the early stage of the nucleus--nucleus collision. 
Open and hidden heavy-flavour production can be affected by the following CNM effects: 
\begin{itemize} 
\item {\it Modification of the effective partonic luminosity} in colliding nuclei, with respect to colliding protons. This effect is due to the different dynamics of partons within free protons with respect those in nucleons, mainly as a consequence of the larger resulting density of partons. 
These effects depend on $x$ and on the scale of the parton--parton interaction $Q^2$ (the square of the four-momentum transfer). 
In collinearly-factorised pQCD calculations the nuclear effects on the parton dynamics are described in terms of nuclear-modified PDFs (hereafter indicated as nPDF). Quite schematically three regimes can be identified for the nPDF to PDF ratio of parton flavour $i$, $R_i(x,Q^2)$, depending on the values of $x$: a depletion ($R_i<1$) ---often referred to as \emph{shadowing} and related to phase-space saturation--- at small $x \lesssim 10^{-2}$, a possible enhancement $R_i>1$ (\emph{anti-shadowing}) at intermediate values $10^{-2} \lesssim x \lesssim 10^{-1}$, and the EMC effect, a depletion taking place at large $x \gtrsim 10^{-1}$. 
The $R_i(x,Q^2)$ parametrisations are determined from a global fit analyses of lepton--nucleus and proton--nucleus data (see \sect{sec:npdf}). 

\item 
The physics of {\it parton saturation} at small $x$ can be also described within the {\it Colour Glass Condensate (CGC)} theoretical framework. Unlike the nPDF approach, which uses DGLAP linear evolution equations, the CGC framework is based on the Balitsky-Kovchegov or JIMWLK non-linear evolution equations (see \sect{sec:saturation}).  
\item {\it Multiple scattering of partons} in the nucleus before and/or after the hard scattering, leading to parton energy loss (either radiative or collisional) and transverse momentum broadening (known as the Cronin effect). In most approaches (see \sect{sec:eloss}) it is characterized by the transport coefficient of cold nuclear matter, $\hat{q}$. 
\item Final-state inelastic interaction, or {\it nuclear absorption}, of \QQbar bound states when passing through the nucleus.   
The important parameter of these calculations is the ``absorption'' (or break-up)  cross section \sabs, namely the inelastic cross section of a heavy-quarkonium state with a nucleon.  
\item On top of the above genuine CNM effects, the large set of particles (partons or hadrons) produced in \pA\ collisions at high energy may be responsible for a modification of open heavy flavour or quarkonium production. It is still highly-debated whether this set of particles could form a ``medium" with some degree of collectivity. If this was the case, this medium could impart a flow to heavy-flavour hadrons. 
Moreover, heavy quarkonia can be dissociated by  {\it comovers}, \ie, the partons or hadrons produced in the collision in the vicinity of the heavy-quarkonium state (see \sect{sec:absorption}). 
\end{itemize} 
 
Assuming factorisation, and neglecting isospin effects, the hadroproduction cross section of a heavy-quark pair \QQbar is given by 
\begin{eqnarray} 
\label{cross-section-factorization} 
\sigma_{{\rm pA}\to \QQbar+X}[\snn] = A \sum_{i,j} \int_0^1 \dd x_i \int_0^1 \dd x_j\,f_i^{\rm N}(x_i,\mu_F^2)\,f_j^{\rm N}(x_j,\mu_F^2)\,\hat{\sigma}_{ij\to\QQbar+X}[x_i, x_j, \snn, \mu_F^2,\mu_R^2] \, , 
\end{eqnarray} 
where $f_i^{\rm N}$ are the nucleon parton distributions, $i$ ($j$) denotes all possible partons in the proton (nucleus) carrying a fraction $x_i$ ($x_j$) of the nucleon momentum, 
$\hat{\sigma}_{ij\to\QQbar+X}$ is the partonic cross section, $\snn$ is the nucleon--nucleon centre-of-mass energy of the collision, and $\mu_F$ ($\mu_R$) is the factorization (renormalisation) scale of the process. 
In high energy hadron collisions (especially at RHIC and LHC), heavy-quarks are mainly produced by gluon fusion~\cite{Mangano:1997ri}. 
 
For a $2 \to 1$ partonic process giving a particle of mass $m$, at leading order there is a direct correspondence between the momentum fractions and the rapidity $y$ of the outgoing particle in the nucleon--nucleon centre-of-mass (CM) frame, 
\begin{eqnarray} 
\label{2-1_x-Bjorken} 
x_1 = \frac{m}{\snn}\exp(y) & \text{and} & x_2 = \frac{m}{\snn}\exp(-y)\,, 
\end{eqnarray} 
where we have indicated as $x_2$ the smallest of the two $x$ values probed in the colliding nucleons. 
For a $2 \to 2$ partonic process, the extra degree of freedom coming from the transverse momentum results  
 in a less direct correspondence leading to the following useful relations 
\begin{eqnarray} 
\label{OHF_x-Bjorken} 
\text{open heavy-flavour (D and B mesons...)} & & x_2 \approx \frac{2 \mt}{\snn}\exp(-y), \\ 
\label{QQ_x-Bjorken} 
\text{quarkonia ($\jpsi$, $\Upsilon$...)} & & x_2 \approx \frac{\mt+\pt}{\snn}\exp(-y). 
\end{eqnarray} 
where $\mt = \sqrt{m^2+\pt^2}$ is the transverse mass of the outgoing particle of mass $m$, transverse momentum $\pt$ and rapidity $\ycm$ in the centre-of-mass frame. 
So, the typical resolution scale should be of the order of the transverse mass of the particle produced. 
 
The typical range for the momentum fractions probed is therefore a function of both the acceptance of the detector (rapidity coverage), and the nature of the particles produced and their associated energy scale. 
Moreover, assuming different underlying partonic production processes can end up in average values of $x$ that can differ from one another. 
 
 
 
Studies of \pA\ collisions since 1980 were first performed on fixed-target experiments at SPS, Tevatron and HERA, and more recently at colliders, RHIC and LHC. 
Current available data are summarised in \tab{Tab-pA_data_collider} for collider experiments and in \tab{Tab-pA_data_fixedtarget} for fixed-target experiments. 
This section is focused on the most recent results from the RHIC and LHC experiments, and their theoretical interpretation. 
 
\begin{table}[!t] 
\centering 
\caption{Available \pA\ data in collider: the probes, the colliding system, $\snn$, the kinematic range (with $\ycm$ the rapidity in the centre-of-mass frame), the observables (as a function of variables) are given as well as the references.} 
\label{Tab-pA_data_collider} 
\begin{tabular}{p{0.12\textwidth}p{0.10\textwidth}p{0.07\textwidth}p{0.21\textwidth}p{0.30\textwidth}p{0.10\textwidth}} 
\hline 
Probes & Colliding system & \snn (TeV) & \ycm & Observables (variables) & Ref. \\ 
\hline 
& & & PHENIX & & \\ 
\hline 
\hfe & \dAu & 0.2 & $|\ycm|<0.35$ & \rdau(\pt,\Ncoll), $\langle\pt^2\rangle$ & \cite{Adare:2012yxa} \\ 
\hfm & & & $1.4<|\ycm|<2$ & \rdau(\Ncoll,\pt) & \cite{Adare:2013lkk} \\ 
\bbbar & & & $|\ycm|<0.5$ & $\sigma(\ycm)$ & \cite{Adare:2014iwg} \\ 
$e^\pm,\mu^\pm$ & & & $|\ycm|<0.5$ \& $1.4<\ycm<2.1$ & $\Delta\phi$, $J_{\rm dAu}$ & \cite{Adare:2013xlp} \\ 
\jpsi & & & $-2.2<\ycm<2.4$ & \rdau, \rcp(\Ncoll,\ycm,$x_2$,\xf,\pt), $\alpha$ & \cite{Adler:2005ph,Adare:2007gn,Adare:2010fn} \\ 
& & & $-2.2<\ycm<2.2$ & \rdau(\pt,\ycm,\Ncoll), $\langle\pt^2\rangle$ & \cite{Adare:2012qf} \\ 
\jpsi, \psiP, \chic & & & $|\ycm|<0.35$ & \rdau(\Ncoll), double ratio & \cite{Adare:2013ezl} \\ 
\ups & & & $1.2<|\ycm|<2.2$ & \rdau(\ycm,$x_2$,\xf), $\alpha$ & \cite{Adare:2012bv} \\ 
\hline 
& & & STAR & & \\ 
\hline 
\Dzero, \hfe & \dAu & 0.2 & $|\ycm|<1$ & yield(\ycm,\pt) & \cite{Adams:2004fc} \\ 
\ups & & & $|\ycm|<1$ & $\sigma$, \rdau(\ycm,\xf), $\alpha$ & \cite{Adamczyk:2013poh} \\ 
\hline 
& & & ALICE & & \\ 
\hline 
$D$ & \pPb & 5.02 & $-0.96<\ycm<0.04$ & $\sigma$, \rppb(\pt,\ycm) & \cite{Abelev:2014hha} \\ 
\jpsi & & & $-4.96<\ycm<-2.96$ \& $2.03<\ycm<3.53$ & $\sigma$, \rppb(\ycm), \rfb & \cite{Abelev:2013yxa} \\ 
\jpsi, \psiP & & & & $\sigma$, \rppb(\ycm,\pt), double ratio & \cite{Abelev:2014zpa} \\  
\jpsi & & & \& $-1.37<\ycm<0.43$ & $\sigma$(\ycm,\pt), \rppb(\ycm,\pt), [\rppb(+\ycm)$\cdot$\rppb(-\ycm)] (\pt) & \cite{Adam:2015iga} \\  
\upsa, \upsb & & & & $\sigma$, \rppb(\ycm), \rfb, ratio & \cite{Abelev:2014oea} \\ 
\hline 
& & & ATLAS & & \\ 
\hline 
\jpsi (from B) & \pPb & 5.02 & $-2.87< \ycm <1.94$ & $\sigma(\ycm,\pt)$, ratio($y$,\pt), \rfb($|y|$,\pt) & \cite{Aad:2015ddl} \\ 
\hline 
& & & CMS & & \\ 
\hline 
\upsn & \pPb & 5.02 & $|\ycm|<1.93$ & double ratio ($E_{\rm T}^{\eta>4},N_{\rm tracks}^{|\eta|<2.4})$ & \cite{Chatrchyan:2013nza} \\ 
\hline 
& & & LHCb & & \\ 
\hline 
\jpsi (from B) & \pPb & 5.02 & $-5.0<\ycm<-2.5$ \& $1.5<y<4.0$ & $\sigma(\pt,\ycm)$, \rppb(\ycm), \rfb(\ycm,\pt) & \cite{Aaij:2013zxa} \\ 
\upsn & & & & $\sigma(\ycm)$, ratio(\ycm), \rppb(\ycm), \rfb & \cite{Aaij:2014mza} \\ 
\hline 
\end{tabular} 
\end{table} 
 
\begin{table}[!htp] 
\centering 
\caption{Available \pA\ data in fixed target: the probes, the target, $\snn$, the kinematic range (with $\ycm$ the rapidity in the centre-of-mass frame), the observables (as a function of variables) are given as well as the references.  
The flag ($\dagger$) means that a cut on $|\cos\theta_{\rm CS}|<0.5$ is applied in the analysis, where $\theta_{\rm CS}$ is the decay muon angle in the Collins-Soper frame. 
Feynman-$x$ variable $\xf=\frac{2p_{\rm L,CM}}{\sqrt{s_{\rm NN}}}$, where $p_{\rm L,CM}$ is the longitudinal momentum of the partonic system in the CM frame, is connected to the momentum fraction variables by $\xf\approx x_1-x_2$, in the limit $p_{\rm T} \ll p$.} 
\label{Tab-pA_data_fixedtarget} 
\begin{tabular}{p{0.13\textwidth}p{0.14\textwidth}p{0.10\textwidth}p{0.19\textwidth}p{0.23\textwidth}p{0.09\textwidth}} 
\hline 
Probes & Target & \snn (GeV) & \ycm (or \xf) & Observables (variables) & Ref. \\ 
\hline 
& & & NA3 & & \\ 
\hline 
\jpsi & H$_2$, Pt & 16.8 - 27.4 & $0<\xf<0.9$ & $\sigma(\xf,\pt)$ & \cite{Badier:1983dg,Badier:1985ri} \\ 
\hline 
& & & NA38 & & \\ 
\hline 
\jpsi, \psiP & Cu, U & 19.4 & $-0.2<y<1.1$ & $\sigma(E_{\rm T},A)$, $\langle \pt^{(2)}\rangle(\epsilon)$, ratio($\epsilon,E_{\rm T},A,L)$ & \cite{Baglin:1991gy,Baglin:1994ui,Baglin:1991vb} \\ 
\jpsi, \psiP, \ccbar & W & 19.4 & $0<y<1$ & ratio($\epsilon$), $\sigma_{\ccbar}(p_{\rm lab})$ & \cite{Lourenco:1993mr} \\ 
\jpsi, \psiP & C, Al, Cu, W & 29.1 & $-0.4<\ycm<0.6$ ($\dagger$) & $\sigma(A)$ and ratio$(A)$ & \cite{Abreu:1998ee} \\ 
\jpsi, \psiP, DY & O, S & 19.4 (29.1) & $0(-0.4)<\ycm<1(0.6)$ & $\sigma(A,L)$, ratio($A,L$) & \cite{Abreu:1999nn} \\ 
\hline 
& & & NA38/NA50 & & \\ 
\hline 
\ccbar & Al, Cu, Ag, W & 29.1 & $-0.52<\ycm<0.48$ ($\dagger$) & $\sigma_{\ccbar}$ & \cite{Abreu:2000nj} \\ 
\hline 
& & & NA50 & & \\ 
\hline 
\jpsi, \psiP, DY & Be, Al, Cu, Ag, W & 29.1 & $-0.4<\ycm<0.6$ & $\sigma(A)$, ratio$(A,E_{\rm T},L)$, $\sabs$ & \cite{Alessandro:2003pi,Alessandro:2006jt} \\ 
\jpsi, \psiP & & & $-0.1<\xf<0.1$ ($\dagger$) & $\sigma(A,L)$, \sabs($\xf$) & \cite{Alessandro:2003pc} \\ 
\ups, DY & & & $-0.5<\ycm<0.5$ ($\dagger$) & $\sigma(A)$, $\langle\pt^2\rangle(L)$, $\langle\pt\rangle$ & \cite{Alessandro:2006dc} \\ 
\hline 
& & & NA60 & & \\ 
\hline 
\jpsi & Be, Al, Cu, In, W, Pb, U & 17.3 (27.5) & $0.3(-0.2)<y<0.8(0.3)$ ($\dagger$) & $\sigma$, \sabs, ratio($L$), $\alpha(\xf,x_2)$ & \cite{Arnaldi:2010ky} \\ 
\hline 
& & & E772 & & \\ 
\hline 
\jpsi,\psiP & H$_2$, C, Ca, W & 38.8 & $0.1<\xf<0.7$ & ratio($A$,\xf,\pt), $\alpha(\xf,x_2,\pt)$ & \cite{Alde:1990wa} \\ 
\ups & & & $-0.15<\xf<0.5$ & $\sigma(\pt,\xf)$, ratio($A$), $\alpha(\xf,x_2,\pt)$ & \cite{Alde:1991sw} \\ 
\hline 
& & & E789 & & \\ 
\hline 
\Dzero & Be, Au & 38.8 & $0<\xf<0.08$ & $\sigma(\pt)$, $\alpha(\xf,\pt)$, ratio & \cite{Leitch:1994vc} \\ 
\bbbar & & & $0<\xf<0.1$ & $\sigma(\xf,\pt)$ & \cite{Jansen:1994bz} \\ 
\jpsi & Be, Cu & 38.8 & $0.3<\xf<0.95$ & $\sigma(\xf)$, $\alpha(\xf)$ & \cite{Kowitt:1993ns} \\ 
\jpsi,\psiP & Be, Au & & $-0.03<\xf<0.15$ & $\sigma(\pt,\xf,y)$, ratio(\pt,\xf) & \cite{Schub:1995pu} \\ 
\jpsi & Be, C, W & & $-0.1<\xf<0.1$ & $\alpha$ (\xf,$x_{\rm target}$,\pt) & \cite{Leitch:1995yc} \\ 
\hline 
& & & E866/NuSea & & \\ 
\hline 
\jpsi, \psiP & Be, Fe, W & 38.8 & $-0.1<\xf<0.93$ & $\alpha$ (\pt,\xf) & \cite{Leitch:1999ea} \\ 
\jpsi & Cu & & $0.3<\xf<0.9$ & $\lambda_\theta(\pt,\xf)$ & \cite{Chang:2003rz} \\ 
\upsn, DY &  & & $0<\xf<0.6$ & $\lambda_\theta(\pt,\xf)$ & \cite{Brown:2000bz} \\ 
\hline 
& & & HERA-B & & \\ 
\hline 
D & C, Ti, W & 41.6 & $-0.15<\xf<0.05$ & $\sigma(\xf,\pt^2)$ & \cite{Abt:2007zg} \\ 
\bbbar, \jpsi & & & $-0.35<\xf<0.15$ & $\sigma$, ratio & \cite{Abt:2005qs} \\ 
\bbbar & & & $-0.3<\xf<0.15$ & $\sigma$ & \cite{Abt:2006xa} \\ 
\bbbar & C, Ti & & $-0.25<\xf<0.15$ & $\sigma$ & \cite{Abt:2002rd} \\  
\jpsi & C, Ti, W & & $-0.225<\xf<0.075$ & $\sigma(A,\ycm)$ & \cite{Abt:2005qr} \\ 
& & & $-0.34<\xf<0.14$ & $\langle\pt^2\rangle(A)$, $\alpha(\pt,\xf)$ & \cite{Abt:2008ya} \\ 
& C, W & & $-0.34<\xf<0.14$ & $\lambda_\theta,\lambda_\phi,\lambda_{\theta\phi}(\pt,\xf)$ & \cite{Abt:2009nu}\\ 
\jpsi, \psiP & C, Ti, W & & $-0.35<\xf<0.1$ & ratio$(\xf,\pt,A)$, $\alpha^\prime$-$\alpha$(\xf) & \cite{Abt:2006va} \\ 
\jpsi, \chic & & & & ratio(\xf,\pt) & \cite{Abt:2008ed} \\ 
\ups & C, Ti, W & & $-0.6<\xf<0.15$ & $\sigma(\ycm)$ & \cite{Abt:2006wc} \\ 
\hline 
\end{tabular} 
\end{table} 
 
\begin{figure}[!tp] 
        \centering 
        \includegraphics[width=0.55\columnwidth]{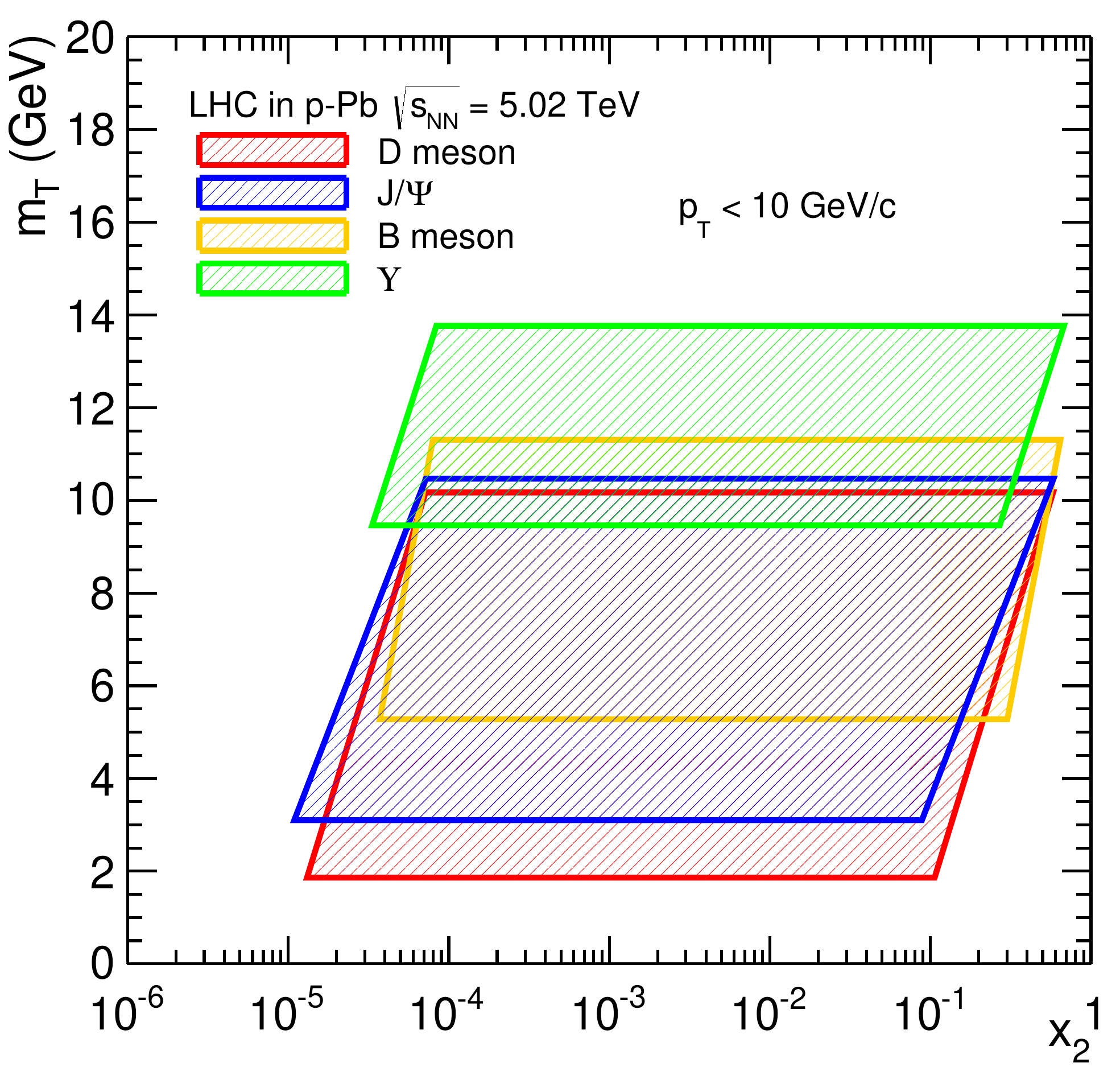} 
        \caption{Accessible $x_2$ and $m_{\rm T}$  range at the LHC ($|\ylab| < 4.5$) in p--Pb collisions at $\snn=5.02$ TeV 
for         different heavy-flavour probes (D and B mesons, $\jpsi$ and $\Upsilon$) with $0<\pt < 10$~GeV/$c$.} 
        \label{Fig-LHCx_range}   
\end{figure}

In LHC \RunOne p--Pb collisions, protons have an energy of 4 TeV and the Pb nuclei an energy $Z/A(4 {\rm TeV})=1.58$~TeV ($Z=82$, $A=208$), leading to $\snn=5.02$~TeV and a relative velocity of the CM with respect to the laboratory frame $\beta=0.435$ in the direction of the proton beam. 
The rapidity of any particle in the CM frame is thus shifted, $y=\ylab-0.465$. 
Applying those experimental conditions to heavy-flavour probes such as D and B mesons and quarkonia, and according to \eq{OHF_x-Bjorken} and~\eq{QQ_x-Bjorken}, leads to a large coverage of $x_2$ from $10^{-5}$ for the D meson at forward rapidity, to 0.5 for 10~GeV/$c$ $\Upsilon$ at backward rapidity, as reported in \fig{Fig-LHCx_range}. 
 
%
 
%

\subsection{Theoretical models for CNM effects}
\label{CNM_theory}

 
We discuss in this section various theoretical approaches to treat CNM effects,  
with emphasis on heavy-quark and quarkonium production at the LHC. 
 
\subsubsection{Typical timescales}\label{sec:timescales} 
 
Before discussing the various theoretical approaches on cold nuclear matter effects, it is useful to recall the  typical time-scales entering the process of heavy-quark hadron and quarkonium production in \pA\ collisions: 
\begin{itemize} 
 
\item The typical time to produce a heavy-quark pair \QQbar, sometimes referred to as the coherence time, which is of the order of $\tau_c \sim 1/m_{\QQbar} \lesssim 0.1$~fm/$c$ in the \QQbar\ rest frame. In the rest frame of the target nucleus, however, this coherence time, $t_{c}=E_{\QQbar}/m^2_{\QQbar}$  (where $E_{\QQbar}$ is the \QQbar\ energy in the nucleus rest frame), can be larger than the nuclear size, leading to shadowing effects due to the destructive interferences from the scattering on different nucleons. 
 
\item The time needed to produce the quarkonium state, also known as the formation time, is much larger than the coherence time. It corresponds to the time interval taken by the \QQbar\ pair to develop the quarkonium wave function. Using the uncertainty principle, it should be related to the mass splitting between the 1S and 2S states~\cite{Kopeliovich:1991pu}, \ie $\tau_{\rm f}\sim\left(m_{2S}-m_{1S} \right)^{-1}\sim$0.3--0.4~fm/$c$. Because of the Lorentz boost, this formation time  in the nucleus rest frame, $t_{\rm f}$, becomes much larger than the nuclear size at the LHC. Consequently the quarkonium state is produced far outside the nucleus and should not be sensitive to nuclear absorption. The time to produce a heavy-quark hadron is longer than for quarkonium production, of the order of ${\Lambda_{\rm QCD}}^{-1} \simeq 1$~fm /$c$ in its rest frame. 
 
\item Another important time-scale is the typical  time needed for the \QQbar\ pair to neutralise its colour. In the colour singlet model, this process occurs through the emission of a perturbative gluon and should thus occur in a time comparable to $\tau_{c}$. In the colour octet model (or colour evaporation model), colour neutralisation happens through a soft process, \ie on ``long'' time-scales, typically of the order the quarkonium formation time $\tau_{\rm f}$. 
\end{itemize} 
 
When discussing the possible nuclear absorption of a quarkonium state in the nucleus, it is common to compare the crossing time of the nucleus, $\tau_{\rm cross}$, which is the time spent by the  state in the nucleus~\cite{Adare:2013ezl} to its formation time $\tau_f$. It is given by $\tau_{\rm cross} = L  / (\beta_z\ \gamma)$, where $L$ is the  longitudinal path of the \QQbar\ pair through the nucleus, $\beta_z$ and $\gamma = \sqrt{1-\beta_z^2}$ are the velocity and Lorentz factor of the quarkonium along the beam direction, both given in the nuclear rest frame. 
 
 
\subsubsection{Nuclear PDFs}
\label{sec:npdf} 
 

The modification of parton densities in nuclei affects the yields of heavy-quark and quarkonium production. In this section, the effects of nPDF on $\jpsi$ and $\Upsilon$ production in p--Pb collisions at the LHC are first presented. The production of open beauty (through its decay into non-prompt $\jpsi$) is then discussed. 
 
\paragraph{$\jpsi$ and $\Upsilon$ production} 
 
The predictions for $\jpsi$ suppression due to the nuclear modifications 
of the  parton densities are described in this section and discussed by Vogt in~\cite{Vogt:2010aa}.   
Here we show results for the rapidity dependence of nPDF effects on $\jpsi$ 
and $\Upsilon$ production in \pPb\ collisions  at $\snn = 5.02$~TeV 
and neglecting any other CNM effect.

The results are 
obtained in the colour evaporation model (CEM) at next-to-leading order in the 
total cross section. 
In the CEM, the quarkonium \Qcal\  
production cross section in p--Pb collisions is some fraction, $F_\Qcal$, of  
all $Q \overline Q$ pairs below the $H \overline H$ threshold where $H$ is 
the lowest mass heavy-flavour hadron, 
\begin{eqnarray} 
\sigma_{{\rm pPb}\to\Qcal+X}^{\rm CEM}[\s]  = A \cdot F_\Qcal \sum_{i,j}  
\int_{4m_{\rm Q}^2}^{4m_H^2} \dd\hat{s} 
\int_0^1 \dd x_i \int_0^1 \dd x_j~ f_i(x_i,\mu_F^2)~ R_j^{\rm Pb}(x_j,\mu_F^2) \,f_j(x_j,\mu_F^2)~  
{\cal J}~\hat\sigma_{ij\to\QQbar+X}[\hat{s},\mu_F^2, \mu_R^2] \,  
\, , \label{sigtil} 
\end{eqnarray}  
where $A$ is the Pb mass number,  $ij = \qqbar$ or $gg$, and $\hat\sigma_{ij\to\QQbar+X}$ is the 
$ij\rightarrow \QQbar+X$ sub-process cross section of centre-of-mass energy $\hat s$.
${\cal J}$ is an appropriate Jacobian with dimension $1/\hat{s}$. 
$f_{i,j}$ is the proton PDF for the parton species $i$, while $R_j^{\rm Pb}$ is 
a nuclear PDF parametrisation for the parton species $j$
 (EPS09~\cite{Eskola:2009uj} for the results shown in this section). 
The normalisation factor $F_\Qcal$ is fitted  
to the forward (integrated over $\xf > 0$)  
$\jpsi$ cross section data on p, Be, Li, 
C, and Si targets (see~\cite{Nelson:2012bc} for details).  
In this way, uncertainties due to  
ignoring any cold nuclear matter effects which are on the order of a few percent 
in light targets are avoided.  The fits are restricted to the forward cross  
sections only.

The values of the central charm quark 
mass and scale parameters are  $m_c = 1.27 \pm 0.09$~GeV/$c^2$, 
$\mu_F/m_c = 2.10 ^{+2.55}_{-0.85}$, and $\mu_R/m_c = 1.60 ^{+0.11}_{-0.12}$  
\cite{Nelson:2012bc}. 
The normalization $F_\Qcal$ is obtained for the central set, 
$(m_c,\mu_F/m_c, \mu_R/m_c) = (1.27 \, {\rm GeV}/c^2, 2.1,1.6)$.   
The calculations for the estimation of the mass and scale uncertainties are 
multiplied by the same value of $F_\Qcal$ to 
obtain the $\jpsi$ uncertainty band~\cite{Nelson:2012bc}. 
$\Upsilon$ production is calculated in the same manner, with the central 
result obtained for $(m_b,\mu_F/m_b, \mu_R/m_b) = (4.65 \pm 0.09 \, {\rm GeV}/c^2, 1.4^{+0.77}_{-0.49},1.1^{+0.22}_{-0.20})$ 
\cite{Nelson_inprog}.   
In the 
NLO calculations of the rapidity and $\pt$ dependence, instead of $m_Q$, 
the transverse mass, $\mt$, is used with $\mt = \sqrt{m_Q^2 + \pt^2}$ 
where $\pt^2 = 0.5\,(p_{{\rm T}_Q}^2 + p_{{\rm T}_{\overline Q}}^2)$. 
All the  
calculations are NLO in the total cross section and assume that the intrinsic 
$\kt$ broadening is the same in  \pp\ as in \pPb. 
 
The mass and scale uncertainties are calculated  
based on results using the one standard deviation uncertainties on 
the quark mass and scale parameters.  If the central, higher and lower limits 
of $\mu_{R,F}/m$ are denoted as $C$, $H$, and $L$ respectively, then the seven 
sets corresponding to the scale uncertainty are  $\{(\mu_F/m,\mu_R/m)\}$ = 
$\{$$(C,C)$, $(H,H)$, $(L,L)$, $(C,L)$, $(L,C)$, $(C,H)$, $(H,C)$$\}$.     
The uncertainty band can be obtained for the best fit sets by 
adding the uncertainties from the mass and scale variations in  
quadrature.  
The uncertainty band associated to the EPS09 NLO set is obtained by calculating the deviations  
from the central EPS09 set 
for the 15 parameter variations on either side of the central set and adding 
them in quadrature.  The uncertainty on $R_{\rm pA}$ associated to the EPS09 NLO variations 
turns out to be larger than 
that coming from 
 the mass and scale variation, as it can be seen below. 
 
\begin{figure*}[t] 
\begin{center} 
\begin{tabular}{cc} 
\resizebox{0.47\textwidth}{!}{\includegraphics{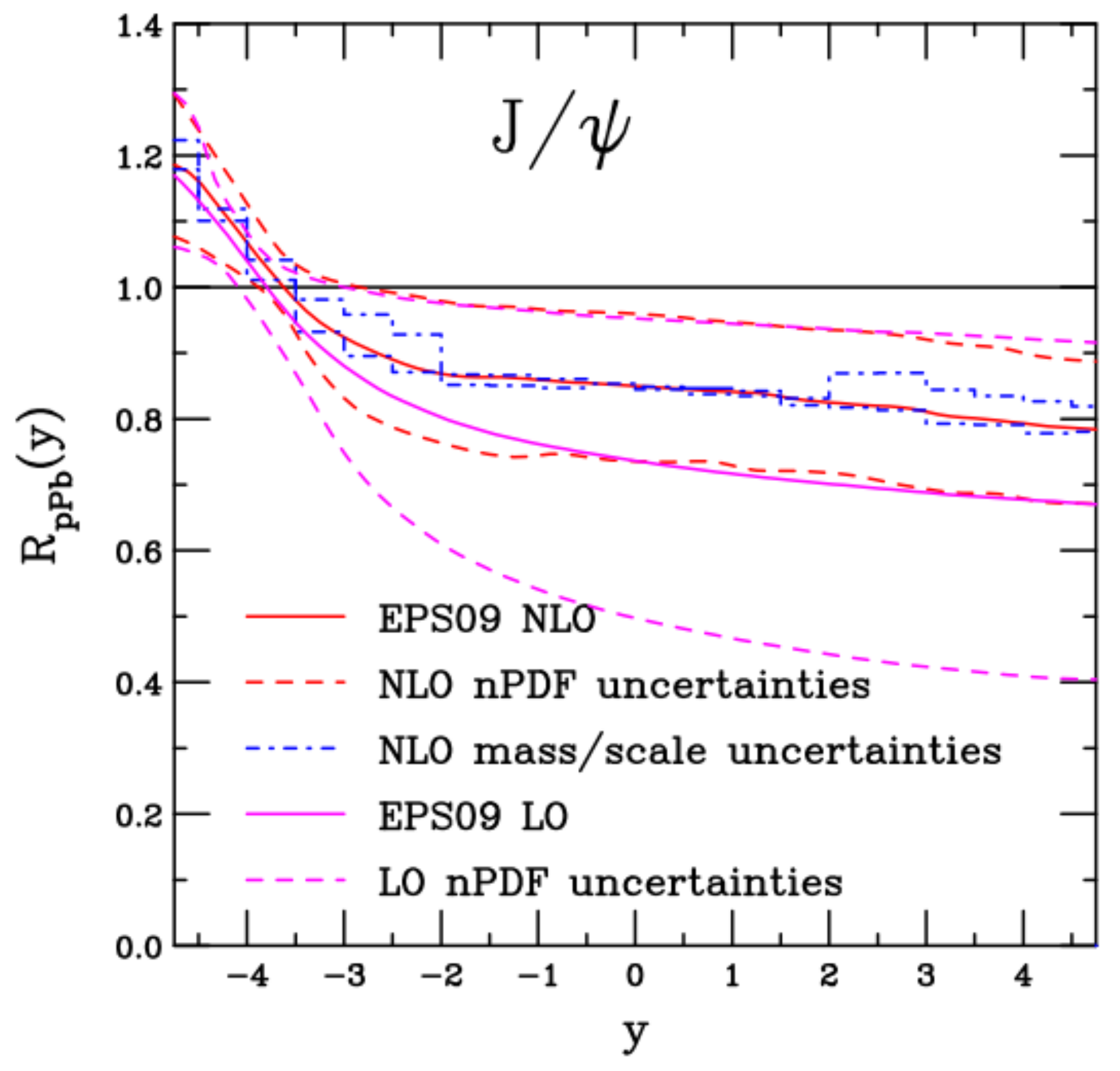}} & 
\resizebox{0.47\textwidth}{!}{\includegraphics{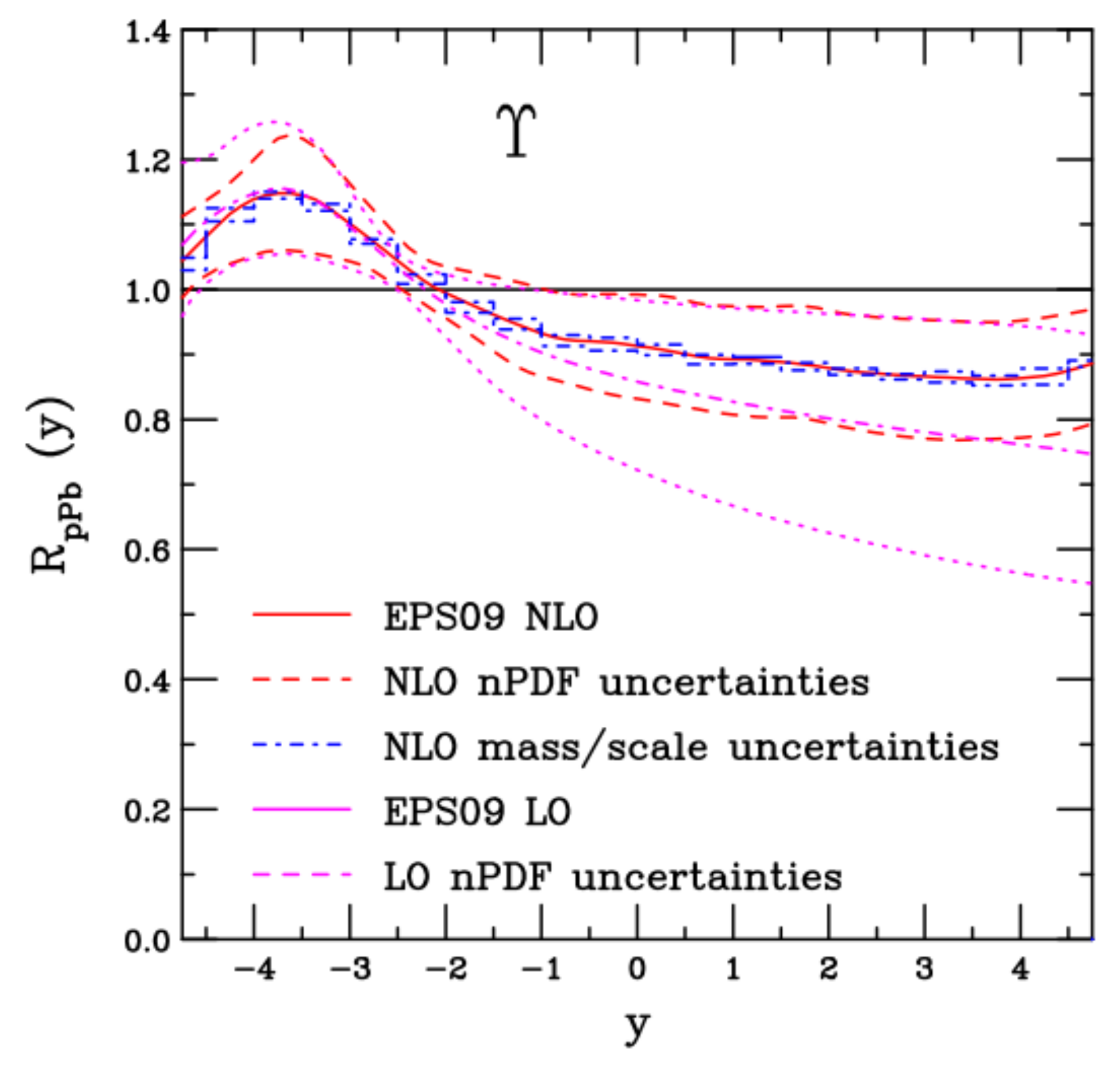}}  
\end{tabular} 
\end{center} 
\caption[]{The nuclear modification factor \rppb\ for $\jpsi$ (left) 
and $\Upsilon$ (right) production calculated using the EPS09 modifications 
 as a function of rapidity. 
The solid red histogram shows the central EPS09 NLO prediction (with its uncertainties shown as red dashed histograms) in \pPb\ collisions at $\snn=5.02$~TeV (integrated over $\pt$) 
while the dot-dashed 
blue histogram shows the dependence on mass and scale.  The magenta curves 
show the LO modification and the corresponding uncertainty band. 
The NLO $\jpsi$ results 
were originally shown in Ref.~\protect\cite{Albacete:2013ei}. 
} 
\label{fig:CNMJpsiUpsypt} 
\end{figure*}

\fig{fig:CNMJpsiUpsypt} (left) shows the uncertainty 
in the shadowing effect on $\jpsi$ due to the variations in the 30 EPS09 NLO 
sets~\cite{Eskola:2009uj} 
(dashed red) as well as those due to the mass and scale uncertainties 
(dashed blue) calculated with 
the EPS09 NLO central set.  The uncertainty band calculated in the CEM  
at LO with the EPS09 LO sets~\cite{Eskola:2009uj} is shown for comparison.   
It is clear that the LO results, represented by the smooth magenta curves in  
\fig{fig:CNMJpsiUpsypt}, exhibit a larger shadowing effect.  This difference 
between the LO results, 
also shown in \ci{Vogt:2010aa}, and the NLO calculations 
arises because the gluon distributions in the proton that the  
EPS09 LO and NLO gluon shadowing parametrisations 
are based on CTEQ61L and CTEQ6M, respectively, which behave very differently 
at low $x$ and moderate values of the factorization scale~\cite{Eskola:2009uj}.  If one uses 
instead the nDS or nDSg parametrisations~\cite{deFlorian:2003qf}, based on the GRV98 LO 
and NLO proton PDFs, the LO and NLO results differ by only a few  
percent.  
The  
right 
panel shows the same calculation for $\Upsilon$ production. 
Here the difference  
between the LO and NLO calculations is reduced because the mass 
scale, and hence the factorization scale, is larger.  The $x$ values 
probed are also correspondingly larger. 
 
The $\pt$ dependence of the nPDF effects 
at forward rapidity for $\jpsi$ and $\Upsilon$ has also been computed in Ref.~\protect\cite{Albacete:2013ei}.   
There is no LO comparison 
because the $\pt$ dependence cannot be calculated in the LO CEM. The effect 
is rather mild and \rppb increases slowly with $\pt$, 
 from roughly $\rppb\simeq  0.7$--$0.9$ for $\jpsi$ at low $\pt$ to $\rppb\simeq  1$ at $\pt= 20$~GeV/c.  There is little difference 
between the $\jpsi$ and $\Upsilon$ results for $\rppb(\pt)$ because, 
for $\pt$ above a few GeV$/c$, the $\pt$ scale dominates over the mass scale. 
The nPDF effects are somewhat similar for open heavy flavour as a function of \pt, yet the effects (estimated using EPS09 NLO) tend to go away faster with \pt\ due to the different production dynamics between quarkonium and open heavy flavour. 
 
\paragraph{Non-prompt $\jpsi$ production} 
 
The nPDF effects on non-prompt \jpsi\ (coming from B decays) has been investigated by Ferreiro et al. in~\cite{delValle:2014wha}. 
Contrary to the more complex case of bottomonium production, it is sufficient to rely  
on LO calculations~\cite{Combridge:1978kx} to deal with open-beauty production data  
integrated in $\pt$ as those of LHCb~\cite{Aaij:2013zxa}. Indeed such computations are sufficient  
to describe the low-$\pt$ cross section up to (1--2)\,$m_b$, where the bulk of the yield lies.  
 
The nPDF effects on non-prompt $\jpsi$ have been evaluated using 
 two parametrisations\footnote{By coherence with the use of LO hard matrix elements, one may prefer to use LO nPDFs.}, namely  
EPS09~LO~\cite{Eskola:2009uj}\footnote{To simplify the comparison, one simply uses the central curve of EPS09 as well as  
four specific extreme curves (minimal/maximal shadowing, minimal/maximal EMC effect),  
which reproduce the envelope of the gluon nPDF uncertainty encoded in EPS09 LO.}  
and nDSg LO~\cite{deFlorian:2003qf}. In addition to the choice of the nPDFs, one also has to fix the value of the  factorization scale  
$\mu_F$ which is set to $\mu_F=\sqrt{m_Q^2+\pt^2}$. One can also consider the spatial dependence of the nPDFs, either by simply assuming an inhomogeneous shadowing 
proportional to the local density~\cite{Klein:2003dj,Vogt:2004dh} or extracting it from a fit~\cite{Helenius:2012wd}.  
These effects would then translate into a 
non-trivial centrality (or impact parameter $b$) dependence of the nuclear modification factor.   
To this end, it is ideal to rely on a Glauber Monte-Carlo which does not factorise the different nuclear effects (such as JIN~\cite{Ferreiro:2008qj} which is used to study the nuclear matter effects on quarkonium production both at RHIC~\cite{Ferreiro:2009ur,Rakotozafindrabe:2012ss} and LHC~\cite{Ferreiro:2011xy,Ferreiro:2013pua} energies.) 
 
\begin{figure}[t] 
\centering 
\includegraphics[height=6.5cm]{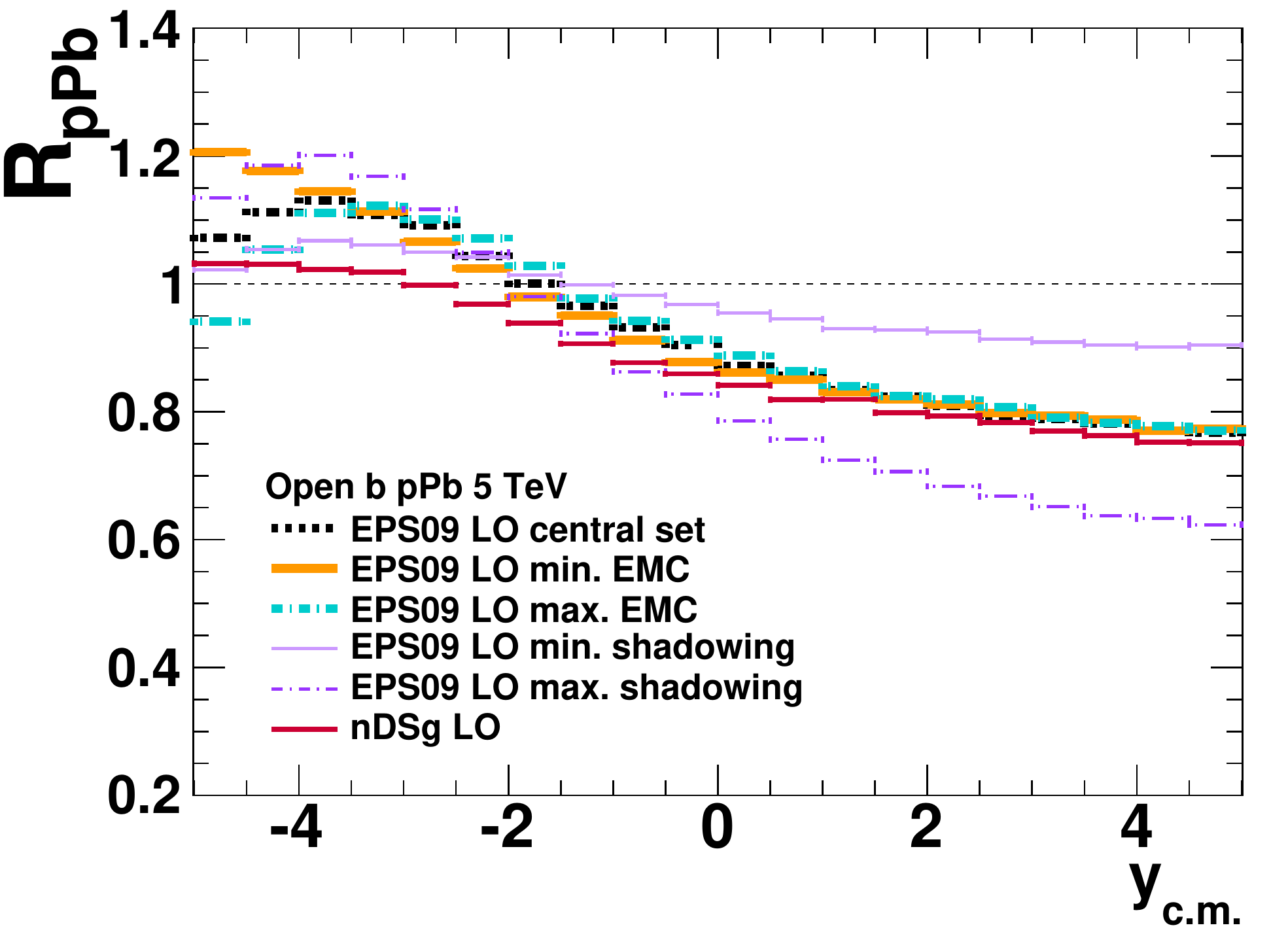}  
\caption{Effect of nPDF as encoded in EPS09 LO on ${\rppb}^{b\to\jpsi}$ at $\snn=5$~TeV.} 
\label{fig:Ferreiro-B} 
\end{figure}

This results in the nuclear modification factor $\rppb$ for  open beauty in  \pPb collisions at 5 TeV shown in \fig{fig:Ferreiro-B}. 
These values will be compared to the data measured by the LHCb collaboration~\cite{Aaij:2013zxa} and shown in \fig{fig:NonPromptJpsi_pPb_LHCb},  at backward and forward rapidity.  
As discussed in~\cite{delValle:2014wha}, the measured values of $\rppb$ slightly favour the nDSg parametrisation, which does not include anti-shadowing. One should, 
however, stress that such a direct theory-data comparison relies on a good control of the interpolated \pp\ cross section, while the forward-over-backward production ratio 
is not affected by any kind of uncertainty on the \pp\ measurements or modelling. In this case, there is no tension  
with EPS09. Finally, one can stress that the nuclear modification factor predicted 
for open beauty is similar to that of inclusive $\Upsilon$(1S). 
 
\subsubsection{Saturation in the Colour Glass Condensate approach}\label{sec:saturation} 
 
Fujii and Watanabe recently computed the heavy quark production cross section in high energy   \pA\ collisions 
in the colour Glass Condensate (CGC) framework~\cite{Fujii:2013yja,Fujii:2013gxa}, 
which is given at the leading order  
in the strong coupling constant $\alpha_s$, 
but includes multiple-scattering effects  
on the gluons and heavy quarks by the dense target~\cite{Blaizot:2004wv}. 
It is expressed in terms of hard matrix elements,  
2-point gluon function in the dilute projectile and  
multi-point gluon functions in the dense target,  
which breaks the $\kt$-factorization~\cite{Fujii:2005vj}. 
The energy dependence in this approach 
is incorporated through the gluon functions 
which obey the non-linear $x$-evolution equation 
leading to the gluon saturation phenomenon.

In the large-$N_c$ approximation (where $N_c=3$ is the number of colours in QCD), 
the multi-point functions reduce to a product of two dipole 
amplitudes in the fundamental representation  
and the evolution equation for the dipole has a closed form, 
called the Balitsky-Kovchegov (BK) equation. 
The BK equation with running coupling corrections (rcBK) is today  
widely exploited for phenomenological studies of 
saturation, and its numerical solution for $x<x_0=0.01$ 
is constrained with HERA DIS data and has been applied  
to hadronic reactions successfully~\cite{Albacete:2009fh}. 
Nuclear dependence is taken into account  
here in the initial condition for the rcBK equation  
by setting larger initial saturation scales, $Q_{s,A}^2(x_0)$ (below which gluon distribution in a nucleus starts to saturate) 
depending on the nuclear thickness. 
 
\begin{figure}[!t] 
\begin{center} 
\resizebox*{!}{7.0cm} 
{\includegraphics[width=15cm,angle=-90]{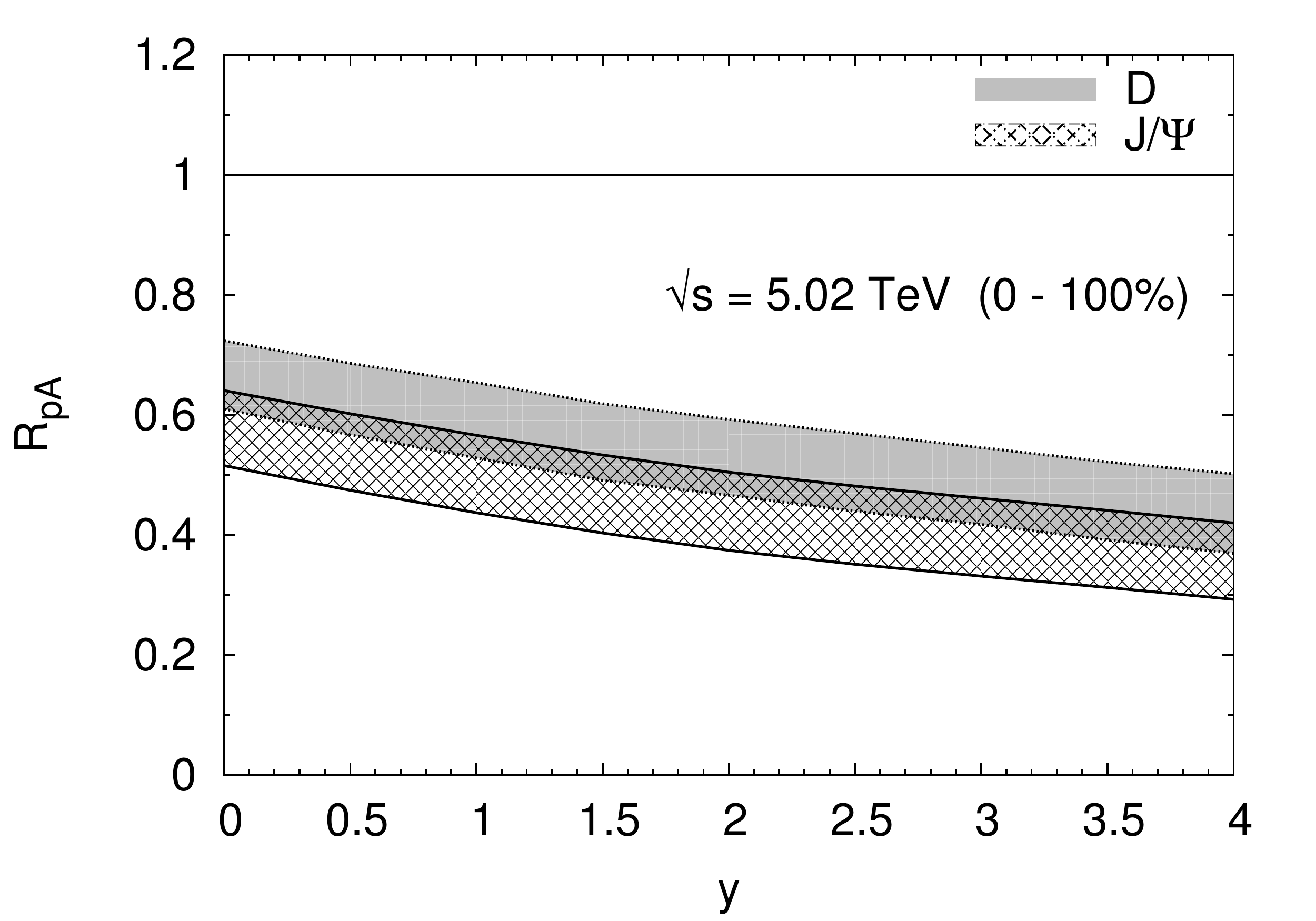}} 
\end{center} 
\caption{ Nuclear modification factor  
$\rpa(y)$ for D and \jpsi\   
in  p--Pb collisions at $\snn=5.02$ TeV in the CGC approach of Fuji and Watanabe~\cite{Fujii:2013yja,Fujii:2013gxa}. 
The bands indicate the uncertainties from the variations $m_c=1.2$--$1.5$~GeV$/c^2$  
and $Q_{s,A}^2(x_0)= (4-6) Q_{s,p}^2(x_0)$. 
} 
\label{fig:HQ-Quarkonium-RpA-y} 
\end{figure}

 
\cis{Fujii:2013yja,Fujii:2013gxa} show the evaluation of  heavy quark production applying the CGC framework in the large-$N_c$ approximation with the numerical solution of the rcBK equation. In hadronisation processes, the colour evaporation model (CEM) is used for $\jpsi$ ($\Upsilon$) and the vacuum fragmentation function for D meson production, assuming that the hadronisation occurs  outside the target as mentioned in \sect{sec:timescales}. 
The rapidity dependence of the nuclear modification factor $\rpa(y)$ of $\jpsi$ is one of the significant observables to investigate the saturation effect and the CGC based model reproduced the RHIC data by setting $Q_{s,A}^2(x_0)=(4 - 6) Q_{s,p}^2(x_0)$.  Extrapolation to the LHC energy predicted a stronger suppression, reflecting stronger saturation effects at the smaller values of $x$ (\fig{fig:HQ-Quarkonium-RpA-y}). Quarkonium suppression in this framework also includes the multiple scattering effects on the quark pair traversing the dense target. The comparison with experimental results will shown in \sect{CNM_ExpData}.

%
 
Several improvements to this approach can be performed. 
The CGC expression for the heavy quark production is derived at LO in the eikonal approximation for the colour sources. The NLO extension should be investigated to be consistent with the use of the rcBK equation. Furthermore, for quarkonium production,  colour channel dependence of the hadronisation process will be important and brings in a new multi-point function, which is simply ignored in CEM. Finally, using a similar approach but with an improved treatment of the nuclear geometry and a different parametrisation of the dipole cross section, Duclou\'e, Lappi and M\"antysaari~\cite{Ducloue:2015gfa} showed that the $\jpsi$ suppression in \pPb\ collisions was less pronounced. 
 
More recently, attempts to compute quarkonium production in \pp\ and \pA\ collisions have been made by implementing small-$x$ evolution and multiple scattering effects in the NRQCD formalism~\cite{Kang:2013hta}.  
Depending on which NRQCD channel dominates the $\jpsi$ production cross section in \pPb\ collisions at the LHC, the $\jpsi$ suppression predicted in this formalism may agree with the current ALICE and LHCb measurements~\cite{Ma:2015sia}. 
 
%

\subsubsection{Multiple scattering and energy loss}\label{sec:eloss} 
 
 
 
In this section various approaches of parton multiple scattering in nuclei are discussed. These effects include \QQbar\ propagation in nuclei, initial and final state energy loss, and coherent energy loss. 
 
\paragraph{\QQbar\ propagation and attenuation in nuclei} 
 
This section summarizes the approach by Kopeliovich, Potashnikova and Schmidt~\cite{Kopeliovich:2010nw,Kopeliovich:2011zz}. At LHC energies, the coherence time, $t_c$, for the production of charm quarks exceeds the typical nuclear size, $t_c\gg R_{\rm A}$. As a consequence, all the production amplitudes from different bound nucleons are in phase. In terms of the dipole description this means that Lorentz time delay ``freezes'' the \ccbar\ dipole separation during propagation through the nucleus, which simplifies calculations compared with the path-integral technique, required at lower energies~\cite{Kopeliovich:1991pu,Kopeliovich:1999am,Kopeliovich:2001ee}.  
 
\begin{figure}[t] 
\centerline{ 
  \scalebox{0.45}{\includegraphics{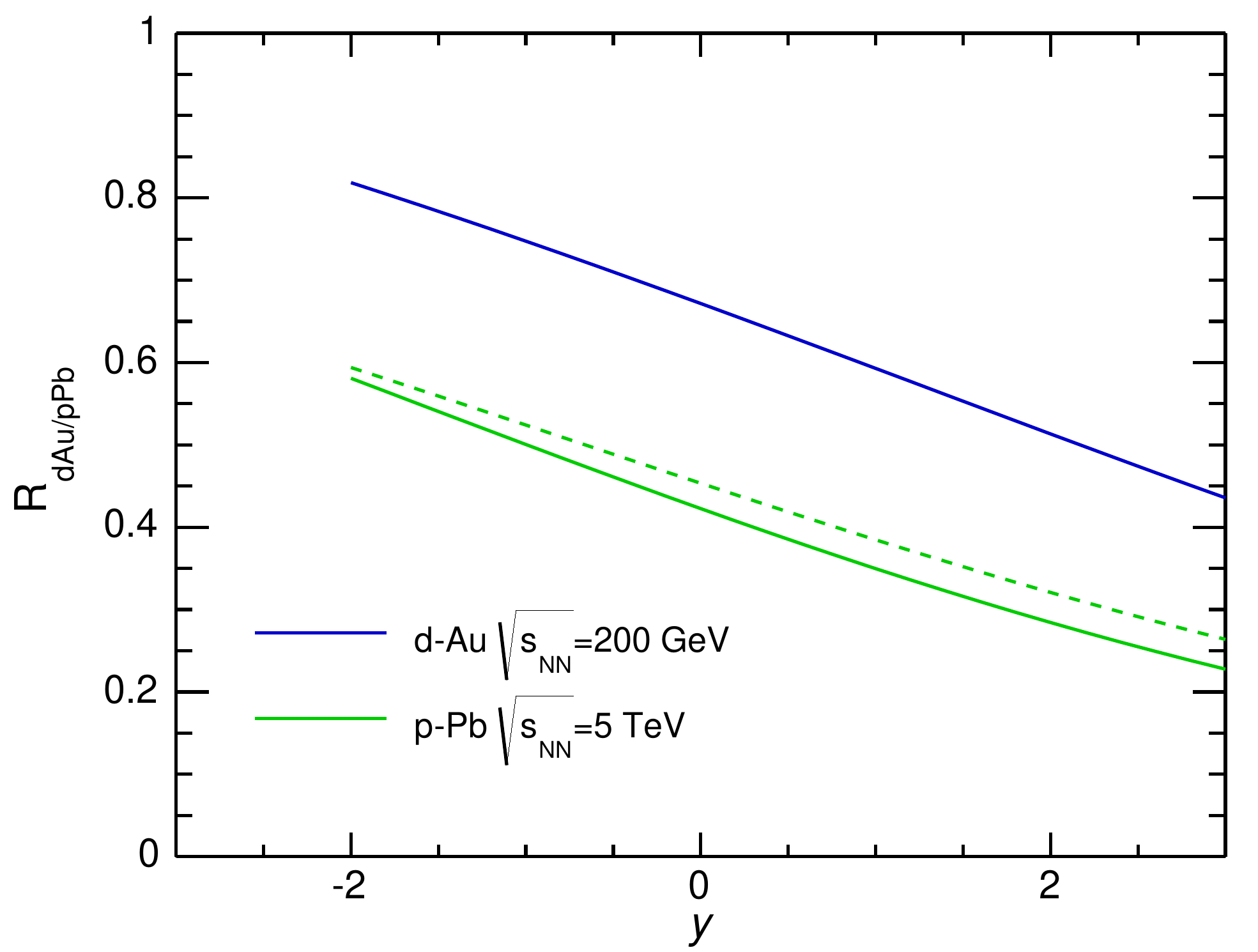}}} 
\caption{\label{dAu} Calculations~\cite{Kopeliovich:2010nw} for the $\pt$-integrated nuclear suppression factor $\rdau(y)$   for $\jpsi$ produced in d--Au collisions with rapidity $y$ at $\snn=200\,$GeV. The upper solid curve presents the result of \eq{120} and \eq{140}, including a small effect from shadowing. The lower solid (dashed) curve shows calculations for LHC at  
$\snn=5\,$TeV, including (excluding) shadowing.} 
 \label{fig:KPS} 
 \end{figure}

Because of the rescattering of the dipole in the nucleus, 
the charmonium suppression in \pA\ collisions with impact parameter $b$ has the form~\cite{Kopeliovich:2001ee,Kopeliovich:2010nw,Kopeliovich:2011zz}, 
\beqn 
\rpa&=&{1\over A}\int \dd^2b\int\limits_{-\infty}^{\infty}\dd z\, 
\rho_{\rm A}(b,z) 
\left|\spa(b,z)\right|^2, 
\label{120}\\ 
\spa(b,z)&=&\int \dd^2r_{\rm T}\,W_{\ccbar}(r_{\rm T})\, 
\exp\left[-{1\over2}\sigma_{\ccbar g}(r_{\rm T})T_-(b,z) 
-{1\over2}\sigma_{\ccbar}(r_{\rm T})T_+(b,z)\right]. 
\label{140} 
\eeqn 
Here $W_{\ccbar}(r_{\rm T})\propto K_0(m_c r_{\rm T})\,r_{\rm T}^2\,\Psi_{f}(r_{\rm T})$ is the distribution function for the dipole size $r_{\rm T}$; $K_0(m_c r_{\rm T})$ describes the $r_{\rm T}$-distribution of the $\ccbar$ dipole in the projectile gluon; $\Psi_f(r_{\rm T})$ is the light-cone wave function of the final charmonium; one factor $r_{\rm T}$ comes from the colour exchange transition $(\ccbar)_8\to(\ccbar)_1$ amplitude, another factor $r_{\rm T}$ originates either from radiation of a gluon (colour-singlet model for $\psi$), or from the wave function of a $P$-wave state ($\chi$).  
The three-body ($g\ccbar$) dipole cross section 
$\sigma_{\ccbar g}(r_{\rm T})={9\over4}\sigma_{\ccbar}(r_{\rm T}/2)-{1\over8}\sigma_{\ccbar}(r_{\rm T})$ is responsible for the $g\to\ccbar$ transition and its nuclear shadowing. The thickness functions are defined as,  
$T_-(b,z)=\int_{-\infty}^z \dd z'\rho_{\rm A}(b,z')$;~ $T_+(b,z)=T_A(b)-T_-(b,z)$, and $T_A(b)=T_-(b,\infty)$ where $\rho_{\rm A}$ is the nuclear density profile. The results of parameter-free calculations~\cite{Kopeliovich:2011zz} of $\rpa$ as a function of rapidity at the energies of RHIC and LHC are shown in \fig{dAu}. 
%
 
At this point one should emphasise that attenuation of $\ccbar$ dipoles in nuclear matter is a  
source of nuclear suppression of $\jpsi$, although it is often not included 
in model calculations. Moreover, independently of model details, 
  the general features of dipole interactions are: 
 {\it (i)} the dipole cross section 
 studied in detail at HERA, which is proportional to the dipole size squared (of the order of $1/m_c^2$) 
and to the gluon density, {\it (ii)} the rise of the dipole cross section (and therefore the magnitude of the nuclear suppression) coming from  
the observed steep rise of the gluon density at small $x$.  
The observed energy independence of nuclear suppression of $\jpsi$ is incompatible with these features, and the only solution would be the presence of a nuclear enhancement mechanism rising with energy. Indeed, such a mechanism was proposed in~\cite{Hufner:1996jw} and developed in 
\cite{Kopeliovich:2001ee}. It comes from new possibilities, compared to a proton target, for $\jpsi$ production due to multiple colour exchange interaction of a $\ccbar$ in the nuclear matter, \eg the relative contribution of double interaction 
is enhanced in nuclei as $A^{1/3}$ and rises with energy proportionally to the dipole cross section~\cite{Hufner:1996jw}. Numerical evaluation of this effect is under way~\cite{Kopeliovich_inprog}. 
This approach for charmonium production cannot be simply extrapolated from  \pA\ to \AAcoll\ collisions~\cite{Kopeliovich:2010nw}. The latter case includes new effects of double colour filtering and a boosted saturation scale~\cite{Kopeliovich:2010nw}.

%
\paragraph{Initial and final state energy loss, power corrections and Cronin effect} 
 
The approach by Sharma and Vitev is now described.  
The basic premise of this approach  
is that CNM can be  evaluated and related to  
the transport properties of large nuclei for quarks and gluons~\cite{Vitev:2006bi}.  
At one extreme, when the scattering from  the medium is largely  incoherent, the parton  
modification is dominated by transverse momentum  broadening. It leads to a Cronin-like  
enhancement  of the cross sections at intermediate $\pt \sim$~few GeV$/c$.  
At the other extreme, when the longitudinal momentum transfer  
is small compared to the  
inverse of the 
path length of the parton  as it propagates through the nucleus,  
the scattering becomes  coherent, which can lead to attenuation, or shadowing. The coherent  
limit is described differently in different  approaches and its  effects are calculated in terms of  
nuclear-enhanced power  corrections to the cross sections. Multiple scattering 
also leads to medium-induced radiative corrections that, in the soft gluon emission limit, have 
the interpretation of energy loss~\cite{Vitev:2007ve}.  
 
The effects are implemented via modifications to the kinematics of hard parton scattering $a+b\rightarrow c+d$. 
For example, in \pA\ collisions  
\begin{eqnarray}   
&{\rm Initial-state \;\; energy \;\; loss }\qquad \qquad &    [ \phi_a({{x}}_{a}) ]_{\rm pA}  
= \left[ \phi_a\left(\frac{ {{x}}_{a}}{1-\epsilon_{a}}\right)  \right]_{\rm NN} \;,   \qquad \epsilon_{a}=\frac{\Delta E_{a}}{E_{a}}   \;,\quad  \qquad \qquad  \label{eq:viteva}\\ 
& {\rm Power \;\; corrections }  \qquad \qquad &   ( {x}_{b})_{\rm pA}  = (x_b)_{\rm NN}\left[1+\frac{\xi_d^2(A^{1/3}-1)}{-\hat{t}+m_d^2}\right]\;,\qquad  \label{eq:vitevb} \\ 
& {\rm Cronin \;\; effect }  \qquad \qquad  & \langle {k}_{{\rm T}a}^2 \rangle_{\rm pA} = \langle {k}_{\rm T}^2 \rangle_{\rm NN} 
+  \langle {k}_{{\rm T}a}^2 \rangle_{\rm IS}\, , \;   \qquad 
\langle {k}_{{\rm T}a}^2 \rangle_{\rm IS} = \left\langle \frac{2 \mu^2 L}{\lambda_{a}}   
\right\rangle   \;  . \label{eq:vitevc} 
\end{eqnarray} 
In \eq{eq:viteva}, $\epsilon_{a}$ is the fractional energy loss for parton $a$ 
prior to the hard collision, which  increases linearly with medium opacity. 
When the inverse longitudinal momentum 
transferred from the nucleus is larger than the Lorentz-contracted longitudinal 
size, the scattering can become coherent. This effect can be included in an 
effective modification of the Bjorken-$x$ variable, as shown in~\eq{eq:vitevb}, in which $\xi_d$ is a parameter monitoring the strength of power corrections. The momentum broadening leading to Cronin effect is given in~\eq{eq:vitevc} in which $\mu$ is the typical transverse momentum transfer in a parton--nucleon scattering and $\lambda_a$ the parton mean free path in the nuclear medium.  
The typical transverse momentum scales and  
scattering lengths  are $\xi_d^2 / 1\; {\rm fm} \sim \mu^2/\lambda 
\simeq  0.1$~GeV$^2$/fm (0.225~GeV$^2$/fm) 
for quarks (gluons) respectively. These yield a quark radiation length $X_0 \sim 50$~fm~\cite{Neufeld:2010dz}.  
For further details, see~\cite{Vitev:2006bi,Neufeld:2010dz,Sharma:2012dy}.  
 This approach has successfully described  
the experimentally observed suppression of light hadron, photon and di-lepton production cross sections. 
\begin{figure}[!t] 
\includegraphics[width=3.2in,angle=0]{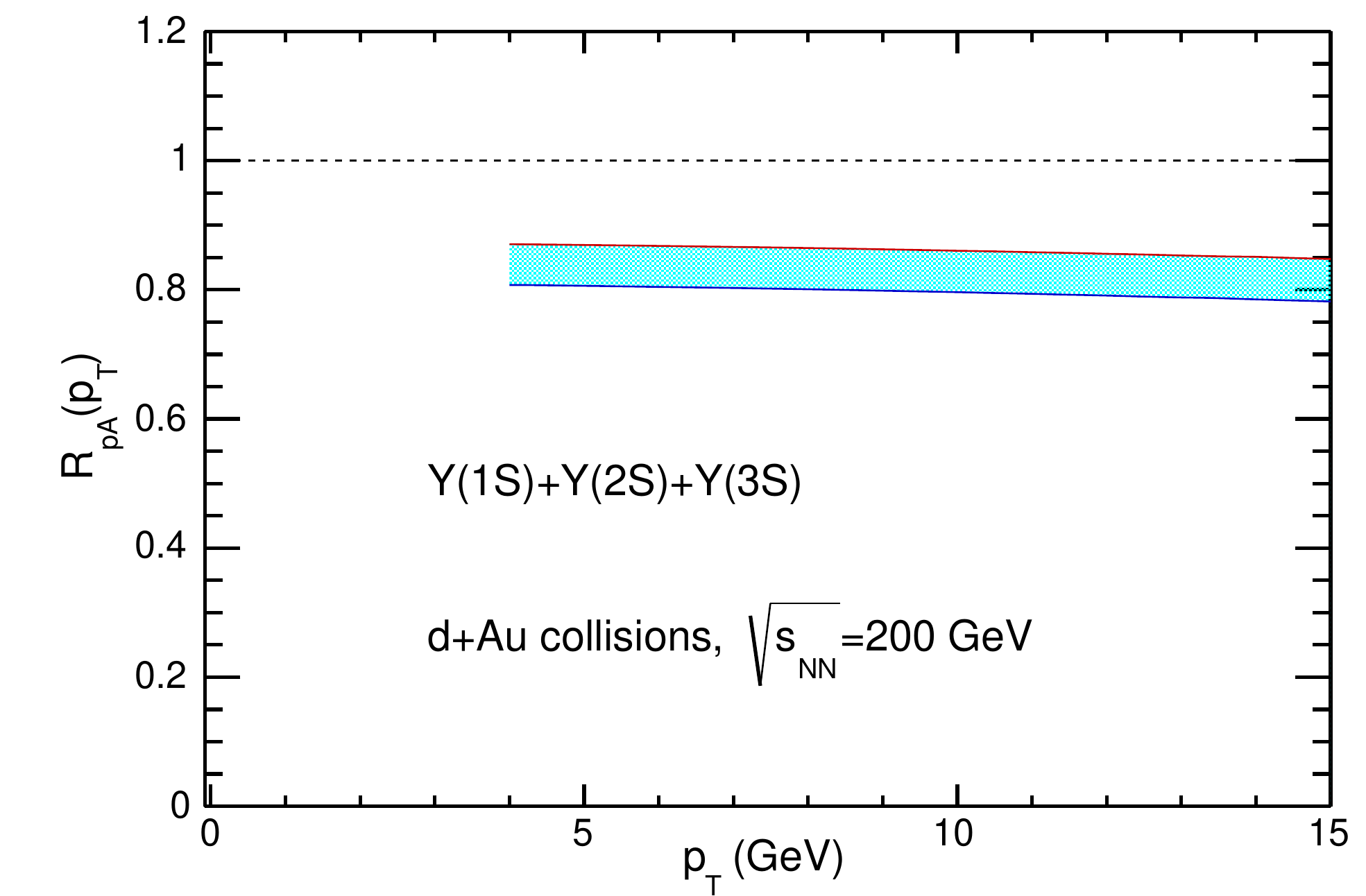}  
\includegraphics[width=3.2in,angle=0]{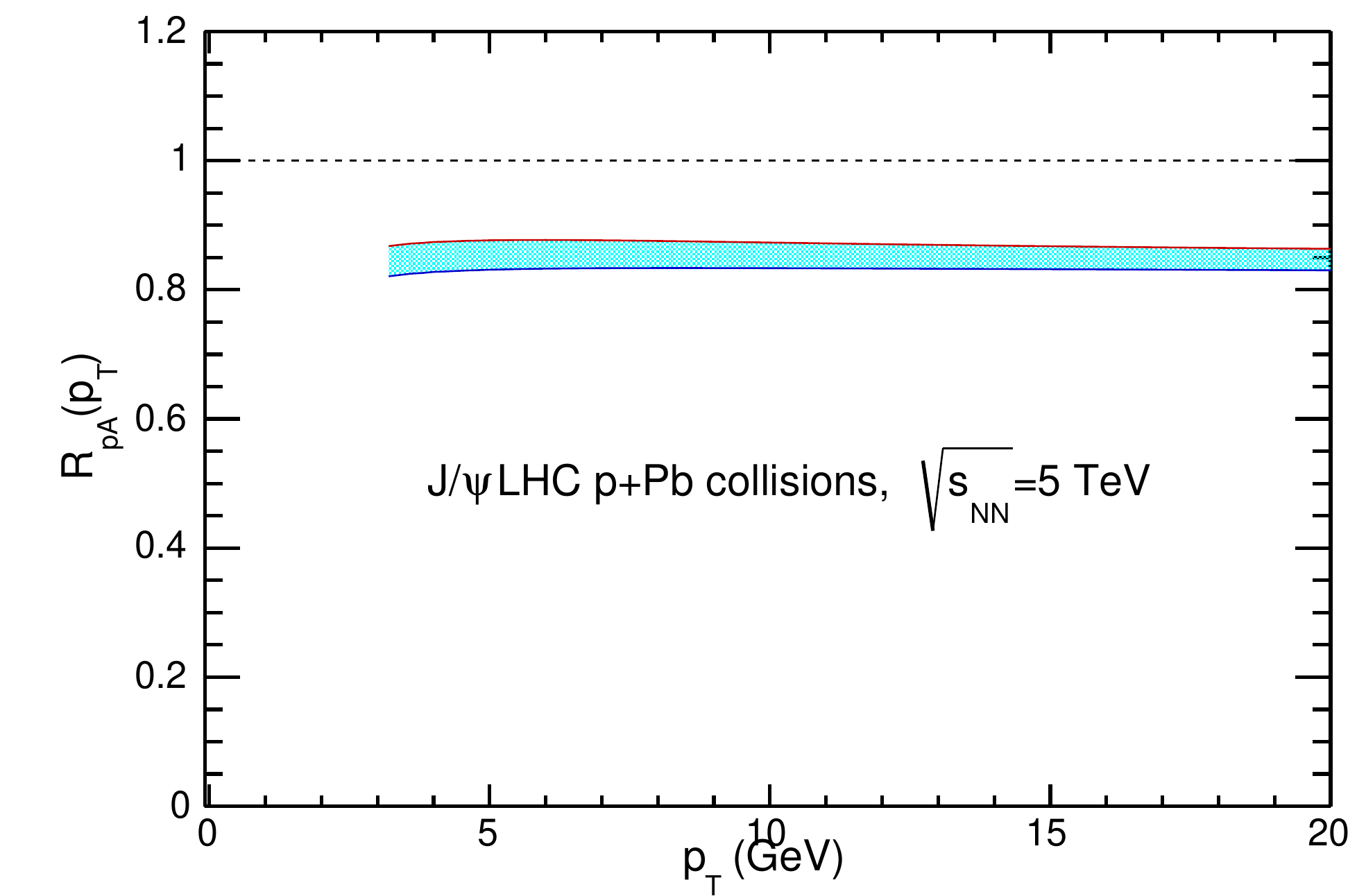}  
\caption{Theoretical calculations~\cite{Sharma:2012dy} for $\Upsilon$ and $\jpsi$  $\rpa$, as a function of \pt and at mid-rapidity, in 
minimum-bias collisions  
with a small (upper curve, red) and large (lower curve, blue) energy loss effect at RHIC and LHC, respectively. } 
\label{fig:RpAJ1}  
\end{figure} 
As the heavy quark introduces a new mass scale, the dependence of CNM corrections on this scale and their relative significance 
needs to be reassessed in light of the experimental data.

For the case of quarkonium production, a large uncertainty arises form the fact that the Cronin effect is not understood~\cite{Sharma:2012dy}, nor have there been attempts to fit it in this approach. Consequently, for $\jpsi$ and $\Upsilon$ results with only CNM energy loss are shown. 
Due to the uncertainties in the magnitude of the Cronin effect and the magnitude of the cold nuclear matter energy loss, the nuclear modification for open heavy flavour can show either small enhancement and small suppression in the region of $\pt \sim$~few GeV$/c$. 
The uncertainties in the magnitude of  $\Delta E / E$ can be quite significant~\cite{Neufeld:2010dz}. Motivated by other multiple parton scattering effects, such as the Cronin and the coherent power corrections, which are both compatible with possibly smaller transport parameters of cold QCD matter we also consider energy loss that is $35\%$ smaller than the one from using the parameters above. The results for quarkonium modification in \pA\ collisions is then presented as a band.         
The left panel of \fig{fig:RpAJ1} shows theoretical predictions for $\Upsilon$  
$\rdau$ at RHIC~\cite{Sharma:2012dy}. 
The right panel of \fig{fig:RpAJ1} shows theoretical  
predictions for  $\jpsi$ $\rppb$ at the LHC~\cite{Sharma:2012dy} that will be compared to data in~\sect{CNM_ExpData}.  
 
 
%
 
\paragraph{Coherent energy loss} 
 
Another approach of parton energy loss in cold nuclear matter has been suggested by Arleo et al. in 
\cis{Arleo:2010rb,Arleo:2012rs,Arleo:2013zua,Peigne:2014uha,Peigne:2014rka}. 
A few years ago it was emphasized that the medium-induced radiative energy loss $\Delta E$ of a high-energy gluon crossing a nuclear medium and being scattered to small angle is proportional to the gluon energy $E$~\cite{Arleo:2010rb,Peigne:2014uha}. The behaviour $\Delta E \propto E$ arises from soft gluon radiation which is {\it fully coherent} over the medium. Coherent energy loss is expected in all situations where the hard partonic process looks like forward scattering of an incoming parton to an outgoing {\it compact} and {\it colourful} system of partons~\cite{Peigne:2014rka}. In the case of $\jpsi$ hadroproduction at low $\pt \lesssim m_{\jpsi}$, viewed in the target rest frame as the scattering of an incoming gluon to an outgoing {\it colour octet} $\ccbar$ pair\footnote{As for instance in the colour Evaporation Model. In the colour Singlet Model for $\jpsi$ production, a {\it colour singlet} $\ccbar$ pair is produced, but in conjunction with a hard gluon, thus making no qualitative difference with the production of a compact colour octet state.}, such an energy loss provides a successful description of $\jpsi$ nuclear suppression in \pA\ as compared to \pp\ collisions, from fixed-target (SPS, HERA, FNAL) to collider (RHIC, LHC) energies~\cite{Arleo:2012rs,Arleo:2013zua}.

\begin{figure}[t] 
\begin{center} 
\includegraphics[width=0.43\textwidth]{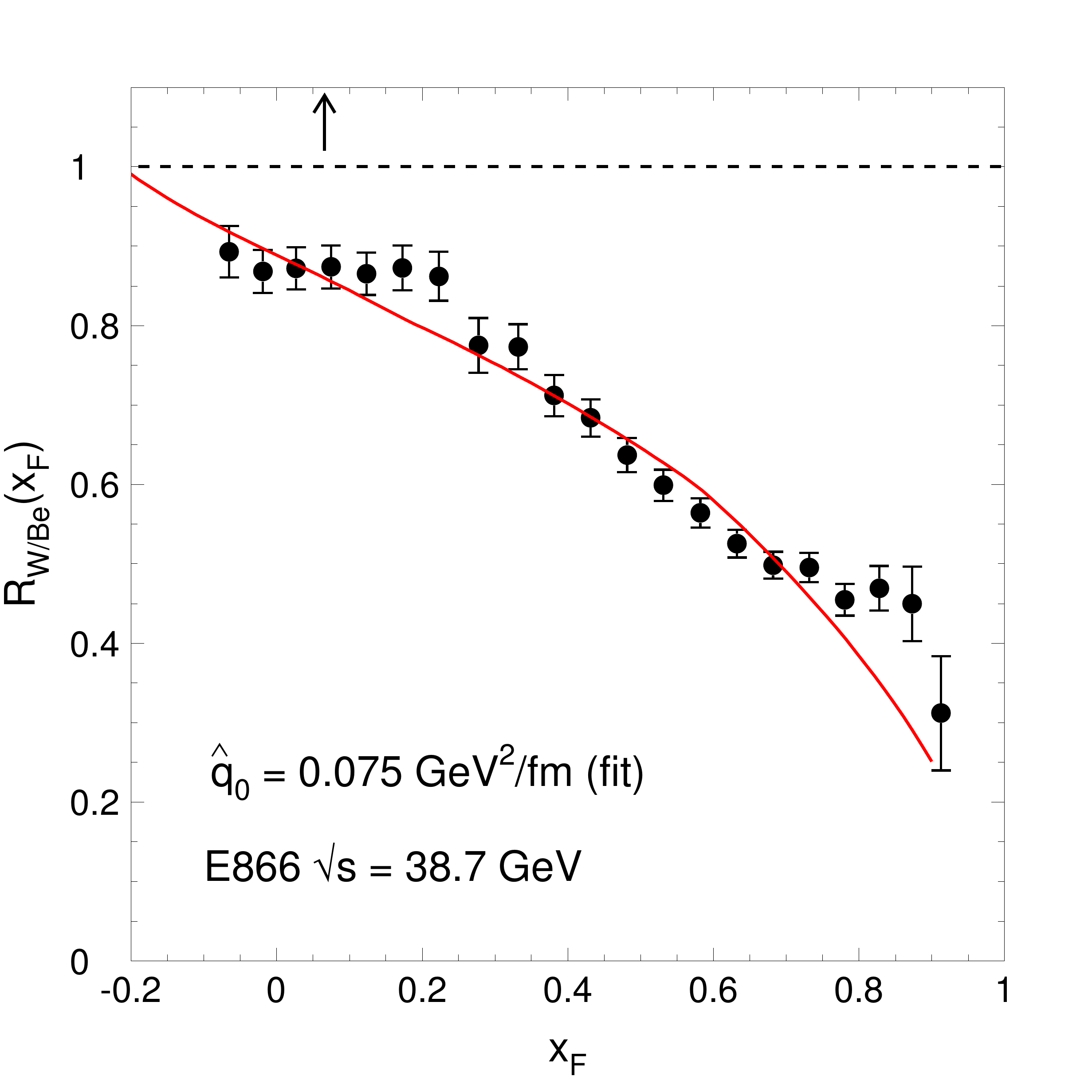}  
\includegraphics[width=0.52\textwidth]{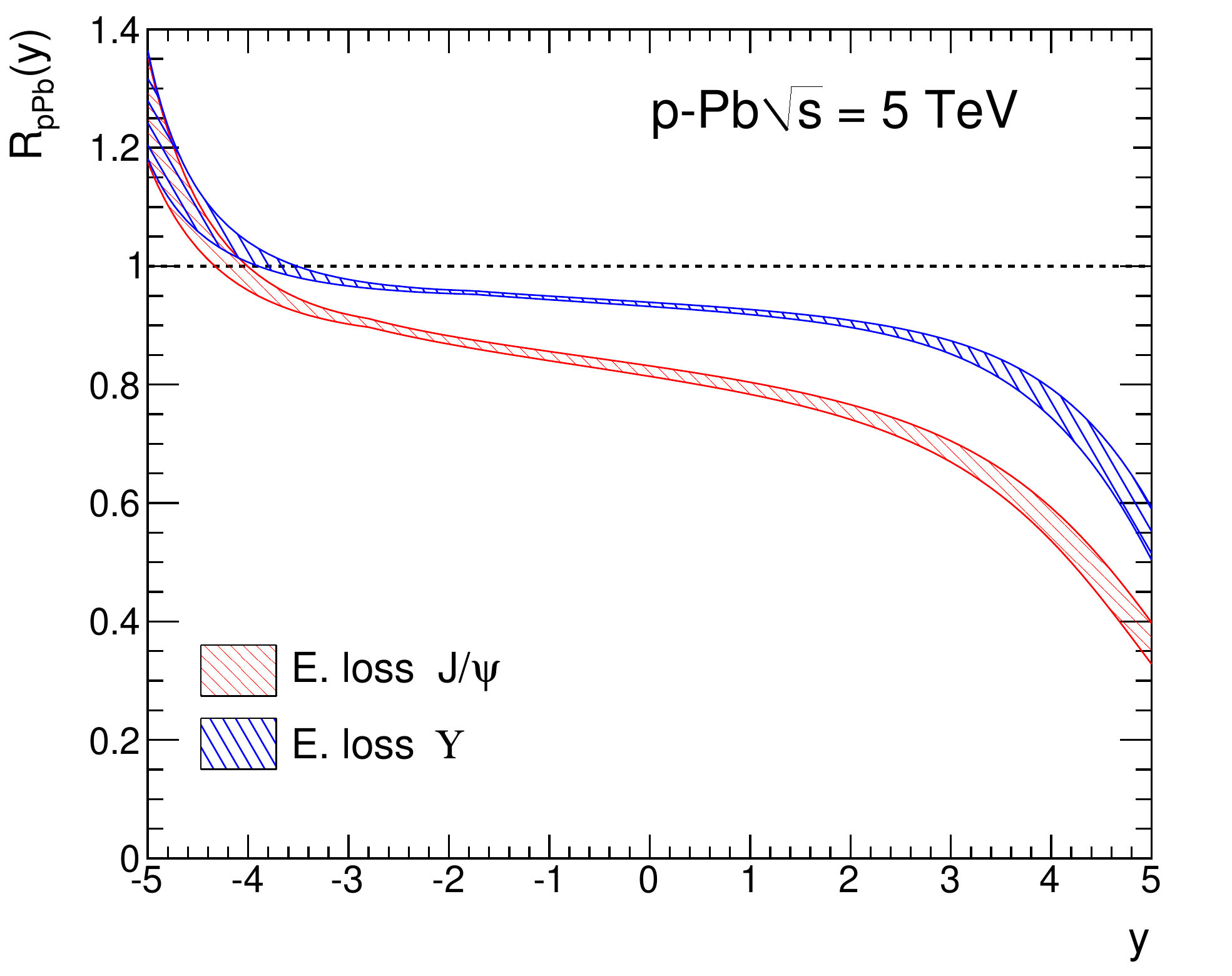}  
\end{center} 
\caption{Left: $\jpsi$ suppression due to coherent energy loss effects, fitted to E866 data in p--W collisions at $\snn=38.7$~GeV, as a function of Feynman-$x$, $x_{\rm F} \simeq 2 p_z^{\jpsi}/\sqrt{s}$. The vertical arrow indicates below which $x_{\rm F}$ values $\jpsi$ production may be sensitive to nuclear absorption. Right: Predictions of $\jpsi$ and $\ups$ suppression in \pPb\ collisions at the LHC. 
From \cis{Arleo:2010rb,Arleo:2012rs,Arleo:2013zua,Peigne:2014uha,Peigne:2014rka}.} 
\label{fig:coherent-loss-vs-LHC} 
\end{figure}  
 
\begin{figure}[t] 
\begin{center} 
\includegraphics[width=0.44\textwidth]{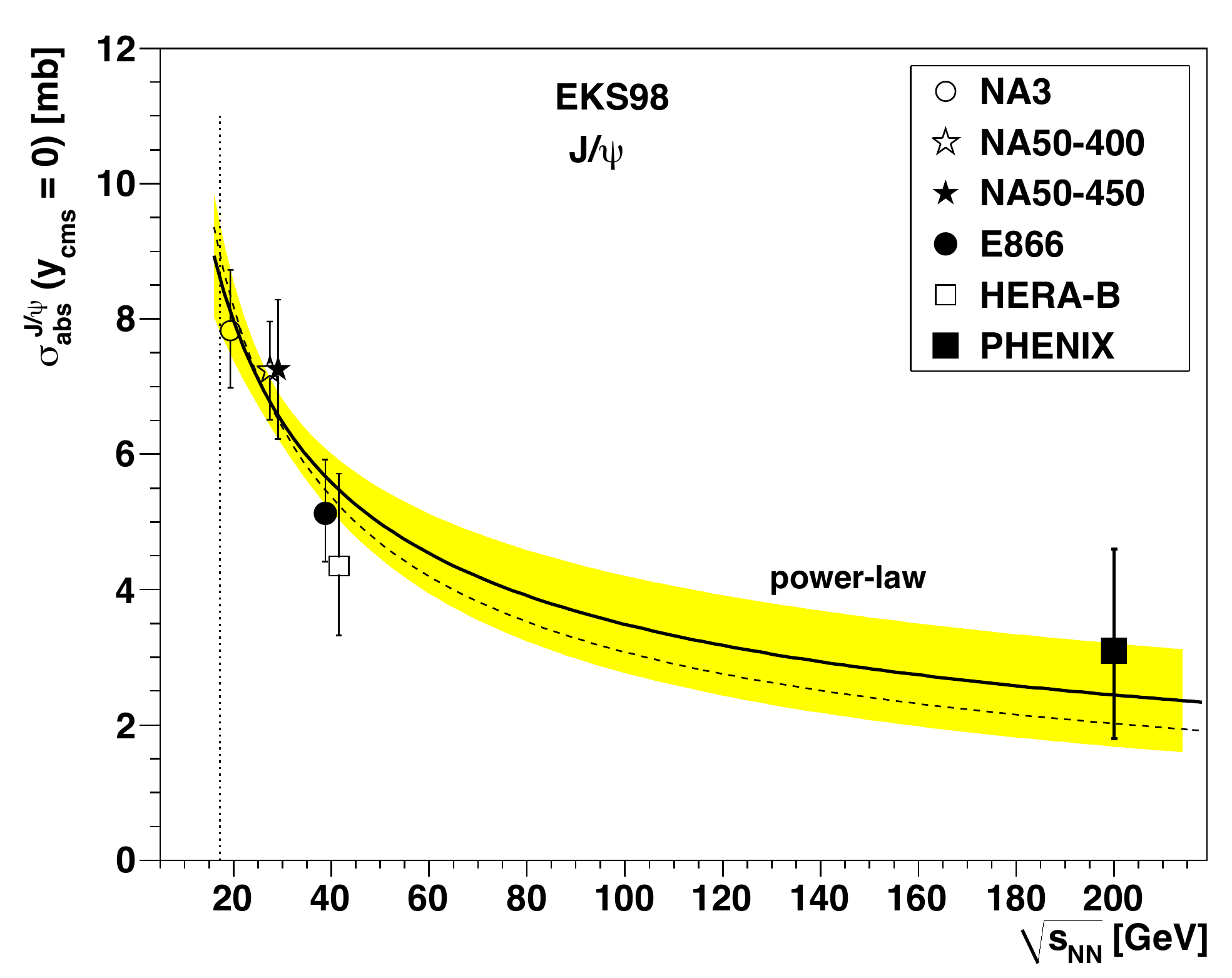} 
\includegraphics[width=0.49\textwidth]{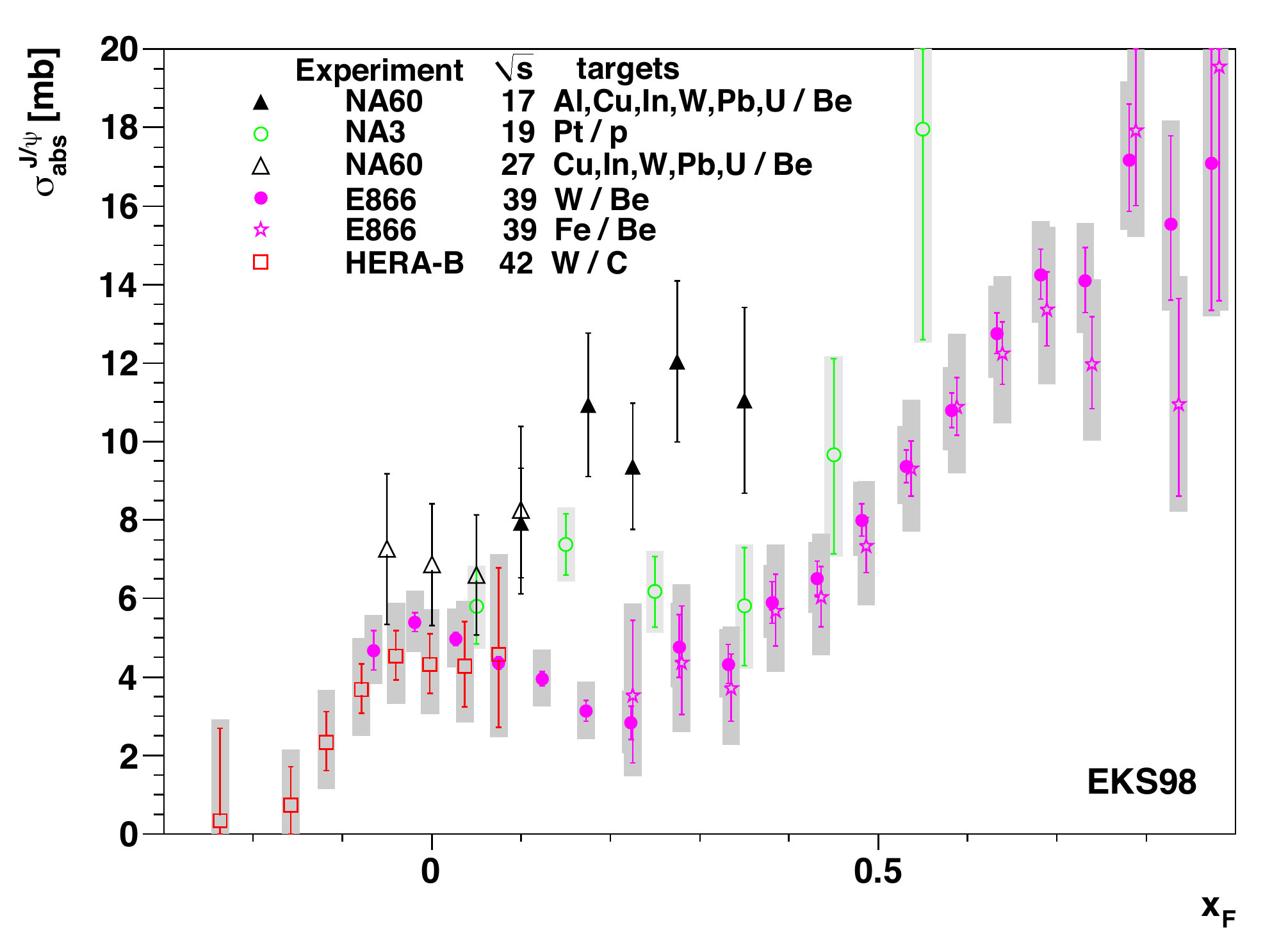}  
\end{center} 
\caption{Left: Energy dependence of $\sigma_{\rm abs}^{\jpsi}$ at mid-rapidity~\cite{Lourenco:2008sk}  
using the EKS98-CTEQ61L nPDFs~\protect\cite{Eskola:1998iy,Eskola:1998df, 
Pumplin:2002vw,Stump:2003yu}. Right: The $\xf$ dependence of $\sigma_{\rm abs}^{\jpsi}$, 
determined~\protect\cite{Lourenco:2008sk} from fixed-target  
measurements 
and using the  
EKS98 nPDFs~\protect\cite{Eskola:1998iy,Eskola:1998df}.} 
\label{fig:LVWfig1} 
\end{figure}

In~\cis{Arleo:2012rs,Arleo:2013zua}, the $\jpsi$ differential cross section $\dd^2\sigma_{\rm pp}/\dd y \, \dd \pt$ is determined from a fit of the \pp\ data, and $\dd^2\sigma_{\rm pA}/\dd y \, \dd \pt$ is obtained by performing a shift in rapidity (and in $\pt$) accounting for the energy loss $\varepsilon$ with probability ${\cal P}(\varepsilon)$ (and for the transverse broadening $\Delta \pt$) incurred by the compact octet state propagating through the nucleus. Independent of the \pp\ production mechanism, the model is thus able to predict $\jpsi$ and $\Upsilon$ nuclear {\it suppression}, $\rpa$, as a function of $y$, $\pt$ and centrality. 
The model depends on a single parameter $\hat{q}_0$, which fully determines both the broadening $\Delta \pt$ and the energy loss probability distribution,  
 ${\cal P}(\varepsilon)$.   
 It is determined by fitting the model calculations to the E866 measurements~\cite{Leitch:1999ea} in p--W collisions at $\snn=38.7$~GeV. The result of the fit, which yields $\hat{q}_0=0.075$~GeV$^2$/fm, is shown in \fig{fig:coherent-loss-vs-LHC} (left) in comparison to the data. 
 
In order to assess the uncertainties of the model predictions,  the parameter entering the \pp\ data parametrisation is varied around its central value, as well as the magnitude of the transport coefficient from 0.07 to 0.09~GeV$^2$/fm~\cite{Arleo:2012rs}. 
The prescription for computing the model uncertainties can be found in~\cite{Arleo:2014oha}. 
The model predictions for $\jpsi$ and $\Upsilon$ suppression  in \pPb\ collisions at the LHC as a function of rapidity are shown in \fig{fig:coherent-loss-vs-LHC} (right). The extrapolation of the model to \AAcoll\ collisions is discussed in \sect{CNM_pAtoAA}. 
 
 
 
 
\subsubsection{Nuclear absorption}\label{sec:absorption} 
 
 

The quarkonium nuclear absorption is characterized by an ``effective'' cross section $\sigma_{\rm abs}$.  
In \ci{Arleo:2006qk}, Arleo and Tram analysed all the $\jpsi$ cross section measurements available at the time, taking into account nuclear absorption and nPDF effects. They found that, within the experimental uncertainties, the absorption cross section does not show a dependence on the $\jpsi$--N centre-of-mass energy, when going from fixed-target to RHIC energy.  
In the approach of \ci{Lourenco:2008sk} discussed below, Louren\c{c}o, Vogt and Woehri  
studied the available fixed-target data   
to  discern a possible dependence of the $\jpsi$ normal absorption at mid-rapidity as a function of the nucleon--nucleon centre-of-mass energy,  
both with 
and without considering nuclear modifications of the parton distributions.

The $\jpsi$ absorption cross section, $\sigma_{\rm abs}^{\jpsi}$, 
was traditionally assumed to be independent of the production kinematics 
until measurements covering broad phase space regions 
showed clear dependences of the nuclear effects on $\xf$ and $\pt$.   
It was further assumed to be independent of collision centre-of-mass energy,  
$\snn$,  neglecting any nuclear effects on the 
parton distributions. However, $\jpsi$ production is sensitive to 
the gluon distribution in the nucleus and the fixed-target measurements 
probe parton momentum fractions, $x$, in the possible anti-shadowing region. 
This effect may enhance the $\jpsi$ production rate at mid-rapidity and  
a larger absorption cross section would be required to match the data. 
 
If one focuses on the behavior of $\jpsi$ production at $\xf \approx 0$, the 
absorption cross section is found to depend on $\snn$, essentially  
independent of the chosen nPDF parametrisation~\cite{Lourenco:2008sk}, 
as shown in \fig{fig:LVWfig1} (left). The yellow band represents the uncertainty corresponding to an empirical power-law fit (solid curve) to all the data points analysed in~\cite{Lourenco:2008sk} from measurements by  
NA3~\cite{Badier:1983dg},  
NA50~\cite{Alessandro:2006jt,Alessandro:2003pc}, E866~\cite{Leitch:1999ea},   
HERA-B~\cite{Abt:2008ya}, NA60~\cite{Scomparin:2009tg} and 
PHENIX~\cite{daSilva:2009yy}.  
The extrapolation of the power-law fit in \fig{fig:LVWfig1} (left) to the current 
LHC \pA\ energy leads to a vanishingly small cross section within the illustrated uncertainties. 
 
%
 
Away from mid-rapidity, the extracted $\sigma_{\rm abs}^{\jpsi}$ grows with $\xf$ up to unrealistically large values,  
as shown in \fig{fig:LVWfig1} (right). This seems to indicate that another mechanism, in addition to absorption and shadowing, such as  
initial-state energy loss, may be responsible for the $\jpsi$ suppression in the forward region  
($\xf > 0.25$). This confirms that the effective parameter $\sigma_{\rm abs}^{\jpsi}$ should not be interpreted as a genuine inelastic cross section. 
It seems that the rise starts closer to $\xf = 0$ for lower collision  
energies~\cite{Brambilla:2010cs}. 
More recent analyses~\cite{McGlinchey:2012bp}, using  
EPS09~\cite{Eskola:2009uj},  
are in general agreement with the results of \ci{Lourenco:2008sk}. 
 
 
Despite different conclusions on the the possible energy dependence of $\sigma_{\rm abs}$ from fixed-target experiments to RHIC energy in~\cite{Arleo:2006qk} and~\cite{Lourenco:2008sk}, one expects nuclear absorption effects to become negligible at the LHC since the quarkonium formation time becomes significantly larger than the nuclear size at all values of the rapidity. 
 This is also confirmed by a more recent analysis. In Ref.~\cite{McGlinchey:2012bp}, the authors show that the $\jpsi$ suppression seems to scale with the crossing time $\tau_{\rm cross}$ (see section~\ref{sec:timescales}), independently of the centre-of-mass energy, above a typical crossing time $\tau_{\rm cross} \gtrsim 0.05$~fm/$c$. Below this scale, however, the lack of scaling indicates that nuclear absorption is probably not the dominant effect. Using the $2\to 1$ kinematics, $\tau_{\rm cross} \simeq 2 m_p\ L\ e^{-y} / \snn$, the condition $\tau_{\rm cross} < 0.05$~fm/$c$ would correspond to $y > - 3.8$ (using $L_{\rm Pb}\simeq 3/4\ R_{\rm Pb} \simeq 5$~fm) at the LHC.
 

%

\subsubsection{Summary of CNM models}\label{sec:cnm-theo-sum} 
 
 A brief summary of these different approaches is given in \tab{tab:cnm-theo-sum}, in which the dominant physical effects and ingredients used in each calculation are given. The model acronyms given in the table match those in the legends of the figures in the next section. 
 
\begin{table}[!htp] 
\centering 
\caption{Summary of the various models of CNM approaches discussed in the text and compared to data in~\sect{CNM_ExpData}. The main physical processes and ingredients used in each calculation are listed.} 
\label{tab:cnm-theo-sum} 
\begin{tabular}{ccccc} 
\hline 
{\it Acronym} & {\it Production mechanism} & {\it Medium effects} & {\it Main parameters} & {\it Ref.}   \\ 
\hline 
\multicolumn{5}{c}{Open heavy flavour} \\ 
\hline 
pQCD+EPS09~LO     & pQCD LO   & nPDF           & 4+1 EPS09~LO sets & \cite{delValle:2014wha} \\ 
SAT             & pQCD LO+CGC   & Saturation               & $Q_{s,p}^2(x_0)$, $Q_{s,A}^2(x_0)$ & \cite{Fujii:2013yja} \\ 
ELOSS           &    pQCD~LO      &            E. loss, power cor., broa.       & $\epsilon_a$, $\xi_d$, $\mu^2$, $\lambda$ & \cite{Vitev:2006bi}\\ 
\hline 
\multicolumn{5}{c}{Quarkonia} \\ 
\hline 
EXT+EKS98LO+ABS & generic $2\rightarrow2$ LO & nPDF and absorption& EKS98~LO, \sabs  & \cite{Ferreiro:2009ur,Rakotozafindrabe:2012ss} \\ 
EXT+EPS09~LO     & generic $2\rightarrow2$ LO   & nPDF           & 4+1 EPS09~LO sets & \cite{Ferreiro:2011xy,Ferreiro:2013pua} \\ 
CEM+EPS09~NLO    & CEM NLO  & nPDF          & 30+1 EPS09~NLO sets & \cite{Vogt:2010aa} \\ 
SAT             & CEM LO+CGC  & Saturation               & $Q_{s,p}^2(x_0)$, $Q_{s,A}^2(x_0)$ & \cite{Fujii:2013gxa} \\ 
ELOSS           &  NRQCD LO        &            E. loss, power cor.      & $\epsilon_a$, $\xi_d$, $\mu^2$, $\lambda$ & \cite{Sharma:2012dy} \\ 
COH.ELOSS       & \pp data  & Coherent E. loss &  $\hat{q}$ & \cite{Arleo:2012rs,Arleo:2013zua} \\ 
KPS             & dipole model         &   Dipole absorption               & $\sigma_{c\bar{c}}$ & \cite{Kopeliovich:2010nw,Kopeliovich:2011zz} \\ 
\hline 
\end{tabular} 
\end{table} 
 

\subsection{Recent RHIC and LHC results}
\label{CNM_ExpData}

In this section we summarise the recent measurements in \pA collisions at RHIC and at the LHC.  
Open heavy-flavour results are described in \sect{sec:CNM:OHF} and hidden heavy-flavour data in  
\sect{sec:CNM:Onia}.  
As described in the previous section, in order to understand the role of the CNM effects,  
the interpretation of these measurements  
is commonly obtained by a comparison with measurements in \pp\ collisions 
at the same centre-of-mass energy as for p--A and in the same rapidity interval.  
At the LHC, so far it has not been possible to carry out \pp measurements 
at the same energy and rapidity as for p--Pb. 
In \sect{sec:CNM:ppRef} the procedures  
to define the pp reference for $R_{\rm pA}$ are described.

\subsubsection{Reference for \pA measurements at the LHC} 
\label{sec:CNM:ppRef} 
 
 
The \pp reference for open heavy-flavour measurements at $\s=5.02$\TeV was obtained either from pQCD calculations or by a pQCD-based  
$\s$-scaling of the measurements performed at $\s=7$\TeV. In some cases, it was also possible to evaluate the $\s$-scaled spectra of  
both the 7\TeV and 2.76\TeV data to $\s=5.02$\TeV and combine them.  
The pQCD-based $\s$-scaling procedure is described in reference~\cite{Averbeck:2011ga}.  
The scaling factor is evaluated as the ratio of the theoretical calculation at the two energies. The scaling uncertainties are determined  
considering the prediction uncertainties, the variation of the renormalisation and factorisation scales, the heavy-quark mass and the PDF  
uncertainties. The assumption behind this calculation is that the values of these parameters remain the same at both energies.  
The scaling factor and uncertainties computed with different heavy-quark production models, FONLL~\cite{Cacciari:2003uh, Cacciari:2012ny}  
and GM-VFNS~\cite{Kniehl:2005st,Kniehl:2005mk}, are in excellent agreement.  
This procedure was verified by comparing the D meson CDF measurements at $\s=1.96\TeV$ at mid-rapidity to a $\s$-scaling of the ALICE data~\cite{Averbeck:2011ga}. 
A different strategy was used in order to evaluate the \pp reference for \jpsi from B decays: the procedure is the same as for \jpsi at forward rapidity and it is described in the following.

In the quarkonium analyses, different strategies have been adopted depending on the precision of the existing measurements. They are mainly based on phenomenological functions and are briefly described in the following.  
 
At mid-rapidity in ALICE,  
the \jpsi \pp integrated cross section reference has been obtained by performing an interpolation based on \jpsi measurements at mid-rapidity  
in \pp collisions  
at \s~=~200\GeV~\cite{Adare:2006kf}, 2.76\TeV~\cite{Abelev:2012kr} and 7\TeV~\cite{Aamodt:2011gj}, and in \ppbar  
collisions at \s~=~1.96\TeV~\cite{Acosta:2004yw}. Several functions (linear, power law and exponential) were used to parametrise the cross  
section dependence as a function of \s. 
The interpolation leads to a total uncertainty of $17\%$ on the integrated cross section. The effect of the asymmetric rapidity coverage,  
due to the shift of the rapidity by 0.465 in the centre-of-mass system in \pPb collisions at the LHC,  
was found to be negligible as compared to the overall uncertainty of the interpolation procedure. Then the same method as described in~\cite{Bossu:2011qe}  
was followed to obtain the \pt-dependent cross section. The method is based on the empirical observation that the \jpsi cross sections measured at different  
energy and rapidity scale with \pt/\meanpt. The \meanpt value was evaluated at \s~=~5.02~TeV by an interpolation of the \meanpt measured at  
mid-rapidity~\cite{Adare:2006kf,Aamodt:2011gj,Acosta:2004yw} using exponential, logarithmic and power law functions.  
 
At forward rapidity, a similar procedure for the \jpsi cross section interpolation has been adopted by ALICE and LHCb and is  
described in~\cite{ALICE:2013spa}.  
In order to ease the treatment of the systematics correlated with energy, the interpolation was limited to results obtained with a single apparatus.  
The inclusive \jpsi cross sections measured at 2.76~\cite{Abelev:2012kr} and 7\TeV~\cite{Abelev:2014qha} were included in the ALICE procedure while   
the inclusive, prompt \jpsi and \jpsi from B-mesons cross sections measured at 2.76~\cite{Aaij:2012asz}, 7~\cite{Aaij:2011jh} and 8\TeV~\cite{Aaij:2013yaa}  
were considered in the LHCb one.  
The interpolation of the cross section with energy is based, as in the mid-rapidity case, on three empirical shapes (linear, power law and exponential).  
The resulting interpolated cross section for inclusive \jpsi obtained  
by ALICE and LHCb in $2.5<\ycm<4$ at \s~=~5.02\TeV were found to be in good agreement with a total uncertainty of $\sim8\%$ and $\sim5\%$  
for ALICE and LHCb, respectively. The interpolation in \s was also performed by ALICE independently for each 
\pt interval and outside  
of the rapidity range of \pp data in order to cope with the \pPb centre-of-mass rapidity shift.   
In that case an additional interpolation with rapidity was carried out by using several empirical functions (Gaussian, second and fourth order polynomials). 
 
In the case of the \ups at forward rapidity, the interpolation procedure results also from a common approach by ALICE and LHCb and  
is described in~\cite{LHCb:08082014sca}. It is based on LHCb measurements in \pp collisions at 2.76\TeV~\cite{Aaij:2014nwa}, 7\TeV~\cite{Abelev:2014qha}  
and 8\TeV~\cite{Aaij:2013yaa}.  
Various phenomenological functions and/or the \s-dependence of the CEM and FONLL models are used for the \s-dependence of the cross section, similarly to the 
\jpsi interpolation procedure at forward rapidity. This interpolation results into a systematic uncertainty that ranges from 8 to $12\%$ depending on the  
rapidity interval.

\subsubsection{Open heavy-flavour measurements} 
\label{sec:CNM:OHF} 
 
Open heavy-flavour production occurs in hard processes at the early stages of the collision (see \sect{sec:pp:Theory:OpenHF} for an introduction to the different calculations).  
As explained in \sect{CNM_theory}, their production in a nuclear environment is affected by the modification of the parton probability density in the nucleus (nPDFs or parton saturation formalisms)  and by the multiple scattering of partons in the nucleus (radiative or collisional parton energy loss, $\kt$ broadening).  
Due to their short lifetimes, open heavy-flavour hadrons are measured via their decay products. Different analyses methods exist:  
{\it (i)} study leptons from heavy-flavour decays; 
{\it (ii)} examine the $\pt$-integrated di-lepton invariant mass distribution, to evaluate the charm and beauty cross sections;  
{\it (iii)} fully reconstruct exclusive decays, such as $\Dzero \to {\rm K}^+ \, \pi^-$ or ${\rm B}^0 \to \jpsi \, {\rm K}^0_{\rm S}$;  
{\it (iv)} select specific (semi-)inclusive decays with a displaced vertex topology, such as beauty decays to leptons or $\jpsi$;
{\it (v)} identify $c$- or $b$-jets from reconstructed jets;  
{\it (vi)} inspect heavy-flavour azimuthal correlations.  
In the analyses where not all the decay products are reconstructed, the correlation between the heavy-flavour hadron kinematics and that of the decay particles has to be considered to properly interpret the measurements.
 
\paragraph{Heavy-flavour decay leptons} 
 
The production of heavy-flavour decay leptons, i.e. leptons from charm and beauty decays, has been studied at RHIC and at LHC energies in \dAu and \pPb collisions at $\snn=200$\GeV and 5.02\TeV respectively.  
The \pA measurements exploit the inclusive lepton $\pt$-spectrum, electrons at mid-rapidity ($|\eta|<0.5$ for PHENIX, $0<\eta<0.7$ for STAR and $|\eta|<0.6$ for ALICE) and muons at forward rapidities ($1.4<|\eta|<2.0$ for PHENIX and $2.5<\eta<4.0$ for ALICE).  
The heavy-flavour decay spectrum is determined by extracting the non-heavy-flavour contribution to the inclusive lepton distribution.  
The photonic background sources are electrons from photon conversions in the detector material and $\pi^0$ and $\eta$ Dalitz decays, which involve virtual photon conversion.  
The contribution of photon conversions is evaluated with the invariant-mass method or via Montecarlo simulations.  
The Dalitz decays contribution can be determined considering the measured $\pi^0$ and $\eta$ distributions.  
Background from light hadrons, hard processes (prompt photons and Drell-Yan) and quarkonia is determined with Montecarlo simulations, based, when possible, on the measured spectrum.  
STAR data is not corrected for the of $\jpsi$ decays contribution, which is non-negligible at high $\pt$.  
Beauty decay electron spectra can be obtained from the heavy-flavour decay electron spectra by a cut or fit of the lepton impact parameter distribution, \ie the distance between the lepton track and the interaction vertex, or exploiting the lepton azimuthal correlation to heavy flavours or charged hadrons. For the latter see the last paragraph of this section.  
 
\begin{figure}[!tbp] 
	\centering 
	\includegraphics[width=0.8\columnwidth]{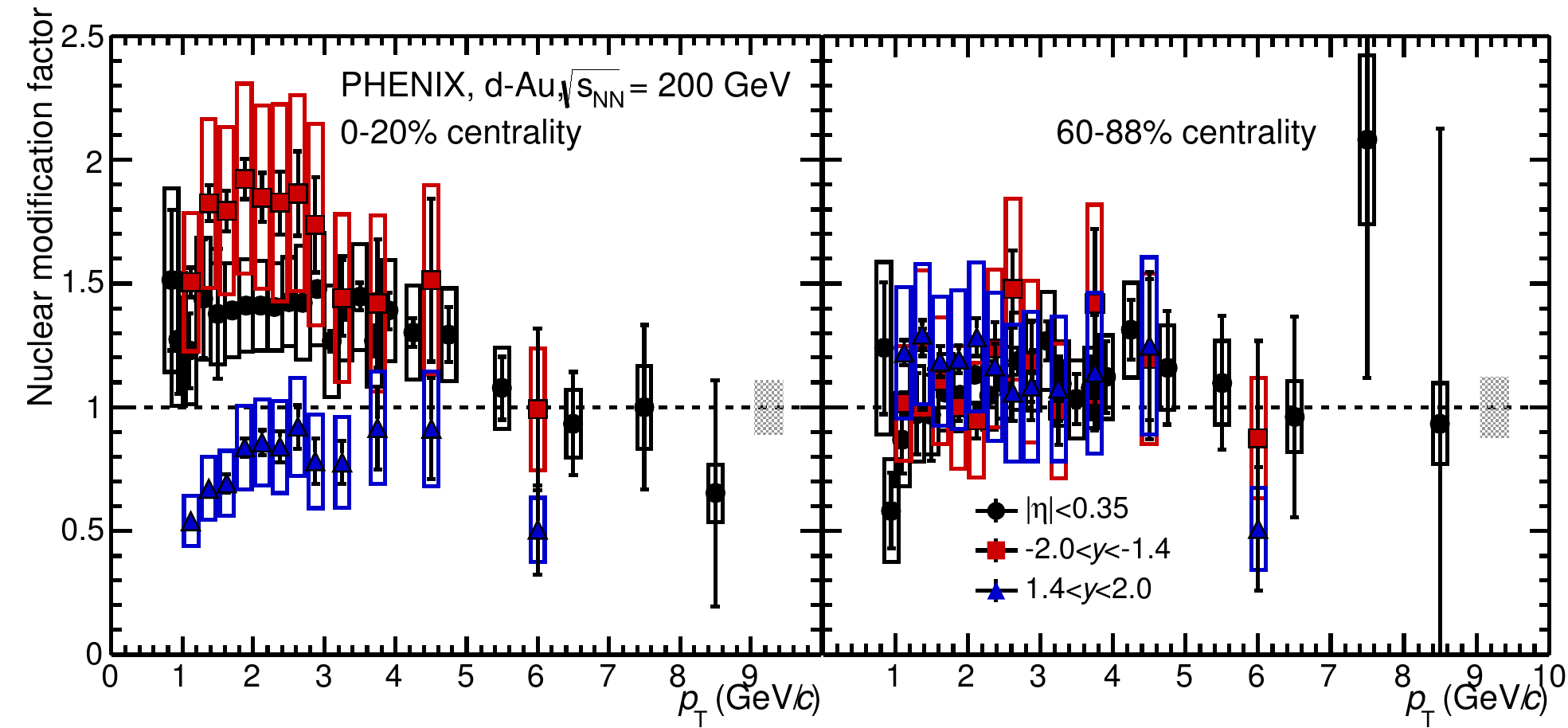} 
	\caption{Nuclear modification factor of heavy-flavour decay leptons in \dAu collisions at $\snn=200$\GeV as a function of transverse momentum in the 0--20\% and 60--88\% centrality classes, as measured with the PHENIX detector~\cite{Adare:2012yxa,Adare:2013lkk}. 
	\label{fig:HFlepton_RdAvspt_RHIC} 
	} 
\end{figure}

\begin{figure}[!tbp] 
	\centering 
	\includegraphics[width=0.95\columnwidth]{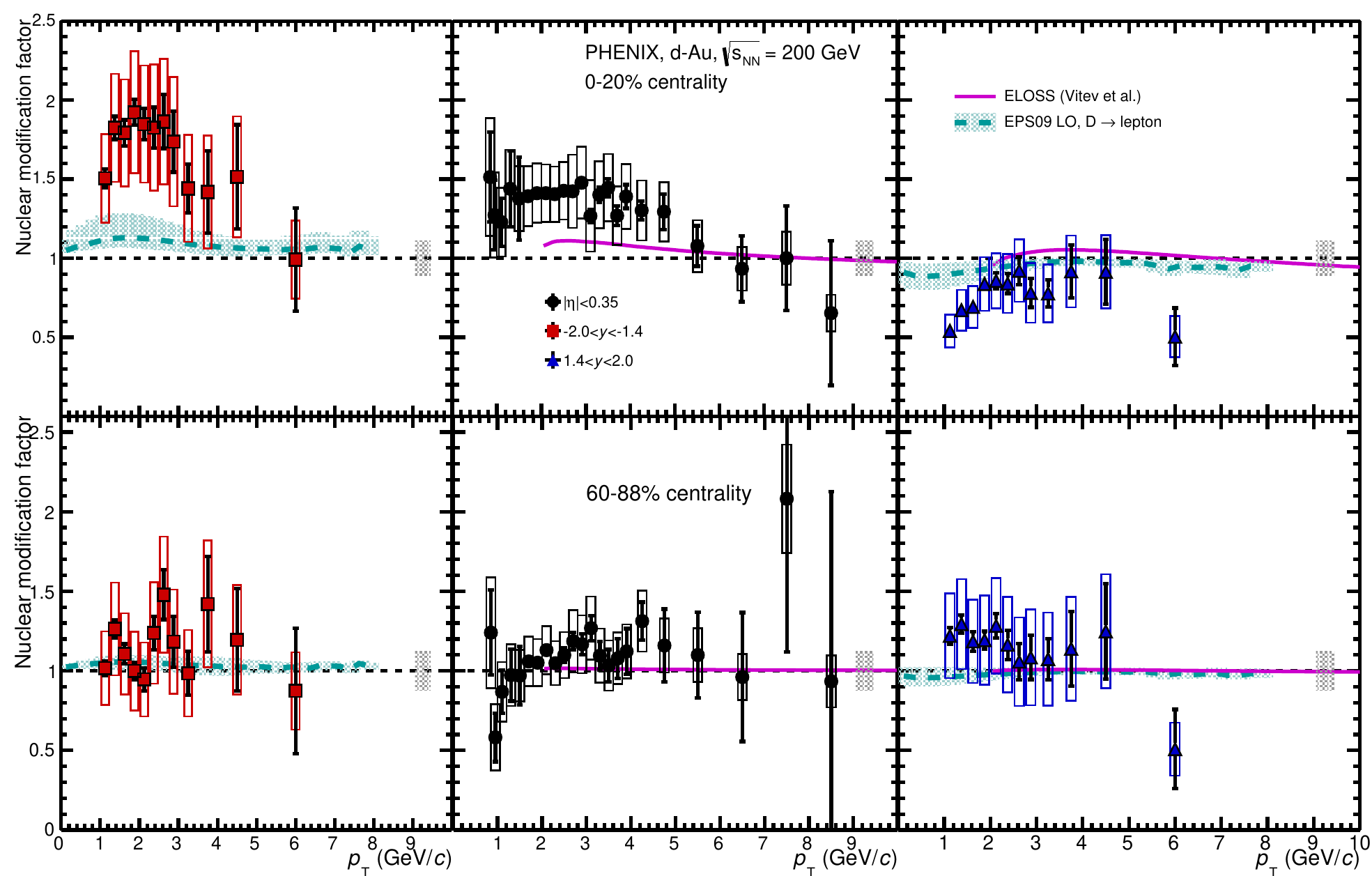} 
	\caption{Nuclear modification factor of heavy-flavour decay leptons in \dAu collisions at $\snn=200$\GeV as a function of transverse momentum in the 0--20\% and 60--88\% centrality classes, as measured with the PHENIX detector~\cite{Adare:2012yxa,Adare:2013lkk}.  
	A PYTHIA calculation considering EPS09 LO is also shown, courtesy of Sanghoon Lim.  
	The calculation by Vitev {\it et al.} considering nPDFs, $\kt$ broadening and CNM energy loss is also shown~\cite{Vitev:2006bi,Vitev:2007ve}.  
	\label{fig:HFlepton_RdAvspt_RHIC_model} 
	} 
\end{figure}

Heavy-flavour decay lepton $\rdau$ measurements at mid-rapidity in minimum-bias d--Au collisions at $\snn=200$\GeV by STAR and PHENIX~\cite{Abelev:2006db,Adare:2012yxa} are consistent and suggest no modification of the multiplicity integrated yields for $1<\pt<10\GeVc$ within uncertainties.  
The $\pt$ dependence of $\rdau$ on the multiplicity and the rapidity was studied by PHENIX~\cite{Adare:2012yxa,Adare:2013lkk} and is reported in \fig{fig:HFlepton_RdAvspt_RHIC}.  
It shows a mild dependence with the multiplicity at mid-rapidity. The results at forward and backward rapidities are similar for peripheral collisions, but evidence a strong deviation for the most central events.  
As shown in \fig{fig:HFlepton_RdAvspt_RHIC_model} and in~\cite{Adare:2012yxa},  
the measurements at forward rapidity are described both by the model of Vitev {\it et al.}~\cite{Vitev:2006bi,Vitev:2007ve}  -- considering nPDFs, $\kt$ broadening and CNM energy loss -- (ELOSS model described in \sect{sec:eloss}) or by nPDFs alone. 
Data at backward rapidity can not be described considering only the nPDFs, suggesting that other mechanisms are at work.  

The preliminary results at LHC energies by the ALICE Collaboration~\cite{Li:2014dha} present $\rppb$ multiplicity-integrated values close to unity at mid-rapidity, as observed at lower energies. The rapidity dependence of the multiplicity-integrated $\rppb$ is also similar to that observed at RHIC. In contrast to RHIC, model calculations with nPDFs present a fair agreement with LHC data.  
The first preliminary measurements of the beauty-hadron decay electron $\rppb$ at mid-rapidity by ALICE are consistent with unity within larger uncertainties~\cite{Li:2014dha}.

The similar behaviour of RHIC and LHC heavy-flavour decay lepton $\rpa$, within the large uncertainties, despite the different $x$-Bjorken ranges covered, suggests that nPDFs might not be the dominant effect in heavy-flavour production. Additional mechanisms like $k_{\rm T}$-broadening, initial or final-state energy loss could be at play.

\paragraph{Dilepton invariant mass} 
 
\begin{figure}[!tbp] 
	\centering 
	\includegraphics[width=0.75\columnwidth]{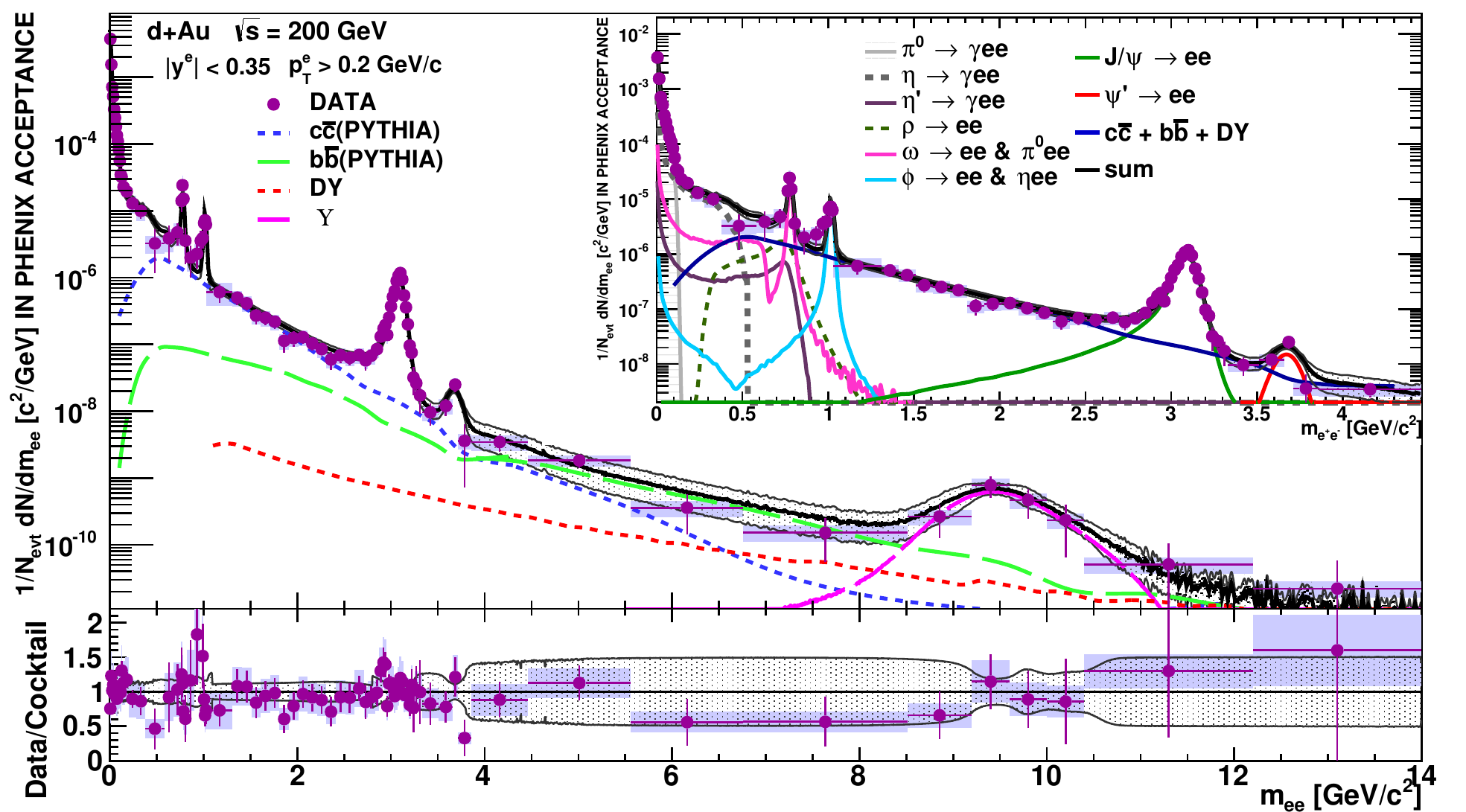} 
	\caption{ 
	Inclusive $\ee$ pair yield from minimum bias \dAu collisions at $\snn=200$\GeV as a function of dilepton invariant mass~\cite{Adare:2014iwg}. The data are compared to the PHENIX model of expected sources. The insert shows in detail the mass range up to $4.5~\GeVc^{2}$. In the lower panel, the ratio of data to expected sources is shown with systematic uncertainties. 
	\label{fig:HFdilepton_dAu_PHENIX} 
	} 
\end{figure}

The $\ccbar$ and $\bbbar$ production cross sections can be obtained by a fit of the $\pt$-integrated dilepton yields as a function of the pair mass. Such measurement has been performed by PHENIX at mid-rapidity in \dAu collisions~\cite{Adare:2014iwg} at $\snn=200$\GeV (see \fig{fig:HFdilepton_dAu_PHENIX}).  
The contributions of pseudo-scalar mesons, $\pi^0$ and $\eta$, and vector mesons, $\omega$, $\phi$, $\jpsi$ and $\Upsilon$ were simulated based on the measured \dAu cross sections. The sources not directly measured  ($\eta^{\prime}$, $\rho$, $\psi^{\prime}$) were studied in simulation and their contribution determined relative to the measured particles. The Drell-Yan mechanism contribution was simulated with the PYTHIA event generator, and its normalisation was one of the fit parameters.  
The resulting $\bbbar$ production cross section is: ${\rm d}\sigma_{\bbbar}/ {\rm d}y = 0.54 \pm 0.11 \, \rm{(stat)} \pm 0.18 \, \rm{ (syst)}$~mb.  
The large model dependence prevents an accurate measurement of $\sigma_{\ccbar}$. 

\paragraph{{\rm D} mesons} 
 
\begin{figure}[!tbp] 
	\centering 
	\includegraphics[width=0.5\columnwidth]{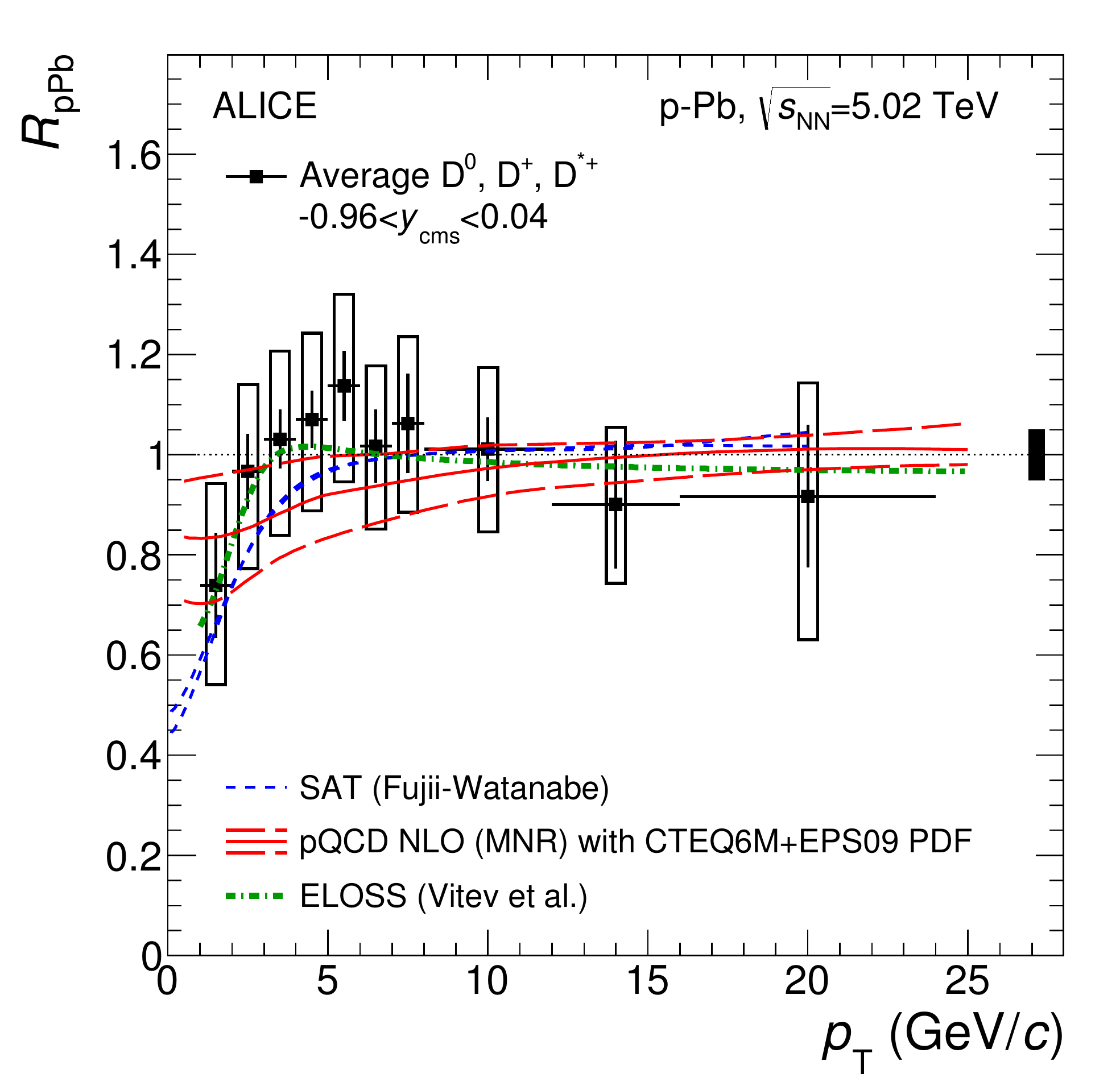} 
	\caption{Nuclear modification factor of prompt D mesons in \pPb collisions at $\snn=5.02$\TeV as a function of transverse momentum as measured with the ALICE detector~\cite{Abelev:2014hha}. The measurements are compared with theoretical calculations including various CNM effects.   
	\label{fig:D_RpAvspt_LHC} 
	} 
\end{figure}

The $\pt$-differential production cross section of $\Dzero$, $\Dplus$, $\Dstarplus$ and $\Ds$ in minimum bias \pPb collisions at $\snn=5.02$~TeV for $|y_{\rm lab}|<0.5$ was published in~\cite{Abelev:2014hha} by ALICE.  
D mesons are reconstructed via their hadronic decays in different $\pt$ intervals from $1\GeVc$ up to $24\GeVc$.  
Prompt D-meson yields are obtained by subtracting the contribution of secondaries from B-hadron decays, determined using pQCD-based estimates~\cite{ALICE:2011aa,Abelev:2014hha}. 
No significant variation of the $\rppb$ among the D-meson species is observed within uncertainties.  
The multiplicity-integrated prompt D (average of $\Dzero$, $\Dplus$ and $\Dstarplus$) meson $\rppb$ is shown in \fig{fig:D_RpAvspt_LHC} together with model calculations.  
$\rppb$ is compatible with unity in the measurement $\pt$ interval, indicating smaller than 10--20\% nuclear effects for $\pt>2\GeVc$. 
Data are described by calculations considering only initial-state effects: NLO pQCD estimates (MNR~\cite{Mangano:1991jk}) considering EPS09 nPDFs~\cite{Eskola:2009uj} or Colour Glass Condensate computations~\cite{Fujii:2013yja} (SAT model described in \sect{sec:saturation}). Predictions including nPDFs, initial or final state energy loss and $\kt$-broadening~\cite{Sharma:2009hn} (ELOSS model discussed in \sect{sec:eloss}) also describe the measurements.   
 
Preliminary measurements of the prompt D meson production as a function of the multiplicity were performed by ALICE~\cite{Russo:2014iia}. The nuclear modification factor of D mesons was evaluated as a function of the event activity, defined in intervals of multiplicity measured in different rapidity intervals. No event activity dependence is observed within uncertainties. D meson production has also been studied as a function of charged-particle multiplicity. The D meson per-event yields increase as a function of the multiplicity at mid-rapidity. The enhancement of the relative D meson yields is similar to that of \pp collisions at $\s=7$\TeV, described in \sect{sec:pp:HadCorrelations}.  The results in \pp collisions favour the scenarios including the contribution of multiple-parton interactions (MPI), parton-percolation or hydrodynamic effects. In \pPb collisions, the cold nuclear matter effects and the contribution of multiple binary nucleon collisions should also be taken into account.

\begin{figure}[!t] 
	\centering 
	\includegraphics[width=0.42\columnwidth]{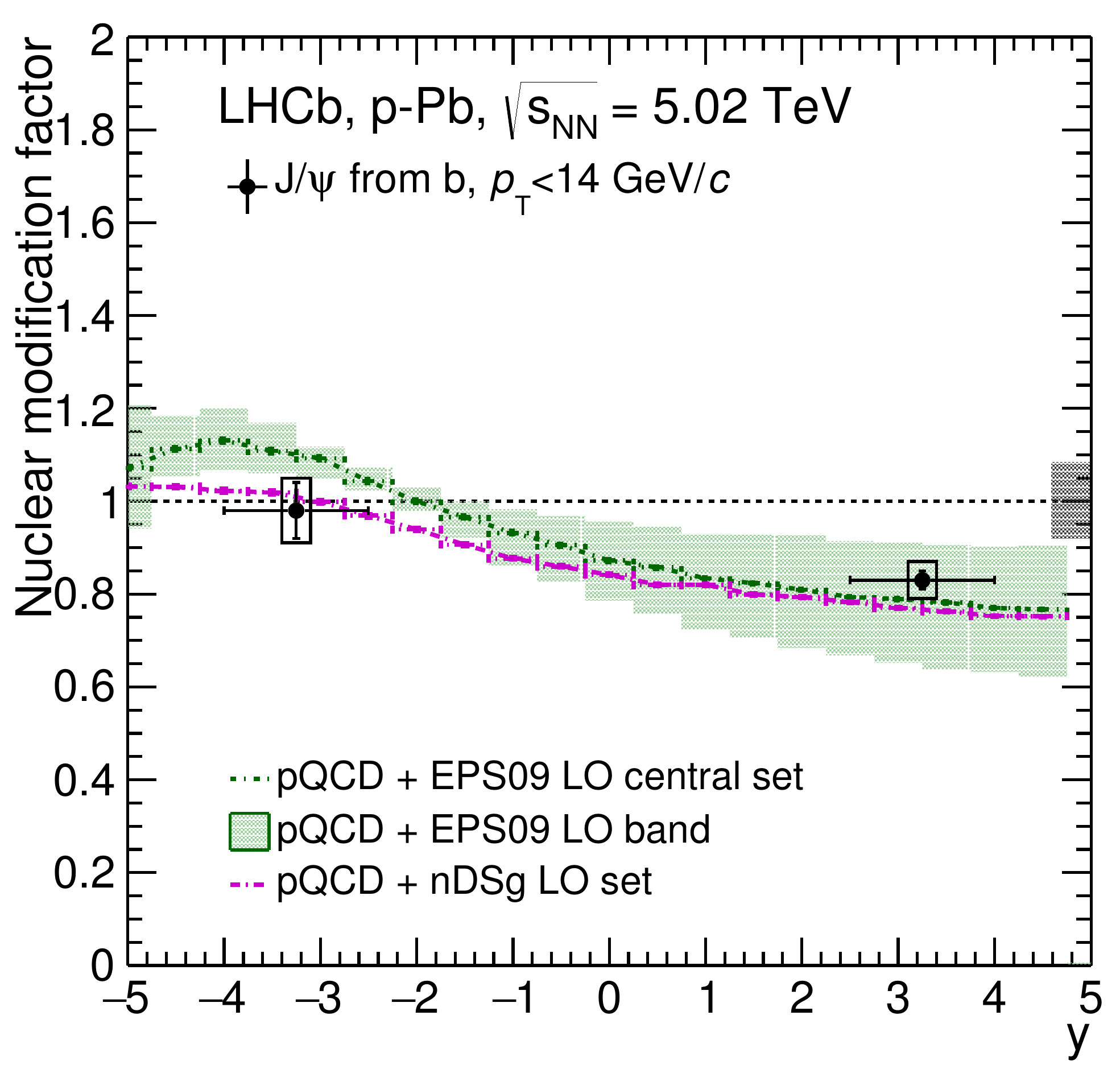} 
	\includegraphics[width=0.42\columnwidth]{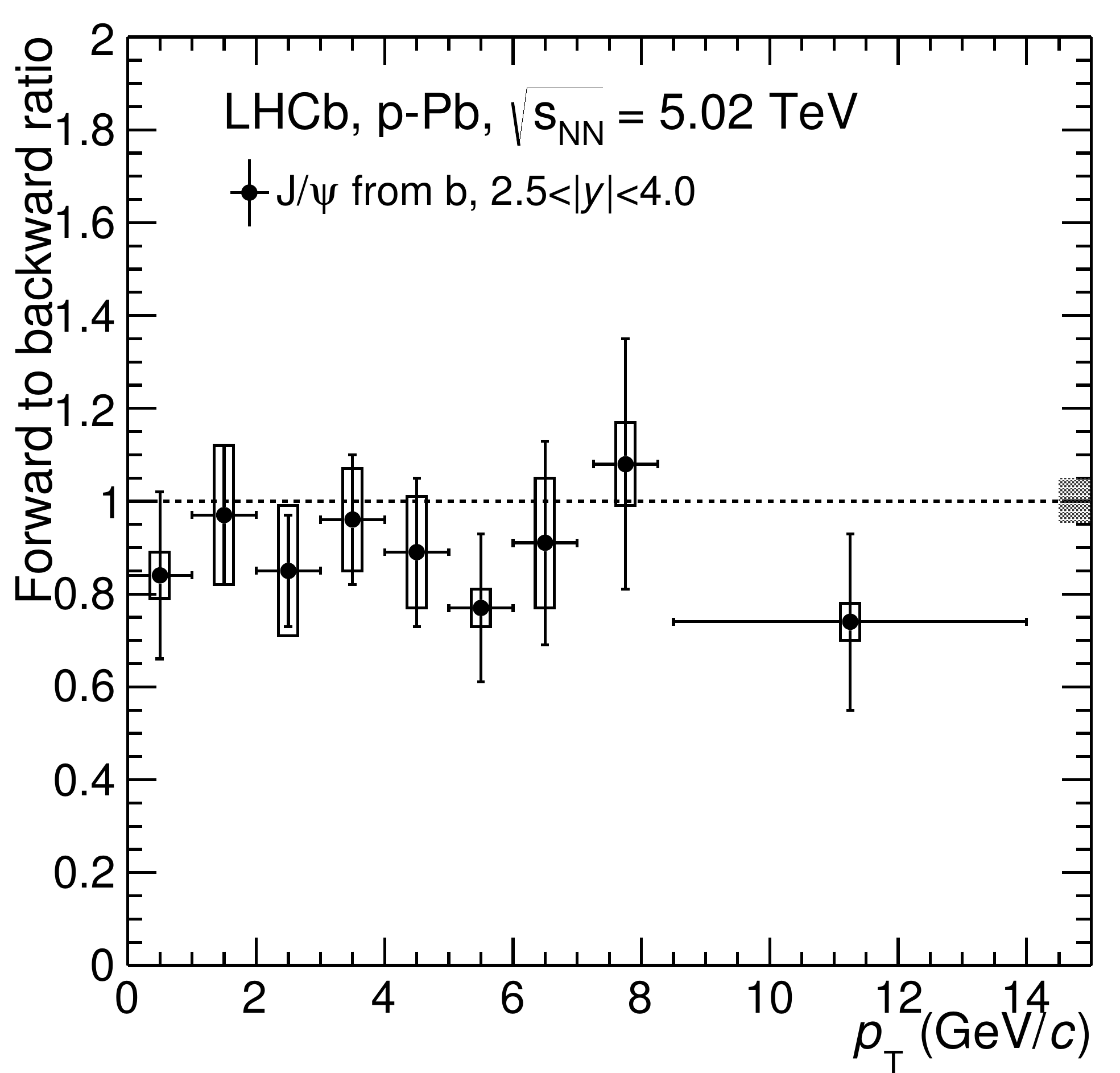} 
	\caption{LHCb measurements of non-prompt $\jpsi$ mesons in \pPb collisions at $\snn=5.02$\TeV~\cite{Aaij:2013zxa}. Left: nuclear modification factor as a function of rapidity, compared to nPDF-based calculations~\cite{delValle:2014wha}. Right: forward to backward rapidity ratio as a function of transverse momentum.  
	\label{fig:NonPromptJpsi_pPb_LHCb} 
	} 
\end{figure}

\paragraph{Open beauty measurements}

The first measurements of the beauty production cross section in \pA collisions down to $\pt=0$ were carried out by LHCb in \pPb collisions at $\snn=5.02$~TeV~\cite{Aaij:2013zxa}. These results were achieved via the analysis of non-prompt $\jpsi$ mesons at large rapidities, $2<\ylab<4.5$.  
$\jpsi$ mesons were reconstructed by an invariant mass analysis of opposite sign muon pairs. The fraction of $\jpsi$ originated from beauty decays, or non-prompt $\jpsi$ fraction, was evaluated from a fit of the component of the pseudo-proper decay time of the $\jpsi$ along the beam direction.  
The $\rppb$ of non-prompt $\jpsi$ was computed considering as \pp reference an interpolation of the measurements performed in the same rapidity interval at $\s=2.76$, 7 and 8\TeV (see \sect{sec:CNM:ppRef}). \fig{fig:NonPromptJpsi_pPb_LHCb}~(left) reports the $\pt$-integrated $\rppb$ as a function of rapidity, whereas \fig{fig:NonPromptJpsi_pPb_LHCb}~(right) presents the double ratio of the production cross section at positive and negative rapidities, $\rfb$, as a function of the $\jpsi$ transverse momentum.  
The $\pt$-integrated $\rppb$ is close to unity in the backward rapidity range, and shows a modest suppression in the forward rapidity region.  
$\rfb$ is compatible with unity within the uncertainties in the measured $\pt$ interval, with values almost systematically smaller than unity. 
These results indicate a moderate rapidity asymmetry, and are consistent with the $\rppb$ ones.  
The results are in agreement with LO pQCD calculations including EPS09 or nDSg nuclear PDF parametrisations.  
The ATLAS Collaboration has also measured the $\rfb$ of non-prompt $\jpsi$ for $8<\pt<30\GeVc$ and $|\ycm|<1.94$~\cite{Aad:2015ddl}. These results are consistent with unity within experimental uncertainties and no significant $\pt$ or $\ycm$ dependence is observed within the measured kinematic ranges.
 
%
A preliminary measurement of the production of B mesons  in \pPb collisions at $\snn=5.02$\TeV was carried out by the CMS collaboration~\cite{CMS:2014tfa,Innocenti:2014hha}. 
B$^0$, B$^+$ and B$_s^0$ mesons are reconstructed via their decays to $\jpsi + {\rm K}$ or $\phi$ at mid-rapidity for $10<\pt<60\GeVc$. The $\dsdpt$ of B$^0$, B$^+$ and B$_s^0$ are described within uncertainties by FONLL predictions scaled by the number of nucleons in the nucleus.  
B$^+$ $\dsdy$ is also described by FONLL binary scaled calculations, and presents no evidence of rapidity asymmetry within the measurement uncertainties.  
These results suggest that B-hadron production for $\pt>10\GeVc$ is not affected, or mildly, by CNM effects.

%
Preliminary results of the $\pt$ and $\eta$ differential cross section of $b$-jets in \pPb collisions at $\snn=5.02$\TeV have been reported by CMS at mid-rapidity~\cite{CMS:2014tca}.  
Jets from $b$-quark fragmentation are identified studying the distribution of secondary vertices, typically displaced by several mm for jets of $\pt \sim 100~\GeVc$.  
The measured $b$-jet fraction for $50< \pt^{b-jet}<400\GeVc$ is consistent with PYTHIA simulations with the Z2 tune~\cite{Sjostrand:2006za,Sjostrand:2007gs}.  
The $\pt$- and $\eta$-differential spectra are also described by binary-scaled PYTHIA simulations within uncertainties. 
$\rppb$ is computed using PYTHIA as \pp reference and is compatible with unity.  
These results conform with the expectations that cold nuclear matter effects are not sizeable at large $\pt$.

\begin{figure}[!t] 
\begin{center} 
\includegraphics[width=0.45\columnwidth,height=0.35\columnwidth]{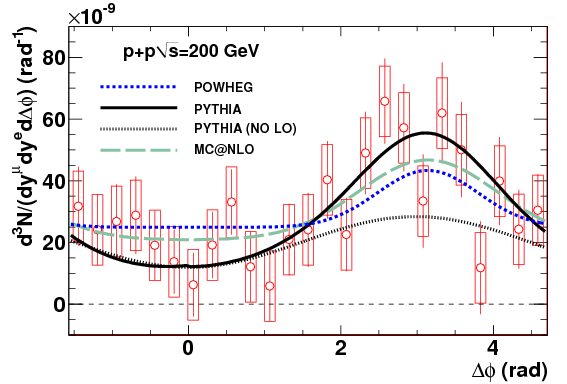} 
\includegraphics[width=0.475\columnwidth,height=0.35\columnwidth]{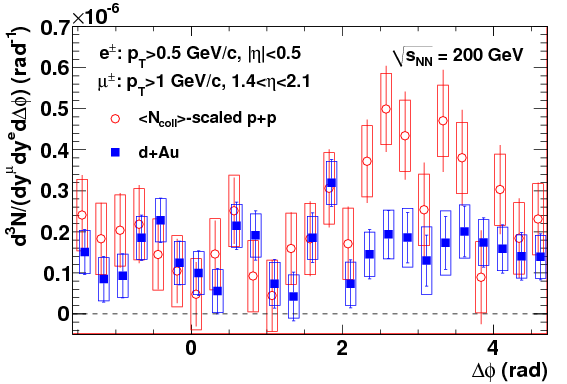} 
 
\caption{ 
Heavy-flavour decay electron  ($\pt>0.5\GeVc$, $|\eta|<0.5$) to heavy-flavour decay muon ($\pt>1\GeVc$, $1.4<\eta<2.1$) $\Delta \phi$ correlations in \pp (left) and \dAu (right) collisions at $\s=200$\GeV~\cite{Adare:2013xlp}.  
The \pp results are compared to POWHEG, PYTHIA and MC\@NLO calculations.  
} 
\label{fig:pA:HFeHFm} 
\end{center} 
\end{figure}

\paragraph{Heavy-flavour azimuthal correlations} 
 
As described in \sect{sec:pp:AngleCorrelations}, heavy-flavour particle production inherits the heavy-quark pair correlation, bringing information on the production mechanisms.  
Heavy-flavour production in \pA collisions is influenced by initial and/or final state effects.  
The modification of the PDFs or the saturation of the gluon wave function in the nucleus predict a reduction of the overall particle yields. 
The CGC formalism also predicts a broadening and suppression of the two-particle away-side azimuthal correlations, more prominent at forward rapidities~\cite{Gelis:2010nm,Marquet:2007vb,Lappi:2012nh}.  
Energy loss or multiple scattering processes in the initial or final state are also expected to cause a depletion of the two-particle correlation away-side yields~\cite{Kang:2011bp}. These effects could also affect heavy-flavour correlations in \pA collisions.  
 
%
%
Heavy-flavour decay electron ($\pt>0.5\GeVc$, $|\eta|<0.5$) to heavy-flavour decay muon ($\pt>1\GeVc$, $1.4<\eta<2.1$) $\Delta \phi$ azimuthal correlations have been studied by PHENIX in \pp and \dAu collisions at $\s=200$\GeV~\cite{Adare:2013xlp}.  
They exploit the forward rapidity muon measurements in order to probe the low-$x$ region in the gold nucleus.  
The analysis considers the angular correlations of all sign combinations of electron-muon pairs. The contribution from light-flavour decays and conversions is removed by subtracting the like-sign yield from the unlike-sign yield.  
\fig{fig:pA:HFeHFm} presents the electron-muon heavy-flavour decay $\Delta \phi$ correlations.  
Model calculations are compared to data for \pp collisions, see \fig{fig:pA:HFeHFm} (left). Calculations from NLO generators seem to fit better the $\Delta \phi$ distribution than LO simulations.   
The corresponding measurement in \dAu collision, see \fig{fig:pA:HFeHFm} (right), shows a reduction of the away-side peak as compared to \pp scaled data, indicating a modification of the charm kinematics due to CNM effects.  
%

Preliminary results of D--hadron azimuthal correlations in \pPb collisions at $\snn=5.02$\TeV were carried out by the ALICE Collaboration~\cite{Bjelogrlic:2014kia}. The measurement uncertainties do not allow a clear  conclusion on a possible modification of heavy-quark azimuthal correlations with respect to pp collisions.

\subsubsection{Quarkonium measurements} 
\label{sec:CNM:Onia} 
 
Quarkonia are mainly measured via their leptonic decay channels.  
In the PHENIX experiment, the Ring Imaging Cherenkov 
 associated with the Electromagnetic Calorimeter (EMCAL) allows one to identify  
electrons at mid-rapidity ($|y| < 0.35$). In this rapidity range where the EMCAL  
can reconstruct the photons, \chic can also be measured from its decay channel  
to \jpsi and photon. At backward and forward rapidity ($1.2 < |y| < 2.2$), 
 two muon spectrometers allow the reconstruction of quarkonia via their muonic  
decay channel. In the STAR experiment, quarkonia are reconstructed at mid-rapidity  
($|y| < 1$) thanks to the electron identification and momentum measurements from  
the TPC. In the ALICE experiment, a TPC at mid-rapidity ($|\ylab| < 0.9$) 
 is used for electron reconstruction and identification and a spectrometer at  
forward rapidity for muon reconstruction ($2.5 < \ylab < 4$). The LHCb 
 experiment is a forward spectrometer that allows for the quarkonium measurement  
via their muonic decay channel for $2 < \ylab < 4.5$. In the CMS experiment,  
quarkonia are reconstructed in a large range around mid-rapidity  ($|\ylab| < 2.4$)  
via the muonic decay channel. In LHCb, CMS and  in ALICE at mid-rapidity, the  
separation of prompt \jpsi from inclusive \jpsi exploits the long lifetime of  
$b$ hadrons, with $c\tau$ value of about 500 $\mu$m, using the good  
resolution of the vertex detector.  
 
\paragraph{Charmonium} 
 
\begin{figure}[!tbp] 
	\centering 
	\includegraphics[width=0.8\columnwidth]{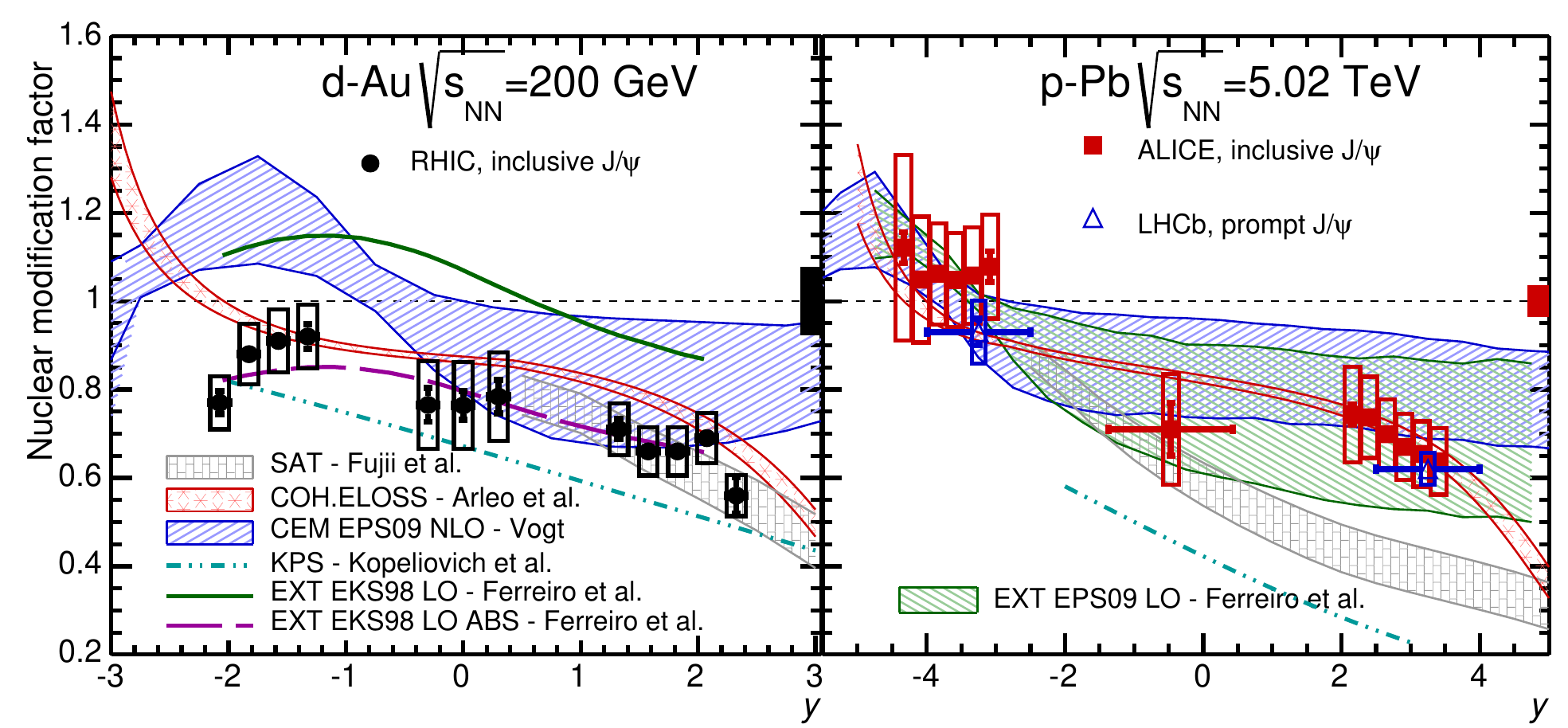} 
	\caption{Left: rapidity dependence of \rdau for inclusive \jpsi in PHENIX~\cite{Adare:2010fn}.  
The error bars represent the uncorrelated uncertainties (statistical and systematic), the open boxes the  
point-to-point correlated systematic uncertainties and the box at unity the correlated one. Right: rapidity  
dependence of \rppb for inclusive \jpsi in ALICE~\cite{Abelev:2013yxa,Adam:2015iga} and prompt \jpsi  
in LHCb~\cite{Aaij:2013zxa}. The error bars represent the statistical uncertainties while the open boxes  
the systematic uncertainties. Other uncertainties are displayed similarly to PHENIX. } 
        \label{fig:fig01}   
\end{figure}

The nuclear modification factor for inclusive and/or prompt \jpsi has been measured for a large range in rapidity  
and is shown in \fig{fig:fig01} for RHIC (left) and LHC (right). It should be emphasized that there  
are no \pp\ measurements at \snn=5.02~TeV at the LHC and the \pp\ cross section interpolation  
procedure described in \sect{sec:CNM:ppRef} results into additional uncertainties.  
 
The measurements from  
PHENIX~\cite{Adare:2010fn} in \dAu collisions at \snn~=~200~\GeV cover four units of rapidity.  
The \jpsi is suppressed with respect to binary-scaled \pp collisions in the full rapidity range with a suppression that can  
reach more than 40\% at $y=2.3$.  
Inclusive \jpsi includes a contribution from prompt \jpsi (direct \jpsi and excited charmonium states,  
\chic and $\rm \psi(2S)$) and a contribution from decays of B mesons. At RHIC energy, the contribution from B-meson  
decays to the inclusive yield is expected to be small, of the order of $3\%$~\cite{Cacciari:2005rk},  
but has not been measured so far in \dAu collisions.  
The contribution from excited states such as \chic and $\rm \psi(2S)$ has been measured at  
mid-rapidity~\cite{Adare:2013ezl} and is discussed  
later in this section. While the inclusive \jpsi \rdau for $ |y| < 0.9$ is found to be  
$0.77\pm0.02\rm{(stat)}\pm0.16\rm{(syst)}$, the correction from \chic and  
$\rm \psi(2S)$ amounts to $5\%$ and leads to a feed-down corrected \jpsi \rdau of  
$0.81\pm0.12\rm{(stat)}\pm0.23\rm{(syst)}$.  
 
At the LHC, the results for inclusive \jpsi from ALICE~\cite{Abelev:2013yxa,Adam:2015iga} and  
for prompt \jpsi from LHCb~\cite{Aaij:2013zxa} show a larger suppression of the \jpsi production with respect  
to the binary-scaled \jpsi production in \pp collisions at forward rapidity (40\% at $\ycm=3.5$).  
In the backward rapidity region the nuclear modification factor is  
slightly suppressed (prompt \jpsi from LHCb) or enhanced (inclusive \jpsi from ALICE) but within the uncertainties  
compatible with unity.  
ATLAS and LHCb have also measured the production of \jpsi from B mesons~\cite{Aaij:2013zxa,Aad:2015ddl}: they contribute to the  
inclusive \jpsi yield integrated over \pt by 8\% at $-4 < \ycm < -2.5$ and 12\% at $1.5<\ycm<4$ with an increase towards mid-rapidity  
and high \pt region. At $\pt>8$ GeV/c, the fraction of B mesons  
can reach up to 34\% at mid-rapidity and up to 26\% in the backward rapidity region covered by LHCb.  
In addition, the nuclear modification factor for \jpsi from B mesons is above 0.8 when  
integrated over \pt, as shown in \fig{fig:NonPromptJpsi_pPb_LHCb},  
At low \pt, a small effect from B mesons on inclusive \jpsi measurements is therefore expected at LHC energy and  
this is confirmed by the comparison of prompt to inclusive \jpsi that shows a good agreement as seen in the  
right panel of \fig{fig:fig01}. 
 
The models based on nuclear PDFs~\cite{Vogt:2010aa,Ferreiro:2008wc,Ferreiro:2013pua} (CEM EPS09 NLO, EXT EKS98 LO and EXT EPS09 LO),  
gluon saturation~\cite{Fujii:2013gxa} (SAT), multiple scattering and energy loss~\cite{Arleo:2010rb,Kopeliovich:2011zz} (COH.ELOSS and KPS)  
described in \sect{CNM_theory} are also shown in \fig{fig:fig01}.  
The uncertainty from the nuclear PDFs on the gluon distribution  
function is large as discussed in \sect{sec:npdf} and is shown by the uncertainty band of the corresponding calculations.  
The models based on nPDFs overestimate the data at RHIC in particular  
at backward rapidity, the anti-shadowing region. At forward rapidity, a strong shadowing with the EPS09 NLO nPDFs  
parametrisation is favoured by the RHIC data. By including a \jpsi absorption cross section,  
$\sigma_{\rm abs}^{\jpsi}=4.2$ mb, the calculation  
from EXT EKS98 LO ABS that uses EKS98 LO nPDFs can describe the \rdau measured at RHIC in the full rapidity range.  
In the latter calculations, since the behaviour of EKS98 is very close to the one of the central set of EPS09 LO,  
the theoretical curves are expected to be similar to those of EPS09 LO nPDFs.   
At the LHC, while the backward rapidity data is well described by the nPDF models, a strong shadowing is  
favoured by the data at forward rapidity. Both EPS09 LO and the lower uncertainty band of EPS09 NLO  
parametrisations provide such a strong shadowing.  
In the COH.ELOSS approach, the rapidity dependence of the nuclear modification factor is well described both at RHIC and LHC energies.  
In the KPS model, the rapidity dependence of the RHIC data is correctly described but the calculations are systematically lower than the measurements.  
At LHC energy, the KPS model overestimates the \jpsi suppression over the full rapidity range.  
Finally, the SAT model is not valid for the full rapidity  
range, see \sect{sec:saturation}. While it describes correctly the data for $y>0.5$ at $\snn=200$~\GeV and the mid-rapidity data at  
$\snn=5.02$~\TeV, it overestimates the \jpsi suppression at forward rapidity at $\snn=5.02$~\TeV.  
 
\begin{figure}[!tbp]
	\centering 
	\includegraphics[width=0.5\columnwidth]{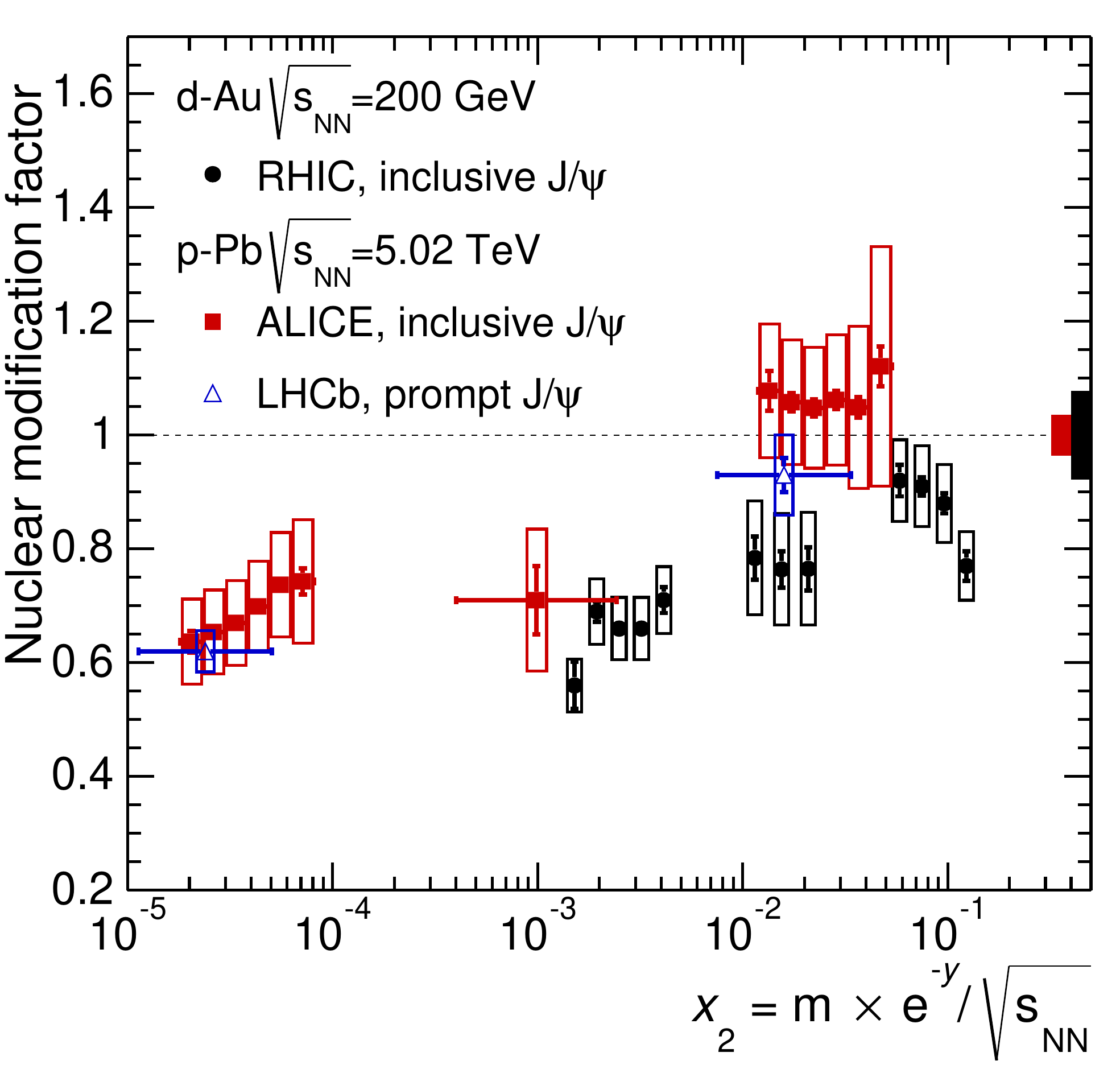} 
	\caption{$x_2 = \frac{m}{\snn}\exp(-y)$ dependence of \rpa for inclusive J/$\psi$  
in PHENIX~\cite{Adare:2010fn} and ALICE~\cite{Abelev:2013yxa,Adam:2015iga} and for prompt \jpsi in LHCb~\cite{Aaij:2013zxa}.  
The uncertainties are described in \fig{fig:fig01}.}  
        \label{fig:fig01x2}   
\end{figure} 
 
It is interesting to check whether a simple scaling exists on $\jpsi$ suppression between RHIC and LHC. The effects of nPDF or saturation are expected to scale with the momentum fraction $x_2$, independently of the centre-of-mass energy of the collision. It is also the case of nuclear absorption, since the $\jpsi$ formation time is proportional to the Lorentz factor $\gamma$, which is uniquely related to $x_2$, $\gamma = m / (2 m_p\ x_2)$, assuming $2\to 1$ kinematics for the production process. 
In order to test the possible  $x_2$ scaling expected in the case of  
nPDF and nuclear absorption theoretical approaches, the data from RHIC and  
LHC~\cite{Adare:2010fn,Abelev:2013yxa,Adam:2015iga,Aaij:2013zxa} of \fig{fig:fig01} are shown together  
in \fig{fig:fig01x2} as a function of $x_2 = \frac{m}{\snn}\exp(-y)$ where the low $x_2$ values correspond to  
forward rapidity data. Note that \eq{QQ_x-Bjorken} for the calculation of $x_2$,  
which refers to a $2 \to 2$ partonic process, can not be used since the \meanpt values for all the data points  
have not been measured. While at $x_2 < 10^{-2}$, the nuclear modification factors are compatible at RHIC and  
LHC energy within uncertainties, the data presents some tension with a $x_2$ scaling at large $x_2$.    
 
\begin{figure}[!t] 
	\centering 
	\includegraphics[width=0.95\columnwidth]{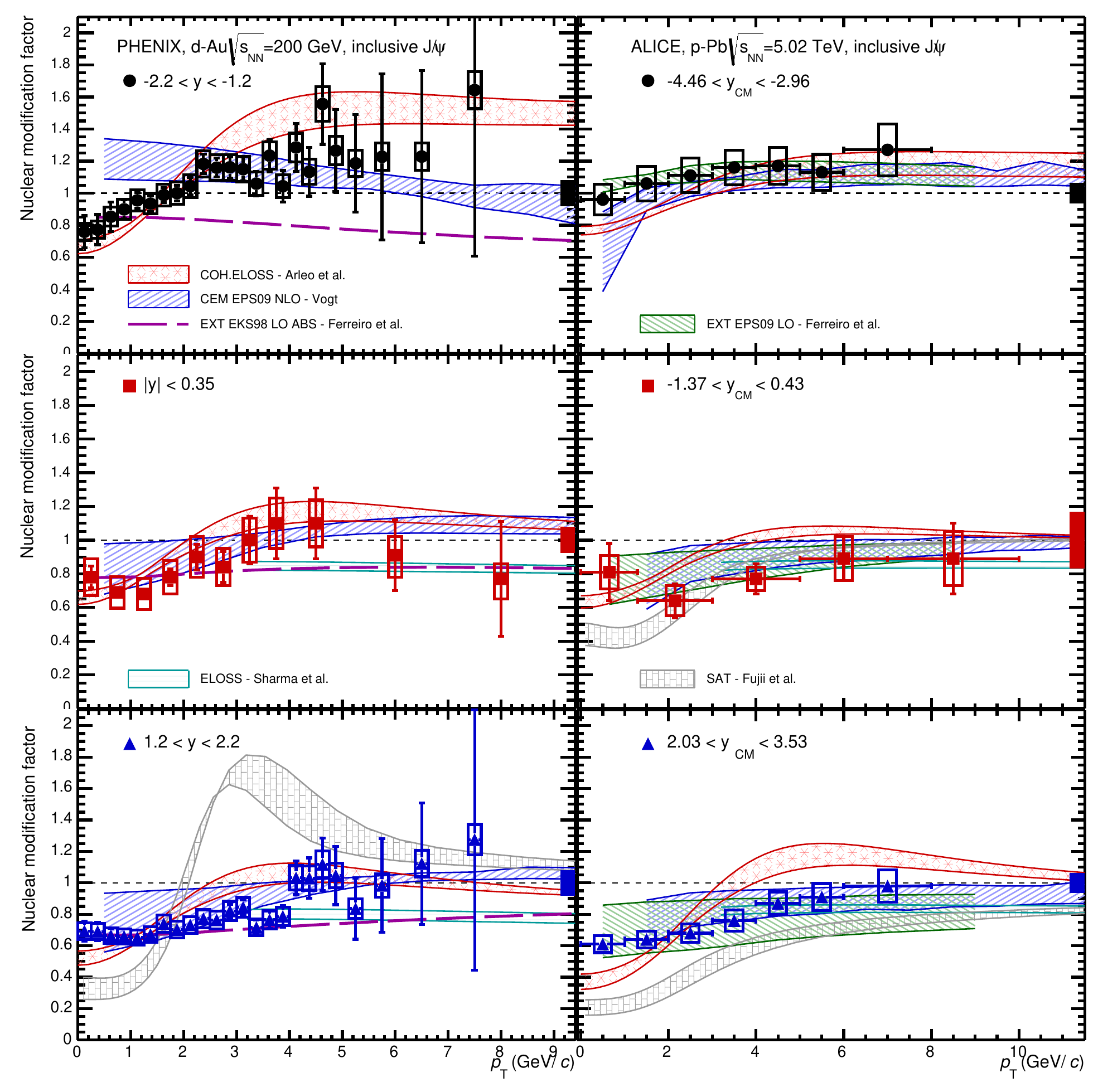} 
	\caption{Left: transverse momentum dependence of \rdau for three rapidity ranges for inclusive \jpsi in PHENIX~\cite{Adare:2012qf}. Right: transverse momentum dependence of \rppb for inclusive \jpsi in ALICE~\cite{Adam:2015iga}. The uncertainties are the same as described in \fig{fig:fig01}.} 
        \label{fig:fig02} 
\end{figure}

The transverse momentum distribution of the nuclear modification factor is shown for different rapidity ranges  
in \fig{fig:fig02} for RHIC (left) and LHC (right) energies. At \snn~=~200~\GeV, the \jpsi \rdau is  
suppressed at low \pt and increases with \pt for the full rapidity range. The mid- and forward rapidity results  
show a similar behaviour: \rdau  
increases gradually with \pt and is consistent with unity at $\pt\gtrsim4$\GeVc. At backward rapidity, \rdau  
increases rapidly to reach 1 at $\pt \approx 1.5$\GeVc and is above unity for $\pt > 2.5$\GeVc. At \snn~=~5.02~\TeV,  
a similar shape and amplitude is observed for \rppb at mid- and forward rapidity: in that case  
it is consistent with unity at $\pt\gtrsim5$\GeVc. The backward rapidity results are consistent throughout the full  
\pt interval with unity.  
 
In addition to the aforementioned models, the calculations based on the energy loss approach from~\cite{Sharma:2012dy} (ELOSS),  
valid for $y \geq 0$ and $\pt > 3$ GeV/c, are also compared to the data. Among these models, only the COH.ELOSS and SAT model includes  
effects from initial- or final-state multiple scattering that may lead to a \pt broadening. The \pt dependence of \rpa is correctly described by the  
CEM EPS09 NLO model except at backward rapidity and $\snn=200$~\GeV. The model based on EXT EKS98 LO ABS  
with an absorption cross section of $4.2$ mb describes the mid- and forward rapidity results at $\snn=200$~\GeV  
but not the \pt dependence at backward rapidity. A good agreement is reached at $\snn=5.02$~\TeV with CEM EPS09 NLO and EXT EPS09 LO calculations.  
The ELOSS model describes correctly the \pt dependence at mid- and forward  
rapidity at both energies. The COH.ELOSS calculations describe correctly the data with however a steeper \pt dependence at forward rapidity  
and $\snn=5.02$~TeV. Finally the SAT model gives a good description of the data at mid-rapidity at $\snn=5.02$~TeV but  
does not describe the \pt dependence at forward rapidity at $\snn=$200~\GeV and overestimates the suppression at forward rapidity at $\snn=5.02$~TeV.  
 
It is also worth mentioning that the ratio \rfb of the nuclear modification factors for a rapidity range  
symmetric with respect to $\ycm \sim 0$ has also been extracted as a function of rapidity and \pt  
at $\snn=5.02$~TeV~\cite{Aaij:2013zxa,Abelev:2013yxa,Aad:2015ddl}. Despite the reduction of statistics (since the rapidity  
range is limited), the \pp cross section and its associated systematics cancels out in the ratio and results on \rfb  
provides additional constraints to the models. 
 
\begin{figure}[!tbp] 
	\centering 
	\includegraphics[width=0.4\columnwidth]{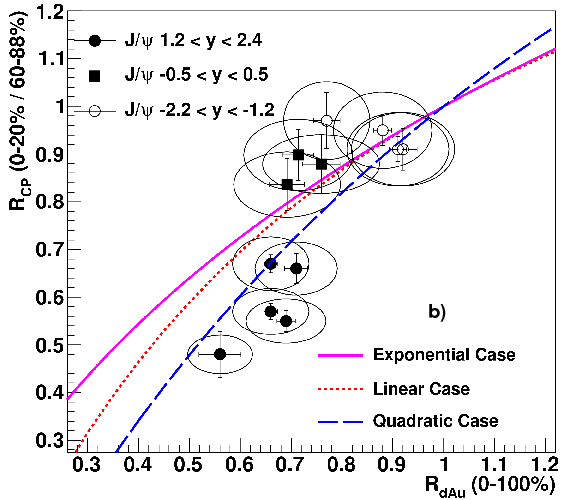} 
	\caption{\rcp as a function of \rdau for inclusive \jpsi in PHENIX~\cite{Adare:2010fn}. The curves are constraint lines for three geometric dependencies of the nuclear modification. The ellipses represent a one standard deviation contour for the systematic uncertainties.} 
        \label{fig:fig03} 
\end{figure}

The dependence of the \jpsi suppression has also been measured in \dAu as a function of the centrality  
of the collision in PHENIX~\cite{Adare:2010fn,Adare:2012qf}. The centrality of the \dAu collision  
is determined thanks to the total energy deposited in the  beam-beam counter (BBC) located in the nucleus direction.  
A larger suppression is observed in central (0--20\%) as compared  
to peripheral (60--88\%) collisions. In order to study the centrality dependence of the nuclear effect, the  
nuclear modification factor between central and peripheral collisions, $R_{\rm CP}$, has been measured.  
\fig{fig:fig03}  
shows \rcp as a function of \rdau: the forward rapidity measurements correspond to the lowest \rcp values.  
The nuclear effect has been parametrised by three functional forms (exponential, linear or quadratic) that depend  
on the density-weighted longitudinal thickness through the nucleus  
$\Lambda(r_\mathrm{T})=\frac{1}{\rho_0}\int \dd z\rho(z,r_\mathrm{T})$. Here $\rho_0$ is the density in the  
centre of the nucleus and $r_\mathrm{T}$ the transverse radial position of the nucleon-nucleon collision relative to the centre  
of the nucleus. While the effect from nuclear absorption is expected to follow an exponential dependence, other models  
like nPDF assume a linear form to describe the centrality dependence of the nuclear effect. While at  
backward and mid-rapidity the data can not discriminate between the functional forms, the forward rapidity data  
suggest that the dependence on $\Lambda(r_\mathrm{T})$ is non-linear and closer to quadratic. 
 
The centrality dependence of the \jpsi nuclear modification factor has also been studied in  
ALICE~\cite{MartinBlanco:2014pia,Lakomov:2014yga}. In  
these analyses, the event activity is determined from the energy measured along the beam line by the Zero Degree  
Neutron (ZN) calorimeter located in the nucleus direction. In the hybrid  
method described in~\cite{Adam:2014qja}, the centrality of the collision in each ZN energy event class is determined  
assuming that the charged-particle multiplicity measured at mid-rapidity is proportional to the number of  
participants in the collision. In the \jpsi case, the data is compatible, within uncertainties, to the binary-scaled  
\pp production for peripheral events at backward and forward rapidity.  
The \jpsi production in \pPb is however significantly modified towards central events.  
At backward rapidity the nuclear modification factor  
is compatible with unity at low \pt and increases with \pt reaching about $1.4$ at \pt$\sim7$~\GeVc.  
At forward rapidity the suppression of \jpsi production shows an increase towards central events  
specially at low \pt. The \jpsi production was also studied as a function of the relative charged-particle multiplicity  
measured at mid-rapidity as it was already done in \pp collisions~\cite{Abelev:2012rz}.  
The results show an increase with  
the relative multiplicity at backward and forward rapidity. At forward rapidity the multiplicity dependence becomes  
weaker than at backward rapidity for high relative multiplicities.  
In \pp collisions~\cite{Abelev:2012rz} this increase is interpreted in terms of the hadronic activity  
accompanying \jpsi production, from contribution of multiple parton-parton interactions or  
in the parton percolation scenario. In \pPb collisions, in addition to the previous contributions, the cold nuclear  
matter effects should be considered when interpreting these results.       
 
\begin{figure}[!tbp] 
	\centering 
	\includegraphics[width=0.8\columnwidth]{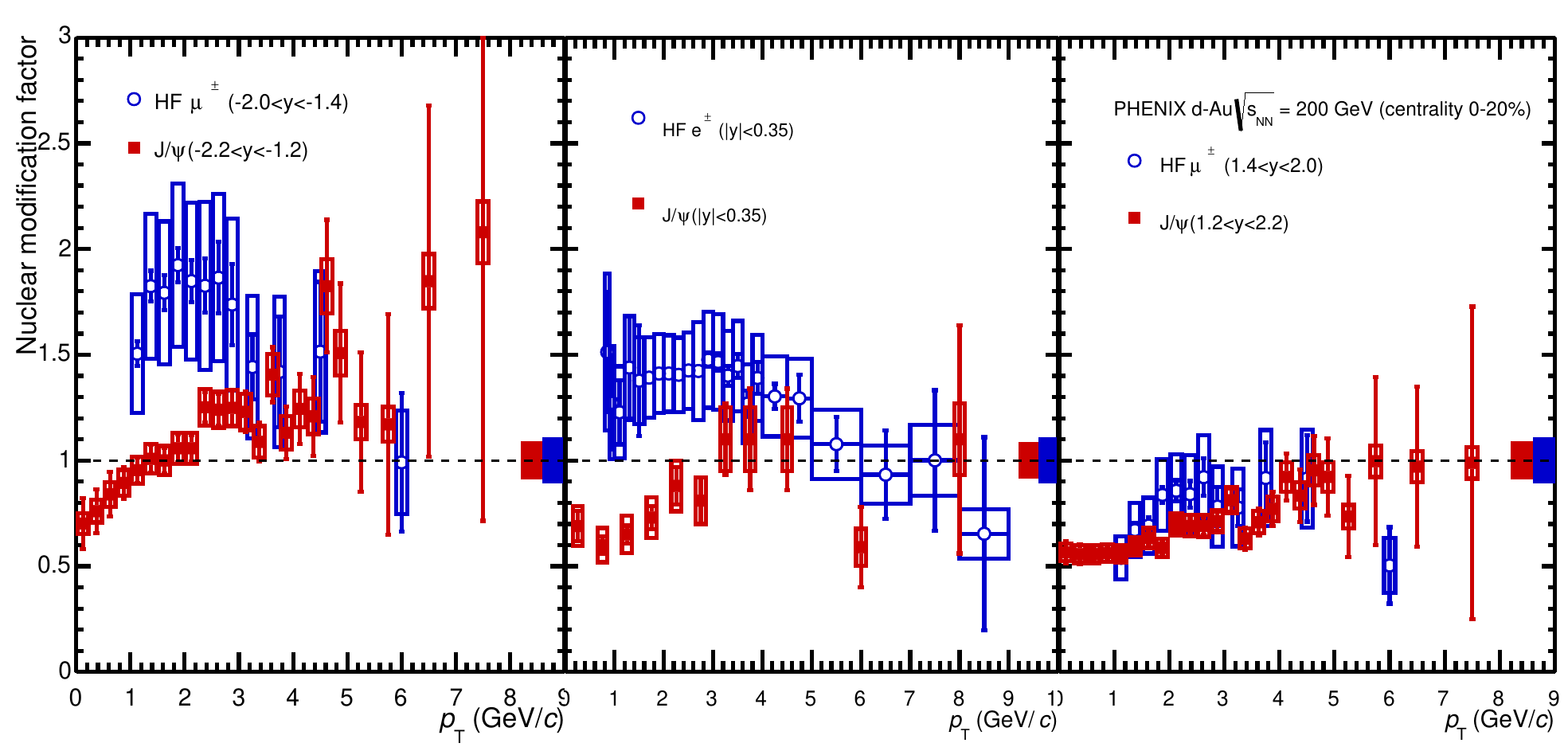} 
	\caption{Transverse momentum of \rdau of inclusive J/$\psi$ for three different rapidity ranges in 0--20\% centrality bin~\cite{Adare:2012qf} and comparison to heavy flavour electron and muon in PHENIX~\cite{Adare:2012yxa,Adare:2013lkk}.} 
        \label{fig:fig05} 
\end{figure}

Since open and hidden heavy flavour hadrons are characterized by the same production process for the heavy quark pair,  
a direct comparison of their productions, if measured over the entire phase space, is expected to single out 
 final-state effects on \jpsi production.  
In \fig{fig:fig05}, the \jpsi \rdau~\cite{Adare:2012qf} is compared to the one of open heavy-flavour  
decay leptons~\cite{Adare:2012yxa,Adare:2013lkk} as measured by PHENIX in central \dAu collisions.  
Despite the fact that the open beauty contribution is not subtracted and  
the measurement is carried out only down to \pt = 1 \GeVc,  
this comparison may already give some hint on the final-state effects on \jpsi production.  
A similar behaviour across the entire \pt range is observed for \rdau at forward rapidity, suggesting that  
the suppression of \jpsi production is related to the suppression of \ccbar pair production.  
At backward and mid-rapidity the \jpsi is clearly more suppressed than leptons from open heavy-flavour decays at  
low \pt, where the charm contributions dominate over those from bottom~\cite{Adare:2009ic}.  
This difference between \jpsi and open charm may originate from additional effects beyond charm quark pair  
production, such as a longer crossing time \tcross of the \ccbar state in the nuclear matter or  
a larger density of comoving medium~\cite{Ferreiro:2014bia}. This comparison suggests that an additional CNM  
final-state effect significantly affects \jpsi production at backward and mid-rapidity at \snn=200~\GeV. One  
should however emphasize that the comparison of the open heavy-flavour and \jpsi production  
is carried out as a function of \pt. The c quark fragments into a charm mesons which  
in turn decays into a lepton and it is not straightforward to relate the decay lepton momentum to the  
parent quark momentum in order to interpret accurately this comparison. 
 
The binding energy of the excited charmonium states is significantly smaller than that of the ground  
state~\cite{Satz:2005hx}: the \psiP  
has the lowest binding energy (0.05~GeV), following by the \chic (0.20~GeV) and the \jpsi (0.64~GeV).  
The excited charmonium  
states are then expected to be more sensitive to the nuclear environment as compared to the \jpsi.  
The relative suppression of the \psiP to \jpsi from earlier measurements at lower energy and at mid-rapidity 
~\cite{Leitch:1999ea,Abt:2006va,Alessandro:2006jt} has been understood as a larger absorption  
of the \psiP in the nucleus since,  
in these conditions, the crossing time \tcross of the \ccbar pair through the nucleus is larger than  
the charmonium formation time \tf.  
At higher energy, \tcross is expected to be always lower than \tf~\cite{Ferreiro:2012mm} except  
maybe for backward rapidity ranges. 
This means that the \ccbar is nearly always in a pre-resonant state when traversing the nuclear matter  
and the nuclear break-up should be the same for the \psiP and \jpsi. 
 
\begin{figure}[!tbp] 
	\centering 
	\includegraphics[width=0.8\columnwidth]{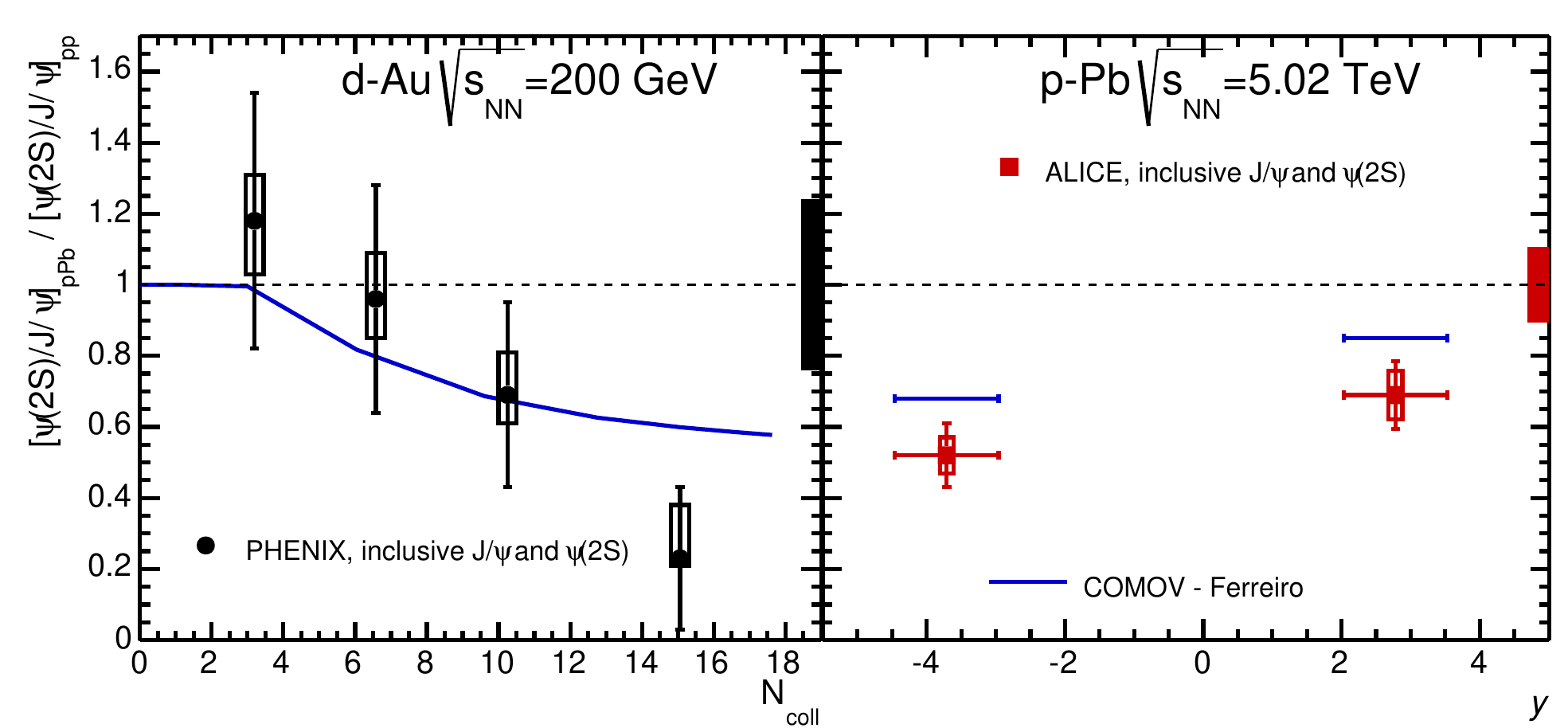} 
	\caption{Left: relative nuclear modification  
($\left[ \psiP/\jpsi \right]_{\rm dAu} / \left[ \psiP/\jpsi \right]_{\rm pp}$) of inclusive  
\jpsi to \psiP as a function of \Ncoll in PHENIX~\cite{Adare:2013ezl}. Right: transverse momentum  
dependence of the relative nuclear modification of \jpsi to $\psi(2S)$ in ALICE~\cite{Abelev:2014zpa}.  
The uncertainties are the same as described in \fig{fig:fig01}.} 
        \label{fig:fig04} 
\end{figure}

The PHENIX experiment has measured $\rdau=0.54\pm0.11\rm{(stat)}^{+0.19}_{-0.16}\rm{(syst)}$ for  
the \psiP and $\rdau=0.77\pm0.41\rm{(stat)}\pm0.18\rm{(syst)}$ for the \chic for $|y|<0.35$~\cite{Adare:2013ezl}.  
While the large uncertainty prevents any conclusion for the \chic, the relative modification factor of the  
\psiP to inclusive \jpsi in d-Au collisions, $\left[ \psiP/\jpsi \right]_{\rm dAu} / \left[ \psiP/\jpsi \right ]_{\rm pp}$ 
equivalent to $\rdau^{\psiP}/\rdau^{\jpsi}$, has been found to be $0.68\pm0.14\rm{(stat)}^{+0.21}_{-0.18}\rm{(syst)}$,  
\ie 1.3 $\sigma$ lower than 1.  
The relative modification factor as a function of \Ncoll is shown in the left panel of \fig{fig:fig04}.  
In the most central collisions, the \psiP is more suppressed than the \jpsi by about $2 \,\sigma$.  
 
ALICE has also measured  
in \pPb collisions at \snn =5.02 TeV the \psiP to \jpsi relative modification factor and has found  
$0.52\pm0.09\rm{(stat)}\pm0.08\rm{(syst)}$ for $-4.46<\ycm<-2.96$ and  
$0.69\pm0.09\rm{(stat)}\pm0.10\rm{(syst)}$ for $2.03<\ycm<3.53$~\cite{Abelev:2014zpa},  
respectively $4 \,\sigma$ and $2 \,\sigma$ lower than unity. In the right panel of \fig{fig:fig04},  
the relative modification factor is shown as a function of rapidity. This double ratio  
has also been measured as a function of \pt~\cite{Abelev:2014zpa} and does not exhibit a significant \pt dependence.  
In addition, preliminary results~\cite{Arnaldi:2014kta} show that the nuclear modification factor of the  
\psiP follows a similar trend as the \jpsi as a function of event activity at forward rapidity but is  
significantly more suppressed at backward rapidity towards central events.  
 
Models based on initial-state effects~\cite{Vogt:2010aa,Ferreiro:2012mm} or coherent energy  
loss~\cite{Peigne:2014uha} do not predict a relative suppression  
of the \psiP production with respect to the \jpsi one. These measurements could indicate that the \psiP production  
is sensitive to final-state  
effects in \pA collisions. A recent theoretical work uses EPS09 LO nPDF and includes  
the interactions of the quarkonium states with a comoving medium~\cite{Ferreiro:2014bia} (COMOV). The COMOV  
calculations are shown in \fig{fig:fig04}. They describe fairly well the PHENIX and  
ALICE results.  
Hot nuclear matter effects were also proposed as a possible explanation for the \psiP relative suppression in central \pPb collisions  
at the LHC~\cite{Liu:2013via}. 

\paragraph{Bottomonium measurements} 
 
\begin{figure}[!tbp] 
	\centering 
	\includegraphics[width=0.8\columnwidth]{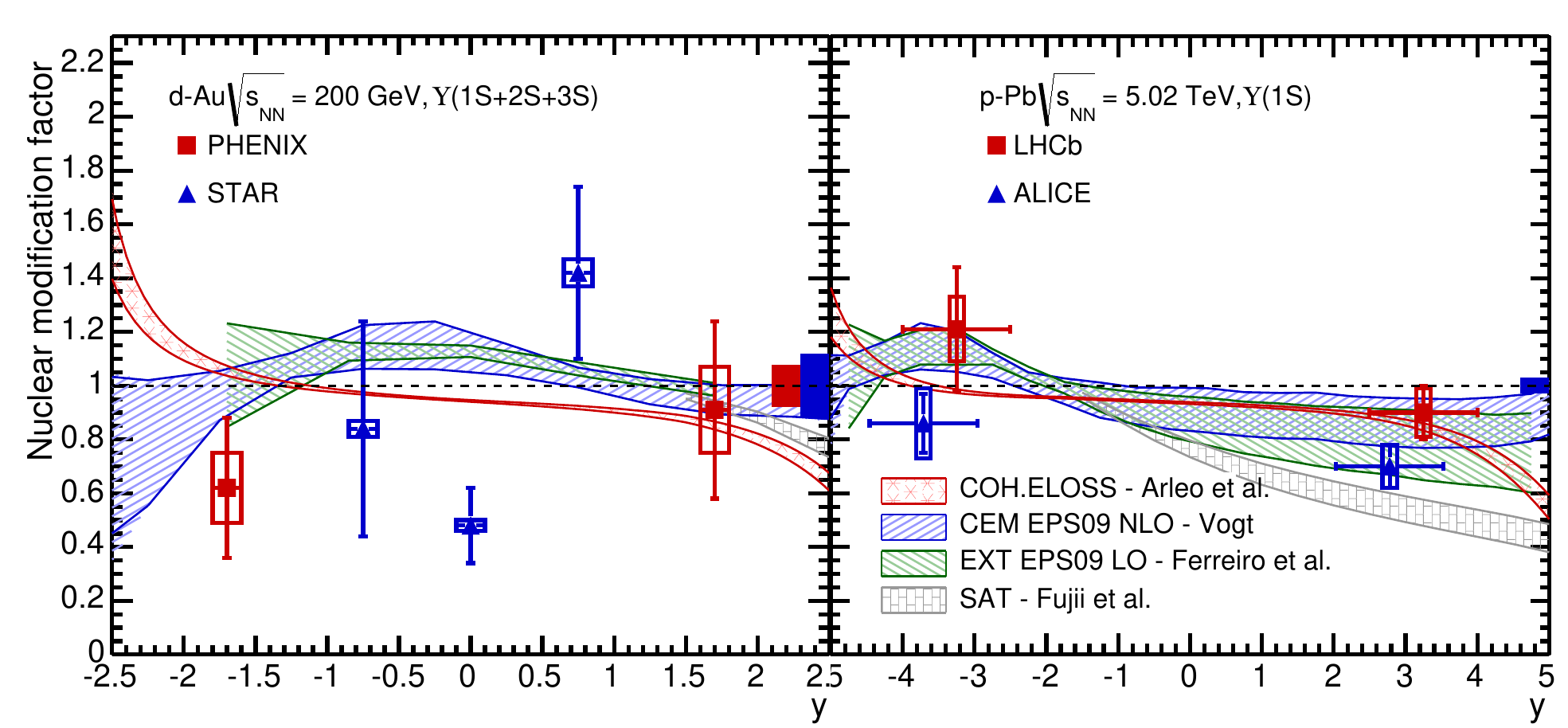} 
	\caption{Left: rapidity dependence of $\upsabc$ in PHENIX~\cite{Adare:2012bv} and STAR~\cite{Adamczyk:2013poh}. Right: rapidity dependence of \rppb for \upsa in ALICE~\cite{Abelev:2014oea} and LHCb~\cite{Aaij:2014mza}} 
        \label{fig:fig06} 
\end{figure}

The nuclear modification factor for bottomonium is shown in \fig{fig:fig06} at  
RHIC~\cite{Adare:2012bv,Adamczyk:2013poh} (left)  
and LHC~\cite{Abelev:2014oea,Aaij:2014mza} (right). At RHIC the 3 \ups states can not be measured separately  
due to the poor statistics and  
invariant mass resolution. At \snn~=~200\GeV, the \rdau is compatible  
with no or a slight suppression over the full rapidity range except at mid-rapidity where a  
suppression by a factor of two is found in \dAu with respect to (binary-scaled) \pp collisions.  
The data suggests a larger suppression by about $40\%$  
at backward rapidity but the uncertainties are large and \rdau is lower than unity by only $1.3 \,\sigma$.  
At \snn~=~5.02\TeV, the \upsa measurements  
from LHCb, despite slightly different rapidity ranges, are systematically higher than those of ALICE but  
the two measurements are consistent within uncertainties. The measured \rppb is consistent with unity  
at backward rapidity and below unity by, at most, $30\%$ at forward rapidity.  
 
The data are compared to models based on nPDFs (CEM EPS09 NLO, EXT EPS09 LO), coherent energy loss (COH.ELOSS) and gluon saturation (SAT). Given the  
limited statistics, the data can not constrain the models in most of the phase space and is in good agreement with the  
theory calculations. Only at mid-rapidity at  
\snn~=~200\GeV, the observed suppression is challenging for all the models, where no suppression is expected. In the  
nPDF based model, the rapidity range where \rpa is higher than unity corresponds to the anti-shadowing region. Clearly  
the data is not precise enough to conclude on the strength of gluon anti-shadowing. 
 
As in the \jpsi case, the ratio \rfb of the nuclear modification factors for a rapidity range  
symmetric with respect to $\ycm \sim 0$ has also been extracted for the \upsa  
at \snn=5.02\TeV~\cite{Abelev:2014oea,Aaij:2014mza}.  
 
Comparison of \upsa \rppb to the one from open beauty from \fig{fig:NonPromptJpsi_pPb_LHCb} can give a hint on  
final-state effects on \upsa. A similar level of suppression is observed for the \upsa and the \jpsi from B mesons.  
Larger statistics data however would be needed to rule out any final-state effect on \upsa production in \pPb. 
 
The study of excited bottomonium states in \pPb collisions  
may indicate the presence of final-state effects in bottomonium production.  
The \upsc has the smallest binding energy (0.2\GeV), followed by the \upsb (0.54 \GeV) and the \upsa (1.10\GeV) 
~\cite{Satz:2005hx}. Since the bottomonium formation time  is expected to be larger than the nuclear size, 
the suppression in \pPb is expected to be the same for all \ups states. 
 
\begin{figure}[!t] 
	\centering 
	\includegraphics[width=0.8\columnwidth]{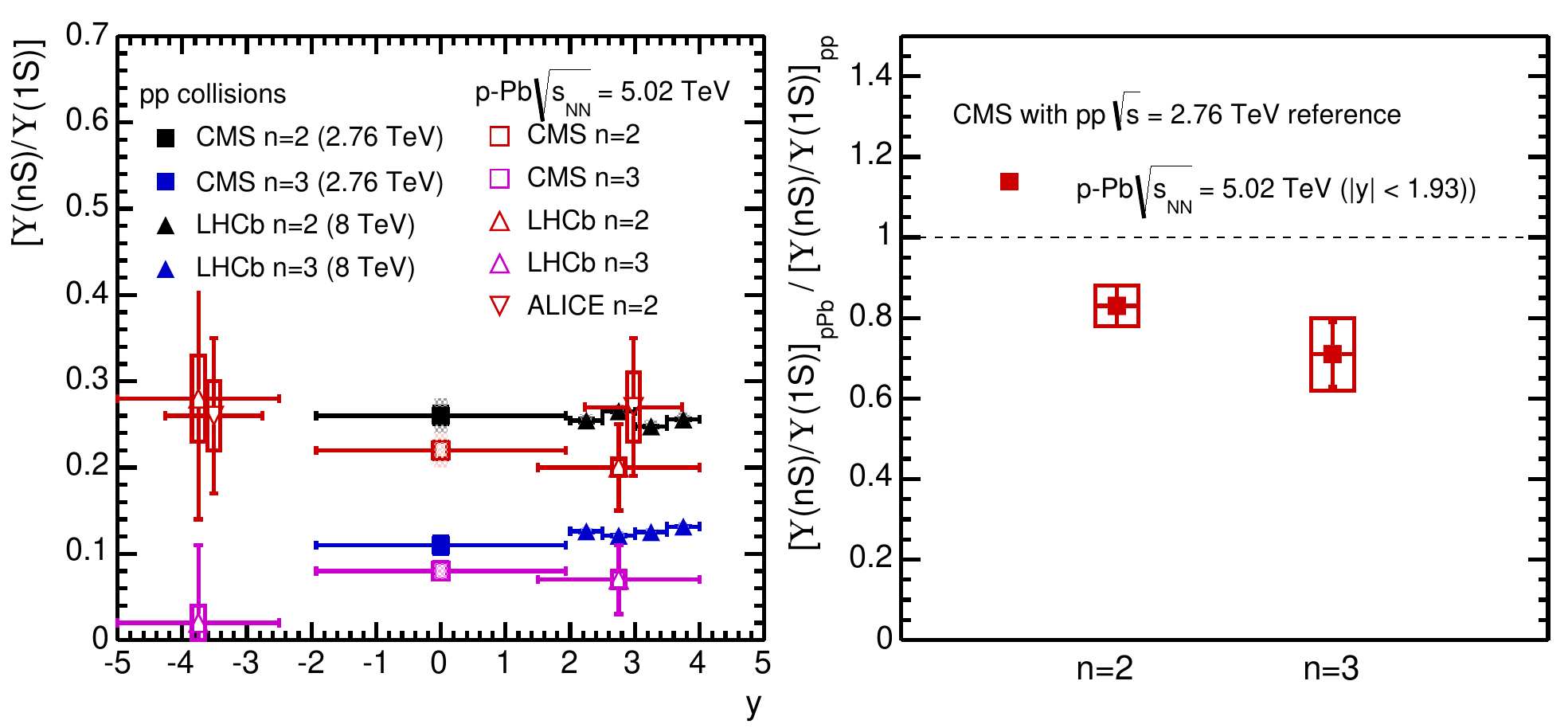} 
	\caption{Left: $\upsn/\upsa$ ratio in \pp and \pPb collisions in ALICE~\cite{Abelev:2014oea}, LHCb~\cite{Aaij:2014mza} and  
CMS~\cite{Chatrchyan:2013nza}. For a better visibility, the ALICE data points are displaced by +0.2 in rapidity. Right: relative modification factor  
$\left[ \upsn/\upsa \right]_{\rm pPb} / \left[ \upsn/\upsa \right ]_{\rm pp}$ in CMS~\cite{Chatrchyan:2013nza}.} 
        \label{fig:fig07} 
\end{figure}

\begin{figure}[!t] 
	\centering 
	\includegraphics[width=0.8\columnwidth]{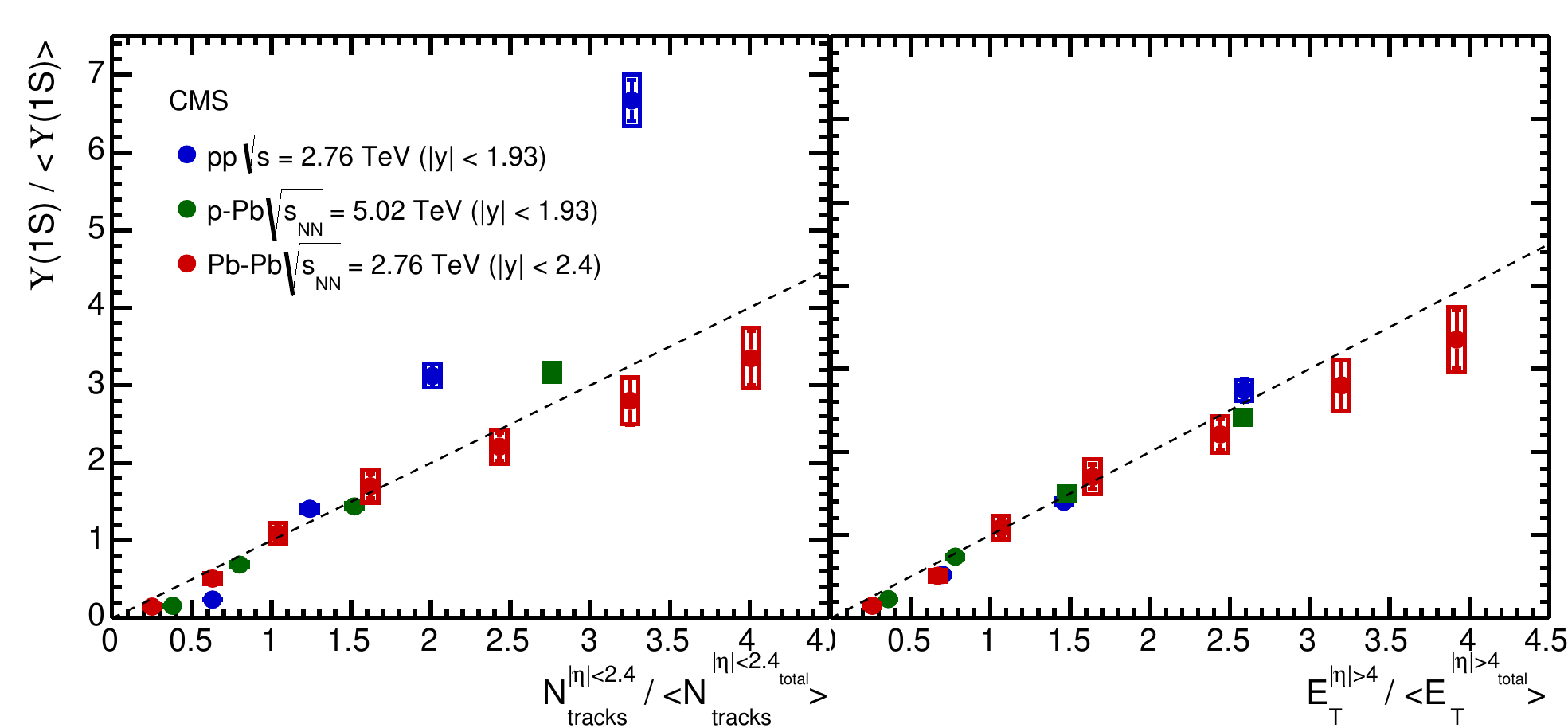} 
	\caption{Self-normalized cross section ratio $\frac{\upsa}{\left < \upsa \right >}$ vs $\frac{N_{\rm tracklets}}{\left < N_{\rm tracklets} \right >}$ (left) and $\frac{E_{\rm T}}{\left < E_{\rm T} \right >}$ (right) in \pp collisions at 2.76\TeV, \pPb collisions at 5.02\TeV and \PbPb collisions at 2.76\TeV in CMS~\cite{Chatrchyan:2013nza}. Here $N_{\rm tracklets}$ is the charged-track multiplicity measured in $|\eta|<2.4$ and $E_{\rm T}$ the transverse energy measured in $4<|\eta|<5.2$. The dotted line is a linear function with a slope equal to unity.} 
        \label{fig:fig08} 
\end{figure}

The CMS experiment has measured the ratio of the excited to the ground state cross section,  
 $\upsn/\upsa$, for $\rm n=2,3$ at mid-rapidity in \pPb collisions.  
ALICE (only for $\rm n=2$) and LHCb have performed similar measurements at backward and forward rapidity.  
The measured ratios $\upsn/\upsa$, shown in the left panel of \fig{fig:fig07} are compared to the ratios measured in  
\pp collisions at, however, different energies (\s~=~2.76 and 8\TeV) and in addition for the backward and  
forward rapidities, in slightly different rapidity ranges.  
It is worth noting that the ratio $\upsn/\upsa$ has been measured for $\rm n=2,3$ at  
\s~=~1.8, 2.76 and 7\TeV~\cite{Abe:1995an,Chatrchyan:2013nza,Chatrchyan:2013yna} at mid-rapidity and  
at \s~=~2.76, 7 and 8\TeV~\cite{Aaij:2014nwa,LHCb:2012aa,Aaij:2013yaa} at forward rapidity.  
The ratio is found to be, within the quoted uncertainties,  
independent of \s, and in the rapidity range $2<\ycm<4$, independent of $\ycm$.   
A stronger suppression than in \pp is observed at mid-rapidity in \pPb collisions for \upsb and \upsc as  
compared to \upsa, which could suggest the presence of final-state effects that affect more the excited states  
as compared to the ground state.  
At forward rapidity, the ratios measured by ALICE and LHCb are similar in \pPb and \pp but the measurements are not precise enough to be sensitive to a difference as observed by CMS.  
 
To better quantify the modification between \pp and \pPb and cancel out some of the systematic uncertainties from  
the detector set-up, the double ratio  
$\left[ \upsn/\upsa \right]_{\rm pPb} / \left[ \upsn/\upsa \right ]_{\rm pp}$ has also been evaluated by CMS  
at mid-rapidity using \pp collisions at 2.76\TeV\cite{Chatrchyan:2013nza} and is displayed in the right panel of  
\fig{fig:fig07}. The double ratio in \pPb is lower than one by $2.4\,\sigma$ for \upsb and \upsc. The double  
ratios signal the presence of different or stronger final-state effect acting on the excited states compared to the  
ground state from \pp to \pPb collisions.  
 
As for the charmonium production, the excited states are not expected to be differentially suppressed by any of the models that  
include initial-state effects nor from the coherent energy loss effect. A possible explanation may come from  
a suppression associated to the comoving medium. Precise measurements in a larger rapidity range, which covers  
different comoving medium density, would help to confirm this hypothesis.

CMS has also performed measurements as a function of the event activity at forward 
($4<|\eta|<5.2$ for the transverse energy $E_{\rm T}$) and mid-rapidity ($|\eta|<2.4$ for the charged-track multiplicity  
$N_{\rm tracklets}$)~\cite{Chatrchyan:2013nza}.  
\fig{fig:fig08} shows the \ups self-normalized cross section ratios \upsa/$\left<\upsa\right>$  
where $\left<\upsa\right>$ is the event-activity integrated value for \pp, \pPb and \pb collisions.  
The self-normalized cross section ratios are found to rise with the  
event activity as measured by these two estimators and similar results are obtained  
for \upsb and \upsc.  
When Pb ions are involved, the increase can be related to the increase in the number of nucleon--nucleon collisions.  
A possible interpretation of the positive correlation between the \ups production yield and the underlying activity  
of the \pp event is related to Multiple-Parton Interactions (MPI) occurring in a single \pp collisions.  
Linear fits performed  
separately for the three collision systems show that the self-normalized ratios have a slope consistent with unity  
in the case of forward event activity. Hence, no significant difference between \pp, \pPb and \pb is observed  
when correlating \ups production yields with forward event activity.  
On the contrary in the case of mid-rapidity event activity, different slopes are found for the three collisions systems.  
These observations are also related to the single cross section ratios $\upsn/\upsa$ as shown in  
\fig{fig:CMS_ups_mult} and discussed in detail in \sect{sec:pp:HadCorrelations}.

\subsection{Extrapolation of CNM effects from {\rm p--A} to {\rm A--A} collisions} 
\label{CNM_pAtoAA}

 
 
It is an important question to know whether cold nuclear matter effects can be simply extrapolated from \pA\ to \AAcoll\ collisions. 
Some of  the CNM effects discussed in \sect{CNM_theory}  can in principle be extrapolated to \AAcoll\ collisions.  
This is the case of  nPDF and coherent energy loss effects, discussed below. Some other approaches, on the contrary, are affected by interference effects between the two nuclei involved in the collision, making delicate an extrapolation to \AAcoll\ collisions. 
 
\paragraph{Nuclear PDF} 
 
Regarding the nPDF effects discussed in \sect{sec:npdf}, 
it is straightforward to make this comparison at leading order in the colour evaporation model (CEM) where the $\pt$ of the $\QQbar$ pair is zero and the $x_1$ and $x_2$ values are related to the quarkonium rapidity by  \eq{2-1_x-Bjorken}.   
As long as the production cross section obeys the factorisation hypothesis, \eq{cross-section-factorization}, the nuclear modification factors (taken at the same energy) also factorize, \ie the following relation  is exact, 
\begin{equation}\label{CNMfactorization} 
\raa^{\rm CNM}(y) = \rpa(+y) \cdot \rpa(-y)\ . 
\end{equation} 
 
\begin{figure}[!htp] 
        \centering 
        \includegraphics[width=0.45\textwidth]{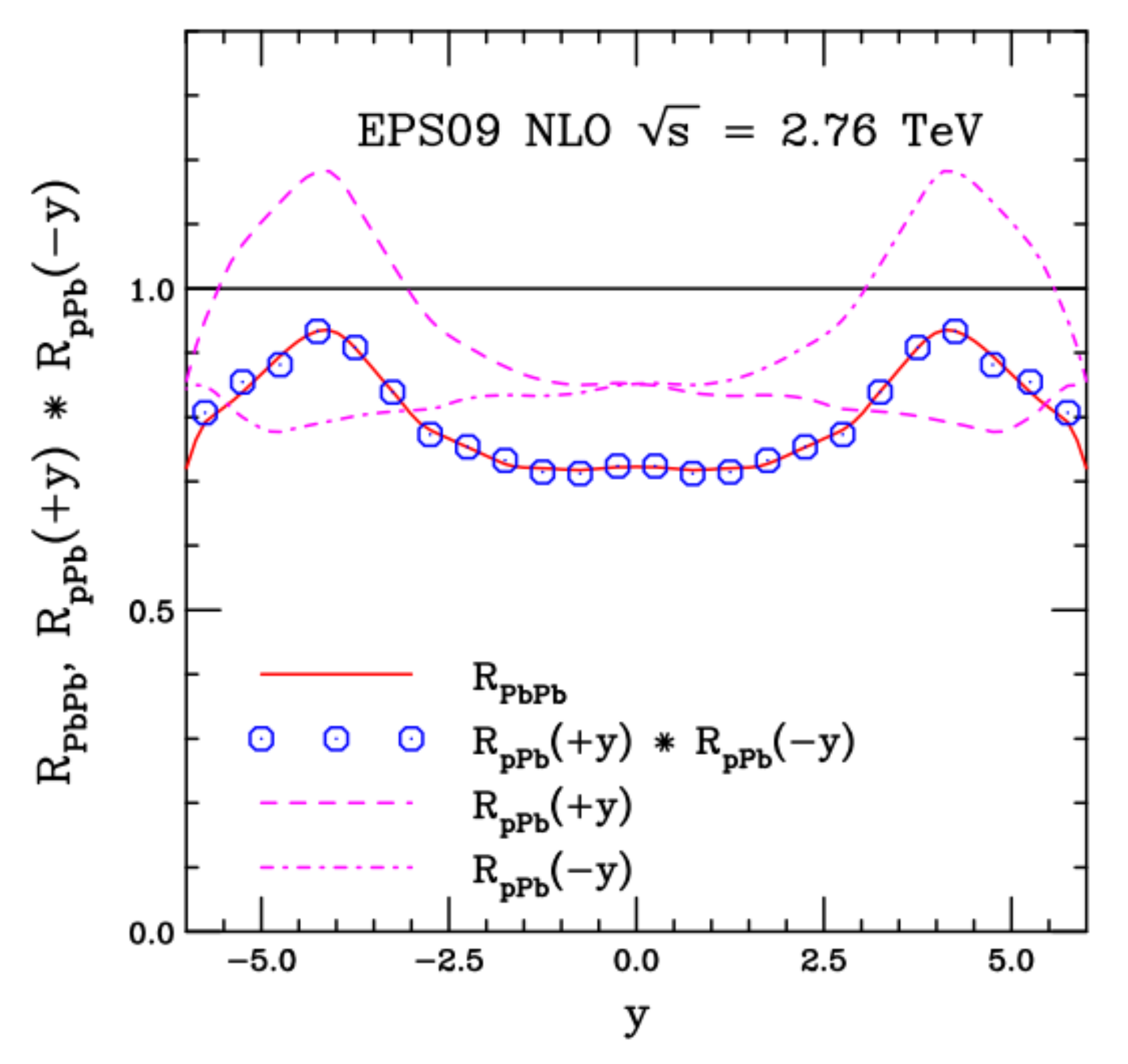}  
        \includegraphics[width=0.45\textwidth]{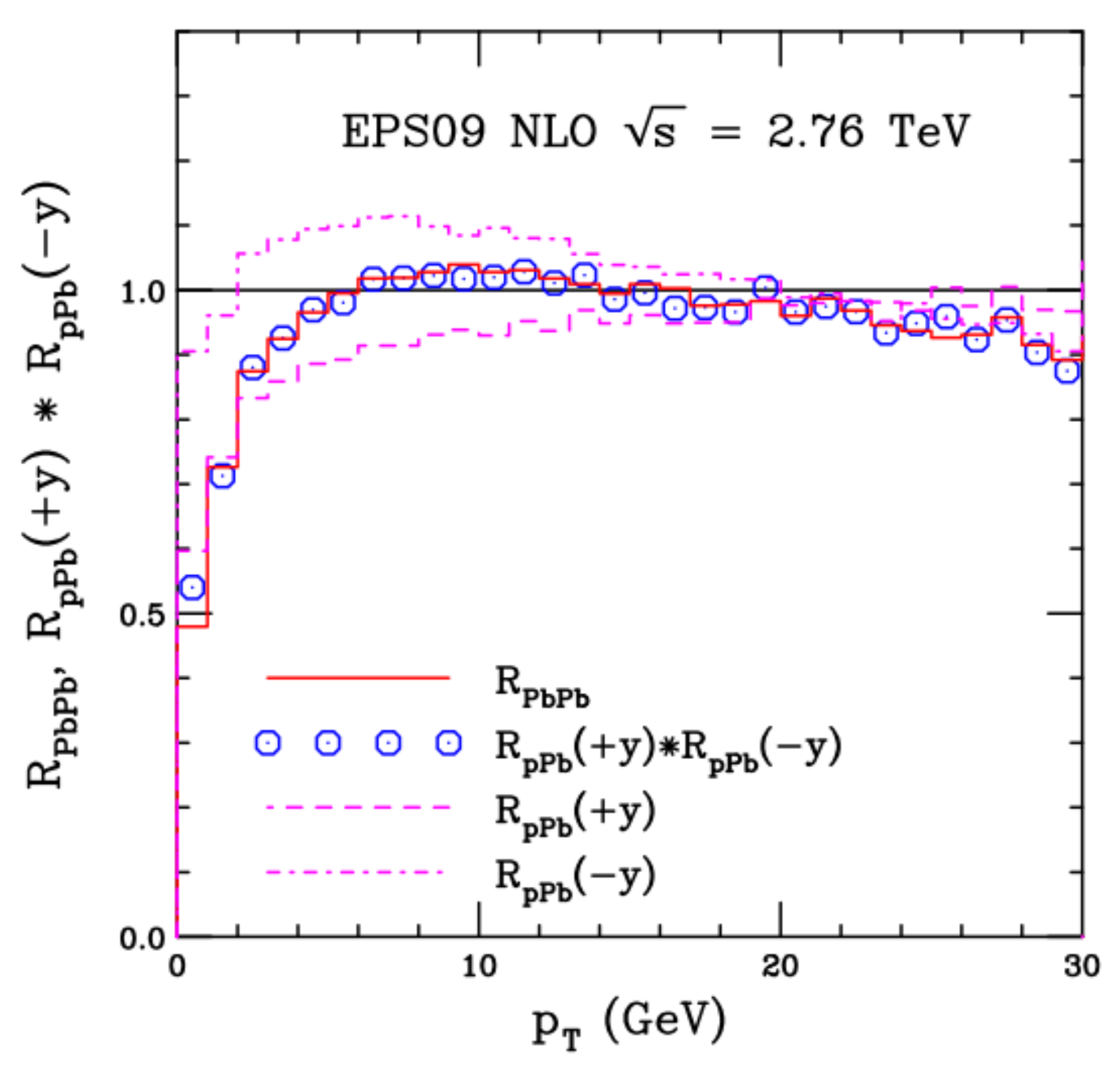} 
        \caption{The $\jpsi$ $\raa$ (red) ratio is compared to the product $\rpa(+y) \cdot \rpa(-y)$ (points) along with the individual \pA\ ratios at forward (dashed) and backward (dot-dashed) rapidity.   
Results are compared for the $y$  (left) and $\pt$  (right) dependencies at NLO, from Ref.~\cite{Vogt:2015uba}.} 
        \label{factpsi} 
\end{figure}

At next-to-leading order in the CEM, however, the assumption of factorization of nPDF effects is less simple to understand because of the large contribution from $2 \rightarrow 2$ diagrams. For such processes, the correlation between the initial momentum fractions $x_1, x_2$ and the rapidity of the  quarkonium state is less straightforward.   
However, the factorisation hypothesis (see \eq{CNMfactorization}) is seen to still hold at NLO, as shown by a calculation using EPS09 NLO central nPDF set at $\snn = 2.76$ TeV in \fig{factpsi} as a function of $y$ (left) and $\pt$ (right)~\cite{Vogt:2015uba}.   
 
 
In principle, this factorization hypothesis can also be applied to open heavy flavour. 
 
\paragraph{Multiple scattering and energy loss} 
 
Let us first discuss how predictions can be extrapolated in the coherent energy loss model. In a generic A--B collision both incoming partons, respectively from the `projectile' nucleus A and the `target' nucleus B, might suffer multiple scattering in the nucleus B and A, respectively. Consequently, gluon radiation off both partons can interfere with that of the final state particle (here, the compact colour octet $\QQbar$ pair), making a priori difficult the calculation of the medium-induced gluon spectrum in the collision of two heavy ions. 
 
\begin{figure}[t] 
\centering 
\includegraphics[width=11cm]{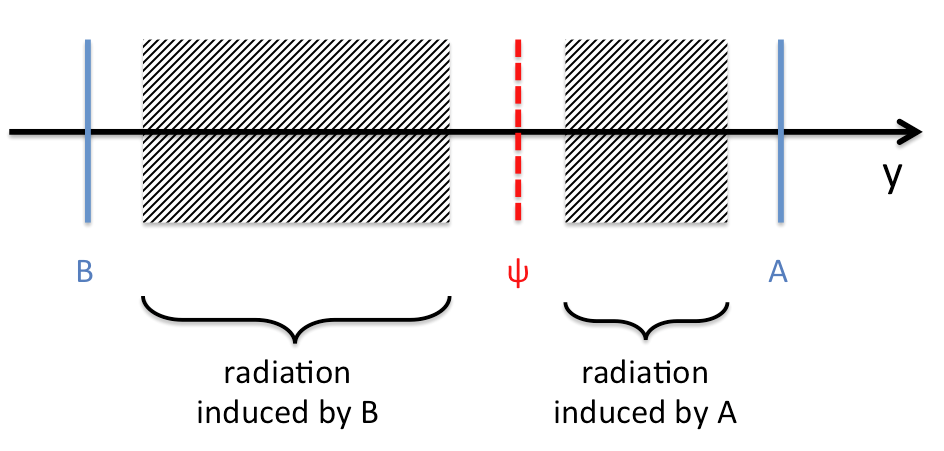} 
\caption{Sketch of the rapidity regions populated by medium-induced radiation in an A--B collision in the coherent energy loss model. The `target' B and `projectile' A move with respectively negative and positive rapidities.} 
\label{fig-rap-regions} 
\end{figure}

However, it was shown in~\cite{Arleo:2014oha} that the gluon radiation induced by rescattering in nuclei A and B occurs in distinct regions of phase space (see \fig{fig-rap-regions}). As a consequence, the energy loss induced by the presence of each nucleus can be combined in a probabilistic manner, making a rather straightforward extrapolation of the model predictions from \pA\ to \AAcoll\ collisions. Remarkably, it is possible to show that the quarkonium suppression in \AAcoll\ collisions follows the factorisation hypothesis (see \eq{CNMfactorization}). 
However, since the energy loss effects do not scale with the momentum fraction $x_2$, the data-driven extrapolation of \pPb\ data at $\snn=5$~TeV to \PbPb\ data at $\snn=2.76$~TeV, discussed below, and which assumes nPDF effects only is not expected to hold~\cite{Arleo:2014oha}.

The model by Sharma and Vitev can also easily be generalized to \AAcoll\ reactions where both incoming and outgoing partons undergo elastic, inelastic and coherent soft interactions in the large nuclei.  
In contrast, the Kopeliovich, Potashnikova and Schmidt approach for charmonium production cannot be simply extrapolated from  \pA\ to \AAcoll\ collisions~\cite{Kopeliovich:2010nw}, because nucleus-nucleus collisions include new effects of double colour filtering and a boosted saturation scale, as explained in detail in~\cite{Kopeliovich:2010nw}.  
 
\paragraph{Data-driven extrapolation} 
 
At RHIC, the \dAu collisions are performed with symmetrical beam energies, so that $y_{\rm CM/lab} = 0$, and at the same nucleon-nucleon centre-of-mass energy than for heavy-ion collisions.  
The direct comparison of \dAu data to heavy-ion data is then easier.  
In this context, the PHENIX experiment has evaluated the \jpsi breakup (\ie absorption) cross section by fitting \rdau as a function of the rapidity, and also as a function of the average number of binary collisions (\Ncoll), and by assuming different shadowing scenarios (EKS and NDSG)~\cite{Adare:2007gn}.  
The two shadowing scenarios with their resulting breakup cross section were applied to \jpsi \raa, both for \CuCu and \AuAu collisions. 
Moreover, an alternative data-driven method~\cite{GranierdeCassagnac:2007aj} was applied to PHENIX data~\cite{Adare:2007gn}.  
This method assumes that all cold nuclear matter effects are parametrised with a modification factor consisting of a function of the radial position in the nucleus. 
 Note that the use of d--Au data in~\cite{GranierdeCassagnac:2007aj} may not be appropriated for peripheral collisions where the size of the deuteron causes significant averaging over impact parameter; on the contrary it should be adequated in central collisions for which the averaging is not so important. An attempt to solve this issue has been proposed in~\cite{Brambilla:2010cs} where an estimate of $R_{\mathrm{pAu}}$ was derived from $\rdau$ using a Glauber 
model including EKS98 nuclear PDF. 

A more recent investigation of RHIC data by Ferreiro et al.~\cite{Ferreiro:2009ur} showed how the use of $2 \to 2$ partonic process instead of the usual $2 \to 1$ can imply a different value of the absorption cross section~\cite{Rakotozafindrabe:2010su}, since the anti-shadowing peak is systematically shifted towards larger rapidities in \dAu.  
The other noticeable consequence is that \rdau versus $y$ is not symmetric anymore around $y \approx 0$.  
This implies that the CNM effects in \raa at RHIC will also show a rapidity dependence, with less suppression from CNM effects at mid-rapidity than at forward rapidity, in the same direction as the one exhibited by the \AuAu and \CuCu data from PHENIX (see extensive comparisons in~\cite{Ferreiro:2009ur}).  
This is quite important since this shape of \raa at RHIC was also considered as a possible hint for hot in-medium recombination effects, while it might come from CNM effects only. 
 
At LHC, the \pPb\ results can not be easily compared to \pb collisions. 
Indeed, the nucleon-nucleon centre-of-mass energies are not the same (5.02 versus 2.76 TeV) and moreover the p and the Pb beam energies per nucleon are different, leading to a rapidity shift of the centre-of-mass frame with respect to the lab frame. 
But assuming factorisation and \eq{CNMfactorization}, a data-driven extrapolation of \pA\ data to \AAcoll\ can be performed. 
 
In a given detector acceptance (at fixed $y_{\rm lab}$), the ratio of $x_2$ values probed in a given process in \pb and \pPb collisions is   
\begin{eqnarray}\label{CNMextrapolation} 
\frac{x_2^{\rm PbPb}}{x_2^{\rm pPb}} = \frac{\sqrt{s_{\rm NN}^{\rm pPb}}}{\sqrt{s_{\rm NN}^{\rm PbPb}}} \exp(-y_{\rm CM/lab}^{\rm pPb}) \, . 
\end{eqnarray} 
At the LHC, the rapidity shift is $y_{\rm CM/lab}^{\rm pPb} = 0.465$. In \RunOne conditions, one has $\sqrt{s_{\rm NN}^{\rm PbPb}} = 2.76$~TeV and $\sqrt{s_{\rm NN}^{\rm pPb}} = 5.02$~TeV and the ratio is $\frac{x_2^{\rm PbPb}}{x_2^{\rm pPb}} = 8\ {\rm TeV} / 7\ {\rm TeV}\simeq 1.14$.  
The typical momentum fraction ranges involved in \pPb collisions are shown in \fig{Fig-LHCx_range}. 
 
This data-driven extrapolation of \pA\ collisions to \AAcoll\ collisions applied by the ALICE collaboration to $\jpsi$ production lead to~\cite{Abelev:2013yxa}: $[\rppb(2<y<3.5) \cdot \rppb(-4.5<y<-3)]^{\jpsi} = 0.75 \pm 0.10 \pm 0.12$, the first uncertainty being the quadratic combination of statistical and uncorrelated systematic uncertainties and the second one the linear combination of correlated uncertainties. 
The application of this result to the interpretation of \PbPb data is discussed in \sect{sec:other_ref}. 
 
 
 
In summary, according to the theoretical and data driven extrapolation approaches, one can conclude that there are non negligible CNM effects on \AAcoll\ results at the LHC (up to 50\% at low \pt). 
A \pt dependence of \jpsi \rppb factorization will be presented in \sect{sec:npdf_aa}.

%

\subsection{Summary and outlook}
\label{CNM_status}

 
 
The LHC \pPb \RunOne has opened a new window on the study of the CNM effects. 
The broad kinematical range probed by the different LHC experiments and the comparison to RHIC \dAu results  
bring new constraints on theoretical models. 
The main observations resulting from the open and hidden heavy-flavour data can be summarised in the following way: 
\begin{itemize} 
\item 
The nuclear modification factor of open heavy-flavour decay leptons 
in d--Au collisions at RHIC shows a dependence on centrality and on rapidity, with 
values smaller than unity at forward rapidity and larger than unity at mid- and backward rapidity in the most central collisions.  
\item  
In p--Pb collisions at the LHC, the D meson nuclear modification factor at mid-rapidity 
and $1<\pt<16$\GeVc is consistent with unity within uncertainties of about 20\%. 
\item The \rpa of \jpsi from B mesons at the LHC shows a modest suppression at forward rapidity and is consistent with unity at backward rapidity.  
\item 
A rapidity dependence of \jpsi suppression has been measured at RHIC and LHC.  
At both energies the suppression is more pronounced at forward than at mid-rapidity. At backward rapidity, $\jpsi$ production is slightly suppressed at RHIC but is compatible with no suppression at the LHC.  
\item  
There is no evidence of $\jpsi$ suppression at large $\pt$ in the full rapidity range at RHIC and LHC.  
\item  
At RHIC, open heavy flavour from lepton decay and \jpsi suppression for $\pt> 1 \GeVc$ are of the same order at forward rapidity but not at backward and mid-rapidity: suppression mechanisms from final-state effects may be at play on \jpsi production at backward and mid-rapidity. 
\item 
\upsa \rpa measurements are compatible with unity except at mid-rapidity at RHIC and forward rapidity at the LHC. Similar level of suppression is observed for the \upsa and the \jpsi from B-mesons at the LHC. However the \upsa \rpa measurements have large statistical uncertainties.  
\item 
Excited states are more suppressed than 1S states at RHIC and LHC suggesting the presence of final-state CNM effects. 
\end{itemize} 
For the theoretical interpretation of the data, the following conclusions can be drawn: 
\begin{itemize} 
\item 
Open heavy-flavour current data do not allow one to favour specific models based on nuclear PDF, parton saturation,  
or initial-parton energy loss. 
\item Regarding \jpsi production, one can conclude the following: 
\begin{itemize} 
\item  
The nuclear PDFs describe well the \rpa despite large theoretical uncertainties at forward rapidity where the data would require strong shadowing effects. At backward rapidity, while the nPDF models describe correctly the LHC data, they do not describe the RHIC data without considering additional effects such as nuclear absorption.  
\item The early CGC prediction of $\jpsi$ \rpa by Fuji and Watanabe is ruled out by the present LHC data at forward rapidity. The calculations do not describe either the \pt dependence of the RHIC data at forward rapidity. Refinements of the model have now been proposed, leading to lesser disagreement with data. 
\item The predictions of the coherent energy loss model describes well the rapidity dependence of \jpsi \rpa both at RHIC and at LHC. Regarding the $\pt$ dependence, the shape of the data is also rather well captured, although the dependence is slightly more abrupt in the model than in the data, especially at forward rapidity. The predicted $\jpsi$ suppression expected in the dipole propagation model by Kopeliovich, Potashnikova 
and Schmidt seems much larger than seen in data, suggesting the need for additional effects to compensate the suppression. Finally, the approach based on energy loss and power corrections by Sharma and Vitev  predicts a moderate and flat  $\jpsi$ and $\Upsilon$ suppression as a function of \pt, above $\pt=4$~GeV/$c$, somehow in contradiction with data. 
\end{itemize} 
\item 
The suppression of excited states relative to 1S state is described so far only by considering the effect from a comoving medium.   
\end{itemize} 
 
The main limitations for the interpretation of the current experimental results are, on the one hand, 
the sizeable experimental uncertainties, on the other hand, the large uncertainties on the  
nuclear modification of the PDFs in the low-$x$ region. 
 
Regarding the experimental uncertainties, in the case of rare probes like  
B mesons, \psiP and \ups, but also high \pt yields, the experimental data suffer from limited statistics.  
For more abundant probes, like heavy-flavour decay leptons.  D mesons,  
B from \jpsi and prompt \jpsi, the size of the systematic uncertainties is the main limitation. 
 
To address part of these issues, a reference \pp period at $\s=5.02$~\TeV and a higher-statistics \pPb period  
at $\s=5.02$~\TeV,  
which will allow a better control of the systematics, during the LHC \RunTwo would be very helpful to improve  
the precision of the current measurements.  
However, for the probes which are using the full LHC luminosity and  
have already a \pp reference at 8~\TeV from the 2012 \RunOne data taking period,  
it would be more interesting to get a new \pPb run at $\snn=8$~\TeV in order  
to study the CNM effects at higher energy.  
These aspects have to be balanced in order to choose the energy for the \pPb run in \RunTwo. 
 
New observables could help to disentangle the various CNM effects. 
First studies of the heavy-flavour azimuthal correlations at RHIC and LHC were carried out and (at RHIC) suggest a modification of charm production kinematics in d--Au.  
A comparison of open to hidden heavy flavour production from  
$\pt =0$ would allow to separate initial- from final-state effects on quarkonia. 
Another open question is related to quarkonium polarisation: can the CNM effects modify  the polarisation of quarkonia? 
In addition, the RHIC capability to collide a polarised-proton beam with nuclei can be used to explore new observables.  
 
Finally, it is not clear whether the CNM effects can be extrapolated from \pA to \AAcoll collisions. 
At present, 
 only  
phenomenological works based on nPDF and coherent energy loss effects have shown that this extrapolation is possible, although with some caveats.  
On the one hand, 
in the nPDF-based models, the main parameter is the probed momentum fraction $x$.  
Since there is a rapidity shift of the centre-of-mass in \pPb collisions at LHC, the optimal strategy would be  to choose the LHC beam energy  
according to \eq{CNMextrapolation}.  
On the other hand, in the coherent energy loss model, the relevant parameter is the \snn value 
and \pA collisions can be directly related to \AAcoll collisions only if taken at the same energy.

%


\newpage


\section{Open heavy flavour in nucleus--nucleus collisions}
\label{OHF}

Heavy-flavour hadrons 
are effective probes of the conditions of the high-energy-density QGP medium formed  
in ultra-relativistic nucleus--nucleus collisions.  
 
Heavy quarks, 
are produced in primary hard QCD scattering processes  
in the early stage of the nucleus--nucleus collision 
and the time-scale of their production (or coherence time)  
is, generally, shorter than the formation time of the QGP, 
$\tau_0\sim 0.1$--1~fm$/c$. 
More in detail, the coherence time of the heavy quark-antiquark pair 
is of the order of the inverse of the virtuality $Q$ of the hard scattering,  
$\Delta\tau\sim 1/Q$.  
The minimum virtuality $2\,m_{c,b}$  
in the production of a \ccbar or \bbbar pair implies a space-time scale  
of $\sim 1/3\GeV^{-1}\sim 0.07~{\rm fm}$ and $\sim 1/10\GeV^{-1}\sim 0.02~{\rm fm}$ for  
charm and for beauty, respectively. 
One exception to this picture are configurations where the quark and antiquark are 
produced with a small relative opening angle in the so-called gluon splitting processes $g\to \qqbar$.  
In this case, the coherence time is increased by a boost factor $E_g/(2\,m_{c,b})\sim E_{c,b}/m_{c,b}$ and becomes  
$\Delta t \sim E_{c,b}/(2\,m_{c,b}^2)$.   
This results, for example, in a coherence time of about 1~fm/$c$ (0.1~fm/$c$) for charm (beauty) quarks with energy of 15\GeV, and of about 1~fm/$c$ 
for beauty quark jets with energy of about 150\GeV. 
The fraction of heavy quarks produced in gluon splitting processes has been estimated using  
perturbative calculations and Monte Carlo event generators, resulting in moderate  
values of the order of 10--20\% for charm~\cite{Mueller:1985zz,Mangano:1992qq} and large values of the order of 50\% for beauty~\cite{Banfi:2007gu}. 
Given that the gluon splitting fraction is moderate for charm and the coherence time is small for beauty from gluon splitting 
when $\pt$ is smaller than about 50\GeVc, 
it is reasonable to conclude that heavy-flavour hadrons in this \pt range probe the heavy quark in-medium interactions. 
 
During their propagation through the medium, heavy quarks interact with its constituents 
 and lose a part of their energy, thus being sensitive to the medium properties. 
 Various approaches have been developed to describe the interaction of the heavy quarks with the surrounding plasma. 
In a perturbative treatment, QCD energy loss is expected to occur 
via both inelastic (radiative energy loss, via medium-induced gluon radiation)~\cite{Gyulassy:1990ye,Baier:1996sk} and elastic (collisional energy loss)~\cite{Thoma:1990fm,Braaten:1991jj,Braaten:1991we} 
processes. However, this distinction is no longer meaningful in strongly-coupled approaches relying for instance on the AdS/CFT conjecture~\cite{Herzog:2006gh,Gubser:2006bz}.  
In QCD, quarks have a smaller colour coupling factor with respect to gluons,  
so that the energy loss for quarks is expected to be smaller than for gluons.  
In addition, the ``dead-cone effect'' should reduce small-angle gluon radiation  
for heavy quarks with moderate  
energy-over-mass values, thus further attenuating  
the effect of the medium. This idea was first introduced in~\cite{Dokshitzer:2001zm}.  
Further theoretical studies confirmed the reduction of the total induced gluon radiation~\cite{Armesto:2003jh,Djordjevic:2003zk,Zhang:2003wk,Wicks:2007am}, 
although they did not support the expectation of a ``dead cone''. 
Other mechanisms such as in-medium hadron formation and dissociation~\cite{Adil:2006ra,Sharma:2009hn}, would determine a 
stronger suppression effect on heavy-flavour hadrons than light-flavour hadrons, because of their smaller formation  
times. 
 
In contrast to light quarks and gluons, which can be produced or annihilated  
during the entire evolution of the medium, heavy quarks are produced  
in initial hard scattering processes  
and their annihilation rate  
is small~\cite{BraunMunzinger:2007tn}. 
Secondary ``thermal" charm production from processes like $gg\to \ccbar$ in the QGP  
is expected to be negligible, unless initial QGP temperatures much larger than  
that accessible at RHIC and LHC are assumed~\cite{Zhang:2007dm}. 
Therefore, heavy quarks preserve their flavour and mass identity while traversing the medium and can be tagged throughout all momentum ranges,  
from low to high \pt, through the measurement of heavy-flavour hadrons in the final state of the collision.  
 

The nuclear modification factor 
\begin{equation} 
\label{eq:Raa} 
\raa(\pt)= 
\frac{1}{\av{\taa}} \cdot  
\frac{\dd N_{\rm AA}/\dd\pt}{\dd\sigma_{\rm pp}/\dd\pt} 
\end{equation} 
---a detailed definition is given in \sect{Cold nuclear matter effects}--- 
is well-established as a sensitive observable  
for the study of the interaction of hard partons  
with the medium.  
At large \pt,  \raa is 
expected to be mostly sensitive to the average energy loss of 
heavy quarks in the hot medium.   
The study of more differential observables can provide important insights 
into the relevance of the various interaction mechanisms and the properties of the medium. In particular, the dependence of  
the partonic energy loss on the in-medium path length is expected to be different for each mechanism (linear for collisional processes~\cite{Thoma:1990fm,Braaten:1991jj,Braaten:1991we} and close 
to quadratic for radiative processes in a plasma~\cite{Baier:1996sk}). 
Moreover, it is still unclear if low-momentum heavy quarks can reach thermal equilibrium with the medium constituents and participate in the collective expansion of the system~\cite{Batsouli:2002qf,Greco:2003vf}. 
It was also suggested that low-momentum heavy quarks could hadronise 
not only via fragmentation in the vacuum, but also via the mechanism of recombination with other quarks from the medium~\cite{Greco:2003vf,Andronic:2003zv}. 
 
These questions can be addressed both with the study of the \raa at low and intermediate \pt (smaller than about five times 
the heavy-quark mass) and with 
azimuthal anisotropy measurements of heavy-flavour hadron production with respect to the reaction plane, 
defined by the beam axis and the impact parameter of the collision. 
For non-central collisions, the two nuclei overlap in an approximately lenticular region, the short axis of which lies in the reaction plane. 
Hard partons are produced at an early stage, when the geometrical anisotropy is not yet reduced by the system expansion. Therefore, partons emitted 
in the direction of the reaction plane (in-plane) have, on average, a shorter in-medium path length than partons emitted orthogonally (out-of-plane), 
leading {\it a priori} to a stronger high-\pt suppression in the latter case. 
In the low-momentum region, the in-medium interactions can also modify the parton emission directions, thus translating the initial spatial anisotropy into a momentum anisotropy of the final-state particles.  
Both effects cause a momentum anisotropy that can be characterised with the coefficients  
$v_n$ and the symmetry planes $\Psi_n$ of the Fourier expansion of the \pt-dependent particle distribution $\dd^2N/\dd\pt\dd\phi$  
in azimuthal angle $\phi$. 
The elliptic flow is the second Fourier coefficient \vtwo. 

The final ambitious goal of the heavy-flavour experimental programmes in nucleus--nucleus collisions is the characterisation of the properties of the produced QCD matter, in particular getting access to the transport coefficients of the QGP.  Theoretical calculations 
encoding the interaction of the heavy quarks with the plasma into a few transport coefficients (see \eg~\cite{Rapp:2009my}) provide the tools to achieve this goal: through a comparison of the experimental data with the numerical outcomes obtained with different choices of the transport coefficients it should be possible, in principle, to put tight constraints on the values of the latter.  
This would be the analogous of the way of extracting information on the QGP viscosity through the comparison of soft-particle spectra with predictions from fluid dynamic models. 
An even more intriguing challenge would be to derive the heavy-flavour transport coefficients through a first principle QCD calculation and confront them with experimental data, via model implementations that describe the medium evolution. 
This chapter reviews the present status of this quest, from the experimental and theoretical viewpoints. 
 
The chapter is organised as follows. The first part of the chapter presents a brief overview of the available data of heavy-flavour production  
in nucleus--nucleus collisions at the RHIC and LHC colliders: in particular, \sect{sec:OHFRAA} describes the measurements of the nuclear modification factor  
\raa, while \sect{sec:OHFv2} focuses on the azimuthal anisotropy. The published RHIC and LHC data are summarised in  
\tabs~\ref{tab:ohf_expSummary_RHIC} and~\ref{tab:ohf_expSummary_LHC}, respectively. 
The second part of the chapter includes a review of the theoretical models for heavy-quark interactions and energy loss in the medium,  
with a detailed description of the model ingredients in terms of the quark--medium interaction (\sect{sec:OHFmodelsInt}) 
and of the medium modelling (\sect{sec:OHFmodelsMed}). A comparative overview of the models  
and comprehensive comparison with data from RHIC and LHC are presented in \sect{sec:modelcomparison}. 
Finally, the theoretical and experimental prospects for the study of heavy-flavour correlations are discussed in \sect{sec:OHFcorr}.

\begin{table} 
 \caption{Open heavy flavour published measurements in \AuAu and \CuCu collisions at RHIC. The nucleon--nucleon energy in the centre-of-mass system (\snn), the covered kinematic ranges and the observables are indicated.} 
 \label{tab:ohf_expSummary_RHIC} 
 \centering 
\begin{tabular}{*{5}{c}lc} 
 \hline 
Probe & Colliding & \snn &  $y_{\rm cms}$ (or $\eta_{\rm cms}$) & \pt & Observables & Ref. \\ 
 & system & (\TeV) &  & (\GeVc) & \\ 
 \hline 
 \multicolumn{7}{c}{PHENIX} \\ 
 \hline 
 \hfe & \AuAu & 62.4 & $|y|<0.35$ & 1 -- 5 & yields~(\pt,centrality) & \cite{Adare:2014rly} \\ 
 & & & & & $\rcp$(\pt) & \\ 
 & & & & & $\raa$(\pt,centrality) & \\ 
 & & & & & $\raa$(\Ncoll,\pt) & \\ 
 & & & & 1.3 -- 3.5 & $\vtwo$(\pt,centrality) & \\ 
 & & & & 1.3 -- 2.5 & $\vtwo$(\snn,centrality) & \\  
 \cline{3-7} 
  & & 130 & $|y|<0.35$ & 0.4 -- 3 & yields~(\pt,centrality) & \cite{Adcox:2002cg} \\ 
  \cline{3-7} 
  & & 200 & $|\eta|<0.35$ & 0.3 -- 9 & yields~(\pt,centrality) & \cite{Adare:2010de,Adare:2006nq,Adler:2005xv,Adler:2004ta} \\ 
 & & & & & $\raa$(\pt,centrality) & \\ 
 & & & & & $\raa$(\Npart,\pt) & \\ 
 & & & & $>0.4$ & $\frac{\dd \sigma_{NN}}{\dd y}$(\Ncoll) & \\ 
 & & & & $>0$ & $\frac{\dd \sigma_{NN}}{\dd y}$(centrality) & \\ 
 & & & & & $\sigma_{NN}^{\ccbar}$(centrality) & \\ 
 & & & & 0.3 -- 5 & $\vtwo$(\pt,centrality) & \\ 
 \cline{3-7} 
  & & 200 & $|y|<0.35$ & 2 -- 4 & $\frac{1}{N_{\mathrm{trig}}^{e_{HF}}}\frac{\dd N_{\mathrm{assoc.}}^{\mathrm{h}}}{\dd \pt}(\pt^{\mathrm{h}},\Delta\phi)$ & \cite{Adare:2010ud} \\ 
 & & & & & $\iaa^{e_{HF}-h}(\pt^{\mathrm{h}},\Delta\phi)$ & \\ 
 & & & & 2 -- 3 & $R_{\mathrm{HS}}(\pt^{\mathrm{h}})$ & \\ 
 \cline{2-7} 
  & \CuCu & 200 & $|y|<0.35$ & 0.5 -- 7 & yields~(\pt,centrality) & \cite{Adare:2013yxp} \\ 
 & & & & & $\raa$(\pt,centrality) & \\ 
 & & & & & $\raa$(\Ncoll,\pt) & \\ 
 & & & & & $\raa$(\Npart,\pt) & \\ 
 & & & & 0.5 -- 6 & $\rcp$(\pt) & \\ 
 \hline 
 \hfm & \CuCu & 200 & $1.4<|y|<1.9$ & 1 -- 4 & yields~(\pt,centrality) & \cite{Adare:2012px} \\ 
 & & & & & $\raa$(\pt,centrality) & \\ 
 & & & & & $\raa$(\Npart) & \\ 
 \hline 
 \multicolumn{7}{c}{STAR} \\ 
 \hline 
 \Dzero& \AuAu & 200 & $|y|<1$ & 0 -- 6 & yields~(\pt,centrality) & \cite{Adamczyk:2014uip} \\ 
 & & & & & $\raa$(\pt,centrality) & \\ 
 & & & & 0 -- 8 & $\raa$($\langle \Npart \rangle$,\pt) & \\ 
 \hline 
 \hfe & \AuAu & 200 & $0<\eta<0.7$ & 1.2 -- 8.4 & yields~(\pt,centrality) & \cite{Abelev:2006db}\\ 
 & & & & & $\raa$(\pt,centrality) & \\ 
 \cline{3-7} 
 & & 39, 62.4, 200 & $|\eta|<0.7$ & 0 -- 7 & $\vtwo$(\pt) & \cite{Adamczyk:2014yew}\\  
 \hline 
 \end{tabular} 
\end{table} 
 
\begin{table} 
 \caption{Open heavy flavour published measurements in \pb collisions at LHC. The nucleon--nucleon energy in the centre-of-mass system (\snn), the covered kinematic ranges and the observables are indicated.} 
 \label{tab:ohf_expSummary_LHC} 
\centering 
\begin{tabular}{*{5}{c}lc} 
 \hline 
 Probe & Colliding & \snn &  $y_{\rm cms}$ (or $\eta_{\rm cms}$) & \pt & Observables & Ref. \\ 
 & system & (\TeV) &  & (\GeVc) & \\ 
 \hline 
 \multicolumn{7}{c}{ALICE} \\ 
 \hline 
\Dzero, \Dplus, \Dstarplus & \pb & 2.76 & $|y|<0.5$ & 2 -- 16 & yields\,(\pt) & \cite{ALICE:2012ab}\\ 
 & & & & & $\raa$(\pt) & \\ 
 & & & & 2 -- 12 & $\raa({\rm centrality})$ & \\  
 & & & & 6 -- 12 & $\raa({\rm centrality})$ & \\ 
 \cline{4-7} 
 & & & $|y|<0.8$ & 2 -- 16 &  $\vtwo$(\pt) & \cite{Abelev:2014ipa,Abelev:2013lca}\\ 
 & & & & & $\vtwo$(centrality,\pt) & \\ 
 & & & & & $\raa^{\text{in/out plane}}$(\pt) & \\ 
 \hline 
 \hfm & \pb & 2.76 & $2.5<y<4$ & 4 -- 10 & $\raa$(\pt) & \cite{Abelev:2012qh} \\ 
 & & & & 6 -- 10 &  $\raa({\rm centrality})$ & \\  
 \hline 
 non-prompt \jpsi & \pb & 2.76 & $|y|<0.8$ & 1.5 -- 10 & \raa(\pt) & \cite{Adam:2015rba} \\ 
 \hline 
 \multicolumn{7}{c}{CMS} \\ 
 \hline 
 $b$-jets & \pb & 2.76 & $|\eta|<2$ & 80 -- 250 & yields~(\pt) & \cite{Chatrchyan:2013exa} \\ 
 & & & & & $\raa$(\pt) & \\ 
 & & & & 80 -- 110 & $\raa({\rm centrality})$ & \\  
 \hline 
 non-prompt \jpsi & \pb & 2.76 & $|y|<2.4$ & 6.5 -- 30 & yields~({\rm centrality}) & \cite{Chatrchyan:2012np} \\ 
 & & & & & $\raa$({\rm centrality}) & \\  
 \hline 
 \end{tabular} 
\end{table} 

\subsection{Experimental overview: production and nuclear modification factor measurements}
\label{sec:OHFRAA} 

\subsubsection{Inclusive measurements with leptons}\label{sec:HFelectrons}
 
Heavy-flavour production can be measured inclusively via the semi-leptonic decay channels. 
The key points of the measurement are the lepton identification and background subtraction. 
 
In the STAR experiment, electrons are identified using the specific energy loss (\dEdx) measurement from the Time Projection Chamber (TPC) together with the Time of Flight information at $\pt < 1.5$\GeVc, and energy and shower shape measurements in the Electro-Magnetic Calorimeter (EMCal) at $\pt > 1.5$\GeVc.  
The background contribution to the electron yield from photonic sources (mainly from photon conversion in the detector material and $\pi^0$ and $\eta$ Dalitz decays) are subtracted statistically using the invariant mass method~\cite{Agakishiev:2011mr, Aggarwal:2010xp}. 
The electron identification in the PHENIX experiment is based on the Ring Imaging Cherenkov detector in conjunction with a highly granular EMCal.  
The subtraction of the electron background is performed by the converter and cocktail methods~\cite{Adare:2006hc, Adler:2005xv}. 
%
%
In the ALICE experiment, electrons are identified in the central pseudo-rapidity region ($|\eta|<0.9$) using four detector systems: the Time Projection Chamber, the Time Of Flight, the EMCal and the Transition Radiation Detector.  
Background electrons are subtracted using both the invariant mass and cocktail methods~\cite{Abelev:2012xe}. 
 
In PHENIX, muons are measured with two muon spectrometers that provide pion rejection at the level of  
$2.5 \times 10^{-4}$ in the pseudo-rapidity range $-2.2 < \eta < -1.2$ and $1.2 < \eta < 2.4$ over the full azimuth. 
%
 
Muons are detected in ALICE 
with the forward muon spectrometer in the pseudo-rapidity range $-4 < \eta < -2.5$.  The extraction of the heavy-flavour contribution to the single muon spectra requires the subtraction of  
muons from the decay in flight of pions and kaons, estimated through the extrapolations of the measurements at mid-rapidity. 
 
In ATLAS, muons are reconstructed in the pseudo-rapidity range $|\eta|<1.05$ by matching the tracks in the Inner silicon Detector (ID) with the ones in the Muon Spectrometer (MS), surrounding the electromagnetic and hadronic calorimeters. 
The background muons arise from pion and kaon decays, muons produced in showers in the calorimeters and mis-association of MS and ID tracks. 
The signal component is extracted through a MC template fit of a discriminant variable that depends on the difference between the ID and MS measurements of the muon momentum, after accounting for energy loss in the calorimeters, and the deflections in the trajectory resulting from decay in flight~\cite{ATLAS:2012ama}. 
 
\begin{figure}[t] 
  \centering 
  \includegraphics[width=0.51\textwidth]{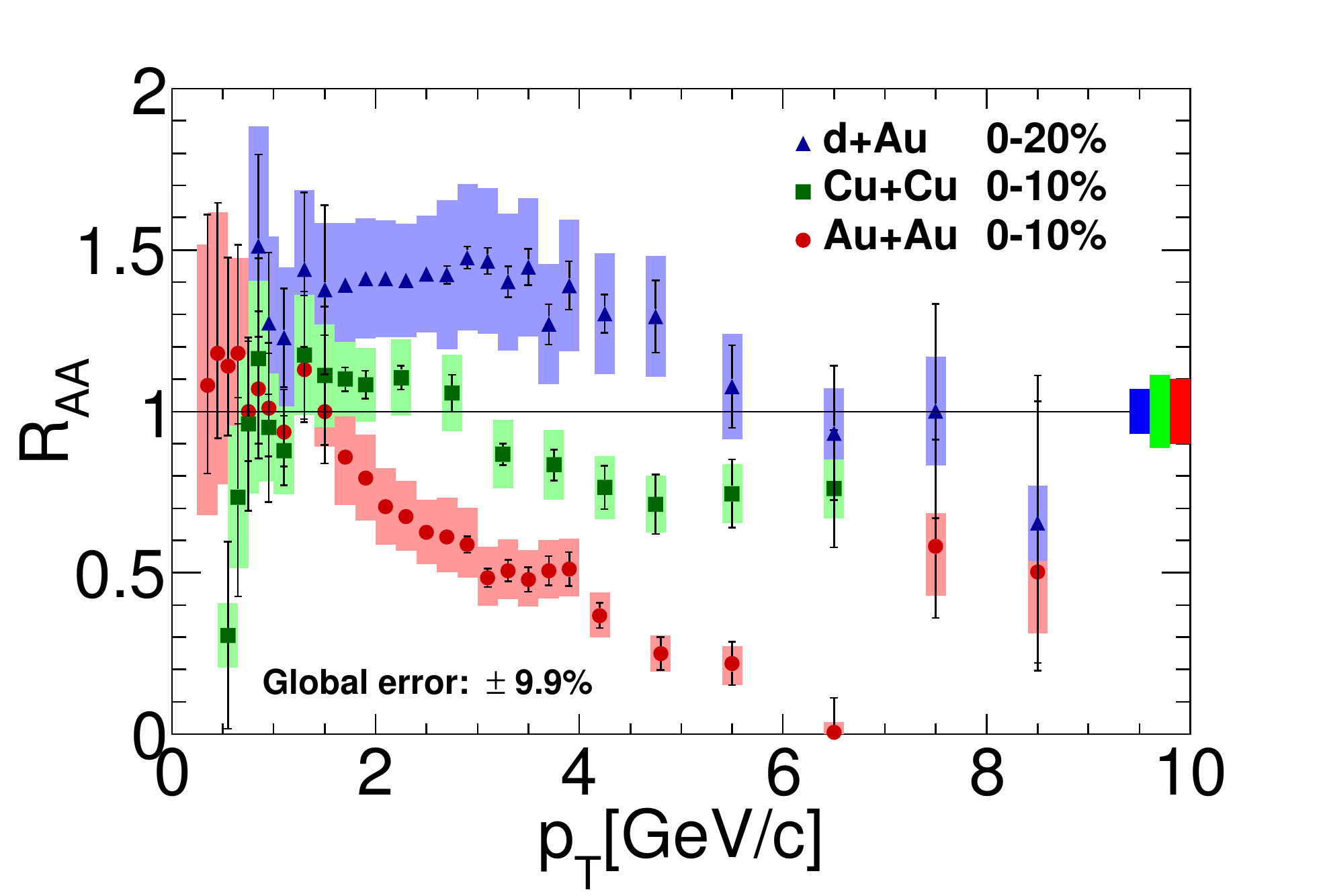}   
  \includegraphics[width=0.42\textwidth]{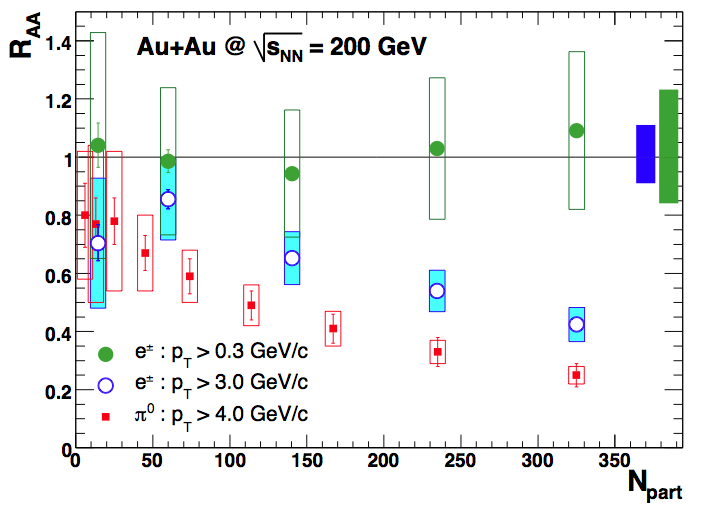} 
  \caption{Left: The transverse momentum dependence of the nuclear modification factors of heavy-flavour decay electrons at mid-rapidity in central \dAu~\cite{Adare:2012yxa}, \CuCu~\cite{Adare:2013yxp} and \AuAu collisions~\cite{Adare:2010de} at $\snn = 200$\GeV.  
%
  Right: \raa of heavy-flavour decay electrons at mid-rapidity with \pt above 0.3 and 3\GeVc and of $\pi^0$ with $\pt > 4$\GeVc as function of number of participants $\Npart$~\cite{Adare:2006nq}.  
  } 
\label{Fig:11} 
\end{figure} 
 
 
%
\begin{figure} 
  \centering 
  \includegraphics[width=0.46\textwidth]{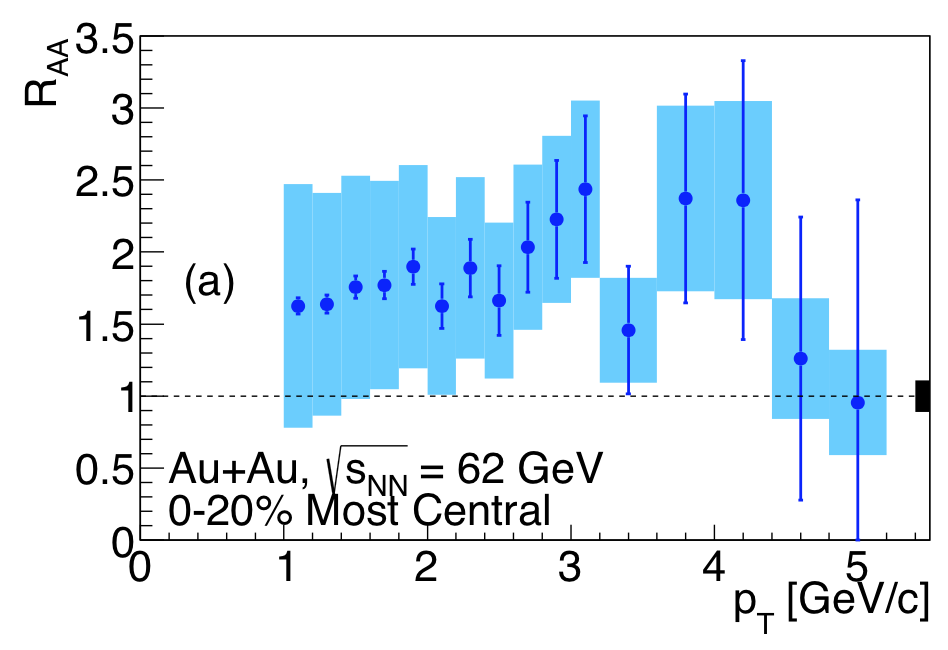} 
  \includegraphics[width=0.47\textwidth]{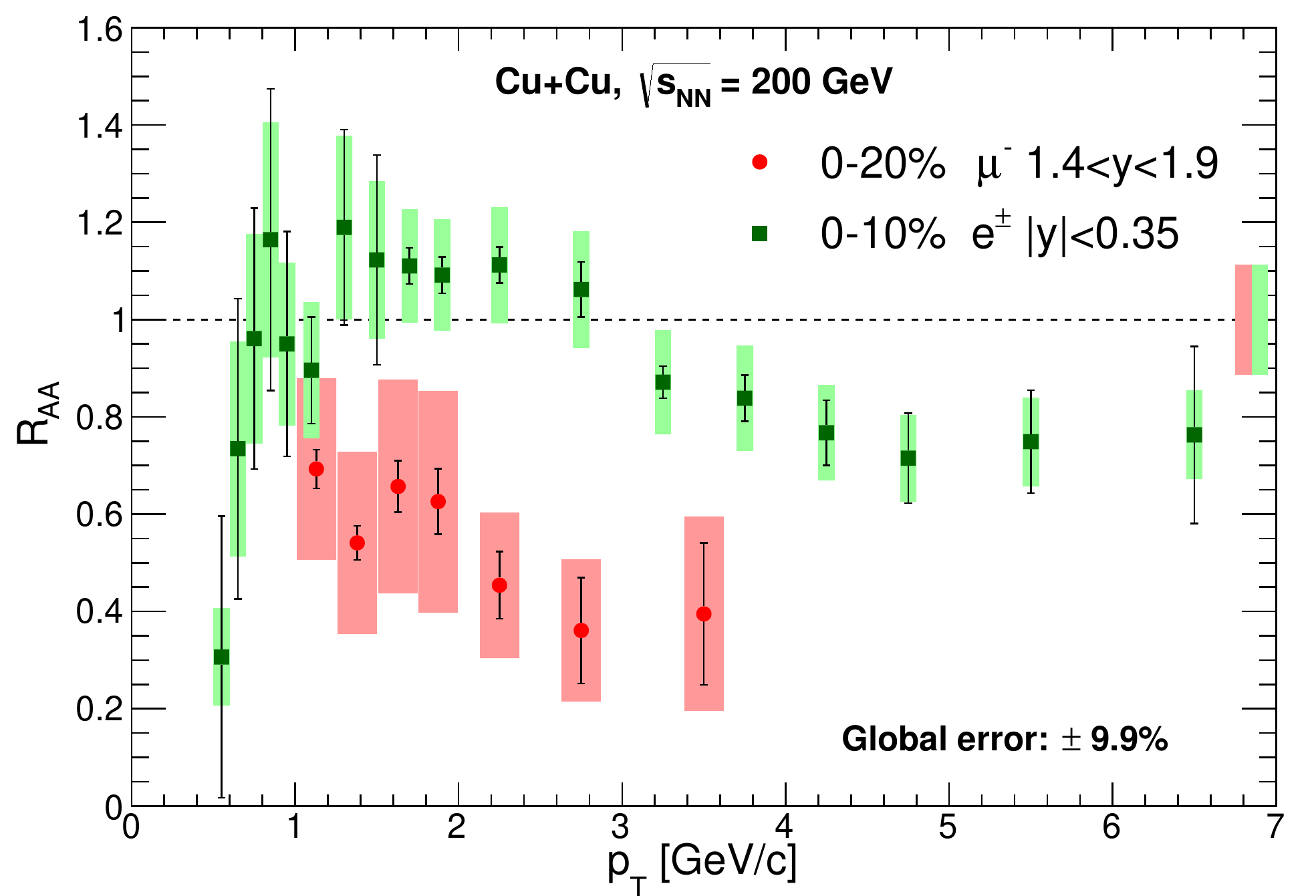} 
  \caption{Left: \raa of heavy-flavour decay electrons at mid-rapidity measured in \AuAu collisions at $\snn = 62.4$\GeV as a function of \pt in the 20\% most central collisions~\cite{Adare:2014rly}. Right: \raa of heavy-flavour decay electrons at mid-rapidity~\cite{Adare:2013yxp} and muons at forward rapidity~\cite{Adare:2012px} for the most central \CuCu collisions at $\snn = 200$\GeV.}   
\label{Fig:12} 
\end{figure} 
 
\begin{figure}[!ht] 
  \centering 
  \includegraphics[width=0.46\textwidth]{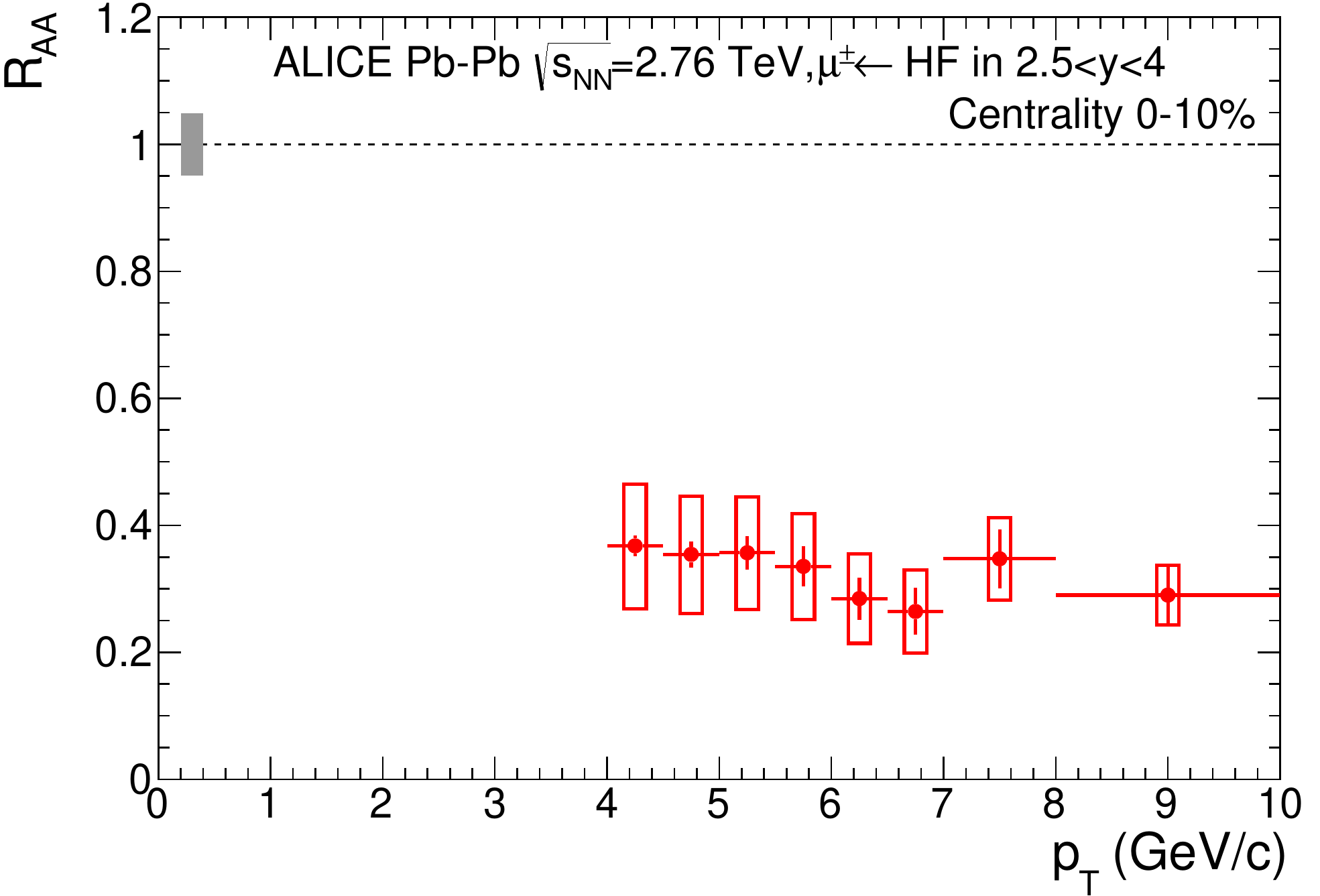} 
  \includegraphics[width=0.46\textwidth]{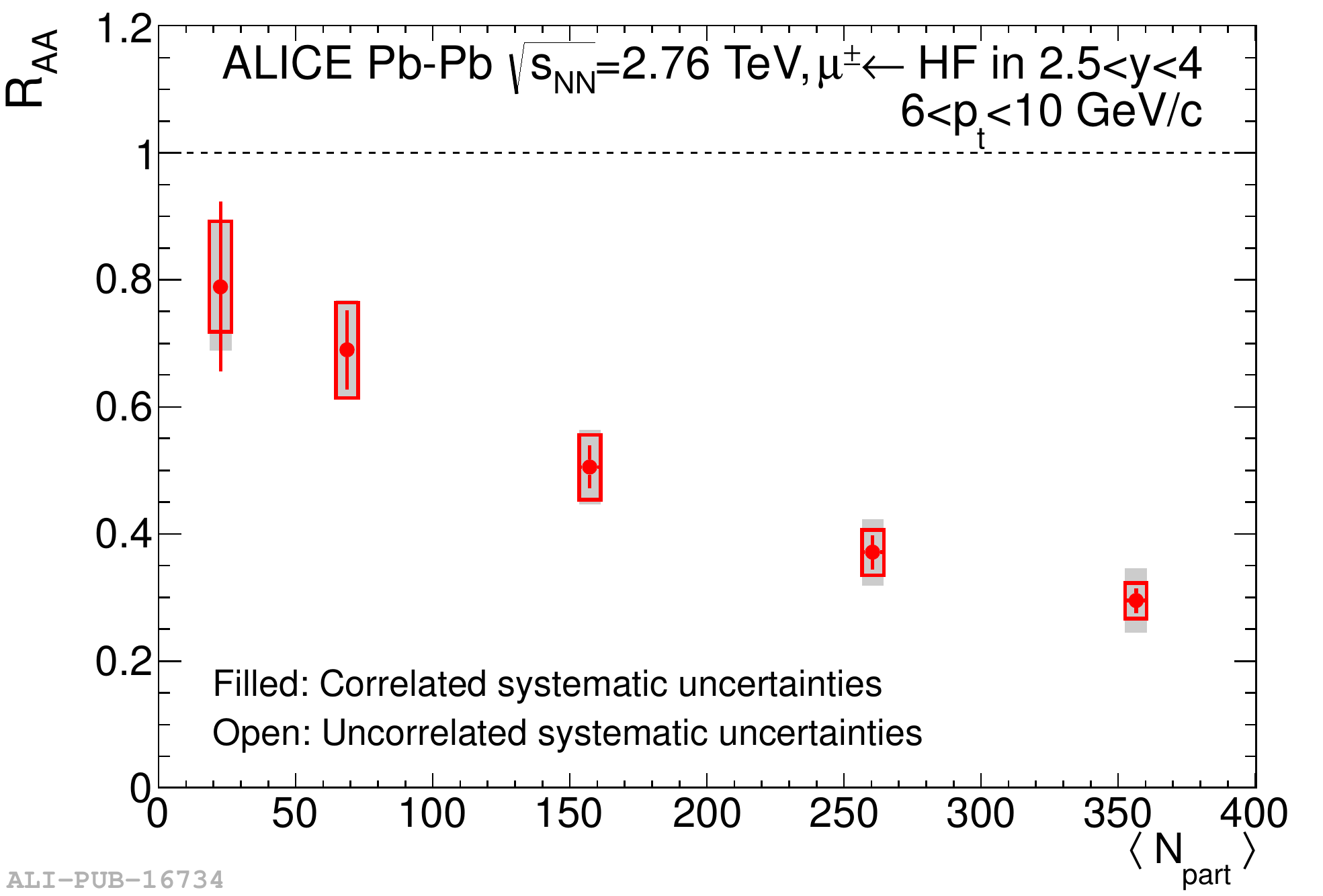} 
  \caption{Nuclear modification factor \raa of heavy-flavour decay muons with $2.5<y<4$ measured in \pb collisions at $\snn$ = 2.76\TeV as a function of \pt in the 10\% most central collisions (left panel) and as a function of the mean number of participating nucleons \av{\Npart} (right panel)~\cite{Abelev:2012qh}. 
  }   
 
\label{Fig:13} 
\end{figure} 
 
%
The STAR~\cite{Abelev:2006db} and PHENIX~\cite{Adare:2010de, Adare:2006nq, Adler:2005xv} Collaborations measured the yield of electrons from heavy-flavour decays at various centre-of-mass energies and in various colliding systems. 
The \pt dependence of the nuclear modification factor measured in the 10\% most central \AuAu collisions at $\snn=200$\GeV is shown in \fig{Fig:11} (left panel). 
The suppression increases with the transverse momentum, reaching a factor of about four for $\pt > 4$\GeVc. 
This strong effect is not observed in \AuAu collisions at $\snn = 62.4$\GeV~\cite{Mustafa:2012jh,Adare:2014rly} for which, however, the \pp reference was not measured at RHIC but taken from ISR data (see left panel of \fig{Fig:12}). 
The left panel of \fig{Fig:11} also shows the comparison with the \raa measured in \dAu and \CuCu collisions at $\snn=200$\GeV: a clear dependence on the colliding system is found. 
In particular, the observation that the nuclear modification factor is consistent or larger than unity in \dAu collisions demonstrates that the high-\pt suppression in nucleus--nucleus collisions 
is induced by the presence of the hot and dense medium. 
The \raa of heavy-flavour decay electrons at mid-rapidity as a function of the collision centrality (represented by the number of participants \Npart)~\cite{Adare:2006nq} is shown in the right panel of \fig{Fig:11}.  
The high-\pt production shows a clear centrality-dependent suppression, reaching a maximum of a factor four in central collisions ($\raa \approx 0.25$). 
At variance, the production of electrons with $\pt>0.3$\GeVc (which measures the charm production yield essentially down to \pt = 0) is consistent with scaling with the number of binary collisions, within experimental uncertainties of about 20\%. 
PHENIX also measured the \raa of heavy-flavour decay muons at forward rapidity~\cite{Adare:2012px} for the most central \CuCu collisions at $\snn =200$\GeV: the observed suppression is stronger than for heavy-flavour decay electrons at mid-rapidity (see \fig{Fig:12}, right).

At the LHC, heavy-flavour production was measured in the leptonic decay channels in \pb collisions at $\snn = 2.76$\TeV. 
\fig{Fig:13} shows the nuclear modification factors of muons from heavy-flavour decays in $2.5<y<4$ measured by ALICE as a function of \pt in the 10\% most central collisions (left panel) and as a function of centrality in $6 < \pt < 10$\GeVc (right panel)~\cite{Abelev:2012qh}. 
The observed suppression increases from peripheral to central collisions, up to a factor of three in central collisions. 
The result is consistent with a preliminary measurement of the \raa of heavy-flavour decay electrons at mid-rapidity (with $4<\pt<18$\GeVc)~\cite{Sakai:2013ata}. 
Moreover, it is also in qualitative agreement with a preliminary measurement of the heavy-flavour decay muon central-to-peripheral nuclear modification factor \rcp at mid-rapidity (with $4<\pt<14$\GeVc), carried out by the ATLAS Collaboration~\cite{Perepelitsa:2012cza}, which shows a suppression of a factor about two, independent of \pt, for the centrality ratio 0--10\%/60--80\%. 
The comparison of the results at forward and mid-rapidity indicate a weak dependence on this variable in the rapidity region $|y|<4$. 

\subsubsection{{\rm D} meson measurements}\label{sec:Dmeson}
 
 
\begin{figure}[!t] 
  \centering 
  \includegraphics[width=0.49\textwidth]{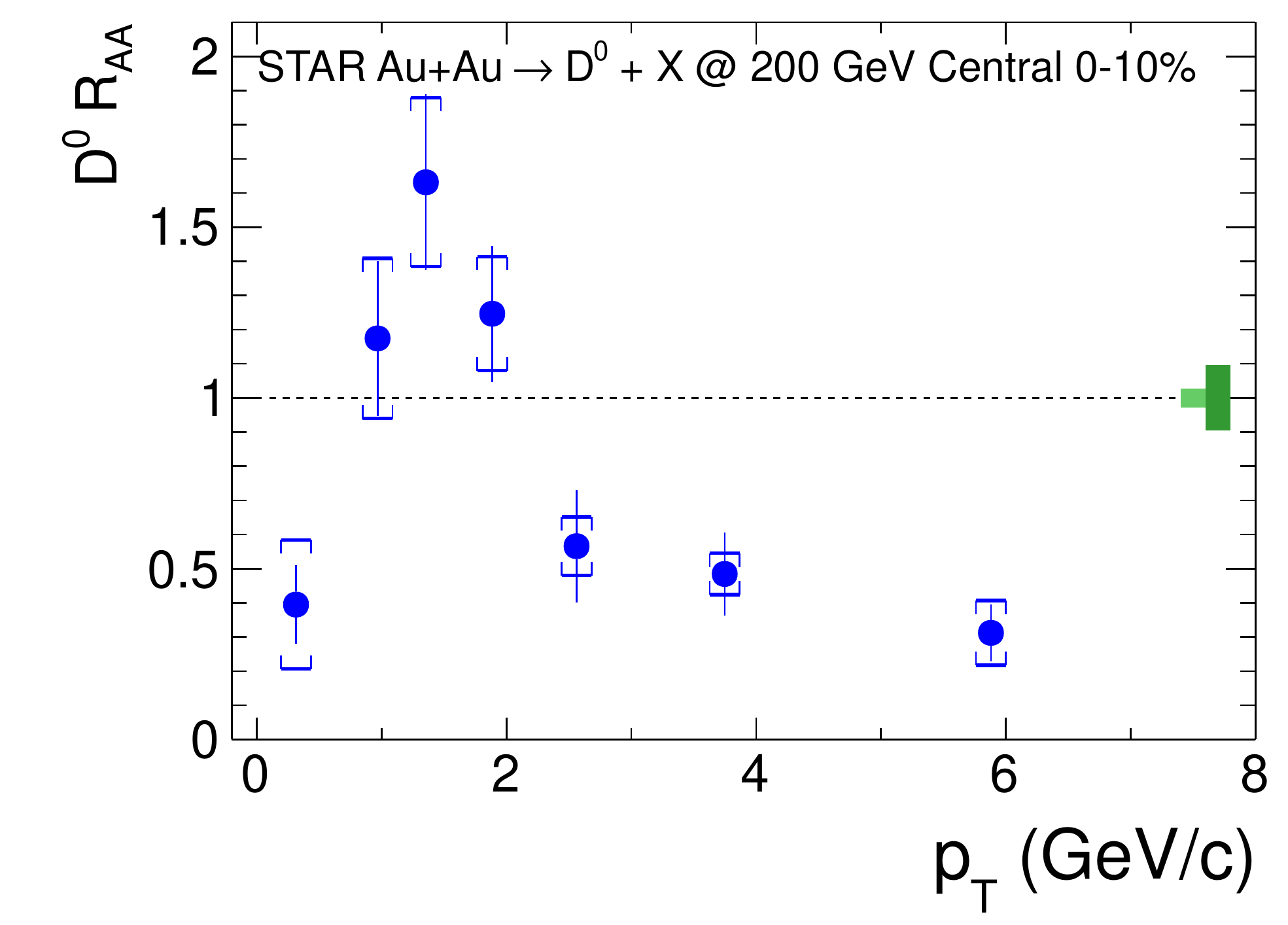}   
  \hfill 
  \includegraphics[width=0.49\textwidth]{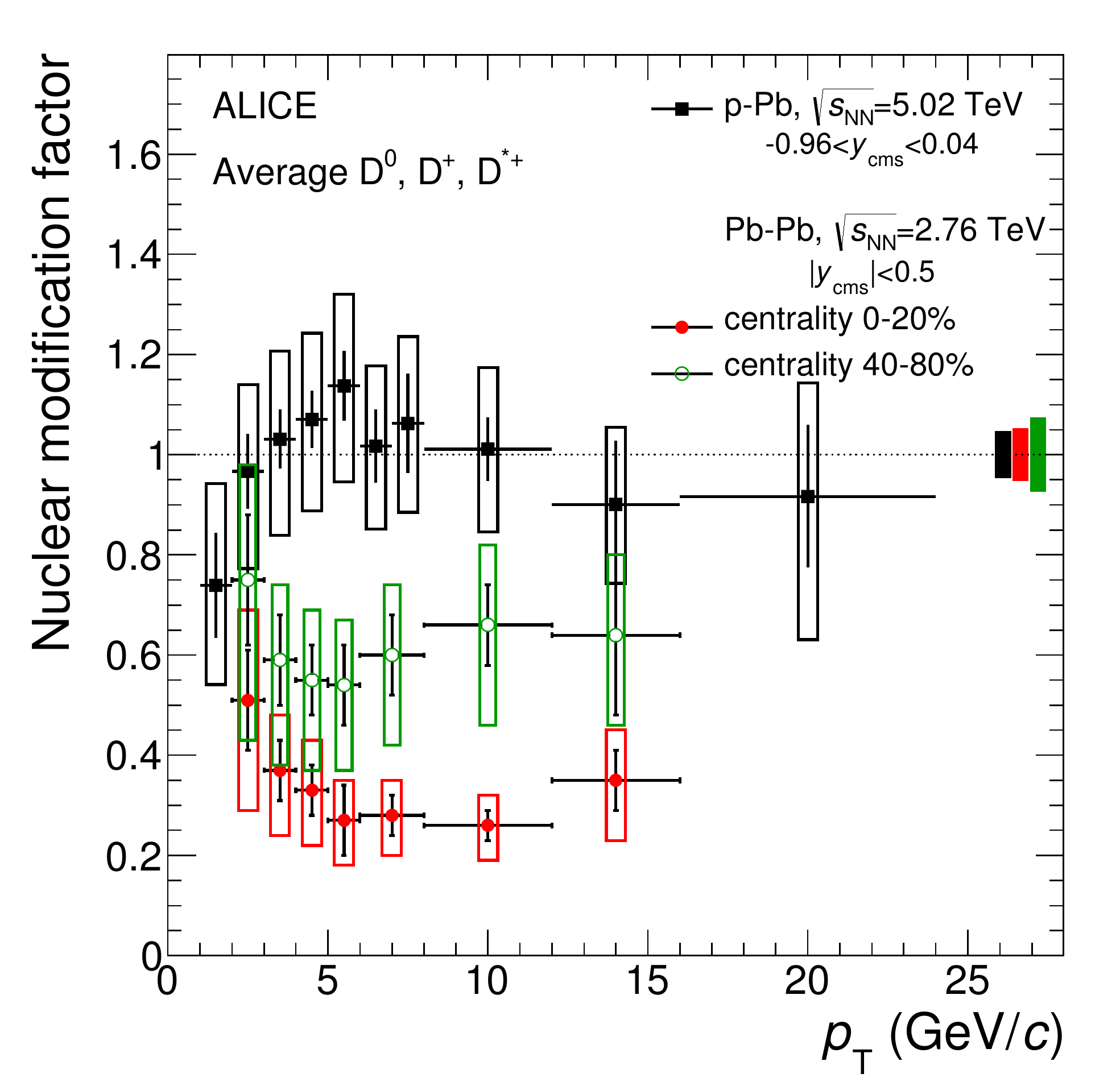} 
  \caption{Left: Transverse momentum (\pt) dependence of the nuclear modification factor \raa of \Dzero mesons in the 10\% most central \AuAu collisions at $\snn = 200$\GeV~\cite{Adamczyk:2014uip}. 
  Right: \raa of prompt D mesons (averaged) versus \pt for the 0--20\% (red discs) and 40--80\% (green circles) centrality classes in \pb collisions at $\snn = 2.76$\TeV~\cite{ALICE:2012ab} and minimum-bias \pPb collisions at $\snn = 5.02$\TeV (black squares)~\cite{Abelev:2014hha}.} 
\label{Fig:22} 
\end{figure}

The differential charm production cross section is determined from measurements of open charm mesons (STAR and ALICE). 
%
D mesons are reconstructed via the hadronic decays  
$\Dzero \to \kaon^-  \pi^+$, 
$\Dplus \to \kaon^-  \pi^+  \pi^+$, and 
$\Dstarplus(2010) \to \Dzero  \pi^+$  
 with $\Dzero \to \kaon^-  \pi^+$, and their charge conjugates.  The mean proper decay lengths of  
$\Dzero$ and $\Dplus$ are of about 120 and 300~$\mu{\rm m}$, respectively, while the $\Dstarplus$ decays strongly with no significant separation from the interaction vertex.  
In the STAR and ALICE experiments, charmed hadrons are measured with an invariant mass analysis of the fully-reconstructed decay topologies in the hadronic decay channels. 
In both experiments, the kaon and pion identification is performed by combining the information of the Time Of Flight and of the specific ionisation energy loss in the TPC~\cite{Adamczyk:2012af, Abelev:2012vra, ALICE:2011aa}. 
The spatial resolution of the ALICE silicon tracker allows, in addition, to reconstruct the decay vertex and apply a topological selection on its separation from the interaction vertex~\cite{Abelev:2012vra}. 
 
%
%
 

%
The left panel of \fig{Fig:22} shows the transverse momentum dependence of the nuclear modification factor \raa for \Dzero mesons in the most central \AuAu collisions at $\snn = 200$\GeV from the STAR experiment~\cite{Adamczyk:2014uip}. The \raa is enhanced at around 1.5\GeVc and shows a strong suppression at $\pt > 3$\GeVc. 
STAR also measured \Dzero mesons in \UU collisions at $\snn = 193$\GeV and observed a similar trend for the \raa as seen in \AuAu collisions~\cite{Ye:2014eia}.  
 
The ALICE experiment measured the production of prompt $\Dzero$, $\Dplus$ and $\Dstarplus$ mesons in \pb collisions at $\snn = 2.76$\TeV~\cite{ALICE:2012ab}. 
The average \raa of D mesons for two centrality classes is shown in the right panel of \fig{Fig:22}. The high-\pt D meson yield for the most central events is strongly suppressed (by factor of about four at 10\GeVc). 
The analysis of the \pb data collected in 2011 allowed to extend the measurement to higher transverse momenta: a similar suppression pattern is observed up to $\pt=30$\GeVc in the 7.5\% most central collisions~\cite{Grelli:2012yv}. 
In addition, the \Ds meson, consisting of a charm and an antistrange quark, was measured for the first time in \pb collisions~\cite{Innocenti:2012ds}. 
The \Ds meson is expected to be sensitive to the possible hadronisation of charm quarks via recombination with light quarks from the medium: the expected abundance of strange quarks in the QGP may lead to an increase of the ratio of strange over non-strange D mesons with respect to \pp collisions in the momentum range where recombination  
can be relevant~\cite{Kuznetsova:2008gr,He:2012df}. 
The observed central value of the \Ds \raa is larger than that of \Dzero, \Dplus and \Dstarplus mesons at low \pt, although the large statistical and systematic uncertainties prevent from drawing any conclusion. 
 
 
Initial-state effects were investigated by the ALICE Collaboration by measuring D production in \pPb collisions~\cite{Abelev:2014hha} (see Section~\ref{sec:CNM:OHF}).  
The nuclear modification factor of prompt D mesons in minimum-bias \pPb at $\snn = 5.02$\TeV is shown in \fig{Fig:22} (right panel). The \raa is compatible with unity within systematic uncertainties. 
 This indicates that the suppression of the D meson yield observed for $\pt>3$\GeVc in central \pb collisions is a final-state effect,  
most likely induced by the interactions of charm quarks within the QGP. 

\subsubsection{Beauty production measurements}
\label{sec:OHFbeauty}

\begin{figure} 
\centering 
\raisebox{8pt}{\includegraphics[width=0.45\textwidth]{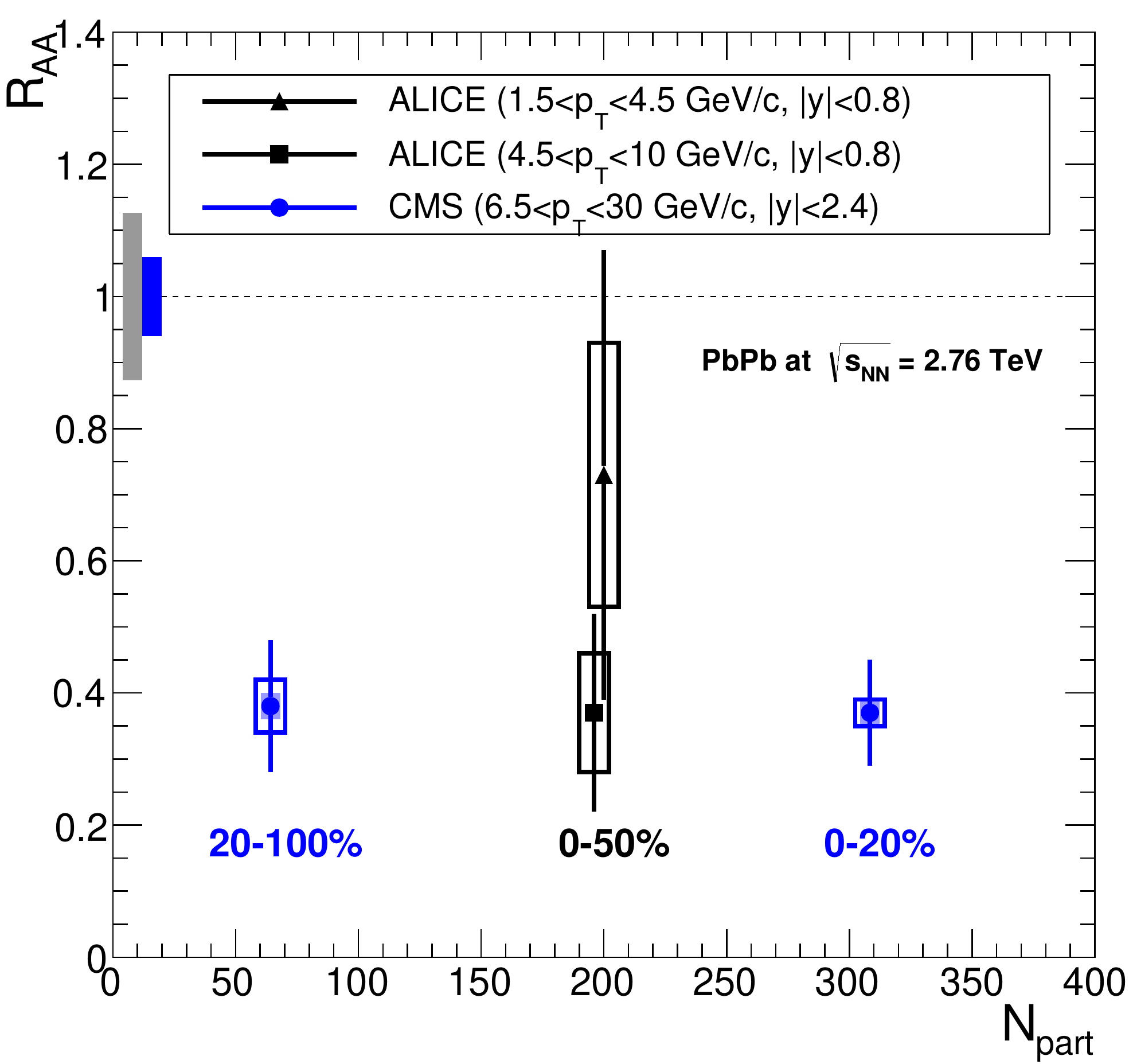}} 
\includegraphics[width=0.45\textwidth,trim=5 12 60 20,clip]{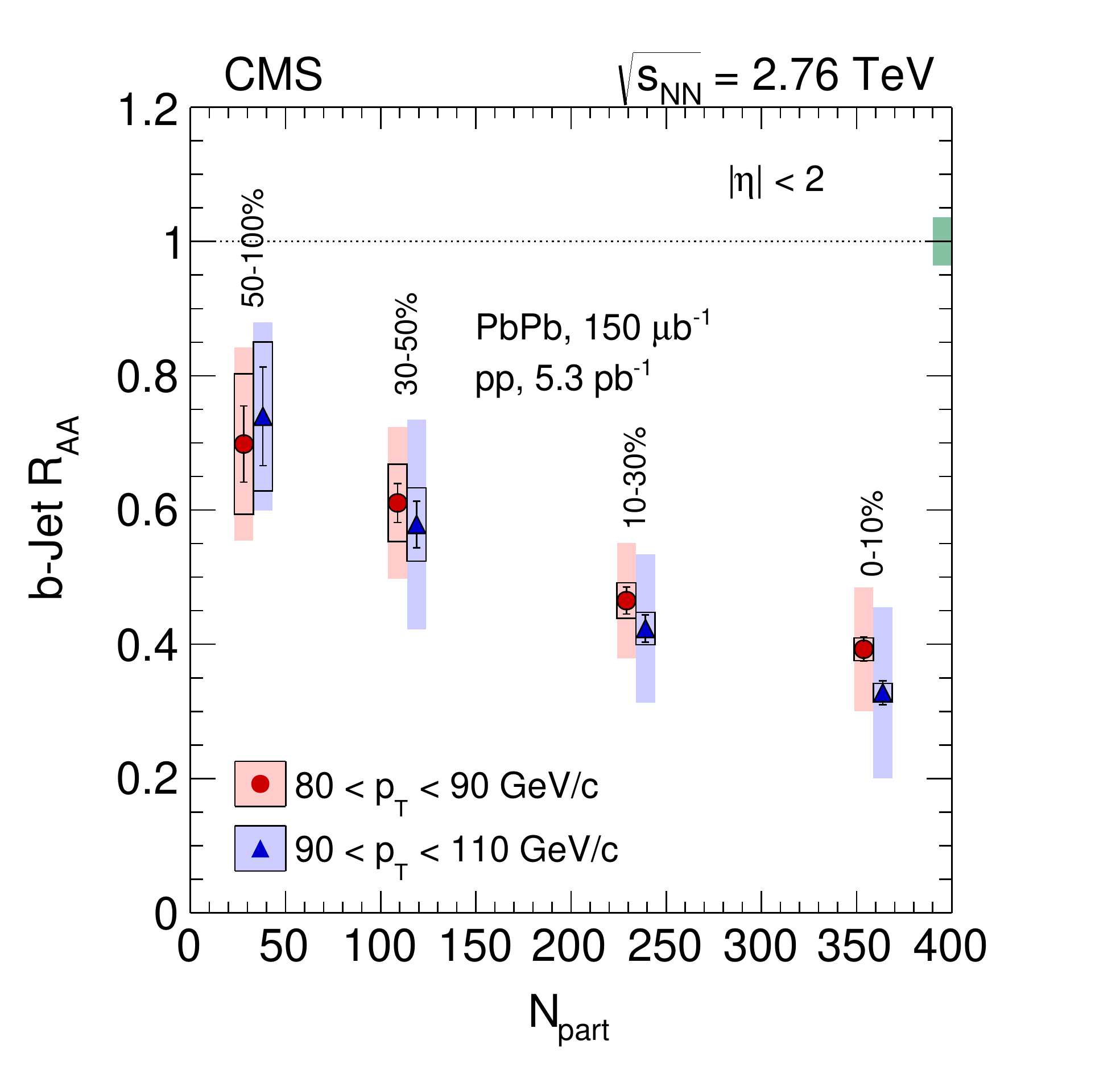} 
\caption[Non-prompt and \bquark-jet \raa from \pb collisions]{\label{fig:bottomPbPbmeas} 
Left:  Non-prompt \jpsi \raa measured in two centrality bins from CMS~\cite{Chatrchyan:2012np} and in one centrality bin for two \pt ranges from ALICE~\cite{Adam:2015rba}. The ALICE points are slightly shifted horizontally for better visibility. The correlated uncertainties are shown as filled box at \raa=1. 
Right:  \raa of \bquark jets, as a function of \Npart from CMS~\cite{Chatrchyan:2013exa}, for two jet \pt selections as indicated in the legend.  Systematic uncertainties are shown as filled boxes, except the \taa uncertainties, depicted as open boxes. The luminosity uncertainty is represented by the green box.} 
\end{figure}

The detection and identification of beauty hadrons usually exploit their long life times, with $c\tau$ values of about 500~$\mu$m. Precise charged particle tracking and vertexing are of crucial importance, with the required resolution of the track impact parameter in the transverse plane being of the order of 100~$\mu$m. Most decay channels proceed as a $b \rightarrow c$ hadron cascade, giving rise to a topology that contains both a secondary and a tertiary decay vertex.   
 
Lepton identification is often exploited in beauty measurements, as the semi-leptonic branching ratio is about 20\%, taking into account both decay vertices. 
The beauty contribution can be extracted from the semi-electronic decays of heavy flavours through a fit of the impact parameter distribution. 
This approach was applied by the ALICE Collaboration in \pp collisions at the LHC~\cite{Abelev:2012sca,Abelev:2014hla} (see \sect{subsubsec:pp:OpenBeauty}) and recently also in \pb collisions~\cite{Festanti:2014foa}, where preliminary results indicate \raa values below unity for electron \pt larger than about 5\GeVc. 
The charm and beauty contribution can be disentangled also by studying the correlations between electrons and associated charged hadrons, exploiting the larger width of the near-side peak for B hadron decays~\cite{Aggarwal:2010xp,Abelev:2014hla,Mischke:2008af}. The main limitation of the beauty measurements via single electrons (or muons) is the very broad correlation between the momentum of the measured electron and the momentum of the parent B meson.  
 
 A more direct measurement is achieved using the inclusive ${\rm B}\to \jpsi + X$ decay mode.  Such decays can be measured inclusively by decomposing the \jpsi yield into its prompt and non-prompt components, using a fit to the lifetime distribution. 
The first measurement with this technique in heavy-ion collisions was performed by the CMS Collaboration, using data from the 2010 \pb run. The \raa of non-prompt  \jpsi in $6.5 < \pt < 30$\GeVc and $|y| < 2.4$ was measured to be $0.37 \pm 0.08 {\rm(stat.)} \pm 0.02 {\rm(syst.)}$ in the 20\% most central collisions (see left panel of \fig{fig:bottomPbPbmeas})~\cite{Chatrchyan:2012np}.  Preliminary measurements from the larger 2011 dataset explore the \pt dependence of \raa~\cite{CMS:2012wba}. 
A recent measurement from the ALICE Collaboration~\cite{Adam:2015rba} (left panel of \fig{fig:bottomPbPbmeas}) shows a similar value of \raa in a close kinematic range ($4.5<\pt<10$\GeVc and $|y|<0.8$).

Further insights into the parton energy loss can be provided through measurements of reconstructed jets and comparison with theory~\cite{Huang:2013vaa}, which is complementary to the studies on B hadrons 
as the reconstructed jet energy is closely related to that of the \bquark quark.  
Assuming that the quark hadronises outside the medium, to first approximation the jet energy represents the sum of the parton energy after its interaction with the medium, as well as any transferred energy that remains inside the jet cone.  CMS has performed a measurement of \bquark jets in \pb collisions by direct reconstruction of displaced vertices associated to the jets~\cite{Chatrchyan:2012np}.  Despite the large underlying \pb event, a light jet rejection factor of about 100 can still be achieved in central \pb events.  The \raa of \bquark jets as function of centrality is shown in \fig{fig:bottomPbPbmeas} (right), for two ranges of jet \pt.  The observed suppression, which reaches a value of about 2.5 in central collisions, does not show any significant difference compared to a similar measurement of the inclusive jet \raa~\cite{CMS:2012rba} within the sizeable systematic uncertainties.  While quark mass effects may not play a role at such large values of \pt, the difference in energy loss between quarks and gluons should manifest itself as a difference in \raa for \bquark jets and inclusive jets, as the latter are dominated by gluon jets up to very large \pt. 
It should be noted, however, that the \bquark jets do not always originate from a primary \bquark quark. 
 As discussed in the introduction to this Section,  at LHC energies, a significant component of \bquark quarks are produced by splitting of gluons into \bbbar pairs~\cite{Banfi:2007gu}.   
For \bquark jets with very large \pt a significant part of the in-medium path-length is likely to be traversed by the parent gluon, as opposed to the \bquark quarks (for example, about 1--2~fm for  
\bquark quarks with \pt of 100--200\GeVc). 
The gluon splitting contribution can be minimised by selecting hard fragments, although this is complicated by the fact that the \bquark-hadron kinematics are not fully measured.  An alternative is to select back-to-back \bquark-tagged jets, a configuration in which the gluon splitting contribution is negligible.  The dijet asymmetry of \bquark jets can then be compared to that of inclusive jets, a measurement that should be feasible with the luminosity expected from the upcoming LHC \RunTwo.  
  

%
%
%
%
%
%

\subsubsection{Comparison of \texorpdfstring{\raa}{R\_AA} for charm, beauty and light flavour hadrons}
 
\begin{figure}[!t] 
  \begin{center} 
\includegraphics[width=0.49\textwidth]{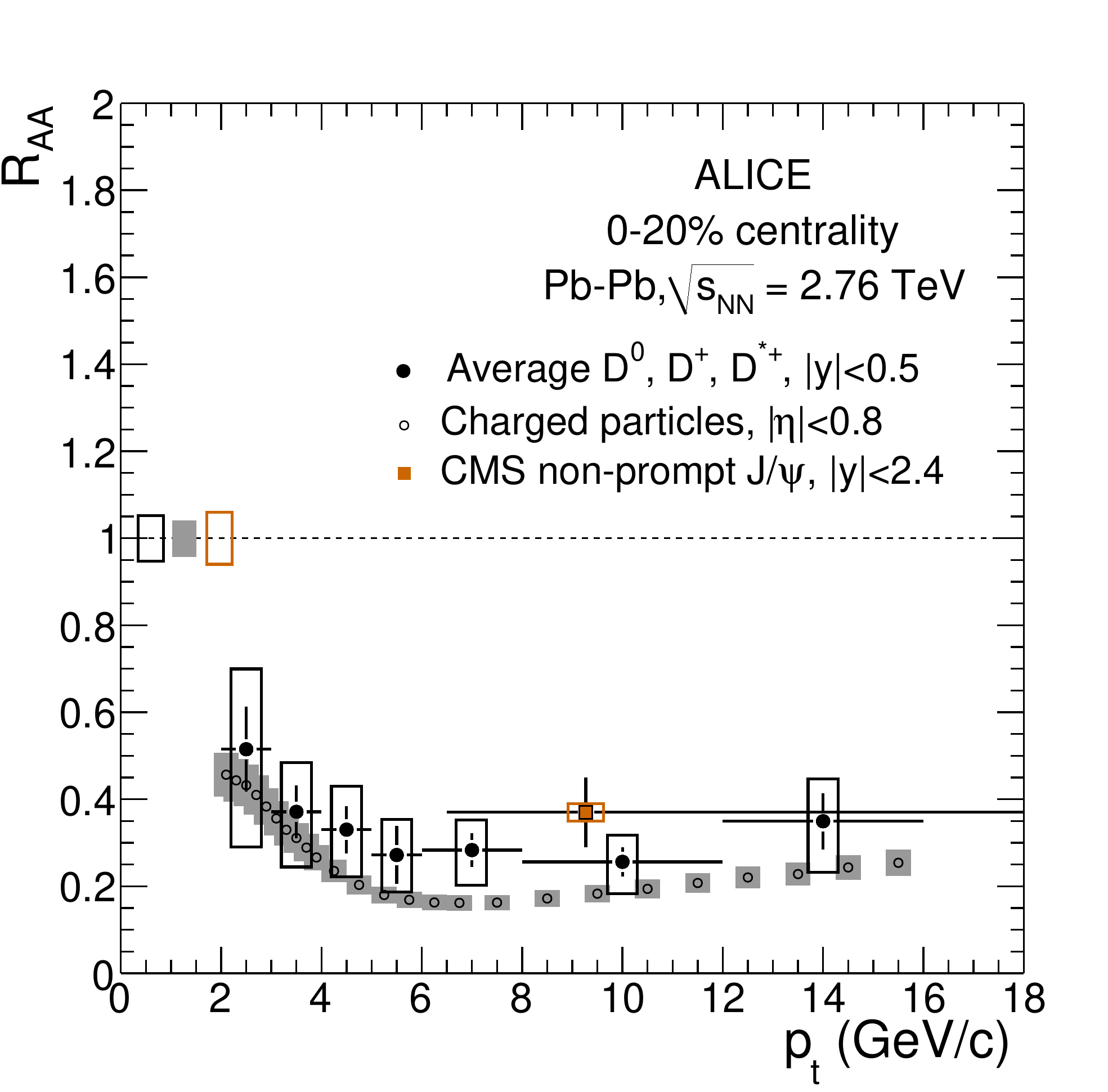} 
  \caption{\raa of D mesons, charged hadrons~\cite{ALICE:2012ab} and non-prompt \jpsi~\cite{Chatrchyan:2012np} in \pb collisions at $\snn=2.76\TeV$ in the 0--20\% centrality class.}  
\label{fig:RAA_LHC_D_charged_npJpsi} 
\end{center} 
\end{figure}

The expected dependence of in-medium energy loss on the parton colour charge and mass can be investigated  
by comparing the nuclear modification factor of charged hadrons, mostly originating from gluon fragmentation at the LHC collision energy, with that of hadrons with charm and beauty. 
The comparison between D meson and charged particle \raa, measured by the ALICE Collaboration~\cite{ALICE:2012ab}  
in \pb collisions at LHC in the centrality class 0--20\% 
 and illustrated in \fig{fig:RAA_LHC_D_charged_npJpsi}, 
shows that the two nuclear modification factors are compatible within uncertainties, although the central values show 
an indication for $\raa^{\rm D}>\raa^{\rm charged}$. 
In the same figure, the nuclear modification factor measured by the CMS Collaboration for non-prompt  
\jpsi mesons (from B decays)  
with $\pt>6.5$\GeVc~\cite{Chatrchyan:2012np} is also shown. Their suppression is clearly weaker 
than that of charged particles, while the comparison with D mesons is not conclusive,  
because of the significant uncertainties of the two measurements. 
In addition, it is worth noting that the \pt of the \jpsi is shifted with respect to the one of the parent B meson (by about 2--3\GeVc on average in the \pt range covered by the CMS measurement), hence the comparison with D mesons is not straight-forward. 
 
Preliminary measurements based on higher-statistics data from the 2011 \pb run at LHC provide a first indication that the nuclear modification factor of B mesons is larger than  
that of D at transverse momentum of about 10\GeVc. The measurements were carried out, as a function of collision centrality, for D mesons with $8<\pt<16$\GeVc and $|y|<0.5$ 
by the ALICE Collaboration~\cite{Bruna:2014pfa} and for non-prompt \jpsi mesons with $6.5<\pt<30$\GeVc and $|y|<1.2$ by the CMS Collaboration~\cite{CMS:2012wba}. 
With these \pt intervals, the average \pt values of the probed D and B mesons are both of about 10--11\GeVc. In central collisions (centrality classes 0--10\% and 10--20\%) 
the \raa values are of about 0.2 and 0.4 for D and non-prompt \jpsi mesons, respectively, and they are not compatible within experimental uncertainties. 
This experimental observation alone does not allow to draw conclusions on the comparison of energy loss of charm and beauty quarks, because several kinematic effects 
contribute to the \raa resulting from a given partonic energy loss. In particular, the shape of the quark \pt distribution (which is steeper for charm than for beauty quarks) 
and the shape of the fragmentation function (which is harder for $b\to \rm B$ than for $c\to \rm D$) have to be taken into account.  Model calculations of heavy-quark production, 
in-medium propagation and fragmentation provide a tool to consistently consider these effects in the comparison of charm and beauty measurements.   
In \sect{sec:modelcomparison} we will show that model calculations including a mass-dependent energy loss result in \raa values significantly larger for \jpsi from B decays 
than for D mesons, consistently with the preliminary results from the ALICE and CMS experiments.

\subsection{Experimental overview: azimuthal anisotropy measurements}
\label{sec:OHFv2}

As mentioned in the introduction to this chapter, the azimuthal anisotropy of particle production in heavy-ion collisions is measured using the Fourier expansion of the  
azimuthal angle ($\phi$) and the \pt-dependent particle distribution $\dd^2 N / \dd\pt \dd\phi$.  
The second coefficient, \vtwo or elliptic flow, which is the dominant component of the anisotropy in non-central nucleus--nucleus collisions,  
is measured using these three 
methods: event plane (EP)~\cite{Poskanzer:1998yz}, scalar product (SP)~\cite{Adler:2002pu} and multi-particle cumulants~\cite{Bilandzic:2010jr}. 
In the following, an overview of the elliptic flow measurements for heavy-flavour particles is presented: the published measurements at RHIC use heavy-flavour decay electrons (\sect{sec:v2leptons}); the published measurements at LHC use D mesons (\sect{sec:v2Dmesons}).

\subsubsection{Inclusive measurements with electrons} 
\label{sec:v2leptons} 
 
The measurement of the production of heavy-flavour decay electrons has been presented in \sect{sec:HFelectrons} .    
In order to determine the heavy-flavour decay electron \vtwo, the starting point is the measurement of \vtwo for inclusive electrons.  Inclusive electrons include, mainly, the so-called photonic (or background) electrons (from photon conversion in the detector material and internal conversions in the Dalitz decays of light mesons), a possible contamination from hadrons, and heavy-flavour decay electrons. Exploiting the additivity of \vtwo, the heavy-flavour decay electron \vtwo is obtained by subtracting from the inclusive electron \vtwo the \vtwo of photonic electrons and hadrons, weighted by the corresponding contributions to the inclusive yield. 
 
\begin{figure}[t] 
\centering 
\includegraphics[width=0.46\textwidth]{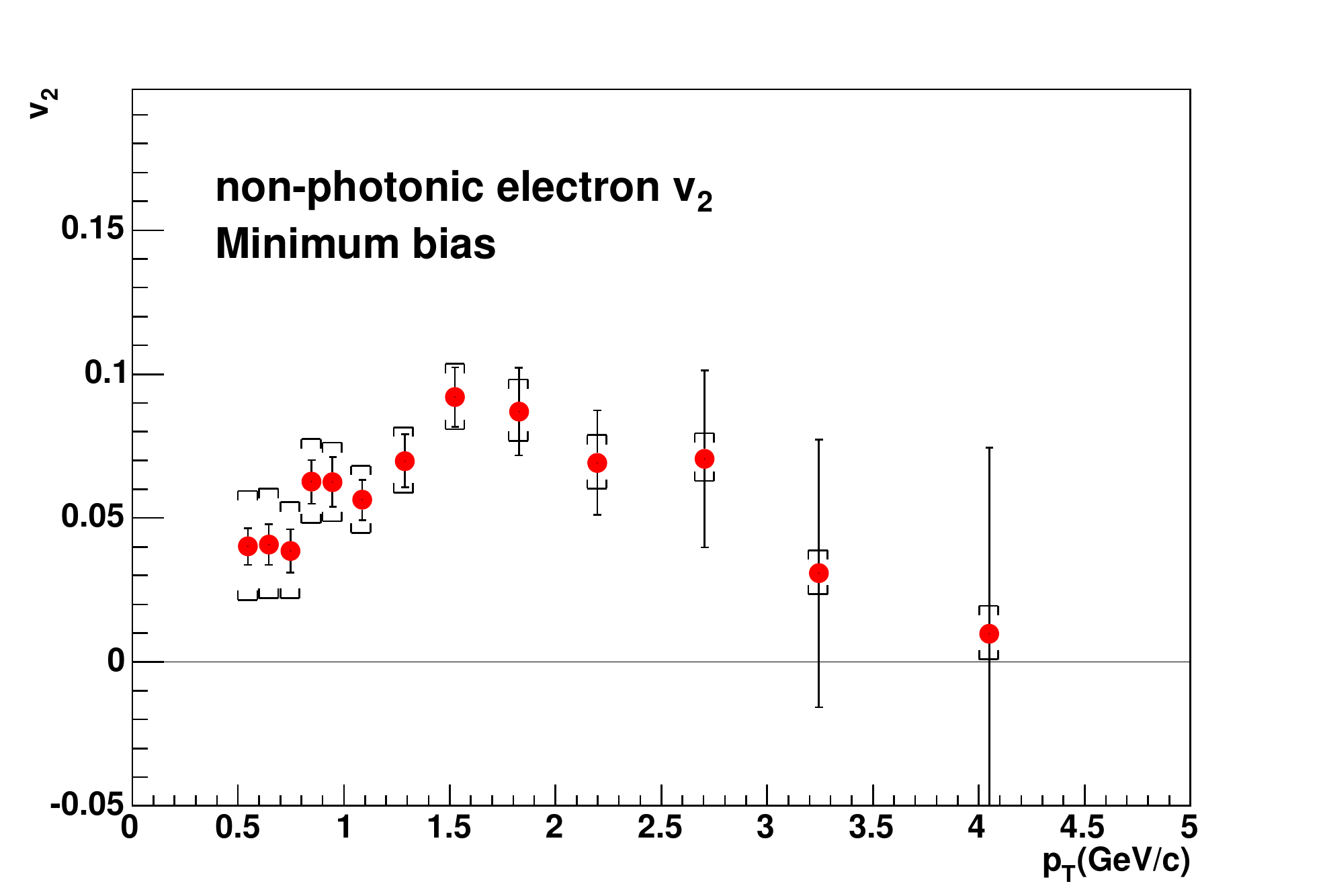} 
\includegraphics[width=0.53\textwidth]{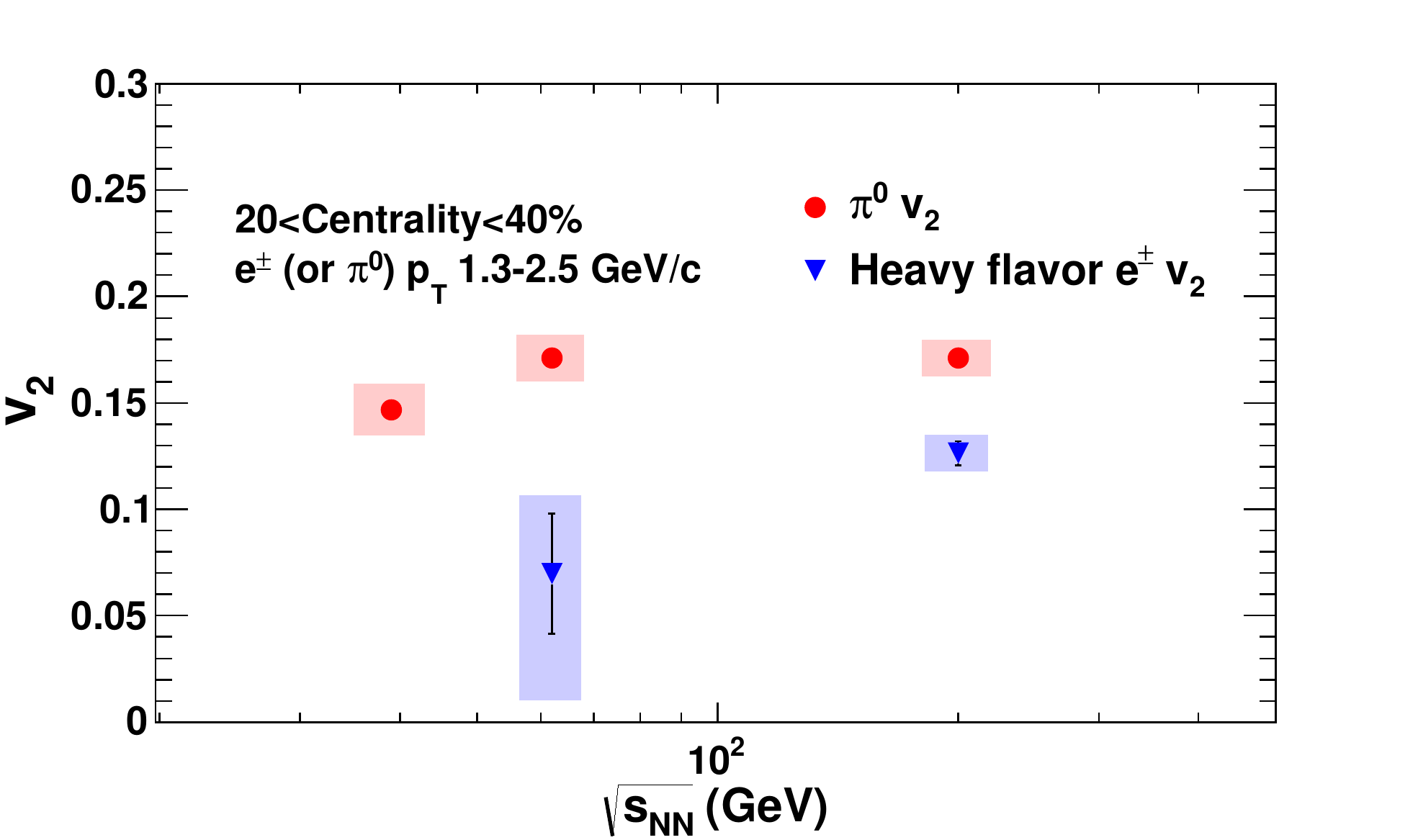} 
 
\caption{Heavy-flavour decay electron \vtwo measured by the PHENIX Collaboration in \AuAu collisions at RHIC. Left: measurement in minimum-bias collisions as a function of \pt at $\snn = 200$\GeV~\cite{Adare:2010de}. Right:  measurements at $\snn=62.4$~and~200\GeV in the 20--40\% centrality class for the interval $1.3<\pt<2.5$\GeVc, compared with the $\pi^0$ \vtwo~\cite{Adare:2014rly}.} 
\label{fig:HFePHENIX} 
\end{figure} 
 
\begin{figure}[t] 
\begin{center} 
\includegraphics[width=0.46\textwidth]{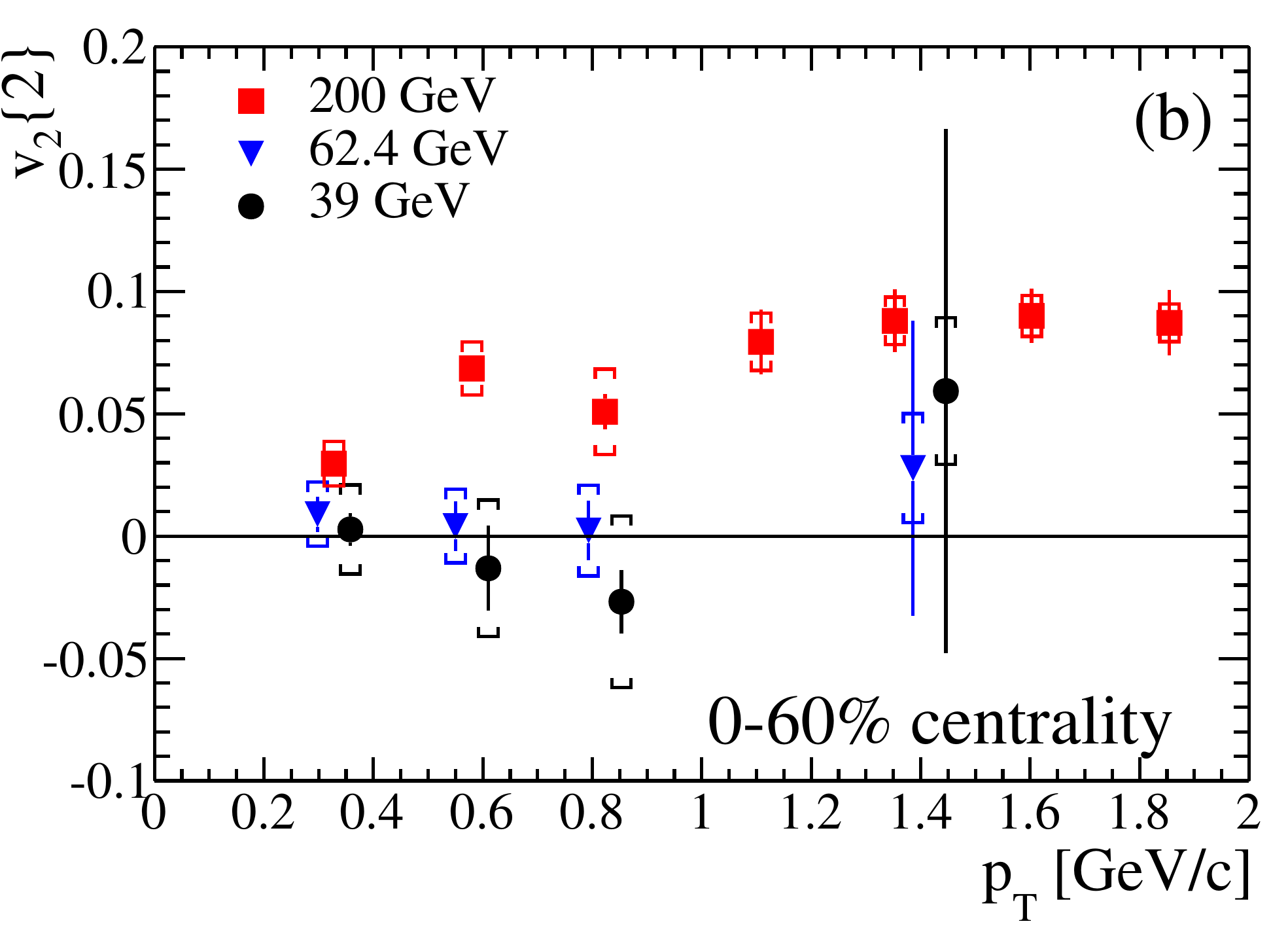} 
\end{center} 
\caption{Heavy-flavour decay electron \vtwo measured by the STAR Collaboration in \AuAu collisions (0--60\% centrality class) at centre-of-mass energies 39, 62.4 and 200\GeV~\cite{Adamczyk:2014yew}.} 
\label{fig:HFeSTAR} 
\end{figure}

The PHENIX Collaboration measured the heavy-flavour decay electron \vtwo in \AuAu collisions at $\snn = 62.4$ and 200\GeV using the event plane method~\cite{Adare:2010de, Adare:2014rly}.  
Electrons were detected at mid-rapidity $|\eta| < 0.35$ in the interval $0.5 < \pt < 5$\GeVc.  
The event plane was instead determined using charged particles at forward rapidity $3.0 < |\eta| < 3.9$. 
This large $\eta$-gap is expected to reduce the non-flow effects (like auto-correlations) in the \vtwo measurement.  
\fig{fig:HFePHENIX}~(left) shows the heavy-flavour decay electron \vtwo for minimum-bias events (without any selection on centrality)~\cite{Adare:2010de}.  
\vtwo is larger than zero in the interval $0.5<\pt<2.5$\GeVc, with a maximum value of about 0.1 at \pt of about 1.5\GeVc. Towards larger \pt the data suggest a decreasing trend, although the statistical uncertainties prevent a firm conclusion.  The study of the centrality dependence of \vtwo (not shown) indicates a maximum effect in the two semi-peripheral centrality classes (20--40\% and 40--60\%), for which the initial spatial anisotropy is largest~\cite{Adare:2010de}.  
The central value of the heavy-flavour electron \vtwo in \AuAu collisions at $\snn = 62.4$\GeV~\cite{Adare:2014rly} is significantly lower than at 200\GeV 
(see \fig{fig:HFePHENIX}~(right)). However, the statistical and systematic uncertainties are sizeable and do not allow to conclude firmly on the energy dependence of \vtwo. 
In \fig{fig:HFePHENIX}~(right) the measurements for heavy-flavour decay electrons with $1.3<\pt<2.5$\GeVc are compared with those for neutral pions with the same \pt: the pions exhibit a larger \vtwo than the electrons; however, this comparison should be taken with care, because the \pt of the heavy-flavour mesons is significantly larger than that of their decay electrons. 
 
  
The STAR Collaboration measured the heavy-flavour decay electron \vtwo in \AuAu collisions at $\snn = 39, ~62.4$ and 200\GeV~\cite{Adamczyk:2014yew}.  
The two-particle cumulant method was used to measure the elliptic flow for the two lower collision energies. The event plane, and both two- and four-particle cumulant methods were used  
at $\snn = 200$\GeV. 
\fig{fig:HFeSTAR} shows the \vtwo measured with two-particle cumulants at the three centre-of-mass energies.  
At $\snn = 200$\GeV the measurement shows a \vtwo larger than zero for $\pt>0.3$\GeVc, compatible with the measurement by the PHENIX Collaboration in the same centrality class (see comparison in~\cite{Adamczyk:2014yew}).   
At  $\snn = 39$ and 62.4\GeV, the $\vtwo\{2\}$ values are consistent with zero within uncertainties. 
 
Preliminary results by the ALICE Collaboration on the elliptic flow of heavy-flavour decay electrons at central rapidity ($|y|<0.6$) and of heavy-flavour decay muons at forward rapidity ($2.5<y<4$) 
in \pb collisions at the LHC show a \vtwo significantly larger than zero in both rapidity regions and with central values similar to those measured at top RHIC energy~\cite{Bailhache:2014fia}. 
 
\begin{figure}[!ht] 
\begin{center} 
\includegraphics[width=0.48\textwidth]{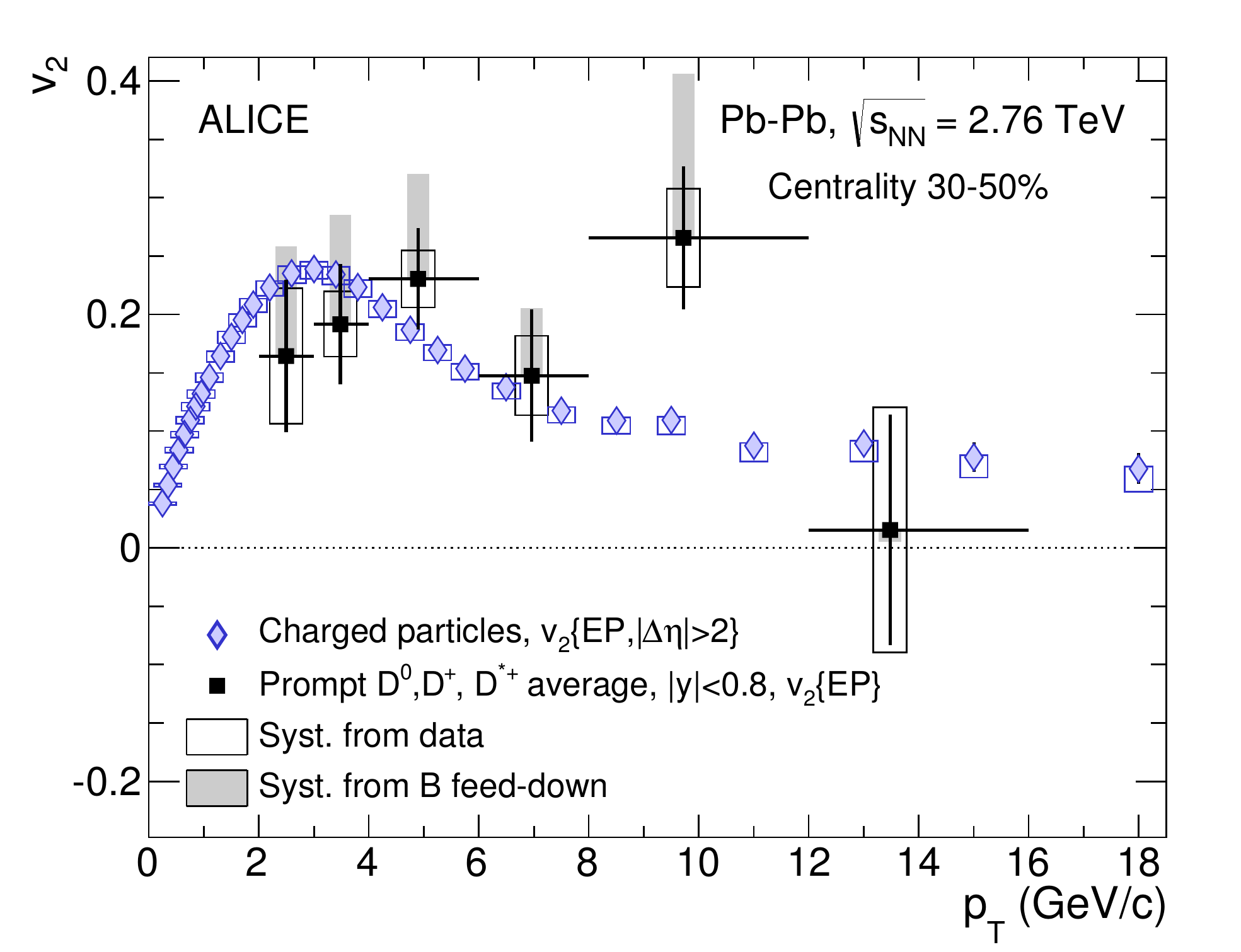} 
\includegraphics[width=0.50\textwidth]{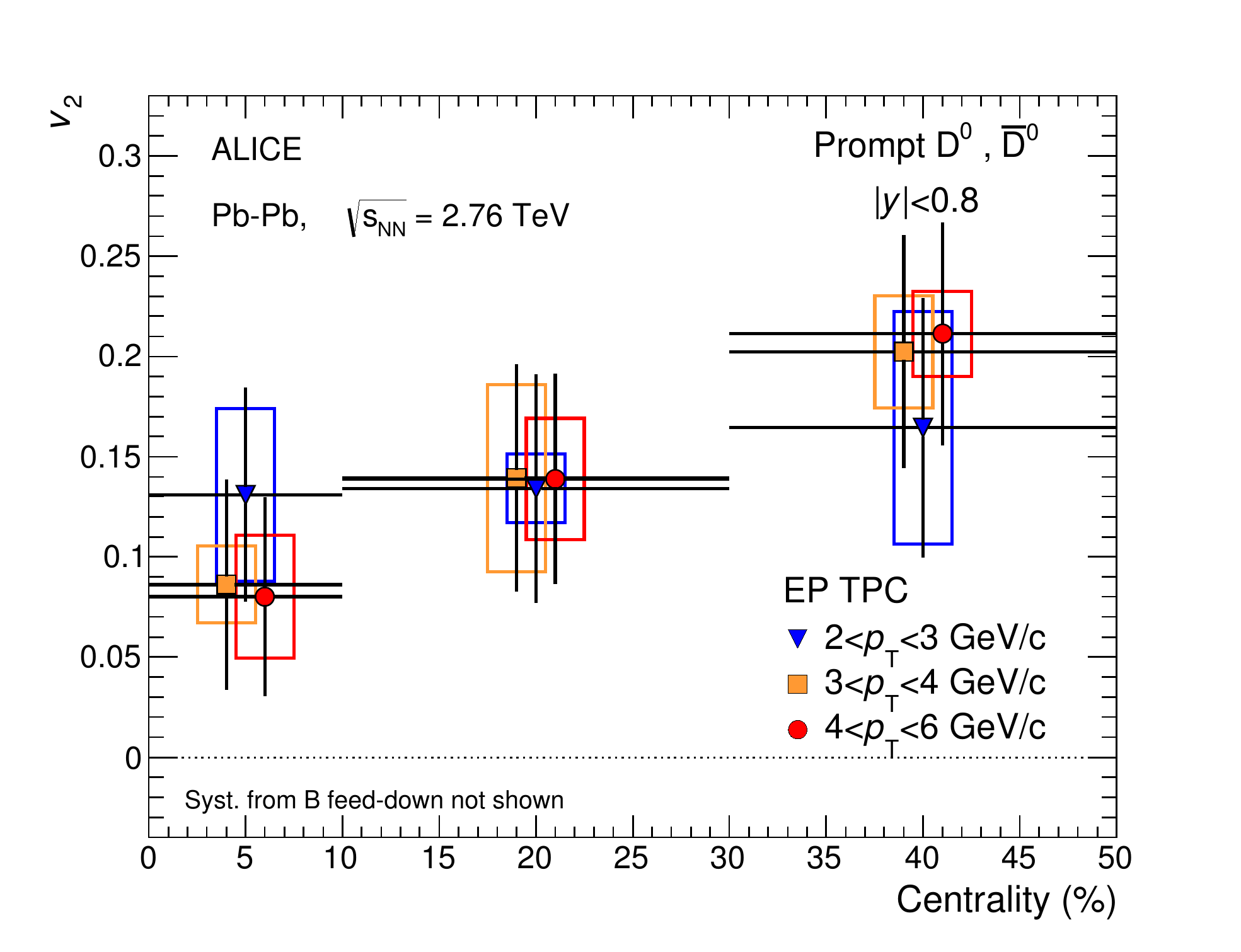} 
\end{center} 
\caption{D meson elliptic flow measured by the ALICE Collaboration in \pb collisions at $\snn=2.76$\TeV~\cite{Abelev:2013lca,Abelev:2014ipa}. Left: average of the \vtwo values for $\Dzero$, $\Dplus$ and $\Dstarplus$ mesons in the centrality class 30--50\% as a function of \pt, compared with the \vtwo of charged particles. Right: centrality dependence of the $\Dzero$ meson \vtwo for three \pt intervals.} 
\label{fig:DmesALICE} 
\begin{center} 
\includegraphics[width=0.45\textwidth]{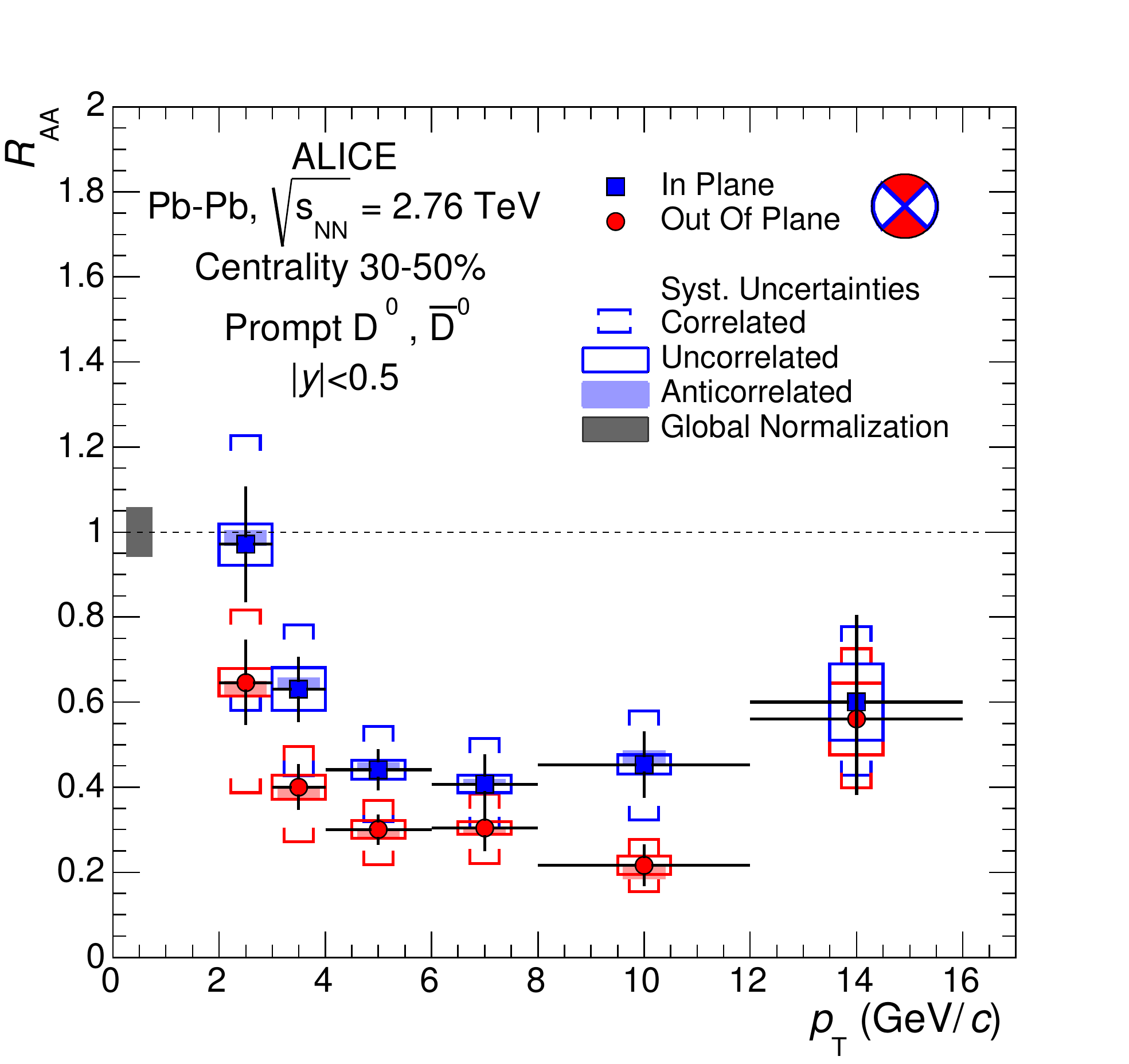} 
\end{center} 
\caption{D meson \raa in the direction of the event plane and in the direction orthogonal to the event plane, measured by the ALICE Collaboration in \pb  
collisions (centrality class 30--50\%) at $\snn=2.76$\TeV~\cite{Abelev:2013lca,Abelev:2014ipa}. } 
\label{fig:DmesRaavsEPALICE} 
\end{figure}

\subsubsection{{\rm D} meson measurements} 
\label{sec:v2Dmesons}

The ALICE Collaboration measured the \vtwo of prompt D mesons in \pb collisions at $\snn=2.76$\TeV~\cite{Abelev:2013lca,Abelev:2014ipa}.   
The D mesons ($\Dzero$, $\Dplus$ and $\Dstarplus$) were measured in $|y| < 0.8$ and $2 < \pt < 16$\GeVc  
using their hadronic decay channels, and exploiting the separation of a few hundred $\mu$m of the decay vertex from the interaction vertex  
to reduce the combinatorial background. 
The measurement of D meson \vtwo was carried out using the event plane, the scalar product and the two-particle cumulant methods. 
 
\fig{fig:DmesALICE}~(left) shows the average of  the \vtwo measurements for $\Dzero$, $\Dplus$ and $\Dstarplus$ in the centrality class 30--50\% as a function of \pt~\cite{Abelev:2013lca}. The measurement shows a \vtwo larger than zero in the interval $2 < \pt< 6$\GeVc with a $5.7\sigma$ significance.  
In the same figure, the \vtwo of charged particles for the same centrality class is reported for comparison: the magnitude of \vtwo is similar for charmed and light-flavour hadrons.  
\fig{fig:DmesALICE}~(right) shows the dependence on collision centrality of the $\Dzero$ meson \vtwo for three \pt intervals. 
An increasing trend of \vtwo towards more peripheral collisions is observed, as expected due to the increasing initial spatial anisotropy.  
 
 
As discussed at the beginning of this Section, the azimuthal dependence of the nuclear modification factor \raa can provide insight into the path length dependence of heavy-quark energy loss. 
The nuclear modification factor of $\Dzero$ mesons in \pb collisions (30--50\% centrality class) was measured by the ALICE Collaboration in the direction of the event plane (in-plane) 
and in the direction orthogonal to the event plane (out-of-plane)~\cite{Abelev:2014ipa}. The results, shown in \fig{fig:DmesRaavsEPALICE}, exhibit a larger high-\pt suppression  
in the out-of-plane direction, where the average path length in the medium is expected to be larger. It is worth noting that the difference between the values of \raa in-plane 
and \raa out-of-plane is equivalent to the observation of $\vtwo > 0$, because the three observables are directly correlated.

\subsection{Theoretical overview: heavy flavour interactions in the medium}
\label{sec:OHFmodelsInt}
 
The approaches describing the heavy-quark--medium interactions 
aim at determining the probability  
${\cal P}_{Q\rightarrow H}(p_Q^{\rm in},p_{H}^{\rm fin})$ that a given heavy quark produced with a  
4-momentum $p_Q^{\rm in}$ escapes the medium as a heavy-flavour hadron of 4-momentum  
$p_{H}^{\rm fin}$.  
 
All the approaches include a description of the interactions that occur between the heavy quarks  
and the partonic constituents of the QGP. 
For ultra-relativistic heavy quarks ($p_Q\gg m_Q$, say $>10\,m_Q$), the dominant source of energy loss is commonly considered to be the 
radiation of gluons resulting from the scattering of the heavy quark on the medium constituents. These are $2\to 3$ processes $q(g)Q\to q(g)Qg$, 
where $q(g)$ is a medium light quark (or gluon). 
As this mechanism proceeds through long formation times, several scatterings contribute 
coherently and quantities like the total energy loss $\Delta E(L)=p_Q^{\rm in}-p_Q^{\rm fin}$ can only be evaluated at the end of the in-medium path length $L$.  
This feature is shared by all schemes that have been developed to evaluate radiative energy loss of 
ultra-relativistic partons~\cite{Armesto:2003jh,Zhang:2003wk,Djordjevic:2003zk}. 
For merely relativistic heavy quarks (say $p_Q<10\,m_Q$), elastic (collisional) processes  
are believed to have an important role as well. These are $2\to 2$ process $q(g)Q\to q(g)Q$. 
The in-medium interactions are gauged by the following, closely related, variables: the \textit{mean free path} $\lambda=1/(\sigma\rho)$ is related to the medium density $\rho$  
and to the cross section $\sigma$ of the parton-medium interaction (for $2\to 2$ or $2\to 3$ processes); 
the \textit{Debye mass} $m_D$ is the inverse  
of the screening length of the colour electric fields in the medium and it is proportional to the temperature $T$ of the medium; the \textit{transport coefficients}  
encode the momentum transfers with the medium (more details are given in the next paragraph). 
 
In the relativistic regime, the gluon formation time for radiative processes becomes small enough that the energy loss probability 
${\cal P}(\Delta E)$ can be evaluated as the result of some local transport equation 
-- like the Boltzmann equation, relying on local cross sections -- evolving from initial to final time.  
This simplification can be applied also to collisional processes.  
When the average momentum transfer is small with respect to the quark mass\footnote{This is the so called 
``grazing expansion''~\cite{balescu1975equil}, well justified for non-relativistic heavy quarks.}, 
the Boltzmann equation can be reduced to the Fokker-Planck equation, which is often further simplified  
to the Langevin stochastic equation (see~\cite{Rapp:2009my} for  
a recent review).  These linear partial-differential equations describe the time-evolution of the momentum distribution $f_Q$ of heavy quarks.  
The medium properties are encoded in three transport coefficients: a) the \textit{drift coefficient} 
-- also called \textit{friction} or \textit{drag coefficient} -- which represents the fractional momentum loss per unit of time 
in the absence of fluctuations and admits various equivalent symbolic representations ($\eta_D,\,A_Q,\,\ldots$) 
and b) the \textit{longitudinal} and \textit{transverse momentum diffusion coefficients} $B_{\rm L}$ 
and $B_{\rm T}$ (or $B_1$ and $B_0$, $\kappa_{\rm L}$ and $\kappa_{\rm T}$,\ldots, depending on the authors),  
which represent the increase of the variance of $f_Q$ per unit of time. For small momentum, the drift and diffusion coefficients are linked through the Einstein relation  
$B=m_Q\,\eta_D\,T$ 
and  
also uniquely related to the spatial diffusion coefficient $D_s$, which describes the spread of the distribution in  
physical space with time. Although the Fokker-Planck approach has some drawbacks\footnote{The Einstein relation is not necessarily satisfied for all momenta $p_Q$  
in an independent calculation of $B$ and $\eta_D$ 
and hence has to be enforced.}, it can also be deduced from more general  
considerations~\cite{risken1989fp}, so that it may still be considered as a valid approach for describing heavy-quark transport even when the Boltzmann equation does not apply, as for instance in the strong-coupling limit.  
 
Some of the approaches consider only partonic interactions and define the  
${\cal P}_{Q\rightarrow H}$ probability as a convolution  
of ${\cal P}_{Q\rightarrow Q'}(p_Q^{\rm in},p_Q^{\rm fin})$ -- the probability for the heavy quark 
to lose $p_Q^{\rm in}-p_Q^{\rm fin}$ in the medium -- with the unmodified fragmentation function. 
A number of approaches also include, for low-intermediate momentum heavy quarks, a contribution of  
hadronisation via recombination (also indicated as coalescence). Finally, some approaches  
consider late-stage interactions of the heavy-flavour hadrons with the partonic or hadronic medium.

In this section, we summarise the various approaches for the calculation of the heavy-quark interactions within the medium.  
\begin{itemize} 
\item \sects~\ref{sec:pQCDEloss} and \ref{sec:mcatshq} are devoted to pQCD and pQCD-inspired calculations 
of radiative and collisional energy loss, as developed 
by Gossiaux \etal (MC$@_s$HQ), Beraudo \etal (POWLANG), Djordjevic \etal, Vitev \etal and Uphoff \etal (BAMPS); 
examples of the relative energy loss ($\Delta E/E$) and the momentum loss per unit length ($\dd P/\dd t$) for \cquark and \bquark quarks are shown. 
\item \sect{sec:sharmaVitev_dissociation} focuses on the calculation by Vitev \etal of in-medium formation and dissociation of heavy-flavour hadrons; 
this proposed mechanism is expected to effectively induce an additional momentum loss with respect to radiative and collisional heavy-quark in-medium interactions alone. 
\item \sects~\ref{sec:Tmatrix} and~\ref{sec:lQCDeloss} describe the calculation of transport coefficients 
through $T$-matrix approach supplemented with a non-perturbative potential extracted from lattice QCD (Rapp \etal, TAMU) or  
through direct \textit{ab initio} lattice-QCD calculations (Beraudo \etal, POWLANG); the transport coefficients that are discussed are  
the spatial diffusion coefficient (or friction coefficient), for which examples are shown, and the momentum diffusion coefficient.  
\item \sect{sec:AdSCFTmodel} presents the AdS/CFT 
approach for the calculation of the transport coefficients, developed by Horowitz \etal. 
\end{itemize} 
 
The implementation of these various approaches in full models that allow to compute the final heavy-flavour hadron  
kinematic distributions will be described in Section~\ref{sec:OHFmodelsMed}, with particular emphasis on  
the modelling of the QGP and its evolution. 

Given the focus of this review, we have chosen not to discuss the theoretical approaches that were not yet applied to LHC energies at the time of writing the document. For example, the modelling of heavy quark energy loss within the Dynamical Quasi-Particle Model (DQPM) approach in 
\cite{Berrehrah:2013mua, Berrehrah:2014kba}, recently integrated in the PHSD transport theory \cite{Song:2015sfa}, appears to be quite promising.

\subsubsection{pQCD energy loss in a dynamical QCD medium}
\label{sec:pQCDEloss}

\begin{figure} 
\begin{center} 
\includegraphics[clip,width=0.48\textwidth]{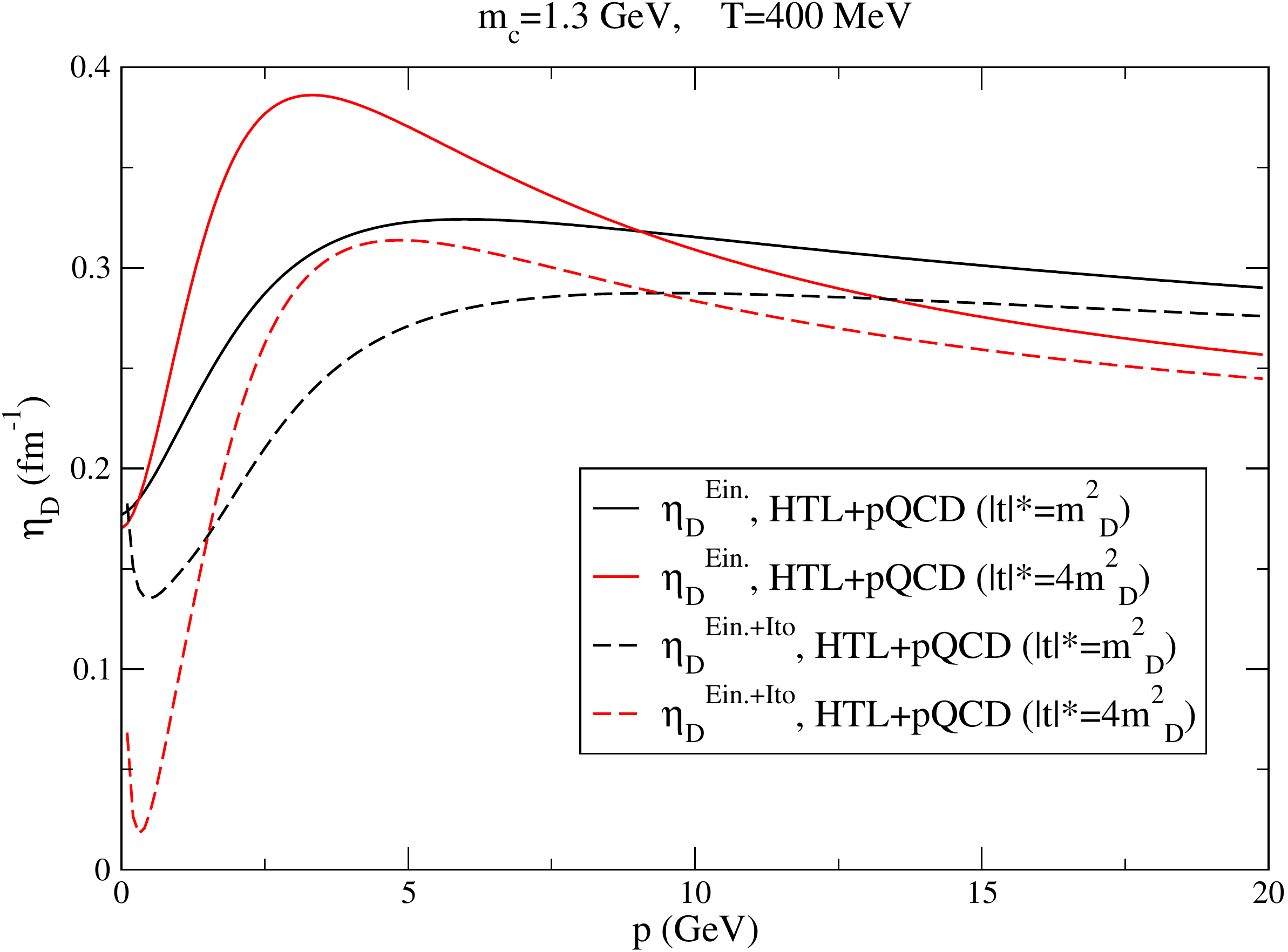} 
\includegraphics[clip,width=0.48\textwidth]{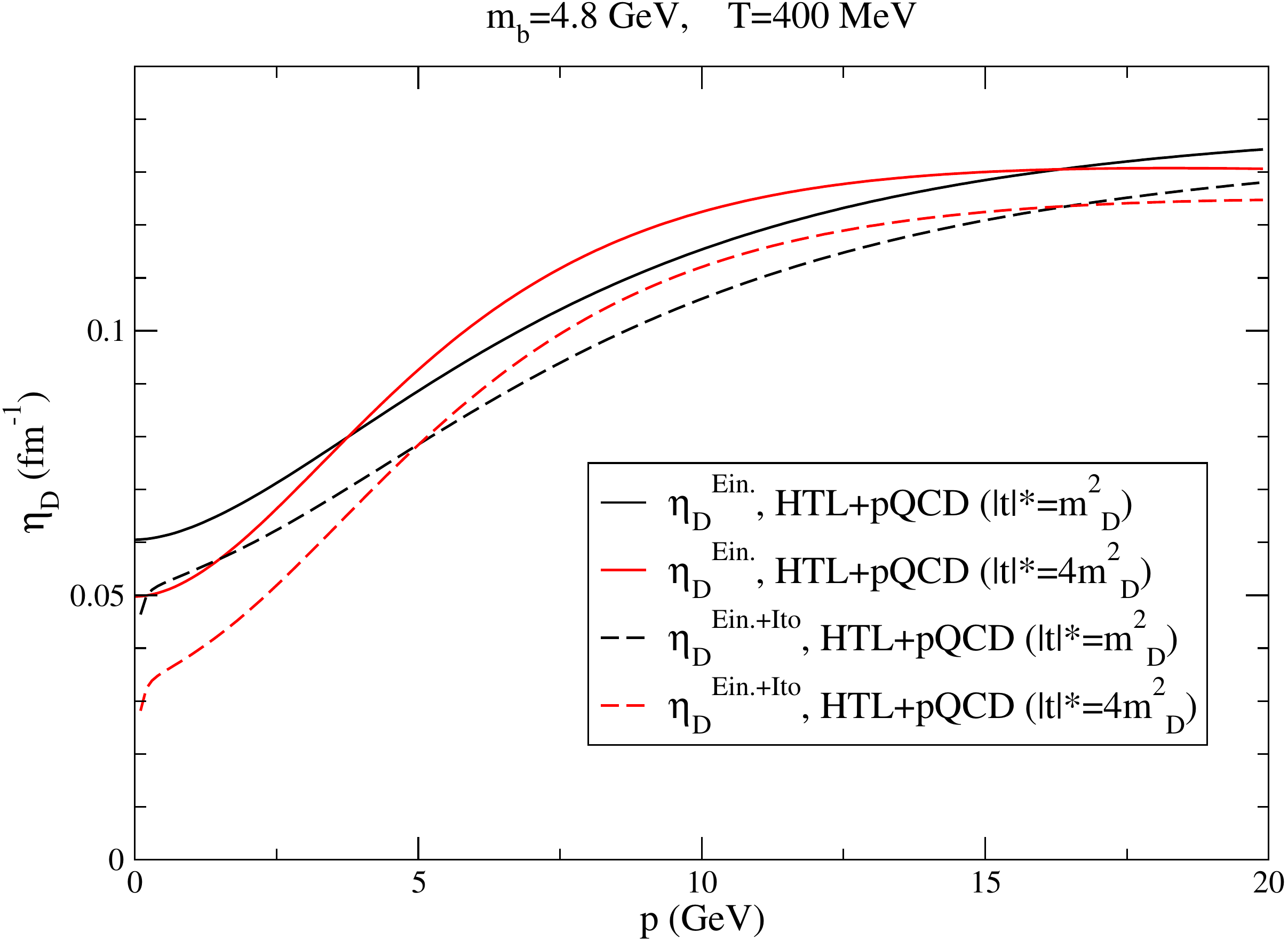} 
\caption{The charm (left panel) and beauty (right panel) friction coefficients in a Quark Gluon Plasma at a temperature $T = 400$\MeV. Continuous curves refer to the results of the Einstein relation $\eta_D=\kappa_{\rm L}/2ET$, while dashed curves include the discretisation correction in \eq{eq:friction} within the Ito scheme~\cite{Ito:1951}. The sensitivity of the results to the intermediate cut-off $|t|*$ separating hard and soft collisions is also displayed.} 
\label{fig:friction} 
\end{center} 
\end{figure}

Within a weak-coupling approach the interaction of heavy quarks with the medium can be described in terms of the uncorrelated scatterings with the light quarks and gluons of the hot deconfined plasma. Neglecting radiative processes one can then attempt an evaluation of the heavy-flavour transport coefficients arising from the $2\to 2$ elastic collisions suffered in the medium: this was the approach followed in \cis{Alberico:2013bza,Alberico:2011zy}, which we briefly summarise. The approach developed by the authors to simulate the propagation of the heavy quarks in the QGP was based on the relativistic Langevin equation (here written in its discretised form) 
\begin{equation} 
  \frac{\Delta\vec{p}}{\Delta t}=-\eta_D(p)\vec{p}+\vec{\xi}(t) 
 \quad{\rm with}\quad \langle\xi^i(t)\xi^j(t')\rangle=b^{ij}(\vec{p})\delta_{tt'}/\Delta t
\label{eq:lange_r_d} 
\end{equation}  
where   
\begin{equation}  
b^{ij}(\vec{p})= \kappa_{\rm L}(p)\hat{p}^i\hat{p}^j+\kappa_{\rm T}(p) 
(\delta^{ij}-\hat{p}^i\hat{p}^j). 
\end{equation} 
The right-hand side is given by the sum of a deterministic friction force and a stochastic noise term. The interaction with the background medium is encoded in the transport coefficients $\kappa_{T/L}$ describing the average squared transverse/longitudinal momentum per unit time exchanged with the plasma. In \cis{Alberico:2013bza,Alberico:2011zy} the latter were evaluated within a weak-coupling set-up, accounting for the collisions with the gluons and light quarks of the plasma. In particular, hard interactions were evaluated through kinetic pQCD calculation; soft collisions, involving the exchange of long-wavelength gluons, required the resummation of medium effects, the latter being provided by the Hard Thermal Loop (HTL) approximation. The friction coefficient $\eta_D(p)$ appearing in \eq{eq:lange_r_d} has to be fixed in order to ensure the approach to thermal equilibrium through the Einstein fluctuation-dissipation relation 
\begin{equation} 
\eta_D^{\rm (Ito)}(p)\equiv\frac{\kappa_{\rm L}(p)}{2TE_p}-\frac{1}{2p}\left[\partial_p\kappa_{\rm L}(p)+\frac{d-1}{2}(\kappa_{\rm L}(p)-\kappa_{\rm T}(p))\right].\label{eq:friction} 
\end{equation} 
In the above, the second term in the right-hand side is a correction (sub-leading by a $T/p$ factor) depending on the discretisation scheme employed in the numerical solution of the stochastic differential equation~(\eq{eq:lange_r_d}) in the case of momentum dependent transport coefficients (for more details see \ci{Beraudo:2009pe}): it ensures that, in the continuum $\Delta t\to 0$ limit, one recovers a Fokker-Planck equation with a proper Maxwell-J\"uttner equilibrium distribution as a stationary solution. Here we have written its expression in the so-called Ito scheme~\cite{Ito:1951}, which is the most convenient for a numerical implementation. Results for the friction coefficient $\eta_D(p)$ of \cquark and \bquark quarks are displayed in \fig{fig:friction}. 
 
The radiative processes, which are neglected in the model described above, are taken into account in other approaches. 
Djordjevic \etal developed a state-of-the-art dynamical energy loss formalism, which \textit{(i)} is applicable for both light and heavy partons, \textit{(ii)} computes both radiative~\cite{Djordjevic:2009cr,Djordjevic:2008iz} and collisional~\cite{Djordjevic:2006tw} energy loss in the \emph{same} theoretical framework, \textit{(iii)} takes into account recoil of the medium constituents, \ie the fact that medium partons are moving (\ie dynamical) particles,  \textit{(iv)} includes realistic finite size effects, \ie the fact that the partons are produced inside the medium and that the medium has finite size. Recently, the formalism was also extended to include \textit{(v)} finite magnetic mass effects~\cite{Djordjevic:2011dd}  and \textit{(vi)} running coupling (momentum dependence of $\alpha_s$)~\cite{Djordjevic:2013xoa}. 
 
In this formalism, radiative and collisional energy losses are calculated for an optically thin dilute QCD medium. Consequently, both collisional and radiative energy losses are computed to the leading order. That is, for collisional energy loss, the loss is calculated for one collisional interaction with the medium, while for radiative energy loss, the loss is calculated for one interaction with the medium accompanied by the emission of one gluon.  
The medium is described as a thermalised QGP~\cite{Kapusta:2006pm,bellac2000thermal} at temperature $T$ and zero baryon density, with $n_f$ effective  
massless quark flavours in equilibrium with the gluons. The Feynman diagrams contributing to the collisional and  the radiative quark energy loss are presented in \cis{Djordjevic:2006tw,Djordjevic:2009cr}. A  
full account of the calculation is presented in \ci{Djordjevic:2006tw} for collisional energy loss, and in~\ci{Djordjevic:2009cr} for radiative energy loss. Since the expression for collisional energy loss is lengthy, it will not be presented here, while the expression for radiative energy loss is given by  
\begin{equation} 
\frac{\Delta E_{\mathrm{dyn}}}{E}=\int \dd x \dd^2k\, x \frac{\dd^3N^g}{\dd x \dd^2k}   
\end{equation} 
with the radiation spectrum  
\begin{eqnarray} 
\frac{\dd^3N^g}{\dd x \dd^2k}   
= \frac{C_R \alpha_s}{\pi}\,\frac{L}{\lambda_\mathrm{dyn}}   
    \int \frac{\dd^2q}{x \pi^2} \, v_\mathrm{dyn}(\bq) 
\,\left(1-\frac{\sin{\left(\frac{(\bk{+}\bq)^2+\chi}{x E^+} \, L\right)}}  
    {\frac{(\bk{+}\bq)^2+\chi}{x E^+}\, L}\right)  
    \frac{2(\bk{+}\bq)}{(\bk{+}\bq)^2{+}\chi} 
    \left(\frac{(\bk{+}\bq)}{(\bk{+}\bq)^2{+}\chi} 
    - \frac{\bk}{\bk^2{+}\chi} 
    \right), 
    \label{DeltaEDyn} 
\end{eqnarray} 
where $\bq$ and $\bk$ are respectively the momentum of the radiated gluon and the momentum of the  
exchanged virtual gluon with a parton in the medium, with both $\bq$ and $\bk$ transverse to the jet direction. 
Here $\lambda_\mathrm{dyn}^{-1} \equiv C_2(G) \alpha_s T$  -- where $C_2(G)=3$ is the gluon quadratic Casimir invariant --  defines the ``dynamical mean free path''~\cite{Djordjevic:2008iz}, $\alpha_s$ is the strong coupling constant, and $C_R=4/3$ is the Casimir factor. Further, $v_\mathrm{dyn}(\bq)=\frac{\mu_E^2}{\bq^2 (\bq^2{+}\mu_E^2)}$ is the effective potential.  
$\chi$ is defined as $m_Q^2 x^2 + m_g^2$, where $m_Q$ is the heavy-quark mass, $x$ is the longitudinal momentum  
fraction of the heavy quark carried away by the emitted gluon and  
$m_g=\frac{\mu_E}{\sqrt{2}}$ is the effective mass for gluons with  
hard momenta $k\gtrsim T$ and $\mu_E$ is the Debye mass. It can be noted that the $C_R$ term encodes the colour charge dependence of energy loss (for radiative energy loss 
off a gluon $C_R$ is 3 instead of $4/3$). 
The $\chi$ term encodes the quark mass dependence of energy loss, which is reduced for increasing values of $m_Q/(\bk+\bq)$. 
 
Note that this dynamical energy loss presents an extension of the well-known static DGLV~\cite{Gyulassy:2000er,Djordjevic:2003zk} energy loss formalism to the dynamical QCD medium.  
The connection between dynamical and static energy losses was discussed in~\cis{Djordjevic:2009cr,Djordjevic:2008iz}. That is, static energy loss can be obtained from the above dynamical energy loss expression by replacing the dynamical mean-free path and effective potential  by equivalent expressions for a static QCD medium:  
$v_\mathrm{dyn}(\bq)\rightarrow v_\mathrm{stat}(\bq)=\frac{\mu_E^2}{(\bq^2{+}\mu_E^2)^2}$ 
and $\lambda_\mathrm{dyn}^{-1}\rightarrow \lambda_\mathrm{stat}^{-1}= 6 \frac{1.202}{\pi^2} \frac{1{+}\frac{n_f}{4}}{1{+}\frac{n_f}{6}} \lambda_\mathrm{dyn}^{-1}$. 
Note that the static DGLV formalism was also used in the WHDG model~\cite{Wicks:2005gt,Wicks:2007am}, 
as well as for the quark energy loss calculation by  Vitev \etal (see \sect{sec:sharmaVitev_dissociation}).

\begin{figure}[tb] 
\centering 
\includegraphics[width=0.48\textwidth]{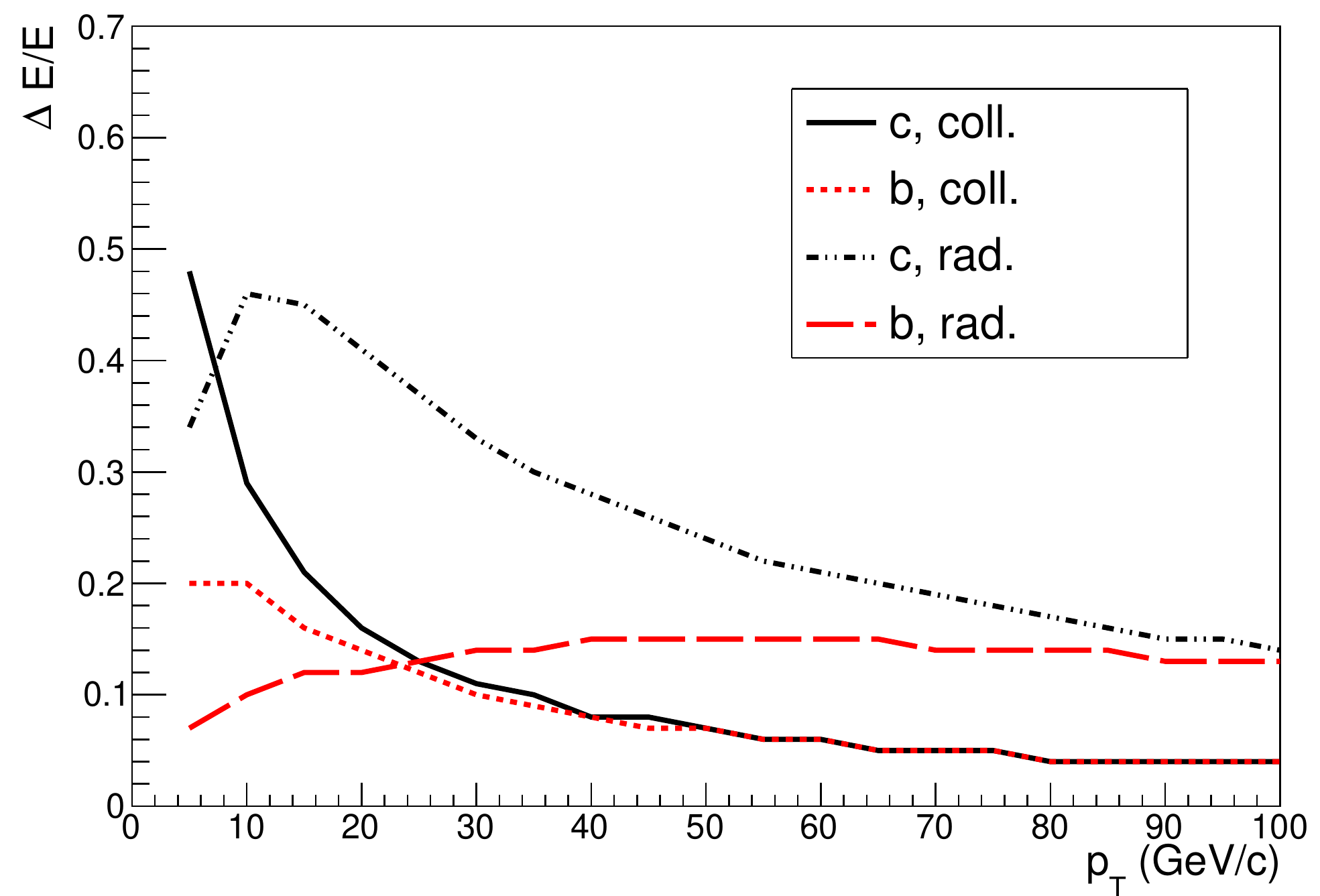} 
\caption{Fractional energy loss (\eq{DeltaEDyn}) evaluated for collisional and radiative processes and  
for charm and beauty quarks, at $T=304~{\rm MeV}$.} 
\label{fig:eloss_djordjevic} 
\end{figure}

The dynamical energy loss formalism was further extended to the case of finite magnetic mass, since various non-perturbative approaches suggest a non-zero magnetic mass at RHIC and LHC collision energies (see \eg~\cis{Maezawa:2010vj,Maezawa:2008kh,Nakamura:2003pu,Hart:2000ha,Bak:2007fk}). The finite magnetic mass is introduced through generalised sum-rules~\cite{Djordjevic:2011dd}.  
The main effect of the inclusion of finite magnetic mass turns out to be the modification of effective cross section $v_\mathrm{dyn}(\bq)$ in \eq{DeltaEDyn} to $v(\bq)=\frac{\mu_E^2-\mu_M^2}{(\bq^2+\mu_M^2) (\bq^2{+}\mu_E^2)}$, where $\mu_M$ is the magnetic mass. 
In \fig{fig:eloss_djordjevic}, the fractional energy loss $\frac{\Delta E}{E}$ corresponding to the  
full model described above is shown, for a path length $L=5~{\rm fm}$ and an effective constant temperature of $T=304\MeV$. 
For charm quarks, radiative energy loss starts to dominate for $\pt> 10$\GeVc, 
while this transition happens for $\pt>25$\GeVc for beauty quark. 
The comparison of radiative energy loss for the two quark species clearly illustrates the dead cone effect, as well as its disappearance when $\pt\gg m_Q$. 

In \cite{Huang:2013vaa}, a calculation for the $b$-jet production in \AAcoll was performed following  
very similar ingredients for the energy loss. The medium-induced gluon spectrum in the soft gluon approximation was evaluated as in~\cite{Vitev:2007ve} in a medium which incorporates Glauber geometry and Bjorken expansion. With the QGP-induced distribution of gluons $\frac{\dd^3 N^g}{\dd x \dd^2k}$ -- of the type of \eq{DeltaEDyn} -- and the related $\frac{\dd^2 N^g}{\dd \omega \dd r}$ ($\omega$ is the energy and $r$ is the angle) of gluons at hand, the fraction $f$ of the in-medium  parton  
shower energy that is contained in the jet cone of radius $R$ was evaluated as:   
\begin{equation} 
 f(R,\omega^{\rm coll})_{(s)}= \frac{\int_0^{R} \dd r 
\int_{\omega^{\rm coll}}^E  \dd\omega \, 
\frac{ \omega \dd^2N^{g}_{(s)}} {\dd\omega  \dd r }} 
{\int_0^{{R}^{\infty}} \dd r \int_{0}^E  \dd\omega \, 
\frac{ \omega \dd^2N^{g}_{(s)}}{\dd \omega \dd r} } \; . 
\label{fractionbjet} 
\end{equation} 
In \eq{fractionbjet} $f(R,0)_{(s)} $ takes into account  medium-induced  
parton splitting effects. On the other hand $f(R^\infty, \omega^{\rm coll})_{(s)}   
= \Delta E^{\rm coll} / E $ is the energy dissipated by the medium-induced  
parton shower into the QGP due to collisional processes.   
$\Delta E^{\rm coll}$ is evaluated as in Ref.~\cite{Neufeld:2011yh,Neufeld:2014yaa} and helps to  
solve for $\omega^{\rm coll}$. Then, for any $R$, \eq{fractionbjet} allows to treat the radiative and collisional energy loss effects on the same footing.   
Writing down explicitly the phase space Jacobian 
$|J(\epsilon)|_{(s)} = 1/\left(1 - [1-f(R,\omega^{\rm coll})_{(s)}]  \epsilon \right)$ for the case of $b$-jets the cross section 
per elementary nucleon-nucleon collision writes:  
\begin{equation} 
\frac{1}{\langle  N_{\rm bin}  \rangle} 
 \frac{\dd^2 \sigma_{AA}^{b-{\rm jet}}(R)}{\dd y \dd\pt} =\sum_{(s)} 
\int_{0}^1  \! \dd\epsilon \;   
\frac{ P_{(s)}(\epsilon) }{  \left(1 - [1-f(R,\omega^{\rm coll})_{(s)}]  \epsilon \right)  }  
\frac{\dd^2\sigma^{\rm CNM,LO+PS}_{(s)} \left(|J(\epsilon)|_{(s)} \pt \right)} {\dd y \dd\pt}. 
\label{incl} 
\end{equation} 
Here, the sum runs over the set of final states (s).  
$\dd^2\sigma^{\rm CNM,LO+PS} / {\dd y \dd\pt} $ includes cold nuclear matter effects.  
 
The same group recently published predictions for photon-tagged and B-meson-tagged $b$-jet production at \linebreak LHC~\cite{Huang:2015mva}. 

\subsubsection{A pQCD-inspired running \texorpdfstring{$\alpha_s$}{alpha\_s} energy loss model in \texorpdfstring{MC$@_s$HQ}{MC@sHQ} and BAMPS} 
\label{sec:mcatshq}
 
In the Monte Carlo at Heavy Quark approach~\cite{Gossiaux:2008jv,Gossiaux:2009mk,Nahrgang:2013saa} (MC$@_s$HQ),  
heavy quarks lose and gain energy by interacting with light partons from the medium (assumed to be in thermal equilibrium) according to rates which include both collisional and radiative types of processes. 
 
For the collisional energy loss, the elements of the transition matrix are calculated from the pQCD Born 
approximation~\cite{Combridge:1978kx,Svetitsky:1987gq}, supplemented by a running coupling constant $\alpha_s(Q^2)$ evaluated  
according to 1-loop renormalisation for $|Q^2|\gg \Lambda_{QCD}^2$ and chosen to saturate at small $Q^2$ in order to satisfy the universality relation~\cite{Dokshitzer:1995qm,Peshier:2008bg}: 
\begin{align} 
\label{alpha_s_continued} 
 \alpha_s(Q^2)= \frac{4\pi}{\beta_0} \begin{cases} 
  L_-^{-1}  & Q^2 < 0\\ 
  \frac12 - \pi^{-1} {\rm arctan}( L_+/\pi ) &  Q^2 > 0 
\end{cases} 
\end{align} 
with $\beta_0 = 11-\frac23\, n_f$  and $L_\pm = \ln(\pm Q^2/\Lambda^2)$ with $\Lambda=200$\MeV and $n_f = 3$.  
The $t$ channel requires infra-red regularisation which describes the physics of the screening at long  
distances~\cite{Weldon:1982aq}. For this purpose one adopts, in a first stage, a similar HTL polarisation as in the usual weak-coupling calculation of the energy loss~\cite{Braaten:1991jj,Braaten:1991we} for the small momentum-transfers, including the running $\alpha_s$ (\eq{alpha_s_continued}), while a semi-hard propagator is adopted for the large momentum-transfers. 
Then the model is simplified by resorting to an effective scalar propagator $\frac{1}{t-\kappa\tilde{m}_D^2(T)}$ 
for the exchanged thermal gluon, with a self-consistent Debye mass evaluated as $\tilde{m}_D^2(T)= \frac{N_c}{3} \left(1+\frac{n_f}{6}\right)4\pi\,\alpha_s(-\tilde{m}_D^2(T))\,T^2$~\cite{Peshier:2006hi} and  
an optimal value of $\kappa$ fixed to reproduce the value of the energy loss obtained at the first stage. 
The resulting model leads to a stronger coupling than previous calculations  
performed with fixed-order $\alpha_s=0.3$. It is also found to be compatible with the calculation of \ci{Peigne:2008nd} -- where the running of $\alpha_s$ is rigorously implemented -- in the region where the latter is applicable. 
 
A similar model is implemented in BAMPS~\cite{Uphoff:2010sh,Uphoff:2011ad,Uphoff:2013rka,Uphoff:2014hza}, although  
with some variations. In BAMPS the Debye mass $m_D^2$ is calculated dynamically from the non-equilibrium distribution functions $f$ of gluons and light quarks via~\cite{Xu:2004mz}   
$m_D^2 = \pi \alpha_s \nu_g \int \frac{\dd ^3p}{(2\pi)^3} \frac{1}{p}  
( N_c f_g + n_f f_q) 
$, 
where $N_c=3$ denotes the number of colours and $\nu_g = 16$ is the gluon degeneracy. While 
MC@$_s$HQ applies the equilibrium Debye mass with quantum statistics for temperatures extracted from the fluid dynamic background, BAMPS treats all particles as Boltzmann particles, due to the non-equilibrium nature of the cascade. Moreover, in BAMPS the scale of the running coupling in the Debye mass is evaluated at the momentum transfer of the process, \eg $\alpha_s(t)$.  
The differences in the treatment lead to a larger energy loss of about a factor of two in MC@$_s$HQ compared to BAMPS. 
 
As for the radiative energy loss, the model mostly concentrates on the case of intermediate energy for which coherence effects do not play the leading role. Exact momentum conservation and scattering on dynamical partons have however to be implemented exactly. In the MC@$_s$HQ approach~\cite{Gossiaux:2010yx,Aichelin:2013mra}, the calculations of \ci{Gunion:1981qs} are 
thus extended for incoherent radiation off a single massless parton to the case of massive quarks. For the central ``plateau'' of radiation, one obtains that the cross section 
$\dd \sigma(Qq\rightarrow Qqg)$ is dominated by a gauge-invariant subclass of diagrams. It can be factorised as the product of the elastic cross section $\dd \sigma(Qq\rightarrow Qq)$ and a factor $P_g$ representing the conditional probability of radiation 
per elastic collision, which is collinear-safe thanks to the heavy-quark mass $m_Q$. Moreover, it was shown  
in~\ci{Aichelin:2013mra} that a fair agreement with the exact power spectra can be achieved by considering the eikonal limit in $P_g$ and preserving the phase-space condition. The ensuing relation reads $\dd \sigma(Qq\rightarrow Qqg)=\dd \sigma(Qq\rightarrow Qq) P_g^{\rm eik}$, 
with   
\begin{equation} 
P_g^{\rm eik}(x,\vec{k}_t,\vec{l}_t)=\frac{3\alpha_s}{\pi^2}\frac{1-x}{x} 
  \bigg(\frac{\vec{k}_t}{k_t^2+x^2m_Q^2} 
  -\\ 
  \frac{\vec{k}_t-\vec{l}_t}{(\vec{k}_t-\vec{l}_t)^2 +x^2 m_Q^2}\bigg)^2\,, 
  \label{eq:OHF_gossiaux_pg} 
  \end{equation} 
where $x$ is the fraction of 4-momentum carried by the radiated gluon, $\vec{k}_t$ its transverse momentum and $\vec{l}_t$ the 
momentum exchanged with the light parton. For the radiation in a medium at finite temperature, the radiated gluon acquires 
a thermal mass, which leads to a modification $x^2 m_Q^2 \rightarrow x^2 m_Q^2 +(1-x) m_g^2$ in \eq{eq:OHF_gossiaux_pg}. 
As a consequence, the power spectra are vastly reduced. In MC@$_s$HQ, an explicit realisation of  
the elastic process is achieved first, and the radiation factor $P_g$ is then sampled along the variables  
$x$ and $\vec{k}_t$. In \ci{Gossiaux:2012cv}, the implementation of radiative processes was generalised to include the coherent radiation, through an interpolation between single and multiple scatterings 
matched to the BDMPS result \cite{Baier:1996kr}. However it neglects the finite path length effects which are important for thin plasmas.  Hereafter, this will be referred to as ``LPM-radiative''. For further description of the model, the reader is referred to~\ci{Nahrgang:2013saa}. 
 
Similar considerations apply for radiative energy loss in BAMPS~\cite{Abir:2011jb,Fochler:2013epa}. Due to the semi-classical transport nature of BAMPS, the LPM effect is included effectively by comparing the formation time of the emitted gluon to the mean free path of the jet \cite{Fochler:2010wn}. Furthermore, the emitted gluon is treated as a massless particle. 
 
 \begin{figure}[tb] 
\includegraphics[width=0.48\textwidth]{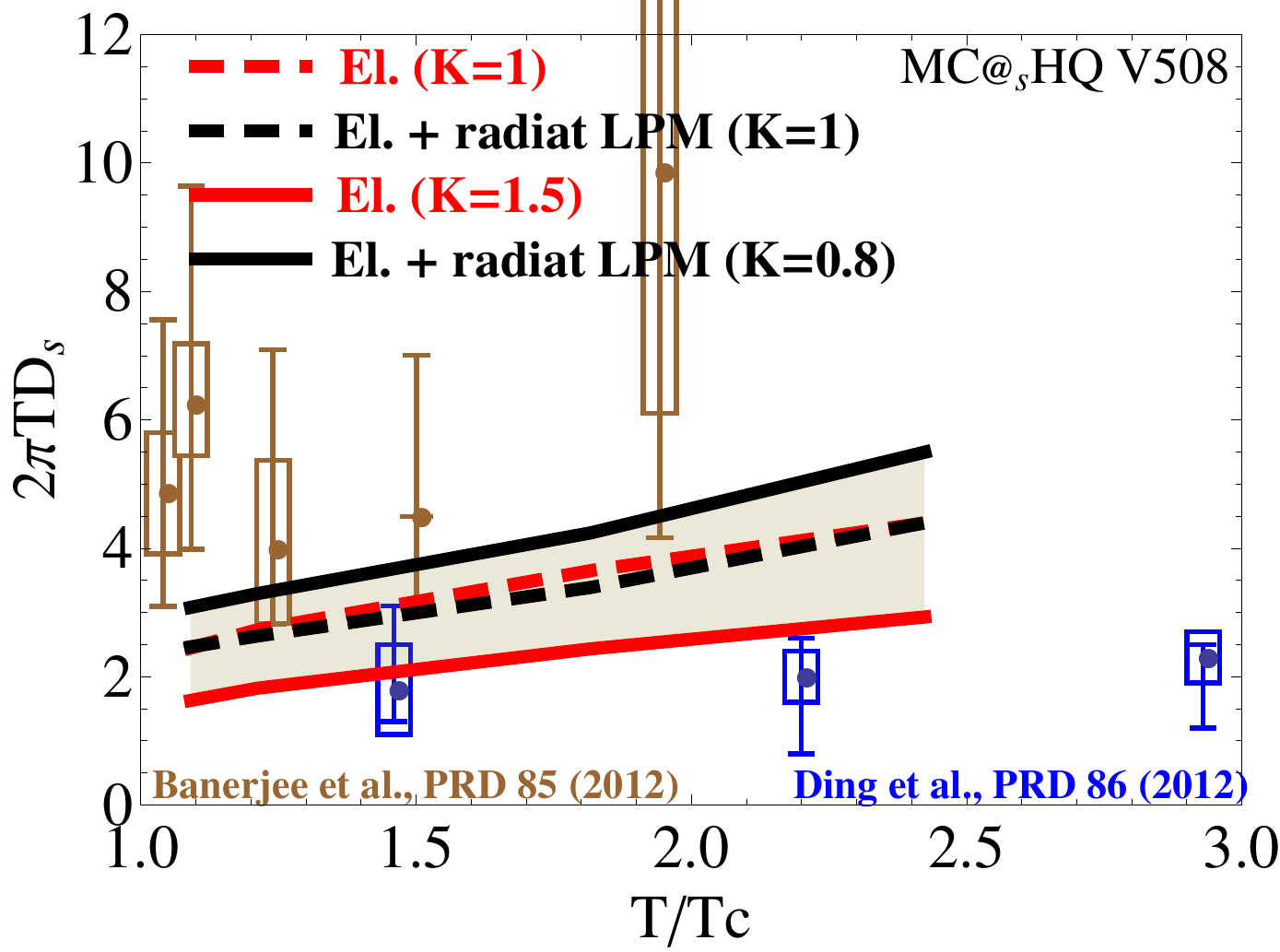} 
\includegraphics[width=0.48\textwidth]{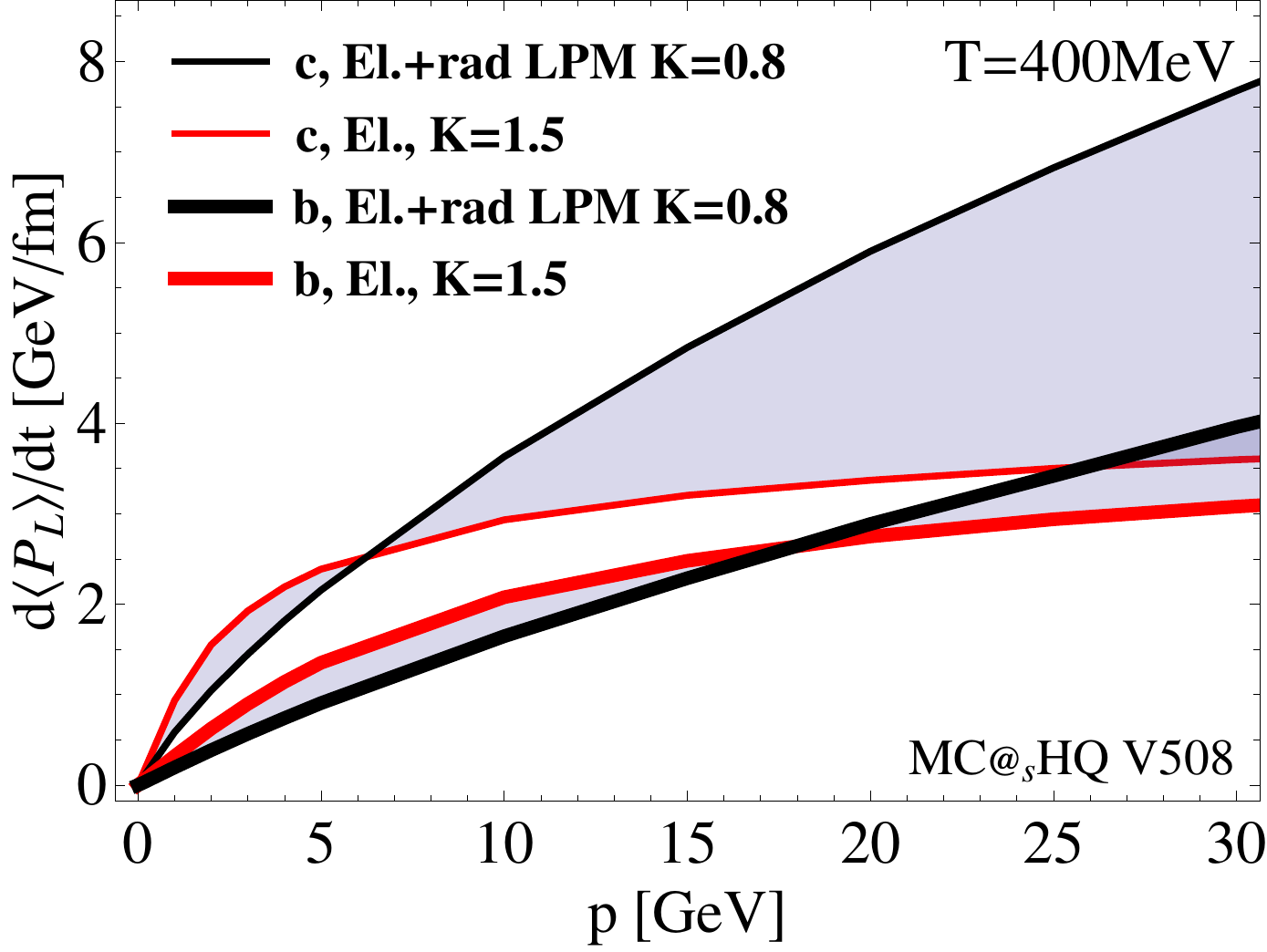} 
 \caption{Macroscopic properties for both elastic and elastic plus LPM-radiative model. On the left panel, the diffusion  
 coefficient $2\pi T D_s$ is plotted vs $T/T_c$ and compared to the l-QCD calculations of~\cite{Ding:2012sp,Banerjee:2011ra}. On the right panel, 
 the average momentum loss per unit of time is plotted vs heavy-quark momentum both for \cquark and \bquark quarks.} 
 \label{fig:gossiaux_properties} 
 \end{figure} 
\fig{fig:gossiaux_properties} illustrates two properties of this energy loss model as implemented in  
MC$@s$HQ model. Both the pure elastic case as well 
as a combination of the elastic and LPM-radiative energy loss are considered. In both cases, the model is calibrated by applying a multiplicative  
$K$-factor to the interaction cross sections, in order to describe the \raa of D mesons for intermediate \pt range in \pb collisions  
at $\snn=2.76$\TeV in the 0--20\% centrality class.  This leads to $K_{\rm el}=1.5$ and  
$K_{\rm el+LPM-rad}=0.8$, while one obtains $K_{\rm el}=1.8$ and  
$K_{\rm el+LPM-rad}=1$ following a similar procedure at RHIC. For the spatial diffusion coefficient $D_s$, one sees that both combinations are compatible with the l-QCD 
calculations of~\cis{Ding:2012sp,Banerjee:2011ra} and thus provide some systematic ``error band'' of the approach. The  
corresponding average momentum loss per unit of time (or length), shown on the right panel of \fig{fig:gossiaux_properties}, illustrates 
the mass-hierarchy, found to be stronger for the radiative component (black lines in the figure). 

\subsubsection{Collisional dissociation of heavy mesons and quarkonia in the QGP}
\label{sec:sharmaVitev_dissociation}
 
Heavy flavour dynamics in dense QCD matter critically depends on the time-scales  
involved in the underlying reaction. Two of these time-scales, the formation time  
of the QGP  $\tau_0$ and its transverse size $L_{QGP}$,   
can be related to the nuclear geometry, the QGP expansion,  and the bulk particle  
properties.  The formation time  $\tau_{\rm form}$ of heavy mesons and quarkonia, on the other hand,  
can be evaluated from the virtuality of the  heavy  quark $Q$  decay  into  
D, B mesons~\cite{Adil:2006ra,Sharma:2009hn} or the time for the \QQbar pair to  
expand to the size of the \jpsi or   
$\Upsilon $ wave function~\cite{Sharma:2012dy}.  For a $\pi^0$ with an energy of 10\GeV,  
$\tau_{\rm form} \sim 25$~fm $\gg L_{QGP}$ affords a relatively simple interpretation 
of light hadron  quenching in terms of radiative and collisional parton-level energy loss~\cite{Kang:2014xsa}.  
On the other hand, for D, B, \jpsi and $\Upsilon(1{\rm S})$,  
one obtains $ \tau_{\rm form} \sim$~1.6, 0.4, 3.3 and 1.4~fm $\ll L_{QGP}$. 
Such short formation times necessitate understanding of heavy meson and quarkonium  
propagation and dissociation in strongly interacting matter. 
 
The Gulassy-Levai-Vitev (GLV) reaction operator formalism 
was developed for calculating the interactions of parton systems 
as  they pass through a dense  strongly-interacting medium.  It was  generalised to the dissociation  
of mesons (quark-antiquark binaries), as long as the momentum exchanges from the medium  
$\mu = g T$  can resolve their internal structure.  The dissociation probability  and dissociation time 
\begin{equation} 
P_d(\pt,m_Q,t) = 1 -   \left| \int \dd^{2}{\vec \Delta k} \dd x \, 
\psi_{f}^* (\Delta {\vec k},x)\psi_{0}(\Delta {\vec k}, x) \right|^{2}  \;,  \qquad  \frac{1}{\langle \tau_{\rm diss}(\pt, t) \rangle} =  
\frac{\partial}{\partial t} \ln  P_d(\pt,m_Q,t) \; ,  
\end{equation} 
can be obtained from the overlap between the  medium-broadened time-evolved and  vacuum  initial 
meson wave functions, $\psi_f$ and  $\psi_0$, respectively. Here, $\psi_f$ has the resummed collisional  
interactions in the QGP. Let us denote by 
\begin{equation} 
\label{convent1} 
f^{Q}({p}_{T},t)= \frac{\dd\sigma^Q(t)}{\dd y \dd^2\pt} \;,   
 \;   f^{Q}({p}_{T},t=0) =  
\frac{\dd \sigma^Q_{PQCD}}{\dd y \dd^2\pt} \;, \qquad      f^{H}({p}_{T},t)= \frac{\dd\sigma^H(t)}{\dd y\dd^2\pt} \;, 
 \;   f^{H}({p}_{T},t=0) = 0 \; ,    
\end{equation} 
the double differential production cross sections for the heavy quarks and 
hadrons.  Initial conditions are also specified  above, in particular the heavy quark distribution is given by  
the perturbative QCD \cquark and \bquark quark cross section. Energy loss in the partonic state can be  
implemented as quenched initial conditions~\cite{Sharma:2009hn,Sharma:2012dy}.  
 Including the loss and  
gain terms one obtains: 
\begin{eqnarray} 
\label{rateq1} 
\partial_t f^{Q}({p}_{\rm T},t)&=&  
-  \frac{1}{\langle \tau_{\rm form}(\pt, t) \rangle} f^{Q}({p}_{\rm T},t)  
 + \, \frac{1}{\langle  
\tau_{\rm diss}(\pt/\bar{x}, t) \rangle} 
\int_0^1 \dd x \,  \frac{1}{x^2} \phi_{Q/H}(x)   
f^{H}({p}_{\rm T}/x,t) \;, \qquad \\ 
\partial_t f^{H}({p}_{\rm T},t)&=&  
-  \frac{1}{\langle \tau_{\rm diss}(\pt, t) \rangle} f^{H}({p}_{\rm T},t)  
 +\, \frac{1}{\langle  
\tau_{\rm form}(\pt/\bar{z}, t) \rangle} 
\int_0^1 \dd z \,  \frac{1}{z^2} D_{H/Q}(z)  
 f^{Q}({p}_{\rm T}/z,t) \;. \qquad  
\label{rateq2} 
\end{eqnarray} 
In \eqs~(\ref{rateq1}) and (\ref{rateq2})   $\phi_{Q/H}(x)$  and  $D_{H/Q}(z)$  are the distribution function of heavy quarks in  
a heavy meson and the fragmentation function of a heavy quark into a heavy mesons, respectively, and  $\bar{z}$ and $\bar{x}$ are 
typical fragmentation and dissociation momentum  fractions. It was checked that   
in the absence of a medium, $\tau_{\rm diss}(\pt, t) \rightarrow \infty$, 
so that the pQCD spectrum of heavy hadrons from vacuum jet fragmentation are recovered. Details for the rate equation relevant to quarkonium formation and dissociation are given in~\ci{Sharma:2012dy}.  
 Solving the above equations in the limit $t \rightarrow \infty$ in the absence and presence of a medium 
allows to evaluate the nuclear modification factor  for heavy-flavour  mesons.

\subsubsection{\texorpdfstring{$T$}{T}-Matrix approach to heavy-quark interactions in the QGP}
\label{sec:Tmatrix}

The thermodynamic $T$-matrix approach is a first-principles framework to  
self-consistently compute one- and two-body correlations in hot and dense matter.  
It has been widely applied to, \eg, electromagnetic plasmas~\cite{Redmer:1997} 
and the nuclear many-body problem~\cite{Jeukenne:1976uy,Brockmann:1990cn}. Its  
main assumption is that the basic two-body interaction can be cast into the form  
of a potential, $V(t)$, with the 4-momentum transfer approximated as   
$t=q^2=q_0^2-\vec q^{\,2}\simeq -\vec q^{\,2}$. This relation is satisfied for charm and  
beauty quarks ($Q = c,b$) in a QGP up to temperatures of  
2-3\,$T_{\rm c}$, since their large masses imply  
$q_0^2\simeq (\vec q^{\,2}/2m_Q)^2\ll \vec q^{\,2}$ with typical momentum transfers  
of $\vec q^{\, 2} \sim T^2$. Therefore, the $T$-matrix formalism is a promising framework  
to treat the non-perturbative physics of heavy-quark (HQ) interactions in the  
QGP~\cite{Mannarelli:2005pz,Cabrera:2006wh,Riek:2010fk}.  
It can be applied to both hidden and open heavy-flavour states, and provides 
a comprehensive treatment of bound and scattering states~\cite{Riek:2010fk}. It  
can be systematically constrained by lattice data~\cite{Riek:2010py}, and  
implemented to calculate heavy-flavour observables in heavy-ion  
collisions~\cite{vanHees:2007me,He:2011qa}.       
  
The potential approximation allows to reduce the 4-dimensional Bethe-Salpeter  
into a 3-dimensional Lipp\-mann-Schwinger equation, schematically given by  
\begin{equation} 
T(s,t) = V(t) + \int \dd^3k  \ V(t') \ G_2(s,k) \ T(s,t'') \   ,  
\end{equation} 
where $G_2$ denotes the in-medium 2-particle propagator. Using the well-known 
Cornell potential in vacuum~\cite{Eichten:1978tg}, heavy quarkonium spectroscopy and heavy-light meson 
masses can be reproduced, while relativistic corrections (magnetic interactions) 
allow to recover perturbative results in the high-energy limit for HQ  
scattering~\cite{Riek:2010fk}.  
 
\begin{figure}[t] 
\includegraphics[width=0.5\columnwidth]{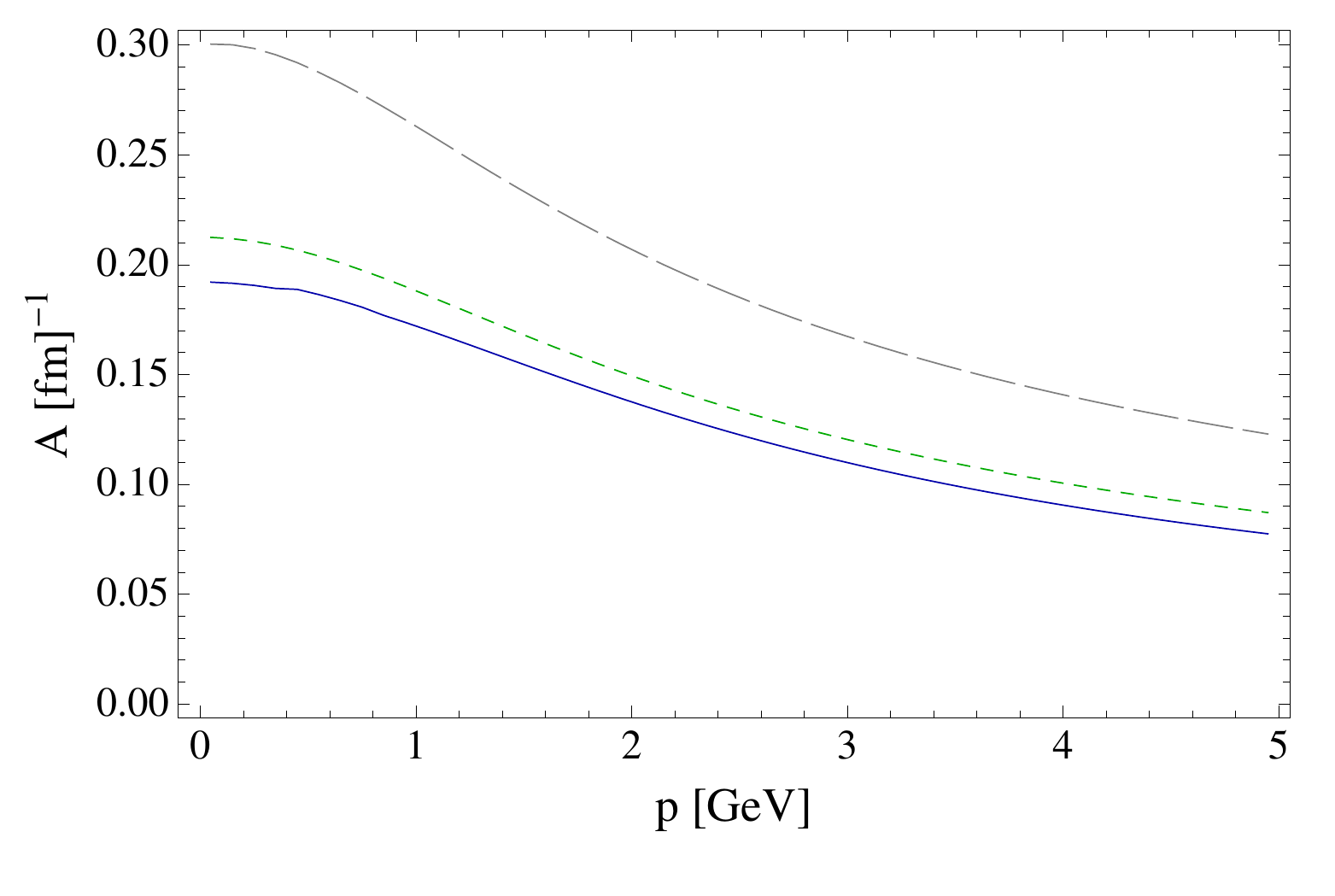} 
\includegraphics[width=0.5\columnwidth]{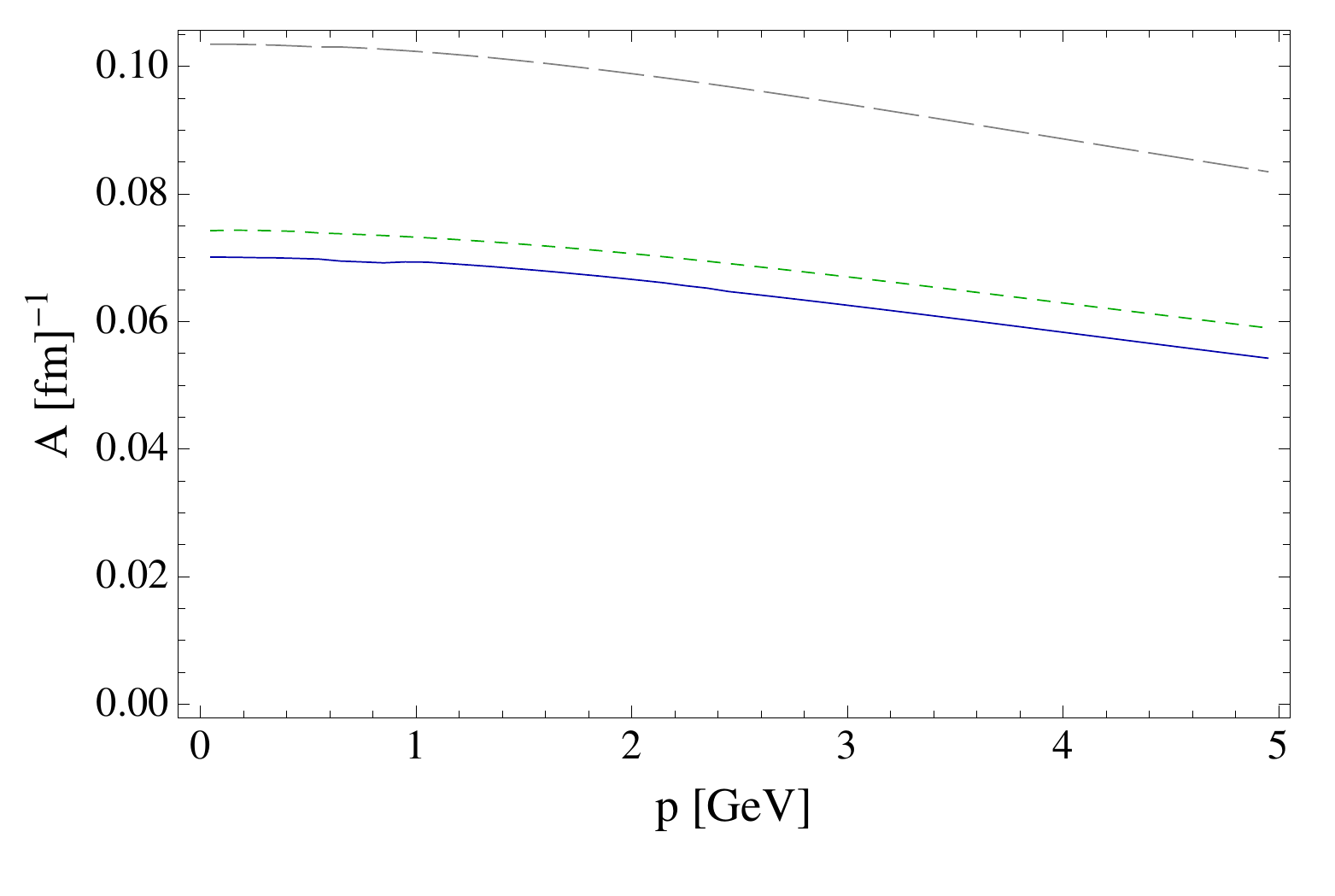} 
\caption{Left: Charm-quark friction coefficient, $A_c$, as a function of momentum 
in the QGP from non-pertubative $T$-matrix scattering amplitudes off thermal light and strange  
quarks~\cite{Riek:2010fk}, as well as gluons~\cite{Huggins:2012dj}; the curves correspond to  
temperatures $T$=1.2, 1.5 and 2\,$T_{\rm c}$ (bottom to top);  
Right: Same as left panel but for bottom quarks. 
Figures are taken from~\ci{Huggins:2012dj}. 
} 
\label{fig_A} 
\end{figure}

\begin{figure} 
\begin{center} 
\includegraphics[width=0.7\columnwidth]{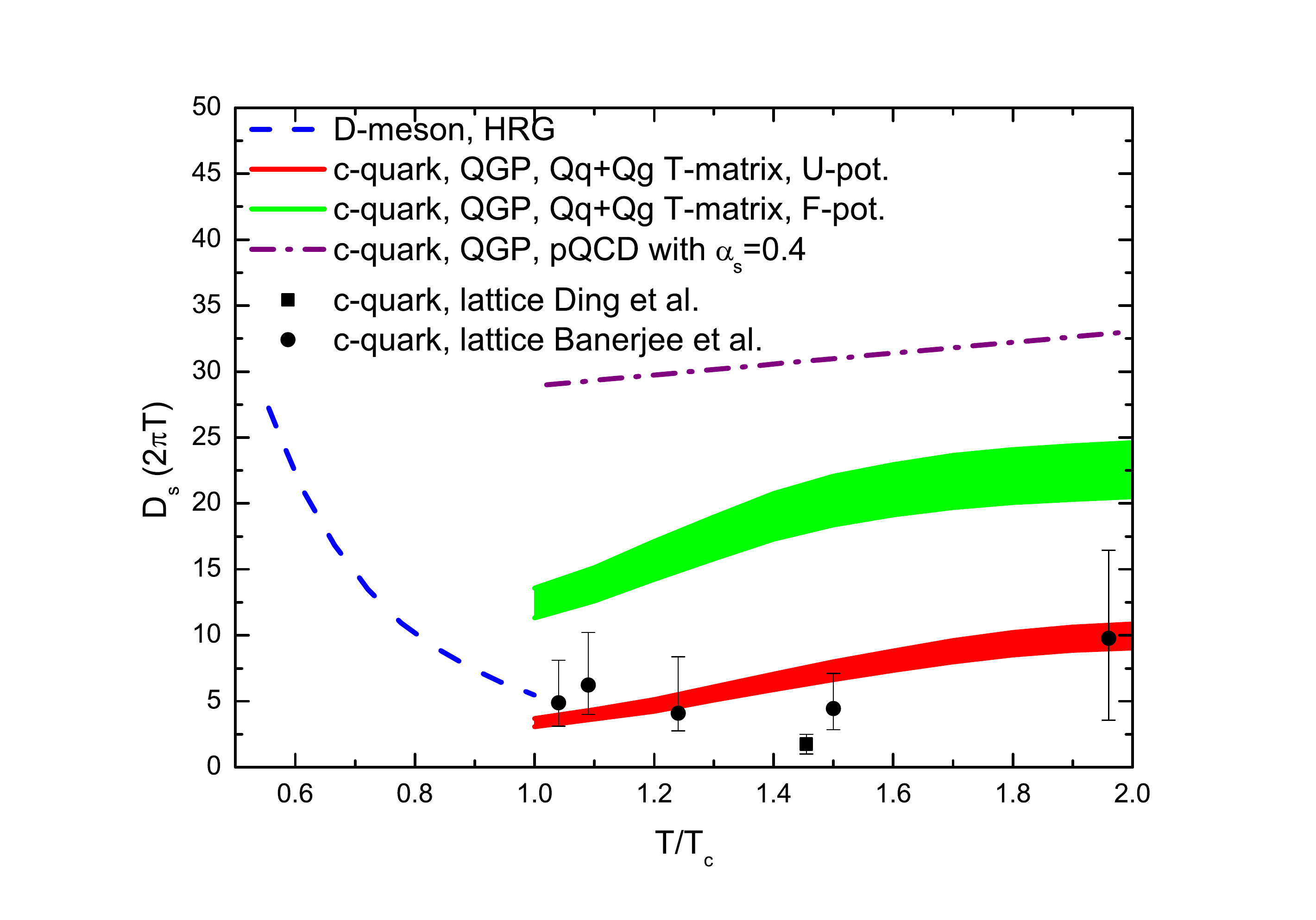} 
\end{center} 
\vspace{-0.5cm} 
\caption{ 
Spatial heavy-flavour diffusion coefficient (defined via the 
relaxation rate at zero momentum) for the $T$-matrix approach in the QGP using the $U$ (lower 
red band) or $F$ potential (upper green band)~\cite{Riek:2010fk}, or pQCD with $\alpha_s$=0.4 
(dash-dotted line), in hadronic matter (dashed line)~\cite{He:2011yi}, and from quenched 
lattice QCD~\cite{Ding:2011hr,Banerjee:2011ra} (data points); figure taken from~\cite{He:2012df}.} 
\label{fig_Ds} 
\end{figure}

The pertinent transport coefficients for a heavy quark of momentum $\vec p$ are  
given by   
\begin{equation} 
A_Q(p)  =  \frac{1}{2\omega_Q(p)(2\pi)^9} \sum\limits_{j=q,\bar{q},g}  
\int \frac{\dd^3k}{2\omega_k}  \frac{\dd^3k'}{2\omega_{k'}}  
\frac{\dd^3p'}{2\omega_Q(p')}  f^j(\omega_k) \ \delta^{(4)}(P_i-P_f) \  
|{\cal M}_{Qj}(s,t)|^2 \left( 1-\frac{\vec p \cdot \vec p\,'}{\vec p^{\,2}} \right) 
\end{equation}  
for the friction coefficient (or relaxation rate) and analogous expressions for momentum diffusion~\cite{Rapp:2009my}.  
The invariant HQ-parton scattering amplitude, ${\cal M}_{Qj}$, is directly proportional to  
the $T$-matrix. An important ingredient is how the HQ potential $V$ is modified in medium.  
This is currently an open question. As limiting cases the HQ free ($F$) and internal ($U$)  
energies computed in lattice-QCD (l-QCD) have been employed~\cite{Kaczmarek:2007pb}. The  
internal energy produces a markedly stronger interaction, and, when employed in the  
$T$-matrix approach, generally leads to better agreement with other quantities computed on 
the lattice (\eg, quarkonium correlators, HQ susceptibility, etc.~\cite{Riek:2010py}). The  
resulting \cquark-quark relaxation rates, including scattering off thermal \uquark, \dquark, \squark quarks  
and gluons, are enhanced over their perturbatively calculated counterparts by up  
to a factor of $\sim$6 at low momenta and temperatures close to $T_{\rm c}$, cf.~left panel  
of \fig{fig_A}. A similar enhancement is found for \bquark quarks, although the  
absolute magnitude of the relaxation rate is smaller than for \cquark quarks by about a  
factor of $m_b/m_c\simeq3$, cf.~right panel of \fig{fig_A}.  
The non-perturbative enhancement is mostly caused by resonant D/B-meson and  
di-quark states which emerge in the colour-singlet and anti-triplet channels as $T_{\rm c}$ is  
approached from above. These states naturally provide for HQ coalescence processes  
in the hadronisation transition, \ie, the same interaction that drives non-perturbative  
diffusion also induces hadron formation. The resummations in the $T$-matrix, together with  
the confining interaction in the potential, play a critical role in this framework. At  
high momenta, both confining and resummation effects become much weaker and the diffusion  
coefficients approach the perturbative (colour-Coulomb) results, although at $p\simeq5$~GeV,  
the enhancement is still about a factor of 2 for charm quarks. With increasing temperature, the  
colour screening of the l-QCD-based interaction potentials leads to an increase in the  
(temperature-scaled) spatial diffusion coefficient, $D_s(2\pi T) = 2\pi T^2/(m_Q\,A_Q)$,  
see \fig{fig_Ds}. 
 
After coalescence into open-charm mesons, the approach also accounts for the diffusion of  
heavy-flavour mesons in the hadronic phase. Pertinent transport coefficients have been worked  
out in~\cite{He:2011yi}, based on effective D-meson scattering amplitudes off light  
hadrons as available from the literature. These include $\pi$, K, $\eta$, $\rho$, $\omega$,  
as well as nucleons and $\Delta$(1232) and their anti-particles. The combined effect of  
these scatterings is appreciable, leading to a hadronic diffusion coefficient comparable  
to the $T$-matrix calculations in the QGP close to $T_{\rm c}$. As first pointed out in~\cite{He:2011yi,He:2012df}, this suggests a minimum of the ($T$-scaled) heavy-flavour  
diffusion coefficient via a smooth transition through the pseudo-critical region, as to  
be expected for a cross-over transition  (see \fig{fig_Ds}). 
 
\subsubsection{Lattice-QCD}
\label{sec:lQCDeloss}
  
First principle non-perturbative results for the transport coefficients can be obtained, although within a limited kinematic domain and with sizeable systematic uncertainties, from lattice QCD (l-QCD) calculations. The theoretical set-up employed to extract the momentum diffusion coefficient $\kappa$~\cite{CasalderreySolana:2006rq} is described in the following. This approach is valid in the non-relativistic limit (for this calculation heavy quarks on the lattice are taken as static colour sources) where the transport of heavy quarks reduces to the Langevin equation 
\begin{equation} 
\frac{\dd p^i}{\dd t}=-\eta_D p^i+ \xi^i(t), 
\end{equation} 
where $\eta_D$ and $\kappa$ are the friction and diffusion coefficients and where $\boldsymbol \xi$ are stochastic forces 
auto-correlated according to $\langle\xi^i(t)\xi^j(t')\rangle\!=\!\delta^{ij}\delta(t-t')\kappa$. 
Hence, in the $p\!\to\!0$ limit, $\kappa$ is given by the Fourier transform of the following force-force correlator  
\begin{equation} 
\label{eq:ber1} 
\kappa=\frac{1}{3}\int_{-\infty}^{+\infty}\!\!\!\dd t\langle\xi^i(t)\xi^i(0)\rangle_{\rm HQ} 
\approx\frac{1}{3}\int_{-\infty}^{+\infty}\!\!\!\dd t{\langle F^i(t)F^i(0)\rangle_{\rm HQ}}\equiv\frac{1}{3}\int_{-\infty}^{+\infty}\!\!\!\dd t D^>(t)=\frac{1}{3}D^>(\omega\!=\!0), 
\end{equation} 
where the expectation value is taken over an ensemble of states containing, besides thermal light partons, a static ($m_Q\!=\!\infty$) heavy quark. 
In a thermal bath, correlators are related by the Kubo-Martin-Schwinger condition, entailing for the spectral function the relation ${\sigma(\omega)}\!\equiv\!D^>(\omega)\!-\!D^<(\omega)={(1-e^{-\beta\omega})D^>(\omega)}$, so that 
\begin{equation} 
{\kappa}\equiv\lim_{\omega\to 0}\frac{{D^>(\omega)}}{3}=\lim_{\omega\to 0}\frac{1}{3}{\frac{\sigma(\omega)}{1-e^{-\beta\omega}}}\underset{\omega\to 0}{\sim}\frac{1}{3}\frac{T}{{\omega}}{\sigma(\omega)}.\label{eq:kappa} 
\end{equation} 
In the static limit magnetic effects vanish and the force felt by the heavy quark can only be due to the chromo-electric field 
\begin{equation} 
{\vec F}(t)=g\int \!\dd{\vec x} \,Q^\dagger(t,{\vec x})t^aQ(t,{\vec x})\,{\vec E}^a(t,{\vec x}). 
\end{equation} 
\eq{eq:kappa} shows how $\kappa$ depends on the low-frequency behaviour of the spectral density $\sigma(\omega)$ of the electric-field correlator in the presence of a heavy quark 
 in the original thermal average in \eq{eq:ber1}. The latter can be evaluated on the lattice for imaginary times $t\!=\!-i\tau$~\cite{CaronHuot:2009uh}:  
\begin{equation} 
{D_E(\tau)}=-\frac{\langle{\rm Re\,Tr}[U(\beta,\tau)gE^i(\tau,{\bf 0})U(\tau,0)gE^i(0,{\bf 0})]\rangle}{\langle{\rm Re\,Tr}[U(\beta,0)]\rangle}. 
\end{equation} 
In the above equation the expectation value is now taken over a thermal ensemble of states of gluons and light quarks, with the Wilson lines $U(\tau_1,\tau_2)$ reflecting the presence of a static heavy quark. As it is always the case when attempting to get information on real-time quantities from l-QCD simulations, the major difficulty consists in reconstructing the spectral density $\sigma(\omega)$ from the inversion of 
\begin{equation} 
{D_E(\tau)}=\int_0^{+\infty}\frac{\dd\omega}{2\pi}\frac{\cosh(\tau-\beta/2)}{\sinh(\beta\omega/2)}{\sigma(\omega)}, 
\end{equation} 
where one knows the correlator $D_E(\tau)$ for a limited set of times $\tau_i\!\in\!(0,\beta)$. Lattice results for the heavy quark diffusion coefficient are currently available for the case of a pure $SU(3)$ gluon plasma~\cite{Banerjee:2012zj,Francis:2011gc}. In transport calculations, depending on the temperature, one relied on the values $\kappa/T^3\!\equiv\!\overline{\kappa}\!\approx\!2.5-4$ obtained in \ci{Francis:2011gc}, which the authors are currently trying to extrapolate to the continuum (\ie zero lattice-spacing) limit. 
 
Being derived in the static $m_Q\!=\!\infty$ limit and lacking any information on their possible momentum dependence, the above results for $\kappa$ have to be taken with some grain of salt when facing the present experimental data (mostly referring to charm at not so small \pt); however they could represent a really solid theoretical benchmark when beauty measurements, for which $M\!\gg\!T$, at low $\pt$ will become available. Bearing in mind the above caveats and neglecting any possible momentum dependence of $\kappa$, the above l-QCD transport coefficients (the friction coefficient $\eta_D=\kappa/2ET$ being fixed by the Einstein relation) were implemented into POWLANG code~\cite{Alberico:2013bza} in order to provide predictions for D mesons, heavy-flavour electrons and \jpsi from B decays. One can estimate what the above results would entail for the average heavy-quark energy-loss:  
\begin{equation} 
\langle{\dd E}/{\dd x}\rangle=\langle{\dd p}/{\dd t}\rangle=-\eta_D\,p=-(\kappa/2ET)\,p=-(\overline{\kappa}T^2/2)\,v. 
\end{equation}    
with $v$, the heavy-quark velocity. Numerically, this would imply a stopping power $\langle -\dd E/\dd x\rangle\approx\overline{\kappa}\cdot 0.4\cdot v$\GeV/fm at $T=400$\MeV and  $\langle -\dd E/\dd x\rangle\approx\overline{\kappa}\cdot 0.1\cdot v$\GeV/fm at $T=200$\MeV. 

\subsubsection{Heavy-flavour interaction with medium in AdS/CFT}
\label{sec:AdSCFTmodel}

The anti-de-Sitter/conformal field theory (AdS/CFT) correspondence~\cite{CasalderreySolana:2011us,DeWolfe:2013cua} is a conjectured dual between field theories in $n$ dimensions and string theories in $n+1$ dimensions (times some compact manifold).  The correspondence is most well understood between $\mathcal{N} = 4$ super Yang-Mills (SYM) and Type IIB string theory; these two theories are generally considered exact duals of one another.  The calculational advantage provided by the conjecture is that there is generally speaking an inverse relationship between the strength of the coupling in the dual theories: when the field theory is weakly-coupled the string theory is strongly-coupled, and vice-versa.  The advantage for QCD physics accessible at current colliders is clear: the temperatures reached at RHIC and LHC are at most a few times $\Lambda_{\mathrm{QCD}}$; it is therefore reasonable to expect that much of the dynamics in these collisions is dominated by QCD physics that is strongly-coupled and hence theoretically accessible only via the lattice, which is then generally restricted to imaginary time correlators, or via the methods of AdS/CFT.  The leading order contribution to string theory calculations (corresponding to a very strong coupling limit in the field theory) comes from classical gravity; one uses the usual tools of Einsteinian General Relativity.  Although much research is focused on finding dual string theories ever closer to QCD, no one has yet found an exact dual; nevertheless, one hopes to gain non-trivial insight into QCD physics by investigating the relevant physics from known AdS/CFT duals.  An obvious limitation of the use of AdS/CFT is that it is difficult to quantify the corrections one expects when going from the dual field theory in which a derivation is performed to actual QCD. 
 
\begin{figure} 
	\centering 
    \includegraphics[width=2.15in]{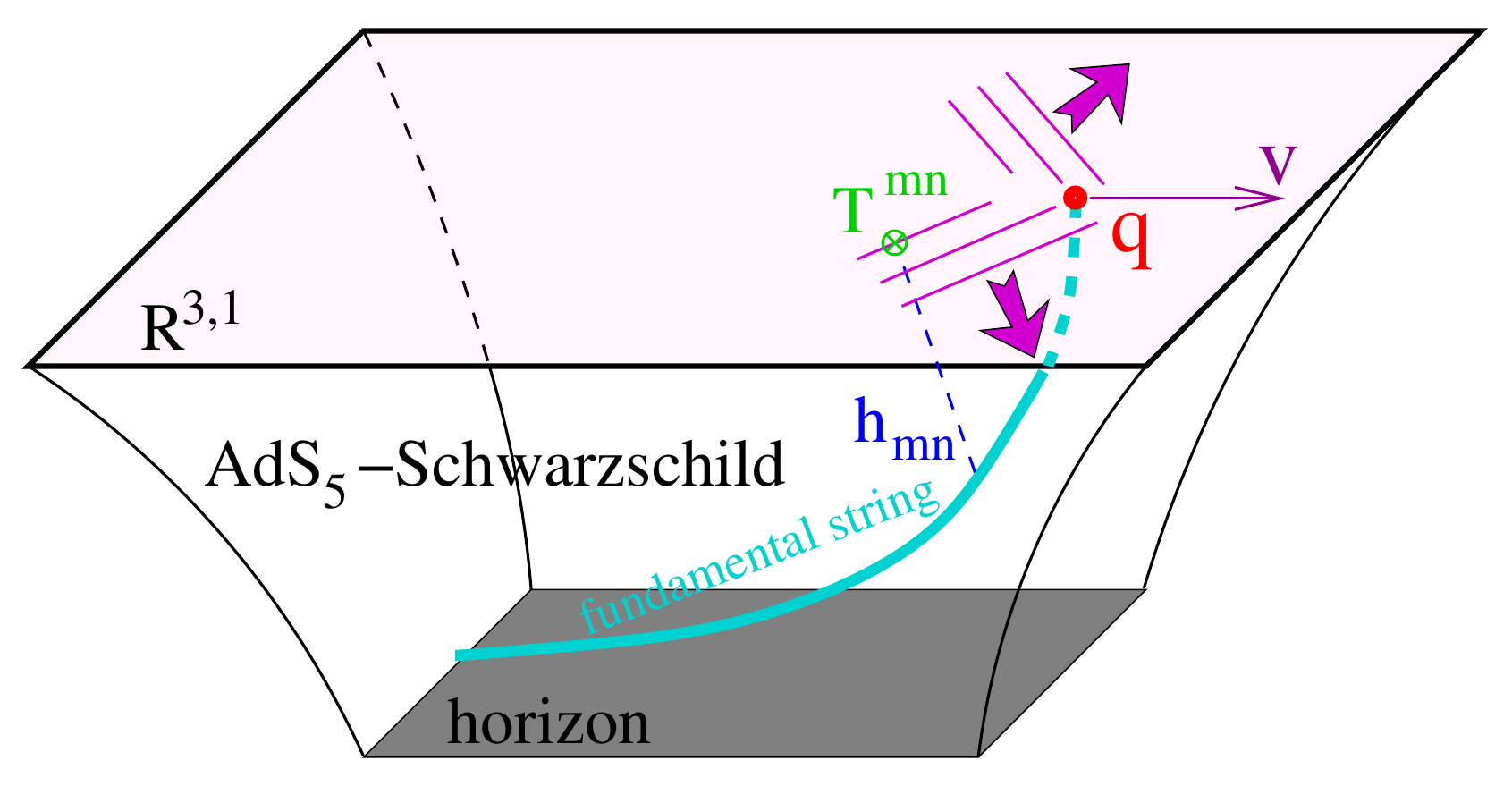} 
\caption{Schema for the heavy quark drag calculation~\cite{DeWolfe:2013cua}.} 
\label{horowitz:f:setup} 
\end{figure}

The main thrust of open heavy flavour suppression research that uses the AdS/CFT correspondence assumes that all couplings are strong, regardless of scale (calculations for light quarks with all couplings assumed strong and calculations for which some couplings are strong and some are weak have also been performed; see~\cite{CasalderreySolana:2011us,DeWolfe:2013cua} and references therein for a review).  For reasons soon to be seen, the result is known as ``heavy quark drag''; 
%
see~\fig{horowitz:f:setup} 
for a picture of the set-up.  The heavy quark is modelled as a string with one endpoint near (or, for an infinitely massive quark, at) the boundary of the AdS space; the string hangs down in the fifth dimension of the space-time towards a black hole horizon (the Hawking temperature of the black hole is equal to the temperature of the Yang-Mills plasma).  As the string endpoint near the boundary moves, momentum flows down the string; this momentum is lost to the thermal plasma.  For a heavy quark moving with constant velocity $v$ in $\mathcal{N} = 4$ SYM, one finds~\cite{Herzog:2006gh,Gubser:2006bz} 
\begin{equation} 
	\label{horowitz:e:adsdrag} 
	\frac{\dd p}{\dd t} = -\frac{\pi\sqrt{\lambda}}{2}T_{SYM}^2\frac{v}{\sqrt{1-v^2}} \;\; \Longrightarrow \;\; \frac{\dd p}{\dd t} = -\mu_Q\,p, 
\end{equation} 
where $\mu_Q = \pi\sqrt{\lambda}T_{SYM}^2/2m_Q$, and the energy loss reduces to a simple drag relationship in the limit of a very heavy quark, for which corrections to the usual dispersion relation $p/m_Q \, = \, v/\sqrt{1-v^2}$ are small.   
 
As the string is dragged, an induced black hole horizon forms in the induced metric on the worldsheet of the string.  This horizon emits Hawking radiation that is dual in the field theory to the influence of the thermal fluctuations of the plasma on the motion of the heavy quark.  The diffusion coefficients have been derived in the large mass, constant motion limit~\cite{Gubser:2006nz,Son:2009vu} as 
\begin{equation} 
	\label{horowitz:e:adsdiffusion} 
	\kappa_{\rm T} = \pi\sqrt{\lambda}\gamma^{1/2}T_{SYM}^3,\;\;\kappa_{\rm L} = \pi\sqrt{\lambda}\gamma^{5/2}T_{SYM}^3. 
\end{equation} 
Note that for $v\,\ne\,0$ the above diffusion coefficients deviate from the usual ones found via the fluctuation-dissipation theorem, which, for the $\mu_Q$ of \eq{horowitz:e:adsdrag}, yields $\kappa_{\rm T}\,=\,\kappa_{\rm L}\,=\,\pi\sqrt{\lambda}\gamma \, T_{SYM}$.  The entire set-up breaks down at a characteristic ``speed limit,'' $\gamma_{crit}^{SL} = (1 + 2 \, m_Q / \sqrt{\lambda}\,T_{SYM})^2 \, \approx 4\,m_Q^2 / \lambda \, T^2$, which corresponds to the velocity at which the induced horizon on the worldsheet moves above the string endpoint (equivalently, if an electric field maintains the constant velocity of the heavy quark, at the critical velocity the field strength is so large that it begins to pair-produce heavy quarks)~\cite{Gubser:2006nz}.  Above the critical velocity, it no longer makes sense to treat the heavy quark as heavy, and one must resort to light flavour energy loss methods in AdS/CFT.  
 
 
 
 
\subsection{Theoretical overview: medium modelling and medium-induced modification of heavy-flavour production}
\label{sec:OHFmodelsMed}

Besides modelling the energy loss as described in \sect{sec:OHFmodelsInt}, 
each model aiming at explaining open heavy flavour observables in \AAcoll collisions needs to include 
several key ingredients. These are: the ``initial'' production of heavy flavour (see \sect{sec:pp:Theory:OpenHF}) possibly affected by cold nuclear matter effects (see \sect{Cold nuclear matter effects}), a space-time description of the QGP evolution up to 
the freeze out, mechanisms for hadronisation (including specific processes like the so-called ``coalescence'') 
and, ultimately, D meson and B meson interactions in the ensuing hadronic matter. 
For a given energy loss model, it has been shown that various choices of these auxiliary ingredients 
could generate a factor of 2 in the observables~\cite{Gossiaux:2011ea}. In this section, the solutions adopted 
in the various models are described in order to better understand their predictions for the modification of heavy-flavour production in \AAcoll.  

\subsubsection{pQCD energy loss in a static fireball (Djordjevic \etal)}
 
The dynamical energy loss formalism discussed in section \ref{sec:pQCDEloss} (Djordjevic \etal) was  
incorporated by the same authors into a numerical procedure in order to calculate medium-modified heavy-flavour hadron momentum distributions. 
This procedure includes \textit{(i)} production of light and heavy-flavour partons based on 
the non-zero-mass variable flavour number scheme VFNS~\cite{Vitev:2006bi}, 
 NLO~\cite{Sharma:2009hn} 
 and FONLL calculations~\cite{Cacciari:2012ny}, respectively, \textit{(ii)} multi-gluon~\cite{Gyulassy:2001nm} and path-length~\cite{Wicks:2005gt,Dainese:2003wq} fluctuations, \textit{(iii)} light~\cite{deFlorian:2007aj} and heavy~\cite{Cacciari:2003zu,Braaten:1994bz,Kartvelishvili:1977pi} flavour fragmentation functions, and \textit{(iv)} decay of heavy-flavour mesons to single electrons/muons and \jpsi~\cite{Cacciari:2012ny}. This model does not include hadronisation via recombination. 
In-medium path length is sampled from a distribution corresponding to a static fireball at fixed effective temperature.  
The \raa predictions are provided for both RHIC and LHC energies, various light and heavy-flavour probes and different collision centralities. This model does not include any free parameter. 
All implementation details are provided in~\ci{Djordjevic:2014tka}. A representative set of these predictions will be presented in \sect{sec:modelcomparison}, while other predictions and detailed comparison with experimental data are provided in~\ci{Djordjevic:2013xoa,Djordjevic:2013pba,Djordjevic:2014tka}.  
In summary, this formalism provides a robust agreement with experimental data, across diverse probes, experiments and experimental conditions. 
 

\subsubsection{pQCD embedded in viscous hydro (POWLANG and Duke)}
 
The starting point of the POWLANG set-up~\cite{Alberico:2013bza} is the generation of the \QQbar pairs in elementary nucleon-nucleon collisions. For this purpose the POWHEG-BOX package~\cite{Frixione:2007nw,Alioli:2010xd} is employed: the latter deals with the initial production in the hard pQCD event (evaluated at NLO accuracy), interfacing it to PYTHIA 6.4~\cite{Sjostrand:2006za} to simulate the associated initial and final-state radiation and other effects, like \eg the intrinsic-\kt broadening. In the \AAcoll case, EPS09 nuclear corrections~\cite{Eskola:2009uj} are applied to the PDFs and the \QQbar pairs are distributed in the transverse plane according to the local density of binary collisions given by the geometric Glauber model. Furthermore, a further \kt broadening depending on the crossed nuclear matter thickness is introduced, as described in~\ci{Alberico:2011zy}. Both in the \pp benchmark and in the \AAcoll case (at the decoupling from the fireball) hadronisation is modelled through independent fragmentation of the heavy quarks, with in-vacuum fragmentation functions tuned by the FONLL authors~\cite{Cacciari:2005rk}. 
Concerning the modelling of the fireball evolution, the latter is taken from the output of the (2+1)d viscous fluid dynamics code of~\ci{Romatschke:2007mq}. At each time step, the update of the heavy-quark momentum according to the Langevin equation is performed in the local rest frame of the fluid, boosting then back the results into the laboratory frame. 
In setting the transport coefficients entering into the Langevin equation, the approach adopted in~\ci{Alberico:2013bza} was to derive the momentum broadening $\kappa_{\rm T/L}(p)$ and to fix the friction coefficient $\eta_D(p)$ so to satisfy the Einstein relation. For the former, two different sets of values were explored: the ones from a weak-coupling calculations described in \sect{sec:pQCDEloss} 
and the ones provided by the lattice QCD calculations described in \sect{sec:lQCDeloss}. 
The local character of these energy-loss models indeed allows their implementation with fluid dynamics as a background. 
 
In first phenomenological studies performed with the POWLANG set-up~\cite{Alberico:2013bza,Alberico:2011zy} hadronisation was modelled as occurring in the vacuum, neglecting the possibility of recombination of the heavy quarks with light thermal partons from the medium. 
Hence no modification of the heavy flavour spectra or hadrochemistry at hadronisation was considered, charm and beauty going into hadrons with the same fragmentation fractions as in the vacuum. 
A medium-modified hadronisation scheme has been recently developed in \cis{Beraudo:2014iva,Beraudo:2014boa}. 
Note that the recombination with light thermal quarks would make the final charmed hadrons inherit part of the flow of the medium, moving present POWLANG results closer to the experimental data. 
First numerical results~\cite{Beraudo:2014iva,Beraudo:2014boa} show that this is actually the case, in particular for what concerns the elliptic flow of D mesons at LHC and  
their \raa at low \pt at RHIC. 
 
 
In the Duke model~\cite{Cao:2013ita}, the Langevin approach was generalized by Cao, Qin and Bass in order to include the contribution of radiative energy loss, thus offering a complementary perspective both with  
respect to the approach of Djordjevic (where static medium is considered) and to POWLANG (where no 
radiative energy loss is implemented). The generalized Langevin equation reads  
\begin{equation}  
\frac{\dd\vec{p}}{\dd t}=-\eta_D(p) \vec{p} +\vec{\xi} + \vec{f}_g 
\end{equation} 
where $\vec{f}_g$ is the semi-classical recoil force exerted on heavy quark due to medium induced gluon radiation. In a (Ito)-discretised scheme, the associated recoil momentum  
$\Delta \vec{p}_g$ is obtain, at each time step 
$\Delta t$, by sampling the radiated gluon radiation spectrum 
$\frac{\dd N_g}{\dd x \dd k_\perp^2 \dd t}$, with an absolute probability of 
radiation $P_{\rm rad}=\int_t^{t+ \Delta t} \dd t 
\int \dd x \dd k_\perp^2 \frac{\dd N_g}{\dd x \dd k_\perp^2 \dd t}$.  
In~\cite{Cao:2013ita}, the ``usual" stochastic forces $\vec{\xi}$ associated 
to the collisional processes are chosen to be autocorrelated in time  
according to $\langle \xi^i(t) \xi^j(t')\rangle =\kappa \delta^{i j} 
\delta(t-t')$, with a spatial diffusion coefficient $D_s=\frac{2 T^2}{\kappa}$ set to $\frac{6}{2\pi T}$, while the gluon radiation spectrum is computed with the pQCD higher-twist approach \cite{Zhang:2003wk}. 
 
The space-time evolution of the temperature and collective flow profiles of the thermalised medium 
are described with a (2+1)d viscous fluid dynamics~\cite{Song:2007fn,Song:2007ux,Qiu:2011hf}. At the  
end of the QGP phase, the hadronisation of heavy quarks is modelled with a hybrid fragmentation plus recombination scenario. Fragmentation  
processes are simulated by PYTHIA~6.4~\cite{Sjostrand:2006za} while the heavy quark coalescence with light quarks is treated with the ``sudden recombination" approach developed  
in~\cite{Oh:2009zj}.

\label{sec:pQCD_hydro}

\subsubsection{pQCD-inspired energy loss with running \texorpdfstring{$\alpha_s$}{alpha\_s} in a fluid-dynamical medium and in 
Boltzmann transport}
 
The implementation of the microscopic models based on a running coupling constant -- described in \sect{sec:mcatshq} -- in the MC@$_s$HQ and BAMPS frameworks is presented here. In its latest version, MC@$_s$HQ couples a Boltzmann transport of heavy quarks to the (3+1)d ideal fluid-dynamical evolution from EPOS2 initial conditions. In its integral version, which includes the hadronic final state interactions, the EPOS2 model describes very well a large variety of observables measured in the light-flavour sector in nucleus--nucleus, proton--nucleus and proton--proton collisions at RHIC and LHC~\cite{Werner:2010aa,Werner:2010ny,Werner:2012xh,Werner:2013ipa}.  Therefore, it provides as a reliable description of the medium from which the thermal scattering partners of the heavy quarks are sampled. 
Due to the fluctuating initial conditions of the fluid dynamics, the heavy-quark evolution can be treated in an event-by-event set-up. 
Initially, the heavy quarks are produced at the spatial scattering points of the incoming nucleons with a momentum distribution from either FONLL~\cite{Cacciari:1998it,Cacciari:2001td,Cacciari:2012ny} or MC@NLO~\cite{Frixione:2003ei,Frixione:2002ik}. The latter combines next-to-leading order pQCD cross sections with a parton shower evolution, which provides more realistic distributions for the initial correlations of heavy quark-antiquark pairs than the back-to-back initialisation applied to single inclusive spectra obtained with FONLL. In recent implementations~\cite{Nahrgang:2014ila}, a convolution of the initial \pt spectrum was applied in order to include (anti-)shadowing at (high) low \pt in central collisions at the LHC according to the EPS09 nuclear shadowing  
effect~\cite{Eskola:2009uj}. After propagation in the deconfined  
medium heavy quarks hadronise at a transition temperature of $T=155$\MeV, which is well in the range of transition temperatures given by lattice QCD calculations \cite{Borsanyi:2010bp}. As described in~\cite{Gossiaux:2009mk}, hadronisation of heavy quarks into D and B mesons can proceed through coalescence (predominant at low \pt) or fragmentation (predominant at large \pt). 
Recently, MC@$_s$HQ+EPOS2 has also been used to study heavy-flavour correlation observables~\cite{Nahrgang:2013saa} (see \sect{sec:OHFcorr}) and higher-order flow coefficients~\cite{Nahrgang:2014vza}. 
 
 
In the BAMPS model~\cite{Xu:2004mz,Xu:2007aa}, the initial heavy quark distribution is obtained from MC@NLO~\cite{Frixione:2002ik,Frixione:2003ei} for \pp collisions through scaling with the number of binary collisions to heavy-ion collisions without taking cold nuclear matter effects into account. After the QGP evolution (that is, after the local energy density has fallen below $\epsilon = 0.6$\GeV/fm$^3$) heavy quarks are fragmented via Peterson fragmentation~\cite{Peterson:1982ak} to D and B mesons. Recombination processes are not considered for the hadronisation. 
 
RHIC heavy-flavour decay electron data can be reproduced with only collisional interactions if their cross section is increased by a $K$-factor of 3.5~\cite{Uphoff:2012gb}. 
 With this parameter fixed, BAMPS predictions~\cite{Uphoff:2012gb} for \vtwo at LHC for various heavy-flavour particles can describe the data, but the \raa is slightly underestimated.  
However, the need of the phenomenological $K$-factor is rather unsatisfying from the theory perspective,  
especially if $K$ is found to deviate vastly from unity. 
Therefore, radiative processes were recently included in BAMPS~\cite{Uphoff:2014hza} and the $K$-factor mocking higher order effects abandoned\footnote{Strictly speaking, the radiative processes include some phenomenological parameter named ``$X$" accounting for the 
LPM effect and calibrated on the $\pi$ production in central \pb collisions.}. The ensuing predictions are in satisfactory agreement with the data, which seems to favour this recent development of the BAMPS model. 


\subsubsection{Non-perturbative \texorpdfstring{$T$}{T}-matrix approach in a fluid-dynamic model (TAMU) and in UrQMD transport}
 
The $T$-matrix approach for heavy-flavour diffusion through QGP, hadronisation  
and hadronic matter~\cite{Riek:2010fk,He:2011yi} described in \sect{sec:Tmatrix}  
has been  
implemented into a fluid-dynamic background medium~\cite{He:2011qa}.   
The latter is based on the 2+1 dimensional ideal fluid dynamics code of Kolb  
and Heinz~\cite{Kolb:2000sd}, but several amendments have been implemented to 
allow for an improved description of bulk-hadron observables 
at RHIC and LHC~\cite{He:2011zx}. First, the quasi-particle QGP equation of 
state (EoS) with first-order transition has been replaced by a lattice-QCD  
EoS which allows for a near-smooth matching into the hadron-resonance gas. 
Second, the initialisation at the thermalisation time has been augmented to  
account for a non-trivial flow field, in particular a significant radial  
flow~\cite{Kolb:2002ve}. Third, the initial energy-density profile has been  
chosen in more compact form, close to a collision profile that turns out to  
resemble initial states from saturation models. 
All three amendments generate a more violent transverse expansion of the medium, which, \eg, have been identified as important ingredients to solve the discrepancy between the fluid dynamics predictions and the measured HBT radii at RHIC (the so-called HBT puzzle~\cite{Pratt:2008qv}). 
These features furthermore lead to an ``early" saturation of the bulk-medium \vtwo~\cite{He:2011zx}, close to the  
phase transition region. Consequently, multi-strange hadrons ($\phi$, $\Xi$  
and $\Omega^-$) need to freeze-out at this point to properly describe their 
\pt spectra and \vtwo. This provides a natural explanation of the  
phenomenologically well established universal kinetic-energy scaling of  
hadron \vtwo at RHIC. For the medium evolution at LHC, an initial radial  
flow is phenomenologically less compelling, and has not been included in  
the tune of fluid dynamics for \pb collisions at $\snn=2.76$\TeV, while the thermalisation time  
($\tau_0=0.4$~fm/$c$) is assumed to be shorter than at RHIC  
(0.6~fm/$c$).  Representative bulk-hadron observables at LHC are  
reasonably well described as a function of both \pt and  
centrality~\cite{He:2014cla}. 
 
 
Heavy-flavour diffusion is implemented into the fluid-dynamical medium employing relativistic  
Langevin simulations of the Fokker-Planck equation. The pertinent non-perturbative  
transport coefficients from the heavy-light $T$-matrix in the QGP and effective  
hadronic theory for D mesons in the hadronic phase are utilised in the local  
rest frame of the expanding medium. The initial heavy-quark  
momentum distributions are taken from FONLL pQCD calculations~\cite{Cacciari:2001td}, which  
describe \pp spectra with suitable fragmentation functions. After diffusion 
through the QGP the HQ distributions are converted into D/D$^*$ mesons using  
the resonance recombination model (RRM)~\cite{Ravagli:2007xx} with \pt-dependent  
formation probabilities from the heavy-light $T$-matrices in the colour-singlet  
channels. The hadronisation is carried out on the hyper-surface corresponding to  
$T_{\rm pc}=170$\MeV.  
The heavy quarks  
that do not recombine are hadronised via the same fragmentation function as used  
in \pp collisions. The resulting D meson distributions are further evolved  
through the hadronic phase until kinetic freeze-out of the fluid-dynamical medium.  
However, the distributions of $\Ds=(c{\overline s})$ mesons, which do not contain any  
light quarks, are frozen out right after hadronisation, in line with the early  
freeze-out of multi-strange mesons.

In recent years, another model~\cite{Lang:2012cx,Lang:2013cca,Lang:2013wya} implementing the non-perturbative $T$-matrix approach described at ~\sect{sec:Tmatrix} has been put forward. 
It was motivated by the necessity of a realistic description for the bulk evolution of the fireball created in ultra-relativistic heavy-ion collisions.  
For this purpose, a transport fluid-dynamics hybrid model of the bulk has been  
developed~\cite{Petersen:2008dd}. It combines the 
Ultra-relativistic Quantum Molecular Dynamics (UrQMD) to describe the 
initial and final stages and ideal fluid dynamics for the intermediate 
stage of the evolution. In this model, the initial collision of the two nuclei is  
simulated with the 
UrQMD cascade model~\cite{Bass:1998ca,Bleicher:1999xi}. After a time 
$t_{\text{start}}=2 R/\sqrt{\gamma_{\text{cm}}^2-1}$,  
 ($R$: radius of the colliding nuclei), when the Lorentz-contracted nuclei have passed 
 through each other ($\gamma_{\text{cm}}$: Lorentz-contraction factor in 
 the centre-of-mass frame) 
the evolution is switched to a relativistic 
ideal-fluid simulation using the full (3+1)-dimensional SHASTA algorithm~\cite{Rischke:1995mt,Rischke:1995ir,Steinheimer:2007iy} by mapping the 
energy, baryon number, and momenta of all particles within UrQMD onto a 
spatial grid. Thermal freeze-out is assumed to occur approximately on 
equal proper-time hyper-surfaces and performed in terms of the usual 
Cooper-Frye prescription~\cite{Cooper:1974mv}. 
 
The diffusion of heavy quarks is described during the fluid-dynamics 
stage of the simulation using a Fokker-Planck description~\cite{Svetitsky:1987gq,GolamMustafa:1997id,Gossiaux:2004qw,vanHees:2004gq,Moore:2004tg,vanHees:2005wb,vanHees:2007me,He:2013zua} 
(``Brownian motion'') employing a relativistic Langevin-Monte-Carlo approach, with  
quark-$Q$ (light-quark--heavy-quark) drag and diffusion coefficients calculated as explained in \sect{sec:Tmatrix}\footnote{An alternative consists in using an effective model for quark-$Q$ 
scattering via D meson like resonance excitations in the QGP based on 
heavy-quark effective theory (HQET) and chiral symmetry in the 
light-quark sector~\cite{vanHees:2004gq}.}. The elastic gluon-$Q$ interaction is computed using  
a leading-order pQCD cross section~\cite{Combridge:1978kx} with a Debye screening mass of 
$m_{Dg}=g T$ in the gluon propagators, which regularises the $t$-channel 
singularities in the matrix elements. The strong coupling constant is 
set to the constant value $\alpha_s=g^2/(4 \pi)=0.4$. 
 
Heavy-quark production is evaluated 
perturbatively on the time-dependent background by UrQMD/hybrid. A first 
UrQMD run is used to determine the collision coordinates of the nucleons 
within the nuclei according to a Glauber initial-state geometry. The 
corresponding space-time coordinates are saved and used in a second full 
UrQMD run as possible production coordinates for the heavy quarks.  
The initial \pt distributions of heavy quarks at $\s=200$\GeV is an ad-hoc  
parametrisation, such that the decay-electron \pt distribution from the calculation 
describes the distribution measured in pp collisions at RHIC~\cite{vanHees:2005wb,vanHees:2007me}.  
For LHC energy, heavy-quark \pt distributions obtained from the PYTHIA event generator are used. 
Finally, at freeze-out temperature the heavy quarks decouple and 
are hadronised either via Peterson \cite{Peterson:1982ak} fragmentation or coalescence.

\subsubsection{lattice-QCD embedded in viscous fluid dynamics (POWLANG)}

In the POWLANG framework (see section \ref{sec:pQCD_hydro}) a set of diffusion 
coefficients $\kappa$ provided by the lattice QCD calculations and described in \sect{sec:lQCDeloss} 
was also implemented. 
 
The main limitation of the lattice QCD approach, providing in principle a non-perturbative result, is the absence of any information on the momentum dependence of $\kappa$.  
The authors of POWLANG make the choice of keeping $\kappa$ constant. 
On the contrary, in the weak-coupling pQCD calculation the longitudinal momentum broadening coefficient $\kappa_{\rm L}(p)$, although starting from a much lower value than the l-QCD one, displays a steep rise with the heavy-quark momentum, which for high enough energy makes it overshoot the lattice-QCD result, taken as constant. Experimental data on the \raa of D mesons and heavy-flavour decay electrons seem to favour an intermediate scenario. 
 
\label{sec:POWLANG}

\subsubsection{AdS/CFT calculations in a static fireball}
 
In the model of \cis{Horowitz:2007su,Horowitz:2011wm}, FONLL~\cite{Cacciari:2012ny} provides both the heavy-flavour production and the fragmentation to D and B mesons.  The medium is described with a static fireball with a transverse profile $T(\vec{x},\,\tau)\propto\rho_{part}(\vec{x})\tau^{-1/3}$ based on the Glauber model.  The energy loss of a heavy quark propagating through the plasma is then given by the AdS/CFT drag derivation, \eq{horowitz:e:adsdrag}, starting at an early thermalisation time $\tau_0  =  0.6$~fm and continuing until $T = T_{hadronisation} = 160$\MeV. Path lengths are  
sampled through a participant transverse density distribution taking into account the nuclear  
diffuseness. 
 
It is non-trivial to connect the parameters of QCD to those of the SYM theory in which the AdS/CFT derivations were performed.  Two common prescriptions~\cite{Gubser:2006nz} for determining the parameters in the SYM theory are to take: \textit{(i)} $\alpha_{SYM}  =  \alpha_s$ and $T_{SYM}  =  T_{QCD}$ or \textit{(ii)} $\lambda_{SYM}  =  5.5$ and $e_{SYM}  =  e_{QCD}$ (and hence $T_{SYM}  =  T_{QCD} / 3^{1/4}$).  In the first prescription, the SYM coupling is taken equal to the QCD coupling and the temperatures are equated.  In the second prescription, the energy densities of QCD and SYM are equated and the coupling is fitted by comparing the static quark--antiquark force from AdS/CFT to lattice results.   
 
Comparing with RHIC data, the pure drag energy loss is qualitatively consistent with electrons from the semi-leptonic decays of heavy mesons~\cite{Abelev:2006db,Adare:2010de}.  In the RHIC calculation, the proportionality constant between the medium temperature and the Glauber participant density is set such that, in the Stefan-Boltzmann limit, the rapidity density of gluons in the medium is $\dd N_g/\dd y  =  1000$, which is similar to that required by perturbative energy loss calculations and is not too different from the entropy one expects from the measured hadronic multiplicity~\cite{Wicks:2005gt}.  With that proportionality constant fixed, predictions for the suppression at LHC are performed  assuming that the temperature of the medium scales with the measured hadronic multiplicity~\cite{Aamodt:2010cz}.   
 
The kinematic range of applicability of the model can be estimated through the \pt scale at which including momentum fluctuations becomes important.  By comparing the momentum lost to drag to the potential momentum gain of the fluctuations, one expects that momentum fluctuations become important at a scale $\gamma_{crit}^{\Delta p^2}  =  m_Q^2 / 4 \, T^2$.  One can see from above that the speed limit at which the entire calculational framework breaks down, $\gamma_{crit}^{SL}$, is parametrically in $\lambda$ smaller than $\gamma_{crit}^{\Delta p^2}$; however, numerically for the finite values of $\lambda$ phenomenologically relevant at RHIC and LHC $\gamma_{crit}^{\Delta p^2}  <  \gamma_{crit}^{SL}$.  In particular, one expects non-trivial corrections to the drag results for $e^\pm$ and D mesons from open heavy flavour for $\pt<4$--5\GeVc.  
Other calculations~\cite{Moore:2004tg,Akamatsu:2008ge} have attempted to include the effect of fluctuations; however their diffusion coefficients were set by the Einstein relations and not those derived from AdS/CFT (recall that the derived diffusion coefficients are qualitatively different from those based on the fluctuation-dissipation theorem except in the limit of $v = 0$).

\subsection{Comparative overview of model features and comparison with data}
\label{sec:modelcomparison}
 
The theoretical models described in the previous sections are compared in \tab{tab:HFmodels} in terms of their ``ingredients'' for heavy-quark production, 
medium modelling, quark--medium interactions, and heavy-quark hadronisation.

In this section we compile a comparison of model calculations with heavy-flavour \raa and \vtwo measurements by the RHIC and LHC experiments. 
 
\begin{figure}[!h] 
 \centering 
 \includegraphics[width=0.48\textwidth]{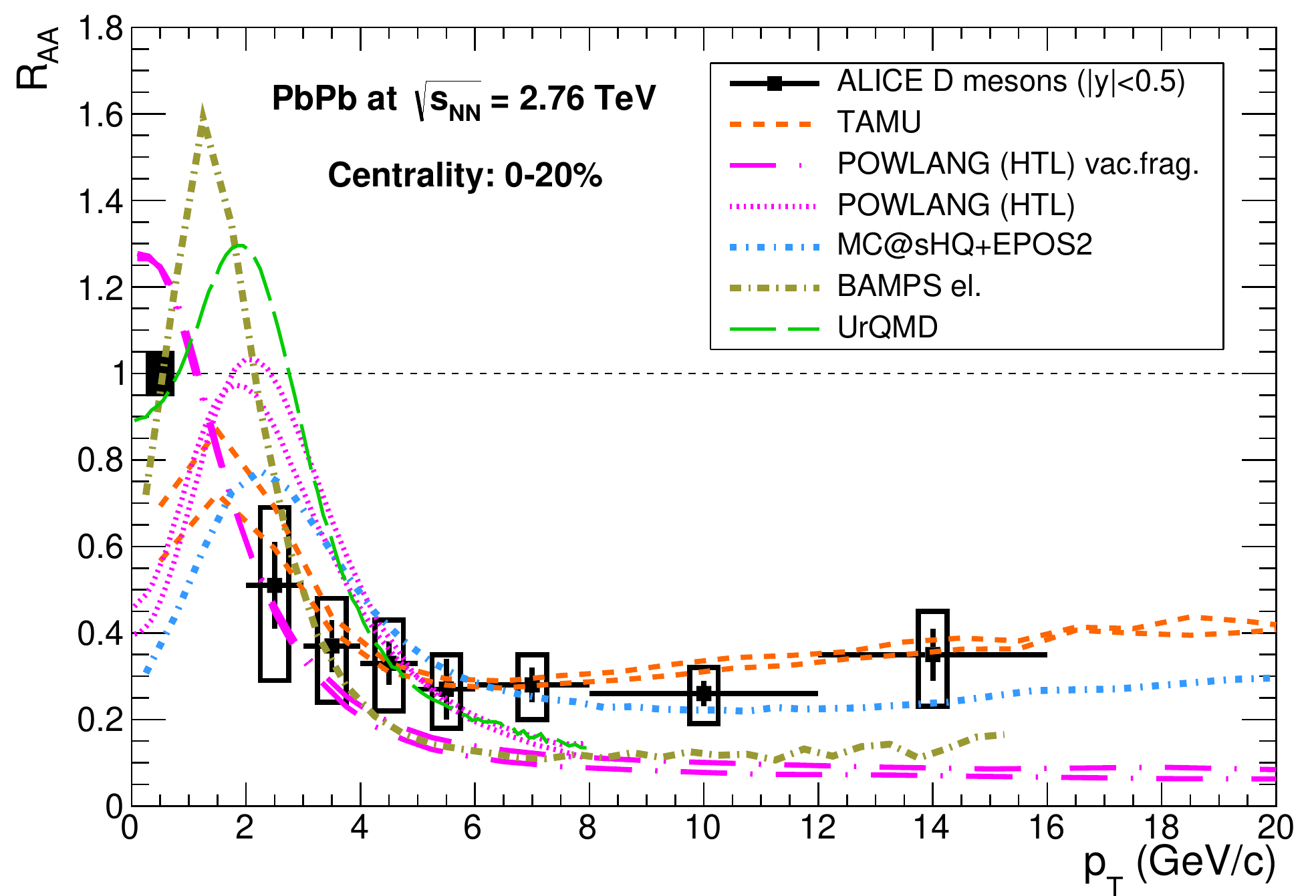} 
 \includegraphics[width=0.48\textwidth]{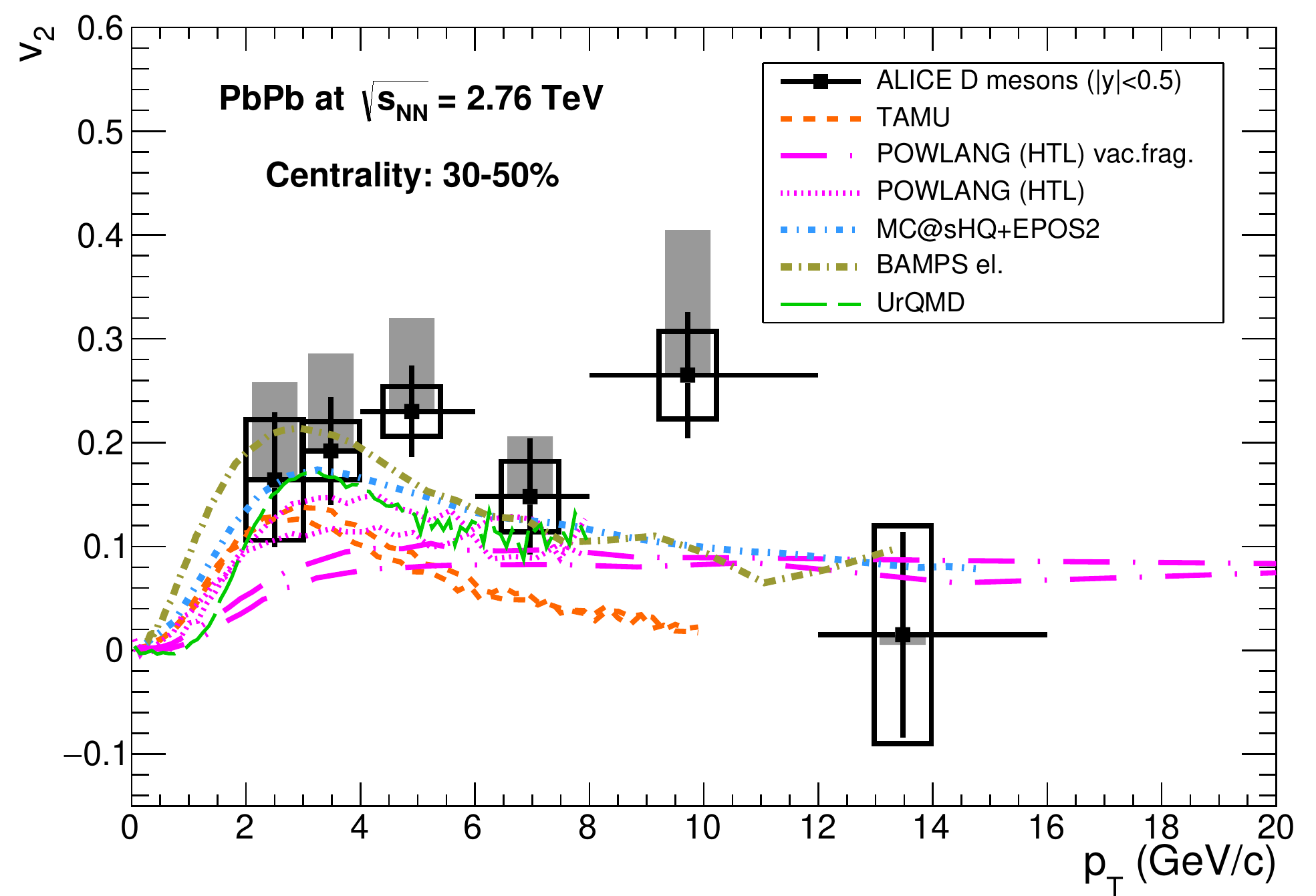} 
  
 \includegraphics[width=0.48\textwidth]{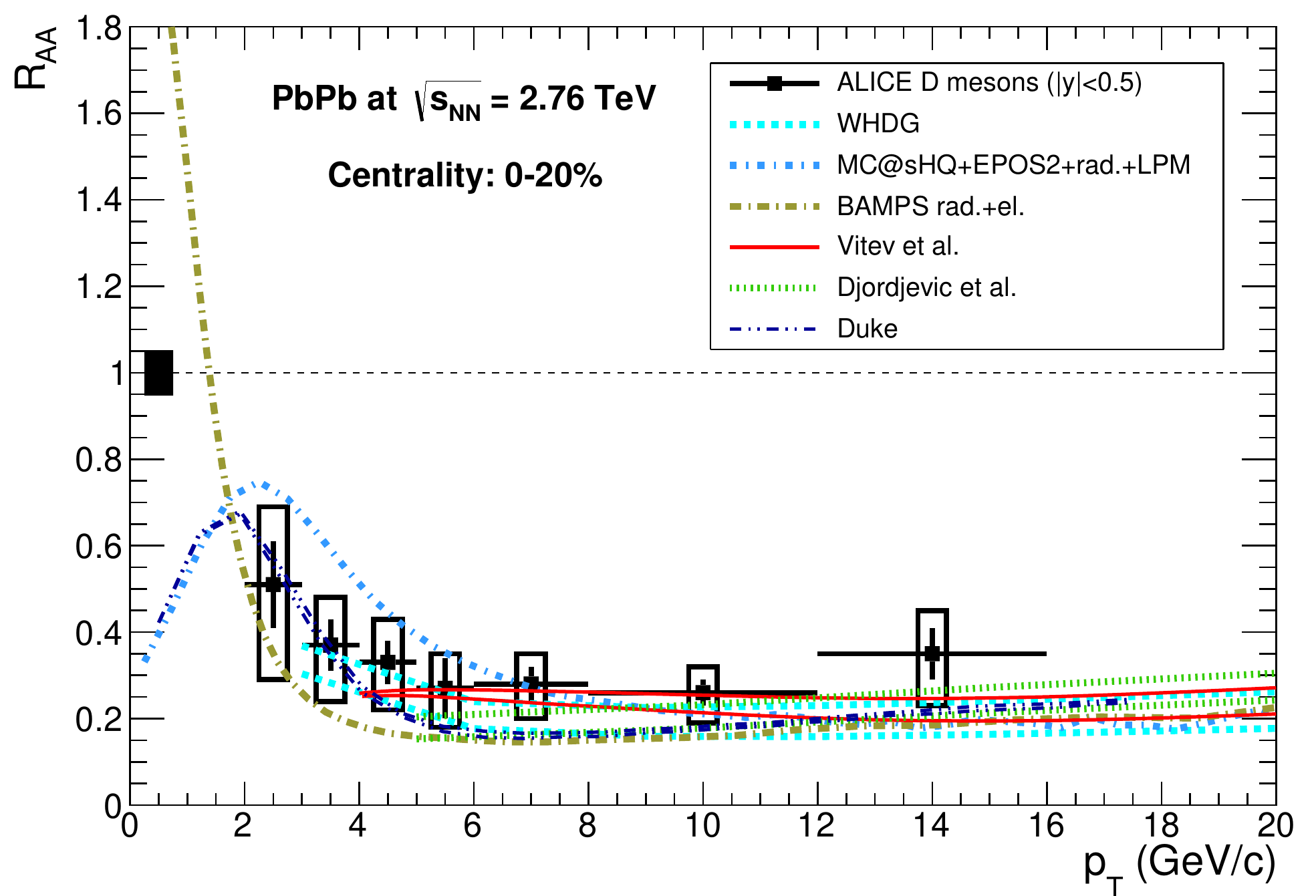} 
 \includegraphics[width=0.48\textwidth]{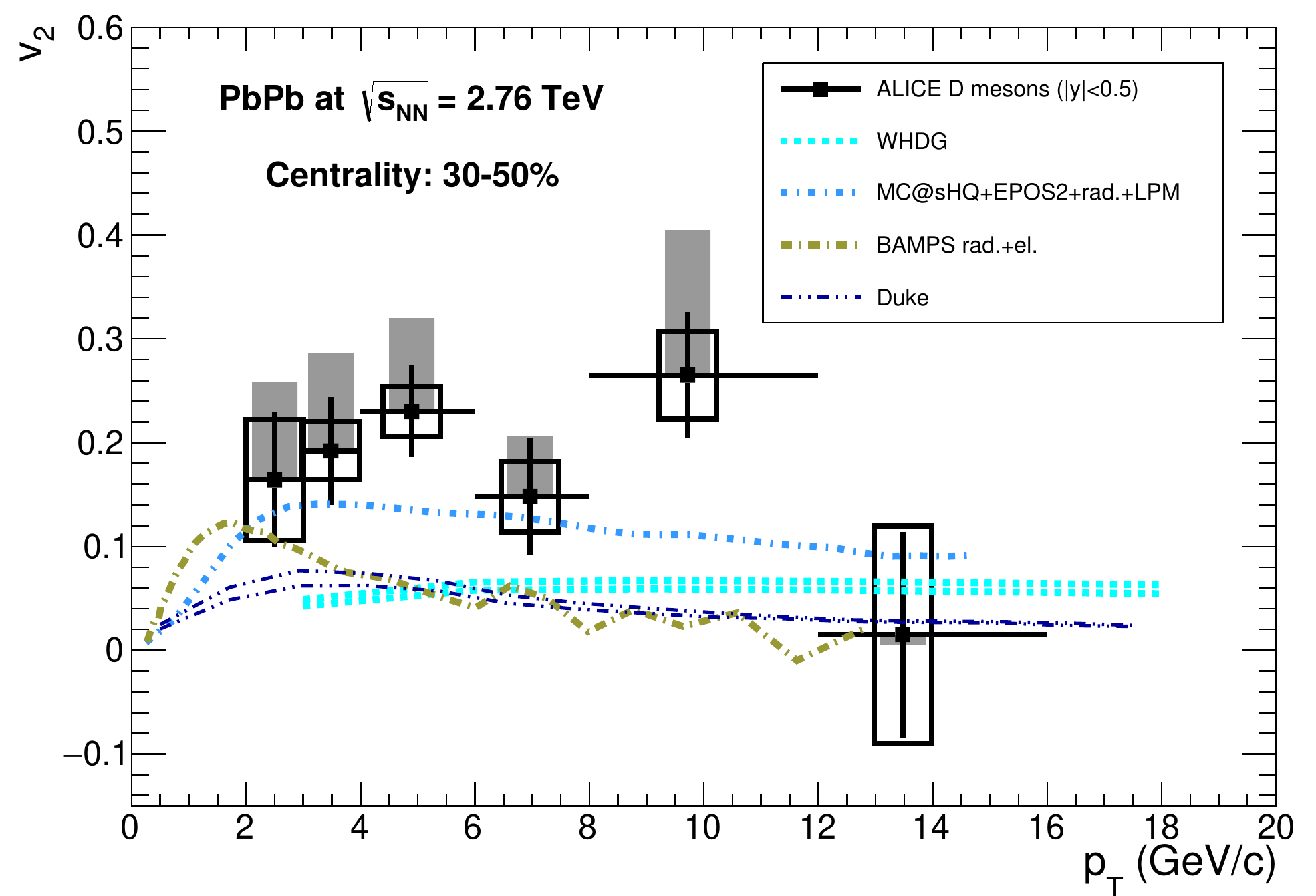} 
  
 \includegraphics[width=0.48\textwidth]{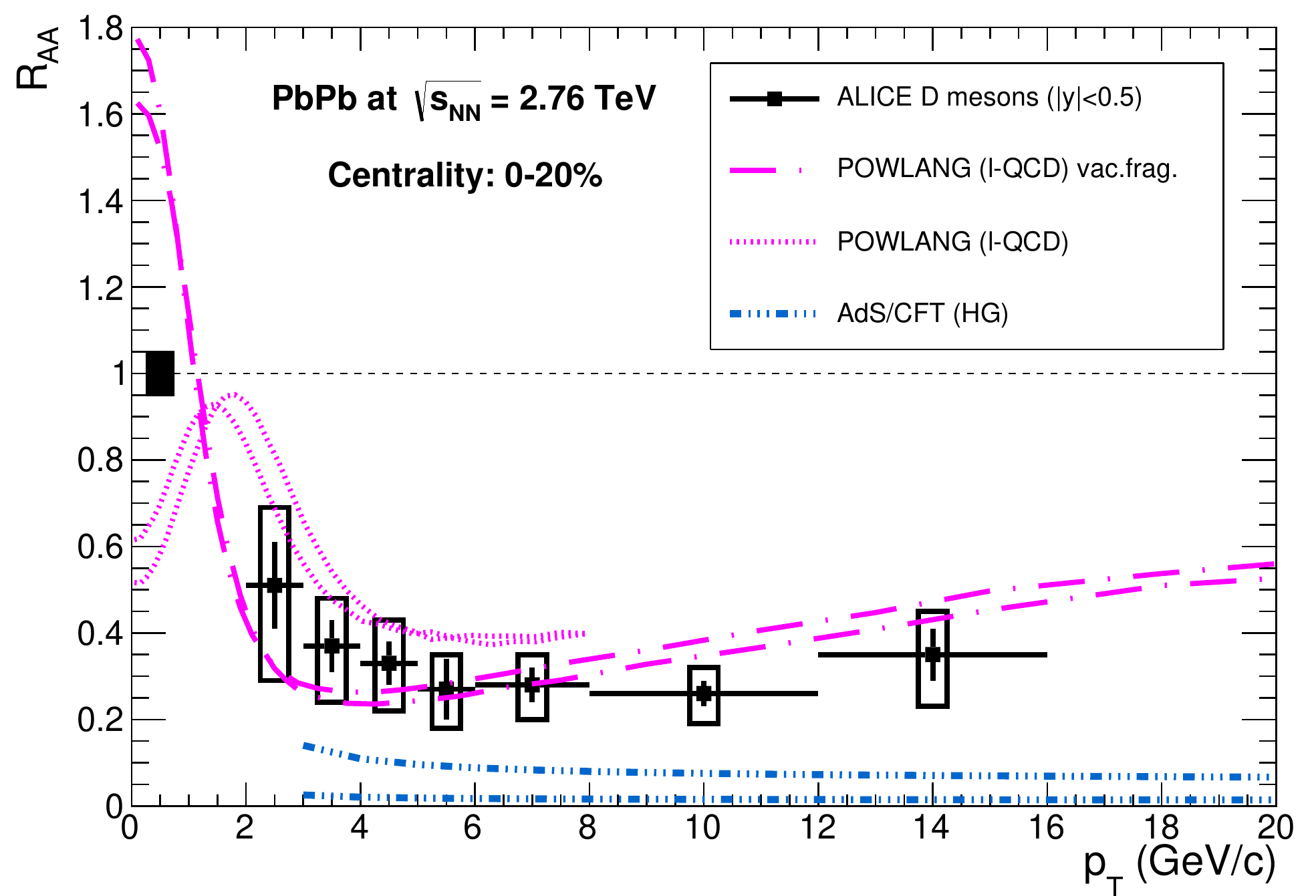} 
 \includegraphics[width=0.48\textwidth]{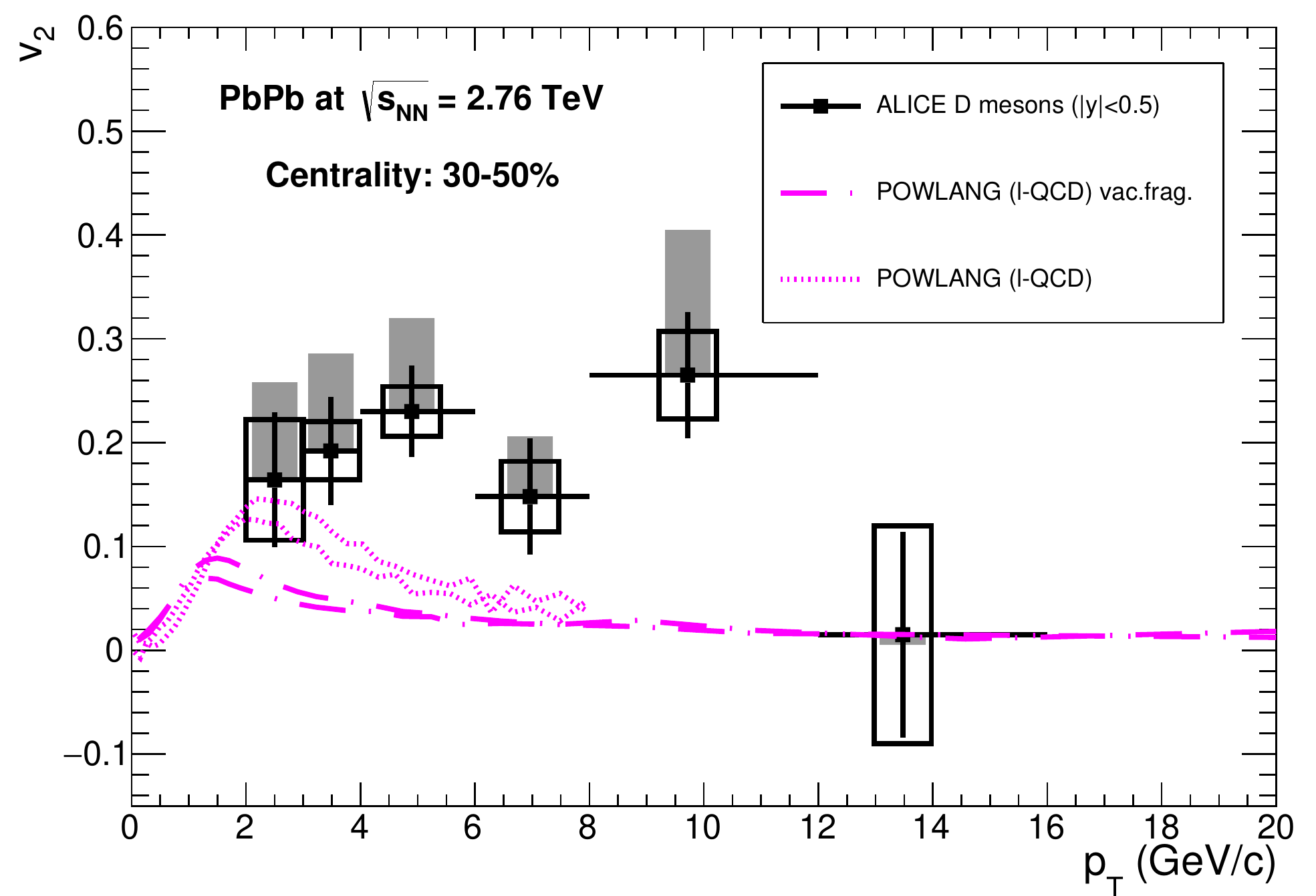} 
 \caption{Left: nuclear modification factor as a function of transverse momentum of averaged prompt D mesons in the 0--20\% most central \pb collisions at \snn=2.76~TeV~\cite{ALICE:2012ab} (the filled box at $\raa=1$ is the systematic uncertainty on the normalisation). Right: \vtwo as a function of transverse momentum of D mesons in the 30--50\% centrality \pb collisions at \snn=2.76~TeV~\cite{Abelev:2013lca} (the filled boxes are the systematic uncertainties on the feed-down subtraction). The results are obtained as an average of the \Dzero, \Dplus and \Dstarplus measurements. The results are compared to model calculations implementing collisional energy loss (top panels), collisional and radiative energy loss (middle panels) and to models which cannot be ascribed to the previous categories (bottom panels).} 
 \label{fig:DmesonALICEcompTheo} 
\end{figure}

\fig{fig:DmesonALICEcompTheo} shows the comparison for D mesons in \pb collisions at $\snn = 2.76~\TeV$, measured by the ALICE Collaboration~\cite{ALICE:2012ab,Abelev:2013lca}. 
The left panels show \raa in the centrality class 0--20\%, the right panels show \vtwo in the centrality class 30--50\%. The models that include only collisional energy loss are shown 
in the upper panels. These models provide in general a good description of \vtwo. The original version of the POWLANG model does not exhibit a clear maximum in \vtwo like the other models, 
which could be due to the fact that it does not include, in such a version, hadronisation via recombination. 
The latter has been recently introduced in the POWLANG model  
and the additional flow inherited by the D mesons from the light quarks moves the calculations to higher values of \vtwo. 
In the TAMU model the decrease of \vtwo towards high \pt is faster than in the other models, which reflects a moderate coupling with the medium, also seen in the  
rise of \raa of D mesons at large \pt.  
In this range, some of the other models over-suppress the \raa and one observes large discrepancies between them, which mostly originate from the medium description as well as from the transport coefficients adopted in each model. At low \pt the models (UrQMD, BAMPS) that do not include PDF shadowing give a \raa value larger than observed in the data. 
The models that include both radiative and collisional energy loss are shown in the central panels.  All these models provide a good description of \raa, but most of them  
underestimate the maximum of \vtwo observed in data. This could be due to the fact that the inclusion of radiative process reduces the weight of collisional process, which  
are more effective in building up the azimuthal anisotropy. In addition, some of these models (Djordjevic \etal, WHDG, Vitev \etal) do not include a fluid-dynamical medium (for this reason, 
the Djordjevic \etal and Vitev \etal models do not provide a calculation for \vtwo), and none of them implements the detailed balance reaction which is mandatory 
to reach thermalisation and then undergo the full drift from the medium. 
The POWLANG model with l-QCD based transport coefficient and the AdS/CFT predictions are plotted in the lower panels. POWLANG provides a good description of \raa,  
while, for what concerns \vtwo, the results depend crucially on the way hadronisation is described, recombination scenarios leading to a larger elliptic flow (although still smaller than the experimental data in the accessible \pt range). 
AdS/CFT, on the other hand, over-predicts the suppression in the full \pt range explored. 
 
\begin{figure} 
 \centering 
 \includegraphics[width=0.48\textwidth]{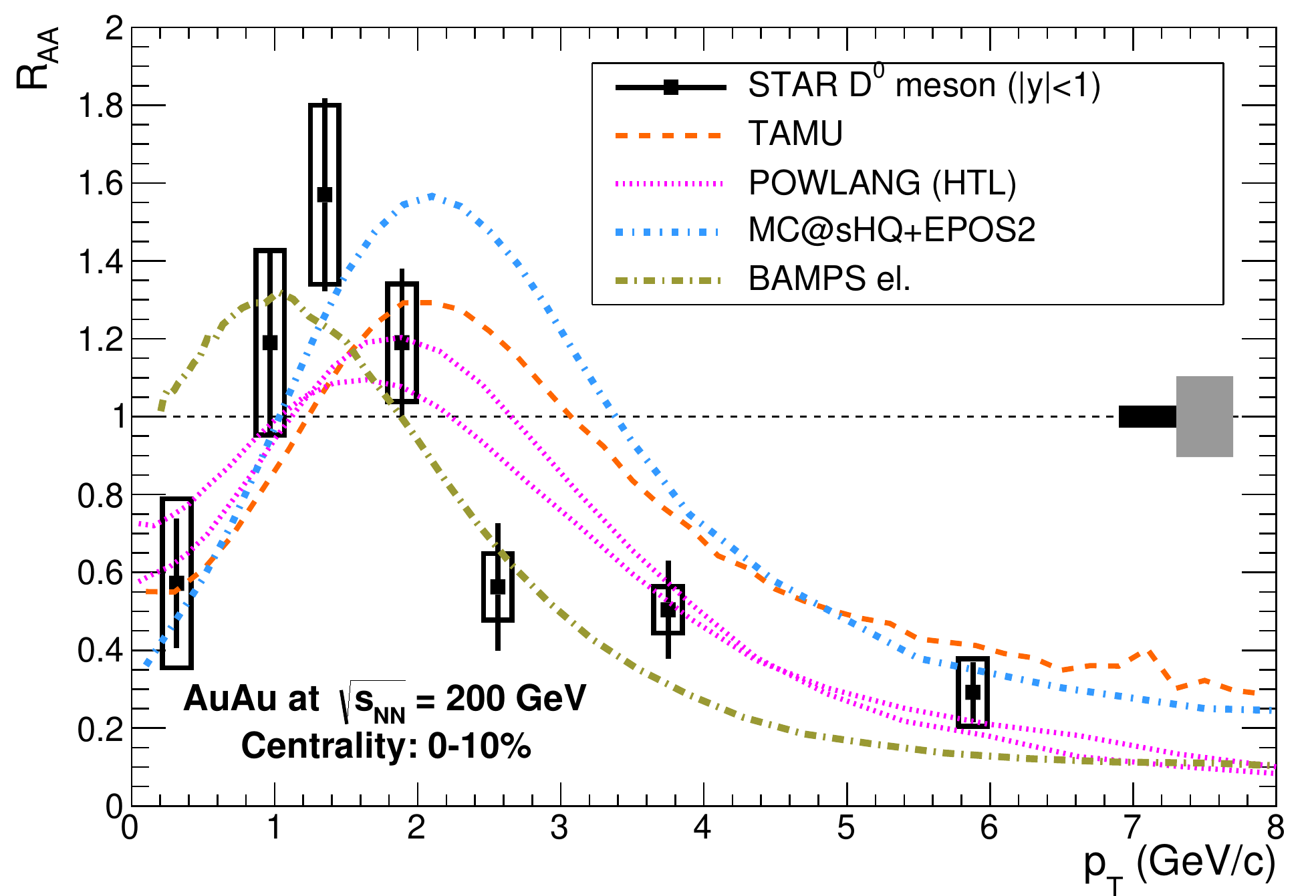} 
 \includegraphics[width=0.48\textwidth]{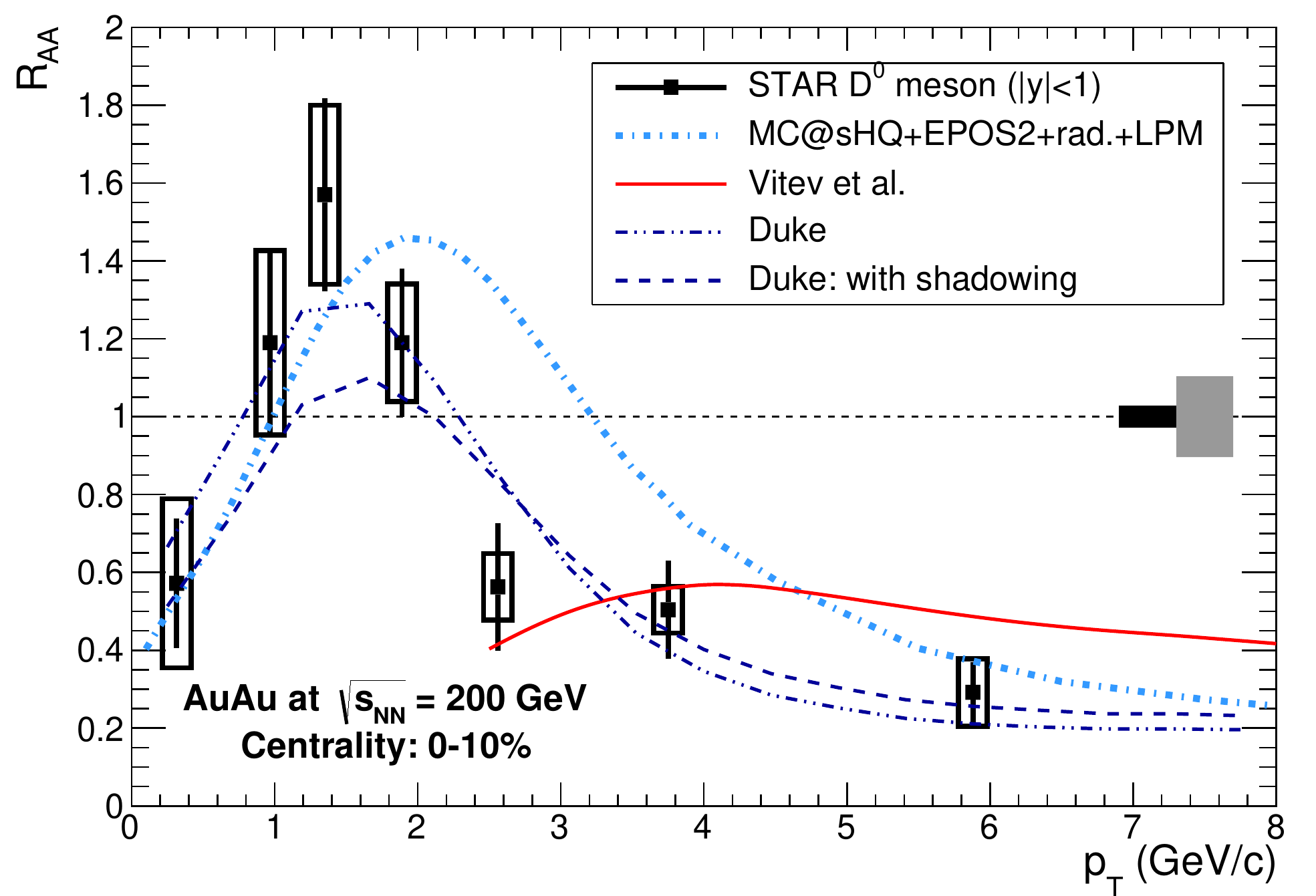} 
  
 \includegraphics[width=0.48\textwidth]{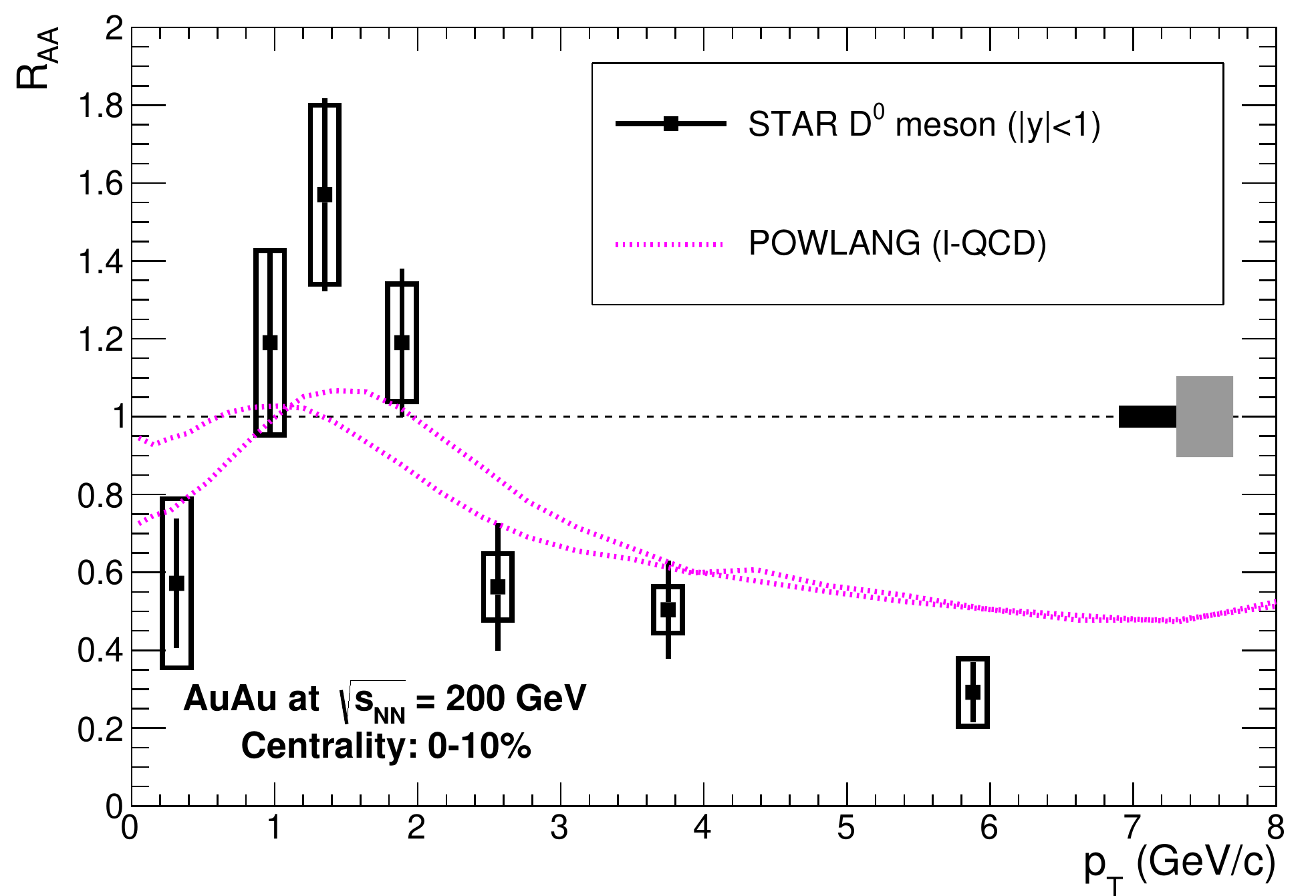} 
 \phantom{\includegraphics[width=0.48\textwidth]{comp_Dmeson_Raa_vs_pt_AuAu200_0_10_coll}} 
 \caption{Nuclear modification factor as a function of transverse momentum of \Dzero mesons in the 0--10\% most central \AuAu collisions at \snn=200~GeV~\cite{Adamczyk:2014uip}. The filled boxes at $\raa=1$ are, from left to right, the systematic uncertainties on the normalisation of \AuAu and \pp data. The results are compared to model calculations implementing collisional energy loss (top left), collisional and radiative energy loss (top right) and to models which cannot be ascribed to the previous categories (bottom left).} 
 \label{fig:DzeroSTARcompTheo} 
\end{figure}

\fig{fig:DzeroSTARcompTheo} shows the comparison for the \Dzero meson \raa in \AuAu collisions at $\snn=200$\GeV, measured by the STAR Collaboration~\cite{Adamczyk:2014uip}. 
The models that include collisional interactions in an expanding fluid-dynamical medium (TAMU, BAMPS, Duke, MC@$_s$HQ, POWLANG) describe qualitatively the shape of \raa in the interval 0--3\GeVc, with a rise, a  
maximum at 1.5\GeVc with $\raa>1$, and a decrease. In these models, this shape is the effect of radial flow on light and charm quarks. The TAMU model also includes flow in the hadronic phase. 
It can be noted that these models predict a similar bump also at LHC energy (left panels of \fig{fig:DmesonALICEcompTheo}): the bump reaches $\raa>1$ for the models that do not include 
PDF shadowing, while it stays below $\raa=0.8$ for the models that include it.  
The present ALICE data for $\pt>2$\GeVc do not allow to draw a strong conclusion. However, the preliminary ALICE data reaching down to 1\GeVc in the centrality class 0--7.5\%~\cite{Grelli:2012yv} do not favour models 
that predict a bump with $\raa>1$. 
 
\begin{figure} 
 \centering 
 \includegraphics[width=0.48\textwidth]{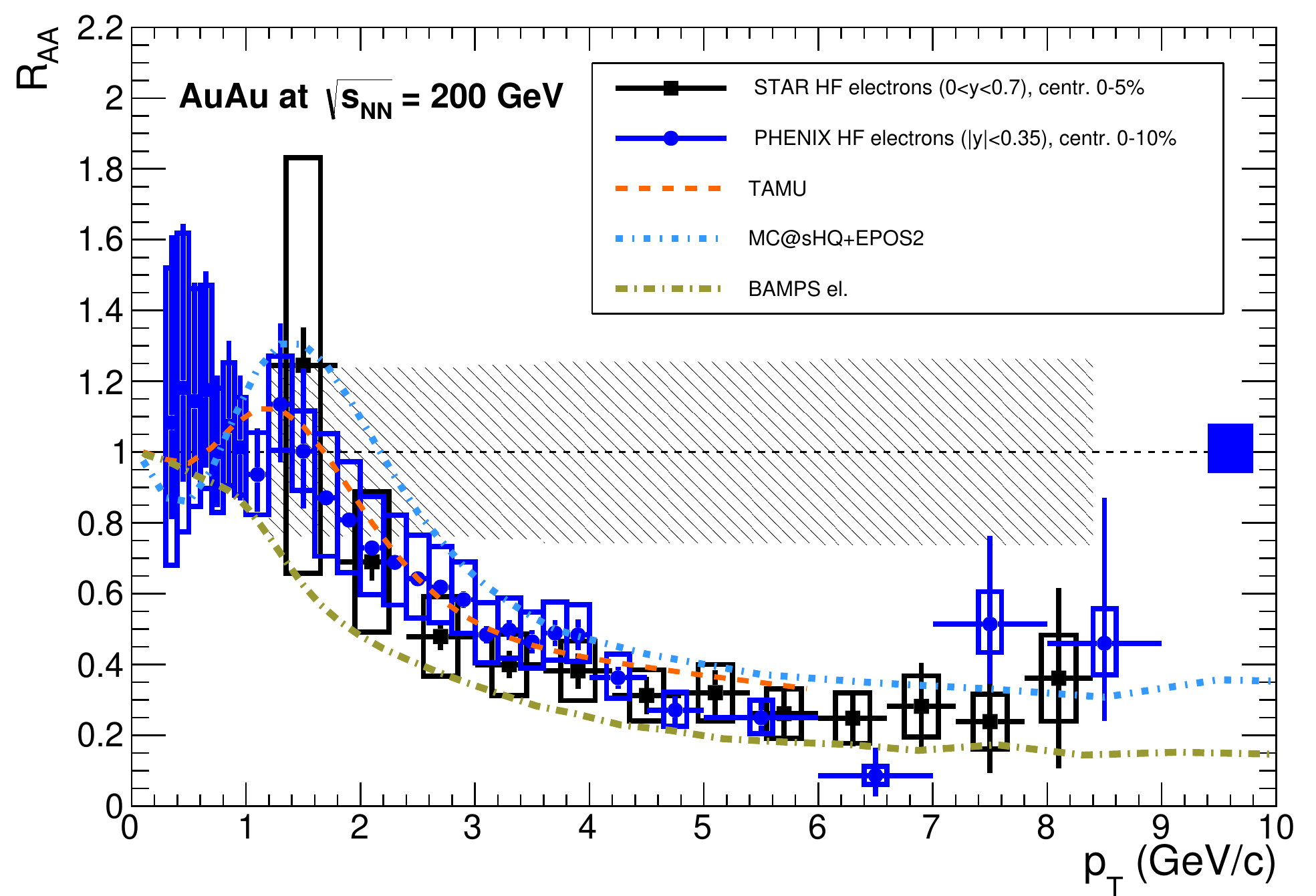} 
 \includegraphics[width=0.48\textwidth]{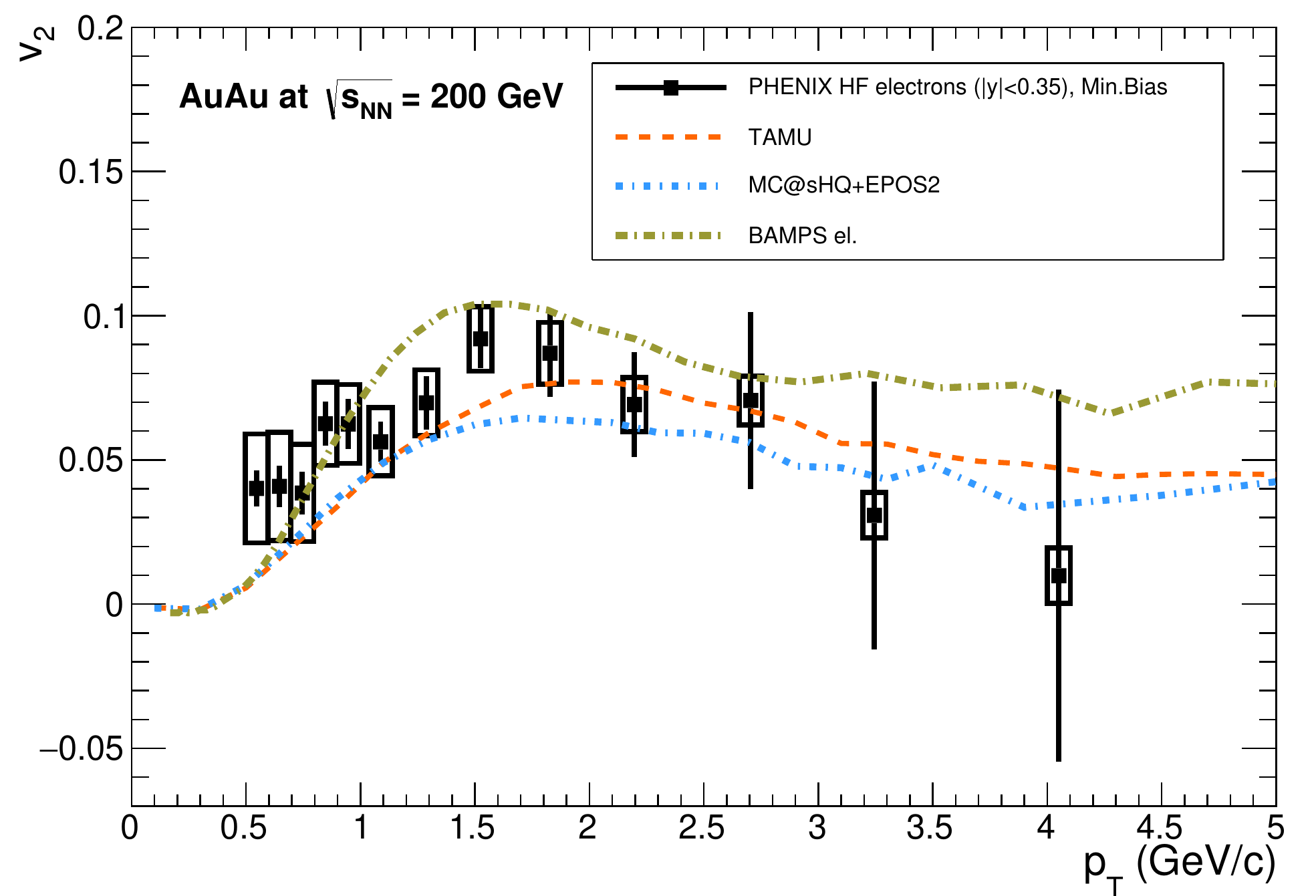} 
  
 \includegraphics[width=0.48\textwidth]{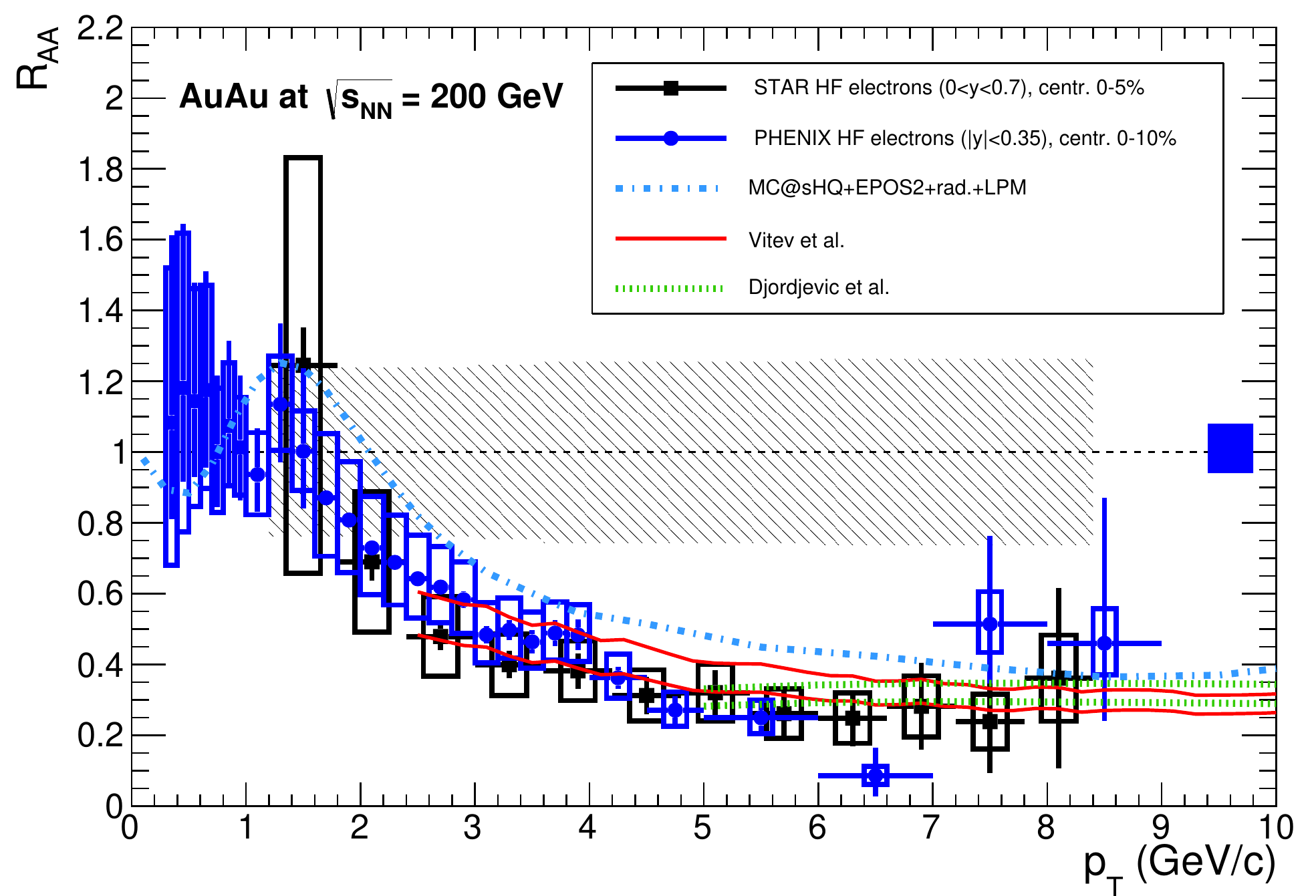} 
 \includegraphics[width=0.48\textwidth]{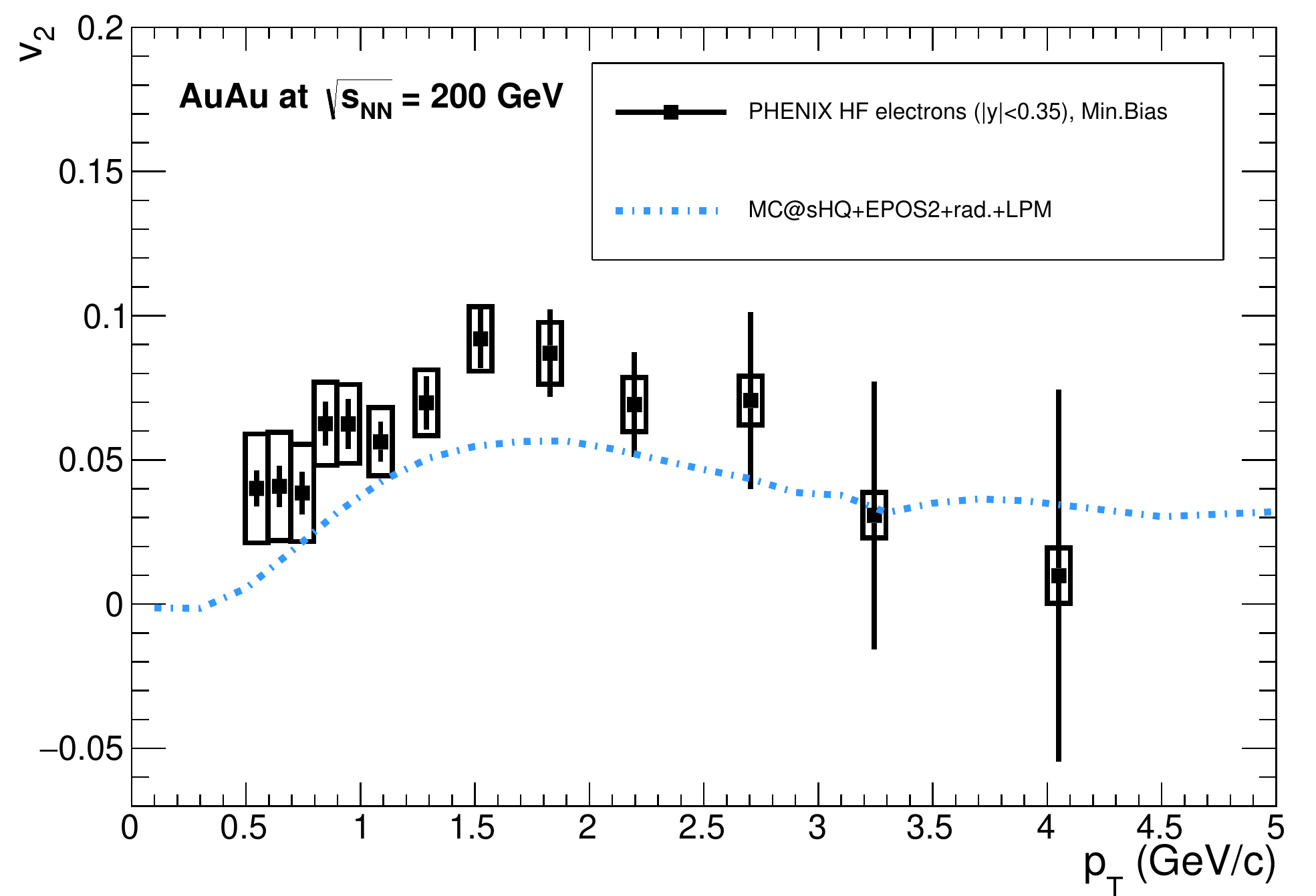} 
 \caption{Left: nuclear modification factor as a function of transverse momentum of heavy flavour electrons in the 0--5\%~\cite{Abelev:2006db} and 0--10\%~\cite{Adare:2010de} most central \AuAu collisions at \snn=200~GeV. The dashed band (filled box) at $\raa=1$ is the normalisation uncertainty for STAR (PHENIX) data. Right: \vtwo as a function of transverse momentum of heavy flavour electrons in \AuAu collisions at \snn=200~GeV~\cite{Adare:2010de}. The results are compared to model calculations implementing collisional energy loss (top panels) and collisional and radiative energy loss (bottom panels).} 
 \label{fig:hfeRHICcompTheo} 
\end{figure}

\begin{figure} 
 \centering 
 \includegraphics[width=0.48\textwidth]{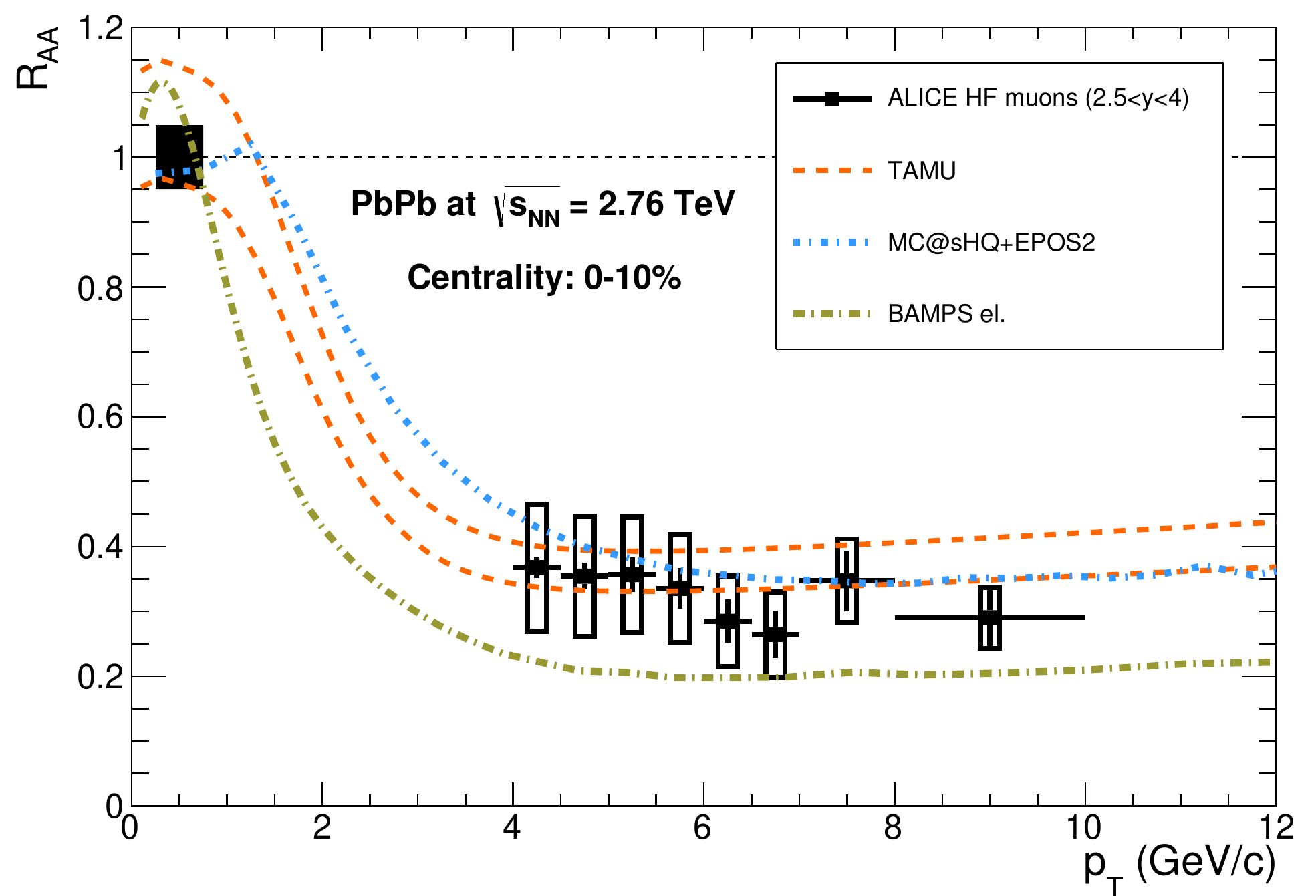} 
 \includegraphics[width=0.48\textwidth]{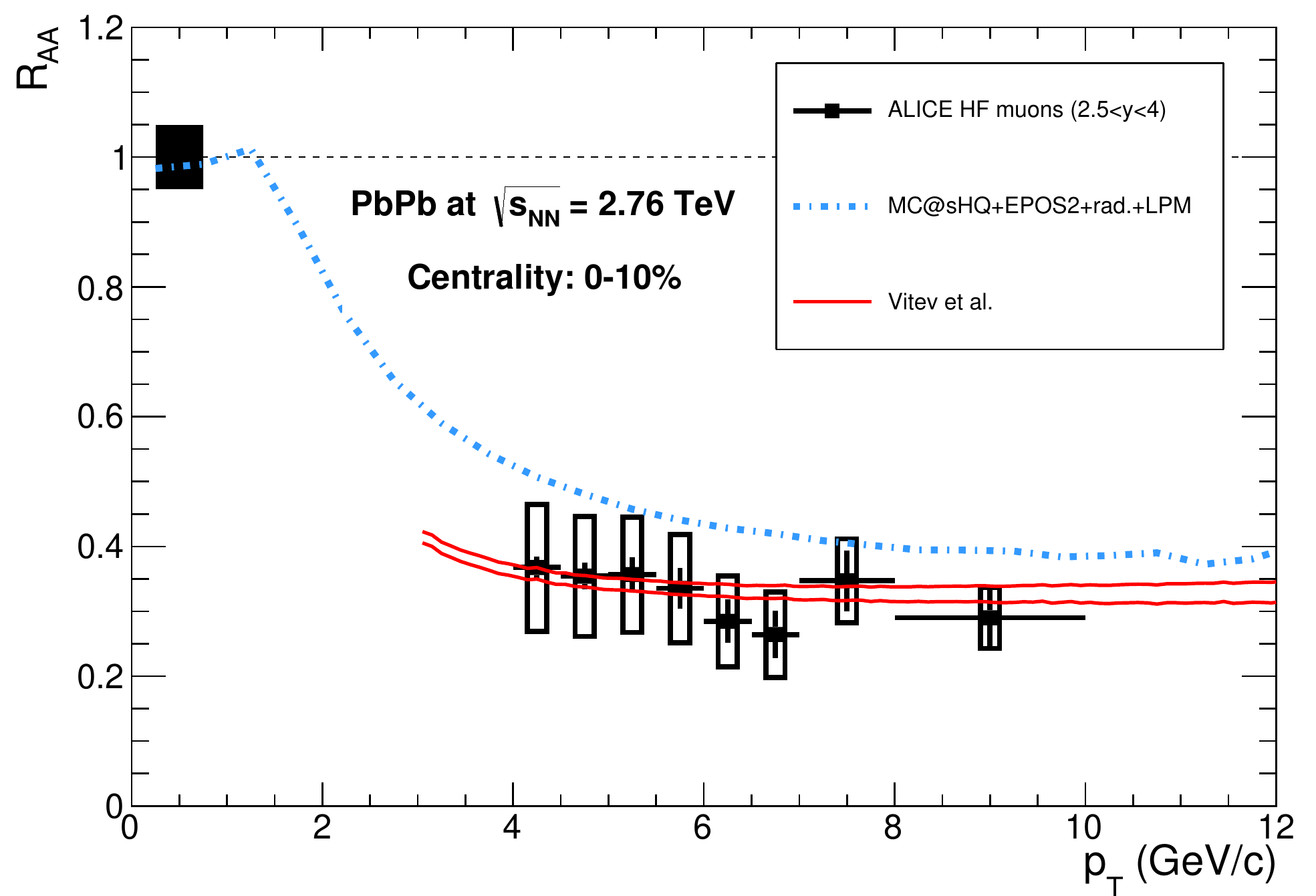} 
 \caption{Nuclear modification factor as a function of transverse momentum of heavy flavour muons with $2.5<y<4$ measured in the 0--10\% most central \pb collisions at \snn=2.76~TeV~\cite{Abelev:2012qh}. The filled box at $\raa=1$ is the systematic uncertainty on the normalisation. The results are compared to model calculations implementing collisional energy loss (left) and collisional and radiative energy loss (right).} 
 \label{fig:hfmuALICEcompTheo} 
\end{figure}

The comparisons with measurements of heavy-flavour decay leptons at RHIC and LHC are shown in ~\figs~\ref{fig:hfeRHICcompTheo} and~\ref{fig:hfmuALICEcompTheo}, respectively. 
The \raa and \vtwo of heavy-flavour decay electrons in \AuAu collisions at top RHIC energy, measured by PHENIX~\cite{Adare:2010de} and STAR~\cite{Abelev:2006db}, are well described by all model calculations. 
Note that in some of the models the quark--medium coupling (medium density or temperature or interaction cross section) is tuned to describe the \raa of pions (Djordjevic \etal, WHDG, Vitev \etal) or electrons (BAMPS) at RHIC. 
The \raa of heavy-flavour decay muons at forward rapidity ($2.5<y<4$), measured by ALICE in central \pb collisions at LHC~\cite{Abelev:2012qh}, is well described by most of the models.  
The BAMPS model tends to over-suppress this \raa, as observed also for the high-\pt \raa of D mesons at RHIC and LHC. The MC@$_s$HQ model describes the data better when  
radiative energy loss is not included. 
In general, it can be noted that the differences between the various model predictions are less pronounced in the case of heavy-flavour decay lepton observables  
than in the case of D mesons. This is due to the fact that the former include a \pt-dependent contribution of charm and beauty decays. In addition, the decay kinematics  
shifts the lepton spectra towards low momentum, reducing the impact on \raa of effects like PDF shadowing, radial flow and recombination. 
 
\begin{figure} 
 \centering 
 \includegraphics[width=0.48\textwidth]{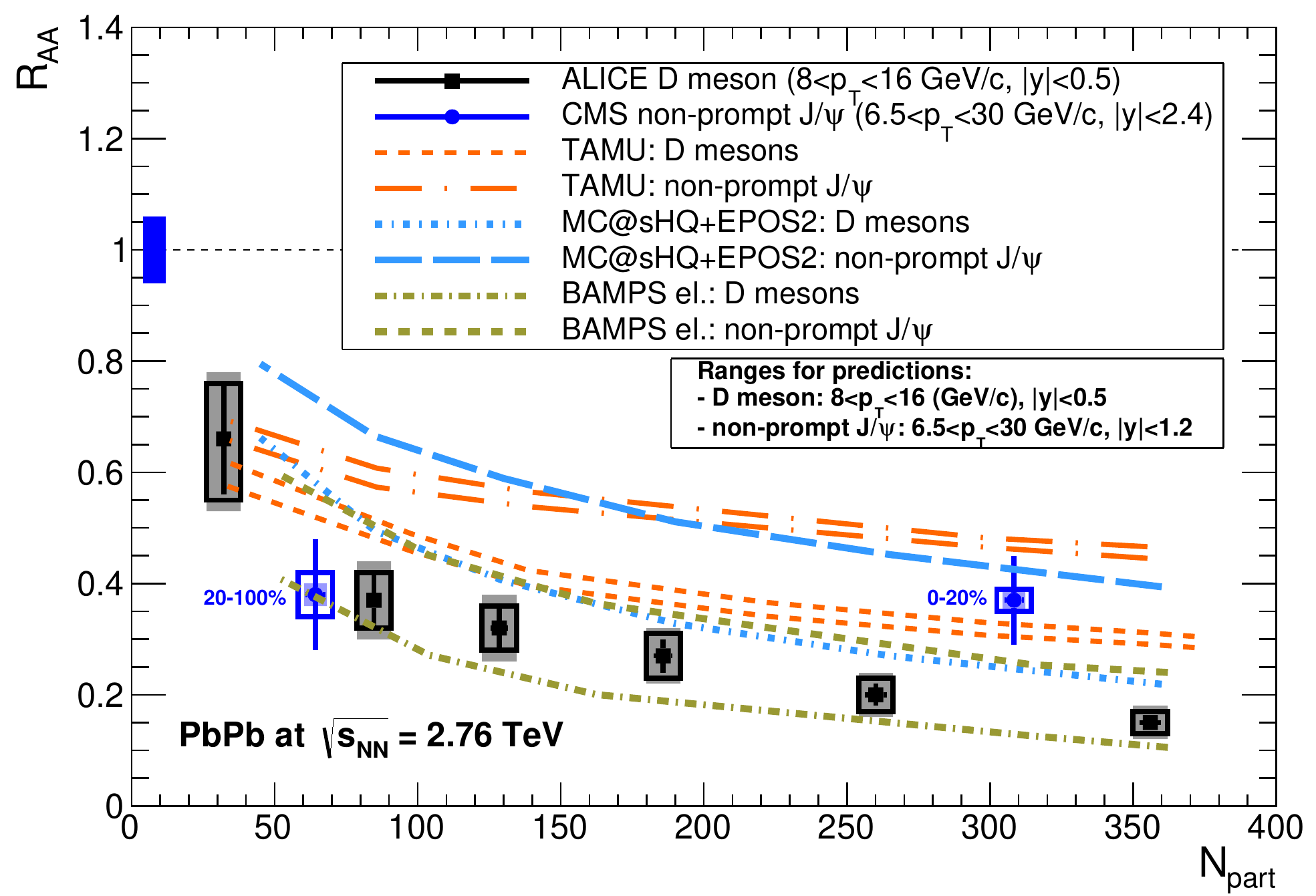} 
 \includegraphics[width=0.48\textwidth]{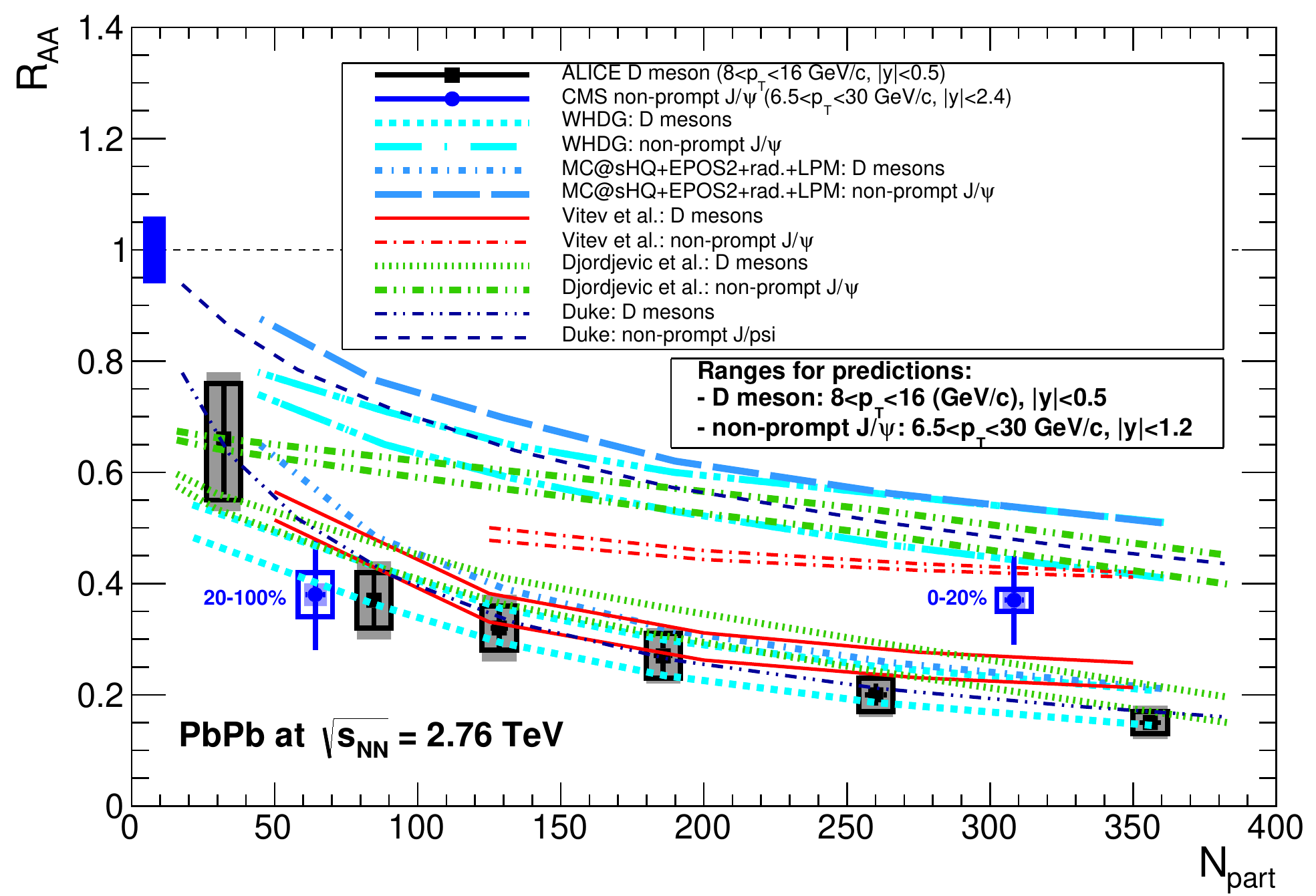} 
 \caption{Nuclear modification factor as a function of the number of participants of averaged prompt D mesons~\cite{ALICE:2012ab} and non-prompt \jpsi~\cite{Chatrchyan:2012np} measured in \pb collisions at \snn=2.76~TeV compared to model calculations implementing collisional (left) and collisional and radiative energy loss (right). The filled box at $\raa=1$ is the systematic uncertainty on the normalisation of non-prompt \jpsi data.  Note that: a) model predictions refer to CMS preliminary data, in a slightly different rapidity range; b) the point at low \Npart for the non-prompt \jpsi \raa refers to a very large centrality interval (20--100\%).} 
 \label{fig:DnonPromptJpsicompTheo} 
\end{figure}

\begin{figure} 
 \centering 
 \includegraphics[width=0.48\textwidth]{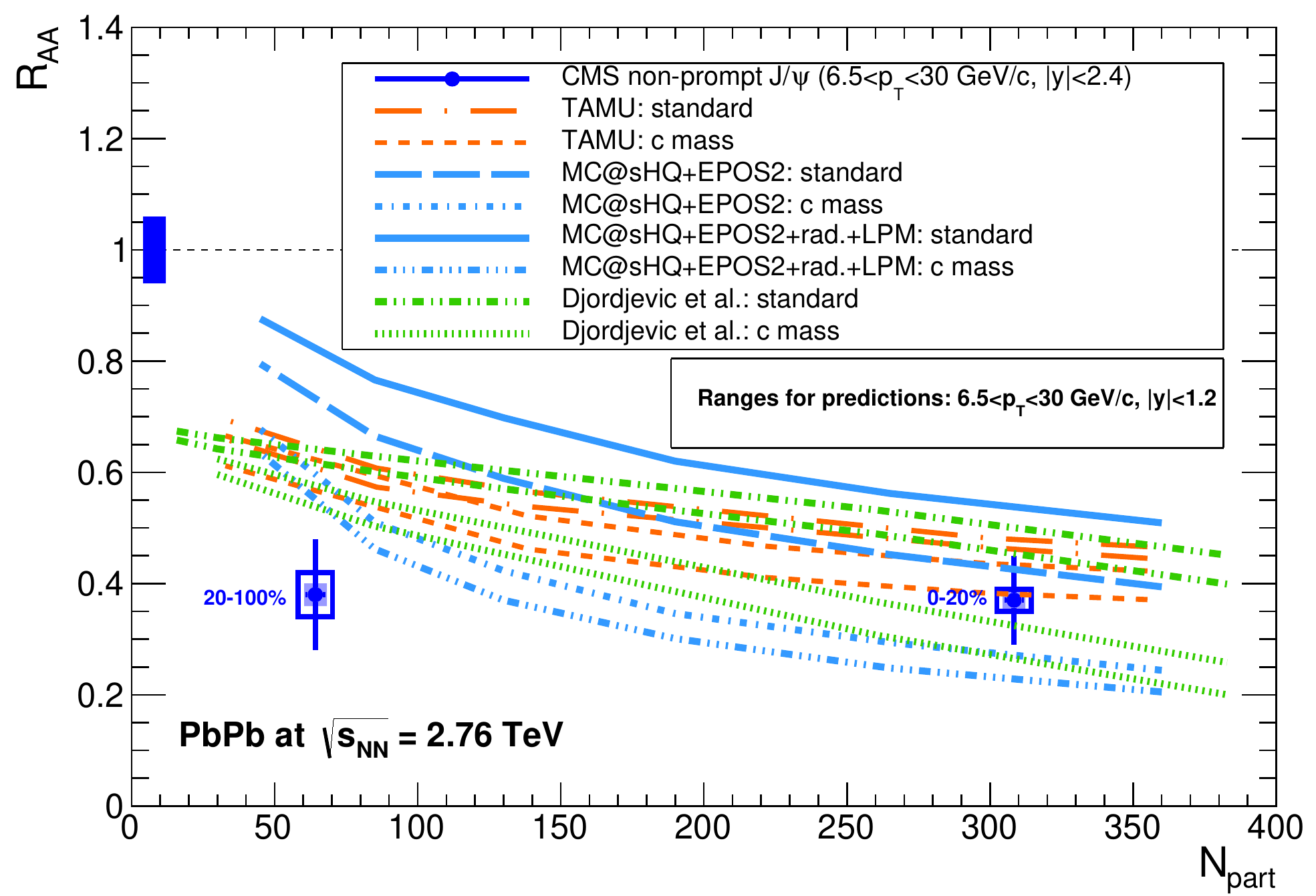} 
 \caption{Quark mass dependence of energy loss. The nuclear modification factor of non-prompt \jpsi~\cite{Chatrchyan:2012np} measured in \pb collisions at \snn=2.76~TeV is compared to model calculations obtained in the same way as in \fig{fig:DnonPromptJpsicompTheo} and assuming that the \bquark quark has the mass of the \cquark quark.} 
 \label{fig:nonPromptJpsiMassEffect} 
\end{figure}

In \fig{fig:DnonPromptJpsicompTheo} we compile the model calculations for the centrality dependence of the nuclear modification factors of D mesons in the interval $8<\pt<16$\GeVc and non-prompt \jpsi mesons  
in the interval $6.5<\pt<30$\GeVc, in \pb collisions at $\snn = 2.76$~\TeV. All models predict the \raa of D mesons to be lower by about 0.2--0.3 units than that of non-prompt \jpsi.  
This difference arises from the mass-dependence of quark--medium interactions. The available published data, from the first limited-statistics \pb run at LHC (in 2010), are reported in the figure: note that the D meson \raa  
measured by the ALICE experiment~\cite{ALICE:2012ab} corresponds to the interval 6--12\GeVc (slightly lower than that of the calculations), while the  
non-prompt \jpsi \raa measured by the CMS experiment~\cite{Chatrchyan:2012np}  corresponds to the large centrality classes 0--20\% and 20--100\%. Due to the large uncertainties and the large centrality  
intervals, the data do not allow for a clear conclusion on the comparison with models. The preliminary ALICE~\cite{Bruna:2014pfa} and CMS~\cite{CMS:2012wba} measurements  
using the higher-statistics 2011 \PbPb sample are well-described by the model calculations. 
The effect of the heavy-quark mass on the nuclear modification factor is illustrated in \fig{fig:nonPromptJpsiMassEffect}, 
where the \raa of non-prompt J/$\psi$ is obtained in the Djordjevic {\it et al.}, MC@$_s$HQ and TAMU 
models using the $c$-quark mass value for the calculation of the  
in-medium interactions of $b$ quarks. In this case, substantially-lower values of \raa are obtained.

\begin{figure} 
 \centering 
 \includegraphics[width=0.48\textwidth]{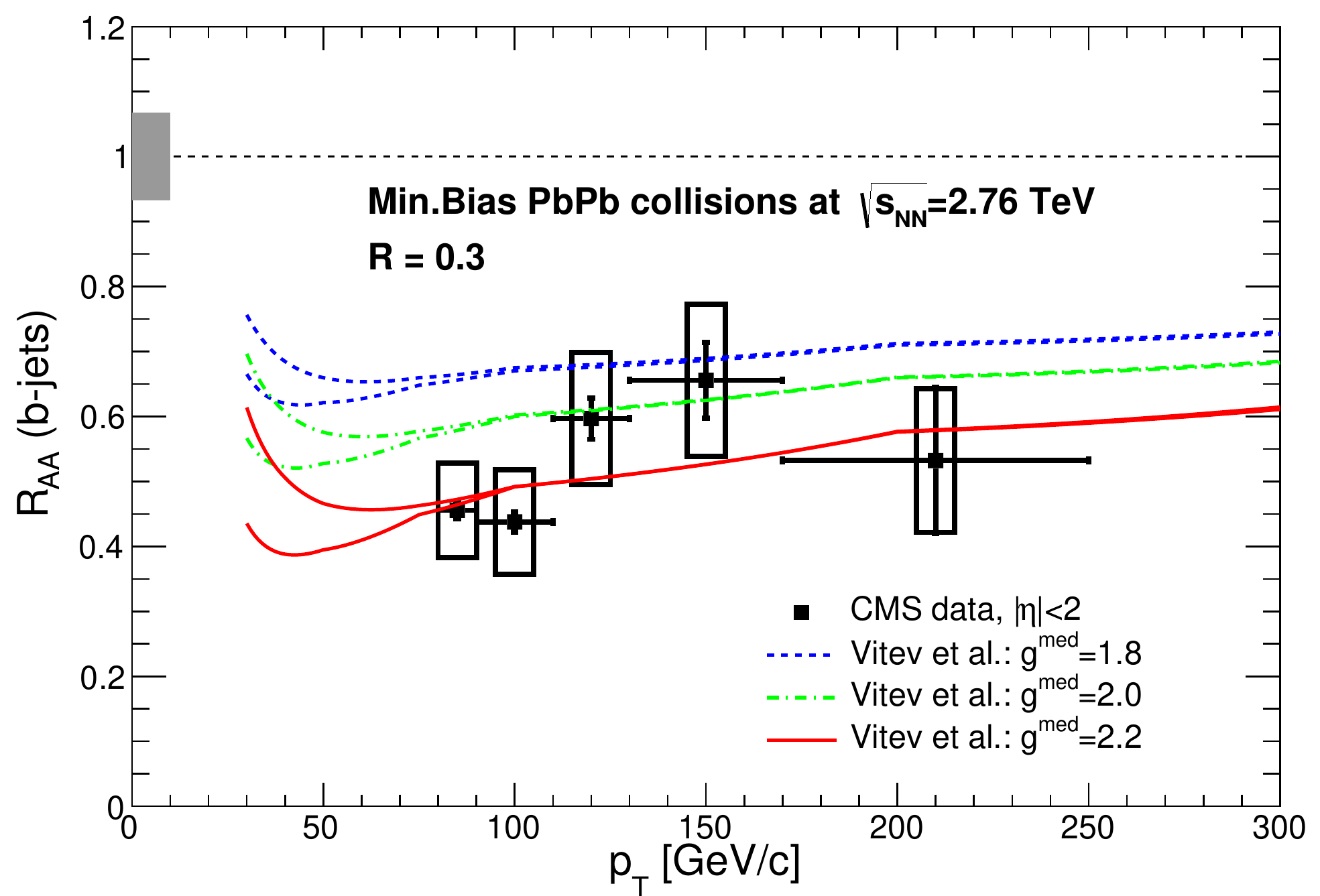} 
 \caption{Nuclear modification factor as a function of transverse momentum of \bquark-jets measured in \pb collisions at \snn=2.76~TeV~\cite{Chatrchyan:2013exa} compared to model calculations. For each value of $g^{\rm med}$, the upper curve is the calculation with the $b$-quark mass and the lower curve is the massless case. The filled box at $\raa=1$ is the systematic uncertainty on the normalisation.} 
 \label{fig:bJetsCMScompTheo} 
\end{figure}

Finally, in \fig{fig:bJetsCMScompTheo} the nuclear modification factor of \bquark-tagged jets measured by the CMS Collaboration in minimum-bias \pb collisions at  $\snn = 2.76$~\TeV is compared with the model described at the end of Section~\ref{sec:pQCDEloss}, including radiative and collisional energy loss. The calculation is shown for three values of the quark--medium coupling parameter $g^{\rm med}$~\cite{Huang:2013vaa}. A precise measurement of this observable in future LHC runs should allow to constrain this parameter to the 10\% level. In addition, an extension of the measurement to transverse momenta lower than 50\GeVc should allow to observe the reduction of suppression due to the mass-dependence of energy loss. 
 
In summary, the comparison of model calculations with currently available data from RHIC and LHC allows for the following considerations: 
\begin{itemize} 
\item the D meson \vtwo measurements at LHC are best described by the models that include 
collisional interactions within a fluid-dynamical expanding medium, as well as hadronisation via recombination; 
\item however, theoretical predictions of the \raa of D mesons from these models are scattered, both at RHIC and LHC, which leaves room for theoretical improvement in the future before reliable conclusions can be drawn; 
\item on the contrary, the models that include radiative and collisional energy loss provide a good description of the D meson \raa, 
but they under-estimate the value of \vtwo at LHC; 
\item the models that include collisional energy loss in a fluid-dynamical expanding medium, hence radial flow, exhibit a bump in the low-\pt D meson \raa,  
which is qualitatively consistent with the RHIC data; 
\item these models predict a bump also at LHC energy, the size of which depends strongly on the nuclear modification of the PDFs (shadowing); the current data 
at LHC are not precise enough to be conclusive in this respect; 
\item most of the models can describe within uncertainties the measurements of \raa and \vtwo for heavy-flavour decay electrons at RHIC (in some models, 
the quark--medium coupling is tuned to describe these data) and of \raa for heavy-flavour decay muons at LHC; 
\item all models predict that the \raa of non-prompt \jpsi from B decays is larger than that of D mesons by about 0.2--0.3 units  
for the \pt region ($\sim 10$\GeVc) for which preliminary data from the LHC experiments exist.  
\end{itemize}

\begin{landscape}
\begin{table}[t]
\caption{Comparative overview of the models for heavy-quark energy loss or transport in the medium described in the previous sections.}
\centering
\begin{tabular}{cccccc}
\hline
{\it Model} & {\it Heavy-quark} & {\it Medium modelling} & {\it Quark--medium} & {\it Heavy-quark} & \it Tuning of medium-coupling \\
 & \it production & & \it interactions & \it hadronisation & \it (or density) parameter(s) \\
\hline
{\bf Djordjevic \etal}    & FONLL & Glauber model  & rad. + coll. energy loss & fragmentation & Medium temperature  \\
   \cite{Djordjevic:2009cr,Djordjevic:2008iz,Djordjevic:2006tw,Djordjevic:2011dd,Djordjevic:2013xoa} &  no PDF shadowing & nuclear overlap & finite magnetic mass & & fixed separately \\
   & & no fl. dyn. evolution&  & &  at RHIC and LHC \\
\hline
{\bf WHDG} & FONLL & Glauber model  & rad. + coll. energy loss & fragmentation & RHIC \\
   \cite{Wicks:2005gt,Wicks:2007am} & no PDF shadowing & nuclear overlap & & & (then scaled with ${\rm d}N_{\rm ch}/{\rm d}\eta$)\\
   &  & no fl. dyn. evolution &  &   &  \\
\hline
{\bf Vitev et al.} & non-zero-mass VFNS & Glauber model  & radiative energy loss & fragmentation & RHIC \\
   \cite{Adil:2006ra,Sharma:2009hn} & no PDF shadowing & nuclear overlap & in-medium meson dissociation & & (then scaled with ${\rm d}N_{\rm ch}/{\rm d}\eta$) \\
  & & ideal fl. dyn. 1+1d  & & & \\
  & & Bjorken expansion & & & \\

\hline
{\bf AdS/CFT (HG)} & FONLL & Glauber model  & AdS/CFT drag & fragmentation & RHIC \\
   \cite{Horowitz:2007su,Horowitz:2011wm} &  no PDF shadowing & nuclear overlap & & & (then scaled with ${\rm d}N_{\rm ch}/{\rm d}\eta$) \\
  &  &  no fl. dyn. evolution &  &  &  \\
 \hline
{\bf POWLANG} & POWHEG (NLO) & 2+1d expansion &  transport with Langevin eq. & fragmentation & assume pQCD (or l-QCD  \\
   \cite{Beraudo:2014iva,Beraudo:2014boa,Alberico:2013bza,Alberico:2011zy,Beraudo:2009pe} &  EPS09 (NLO)  & with viscous  & collisional energy loss & recombination & $U$ potential) \\
   & PDF shadowing &  fl. dyn. evolution &  &  & \\
 \hline 
{\bf MC@$_s$HQ+EPOS2} & FONLL & 3+1d expansion &  transport with Boltzmann eq.  &  fragmentation & QGP transport coefficient \\
   \cite{Gossiaux:2008jv,Gossiaux:2009mk,Nahrgang:2013saa} & EPS09 (LO)  & (EPOS model) & rad. + coll. energy loss & recombination & fixed at LHC, slightly\\
   & PDF shadowing & &  & & adapted for RHIC \\
 \hline 
{\bf BAMPS} & MC@NLO & 3+1d expansion &  transport with Boltzmann eq. &  fragmentation & RHIC \\
   \cite{Uphoff:2010sh,Uphoff:2011ad,Uphoff:2013rka,Uphoff:2014hza} &  no PDF shadowing & parton cascade & rad. + coll. energy loss & & (then scaled with ${\rm d}N_{\rm ch}/{\rm d}\eta$)\\
 \hline 
{\bf TAMU} & FONLL & 2+1d expansion &  transport with Langevin eq.  &  fragmentation & assume l-QCD \\
   \cite{He:2011yi,He:2012df,He:2014cla} & EPS09 (NLO)  & ideal fl. dyn. & collisional energy loss &  recombination & $U$ potential  \\
   & PDF shadowing & & diffusion in hadronic phase  & & \\
 \hline 
{\bf UrQMD} & PYTHIA & 3+1d expansion &  transport with Langevin eq.  &  fragmentation &  assume l-QCD  \\
   \cite{Lang:2012cx,Lang:2013cca,Lang:2013wya} &  no PDF shadowing & ideal fl. dyn. & collisional energy loss &  recombination & $U$ potential \\
 \hline 
{\bf Duke} & PYTHIA & 2+1d expansion &  transport with Langevin eq.  &  fragmentation & QGP transport coefficient \\
   \cite{Cao:2011et,Cao:2013ita} &  EPS09 (LO)  & viscous fl. dyn. & rad. + coll. energy loss &  recombination & fixed at RHIC and LHC \\
   &  PDF shadowing &  &  & & (same value) \\
 \hline 
\end{tabular}
\label{tab:HFmodels}
\end{table}
\end{landscape}

\clearpage

\subsection{Heavy-flavour correlations in heavy-ion collisions: status and prospects}
\label{sec:OHFcorr}
 
Angular correlations of charged hadrons proved to be key observables at RHIC and LHC energies to study energy loss and QGP properties~\cite{Arsene:2004fa,Adcox:2004mh,Back:2004je,Adams:2005dq,Chatrchyan:2011eka,Aamodt:2011vg}, providing measurements that are complementary to single-particle observables like the \raa and \vtwo. 
Two-particle correlation distributions are defined in terms of the ($\Delta\phi$, $\Delta\eta$) distance between a \pt-selected trigger particle and a (set of) associated particles, generally with  
lower \pt than the trigger particle.  
On the near side ($\Delta\phi\sim0$), the correlations provide information on the properties of the jet leaving the medium, while on the away side  ($\Delta\phi\sim \pi$) they reflect the ``survival" probability of the recoiling parton that traverses the medium. Di-hadron correlation measurements typically carry geometrical and kinematical biases~\cite{Renk:2006pk}. Triggering on a high-\pt particle tends to favour the selection of partons produced near the surface of the medium, which lost a small fraction of their energy and could still fragment to hadrons at high \pt (geometrical bias). In addition, when comparing to the vacuum case (\pp collisions) with the same conditions on the trigger particles, one might have different contributions of quark and gluon jets and different partonic energies in nucleus--nucleus and \pp collisions (parton and kinematical biases). 
Together with single particle measurements and fully reconstructed jets, di-hadron correlations can constrain energy loss models by adding information on the path length dependence of the energy loss and the relative contributions of collisional and radiative energy loss. 
 
 
Recent works~\cite{Nahrgang:2013saa,Cao:2014pka,Uphoff:2013rka,Renk:2013xaa,Lang:2013wya,Beraudo:2014boa} have shown that the azimuthal distributions of heavy quark-antiquark pairs are sensitive to the different interaction mechanisms, collisional and radiative. The relative angular broadening of the \QQbar pair does not only depend on the drag coefficient discussed above (see \sect{sec:OHFmodelsInt}) 
but also on the momentum broadening in the direction perpendicular to the initial quark momentum, $\langle p_\perp^2\rangle$, which is not probed 
directly in the traditional \raa and \vtwo observables.  This is one of the motivation for measuring azimuthal correlations of heavy-flavour. 
 
The experimental challenges in measurements like D--$\rm \overline{D}$ correlations in heavy-ion collisions come from the reconstruction of both the hadronic decays of the back-to-back D mesons, which require large statistics to cope with low branching ratios and low signal-to-background in nucleus--nucleus collisions. As an alternative, correlations of D mesons with charged hadrons (D--h), correlations of electrons/muons from decays of heavy-flavour particles with charged hadrons (\electron--h) and correlations of D--\electron, $e^+$--$e^-$,  $\mu^+$--$\mu^-$ and \electron--$\mu$ pairs (where electrons and muons come from heavy-flavour decays) can be studied. 
Such observables might, however, hide decorrelation effects intrinsic to the decay of heavy mesons.  
In addition, in the case of correlations triggered by electrons or muons from heavy-flavour decays, the interpretation of the results is complicated by the fact that the lepton carries only a fraction of the momentum of the parent meson. This makes the understanding of \pp collisions as baseline a very crucial aspect of these analyses (see \sect{sec:pp:AngleCorrelations}).

Heavy-flavour azimuthal correlations are 
being studied in \dAu collisions at RHIC and \pPb collisions both at RHIC and at the LHC to understand how the presence of the nucleus might affect the properties of heavy-flavour pair production. 
The results are discussed in \sect{sec:CNM:OHF}. 
 
%
 
\begin{figure}[!ht] 
 \begin{center} 
  \includegraphics[angle=0,width=0.5\textwidth]{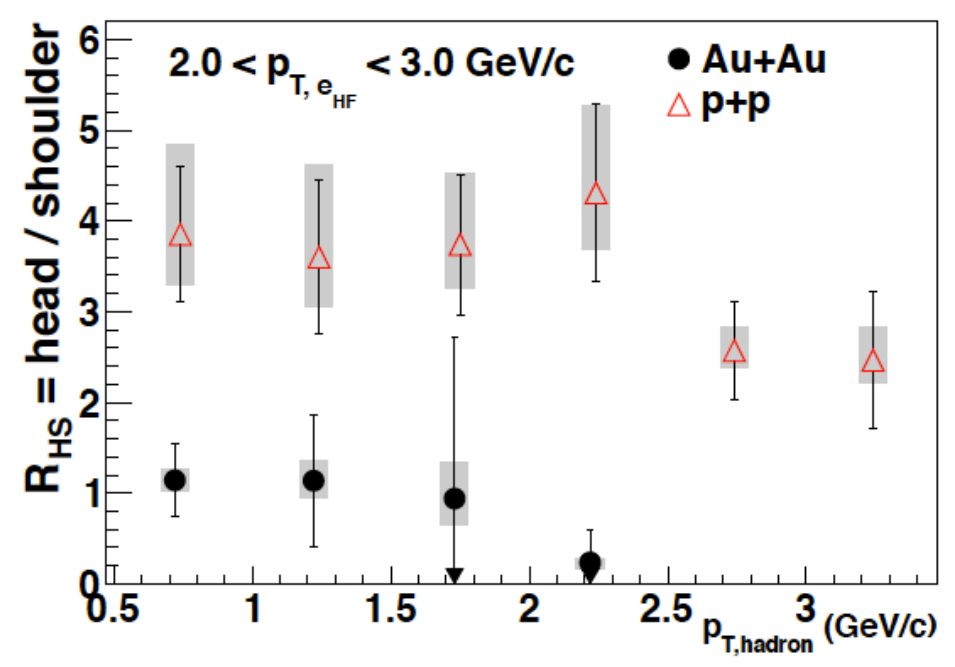} 
  \includegraphics[angle=0,width=0.36\textwidth]{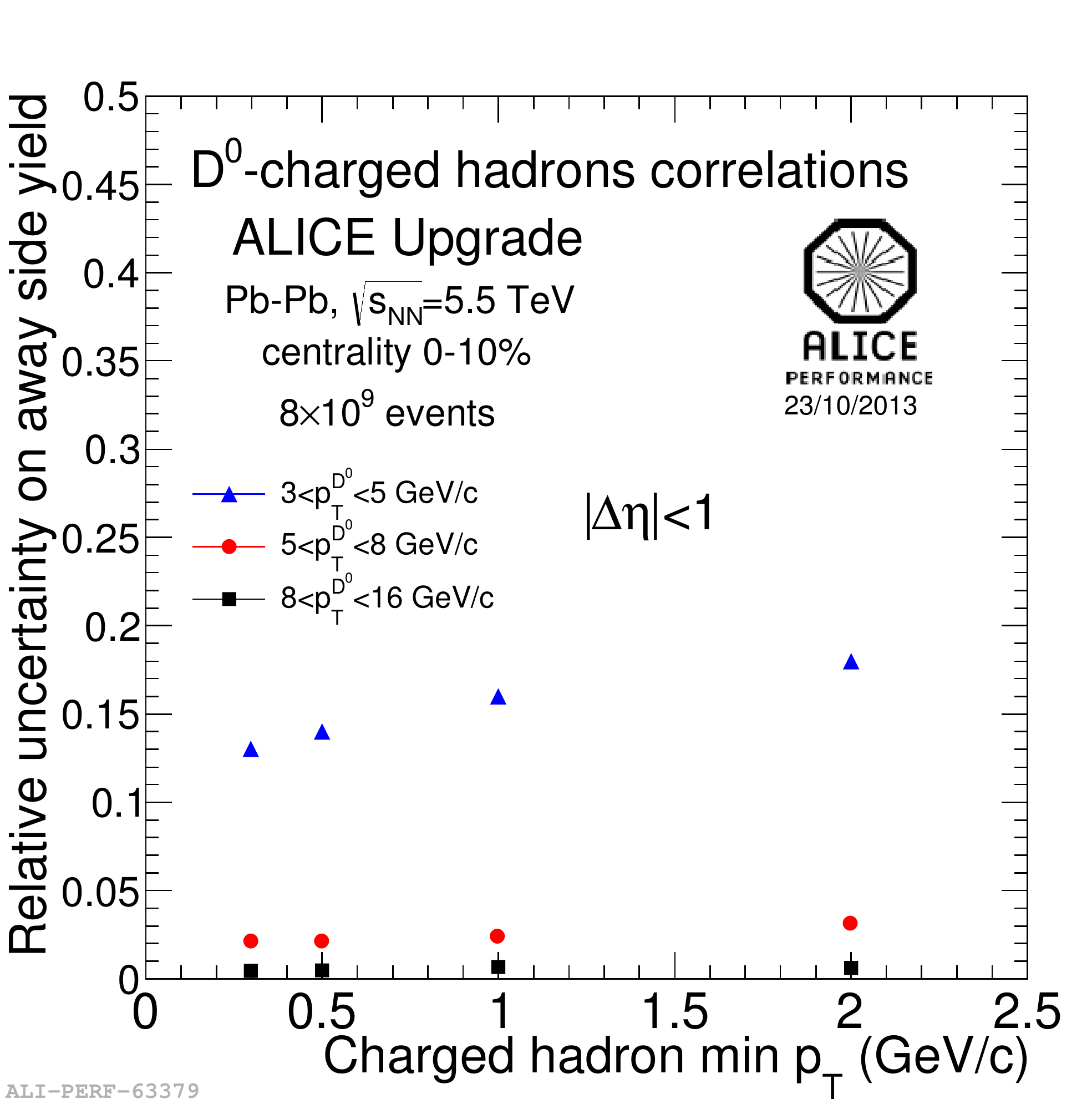} 
     \end{center} 
 \caption{Left: ratio of yields in the away-side region ($2.51<\phi<\pi$) to those in the shoulder region  ($1.25<\phi<2.51$)  in \pp and \AuAu collisions from PHENIX~\cite{Adare:2010ud}. Right: relative uncertainty on the away-side yield in D--h correlations as a function of the charged hadron \pt for three ranges of D meson transverse momenta, with the ALICE and LHC upgrades~\cite{Colamaria:2014uja}.} 
 \label{HFcorrPb}  
\end{figure} 
 
\begin{figure}[!t] 
 \includegraphics[width=0.33\textwidth]{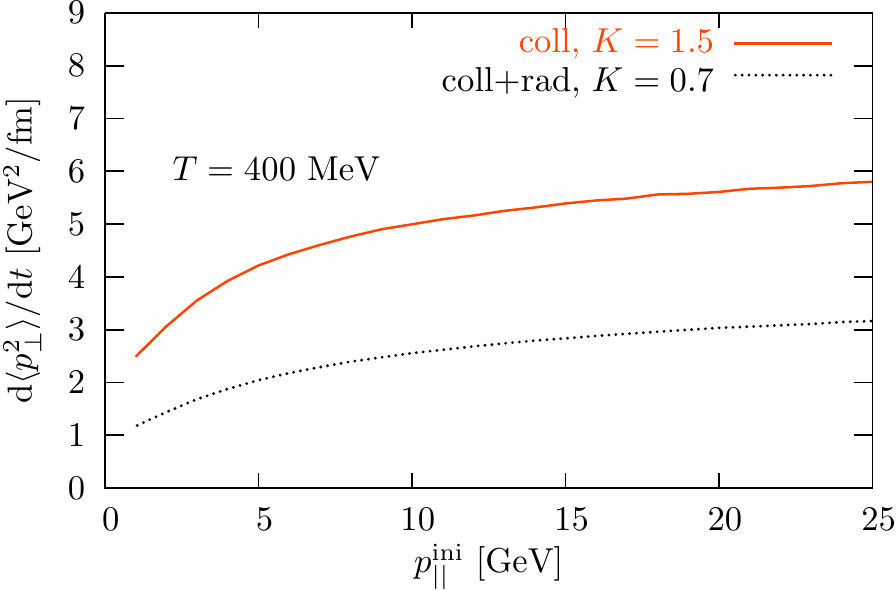} 
 \includegraphics[width=0.33\textwidth]{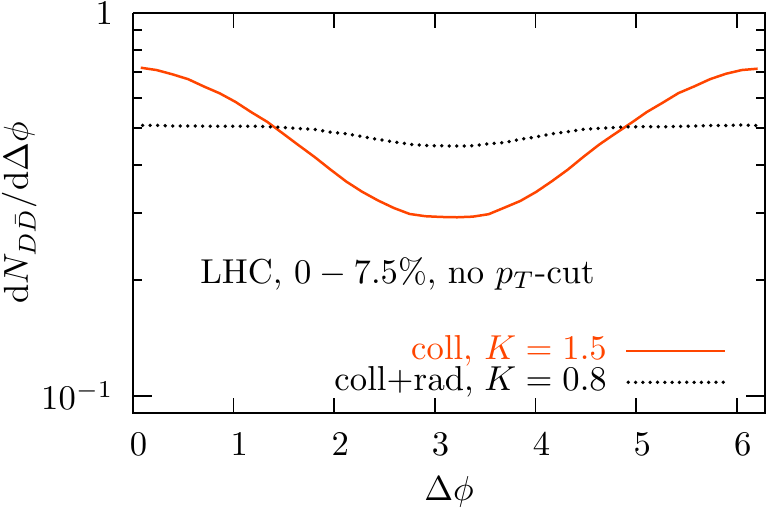} 
 \includegraphics[width=0.32\textwidth]{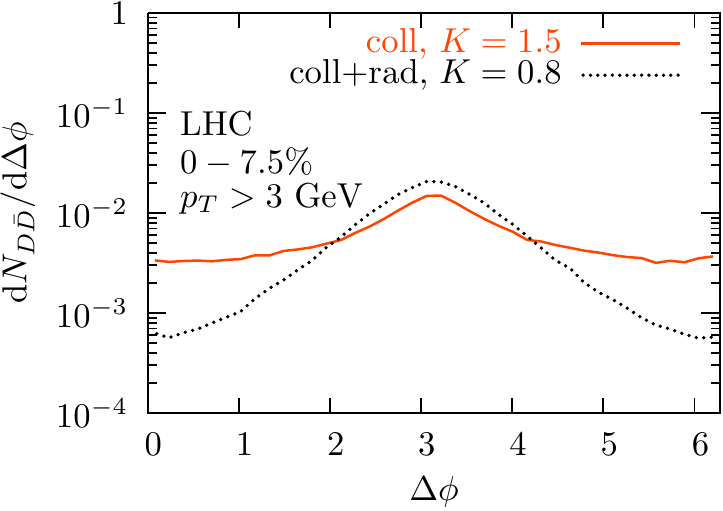} 
 \caption{Perpendicular momentum of charm quarks acquired in a QGP medium at $T=400$\MeV as a function of the initial momentum $p_{||}^{\rm ini}$ (left). 
 Angular correlations of $\mathrm{D\overline{D}}$ pairs in \pb collisions at LHC without (centre) and with (right) a lower momentum cut~\cite{Nahrgang:2013saa}.} 
 \label{fig:correlation} 
 \end{figure}

Measurements of heavy-flavour correlations in nucleus--nucleus collisions were carried out at both RHIC and LHC with \electron--h correlations~\cite{Adare:2010ud,Wang:2008kha,Thomas:2014cwa} (where electrons come from heavy-flavour decays), but the current statistics prevents us from drawing quantitative conclusions. Such measurements are expected to provide more information about how the charm and beauty quarks propagate through the hot and dense medium and how this affects and modifies the correlation structures.  
In particular, PHENIX reported a decrease of the ratio of yields in the away-side region ($2.51<\Delta\phi<\pi$) to those in the shoulder region  ($1.25<\Delta\phi<2.51$)  from \pp to \AuAu collisions (left panel of \fig{HFcorrPb}).  
 
Further measurements of heavy-flavour triggered azimuthal correlations  will be promising in future data takings at both RHIC (with the new silicon tracker detectors) and LHC (with the machine and detector upgrades). As reported in \fig{HFcorrPb}, right, the relative uncertainty on the away-side yield in D--h correlations in central \pb collisions with the ALICE and LHC upgrades will be $\approx 15 \%$ for low-\pt D mesons and only a few percent for intermediate/high \pt.  
 
  
Several theoretical works have recently addressed angular correlations of heavy-flavour particles 
in nucleus--nucleus collisions~\cite{Nahrgang:2013saa,Cao:2014pka,Uphoff:2013rka,Renk:2013xaa,Lang:2013wya,Beraudo:2014boa}. 
However, none of these approaches presently includes the interactions of D and B mesons in the hadronic phase present in the 
late stages of the system evolution. These interactions could add a further smearing on top of QGP-induced modification of the heavy-quark angular correlations.  
For the traditional \raa and \vtwo observables, a first step in this  
direction was made in~\cis{He:2012df,Ozvenchuk:2014rpa,Cao:2014dja}, with effects found of the order of 20\% at most. We now focus on a particular model in order to illustrate the sensitivity of heavy-flavour angular correlations to the type of interaction mechanism~\cite{Nahrgang:2013saa}. In \fig{fig:correlation} (left), the transverse momentum broadening per unit of time is shown as a function of the initial momentum $p_{||}^{\rm ini}$ of charm quarks for the purely collisional and collisional plus radiative (LPM) interactions as applied within the MC@$_s$HQ model (see \sect{sec:mcatshq}). For all initial momenta, $\langle p_\perp^2\rangle$ is larger in a purely collisional interaction mechanism. $\langle p_\perp^2\rangle$ has similar numerical values for charm and for beauty quarks.   
A larger $\langle p_\perp^2\rangle$ leads to a more significant change of the initial relative pair azimuthal angle $\Delta\phi$ during the evolution in the medium. This means that for a purely collisional interaction mechanism one expects a stronger broadening of the initial correlation at $\Delta\phi=\pi$, as seen in the central and right panels of \fig{fig:correlation}. In the central panel, the $\Delta\phi$ distribution of all initially correlated pairs is shown after hadronisation into $\mathrm{D\overline{D}}$ pairs. Since no cut in \pt is applied, these distributions are dominated by low-momentum pairs, while in the right panel a cut of $\pt>3$\GeVc is applied. The low-momentum pairs show the influence of the radial flow of the underlying QGP medium, which tends to align the directions of the quark and the antiquarks toward smaller opening angles. It again happens more efficiently for larger  $\langle p_\perp^2\rangle$ of the underlying interaction mechanism. This effect, which was called ``partonic wind''~\cite{Zhu:2007ne}, 
is thus only seen for the purely collisional interaction mechanism. A \pt threshold reveals clearly the residual correlation around $\Delta\phi\sim\pi$. Here in the purely collisional scenario one sees a larger background of pairs that decorrelated during the evolution in the QGP than for the collisional plus radiative (LPM) scenario.  
  
For these calculations an initial back-to-back correlation has been assumed. Next-to-leading order processes, however, destroy this strict initial correlation already in proton--proton collisions. Unfortunately the theoretical uncertainties on these initial distributions are very large, especially for charm quarks. Here, a thorough experimental study of heavy-flavour correlations in proton--proton and proton--nucleus collisions is very important for validating different initial models. Also enhanced theoretical effort in these reference systems is necessary.


\subsection{Summary and outlook}
\label{sec:OHFconclusion}
 
The LHC \RunOne has provided a wealth of measurements of heavy-flavour production in heavy-ion collisions, 
which have extended and complemented the results from the RHIC programme. The main observations and  
their present interpretation are summarized in the following. 
 
{\it High-$\pt$ region} (above 5--10~GeV/$c$): in this region, heavy-flavour measurements are expected to provide information  
mainly on the  
properties of in-medium energy loss. 
\begin{itemize} 
\item The $\raa$ measurements show a strong suppression with respect to binary scaling in central nucleus--nucleus collisions for D mesons, heavy-flavour decay leptons and J$/\psi$ from B decays. 
The suppression of D mesons and heavy-flavour decay leptons is similar, within uncertainties, at RHIC and LHC energies. 
Given that a suppression is not observed in proton(deuteron)--nucleus collisions, the effect in nucleus--nucleus collisions can be attributed to in-medium energy loss. 
\item The suppression of D mesons with average $\pt$ of about 10~GeV/$c$ is stronger than that of J/$\psi$ decaying from B mesons with similar average $\pt$. This observation, still based on preliminary results, is consistent with the expectation of lower energy loss for heavier quarks and it is described by model calculations that implement radiative and collisional energy loss with this feature. 
\item The suppression of D mesons and pions is consistent within uncertainties at both RHIC and LHC. While there is no experimental evidence of the colour-charge dependence of energy loss, model calculations indicate that similar $\raa$ values can result from the combined effect of colour-charge dependent energy loss and the softer $\pt$ distribution and fragmentation function of gluons with respect to $c$ quarks. 
\item At very high-$\pt$ (above 100~GeV/$c$), a similar $\raa$ is observed for $b$-tagged jets and inclusive jets. This observation is consistent with a negligible effect of the heavy quark mass at these scales.  
\end{itemize} 
 
{\it Low-$\pt$ region} (below 5--10~GeV/$c$): in this region, heavy-flavour measurements  
are expected to provide information on the total production yields (and the role of initial-state effects) 
and on heavy-quark in-medium dynamics (participation to collective expansion, in-medium hadronisation effects). 
\begin{itemize} 
\item The measurements of electrons (in particular) and D mesons at RHIC show that the total production of charm quarks 
is consistent with binary scaling within uncertainties of about 30--40\%. The available measurements at LHC do not extend  
to sufficiently-small $\pt$ to provide an estimate of the total yields. 
\item The D meson $\raa$ at RHIC energy shows a pronounced maximum at $\pt$ of about 1--2~GeV/$c$ (where $\raa$ becomes larger than unity). This feature is not observed in the measurements at LHC energy. Model calculations including collisional (elastic) interaction processes in an expanding medium and a contribution of hadronisation via in-medium quark recombination, as well as initial-state  
gluon shadowing, describe qualitatively the behaviour observed at both energies. In these models the bump at RHIC is due to radial flow  
and the effect on $\raa$ at LHC is strongly reduced because of the harder $\pt$ distributions and of the effect of gluon shadowing. 
\item A positive elliptic flow $v_2$ is measured in non-central collisions for D mesons at LHC and heavy-flavour decay leptons  
at RHIC and LHC. The D meson $v_2$ at LHC is comparable to that of light-flavour hadrons (within uncertainties of about 30\%). 
These measurements indicate that the interaction with the medium constituents transfers information about the azimuthal anisotropy of the system to charm quarks. 
The $v_2$ measurements are best described by the models that include collisional interactions within a  
fluid-dynamical expanding medium, as well as hadronisation via recombination. 
\end{itemize} 
 
The main open questions in light of these observations are: 
\begin{itemize} 
\item Does the total charm and beauty production follow binary scaling or is there a significant gluon shadowing effect? This requires a precise measurement of charm and beauty production down to zero $\pt$, in proton--proton, proton--nucleus and nucleus--nucleus collisions. 
\item Can there be an experimental evidence of the colour-charge dependence of energy loss? This requires a precise comparison of D mesons and pions in the intermediate $\pt$ region, at both RHIC and LHC. 
\item Is the difference in the nuclear modification factor of charm and beauty hadrons consistent with the  
quark mass dependent mechanisms of energy loss? Can it provide further insight on these mechanisms (for example, the gluon 
radiation angular distribution)? This requires a precise measurement of D and B meson (or J/$\psi$ from B) $\raa$ over a wide $\pt$ range and as a function of collision centrality. This will also be mandatory in order to extract the precise path-length 
dependence of energy loss, which cannot be extracted from the actual data. 
\item Does the positive elliptic flow observed for D mesons and heavy-flavour decay leptons result from the charm quark interactions 
in the expanding medium? Are charm quarks thermalised in the medium? Is there a contribution (dominant?) inherited from light quarks 
via the recombination process? What is the contribution from the path length dependence of energy loss? This requires precise measurements of the elliptic flow and of the higher order flow coefficients of charm and beauty hadrons over a wide $\pt$ interval, and their comparison with light-flavour hadrons. 
\item What is the role of in-medium hadronisation and of radial flow for heavy quarks? This requires measurements of $\raa$ and $v_2$ 
of heavy flavour hadrons with different quark composition and different masses, namely D, ${\rm D}_s$, B, ${\rm B}_s$, $\Lambda_c$, $\Xi_c$,  
$\Lambda_b$.      
\item What is the relevance of radiative and collisional processes in heavy quark energy loss? What is the path length dependence of the two types of processes? This requires precise simultaneous measurements of the $\raa$ and $v_2$ and their comparison with model calculations. Heavy-quark correlations are also regarded as a promising tool  
in this context.  
\end{itemize} 
The outlook for addressing these open questions with the future experimental programmes at RHIC and LHC is discussed in \sect{sec:upgrade}. 
 
From the theoretical point of view, 
a wide range of models, also with somewhat different ``ingredients", can describe most of the available data, at least qualitatively. The main challenges in the theory sector is thus to connect the data with the fundamental properties of the QGP and of the theory of the strong interaction. For this purpose, it is important to identify the features of the  quark--medium interaction that are needed for an optimal description of all aspects of the data and to reach a uniform treatment of the ``external inputs" in the models (e.g. using state-of-the-art pQCD baseline, fragmentation functions and fluid-dynamical medium description, and fixing transport coefficients on those that will be ultimately obtained from lattice calculations for finite \pt).


\section{Quarkonia in nucleus--nucleus collisions}
\label{sec:quarkonia}

\label{sec:intro_quarkonia}

Quarkonia are considered important probes of the QGP formed in heavy-ion 
collisions. In a hot and deconfined medium quarkonium production is 
expected to be significantly suppressed with respect to the proton-proton yield, 
scaled by the number of binary nucleon-nucleon collisions, as long as the total 
charm cross section remains unmodified\footnote{As open heavy flavour and 
  quarkonia are produced via the same processes, any modifications of the 
  initial state will not modify the yield ratio of quarkonia to open heavy 
  flavour states.}. The origin of such a suppression, taking place in the QGP, 
is thought to be the colour screening of the force that binds the \ccbar 
(\bbbar) state~\cite{Matsui:1986dk}. In this scenario, quarkonium suppression 
should occur sequentially, according to the binding energy of each meson: 
strongly bound states, such as the \upsa\ or the \jpsi, should melt at higher 
temperatures with respect to the more loosely bound ones, such as the \chib, 
\upsb, or \upsc for the bottomonium family or the \psiP and the \chic for the 
charmonium one. As a consequence, the in-medium dissociation probability of 
these states should provide an estimate of the initial temperature reached in 
the collisions~\cite{Digal:2001ue}. However, the prediction of a sequential 
suppression pattern is complicated by several factors. Feed-down 
decays of higher-mass resonances, and of $b$-hadrons in the case of charmonium,  
contribute to the observed yield of quarkonium states. Furthermore, other hot and cold 
matter effects can play a role, competing with the suppression mechanism. 
 
On the one hand, the production of $c$ and $\overline{c}$ quarks increases with increasing  
centre-of-mass energy. Therefore, at high energies, as at the LHC, 
the abundance of $c$ and $\overline{c}$ quarks might lead to a new charmonium 
production source: the (re)combination of these quarks throughout the collision 
evolution~\cite{Thews:2000rj} or at the hadronisation 
stage~\cite{BraunMunzinger:2000px,Stachel:2013zma}. This additional charmonium 
production mechanism, taking place in a deconfined medium, enhances the \jpsi 
yield and might counterbalance the expected \jpsi suppression. Also the \bbbar 
cross section increases with energy, but, given the smaller number of \bbbar 
pairs, with respect to \ccbar, this contribution is less important for 
bottomonia even in high-\snn collisions. 
 
On the other hand, quarkonium production is also affected by several effects 
related to cold matter (the so-called cold nuclear matter effects, CNM) 
discussed in \sect{Cold nuclear matter effects}. For example, the production 
cross section of the \QQbar pair is influenced by the kinematic parton 
distributions in nuclei, which are different from those in free protons and 
neutrons (the so-called nuclear PDF effects). In a similar way, approaches based 
on the Colour-Glass Condensate (CGC) effective theory assume that a gluon 
saturation effect sets in at high energies. This effect influences the 
quarkonium production occurring through fusion of gluons carrying small values 
of the Bjorken-$x$ in nuclei. Furthermore, parton energy loss in the nucleus may 
decrease the pair momentum, causing a reduction of the quarkonium production at 
large longitudinal momenta. Finally, while the \QQbar pair evolves towards the 
fully-formed quarkonium state, it may also interact with partons of the crossing 
nuclei and eventually break-up. This effect is expected to play a dominant role 
only for low-\snn collisions, where the crossing time of the (pre)-resonant 
state in the nuclear environment is rather large. On the contrary, this 
contribution should be negligible at high-\snn, where, due to the decreased 
crossing time, resonances are expected to form outside the nuclei. 
Cold nuclear matter effects are investigated in proton-nucleus collisions. 
 Since these effects are present also in 
nucleus-nucleus interactions, a precise knowledge of their role is crucial in 
order to correctly quantify the effects related to the formation of hot QCD 
matter. 
 
The in-medium modification of quarkonium production, induced by either hot or 
cold matter mechanisms, is usually quantified through the nuclear modification 
factor \raa, defined as the ratio of the 
quarkonium yield in \AAcoll collisions ($N_{\rm AA}^{\QQbar}$) and the expected 
value obtained by scaling the production cross section in pp collisions ($\sigma_{\rm 
  pp}^{\QQbar}$) by the average nuclear overlap function ($\av{\taa}$), evaluated through a 
Glauber model calculation~\cite{Aamodt:2010cz}: 
\begin{equation} 
  \centering 
  \raa = \frac{N_{\rm AA}^{\QQbar}}{\langle T_{\rm AA} \rangle \times \sigma_{\rm pp}^{\QQbar}}\,. 
\end{equation} 
\raa is expected to equal unity if nucleus--nucleus collisions behave 
as a superposition of nucleon--nucleon interactions. This is, \eg, the case for 
electroweak probes (direct $\gamma$, W, and Z) that do not interact 
strongly~\cite{Afanasiev:2012dg,Chatrchyan:2012vq,Chatrchyan:2012nt,Aad:2012ew,Chatrchyan:2014csa}. 
Such a scaling is assumed to approximately hold for the total charm cross 
section, although an experimental verification has large uncertainties at RHIC 
($\approx 30\%$)~\cite{Adare:2010de,Adamczyk:2014uip} and is still lacking at the LHC 
(see discussion in \sect{OHF}).  A value of \raa 
different from unity implies that the quarkonium production in \AAcoll is 
modified with respect to a binary nucleon-nucleon scaling. Further insight on 
the in-medium modification of quarkonium production can be obtained by  
investigating the rapidity and transverse momentum dependence of the nuclear 
modification factor. 
 
The information from \raa can be complemented by the study of the quarkonium 
azimuthal distribution with respect to the reaction plane, defined by the beam 
axis and the impact parameter vector of the colliding nuclei. The second 
coefficient of the Fourier expansion of the  
azimuthal distribution, \vtwo, is called elliptic flow, as explained in 
\sect{OHF}. Being sensitive to the dynamics of the partonic stages of 
heavy-ion collisions, \vtwo can provide details on the 
quarkonium production mechanisms: in particular, \jpsi produced through a 
recombination mechanism, should inherit the elliptic flow of the charm quarks in 
the QGP, acquiring a positive \vtwo. 
 
Studies performed for thirty years, first at the SPS (\snn = 17\GeV) and then at 
RHIC (\snn = 39--200\GeV)\footnote{References to experimental results 
  are reported in Tables~\ref{tab:expSummary_SPS} 
  and~\ref{tab:expSummary_RHIC}.}, have indeed shown a reduction of the \jpsi 
yield beyond the expectations from cold nuclear matter effects (such as nuclear 
shadowing and \ccbar break-up). 
Even if the centre-of-mass energies differ by a factor of ten, the amount of suppression, with respect to \pp collisions, observed by SPS and RHIC experiments at midrapidity is rather similar.
This observation suggests the existence of an additional contribution to \jpsi production, the previously mentioned (re)combination process, which sets in already at RHIC energies and can compensate for some of the quarkonium suppression due to screening in the QGP. Furthermore, \jpsi suppression at RHIC is, unexpectedly, smaller at midrapidity than at forward rapidity ($y$), in spite of the higher energy density which is reached close to $y \sim 0$. The stronger \jpsi\ suppression at forward-$y$ might be considered a further indication of the role played by (re)combination processes. Note however that the rapidity dependence of the (re)combination contribution is expected to be rather small~\cite{Zhao:2007hh,Zhao:2008pp}. On the other hand, at RHIC energies, cold nuclear matter effects can also explain the observed difference~\cite{Brambilla:2010cs}, at least partially. 
  
The measurement of charmonium production is especially promising at 
the LHC, where the higher energy density reached in the medium and the larger 
number of \ccbar pairs produced in central \PbPb collisions (increased by a 
factor ten with respect to RHIC energies, see \fig{fig:pp:CharmAndBottomXsec}) should help to disentangle suppression 
and (re)combination scenarios. Furthermore, at LHC energies also bottomonium 
states, which were barely accessible at lower energies, are abundantly produced. 
Bottomonium resonances should shed more light on the processes affecting the 
quarkonium behaviour in the hot matter. The \ups mesons are, as previously 
discussed, expected to be less affected by production through (re)combination 
processes, due to the much smaller abundance of $b$ and $\overline{b}$ quarks in 
the medium with respect to $c$ and $\overline{c}$ (in central \PbPb collisions 
at LHC energies, the number of \ccbar is a factor $\sim 20$ higher than the 
number of \bbbar pairs). Furthermore, due to the larger mass of the $b$ quark, 
cold nuclear matter effects, such as shadowing, are expected to be less 
important for bottomonium than for charmonium states. 
 
\begin{figure}[!t] 
  \centering 
   \includegraphics[width=0.4\textwidth]{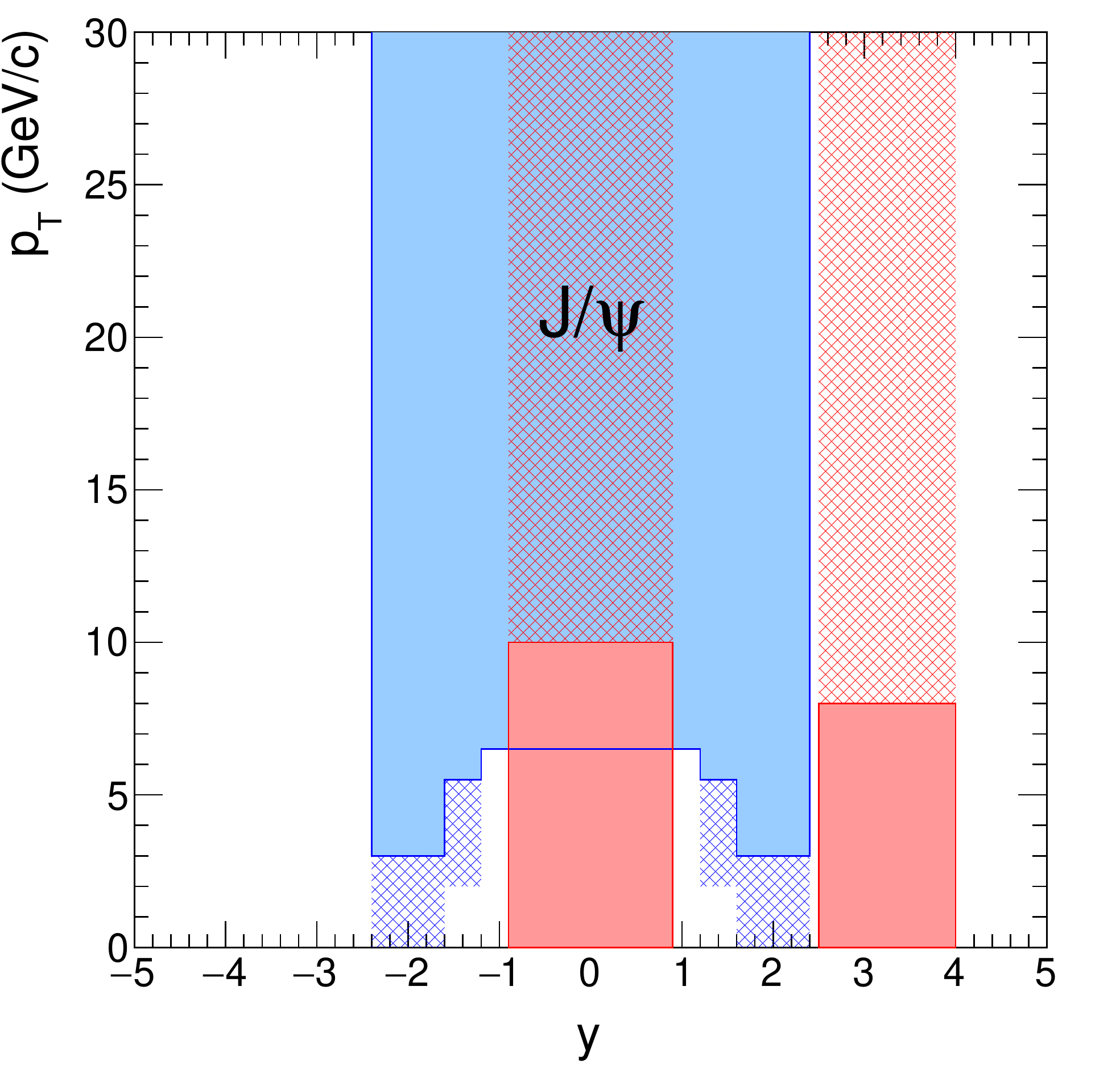} 
   \includegraphics[width=0.4\textwidth]{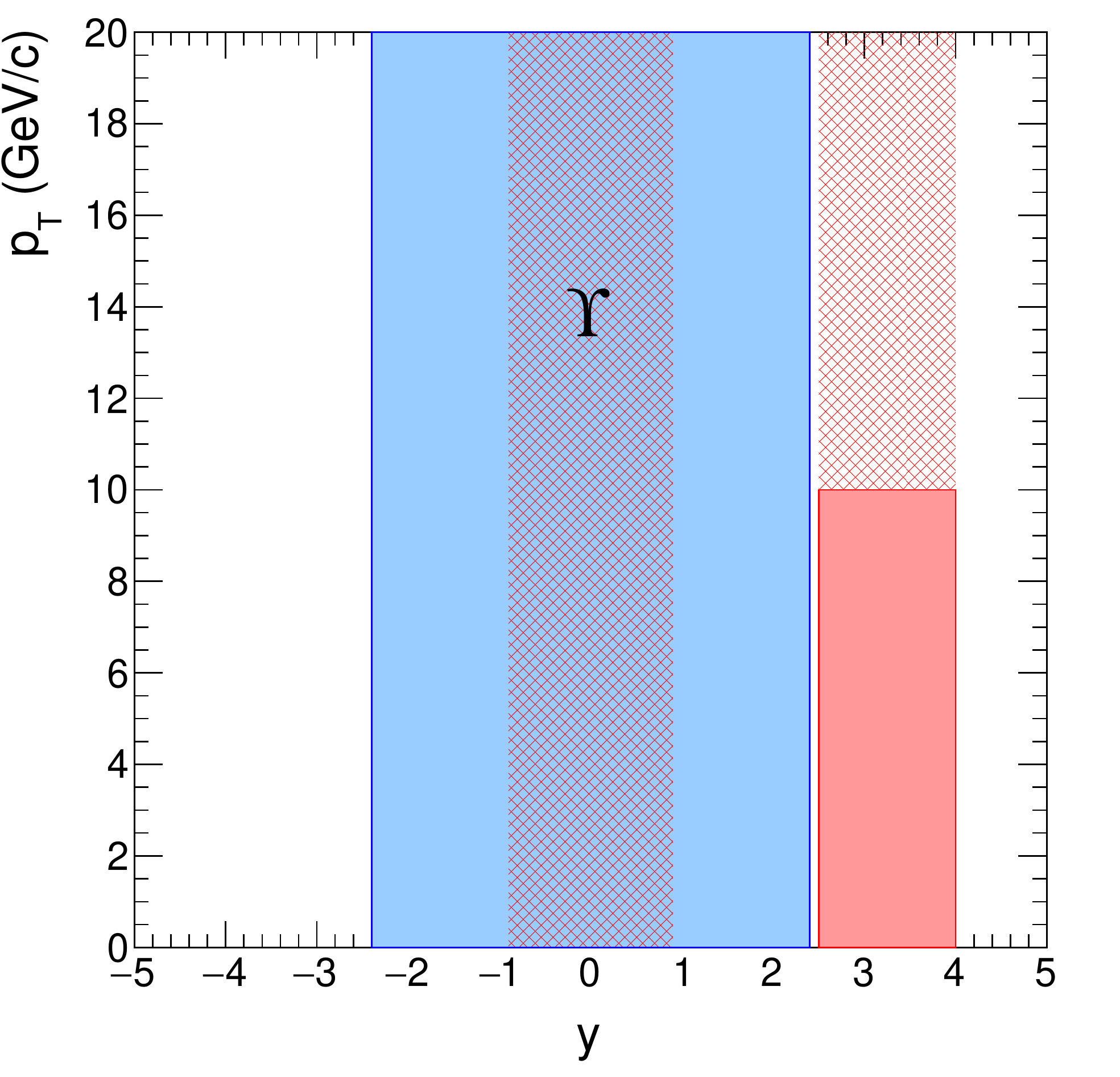} 
   \caption{Left: \pt-$y$ acceptance coverage of the ALICE (red) and CMS (blue) 
     experiments for \jpsi. Right: \pt-$y$ acceptance coverage of the ALICE and 
     CMS experiments for \upsn. Filled areas correspond the the ranges 
     investigated in recent ALICE and CMS quarkonium publications. The hashed 
     areas correspond to the acceptance range which can potentially be covered 
     by the experiments. In fact, while the high-\pt reach in ALICE is limited 
     by statistics, the low-\pt \jpsi coverage by CMS is limited by the muon 
     identification capabilities, affected by the large background in \PbPb 
     collisions.} 
  \label{fig:Acceptance} 
\end{figure} 
 
\begin{table*}[!t] 
  \centering 
  \caption{Quarkonium results obtained in \AAcoll at SPS. The nucleon-nucleon 
    energy in the centre-of-mass frame (\snn), the covered kinematic range, the 
    probes and observables are reported.} 
  \label{tab:expSummary_SPS} 
  \begin{tabular*}{\textwidth}{@{\extracolsep{\fill}}llllll@{~}l@{}} 
    \hline 
    Probe & Colliding  & \snn & $y$ & \pt  & Observables & Ref. \\ 
    & system & (\GeV) &  & (\GeVc)& & \\ 
    \hline 
    \multicolumn{7}{c}{NA38}\\ 
    \hline 
    \jpsi & S--U & 17.2 & $0<y<1$ & $\pt>0$& $\sigma_{\jpsi}$, $\sigma_{\jpsi}/\sigma_{\text{Drell-Yan}}(\text{cent.})$ & \cite{Abreu:1998wx}\\ 
    \psiP & & & & & $\sigma_{\psiP}$, $\sigma_{\psiP}/\sigma_{\text{Drell-Yan}}(\text{cent.})$ & \\ 
    \hline 
    \multicolumn{7}{c}{NA50}\\ 
    \hline 
    \jpsi & \pb & 17.2 & $0<y<1$ & $\pt>0$& yield(\pt), $\sigma_{\jpsi}$ and $\sigma_{\jpsi}/\sigma_{\text{Drell-Yan}}(\text{cent.})$ & \cite{Abreu:1997ji,Abreu:1997jh,Abreu:1999qw,Abreu:2000ni,Abreu:2000xe,Abreu:2001kd,Alessandro:2004ap}\\ 
    \psiP & & & & & yield(\pt), $\sigma_{\psiP}/\sigma_{\text{Drell-Yan}}$ and $\sigma_{\psiP}/\sigma_{\jpsi}(\text{cent.})$ & \cite{Abreu:2000xe,Alessandro:2006ju}\\ 
    \hline 
    \multicolumn{7}{c}{NA60} \\ 
    \hline 
    \jpsi & In--In & 17.2 & $0<y<1$ & $\pt>0$& $\sigma_{\jpsi}/\sigma_{\text{Drell-Yan}}(\text{cent.})$ & \cite{Arnaldi:2007zz,Arnaldi:2009ph}\\ 
    & & & & & polarization & \cite{Arnaldi:2009ph}\\  
    \hline 
  \end{tabular*} 
\end{table*} 
\begin{table*}[!h] 
  \centering 
  \caption{Quarkonium results obtained in \AAcoll from RHIC experiments. The 
    experiment, the probes, the collision energy (\snn), the covered kinematic range 
    and the observables are indicated.} 
 \label{tab:expSummary_RHIC} 
  \begin{tabular*}{\textwidth}{@{\extracolsep{\fill}}llllll@{~}l@{}} 
    \hline 
    Probe  & Colliding  & \snn  & $y$  & \pt  & Observables & Ref. \\ 
     & system & (\GeV) &  & (\GeVc)& & \\ 
    \hline 
    \multicolumn{7}{c}{PHENIX} \\ 
    \hline 
    \jpsi & \AuAu & 200 & $1.2<|y|<2.2$  & $\pt>0$ & yield and $\raa(\text{cent.},\,\pt,\,y)$&\cite{Adler:2003rc,Adare:2006ns,Adare:2011yf} \\ 
    & & & $|y|<0.35$ & & & \\ 
    & & & & $0<\pt<5$ & $\vtwo(\pt,\,y)$ & \cite{Silvestre:2008tw}\\ 
    & & & $1.2<|y|<2.2$ & & & \cite{Atomssa:2009ek}\\ 
    & \CuCu & & $1.2<|y|<2.2$ & $\pt>0$ & yield and $\raa(\text{cent.},\,\pt,\,y)$ & \cite{Adare:2008sh}\\ 
    &  &   & $|y|<0.35$ & & &\\ 
    & \CuAu & & $1.2<|y|<2.2$ & & yield and $\raa(\text{cent.},\,y)$ &\cite{Adare:2014nsa}\\ 
    & \UU & 193 & $1.2<|y|<2.2$ & $\pt>0$ & $\raa(\text{cent.})$ & \cite{daSilva:2014ria}\\
    & \AuAu & 62.4 & & & yield$(\text{cent.},\,\pt)$, $\raa(\text{cent.})$ & \cite{Adare:2012wf}\\ 
    & & 39 & & & &\\ 
    \upsabc & & 200  & $|y|<0.35$ & & yield, $\raa\text{(cent.)}$& \cite{Adare:2014hje}\\ 
    \hline 
    \multicolumn{7}{c}{STAR}\\ 
    \hline 
    \jpsi & \AuAu & 200 & $|y|<1$ & $\pt>0$ & yield and $\raa(\text{cent.},\,\pt)$ & \cite{Adamczyk:2012ey,Adamczyk:2013tvk}\\ 
     & & & & & $\vtwo(\text{cent.},\,\pt)$ & \cite{Adamczyk:2012pw}\\ 
     & \CuCu & & & & yield and $\raa(\text{cent.},\,\pt)$  & \cite{Abelev:2009qaa,Adamczyk:2013tvk}\\ 
     & \UU & 193 & & & $\raa(\pt)$ & \cite{Zha:2014nia}\\ 
     & \AuAu & 62.4 & & & yield, $\raa(\text{cent.},\,\pt)$ & \\ 
     & & 39 & & & & \\ 
     \upsa & & 200 & & & $\sigma$ and $\raa(\text{cent.})$& \cite{Adamczyk:2013poh}\\ 
     \upsabc & & & & & $\raa(\text{cent.})$ & \\ 
     & \UU & 193 & & & & \cite{Zha:2014nia}\\ 
    \hline 
  \end{tabular*} 
\end{table*}

The four large LHC experiments (ALICE, ATLAS, CMS, and LHCb) have carried out 
studies on quarkonium production either in \PbPb collisions at \snn = 
2.76\TeV\footnote{References to experimental results are reported in 
  \tab{tab:expSummary_LHC}.} or in \pPb collisions at \snn = 5.02\TeV. Quarkonium 
production has been also investigated in \pp interactions at \s = 2.76, 7 and 
8\TeV.  
The four 
experiments are characterised by different kinematic coverages, allowing one to 
investigate quarkonium production in $|y|<4$, down to zero transverse momentum. 
 
ATLAS and CMS are designed to measure quarkonium production by reconstructing the 
various states in their dimuon decay channel. They both cover the mid-rapidity 
region: depending on the quarkonium state under study and on the \pt range 
investigated, the CMS rapidity coverage can reach up to $|y|<2.4$, and a similar 
$y$ range is also covered by ATLAS. ALICE measures quarkonium in two rapidity 
regions: at mid-rapidity ($|y|<0.9$) in the dielectron decay channel and at 
forward rapidity ($2.5<y<4$) in the dimuon decay channel, in both cases down to 
zero transverse momentum. LHCb has taken part only in the \pp and \pA LHC 
programmes during Run~1 and their results on quarkonium production, reconstructed through the 
dimuon decay channel, are provided at forward rapidity ($2<y<4.5$), down to zero 
\pt. As an example, the \pt-$y$ acceptance coverages of the ALICE and CMS 
experiments are sketched in \fig{fig:Acceptance} for \jpsi (left) and \ups 
(right). 
 
In Tables ~\ref{tab:expSummary_SPS}--\ref{tab:expSummary_LHC}, a summary of the 
charmonium and bottomonium results obtained in \AAcoll collisions by the SPS, 
RHIC, and LHC experiments are presented, respectively.

\begin{table*}[!t] 
  \centering 
  \caption{Quarkonium results obtained in \AAcoll from LHC experiments. The 
    experiment, the probes, the collision energy (\snn), the covered kinematic range 
    and the observables are indicated.} 
  \label{tab:expSummary_LHC} 
  \begin{tabular*}{\textwidth}{@{\extracolsep{\fill}}llllll@{~}l@{}} 
    \hline 
    Probe & Colliding  & \snn  & $y$ & \pt  & Observables & Ref. \\ 
      & system &  (\TeV) &  & (\GeVc)& & \\ 
    \hline 
    \multicolumn{7}{c}{ALICE} \\ 
    \hline 
    \jpsi & \pb & 2.76 & $|y|<0.9$ & $\pt>0$ & $\raa(\text{cent.,\,\pt})$ & \cite{Abelev:2013ila,Adam:2015rba}\\  
    & & & $2.5<y<4$ & $\pt>0$ & $\raa(\text{cent.},\,\pt,\,y)$ & \cite{Abelev:2012rv,Abelev:2013ila}\\  
    & & & & $0<\pt<10$ & $\vtwo(\text{cent.},\,\pt)$ & \cite{ALICE:2013xna}\\ 
    \psiP & & & & $\pt<3$ & $\frac{(N_{\psiP}/N_{\jpsi})_{\mathrm{Pb-Pb}}}{(N_{\psiP}/N_{\jpsi})_{\mathrm{pp}}}(\text{cent.})$ & \cite{Arnaldi:2012bg}\\ 
    & & & & $3<\pt<8$ & & \\ 
    \upsa & & & & $\pt>0$ & $\raa(\text{cent.},\,y)$ & \cite{Abelev:2014nua} \\ 
    \hline 
    \multicolumn{7}{c}{ATLAS} \\ 
    \hline 
    \jpsi & \pb & 2.76 & $|\eta|<2.5$ & $\pt\gtrsim6.5$ & $\rcp(\text{cent.})$ & \cite{Aad:2010aa}\\ 
    \hline 
    \multicolumn{7}{c}{CMS} \\ 
    \hline 
    \jpsi (prompt) & \pb & 2.76   & $|y|<2.4$ & $6.5<\pt<30$& yield and $\raa(\text{cent.},\,\pt,\,y)$ & \cite{Chatrchyan:2012np}\\  
    & & & & & $\vtwo(\text{cent.},\,\pt,\,y)$ & \cite{CMS:2013dla}\\ 
    & & & $1.6<|y|<2.4$ & $3<\pt<30$ & &\\ 
    & & & $|y|<1.2$ & $6.5<\pt<30$ & yield and \raa & \cite{Chatrchyan:2012np}\\ 
    & & & $1.2<|y|<1.6$ & $5.5<\pt<30$ & & \\ 
    & & & $1.6<|y|<2.4$ & $3<\pt<30$ & &\\ 
    \psiP (prompt) & & & $1.6<|y|<2.4$ & $3<\pt<30$ & \raa, $\frac{(N_{\psiP}/N_{\jpsi})_{\mathrm{Pb-Pb}}}{(N_{\psiP}/N_{\jpsi})_{\mathrm{pp}}}(\text{cent.})$ & \cite{Khachatryan:2014bva}\\ 
    & & & $|y|<1.6$ & $6.5<\pt<30$ & & \\ 
    \upsa & & & $|y|<2.4$ & $\pt>0$ & yield and $\raa(\text{cent.},\,\pt,\,y)$ & \cite{Chatrchyan:2012np}\\ 
    \upsn & & & $|y|<2.4$ & $\pt>0$ & $\raa(\text{cent.})$ & \cite{Chatrchyan:2011pe,Chatrchyan:2012lxa}\\  
    & & & & & $\frac{(N_{\upsb}/N_{\upsa})_{\mathrm{Pb-Pb}}}{(N_{\upsb}/N_{\upsa})_{\mathrm{pp}}}(\text{cent.})$ & \cite{Chatrchyan:2013nza}\\  
    \hline 
  \end{tabular*} 
\end{table*} 

This section is organised as follows. In the first part, a theoretical overview is 
presented, in which the sequential suppression pattern of quarkonia and the 
lattice calculations are introduced. Other effects, such as modifications of the 
parton distribution functions inside nuclei and their influence on 
nucleus-nucleus collisions are discussed. Along with the suppression, the 
enhancement of quarkonia is also considered through two different approaches to 
(re)generation: the statistical hadronisation model and transport models. In the 
context of bottomonium studies,  
non-equilibrium effects on quarkonium suppression in the anisotropic 
hydrodynamic framework are also discussed. Finally, the collisional 
dissociation model and the comover interaction model are briefly 
introduced. 
 
In the second part, experimental quarkonium results are reviewed. The recent LHC results, 
starting with a brief discussion on the quarkonium production cross sections in 
\pp collisions as necessary references to build the nuclear modification 
factors, are presented. The description of the experimental \raa results for 
\jpsi production, both at low and high \pt is then addressed. The LHC 
results are compared to those at RHIC energies and to  
theoretical models. A similar discussion is also introduced for the \jpsi 
azimuthal anisotropy. Results obtained at RHIC from variations of the 
beam-energy and collision-system are also addressed. The charmonium section is 
concluded with a discussion of \psiP production. 
Next, the bottomonium results on ground and excited states at RHIC and LHC energies are discussed.  
 
Finally, other possible 
references for the quarkonium behaviour in nucleus-nucleus collisions, namely 
proton--nucleus collisions and open heavy flavour, production are discussed.


\subsection{Theory overview}

\subsubsection{Sequential suppression and lattice QCD}
\label{sec:seq-supp}

Historically, the large masses of charm and beauty quarks provide the basis for a quarkonium spectroscopy through 
non-relativistic potential theory, introducing a confining potential in terms of 
a string tension~\cite{Satz:2005hx}. 
 
\begin{table*}[!b] 
  \centering 
  \caption{Mass, binding energy, and radius for charmonia and bottomonia~\cite{Satz:2005hx}.} 
  \label{tab:resonances} 
  \begin{tabular*}{\textwidth}{@{\extracolsep{\fill}}ccccccccc@{}} 
    \hline 
    {\rm state}& $J/\psi$ & $\chic\text{(1P)}$ & $\psiP$ & 
    $\upsa$ & $\chib\text{(1P)}$ &  \upsb & $\chib\text{(2P)}$ & $\upsc$ \\ 
    \hline 
    {\rm mass~[GeV$/c^2$]}& 
    3.07 & 3.53 & 3.68 & 9.46 & 9.99 & 10.02 & 10.26 & 10.36 \\ 
    ${\rm binding}$ {\rm[GeV]}& 
    0.64 & 0.20 & 0.05 & 1.10 & 0.67 & 0.54 & 0.31 & 0.20 \\ 
    {\rm radius~[fm]}& 
    0.25 & 0.36 & 0.45 & 0.14 & 0.22 & 0.28 & 0.34 & 0.39 \\ 
    \hline 
  \end{tabular*} 
\end{table*} 
 
The QGP consists of deconfined colour charges, so that the 
binding of a \QQbar pair is subject to the effect of colour screening which 
limits the range of strong interactions. Intuitively, the fate of heavy 
quark bound states in a QGP depends on the size of the colour screening radius 
$r_D$ (which is inversely proportional to the temperature, so that it decreases 
with increasing temperature) in comparison to the quarkonium binding radius 
$r_Q$: if $r_D \gg r_Q$, the medium does not really affect the heavy quark 
binding. Once $r_D \ll r_Q$, however, the two heavy quarks cannot ``see'' each 
other any more and hence the bound state will melt. It is therefore expected 
that quarkonia will survive in a QGP through some range of 
temperatures above $T_c$, and then dissociate once $T$ becomes large enough. 
Recent studies 
have shown that the  Debye-screened potential develops an imaginary part, implying a class of thermal effects that generate a finite width for the
quarkonium peak in the spectral function. These results  can be used to study quarkonium  in a weakly coupled Quark Gluon Plasma
within an EFT (Effective Field Theories) framework \cite{Brambilla:2010vq}.
On the other hand lattice-QCD enables ab initio study of
quarkonium correlation functions in the strongly coupled regime.
%
The sequential dissociation scenario
is confirmed by all these approaches~\cite{Brambilla:2010cs}. 

In vacuum, progress in lattice calculations and effective field theories 
have turned quarkonium physics into a powerful tool to determine the
heavy-quark masses and the strength of the QCD coupling, 
with an accuracy comparable to other techniques.
The measurements of quarkonia in heavy-ion collisions 
provide quantitative inputs for the study of QCD at high density and
temperature, providing an experimental basis for analytical and lattice 
studies to extract
the in-medium properties of heavy-flavor particles and the
implications for the QCD medium~\cite{Brambilla:2004wf,Brambilla:2004jw,Brambilla:2010cs,Brambilla:2014jmp}.
 
Finite-temperature lattice studies on quarkonium mostly consist of calculations 
of spectral functions for temperatures in the range explored by the experiments. 
The spectral function $\rho(\omega)$ is the basic quantity encoding the 
equilibrium properties of a quarkonium state. It characterises the spectral 
distribution of binding strength as a function of energy $\omega$. Bound or resonance 
states manifest themselves as peaks with well defined mass and spectral width. 
The in-medium spectral properties of quarkonia are related to phenomenology, 
since the masses determine the equilibrium abundances, their inelastic widths 
determine formation and destruction rates (or chemical equilibration times) and 
their elastic widths affect momentum spectra (and determine the kinetic 
equilibration times). 
%
%
 
Spectral functions play an important role in understanding how elementary 
excitations are modified in a thermal medium. They are the power spectrum of 
autocorrelation functions in real time, hence provide a direct information on 
large time propagation. In the lattice approach such real time evolution is not 
directly accessible: the theory is formulated in a four dimensional box -- three 
dimensions are spatial dimensions, the fourth is the imaginary (Euclidean) time 
$\tau$. The lattice temperature $T_L$ is realised through (anti)periodic 
boundary conditions in the Euclidean time direction -- $T_L = 1/N_\tau$, where 
$N_\tau$ is the extent of the time direction, and can be converted to physical 
units once the lattice spacing is known. The spectral functions appear now in 
the decomposition of a (zero-momentum) Euclidean propagator $G(\tau)$: $ G(\tau) 
= \int_{0}^\infty \rho(\omega) \frac{\dd\omega}{2\pi}\, K(\tau,\omega)$, with 
$K(\tau,\omega) = \frac{\left(e^{-\omega\tau} + e^{-\omega(1/T - \tau)}\right)} 
{1 - e^{-\omega/T}}$. The $\tau$ dependence of the kernel $K$ reflects the 
periodicity of the relativistic propagator in imaginary time, as well as its $T$ 
symmetry. The Bose--Einstein distribution, intuitively, describes the wrapping 
around the periodic box, which becomes increasingly important at higher 
temperatures. 
 
The procedure is, then, based on the generation of an appropriate ensemble of 
lattice gauge fields at a temperature of choice, on the computation on such an 
ensemble of the Euclidean propagators $G(\tau)$, and on the extraction of the 
spectral functions. 
%
All such quarkonium studies yield qualitatively the same result: a given 
quarkonium state melts at a temperature above, or possibly at, the phase 
transition temperature. There are, however, disagreements between different 
calculations in the precise temperatures for the following reasons. First, 
experiences with lattice calculations have demonstrated that it is extremely 
important to have results in the continuum limit, and with the proper matter 
content. 
This means that the masses of the dynamical quark fields which are used in the 
generation of the gauge ensembles must be as close as possible to the physical 
ones, and the lattice spacing should be fine enough to allow for making contact with 
continuum physics. These systematic effects, which have been studied in detail 
for bulk thermodynamics, are still under scrutiny for the spectral functions. 
Second, the calculation of spectral functions using Euclidean propagators as an 
input is a difficult, possibly ill-defined, problem. It has been mostly tackled 
by using the Maximum Entropy Method (MEM)~\cite{Asakawa:2000tr}, which has 
proven successful in a variety of applications. Recently, an alternative 
Bayesian reconstruction of the spectral functions has been proposed in 
Refs.~\cite{Rothkopf:2011ef,Burnier:2013nla} and applied to the analysis of 
configurations from the HotQCD Collaboration~\cite{Kim:2014iga}. 
 
Most calculations of charmonium spectral functions have been performed in the 
quenched approximation ---neglecting quark loops---, although recently the 
spectral functions of the charmonium states have been studied as a function of 
both temperature and momentum, using as input relativistic propagators with two 
light quarks~\cite{Aarts:2007pk,Kelly:2013cpa} and, more recently, including the 
strange quark, for temperatures ranging between $0.76\,T_c$ and $1.9\,T_c$. The 
sequential dissolution of the peaks corresponding to the S- and P-wave states is 
clearly seen. The results are consistent with the expectation that charmonium 
melts at high temperature, however as of today they lack quantitative precision 
and control over systematic errors. 
 
\begin{figure}[!t] 
  \centering 
  \includegraphics[height=4cm]{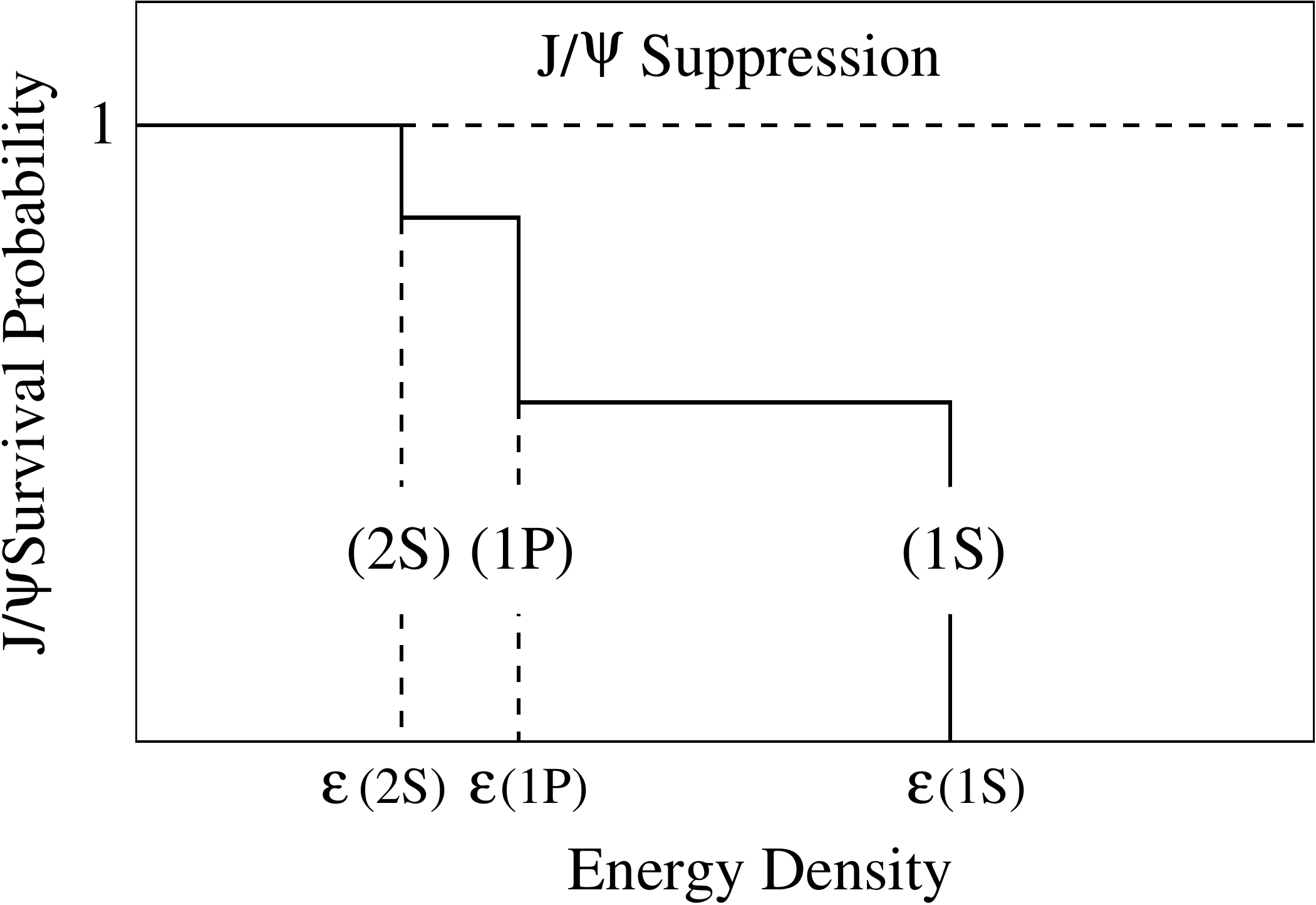}\hskip1cm 
  \includegraphics[height=4cm]{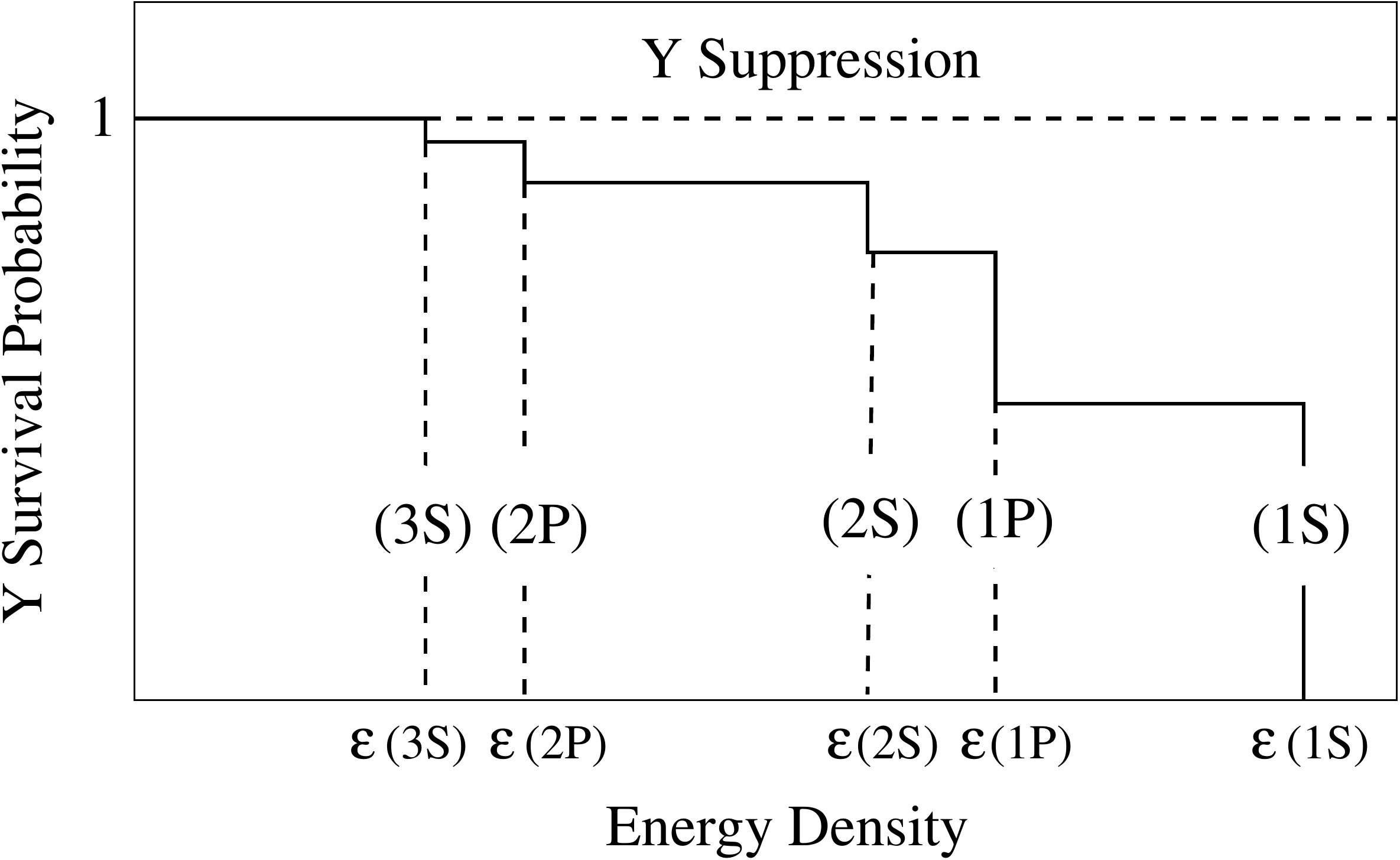} 
  \caption{Sequential quarkonium suppression for \jpsi (left) and \upsa (right) 
    states~\cite{Karsch:2005nk}.} 
  \label{fig:seq} 
\end{figure}

The survival probability for a given quarkonium state 
depends on its size and binding energy (see \tab{tab:resonances} for details\footnote{Note that in this Table
the calculation of the binding energies and radii of the quarkonium  states is made with an arbitrary potential model with
arbitrary parameters, so they do not correspond to the experimental masses of quarkonia but
are model dependent. These values are to be taken as an illustration of the expected Debye screening ordering.}). 
Hence the excited states will be dissolved at a lower initial temperature than 
the more tightly-bound ground states. However, only a 
fraction (about 60\%) of the observed \jpsi is a directly produced $\text{(1S)}$ 
state, the remainder is due to the feed-down of excited states, with about 30\% 
from $\chic\text{(1P)}$ and 10\% from \psiP 
decays~\cite{Antoniazzi:1992af,Antoniazzi:1992iv,Lemoigne:1982jc}. A similar 
decay pattern arises for \ups 
production~\cite{Abe:1995an,Affolder:1999wm,Aaij:2012se,Aad:2011ih,Aaij:2014caa}. 
The decay processes occur far outside the produced medium, so that the medium 
affects only the excited states. As a result, the formation of a hot deconfined 
medium in nuclear collisions will produce a sequential quarkonium suppression 
pattern~\cite{Karsch:2005nk}, as illustrated in \fig{fig:seq}. Increasing the 
energy density of the QGP above deconfinement first leads to \psiP dissociation, 
removing those \jpsi's which otherwise would have come from \psiP decays. Next 
the \chic melts, and only for a sufficiently hot medium also the direct \jpsi 
disintegrate. For the bottomonium states, a similar pattern 
holds~\cite{Aarts:2013kaa,Aarts:2011sm,Aarts:2010ek,Aarts:2014cda}. 
 
\subsubsection{Effect of nuclear PDFs on quarkonium production in nucleus--nucleus collisions}
\label{sec:npdf_aa}

The predictions for quarkonium suppression in \AAcoll collisions, considering 
only modifications of the parton densities in the nucleus, the so called nuclear 
PDFs, are described in this subsection. There are other possible cold matter 
effects on quarkonium production in matter in addition to shadowing: breakup of 
the quarkonium state due to inelastic interactions with nucleons (absorption) or 
produced hadrons (comovers) and energy loss in cold matter, as discussed 
in \sect{Cold nuclear matter effects}. The midrapidity quarkonium 
absorption cross section for breakup by nucleon interactions decreases with 
centre-of-mass energy~\cite{Lourenco:2008sk,McGlinchey:2012bp}, becoming 
negligible at LHC energies. In addition, cold matter suppression due to energy 
loss does not have a strong rapidity dependence. Thus, shadowing is expected to be 
the dominant cold matter effect in what concerns the modification of the shape 
of the quarkonium rapidity distribution. It will also produce a relatively small 
effect on the shape of the quarkonium \pt distribution at low \pt. 
 
\begin{figure*}[!t] 
  \centering 
  \includegraphics[width=0.45\textwidth]{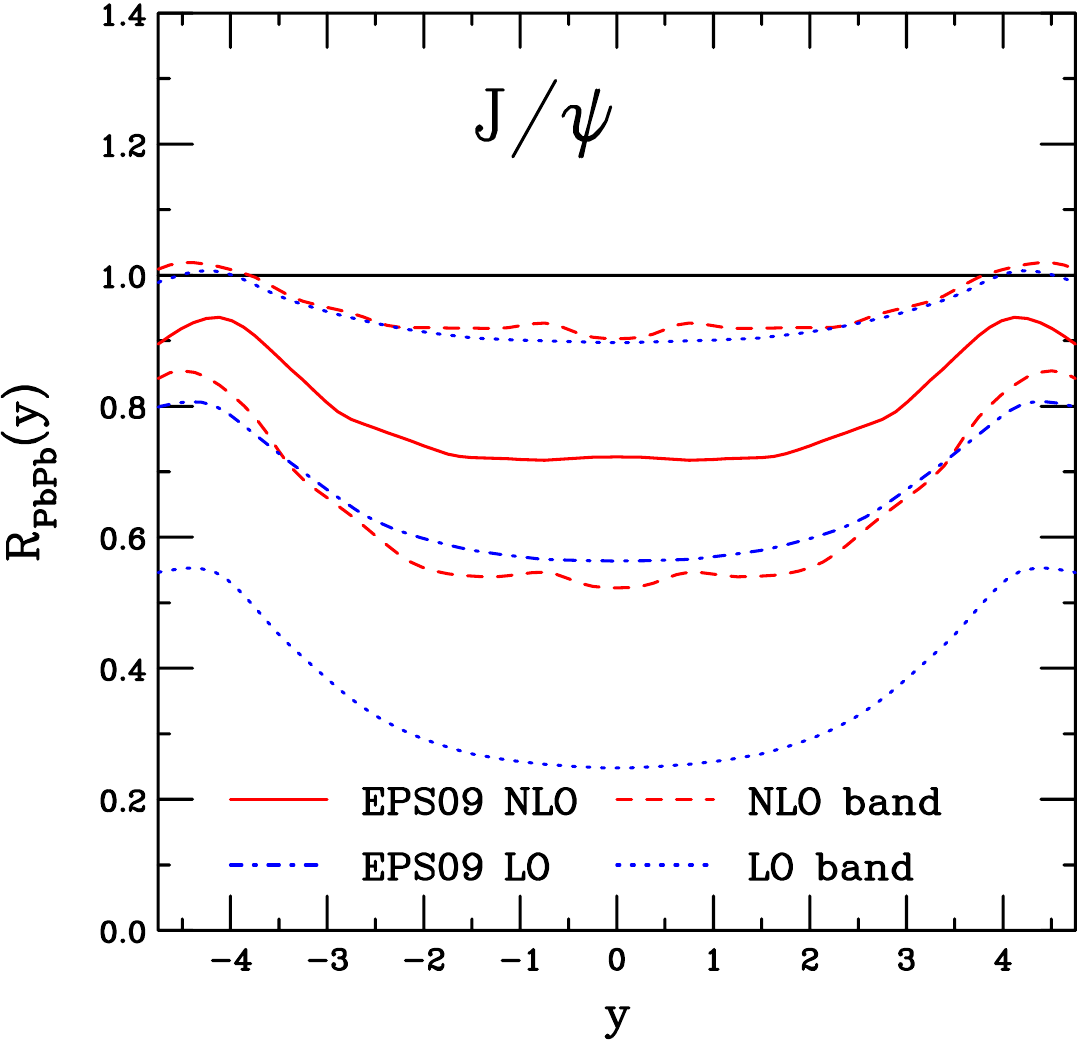} 
  \includegraphics[width=0.45\textwidth]{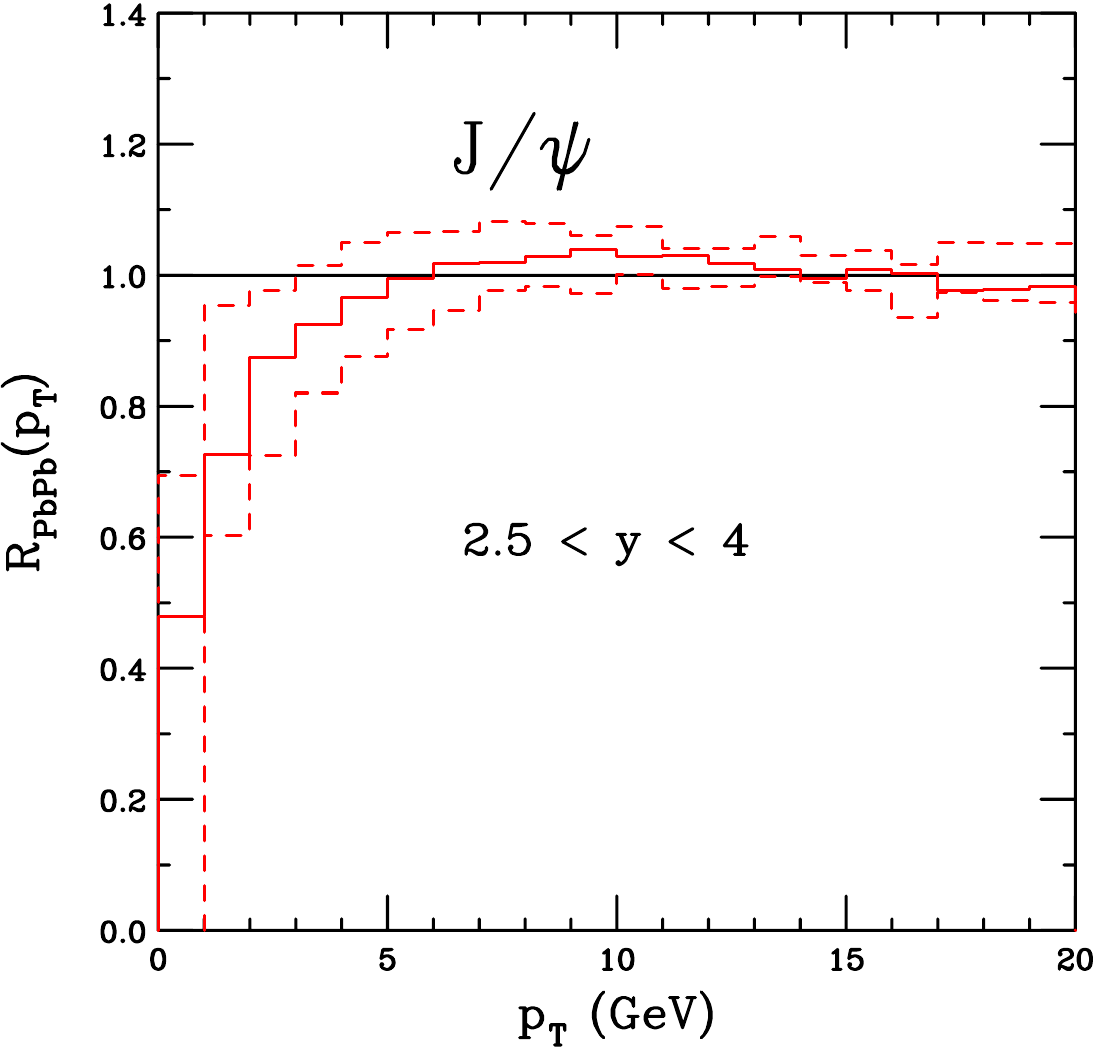}\\ 
  \includegraphics[width=0.45\textwidth]{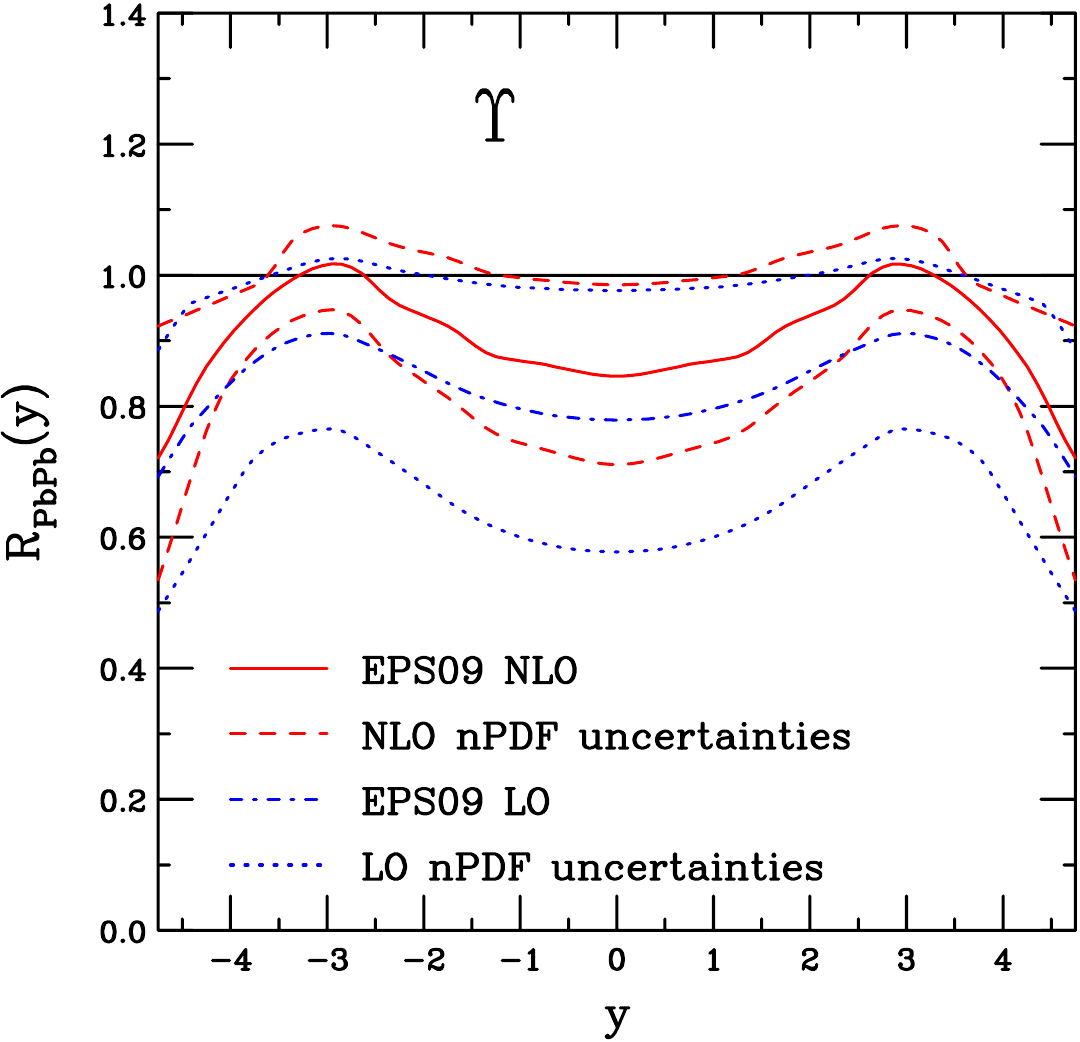} 
  \includegraphics[width=0.45\textwidth]{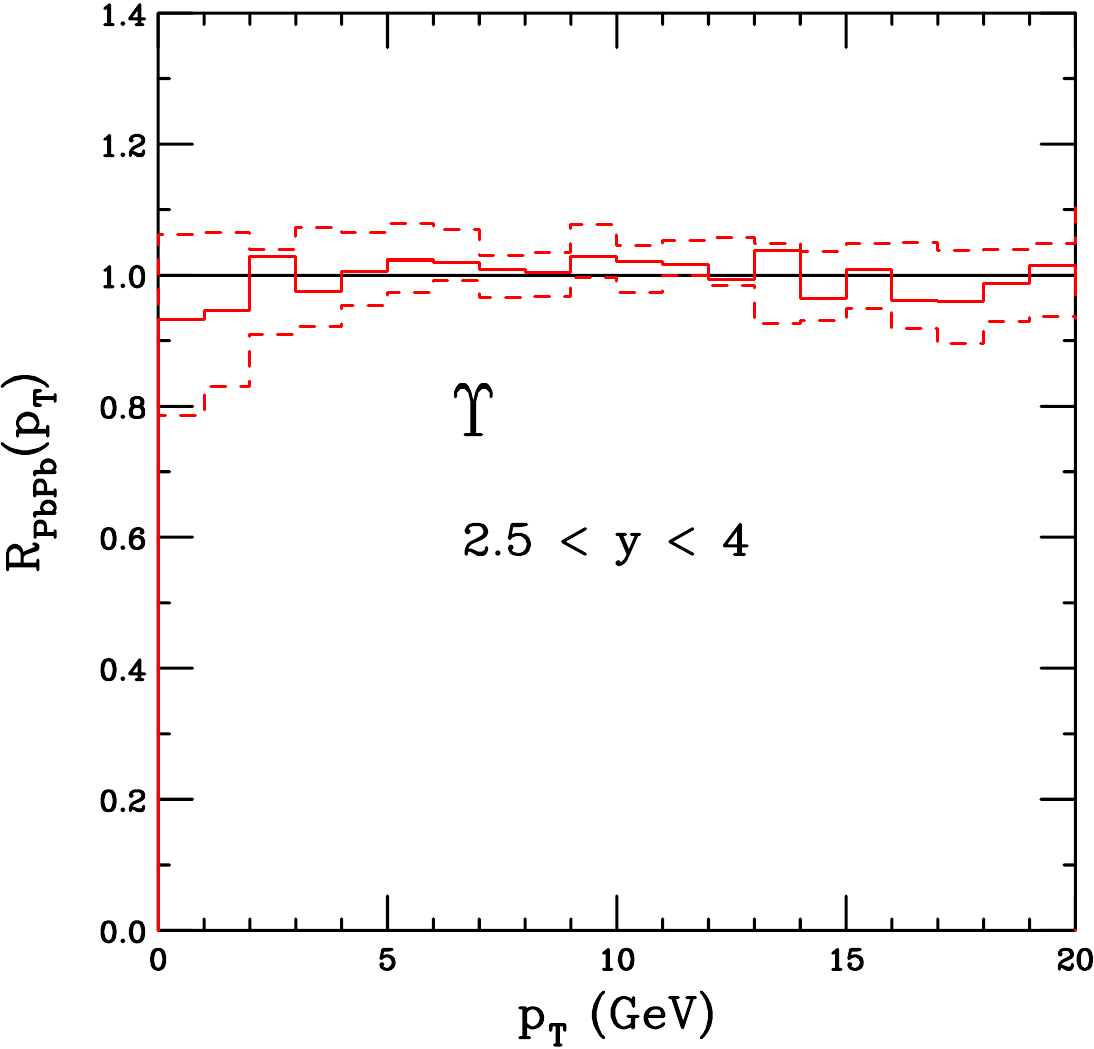} 
  \caption{The nuclear modification factor $R_{\rm AA}$ for \jpsi (upper) 
    and \ups (lower) production, calculated in the CEM model using the EPS09 modifications~\cite{Eskola:2009uj}, is shown for \PbPb collisions at \snn=2.76 \TeV. 
    The results are presented as a function of rapidity (left) and \pt 
    (right)~\cite{Vogt:2010aa}. The dashed red histogram shows the EPS09 NLO uncertainties. The 
    blue curves show the LO modification and the corresponding uncertainty band 
    as a function of rapidity 
    only. 
  } 
    \label{fig:JpsiUpsypt} 
\end{figure*}

\begin{figure}[t!] 
\vskip 0.3cm 
\begin{minipage}[t]{.45\textwidth} 
\begin{center} 
\includegraphics[width=1.0\textwidth]{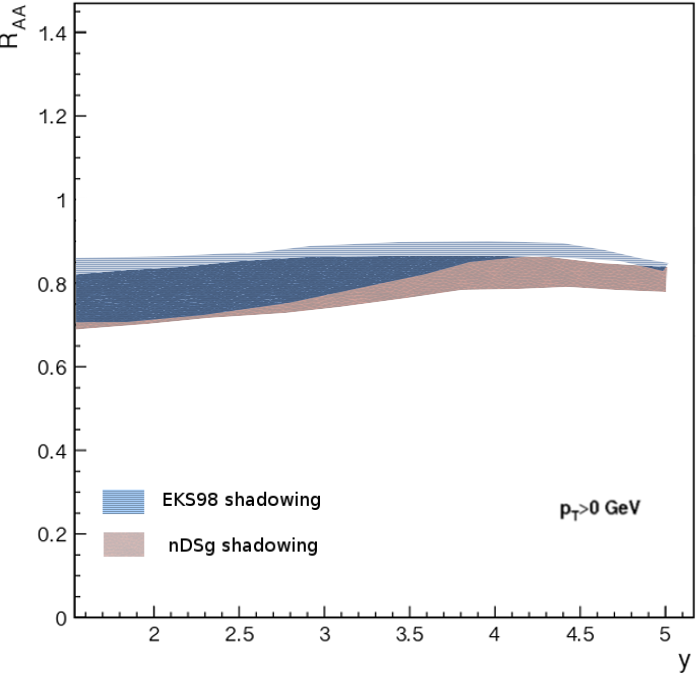} 
\end{center} 
\end{minipage} 
\begin{minipage}[t]{.45\textwidth} 
\begin{center} 
\includegraphics[width=0.86\textwidth]{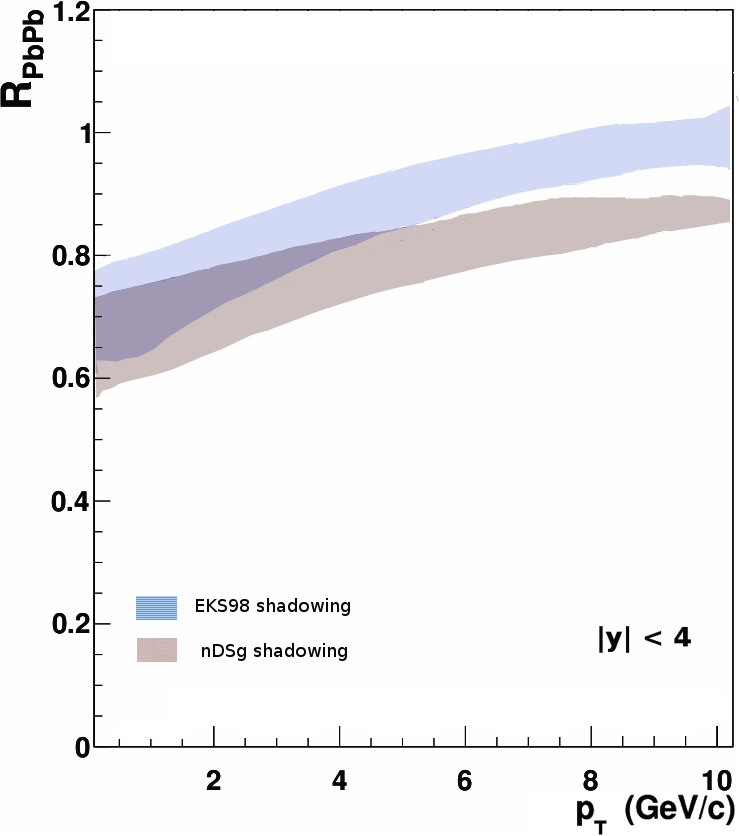} 
\end{center} 
\end{minipage} 
\vskip -0.4cm 
\caption{\jpsi rapidity (left) and \pt dependence (right) of the EKS98 LO and 
  nDSg LO shadowing corrections performed using the CSM model according 
  to~\cite{Ferreiro:2008wc,Rakotozafindrabe:2011rw} in \pb collisions at \snn = 
  2.76\TeV. The bands for the EKS98/nDSg models shown in the figure correspond 
  to the variation of the factorisation scale.} 
\label{fig1and2ferreiro} 
\hfill 
\begin{center} 
\includegraphics[width=1.0\textwidth]{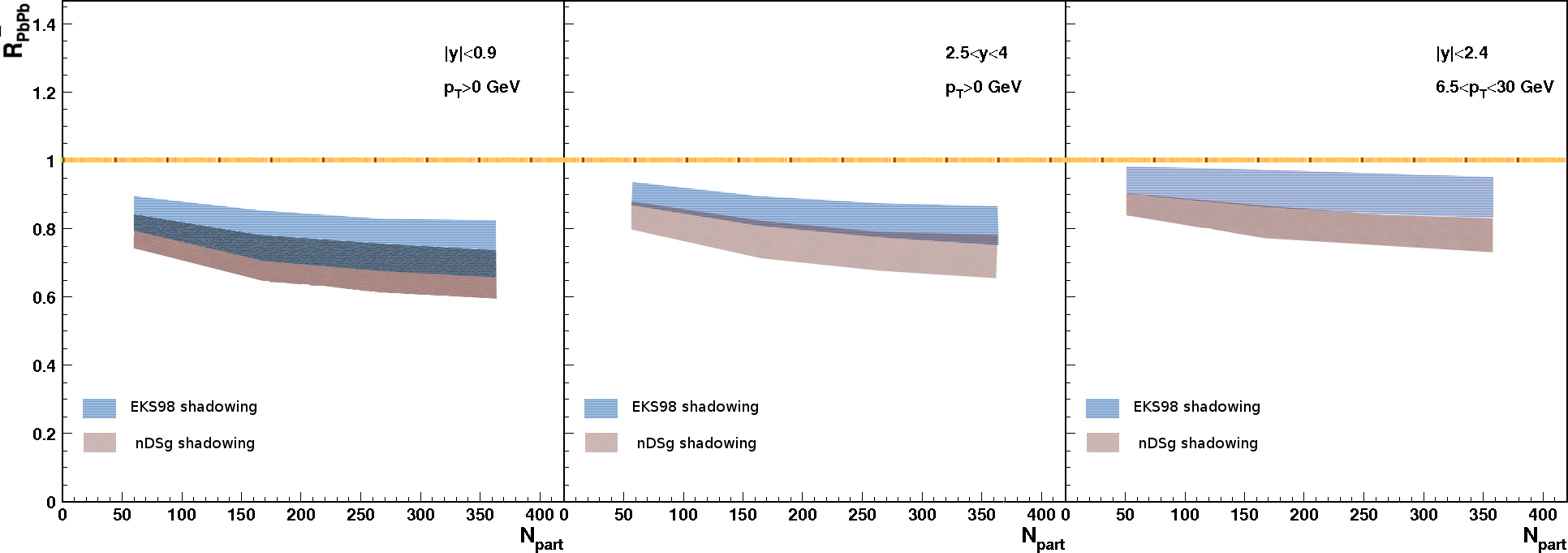} 
\end{center} 
\caption{ \jpsi centrality dependence of the EKS98 LO and nDSg LO shadowing 
  corrections performed using the CSM model according 
  to~\cite{Ferreiro:2008wc,Rakotozafindrabe:2011rw} in \pb collisions at \snn = 
  2.76\TeV. The bands for the EKS98/nDSg models shown in the figure correspond 
  to the uncertainty in the factorisation scale.} 
\label{fig3ferreiro} 
\end{figure}

\fig{fig:JpsiUpsypt} shows the results for the dependence of shadowing on rapidity, 
transverse momentum, and centrality are shown for \jpsi and \ups production in 
\pb collisions at \snn = 2.76\TeV, neglecting absorption. Results 
obtained in the colour evaporation model (CEM) at next-to-leading order (NLO) in 
the total cross section (leading order in \pt) are discussed first, followed by 
results from a leading order colour singlet model (CSM) calculation. 
 
The CEM calculation was described in  \sect{Cold nuclear matter effects}. Here only a 
few pertinent points are repeated. In the CEM, the 
quarkonium production cross section in \pp collisions is some fraction, $F_\Qcal$, 
of all \QQbar pairs below the $H \overline H$ threshold where $H$ is the lowest-mass heavy-flavour hadron, 
\begin{eqnarray} 
\sigma_{{\rm PbPb}\to\Qcal+X}^{\rm CEM}[\s]  = A^2\, F_\Qcal \sum_{i,j}  
\int_{4m_{\rm Q}^2}^{4m_H^2} \dd\hat{s} 
\int_0^1 \dd x_i \int_0^1 \dd x_j~ R_i^{\rm Pb}(x_i,\mu_F^2)\,f_i(x_i,\mu_F^2)~ R_j^{\rm Pb}(x_j,\mu_F^2)\,f_j(x_j,\mu_F^2)~  
{\cal J}~\hat\sigma_{ij\to\QQbar+X}[\hat{s},\mu_F^2, \mu_R^2] \,  
\, , \label{sigtilAA} 
\end{eqnarray}  
where $ij = \qqbar$ or $gg$ and $\hat\sigma_{ij\to\QQbar+X}$ is the $ij\rightarrow 
\QQbar$ sub-process cross section at centre-of-mass energy $\hat{s}$, while 
$\mathcal{J}$ is an appropriate Jacobian with dimension $1/\hat{s}$. The 
normalisation factor $F_\Qcal$ is fitted to an appropriate subset of the 
available data, restricting the fits to measurements on light nuclear targets to 
avoid any significant cold matter effects. For the \jpsi and $\Upsilon$ results shown here, the 
normalisation $F_\Qcal$ is based on the same central mass and scale parameter values 
as those obtained for open charm, $(m_c,\mu_F/m_c, \mu_R/m_c) = 
(1.27\mathrm{\,GeV}/c^2,\,2.1,\,1.6)$ ~\cite{Nelson:2012bc}, and beauty, 
$(m_b,\mu_F/m_b, \mu_R/m_b) = (4.65 \, {\rm GeV}/c^2, 1.4,1.1)$ 
~\cite{Nelson_inprog}. The mass and scale uncertainties on the CEM calculation 
are shown in the previous section. They are smaller than those due to the 
uncertainties of the EPS09 shadowing parametrisation~\cite{Eskola:2009uj}. All 
the CEM calculations are NLO in the total cross section and assume that the 
intrinsic \kt broadening is the same in \pb as in pp.

The upper left-hand panel of \fig{fig:JpsiUpsypt} shows the uncertainty in the 
shadowing effect on \jpsi due to the variations in the 30 EPS09 NLO 
sets~\cite{Eskola:2009uj} (red). The uncertainty band calculated in the CEM at 
LO with the EPS09 LO sets is shown for comparison (blue).  It is clear that the LO 
results 
exhibit a larger shadowing effect. This difference between the LO results, also 
shown in Ref.~\cite{Vogt:2010aa}, and the NLO calculations arises because the 
EPS09 LO and NLO gluon shadowing parametrisations differ significantly at low 
$x$~\cite{Eskola:2009uj}. 
 
In principle, the shadowing results should be the same for LO and NLO. 
Unfortunately, however, the gluon modifications, particularly at low $x$ and 
moderate $Q^2$, are not yet sufficiently constrained. The lower left panel shows 
the same calculation for \ups production. Here, the difference between the LO 
and NLO calculations is reduced because the mass scale, as well as the range of 
$x$ values probed, is larger. Differences in LO results relative to, \eg, the 
colour singlet model arise less from the production mechanism than from the 
different mass and scale values assumed, as we discuss below.

It should be noted that the convolution of the two nuclear parton densities 
results in a $\sim 20$\% suppression at NLO for $|y|\leq 2.5$ with a mild 
decrease in suppression at more forward rapidities. The gluon antishadowing peak 
at $|y| \sim 4$ for \jpsi and $|y| \sim 2$ for \ups with large $x$ in the 
nucleus is mitigated by the shadowing at low $x$ in the opposite nucleus with 
the NLO parametrisation. The overall effect due to NLO nPDFs in both nuclei is 
a result with moderate rapidity dependence and $R_{\rm AA}^{\jpsi} \sim 0.7$ for 
$|y|\leq 5$ and $R_{\rm AA}^{\ups} \sim 0.84$ for $|y| \leq 3$. The nPDF effect 
gives more suppression at central rapidity than at forward rapidity, albeit less 
so for the LHC energies than for RHIC where the antishadowing peak at \snn = 
200\GeV is at $|y| \sim 2$. The difference between the central value of 
$R_{\rm AA}$ at LO and NLO is $\sim 30$\% for the \jpsi and $\sim 10$\% for the 
\ups. If a different nPDF set with LO and NLO parametrisations, such as nDSg~\cite{deFlorian:2003qf}, 
is used, the difference between LO and NLO is reduced to a few percent since the 
difference between the underlying LO and NLO proton parton densities at low $x$ 
is much smaller for nDSg than for EPS09~\cite{RVinprog}. 
 
The uncertainty is larger in the LO CEM calculation for several reasons. First 
and foremost is the choice of the underlying proton parton densities. If the $x$ 
and $Q^2$ dependence at LO and NLO is very different, the resulting nuclear 
parton densities will reflect this~\cite{RVinprog}. Other factors play a smaller 
role. For example, the $x$ values in the $2 \rightarrow 1$ kinematics at LO is 
somewhat lower than the $2 \rightarrow 2$ and $2\rightarrow 3$ kinematics (for 
the LO+virtual and real NLO contributions respectively) of the NLO CEM 
calculation. Next, the \pt scale enters in the complete NLO calculation where it 
does not in the LO, leading to both a slightly larger $x$ value for higher \pt 
as well as a larger scale so that the NLO calculation is on average at a higher 
scale than the LO. 
 
The right panels of \fig{fig:JpsiUpsypt} show the \pt dependence of the effect 
at forward rapidity for \jpsi (upper) and \ups (lower).  
The effect is rather mild and increases slowly with \pt. There is little 
difference between the \jpsi and \ups results for $R_{\text{\pb}}(\pt)$ because, 
for \pt above a few GeV, the \pt scale is dominant. 
There is no LO 
comparison here because the \pt dependence cannot be calculated in the LO CEM. 
 
 
However, the leading order colour single model calculation (LO CSM) of \jpsi 
production, shown to be compatible with the magnitude of the of the 
\pt-integrated cross sections, is a $2 \rightarrow 2$ process, $g + g 
\rightarrow \jpsi + g$, which has a calculable \pt dependence at LO, as in the 
so-called {\it extrinsic} scheme~\cite{Ferreiro:2008wc}. 
  
In this approach, one can use the partonic differential cross section computed 
from any $2 \rightarrow 2$ theoretical model that satisfactorily describes the 
data down to low \pt. Here, a generic $2\to 2$ matrix element which matches the 
\pt dependence of the data has been used and 
%
%
%
%
the parametrisations EKS98 LO~\cite{Eskola:1998df} and nDSg 
LO~\cite{deFlorian:2003qf} have been employed. The former 
coincides with the mid value of EPS09 LO~\cite{Eskola:2009uj}. The error bands 
for the EKS98 and nDSg models shown in ~\fig{fig1and2ferreiro} correspond to the 
variation of the factorisation scale ($0.5 m_{\rm T} < \mu_F < 2 m_{\rm T}$).


The spatial dependence of the nPDF has been included in this approach through a 
probabilistic Glauber Monte-Carlo framework, {\sf JIN}~\cite{Ferreiro:2008qj}, 
assuming an inhomogeneous shadowing proportional to the local 
density~\cite{Klein:2003dj,Vogt:2004dh}. Results are shown in 
\fig{fig3ferreiro}.


\subsubsection{Statistical (re)generation models}
\label{sec:regeneration}

Over the past 20 years thorough evidence has been gathered that production of 
hadrons with $u$, $d$, $s$-valence quarks in heavy-ion collisions can be 
described using 
a statistical model reflecting a hadro-chemical equilibrium 
approach~\cite{Stachel:2013zma,Andronic:2008gu}. Hadron yields from top AGS 
energy ($\sim10\GeV$) up to the LHC are reproduced over many orders of magnitude 
employing a statistical operator that incorporates a complete hadron resonance 
gas. In a grand canonical treatment, the only thermal parameters are the 
chemical freeze out temperature $T$ and the baryo-chemical potential $\mu_b$ 
(and the fireball volume $V$, in case yields rather than ratios of yields are 
fitted). These parameters are fitted to data for every collision system as 
function of collision energy. The temperature  
initially rises with \snn and flattens  at a value of $(159\pm2)\MeV$ close to 
top SPS energy, while the baryo-chemical potential drops smoothly and reaches a 
value compatible with zero at LHC energies. In the energy range where $T$ 
saturates, it has been found to coincide with the (quasi-)critical 
temperature found in lattice QCD. 
 
Deconfinement of quarks is expected in a QGP and for heavy quarks, in particular, 
this has been formulated via modification of the heavy quark potential in a 
process analogue to Debye screening in QED~\cite{Matsui:1986dk} (see 
\sect{sec:seq-supp}). Heavy quarks are not expected to be produced thermally but 
rather in initial hard scattering processes. Even at top LHC energy thermal 
production is only a correction at maximally the 10\% 
level~\cite{BraunMunzinger:2000dv}. Therefore a scenario was proposed, in which 
charm quarks, formed in a high energy nuclear collision in initial hard 
scattering, find themselves colour-screened, therefore deconfined in a QGP, and 
hadronise with light quarks and gluons at the phase 
boundary~\cite{BraunMunzinger:2000px,BraunMunzinger:2000ep,BraunMunzinger:2009ih}. 
At hadronisation open charm hadrons as well as charmonia are formed according to 
their statistical weights and the mass spectrum of charmed hadrons. 
 
Since for each beam energy the values of $T$ and $\mu_b$ are already fixed by 
the measured light hadron yields, the only additional input needed is the 
initial charm production cross section per unit rapidity in the appropriate 
rapidity interval. The conservation of the number of charm quarks is introduced 
in the statistical model via a fugacity $g_c$, where all open charm hadron 
yields scale proportional to $g_c$, while charmonia scale with $g_c^2$ since 
they are formed from a charm and an anticharm quark. A logical consequence of 
this is that at energies below LHC energy, where the charm yield is small, 
charmonium production is suppressed in comparison to scaled pp collisions, while 
for LHC energies, the charm yield is larger and the charmonium yield is 
enhanced 
~\cite{BraunMunzinger:2000px,BraunMunzinger:2000ep,BraunMunzinger:2009ih}. 
 
Already a comparison to first data on \jpsi production from PHENIX at RHIC using 
a charm cross section from perturbative QCD proved 
successful~\cite{Andronic:2003zv}. When more data became available it was found 
that in particular the rapidity and centrality dependence of \jpsi \raa from 
RHIC and the \psiP to \jpsi ratio from NA50 at the SPS were well reproduced by 
this approach~\cite{Andronic:2006ky,Andronic:2007bi}. In order to treat properly 
the centrality dependence, also production in the dilute corona using the pp 
production cross section of \jpsi is 
considered~\cite{Andronic:2006ky,Andronic:2007bi}. While it was clear that for 
LHC energies larger values for \raa of \jpsi are expected than at RHIC, \raa 
depends linearly on the unknown \ccbar cross section. Predictions for an 
expected range were given in~\cite{Andronic:2008gm}. 
 
The comparison of the statistical hadronisation predictions with the LHC data 
require the knowledge of the \ccbar cross section. This quantity has been 
measured in \pp collisions at \s = 7\TeV and is then extrapolated to the lower 
\pb beam energy, \ie \snn = 2.76\TeV. Since the current data are for half the 
LHC design energy, the open charm cross section is at the lower end of the range 
considered in Ref.~\cite{Andronic:2008gm}. The uncertainty on this model 
prediction comes from the uncertainty on the \ccbar cross section and it stems 
from the measurement of the \ccbar cross section itself, \s, and shadowing 
extrapolations. 
 
As it will be discussed in \sect{sec:raa_lowpt}, the statistical model 
reproduces the significant increase observed, for central collisions, in the 
\jpsi \raa from RHIC to the LHC (see \fig{fig:stat_mod_charm}). 
 
The statistical hadronisation picture, and therefore the increase in \raa at 
LHC, applies to thermalised charm quarks and, therefore, is necessarily a low 
\pt phenomenon. This is in line with a drop in \raa for larger \pt observed in 
the data. The statistical hadronisation model in itself makes no prediction of 
spectra without additional input. Given a velocity distribution of the quarks at 
hadronisation, the spectra and their moments are fixed. As examples in 
Ref.~\cite{Andronic:2006ky,Andronic:2007bi}, \jpsi spectra are predicted for 
different $T$ and collective expansion velocity of the medium at hadronisation. 
The narrowing of $\langle p_T \rangle$ and its root-mean-square as compared to 
\pp collisions in the ALICE data are in line with this expectation. A precise 
measurement of the spectral shape is an important test of the model awaiting 
larger data samples. 
 
Another characteristic feature of the statistical hadronisation model is an 
excited state population driven by Boltzmann factors at the hadronisation 
temperature. So far the only successful test of this prediction is the 
\psiP/\jpsi ratio at the SPS. Data for \psiP and \chic at LHC and RHIC will be 
crucial tests of this model and will allow, if measured with sufficient 
precision (10--20\%), to differentiate between transport model predictions (see 
\sect{sec:transport}) and statistical hadronisation at the phase boundary. 
 

\subsubsection{Transport approach for in-medium quarkonia}
\label{sec:transport}

In the transport models, there is continuous dissociation and (re)generation of 
quarkonia over the entire lifetime of the deconfined stage. The space-time 
evolution of the phase-space distribution, $f_{\cal Q}$, of a quarkonium state 
${\cal Q} = \Psi, \Upsilon$ ($\Psi$=$J/\psi, \chi_c, \dots $; 
$\Upsilon$=$\Upsilon(1S),\chi_b, \dots$) in hot and dense matter may be 
described by the relativistic Boltzmann equation, 
\begin{equation} 
  p^\mu \partial_\mu f_{\cal Q}(\vec r,\tau;\vec p) = 
  - E_p \ \Gamma_{\cal Q}(\vec r,\tau;\vec p) \ f_{\cal Q}(\vec r,\tau;\vec p) + 
  E_p \ \beta_{\cal Q}(\vec r,\tau;\vec p)  \  
\label{eq:boltz} 
\end{equation}    
where $p_0=E_p=(\vec p^2 +m_{\cal Q}^2)^{1/2}$, $\tau$ is the proper time, and 
$\vec{r}$ is the spatial coordinate. $\Gamma_{\cal Q}$ denotes the dissociation 
rate\footnote{A possible mean-field term has been neglected.} and the gain term, 
$\beta_{\cal Q}$, depends on the phase-space distribution of the individual 
heavy (anti-)quarks, $Q=c,\,b$ in the QGP (or $\rm D$, $\overline{\rm D}$ mesons in 
hadronic matter). If the open charm states are thermalised, and in the limit of 
a spatially homogeneous medium, one may integrate over the spatial and 
3-momentum dependencies to obtain the rate 
equation~\cite{Grandchamp:2003uw,Grandchamp:2005yw,Rapp:2009my} 
\begin{equation} 
  \frac{dN_{\cal Q}}{d\tau} =  -\Gamma_{\cal Q}(T) [ N_{\cal Q} - N_{\cal Q}^{\rm eq}(T) ] \,.  
  \label{eq:rate} 
\end{equation} 
The key ingredients to the rate equation are the {\it transport 
coefficients}: the inelastic reaction rate, $\Gamma_{\cal Q}$, for both 
dissociation and formation ---{\it detailed balance}---, and the quarkonium 
equilibrium limit, $N_{\cal Q}^{\rm eq}(T)$. 
 
\begin{figure}[t] 
  \centering 
  \includegraphics[width=0.45\textwidth]{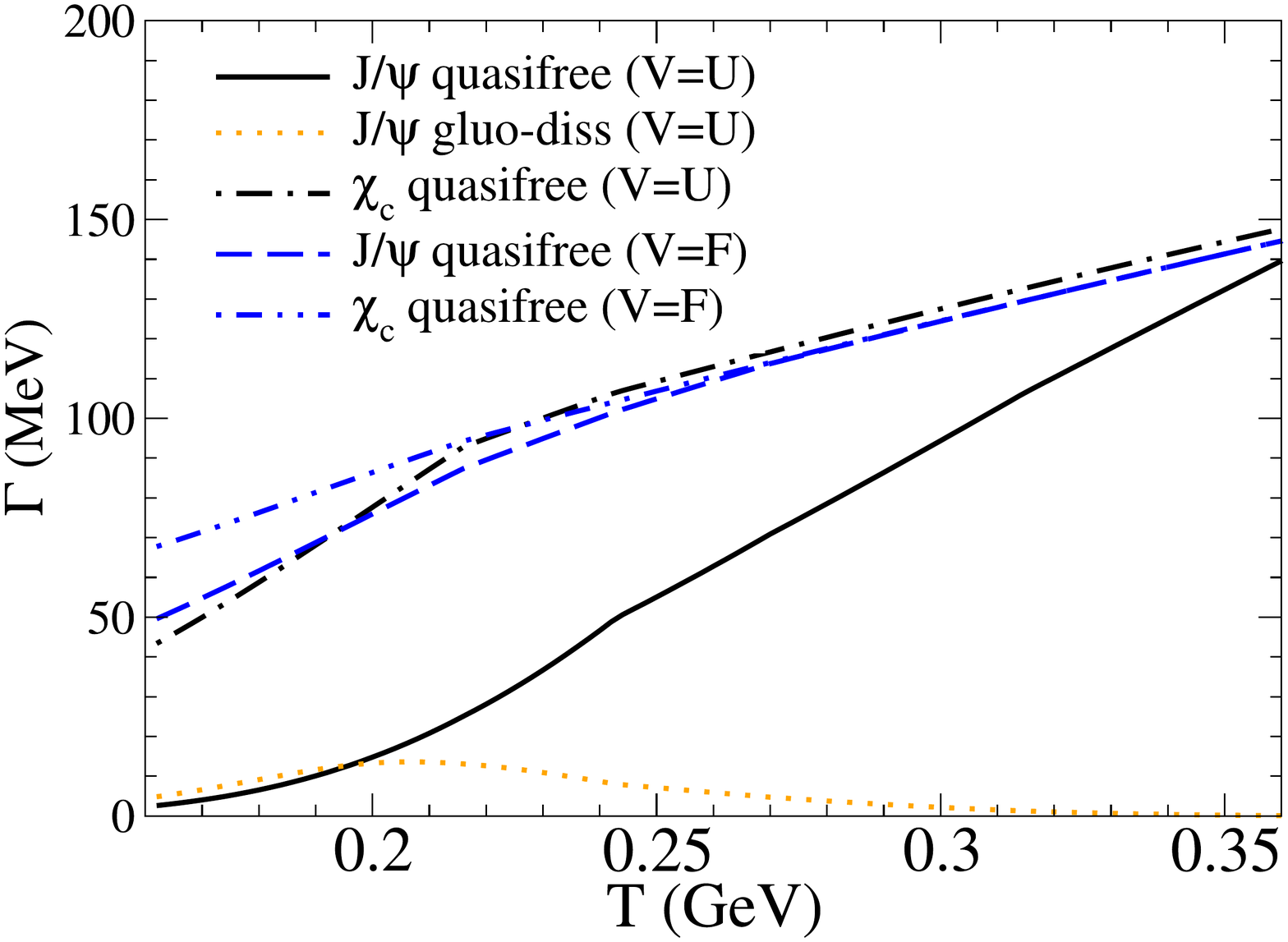} 
  \includegraphics[width=0.45\textwidth]{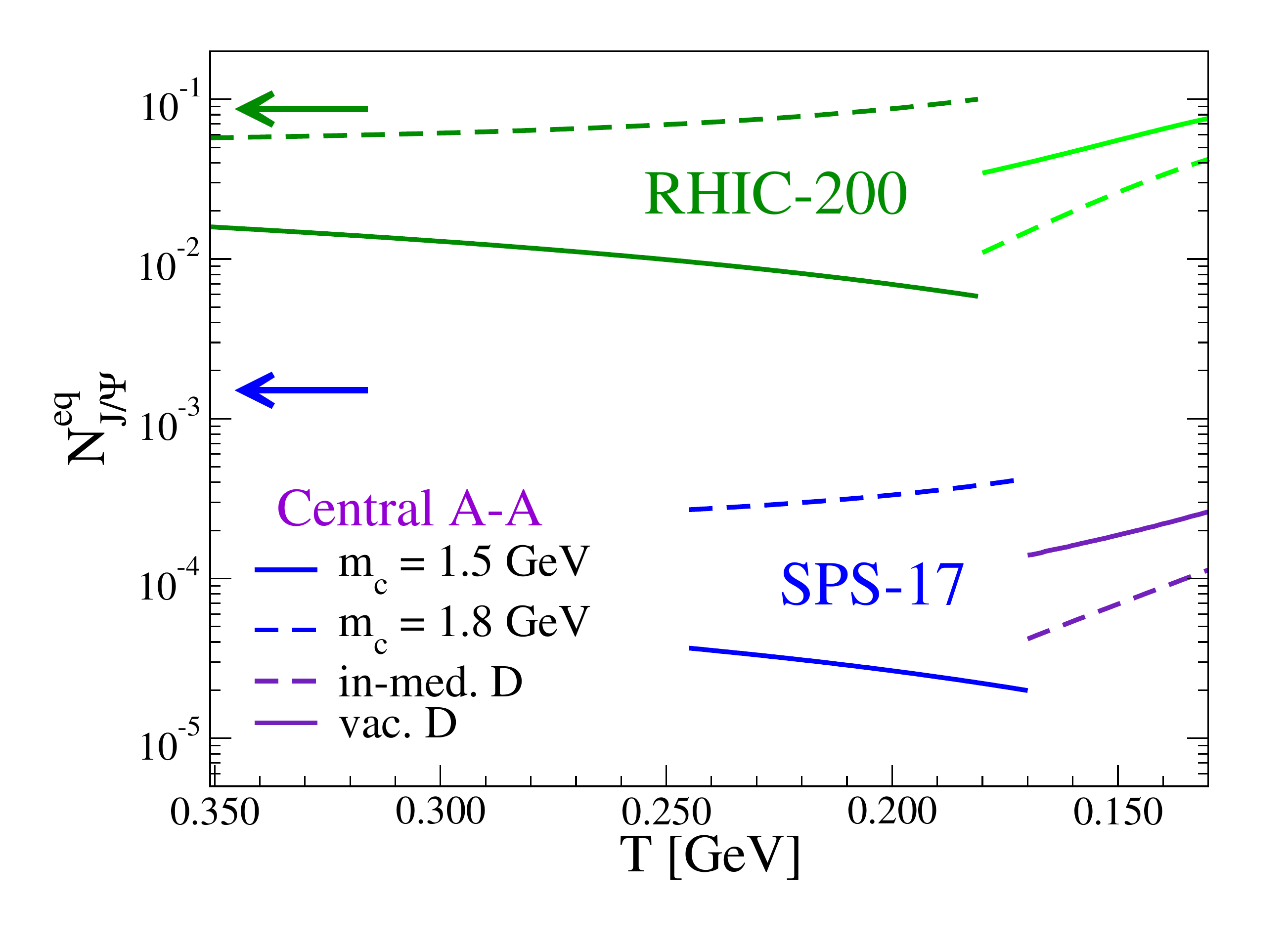} 
  \caption{Transport coefficients of charmonia in the QGP. Left: inelastic 
    reaction rates for J$/\psi$ and $\chi_c$ in strong- ($V$=$U$) and 
    weak-binding ($V$=$F$) scenarios defined in the text. Right: \jpsi 
    equilibrium numbers for conditions in central \PbPb and \AuAu at full SPS 
    and RHIC energies, respectively, using different values of the in-medium 
    $c$-quark mass in the QGP ($T\ge180\MeV$) and for D-mesons in hadronic 
    matter ($T\le180\MeV$); in practice the equilibrium numbers are constructed 
    as continuous across the transition region. } 
  \label{fig:trans} 
\end{figure}

The reaction rate can be calculated from inelastic scattering amplitudes of 
quarkonia on the constituents of the medium (light quarks and gluons, or light 
hadrons). The relevant processes depend on the (in-medium) properties of the 
bound state~\cite{Grandchamp:2001pf}. In the QGP, for a tightly bound state 
(binding energy $E_B\geq T$), an efficient process is 
gluo-dissociation~\cite{Bhanot:1979vb}, $g+{\cal Q}\to Q+\overline{Q}$, where 
all of the incoming gluon energy is available for break-up. However, for loosely 
bound states ($E_B <T$ for excited and partially screened states), the phase 
space for gluo-dissociation rapidly shuts off, rendering ``quasi-free" 
dissociation, $p+{\cal Q}\to Q+\overline{Q}+p$ ($p=q,\overline{q},g$), the 
dominant process~\cite{Grandchamp:2001pf}, cf.~\fig{fig:trans} (left). 
Gluo-dissociation and inelastic parton scattering-dissociation of
quarkonia have also been studied within an EFT approach~\cite{Brambilla:2013dpa}.

The equilibrium number densities are simply those of $Q$ quarks (with 
spin-colour and particle-antiparticle degeneracy $6 \times 2$) and quarkonium 
states (summed over including their spin degeneracies $d_{\cal Q}$). 
 
\noindent The quarkonium equilibrium number is given by: 
\begin{equation} 
  N_{\cal Q}^{\rm eq} =  
  V_{\mathrm{FB}} \sum\limits_{\cal Q}  n_{\cal Q}^{\rm eq}(m_{\cal Q};T,\gamma_Q) 
  = V_{\mathrm{FB}} \ \sum\limits_{\cal Q} d_{\cal Q} \  \gamma_Q^2 \int \frac{d^3p}{(2\pi)^3} f_{\cal Q}^B(E_p;T) \  
  \label{eq:Neq} 
\end{equation} 
where $V_{\rm FB}$ refers to the fireball volume, $d_{\cal Q}$ is the spin 
degeneracy and $f_{\cal Q}^B$ corresponds to the Bose distribution. 
 
\noindent The open heavy-flavour (HF) number, $N_{\rm{op}}$, follows from the 
corresponding equilibrium densities, \eg 
\begin{equation} 
N_{\rm{op}}=N_{\cal Q}+N_{\overline{\cal Q}} = V_{\rm FB} 12\gamma_Q \int 
\frac{d^3p}{(2\pi)^3} f_{Q}^F(E_p;T) 
\end{equation} 
for heavy (anti-)quarks in the QGP. 
 
Assuming relative chemical equilibrium between all available states containing 
heavy-flavoured quarks at a given temperature and volume of the system, the 
number of \QQbar pairs in the fireball ---usually determined by the initial hard 
production--- is matched to the equilibrium numbers of HF states, using a 
fugacity factor $\gamma_Q=\gamma_{\overline{Q}}$, by the condition: 
\begin{equation} 
  N_{\QQbar}=\frac{1}{2} N_{\rm{op}}\frac{I_1(N_{\rm{op}})}{I_0(N_{\rm{op}})}+ 
  V_{\mathrm{FB}} \ \gamma_Q^2\sum\limits_{\cal Q} n_{\cal Q}^{\rm eq}(T) 
  \ . 
  \label{eq:NQQ} 
\end{equation} 
The ratio of Bessel functions above, $I_1/I_0$, enforces the canonical limit for 
small $N_{\rm op} \le1$. 
 
The quarkonium equilibrium limit is thus coupled to the open HF spectrum in 
medium; \eg, a smaller $c$-quark mass increases the $c$-quark density, which 
decreases $\gamma_c$ and therefore reduces $N_{J/\psi}^{\rm eq}$, by up to an 
order of magnitude for $m_c=1.8 \to 1.5$\,GeV/$c^2$, cf.~\fig{fig:trans} 
(right). 
 
In practice, further corrections to $N_{\cal Q}^{\rm eq}$ are needed for more 
realistic applications in heavy-ion collisions. First, heavy quarks cannot be 
expected to be thermalised throughout the course of a heavy-ion collision; 
harder heavy-quark (HQ) momentum distributions imply reduced phase-space overlap 
for quarkonium production, thus suppressing the gain term. In the rate equation 
approach this has been implemented through a relaxation factor ${\cal R} = 
1-\exp(-\int \dd\tau/\tau_Q^{\rm therm})$ multiplying $N_{\cal Q}^{\rm eq}$, where 
$\tau_Q^{\rm therm}$ represents the kinetic relaxation time of the HQ 
distributions~\cite{Grandchamp:2002wp,Grandchamp:2003uw}. This approximation has 
been quantitatively verified in Ref.~\cite{Song:2012at}. Second, since HQ pairs 
are produced in essentially point-like hard collisions, they do not necessarily 
explore the full volume in the fireball. This has been accounted for by 
introducing a correlation volume in the argument of the Bessel functions, in 
analogy to strangeness production at lower energies~\cite{Hamieh:2000tk}. 
 
\begin{figure}[t] 
  \centering 
  \includegraphics[width=0.45\textwidth]{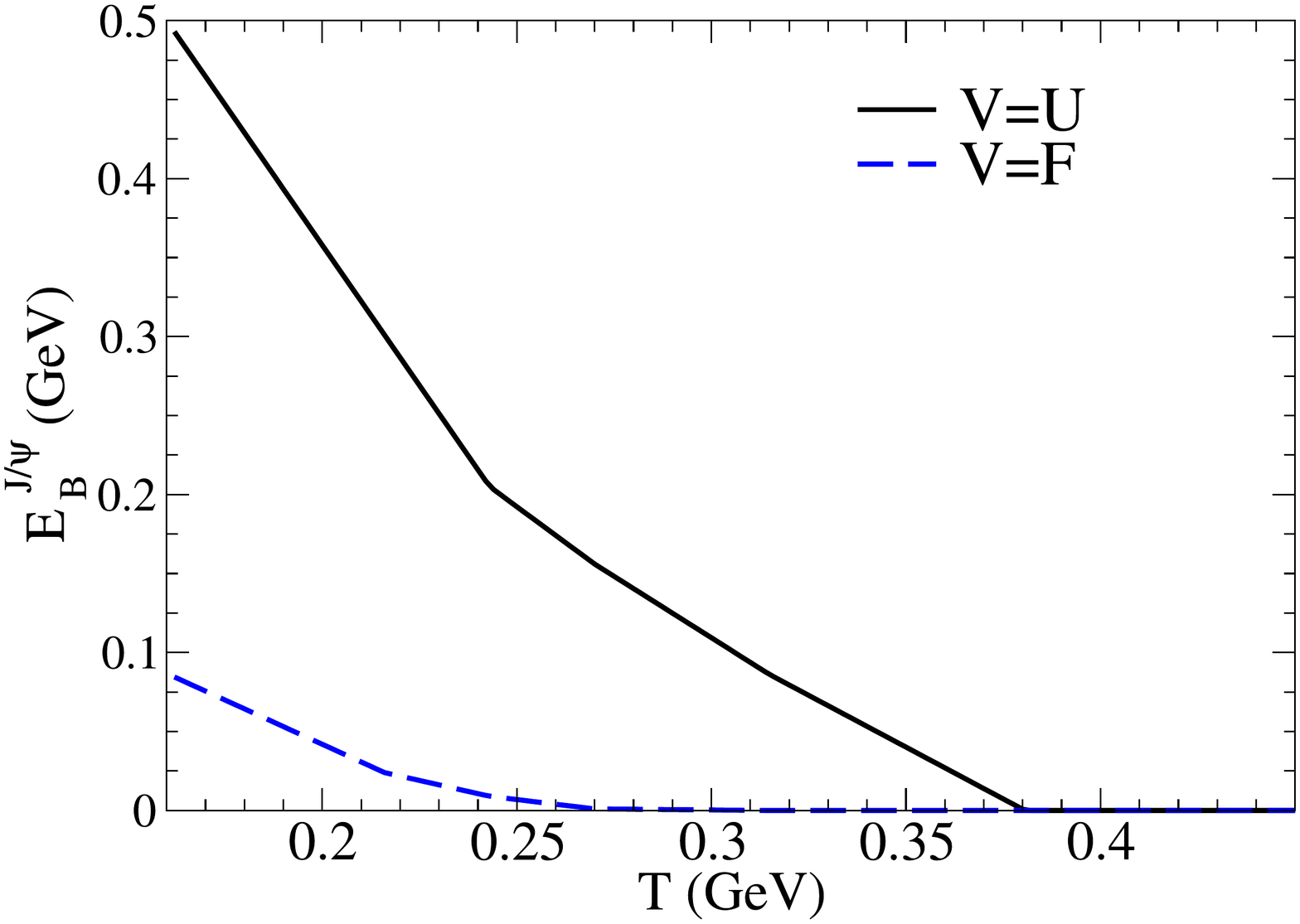} 
  \includegraphics[width=0.45\textwidth]{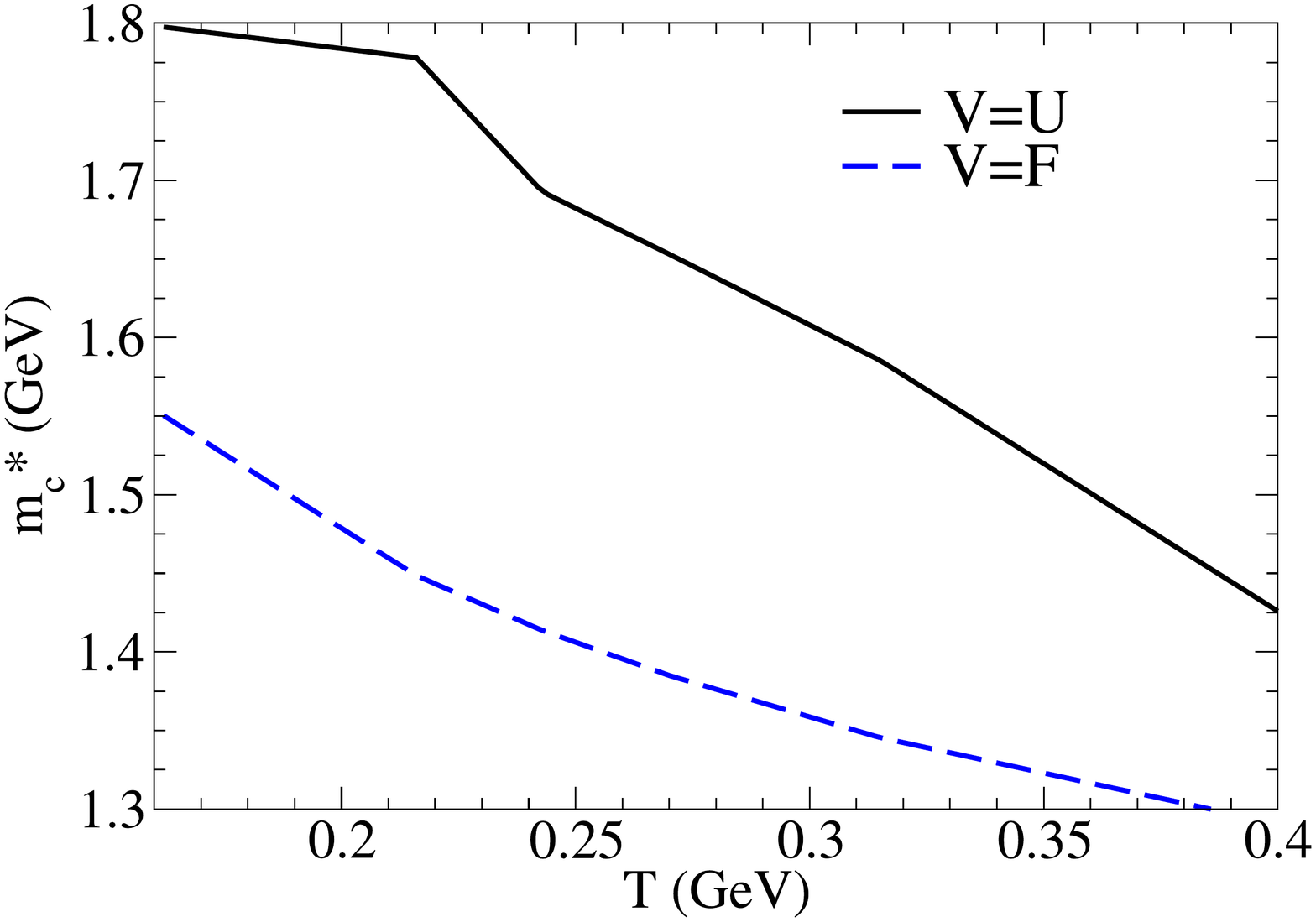} 
  \caption{Temperature dependence of J$/\psi$ binding energy (left panel) and 
    charm-quark mass (right panel) in the QGP in the strong- and weak-binding 
    scenarios (solid ($V$=$U$) and dashed lines ($V$=$F$), respectively) as 
    implemented into the rate equation approach~\cite{Zhao:2010nk}.} 
  \label{fig:med} 
\end{figure}

An important aspect of this transport approach is a controlled implementation of 
in-medium properties of the quarkonia~\cite{Grandchamp:2003uw,Zhao:2010nk}. 
Colour-screening of the QCD potential reduces the quarkonium binding energies, 
which, together with the in-medium HQ mass, $m_Q^*$, determines the bound-state 
mass, $m_{\cal Q} = 2m_Q^*- E_B$. As discussed above, the interplay of $m_{\cal 
  Q}$ and $m_Q^*$ determines the equilibrium limit, $N_{\cal Q}^{\rm eq}$, while 
$E_B$ also affects the inelastic reaction rate, $\Gamma_{\cal Q}(T)$. To 
constrain these properties, pertinent spectral functions have been used to 
compute Euclidean correlators for charmonia, and required to approximately agree 
with results from lattice QCD~\cite{Zhao:2010nk}. Two basic scenarios have been 
put forward for tests against charmonium data at the SPS and RHIC: a 
strong-binding scenario (SBS), where the J$/\psi$ survives up to temperatures of 
about 2\,$T_c$, and a weak-binding scenario (WBS) with $T_{\rm diss}\simeq 
1.2\,T_c$, cf.~\fig{fig:med}. These scenarios are motivated by microscopic 
$T$-matrix calculations~\cite{Riek:2010fk} where the HQ internal ($U_{\QQbar}$) 
or free energy ($F_{\QQbar}$) have been used as potential, respectively. A more 
rigorous definition of the HQ potential, and a more direct implementation of the 
quarkonium properties from the $T$-matrix approach is warranted for future work. 
The effects of the hadronic phase are generally small for \jpsi and bottomonia, 
but important for the \psiP, especially, if its direct decay channel $\psiP\to 
\overline{\rm D}{\rm D}$ is opened (due to reduced masses and/or finite widths 
of the D mesons)~\cite{Grandchamp:2002wp,Grandchamp:2003uw}. 
 
The rate equation approach has been extended to compute \pt spectra of charmonia 
in heavy-ion collisions~\cite{Zhao:2007hh}. Toward this end, the loss term was 
solved with a 3-momentum dependent dissociation rate and a spatial dependence of 
the charmonium distribution function, while for the gain term blast-wave 
distributions at the phase transition were assumed (this should be improved in 
the future by an explicit evaluation of the gain term from the Boltzmann 
equation using realistic time-evolving HQ distributions, see 
Ref.~\cite{Zhao:2010ti} for initial studies)~\cite{Schnedermann:1993ws}. In 
addition, formation time effects are included, which affect quarkonium 
suppression at high \pt~\cite{Zhao:2008vu}. 
 
To close the quarkonium rate equations, several input quantities are required 
which are generally taken from experimental data in \pp and \pa collisions, \eg, 
quarkonia and HQ production cross sections (with shadowing corrections), and primordial nuclear absorption effects 
encoded in phenomenological absorption cross sections. Feed-down effects from 
excited quarkonia (and $b$-hadron decays into charmonium) are accounted for. The 
space-time evolution of the medium is constructed using an isotropically 
expanding fireball model reproducing the measured hadron yields and their \pt 
spectra. The fireball resembles the basic features of hydrodynamic 
models~\cite{vanHees:2014ida}, but an explicit use of the latter is desirable 
for future purposes. 
 
Two main model parameters have been utilised to calibrate the rate equation 
approach for charmonia using the centrality dependence of inclusive \jpsi 
production in \PbPb collisions at the SPS (\snn= 17\,GeV) and in \AuAu 
collisions at RHIC (\snn = 200\GeV): the strong coupling constant $\alpha_s$, 
controlling the inelastic reaction rate, and the $c$-quark relaxation time 
affecting the gain term through the amended charmonium equilibrium limit. With 
$\alpha_s\simeq0.3$ and $\tau_c^{\text{therm}}\simeq$\,4--6 (1.5--2)~fm/$c$ for the 
SBS (WBS), the inclusive \jpsi data at SPS and RHIC can be reasonably well 
reproduced, albeit with different decompositions into primordial and regenerated 
yields (the former are larger in the SBS than in the WBS). The 
$\tau_c^{\text{therm}}$ obtained in the SBS is in the range of values calculated 
microscopically from the $T$-matrix approach using the 
$U$-potential~\cite{Riek:2010fk}, while for the WBS it is much smaller than 
calculated from the $T$-matrix using the $F$-potential. Thus, from a theoretical 
point of view, the SBS is the preferred scenario. 
 
With this set-up, namely the TAMU transport model, quantitative predictions for 
\PbPb collisions at the LHC (\snn= 2.76\,TeV) have been carried out for the 
centrality dependence and \pt spectra of \jpsi~\cite{Zhao:2011cv}, as well as 
for \upsa, \chib, and \upsb production~\cite{Emerick:2011xu}. 
 
Similar results are obtained in the transport approach THU developed by the 
Tsinghua group~\cite{Yan:2006ve,Liu:2009nb}, which differs in details of the 
implementation, but overall asserts the robustness of the conclusions. In the 
THU model, the quarkonium distribution is also governed by the Boltzmann-type 
transport equation. 
The cold nuclear matter effects change the initial quarkonium distribution and 
heavy quark distribution at $\tau_0$. 
The interaction between the quarkonia and the medium is reflected in the loss 
and gain terms and depends on the local temperature $T(\vec{r},\tau)$ and 
velocity $u_\mu(\vec{r},\tau)$, which are controlled by the energy-momentum and 
charge conservations of the medium, $\partial_\mu T^{\mu\nu}=0$ and 
$\partial_\mu n^\mu=0$. 
 
Within this approach, the centrality dependence of the nuclear modification 
factor \raa can be obtained and compared to experimental results at low \pt. In 
contrast to collisions at SPS and RHIC energies, at LHC energies the large 
abundance of $c$ and $\overline{c}$ quarks increases their combining probability 
to form charmonia. Hence this regeneration mechanism becomes the dominant source 
of charmonium production for semi-central and central collisions at the LHC. The 
competition between dissociation and regeneration leads to a flat structure of 
the \jpsi yield as a function of centrality. This flat behaviour should 
disappear at higher energies or, regeneration being a \pt-dependent mechanism, 
with increasing \pt. 
 
The charmonium transverse momentum distribution contains more dynamic 
information on the hot medium and can be calculated within the transport 
approach. The regeneration occurs in the fireball, and therefore the thermally 
produced charmonia are mainly distributed at low \pt, their contribution 
increasing with centrality. On the other hand, those charmonia from the initial 
hard processes carry high momenta and dominate the high \pt region at all 
centralities. This different \pt behaviour of the initially-produced and 
regenerated charmonia can even lead to a minimum located at intermediate \pt. 
Moreover, this particular \pt behaviour will lead to an evolution of the mean 
transverse momentum, $\langle\pt\rangle$, with centrality that would be higher 
for SPS than for LHC nuclear collisions, once normalised to the corresponding 
proton-proton $\langle\pt\rangle$~\cite{Zhou:2009vz,Zhou:2014kka}. 
At the SPS, almost all the measured \jpsi are produced through initial hard 
processes and carry high momentum. At RHIC, the regeneration starts to play a 
role and even becomes equally important as the initial production in central 
collisions. At the LHC, regeneration becomes dominant, and results in a 
decreasing of $\langle\pt\rangle$ with increasing centrality. 
 
\begin{figure}[htb] 
  \centering 
  \includegraphics[width=0.50\textwidth]{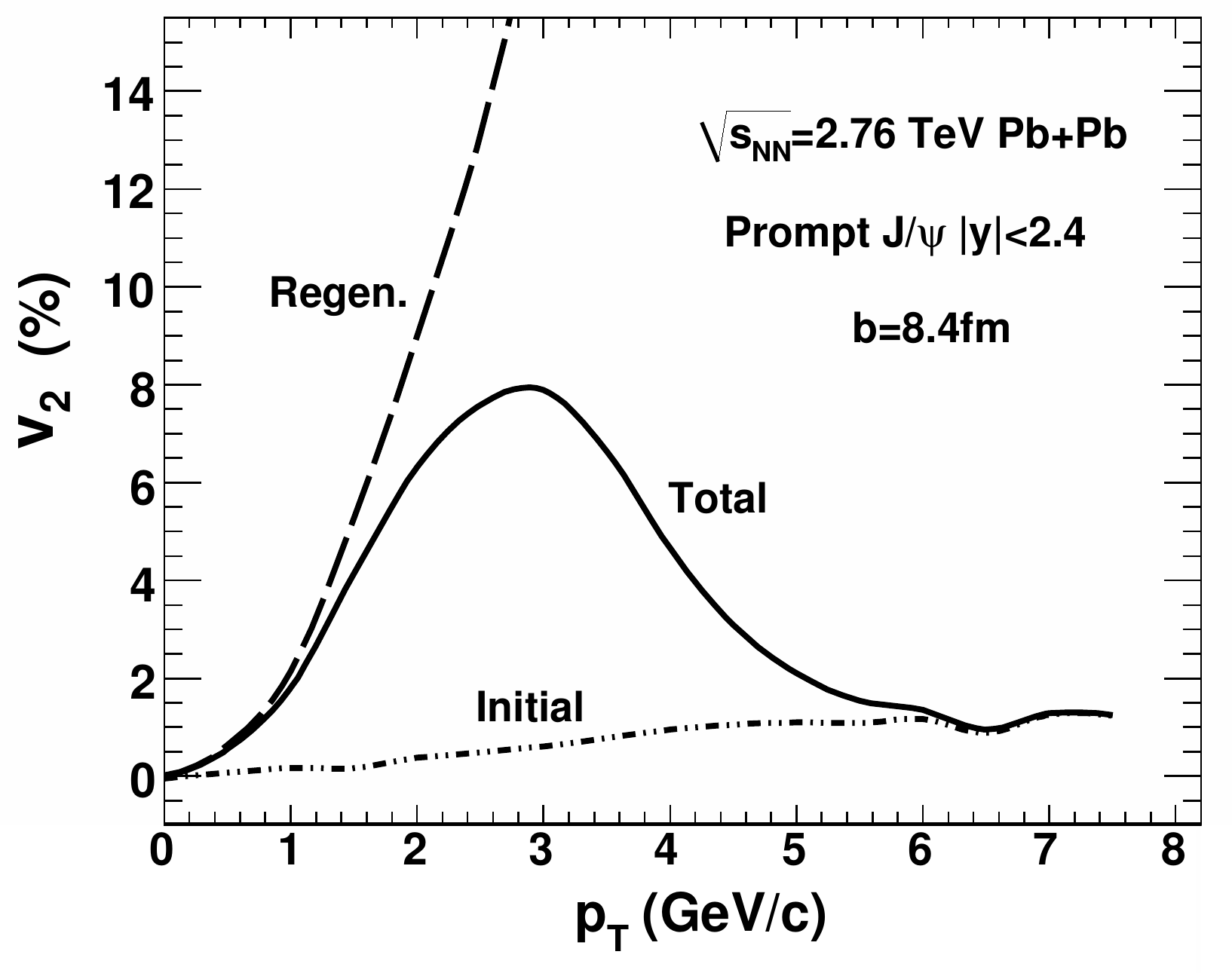} 
  \caption{Elliptic flow \vtwo for prompt \jpsi in \PbPb 
    collisions at \snn = 2.76\TeV as predicted by the THU model. The calculation is with impact parameter b = 8.4\,fm, 
    corresponding to the 0--100\% centrality range. The dot-dashed, dashed and solid 
    lines represent the initial, regeneration, and total contributions, 
    respectively.} 
  \label{fig:zhuang4} 
\end{figure}

Concerning the \jpsi elliptic flow, due to the strong interaction between the 
heavy quarks and the hot medium, the regenerated charmonia inherit collective 
flow from the charm quarks. Furthermore, primordial \jpsi might acquire a \vtwo 
induced by a path-length dependent suppression. As shown in \fig{fig:zhuang4}, 
the \jpsi \vtwo will, therefore, result from the interplay of two contributions, 
a regeneration component, dominant at lower \pt and the primordial \jpsi 
component that takes over at higher \pt. Hence, given the increasing 
regeneration fraction with colliding energy, the \jpsi elliptic flow is expected 
to become sizeable at LHC while it should be almost zero at RHIC. 
 
\subsubsection{Non-equilibrium effects on quarkonium suppression}
\label{sec:non-equilib}

Since heavy quarkonium states have a short formation time in their rest frame 
($< 1~\text{fm}/c$), they are sensitive to the early-time dynamics of the QGP. 
As a consequence, it is necessary to have dynamical models that can accurately 
describe the bulk dynamics of the QGP during the first $\text{fm}/c$ of its 
lifetime. This is complicated by the fact that, at the earliest times after the 
initial nuclear impact, the QGP is momentum-space anisotropic in the local rest 
frame. The existence of large QGP momentum-space anisotropies is found in both 
the weak and strong coupling limits (see e.g. 
Ref.~\cite{Attems:2012js,Berges:2012iw,Heller:2012je,vanderSchee:2013pia}). In 
both limits, one finds that the longitudinal pressure, ${\cal P}_L = T^{zz}$, is 
much less than the transverse pressure, ${\cal P}_T = (T^{xx}+T^{yy})/2$, at 
times smaller than $1~\text{fm}/c$. During the QGP evolution this momentum-space 
anisotropy relaxes to zero, but it does so only on a time-scale of several 
$\text{fm}/c$. In addition, the momentum-space anisotropy grows larger as one 
approaches the transverse edge of the QGP, where the system is colder. The 
existence of such momentum-space anisotropies is consistent with first- and 
second-order viscous hydrodynamics; however, since these approaches rely on 
linearisation around an isotropic background, it is not clear that these methods 
can be applied in a far-from-equilibrium situation. In order to address this 
issue, a non-perturbative framework, called anisotropic hydrodynamics (aHYDRO), 
has been developed. This framework allows the system to be arbitrarily 
anisotropic~\cite{Florkowski:2010cf,Martinez:2010sc,Martinez:2012tu,Bazow:2013ifa}. 
 
The time-evolution provided by aHYDRO has to be folded together with the 
non-equilibrium (anisotropic) quarkonium rates. These were first considered in 
Ref.~\cite{Dumitru:2007hy,Dumitru:2009ni,Burnier:2009yu,Dumitru:2009fy,Margotta:2011ta} 
where the effect of momentum-space anisotropy was included for both the real and 
imaginary parts of potential. In this context, the imaginary part of the 
potential plays the most important role as it sets the in-medium decay rate of 
heavy quarkonium states. The calculations of the resulting decay rates in 
Ref.~\cite{Margotta:2011ta} demonstrated that these in-medium decay rates were 
large with the corresponding lifetime of the states being on the order of 
$\text{fm}/c$. In practice, one integrates the decay rate over the lifetime of 
the state in the plasma as a function of its three-dimensional position in the system and its 
transverse momentum. The result of this is a prediction for the \raa that 
depends on the assumed shear-viscosity to entropy density ratio ($\eta/s$) of 
the QGP since this ratio determines the degree to which the system remains 
isotropic. 
The results obtained for the inclusive \upsa and \upsb 
suppression~\cite{Strickland:2011mw,Strickland:2011aa,Strickland:2012as} have a 
significant dependence on the assumed value of $\eta/s$, in particular for the 
inclusive \upsa. This ratio can be determined from independent collective flow 
measurements and at the energies probed by the LHC one finds that $4\pi\,\eta/s 
\sim 1\text{--}3$~\cite{Strickland:2011aa,Schenke:2011tv}. The upper limit of 
this range seems to be compatible with the CMS data (the comparison will be shown  
in \sect{sec:bottomonium});  
however, since the model 
used did not include any regeneration effects, it is possible that the final 
$\eta/s$ could be a bit lower than three times the lower bound.  
Furthermore, it 
should be pointed out that feed-down fractions based on CDF measurements with 
$\pt>8\GeVc$ are used~\cite{Abe:1995an,Affolder:1999wm}, which with 
$\approx50\%$ is larger than the fraction one would obtain when using the recent 
$\chib\text{nP}\rightarrow\upsa$ measurements by LHCb that extend to slightly 
lower \pt~\cite{Aaij:2014caa}. In the later case the total \upsa feed down 
contribution is $\approx30\%$ for $\pt>6\GeVc$.


\subsubsection{Collisional dissociation of quarkonia from final-state interactions}
\label{sec:finalstate_diss}

The model described in Section~\ref{sec:sharmaVitev_dissociation} can also be 
modified to describe the dissociation dynamics of quarkonia in the QGP. The 
model includes both initial state cold nuclear matter energy loss and final 
state effects, such as radiative energy loss for the colour-octet state and 
collisional dissociation for quarkonia, as they traverse the created hot medium. 
The main differences with respect to the formalism discussed 
in~\sect{sec:sharmaVitev_dissociation} are (a) that once a high-\pt quarkonium is 
dissociated, it is unlikely that it will fragment again to form a new 
quarkonium, (b) the formation time is given not by fragmentation dynamics but by 
binding energies. A self-consistent description of the formation of a quarkonium 
in a thermal QGP is a challenging problem~\cite{Sharma:2012dy} and assumes that 
the formation time lies between $1/(2E_b)$ and $1/(E_b)$, and that the wave 
function does not show significant thermal effects in this short time. 
 
When compared to the \jpsi \raa results, obtained by the CMS experiment, the 
model is consistent with the observations for the peripheral events, but 
underestimates the suppression for the most central events, suggesting that 
thermalisation effects on the wave functions may be substantial. 

\subsubsection{Comover models}
\label{sec:comovers}

The comover interaction model (CIM) was originally developed in the nineties in 
order to explain both the suppression of charmonium yields and the strangeness 
enhancement in nucleus-nucleus collisions at the 
SPS~\cite{Capella:2000zp,Capella:2007jv,Ferreiro:2012rq}. It includes the 
initial-state nuclear effects, the so-called nuclear shadowing. It takes into 
account the quarkonium dissociation due to interactions with the comoving medium 
and the recombination of \QQbar into secondary quarkonium states. It is based on 
the well-known gain and loss differential equations in transport theory for a 
quarkonium state $\mathcal{Q}$: 
\begin{equation}\label{eq:recorateeq} 
  \tau \frac{\dd N_{\cal Q}}{\dd\tau}\left( b,s,y \right) = -\sigma_{\text{co}} \left[ N^{\text{co}}(b,s,y) N_{\cal Q}(b,s,y) \,-\, 
    N_Q(b,s,y) N_{\overline{Q}}(b,s,y) \right], 
\end{equation} 
as a function of impact parameter $b$, centre-of-mass energy squared $s$, and 
rapidity $y$. The first term refers to the quarkonium dissociation and the 
second term takes care of the recombination of \QQbar into secondary quarkonium 
states. The variable $\sigma_{\text{co}}$ denotes the cross section of 
quarkonium dissociation due to interactions with the comoving medium, with 
density $N^{\text{co}}$. 
 
Assuming a dilution in time of the densities due to longitudinal motion, which 
leads to a $\tau^{-1}$ dependence on proper time, the approximate solution of 
\eq{eq:recorateeq} gives the survival probability: 
\begin{equation}\label{eq:fullsupp} 
  S^{\text{co}}(b,s,y) = \exp \left\{-\sigma_{\text{co}} 
  \left[N^{\text{co}}(b,s,y)\,-\, \frac{N_Q(b,s,y) 
  N_{\overline{Q}} (b,s,y)}{N_{\cal Q}(b,s,y)} \right] \, \ln 
\left[\frac{N^{\text{co}}(b,s,y)}{N_{\pp} (y)}\right] \right\}. 
\end{equation} 
Using the inverse proportionality between proper time and densities,\ie 
$\tau_f/ \tau_0= N^{\text{co}}(b, s, y)/N_{\pp}(y)$ ---the interaction stops when 
the densities have diluted, reaching the value of the \pp density at the same 
energy--- 
it can be concluded that the 
result depends only on the ratio $\tau_f/ \tau_0$ of final over initial 
time.


\subsubsection{Summary of theoretical models for experimental comparison}
\label{sec:summ_theo}

Different theoretical models are available for comparison. Among them, the 
statistical hadronisation model, the transport model, the collisional 
dissociation model, and the comover model will be compared to charmonium 
experimental results in the next section. 
Their principal characteristics can be summarised as follows. 
 
In the statistical hadronisation model, the charm (beauty) quarks and 
antiquarks, produced in initial hard collisions, thermalise in a QGP and form 
hadrons at chemical freeze-out. It is assumed that no quarkonium state survives 
in the deconfined state (full suppression) and, as a consequence,  
also CNM effects are not included in this model.   
An important aspect in this scenario is the canonical suppression of open charm 
or beauty hadrons, which determines the centrality dependence of production 
yields in this model. The overall magnitude is determined by the input charm 
(beauty) production cross section. 
 
Kinetic (re)combination of heavy quarks and antiquarks in a QGP provides an 
alternative quarkonium production mechanism. In transport models, there is 
continuous dissociation and (re)generation of quarkonia over the entire lifetime 
of the deconfined state. A hydrodynamical-like expansion of the fireball of 
deconfined matter, constrained by data, is part of such models, alongside an 
implementation of the screening mechanism with inputs from lattice QCD. Other 
important ingredients are parton-level cross sections. Cold nuclear matter 
effects are incorporated by means of an overall effective absorption cross 
section that accounts for (anti-)shadowing, nuclear absorption, and Cronin 
effects. 
 
The collisional dissociation model considers, in addition to modifications of 
the binding potential by the QGP and cold nuclear matter effects, radiative 
energy loss of the colour octet quarkonium precursor and collisional 
dissociation processes inside the QGP. 
 
Similarly, the comover interaction model includes dissociation of quarkonia by 
interactions with the co-moving medium of hadronic and partonic origin. 
Regeneration reactions are also included. Their magnitude is determined by the 
production cross section of \ccbar pairs and quarkonium states. Cold nuclear 
matter effects are taken into account by means of (anti-)shadowing 
models. 
 
Summarising: 
\begin{itemize} 
\item Statistical hadronisation assumes full suppression of primordial quarkonia and 
  regeneration at the phase boundary. 
\item Transport models include cold nuclear absorption, direct suppression, and 
  regeneration. 
\item Collisional dissociation models include initial state cold nuclear matter 
  effects and final state effects based on radiative energy loss and collisional 
  dissociation. 
\item Comover models include shadowing, interaction with co-moving medium, and 
  regeneration. 
\end{itemize} 
 
In transport and comover models, at LHC energies, a large fraction of \jpsi ($> 
50\%$ in most central collisions) is produced by charm quark recombination. In 
the statistical hadronisation model, all \jpsi are generated at the 
hadronisation stage by purely statistical mechanisms. In order to include 
(re)generation, a cross section $\dd\sigma_{\rm pp}^{\ccbar}/\dd y \approx 
0.6\text{--}0.8$~mb at midrapidity at \snn = 2.76\TeV has been considered in the 
transport and comover models. It corresponds to $\sigma_{\rm pp}^{\ccbar}$ 
around 5~mb, which agrees with experimental data (see \fig{fig:pp:CharmAndBottomXsec}). Currently available data, 
however, offer only very little constrains at 2.76\TeV due to the lack of 
D meson measurements at $\pt<2\GeVc$ in Pb--Pb collisions. The used value is about a factor of two higher 
than the one used in the statistical hadronisation model. Note nevertheless 
that there is no contradiction, since in the latter the initial-state shadowing 
is not modelled. The choice of smaller cross section in \pp, 
$\dd\sigma_{\rm pp}^{\ccbar}/\dd y \approx 0.3\text{--}0.4$~mb, takes into account a 
shadowing effect that reduces the charm cross section in \PbPb by up to a factor 
of two. 

In order to compare with experimental data on bottomonium, also the 
hydrodynamical formalism assuming finite local momentum-space anisotropy due to 
finite shear viscosity will be considered. The main ingredients are: screened 
potential, hydrodynamical-like evolution of the QGP, and feed-down from higher 
mass states. Neither cold nuclear matter effects nor recombination are included. 


\subsection{Experimental overview of quarkonium results at RHIC and LHC}
\label{LHC_exp}

\subsubsection{Proton--proton collisions as a reference for \raa at the LHC}
\label{sec:pp_ref}

The medium effects on quarkonia are 
usually quantified via the nuclear modification factor \raa, basically comparing 
the quarkonium yields in \AAcoll to the \pp ones. A crucial ingredient for the 
\raa evaluation is, therefore, $\sigma_{\rm pp}$, the quarkonium production cross 
section in \pp collisions measured at the same energy as the \AAcoll data. 
 
During LHC Run~1, \pp data at \s = 2.76 \TeV were collected in two short 
data taking periods in 2011 and 2013. When the collected data sample was large 
enough, the quarkonium $\sigma_{\rm pp}$ was experimentally measured, otherwise 
an interpolation of results obtained at other  energies was made. 
 
More in detail, the \jpsi\ cross section ($\sigma_{\rm pp}^{\jpsi}$) adopted by 
ALICE for the forward rapidity \raa results is based on the 2011 \pp 
data-taking. The $\lumi_{\text{int}} = 19.9\nbinv$ integrated luminosity, 
corresponding to $1364\pm53$ \jpsi reconstructed in the dimuon decay channel, 
allows for the extraction of both the integrated as well as the \pt and $y$ 
differential cross sections~\cite{Abelev:2012kr}. The statistical 
uncertainty is 4\% for the integrated result, while it ranges between 6\% and 
20\% for the differential measurement. Systematic uncertainties are $\sim$8\%. 
The collected data ($\lumi_{\text{int}} = 1.1\nbinv$) allow for the evaluation 
of $\sigma_{\rm pp}^{\jpsi}$ also in the ALICE mid-rapidity region, where \jpsi are 
reconstructed through their dielectron decay. The measurement is, in this case, 
affected by larger statistical and systematic uncertainties, of about  
23\% and 18\%, respectively. Therefore, the $\sigma_{\rm pp}^{\jpsi}$ reference for 
the \raa result at mid-rapidity was obtained performing an interpolation 
based on mid-rapidity results from PHENIX at \s = 0.2\TeV~\cite{Adare:2006kf}, 
CDF at \s = 1.96\TeV~\cite{Acosta:2004yw}, and ALICE at \s = 
2.76~\cite{Abelev:2012kr} and 7\TeV~\cite{Aamodt:2011gj}. The 
interpolation is done by fitting the data points with several functions assuming 
a linear, an exponential, a power law, or a polynomial \s-dependence. The 
resulting systematic uncertainty is, in this case, 10\%, \ie smaller than the 
one obtained directly from the data at \s = 2.76\TeV. 
 
The \jpsi \pp cross section used as a reference for the \raa measurements 
obtained by CMS is based on the results extracted from the data 
collected at \s = 2.76\TeV in 2011, corresponding to an integrated luminosity 
$\lumi_{\text{int}} = 231\nbinv$~\cite{Chatrchyan:2012np}. The number of prompt 
and non-prompt \jpsi in the range $|y|<2.4$ and $6.5<\pt<30\GeVc$ are $830\pm 
34$ and $206\pm20$, respectively. The systematic uncertainty on the signal 
extraction varies between 0.4\% and 6.2\% for prompt \jpsi and 5\% and 20\% for 
non-prompt \jpsi. Since the adopted reconstruction procedure is the same in \pp 
and \PbPb collisions, many of the reconstruction-related systematic 
uncertainties cancel when \raa is computed. 
 
The limited size of the \pp data sample at \s = 2.76\TeV has not allowed ALICE 
to measure the \ups\ cross section. The reference adopted by ALICE for the \raa 
studies~\cite{Abelev:2014nua} is, in this case, based on the \pp\  
measurement by LHCb~\cite{Aaij:2014nwa}. However, since the LHCb result is obtained in a 
rapidity range ($2<y<4.5$) not exactly matching the ALICE one ($2.5<y<4$), the 
measurement is corrected through a rapidity interpolation based on a Gaussian 
shape. 
 
For the \ups \raa, CMS results are based on the \pp reference cross section 
extracted from \pp data at \s = 2.76\TeV~\cite{Chatrchyan:2012np}. 
The number of \upsa with $|y|<2.4$ 
and $0<\pt<20\GeVc$ is $101\pm12$, with a systematic uncertainty on the signal 
extraction of $\sim10\%$. 
 
In \tab{tab:ppref_summary} the datasets and the approach adopted for the 
evaluation of the \pp reference are summarised. 
 
\begin{table*}[t] 
  \centering 
  \caption{Overview of the \pp datasets and approaches adopted for the 
    evaluation of the $\sigma_{\rm pp}$ production cross section for the quarkonium states under study.} 
  \label{tab:ppref_summary} 
  \begin{tabular*}{\textwidth}{@{\extracolsep{\fill}}ccc@{}} 
    \hline 
    & ALICE & CMS \\ 
    \hline 
    \jpsi & forward-$y$: $\sigma^{\jpsi}_{\pp}$ from \pp data at \s = 2.76\TeV&  $\sigma^{\jpsi}_{\pp}$ from \pp data at \s = 2.76\TeV \\ 
    & mid-$y$: $\sigma^{\jpsi}_{\pp}$ from interpolation of ALICE, CDF and PHENIX data& \\ 
    $\Upsilon$ & $\sigma^{\ups}_{\pp}$ from LHCb \pp data at \s = 2.76\TeV + 
    $y$-interpolation & $\sigma^{\ups}_{\pp}$ from \pp data at \s = 2.76\TeV\\ 
    \hline    
  \end{tabular*} 
\end{table*} 
 

\subsubsection{\texorpdfstring{$\rm J/\psi$}{J/psi} \raa results at low \pt}
\label{sec:raa_lowpt}

\begin{figure}[t] 
\centering 
\includegraphics[width=7cm,clip]{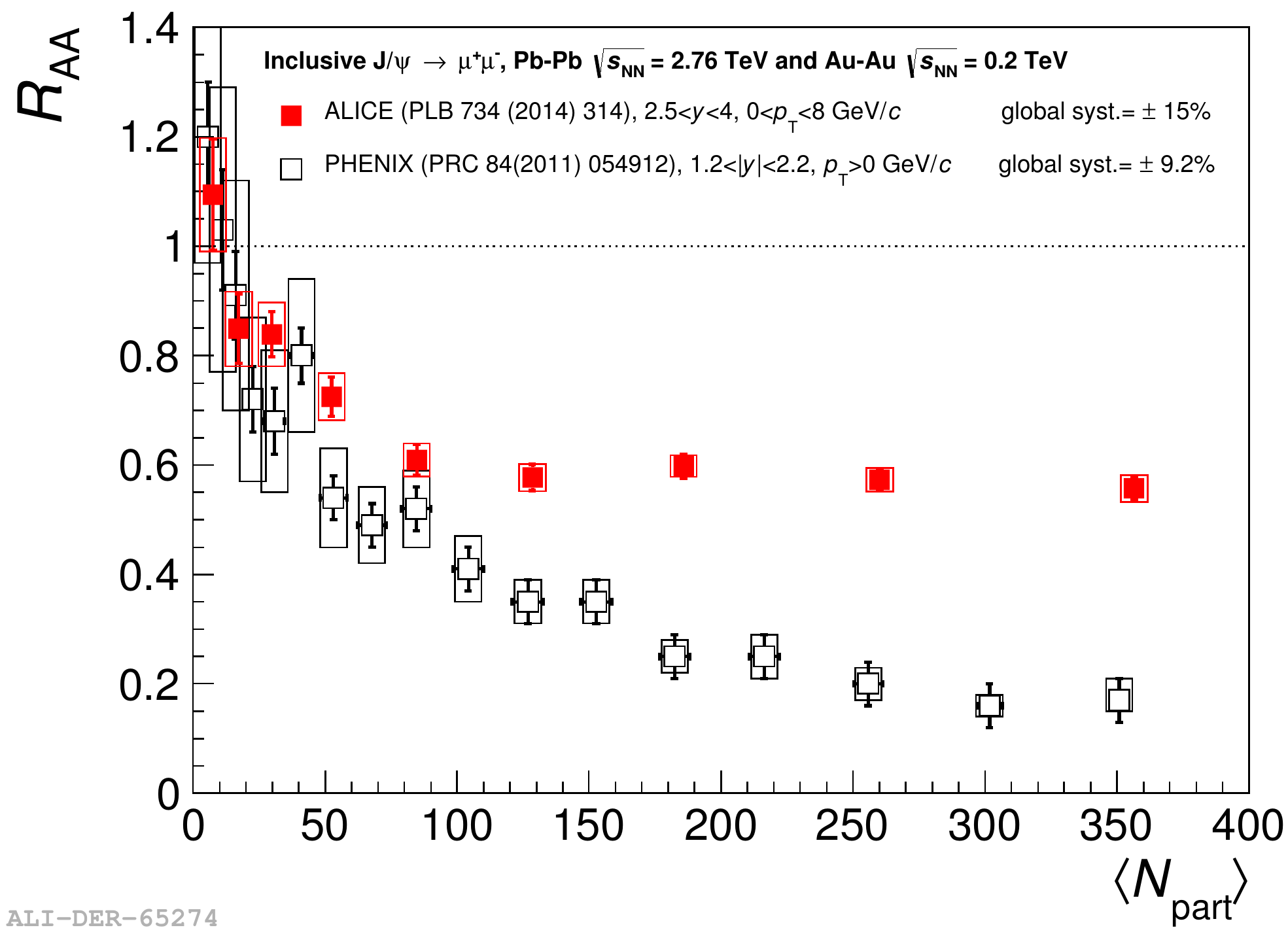} 
\includegraphics[width=7cm,clip]{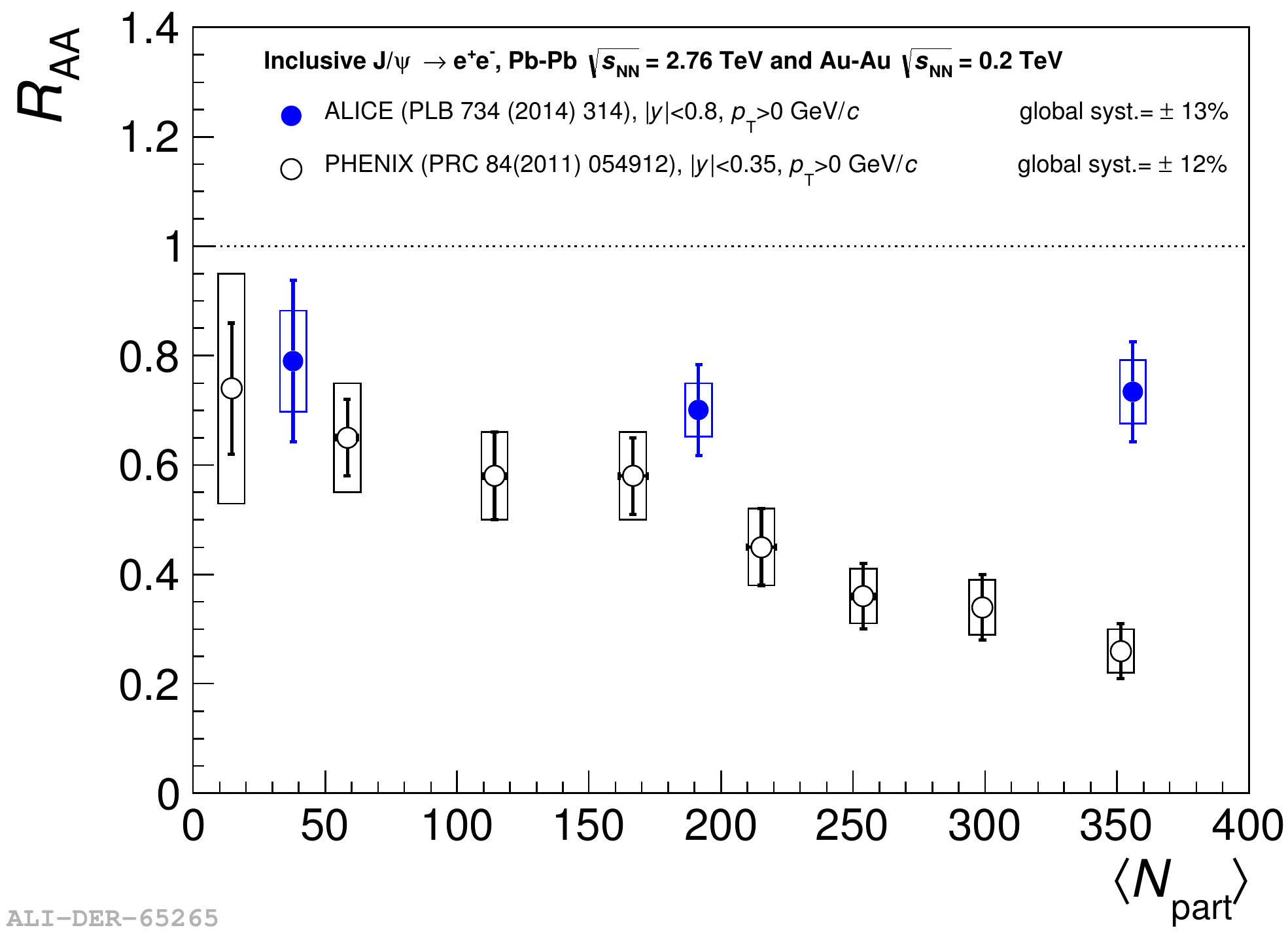} 
\caption{ALICE~\cite{Abelev:2013ila,Abelev:2012rv} (closed symbols) and 
  PHENIX~\cite{Adare:2011yf} (open symbols) inclusive \jpsi nuclear modification 
  factor versus the number of participant nucleons, at forward rapidity (left) 
  and at mid-rapidity (right).} 
\label{fig:Alice_Phenix_Centr}        
\end{figure}

The experiments ALICE at the LHC and PHENIX and STAR at RHIC measure the inclusive \jpsi production (prompt \jpsi plus those coming from $b$-hadron decays) in the low \pt region, down to $\pt = 0$. STAR measures \jpsi reconstructed from their \ee decay at mid-rapidity ($|y| <$ 1), while PHENIX detects charmonia in two rapidity ranges: at mid-rapidity ($|y|<0.35$) in the \ee decay channel and at forward rapidity ($1.2<|y|<2.2$) in the \mumu decay channel. Similarly, ALICE studies the inclusive \jpsi production in the \ee decay channel at mid-rapidity ($|y|<0.9$) and in the \mumu decay channel at forward rapidity ($2.5<y<4$). A summary of the main experimental results, together with their kinematic coverage and references, is given in Tables~\ref{tab:expSummary_RHIC} and ~\ref{tab:expSummary_LHC}. The experiments have investigated the centrality dependence of the \jpsi nuclear modification factor measured in \AAcoll collisions, \ie\ \AuAu at \snn = 200\GeV for PHENIX~\cite{Adare:2011yf} and STAR~\cite{Adamczyk:2013tvk} and \pb at \snn = 2.76\TeV in the ALICE case~\cite{Abelev:2013ila,Abelev:2012rv}. As an example, PHENIX and ALICE results are shown in
\fig{fig:Alice_Phenix_Centr} for the forward (left) and the mid-rapidity (right) regions. While the RHIC results show an increasing suppression towards more central collisions, the ALICE \raa has a flatter behaviour both at forward and at mid-rapidity. In the two $y$ ranges there is a clear 
evidence for a smaller suppression at the LHC than at RHIC. 

Partonic transport models that include a (re)generation process for \jpsi due to 
the (re)combination of \ccbar pairs along the history of the collision indeed 
predict such a behaviour~\cite{Zhao:2011cv,Liu:2009nb,Ferreiro:2012rq}, the 
smaller suppression at the LHC being due to the larger \ccbar pair multiplicity 
which compensates the suppression from colour screening in the deconfined phase. 
\begin{figure}[htb] 
  \centering 
  \includegraphics[width=0.45\textwidth]{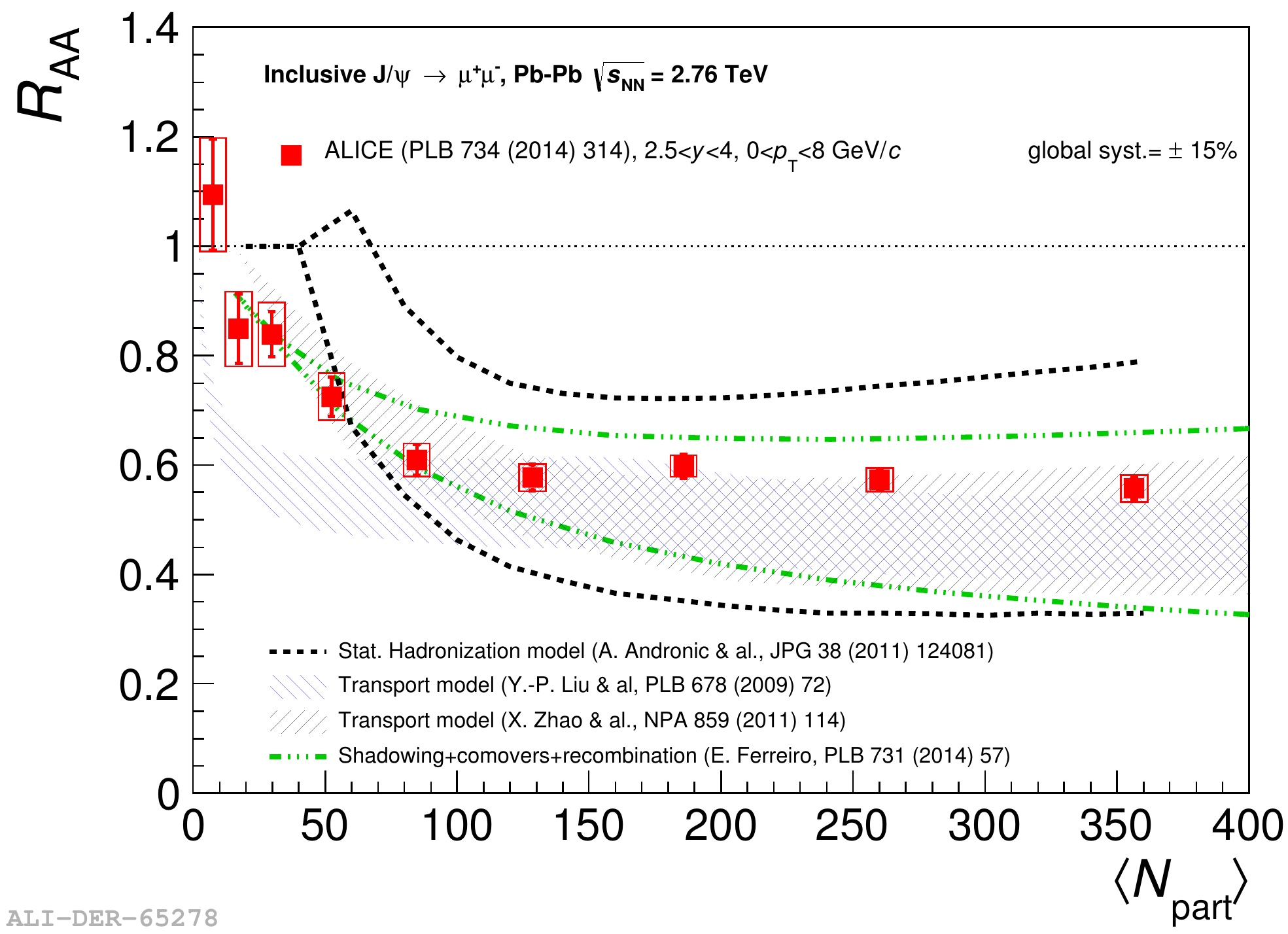} 
  \includegraphics[width=0.45\textwidth]{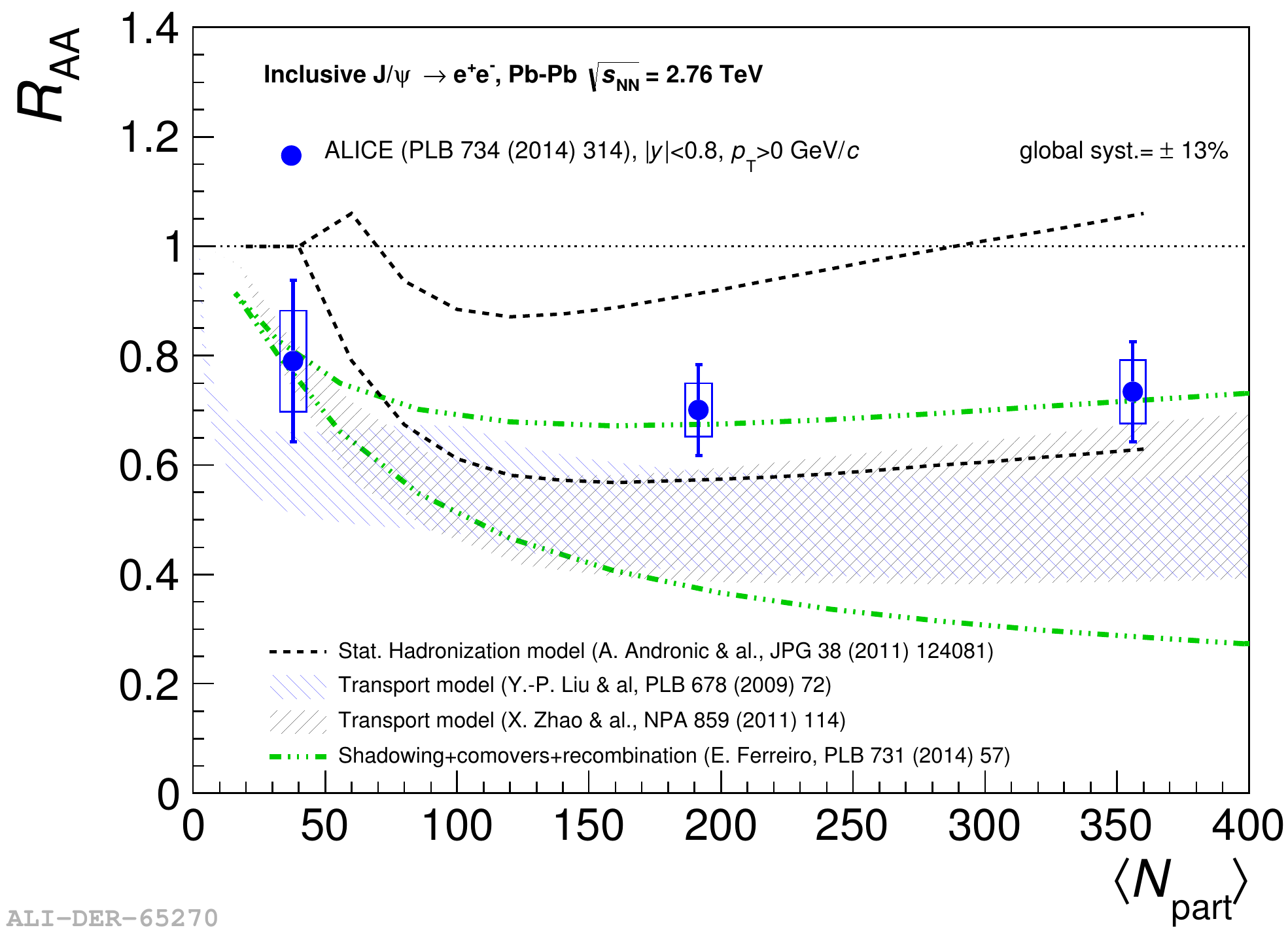} 
  \caption{Comparison of the ALICE \jpsi \raa at forward rapidity (left) and 
    mid-rapidity (right) with the theory predictions based on the TAMU (X. Zhao 
    et al.) and THU (Y.P. Liu et al.) transport models discussed in 
    \sect{sec:transport}. Bands correspond to the uncertainty associated to the 
    model, i.e. to the variation of the charm cross section for the THU model 
    and a variation of the shadowing amount for the TAMU approach. Predictions 
    from the statistical model discussed in \sect{sec:regeneration} (A. Andronic 
    et al.) are also shown. The two curves correspond, in this case, to two 
    assumptions on the values of the $d\sigma_{c\bar{c}}/dy$ cross sections. 
    Calculations based on the comover model (E. Ferreiro), presented in 
    \sect{sec:comovers}, are included in the plot. The lower and upper curves 
    correspond to variations of the charm cross section.} 
  \label{fig:RAA_models_NPart} 
\end{figure} 
The \raa centrality dependence was predicted by the TAMU and THU 
transport models, discussed in \sect{sec:transport}. For both models, 
(re)generation becomes the dominant source for charmonium production for 
semi-central and central collisions and the competition between the dissociation 
and (re)generation mechanisms leads to the observed flat structure of the \jpsi 
\raa as a function of centrality. The comparison of the predictions of the two 
transport models with the ALICE data is shown in \fig{fig:RAA_models_NPart} for 
the forward (left) and mid-rapidity (right) regions. 
 
\begin{figure}[h] 
\vspace*{-0mm} 
  \centering 
  \includegraphics[width=0.42\textwidth]{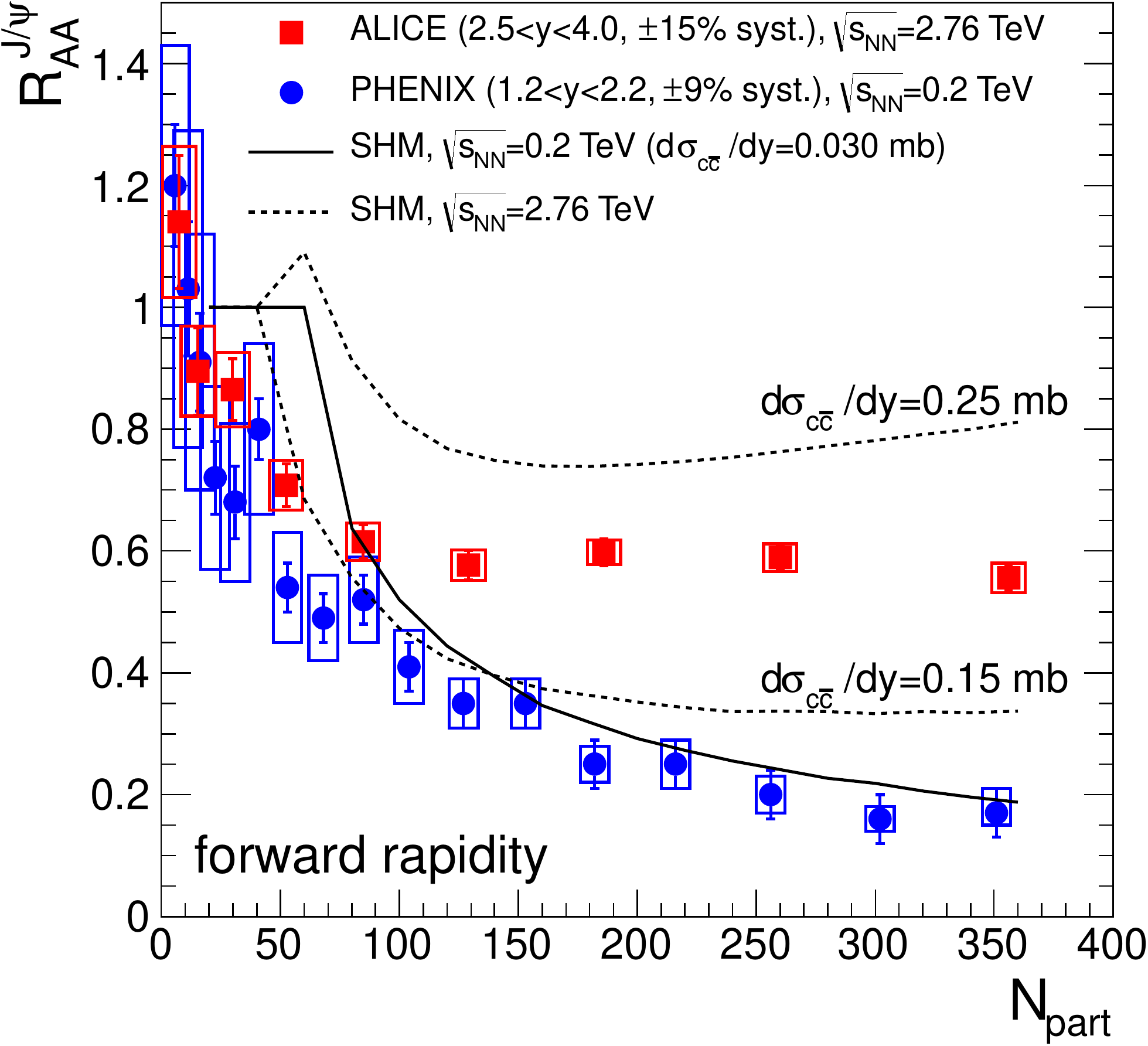} 
  \includegraphics[width=0.42\textwidth]{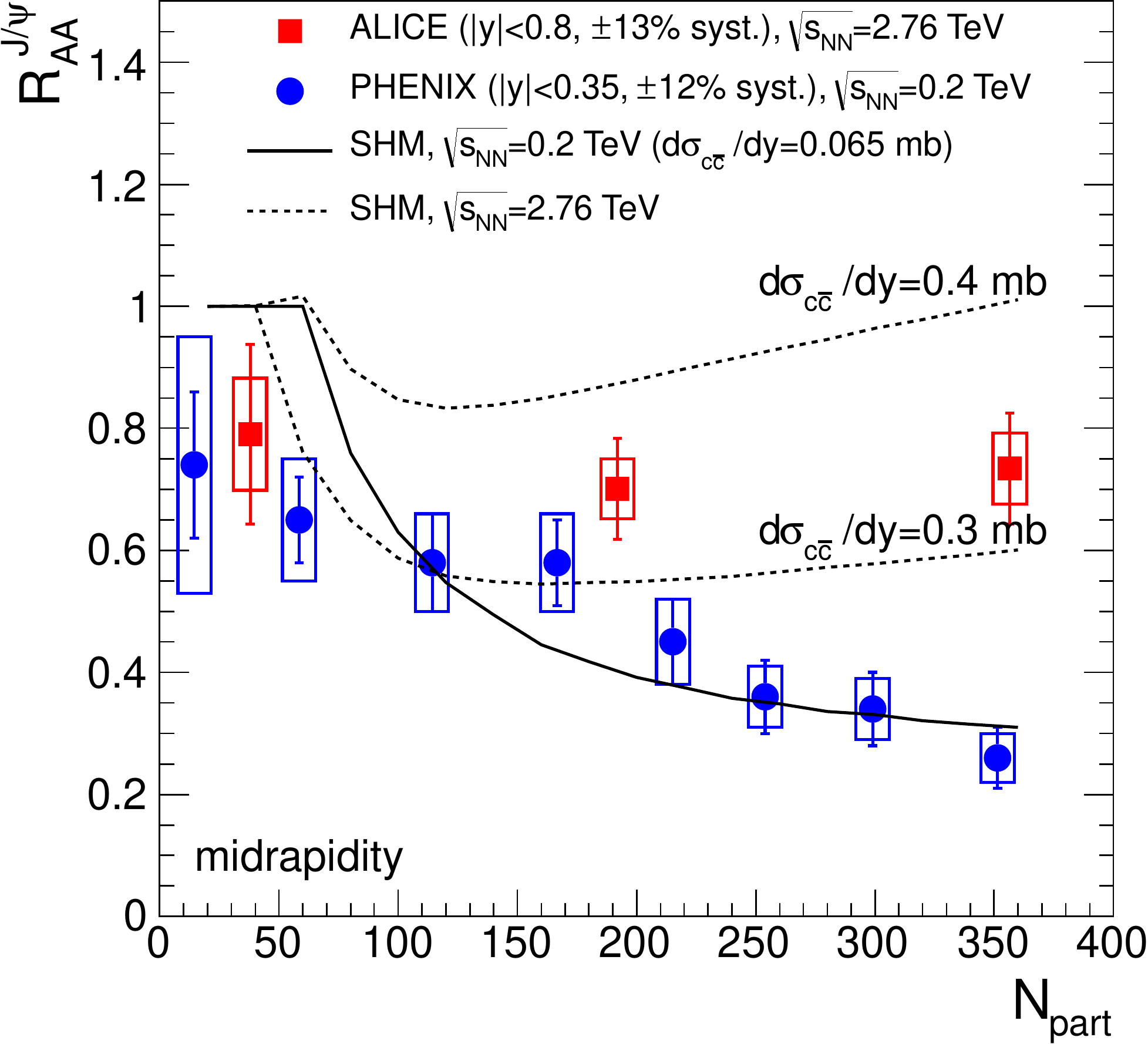} 
  \caption{\jpsi \raa from ALICE~\cite{Abelev:2013ila} and 
    PHENIX~\cite{Adare:2011yf} compared to predictions from the statistical 
    hadronisation model~\cite{Andronic:2011yq}.} 
\label{fig:stat_mod_charm} 
\end{figure} 
 
A similar behaviour is expected by the statistical model~\cite{Andronic:2011yq}, 
discussed in \sect{sec:regeneration}, where the \jpsi yield is completely 
determined by the chemical freeze-out conditions and by the abundance of 
$c\overline{c}$ pairs. In Figures~\ref{fig:RAA_models_NPart} and 
\ref{fig:stat_mod_charm}, the statistical model predictions are compared to the 
ALICE \raa in the two covered rapidity ranges. As discussed in 
\sect{sec:regeneration}, a crucial ingredient in this approach is the \ccbar 
production cross section: the error band in the figures stems from the 
measurement of the \ccbar cross section itself and from the correction 
introduced to take into account the \s extrapolation to 
evaluate the cross section at the \PbPb energy (\snn = 2.76\TeV). In 
\fig{fig:stat_mod_charm} the RHIC data~\cite{Adare:2011yf} and the corresponding 
statistical model calculations are also shown. Inspecting 
\fig{fig:stat_mod_charm}, for central collisions a significant increase in the 
\jpsi \raa at LHC as compared to RHIC is visible and well reproduced by the 
statistical hadronisation model. In particular, as a characteristic feature of 
the model, the shape as a function of centrality is entirely given by the charm 
cross section at a given energy and rapidity and is well reproduced both at RHIC 
and LHC. This applies also to the maximum in \raa at mid-rapidity due to the 
peaking of the charm cross section there. 
 
\begin{figure}[t] 
\centering 
\includegraphics[width=0.42\textwidth]{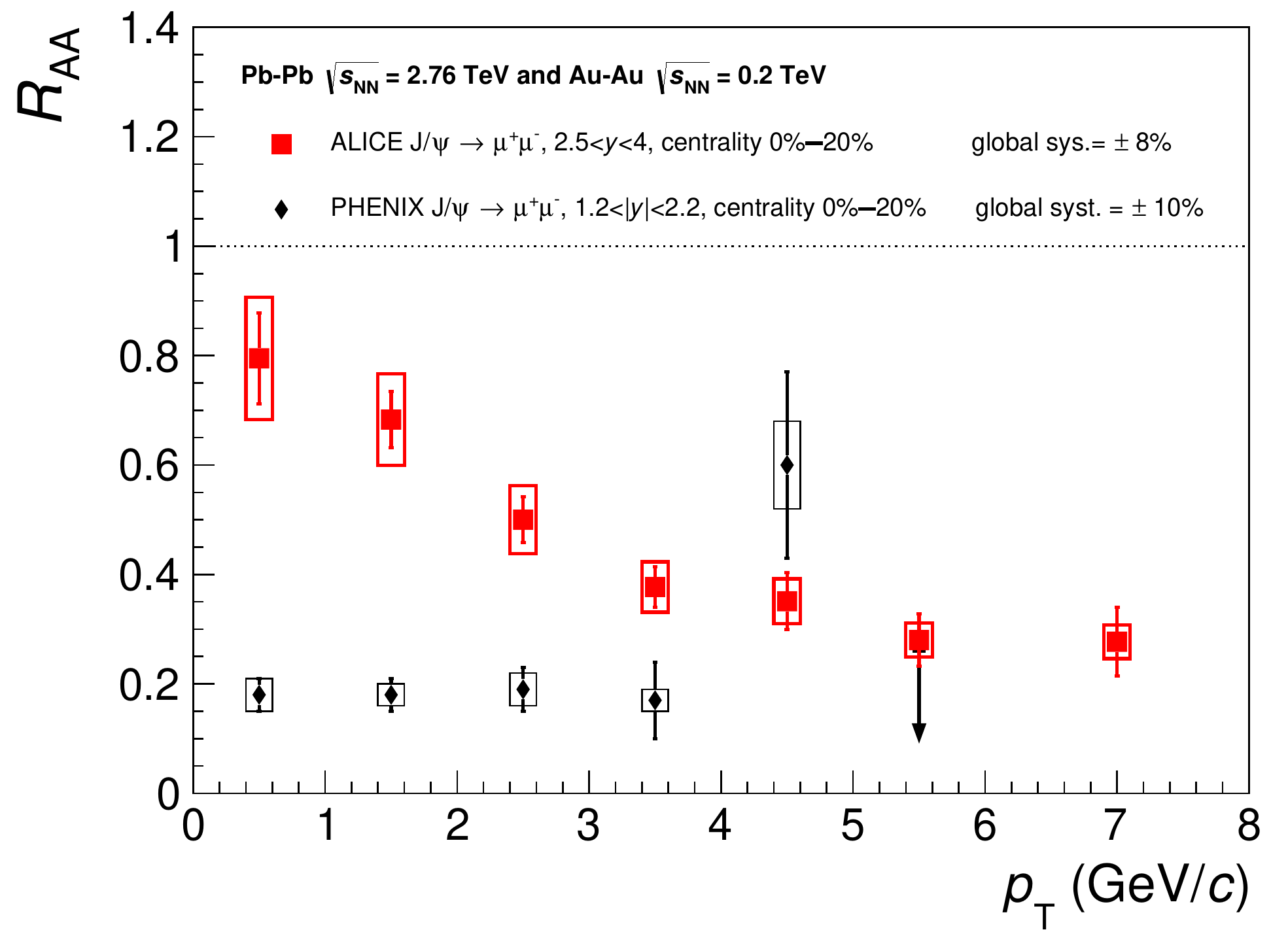} 
\includegraphics[width=0.42\textwidth]{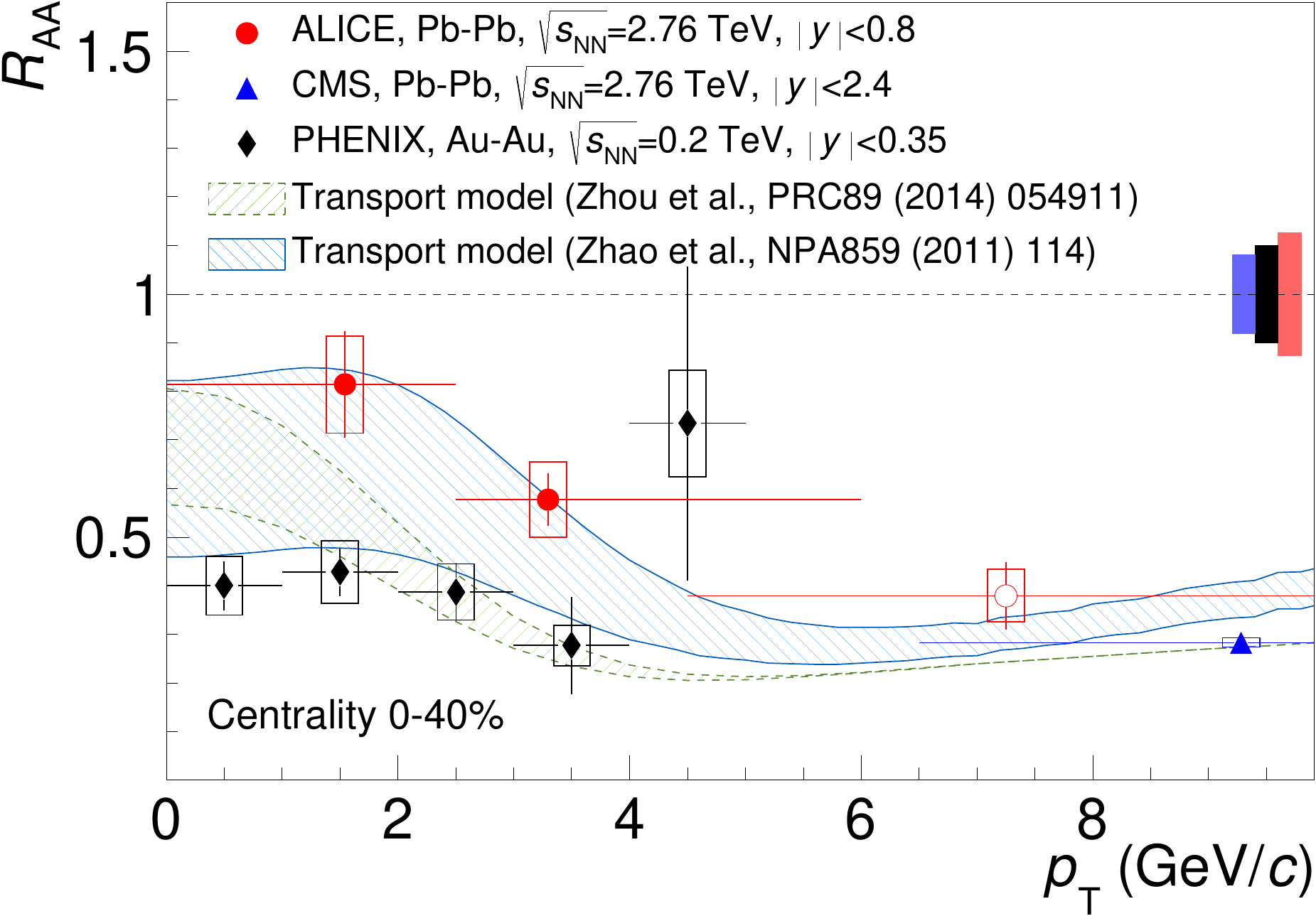} 
\caption{ALICE inclusive \jpsi \raa versus \pt~\cite{Abelev:2013ila}. Left: 
  forward rapidity result compared to the PHENIX result~\cite{Adare:2011yf} in a 
  similar rapidity region. Both results are obtained in the 0-20\% most central 
  collisions. Right: mid-rapidity result~\cite{Adam:2015rba} compared to the 
  PHENIX result~\cite{Adare:2011yf}, both evaluated in the 0-40\% most central 
  collisions} 
\label{fig:ALICE_RAAvspT}        
\end{figure}

The (re)combination or the statistical hadronisation process are expected to be 
dominant in central collisions and, for kinematical reasons, they should 
contribute mainly at low \pt, becoming negligible as the \jpsi \pt increases. 
This behaviour is investigated by further studying the \raa \pt-dependence. In 
\fig{fig:ALICE_RAAvspT}, the ALICE \jpsi \raa(\pt), measured at forward 
rapidity (left) or at mid-rapidity~\cite{Adam:2015rba} (right), are compared to 
corresponding PHENIX results obtained in similar rapidity ranges. The forward 
rapidity result has been obtained in the centrality class 0--20\%, while 
the mid-rapidity one in 0--40\%. In both rapidity 
regions, a striking different pattern is observed: while the ALICE \jpsi \raa 
shows a clear decrease from low to high \pt, the pattern observed at low 
energies is rather different, being almost flat versus \pt, with a suppression 
up to a factor four (two) stronger than at LHC at forward rapidity (mid-rapidity). 
 
\begin{figure}[t] 
\centering 
\includegraphics[width=0.42\textwidth]{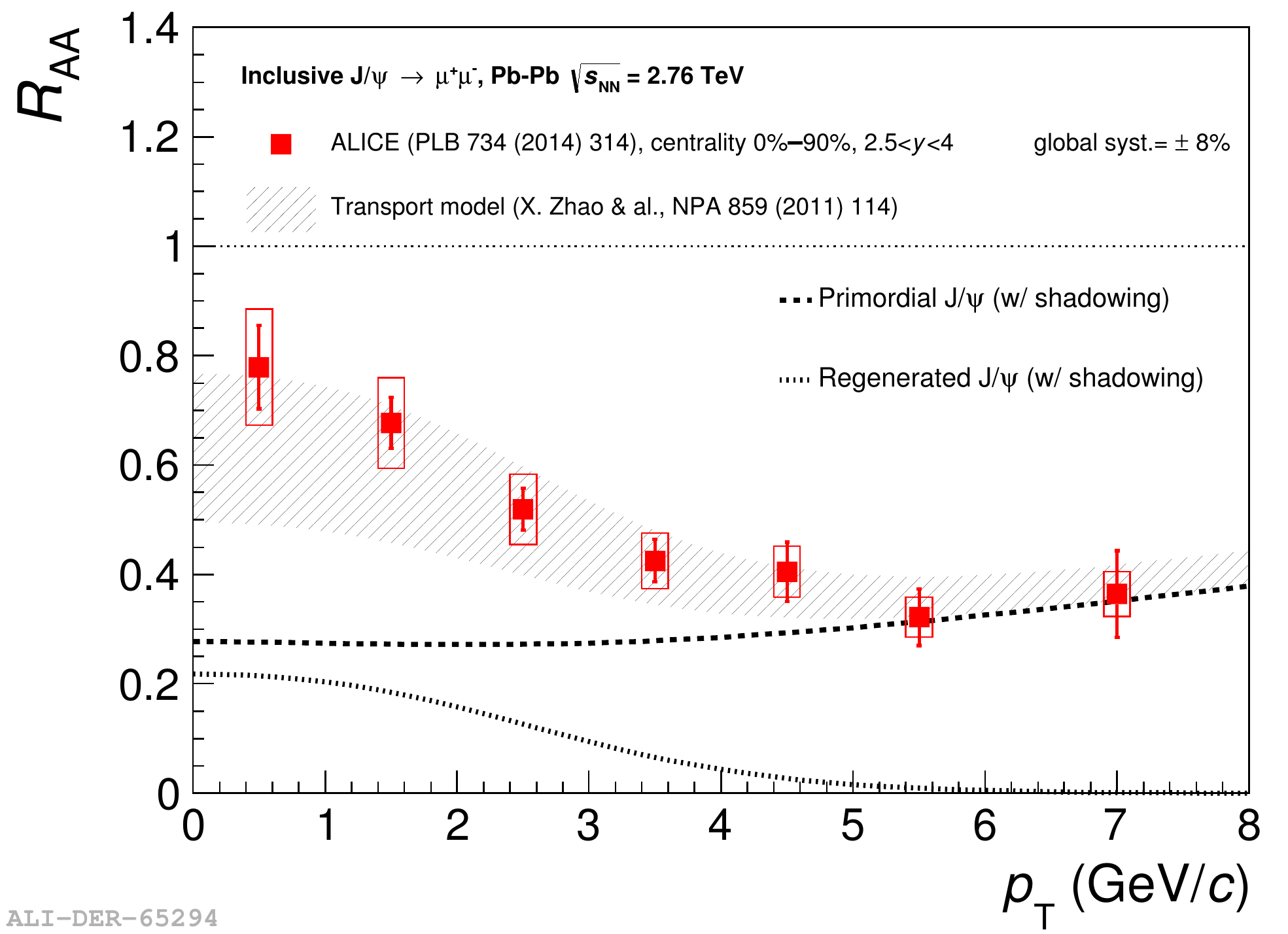} 
\includegraphics[width=0.42\textwidth]{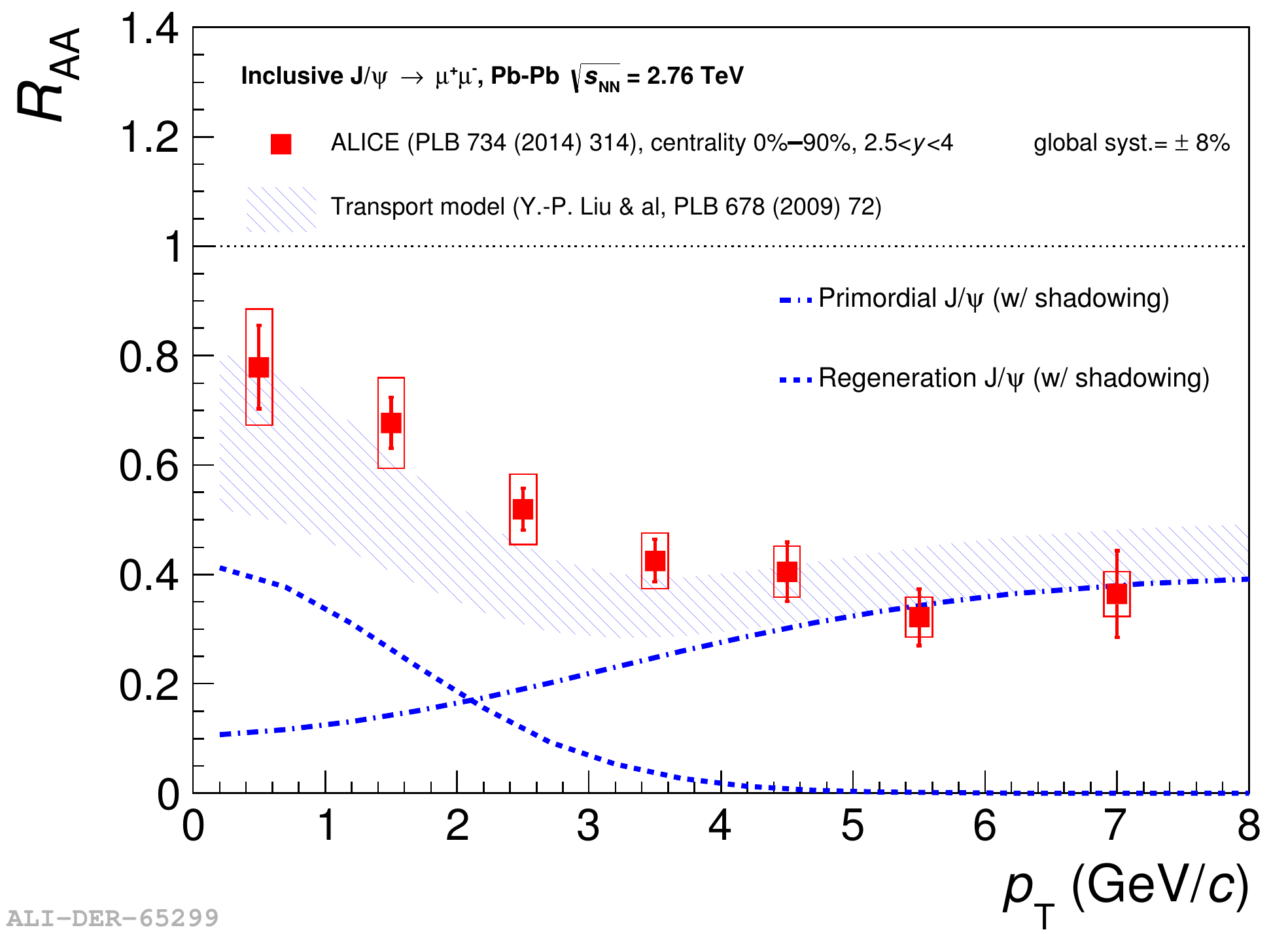} 
\caption{ALICE inclusive \jpsi \raa, measured in the forward rapidity region, 
  versus \pt~\cite{Abelev:2013ila}, compared to the TAMU (left) and THU (right) 
  theoretical transport calculations including a (re)combination component to 
  the \jpsi production.} 
\label{fig:ALICE_RAAvspT_models}        
\end{figure}

Models, such as TAMU and the THU that include a \pt-dependent contribution from 
(re)combination, amounting to $\approx50\%$ at low \pt and vanishing for high 
\pt~\cite{Zhao:2011cv,Liu:2009nb}, are found to provide, also in this case, a 
reasonable description of the data, as it can be observed in 
\fig{fig:ALICE_RAAvspT_models} for the forward rapidity result or in 
\fig{fig:ALICE_RAAvspT} (right) for the mid-rapidity one. 
 
\begin{figure}[t] 
\centering 
\includegraphics[width=0.42\textwidth]{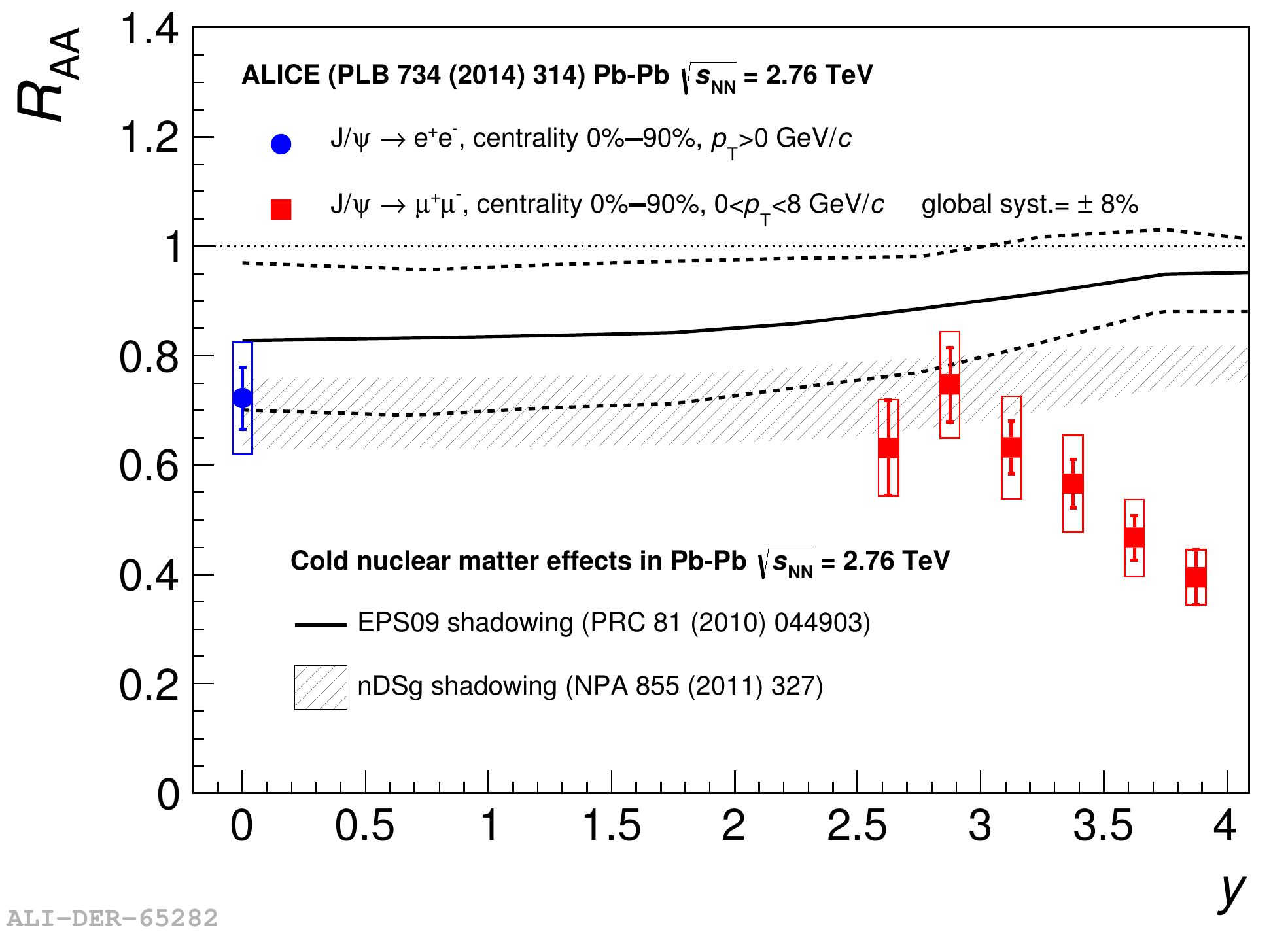} 
\caption{ALICE inclusive \jpsi \raa versus rapidity~\cite{Abelev:2013ila}, 
  compared to nPDF calculations (see \sect{sec:npdf_aa}).} 
\label{fig:ALICE_RAAvsy}        
\end{figure}

Finally, the rapidity dependence of the \jpsi \raa is shown in 
\fig{fig:ALICE_RAAvsy}. At forward-$y$ the \jpsi \raa decreases by about 40\% 
from $y = 2.5$ to $y = 4$. The \raa $y$-dependence is compared to shadowing 
calculations discussed in \sect{sec:npdf_aa}. As expected, the contribution of 
cold nuclear matter alone, such as shadowing, cannot account for the observed 
suppression, clearly indicating the need of the aforementioned hot matter 
effects. 
 
As discussed, the ALICE results are for inclusive \jpsi, therefore including two 
contributions: the first one from \jpsi direct production and feed-down from 
higher charmonium states and the second one from \jpsi originating from 
$b$-hadron decays. Beauty hadrons decay mostly outside the fireball, hence the 
measurement of non-prompt \jpsi \raa is mainly connected to the $b$ quark 
in-medium energy loss, discussed in \sect{sec:OHFbeauty}. Non-prompt \jpsi are, 
therefore, expected to behave differently with respect to the prompt ones. 
In the low-\pt 
region covered by ALICE the fraction of non-prompt \jpsi is smaller than 
15\%~\cite{Book:2014mia} (slightly depending on the $y$ range). Based on this 
fraction, the ALICE Collaboration has estimated the influence of the non-prompt 
contribution on the measured inclusive \raa. At mid-rapidity the prompt \jpsi 
\raa can vary within $-7\%$ and $+17\%$ with respect to the inclusive \jpsi \raa 
assuming no suppression ($\raa^{\text{\rm non-prompt}}=1$) or full suppression 
($\raa^{\text{\rm non-prompt}}=0$) for beauty, respectively. At forward-$y$, the 
prompt \jpsi \raa would be 6\% lower or 7\% higher than the inclusive result in 
the two aforementioned cases~\cite{Abelev:2013ila}.  
 

\subsubsection{\texorpdfstring{$\rm J/\psi$}{J/psi} \raa results at high \pt}
\label{sec:raa_hipt}

The CMS experiment is focused on the study of the \jpsi production at high \pt. 
The limit in the charmonium acceptance at low-\pt is due to the fact that muons 
from the charmonium decay need a minimum momentum ($p\approx3\text{--}5\GeVc$) 
to reach the muon tracking stations, overcoming the strong CMS magnetic field (3.8~T) and the 
energy loss in the magnet and its return yoke. The CMS vertex reconstruction capabilities allow 
the separation of non-prompt \jpsi from $b$-hadron decays from prompt \jpsi, using 
the reconstructed decay vertex of the \mumu pair. The prompt \jpsi include 
directly-produced \jpsi as well as those from decays of higher charmonium states 
(\eg \psiP and \chic), which can not be removed because their decay lengths are 
orders of magnitude smaller compared to those from $b$ decays, and not 
distinguishable in the analysis of the \pb data. 
 
As discussed in \sect{sec:pp_ref}, the \pp reference sample, recorded in 2011 at 
the same centre-of-mass energy per nucleon-nucleon pair as the \pb 
data, was used to evaluate the \PbPb \raa. 
 
\begin{figure}[!ht] 
  \centering 
  \includegraphics[width=0.45\textwidth]{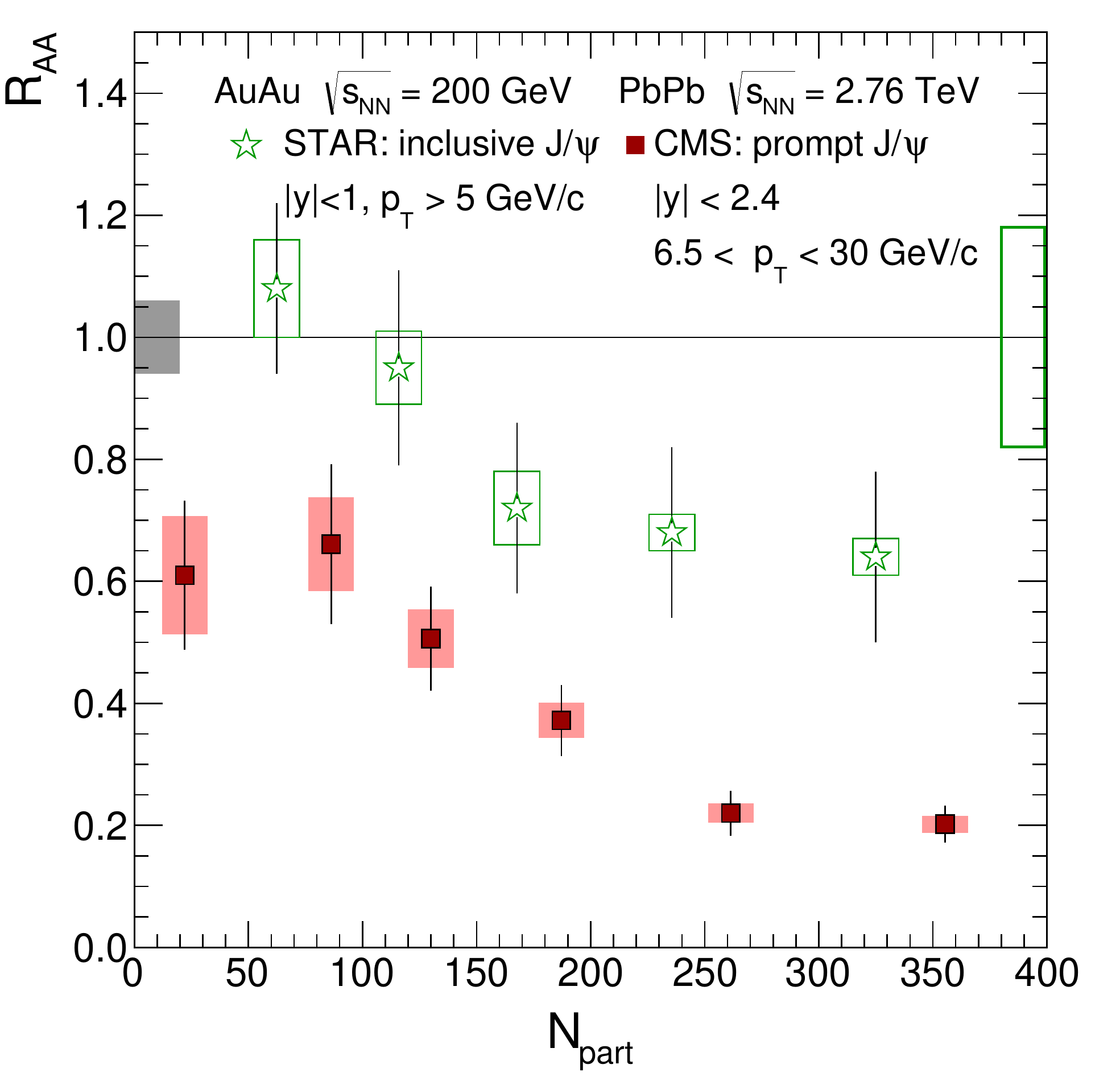} 
  \includegraphics[width=0.45\textwidth]{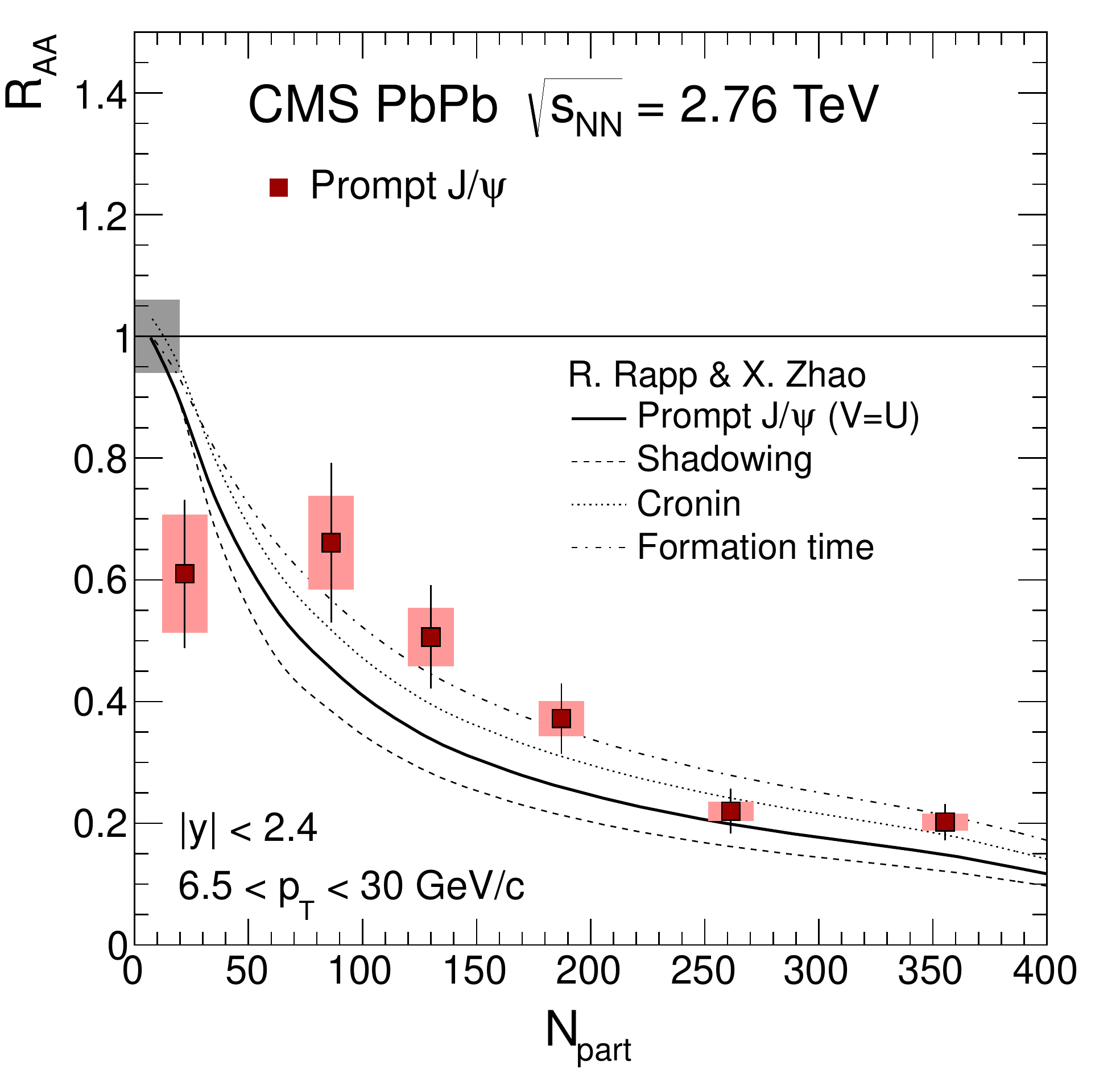} 
  \caption{\jpsi \raa as function of centrality. Left: CMS high-\pt prompt 
    \jpsi, $6.5<\pt<30\GeVc$ and $|y|<2.4$ (squares)~\cite{Chatrchyan:2012np}, 
    and STAR inclusive \jpsi with $\pt>5\GeVc$ measured in $|y|<1$ (open 
    stars)~\cite{Adamczyk:2012ey}; Right: CMS high-\pt prompt \jpsi compared to 
    the TAMU transport model calculation discussed in \sect{sec:transport}.} 
  \label{fig:CMS_JPsi_Raa} 
\end{figure} 
\begin{figure}[!ht] 
  \centering 
  \includegraphics[height=0.4\textwidth]{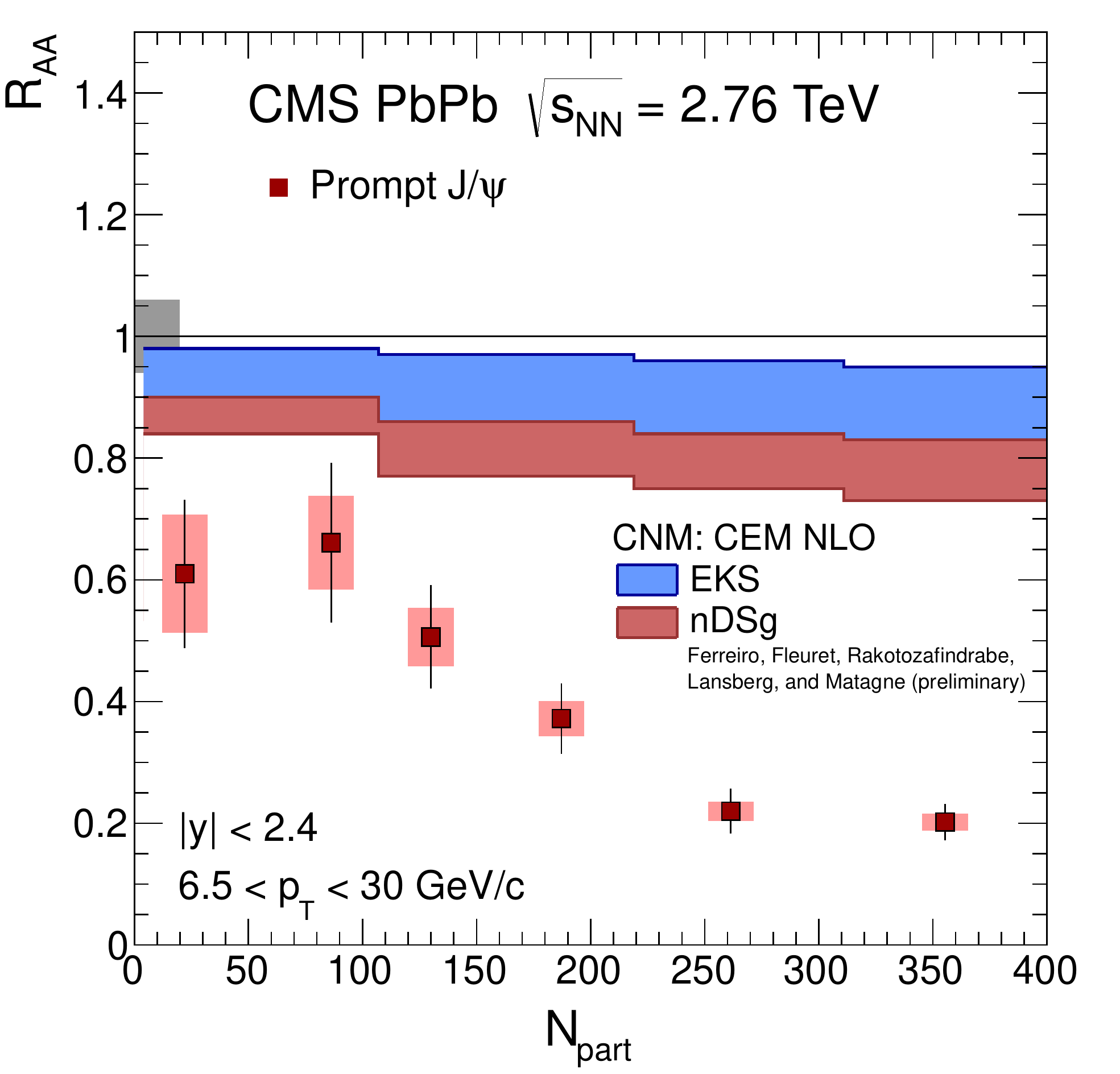} 
  \includegraphics[height=0.4\textwidth]{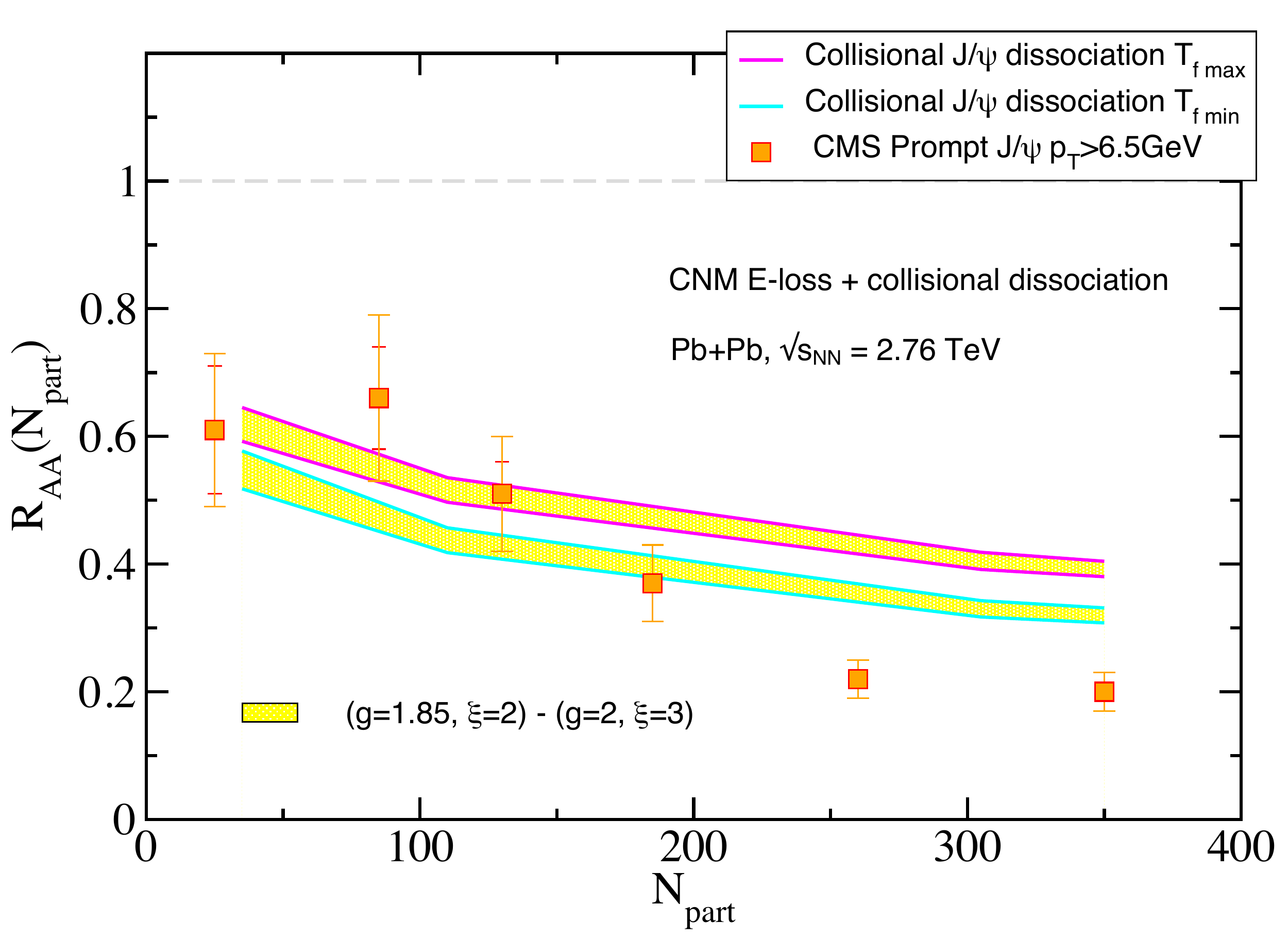} 
  \caption{CMS prompt \jpsi \raa as function of centrality~\cite{Chatrchyan:2012np}, compared to the nPDF 
    calculations discussed in \sect{sec:npdf_aa} (left) and to the collisional 
    dissociation model described in \sect{sec:finalstate_diss} (right).} 
  \label{fig:CMS_jpsi_Raa_shad} 
\end{figure} 
 
The \jpsi \raa was evaluated in the \PbPb data sample collected in 2010, 
corresponding to $\lumi_{\text{int}} = 7.3\mubinv$. The nuclear modification 
factor, integrated over the rapidity range $|y|<2.4$ and \pt range 
$6.5<\pt<30\GeVc$, was measured in six centrality 
bins~\cite{Chatrchyan:2012np}, starting with the 0--10\% bin (most central), up 
to the 50--100\% bin (most peripheral). The \raa obtained for prompt \jpsi, when 
integrating over the \pt range $6.5<\pt<30\GeVc$ and $|y|<2.4$, is shown in 
\fig{fig:CMS_JPsi_Raa} (left). The same centrality dependence, with a smooth 
decrease towards most central collisions, is observed also for inclusive \jpsi, 
even if the suppression is slightly more important for prompt \jpsi. In both 
cases the \jpsi \raa is still suppressed even in the (rather wide) most 
peripheral bin. A more recent analysis, based on the larger 2011 \PbPb data 
sample ($\lumi_{\text{int}} = 150 \mubinv$), has allowed to study the \raa in a 
much narrower centrality binning (12 centrality bins) and confirms the observed 
pattern~\cite{CMS:2012wba}. 
 
In the left panel of \fig{fig:CMS_JPsi_Raa}, a comparison is made with the 
inclusive \jpsi measurement from the STAR Collaboration~\cite{Adamczyk:2012ey}, 
at a more than ten times smaller collision energy, but in a similar high-\pt 
kinematic region: $\pt>5\GeVc$ and $|y|<1$. The rightmost bin corresponds to 
0--10\% centrality, while the leftmost bin to 40--60\% centrality.  
The suppression is smaller at RHIC than at LHC energies, with no 
significant suppression for collisions with a centrality more peripheral than 
30\% in the RHIC case. These results seem to support a higher medium temperature 
reached in Pb--Pb collisions at \snn = 2.76\TeV collisions than in \AuAu collisions at \snn = 
0.2\TeV. 
 
In \fig{fig:CMS_JPsi_Raa} (right) the prompt \jpsi \raa centrality dependence is 
compared with the predictions of the TAMU transport model. The observed 
suppression, increasing as a function of centrality, is due to the melting of 
primordial \jpsi. The TAMU model provided a reasonable description of 
the ALICE low-\pt \jpsi \raa (see \fig{fig:RAA_models_NPart}), with a significant 
recombination contribution. On the contrary, no 
recombination component is needed to describe the high-\pt \jpsi results.

In \fig{fig:CMS_jpsi_Raa_shad} (left), the prompt \jpsi \raa is compared to 
shadowing calculations. As already discussed for low-\pt \jpsi results, 
shadowing, here considered as the only cold nuclear matter effect, cannot 
account for the observed suppression, clearly indicating that other cold or hot 
matter effects are needed to describe the experimental results. 
 
In \fig{fig:CMS_jpsi_Raa_shad} (right), the centrality dependence of the prompt 
\jpsi \raa is compared to the collisional dissociation model, discussed in 
\sect{sec:finalstate_diss}. The model describes the more peripheral events, but 
underestimates the suppression for the most central events. It also 
underestimate the \pt dependence of the \jpsi \raa.

The CMS Collaboration also measured the \raa of 
non-prompt \jpsi, presented in \sect{sec:OHFbeauty}.


\subsubsection{\texorpdfstring{$\rm J/\psi$}{J/psi} azimuthal anisotropy}
\label{sec:jpsi_flow}

Further information on the \jpsi production mechanism can be accessed by studying 
the azimuthal distribution of \jpsi with respect to the reaction plane. 
As discussed in \sect{OHF}, the positive $v_2$ measured for D mesons at LHC and  
heavy-flavour decay electrons at RHIC suggests that charm quarks participate 
in the collective expansion of the medium and do acquire some elliptic flow  
as a consequence of the multiple collisions with the medium constituents.  
\jpsi produced through a recombination mechanism, should inherit the elliptic 
flow of the charm quarks in the QGP and, as a consequence, \jpsi are expected to 
exhibit a large \vtwo. Hence this quantity is a further signature to identify 
the charmonium production mechanism. 
 
\begin{figure}[!t] 
  \centering 
  \includegraphics[width=0.45\textwidth]{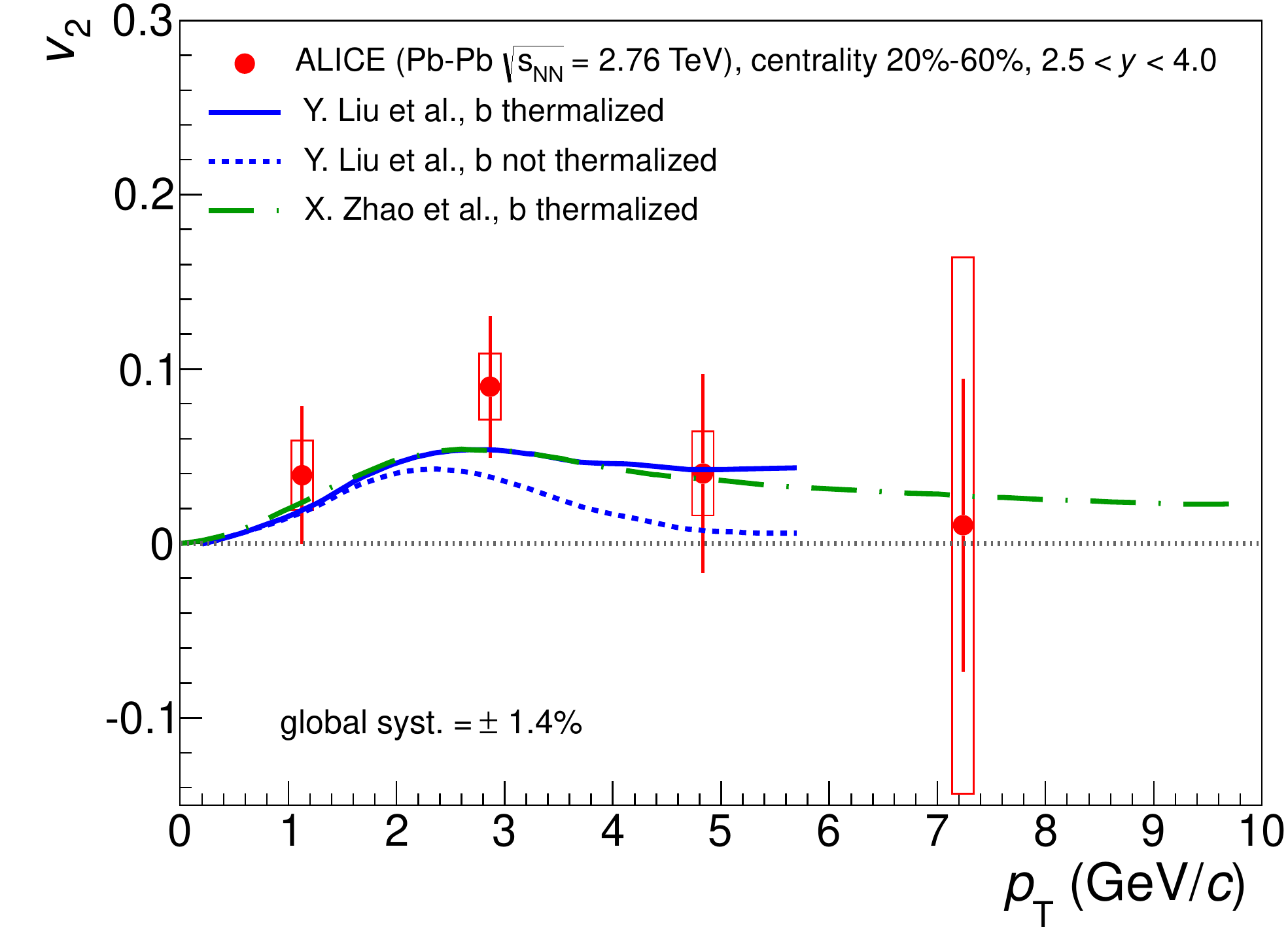} 
  \includegraphics[width=0.485\textwidth]{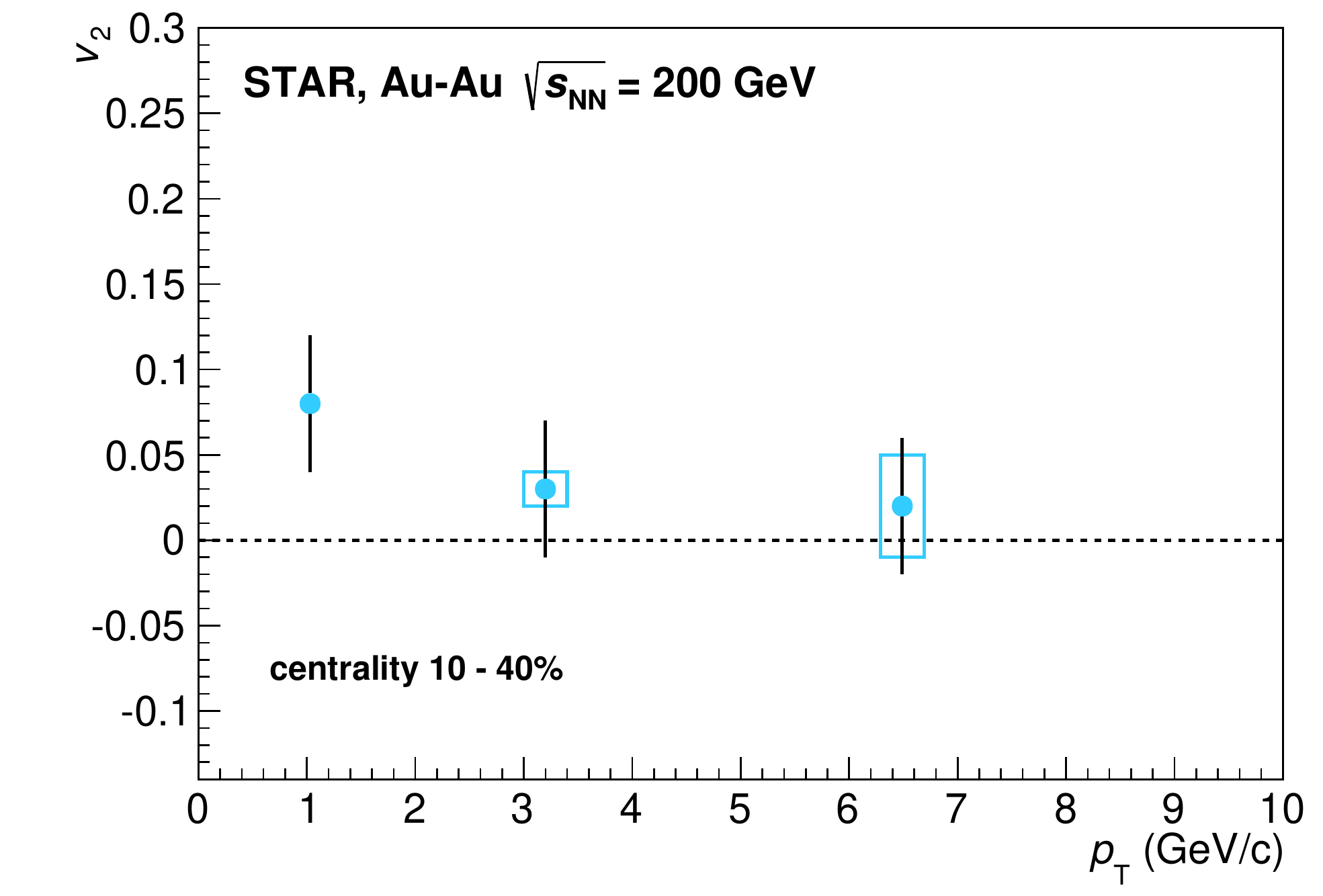} 
  \caption{Left: ALICE inclusive \jpsi measurement as a function of transverse 
    momentum for semi-central \mbox{Pb-Pb} collisions~\cite{ALICE:2013xna}, 
    compared to TAMU~\cite{Zhao:2012gc} and THU~\cite{Liu:2009gx} transport 
    models calculations. Right: STAR inclusive \jpsi measurement as a function 
    of transverse momentum in different centrality bins~\cite{Adamczyk:2012pw}.} 
  \label{fig:jpsi_v2} 
\end{figure}

ALICE measured the inclusive \jpsi elliptic flow in \pb collisions at 
forward rapidity~\cite{ALICE:2013xna}, using the event-plane technique. For 
semi-central collisions there is an indication of a positive \vtwo, reaching 
$\vtwo = 0.116\pm0.046\,\text{(stat.)}\pm0.029\,\text{(syst.)}$ in the 
transverse momentum range $2<\pt<4\GeVc$, for events in the 20--40\% centrality 
class. In \fig{fig:jpsi_v2} (left), the \jpsi \vtwo in the 20--60\% centrality class is compared with the TAMU 
and THU transport model calculations, which also provide a fair description of 
the \raa results, discussed in \sect{sec:raa_lowpt}. Both models, which 
reasonably describe the data, include a fraction ($\approx30\%$ in the 
centrality range 20--60\%) of \jpsi produced through (re)generation mechanisms, 
under the hypothesis of thermalisation or non-thermalisation of the $b$-quarks. 
More in details, charm quarks, in the hot medium created in \pb collisions at 
the LHC, should transfer a significant elliptic flow to regenerated \jpsi. 
Furthermore, primordial \jpsi might acquire a \vtwo induced by a path-length 
dependent suppression due to the fact that \jpsi emitted out-of-plane traverse a 
longer path through the medium than those emitted in-plane. Thus, out-of-plane 
emitted \jpsi will spend a longer time in the medium and have a higher chance to 
melt. The predicted maximum \vtwo at \pt = 2.5\GeVc is, therefore, the result of 
an interplay between the regeneration component, dominant at low \pt and the 
primordial \jpsi component which takes over at high \pt (see 
\fig{fig:zhuang4}). The \vtwo measurement complements the \raa results, 
favouring a scenario with a significant fraction of \jpsi produced by 
(re)combination in the ALICE kinematical range. 
 
At RHIC, measurements by the STAR Collaboration~\cite{Adamczyk:2012pw} of the 
\jpsi \vtwo in \AuAu collisions at \snn = 200\GeV are consistent with zero for 
$\pt>2\GeVc$ albeit with large uncertainties, as shown in \fig{fig:jpsi_v2} 
(right), while a hint for a positive \vtwo might be visible in the lowest \pt 
bin ($0<\pt<2\GeVc$). Results do not show a dependence on centrality. The 
measurement seems to disfavour the \jpsi formation through recombination 
mechanisms at RHIC energies, contrarily to what happens in \PbPb collisions at 
the LHC. 
 
CMS has investigated the prompt \jpsi \vtwo 
as a function of the centrality of the collisions and as a 
function of transverse momentum~\cite{CMS:2013dla}. Preliminary results indicate 
a positive \vtwo. The observed 
anisotropy shows no strong centrality dependence when integrated over rapidity 
and \pt. The \vtwo of prompt \jpsi, measured in the 10--60\% centrality class, 
has no significant \pt dependence either, whether it is measured at low \pt, 
$3<\pt<6.5\GeVc$, in the forward rapidity interval $1.6<|y|<2.4$, or at 
high \pt, $6.5<\pt<30\GeVc$, in the rapidity interval $|y|<2.4$. The preliminary CMS 
result supports the presence of a small anisotropy over the whole \pt range,  
but the present level of precision does not allow for a definitive 
answer on whether this anisotropy is constant or not. In the rapidity interval 
$|y|<2.4$, for $\pt>8\GeVc$, the anisotropy is similar to that observed for 
charged hadrons, the latter being attributed to the path-length dependence of 
 partonic energy loss~\cite{Chatrchyan:2012xq}.


\subsubsection{\texorpdfstring{$\rm J/\psi$}{J/psi} \raa results for various colliding systems and beam energies at RHIC}
\label{sec:rhic_exp}

A unique feature of RHIC is the possibility of accelerating various symmetric or 
asymmetric ion species, allowing for the study of charmonium suppression as a 
function of the system size. Furthermore, since at RHIC it is possible to 
collect data at various \snn, the charmonium production beam-energy dependence 
was also investigated from the top energy \snn = 200\GeV down to \snn = 
39\GeV. 
 
\begin{figure}[t] 
  \centering 
  \includegraphics[width=0.52\textwidth]{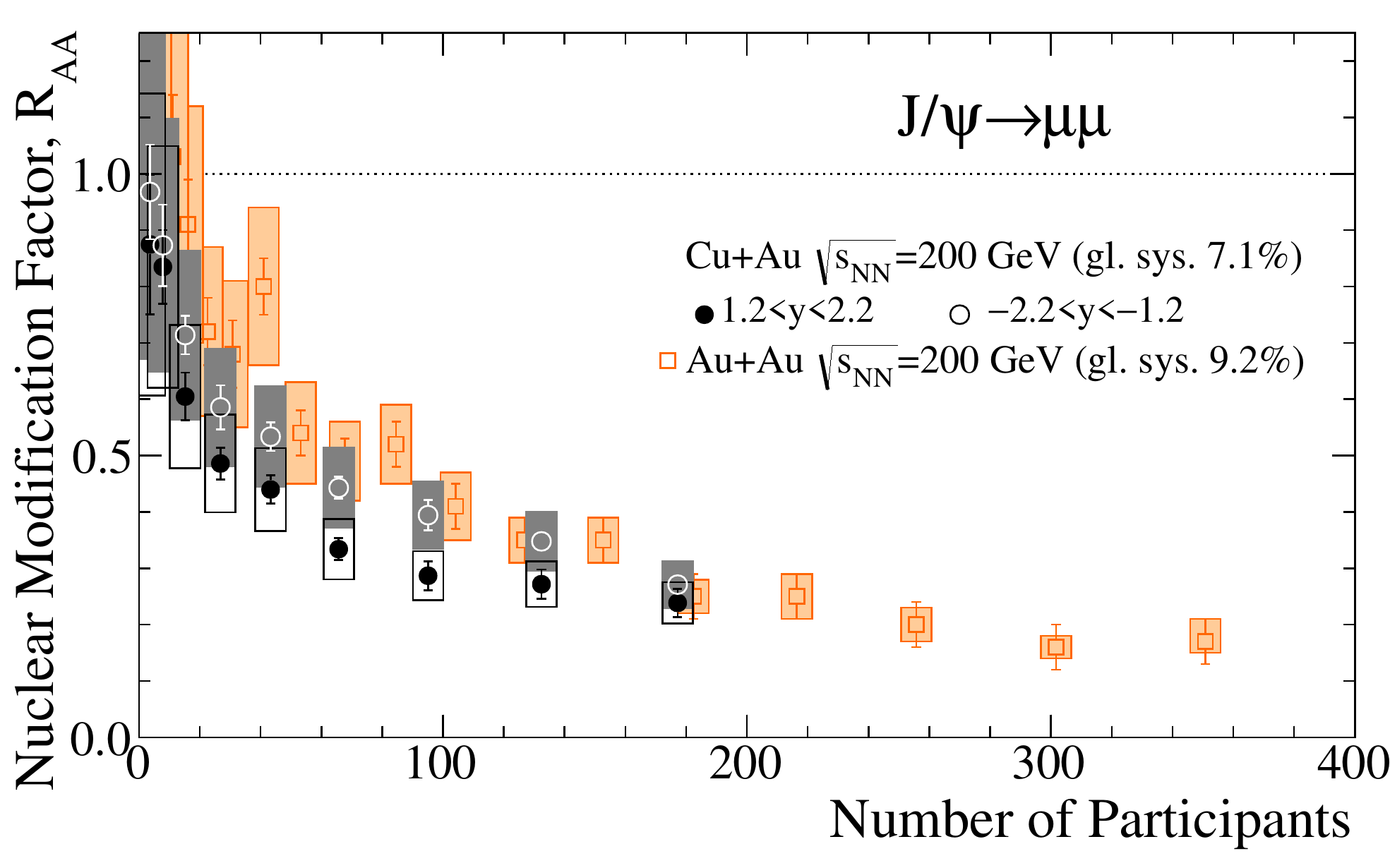} 
  \includegraphics[width=0.42\textwidth]{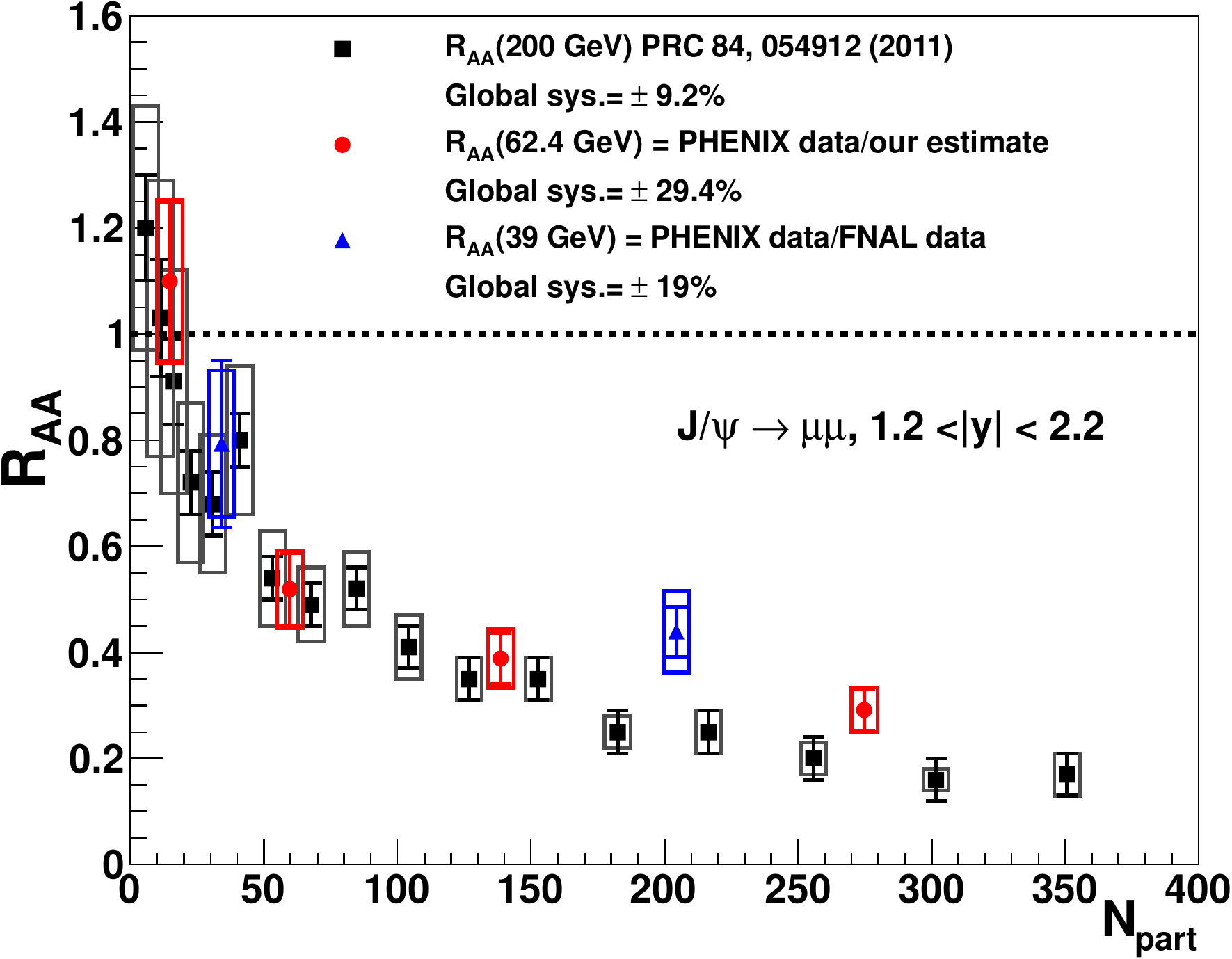} 
  \caption{Left: PHENIX \jpsi \raa measured in \CuAu~\cite{Adare:2014nsa} and \AuAu collisions, shown as a 
    function of the collision centrality. Right: PHENIX \jpsi \raa at various 
    collision energies (\snn= 39, 62.4 and 200 \GeV)~\cite{Adare:2012wf}.} 
  \label{fig:PHENIX_scans} 
\end{figure}

The PHENIX Collaboration measured \jpsi production from asymmetric \CuAu 
heavy-ion collisions at \snn = 200 \GeV at both forward (Cu-going direction) and 
backward (Au-going direction) rapidities~\cite{Adare:2014nsa}. The nuclear 
modification of \jpsi yields in \CuAu collisions in the Au-going direction is 
found to be comparable to that in \AuAu collisions when plotted as a function of 
the number of participating nucleons, as shown in \fig{fig:PHENIX_scans} (left). 
In the Cu-going direction, \jpsi\ production shows a stronger suppression. This 
difference is comparable to  
expectation from nPDF effects due to stronger low-$x$ gluon suppression in the 
larger Au nucleus. 
 
Moreover, the PHENIX Collaboration measured nuclear modification factors also 
by varying the collision energies, studying \AuAu data at \snn = 39 and 
62.4\GeV~\cite{Adare:2012wf}. The observed suppression 
patterns follow a trend very similar to those 
previously measured at \snn = 200\GeV, as shown in \fig{fig:PHENIX_scans} 
(right). Similar conclusions can be drawn also from preliminary STAR 
results~\cite{Zha:2014nia}. 
 
In spite of the large uncertainties associated to these results, up to now, this
similarity presents a challenge to theoretical models that contain competing hot
and cold matter effects with possibly different energy dependencies. 
For example, in the TAMU transport model~\cite{Grandchamp:2002wp}, the larger \jpsi
suppression towards higher collision energies due to higher energy-densities is
counter-balanced by a larger contribution from (re)combination due to the
increase of the total charm cross section, leading to an overall \jpsi
suppression that is nearly independent of the collision energy in the range
probed by the SPS and RHIC.


\subsubsection{Excited charmonium states}
\label{sec:exc_charm}

The measurement of excited charmonium states in heavy-ion collisions is 
experimentally challenging. The \psiP, observed via its \mumu decay, is expected 
to yield 50 times less events than the corresponding \jpsi decay, while being 
subject to similar background rates. The P-wave states decay radiatively into 
\jpsi and a low energy photon that is difficult to find in the background of 
thousands of photons resulting from neutral pion decays. So far, only the \psiP 
was measured in heavy-ion collisions, by NA50 at the 
SPS~\cite{Alessandro:2006ju} and by CMS at the 
LHC~\cite{Khachatryan:2014bva} (preliminary measurements also exist from 
  the ALICE Collaboration~\cite{Arnaldi:2012bg}). NA50 found a suppression of 
\psiP relative to \jpsi that increases with centrality, an observation that is 
consistent with a sequential dissociation of charmonia. At the same time the 
\psiP to \jpsi ratio reached in central \pb collisions is also consistent with 
the prediction of the statistical hadronisation model, leaving open the question 
whether all charmonia melt at the SPS. 
 
\begin{figure}[t] 
\centering 
\includegraphics[width=0.45\textwidth]{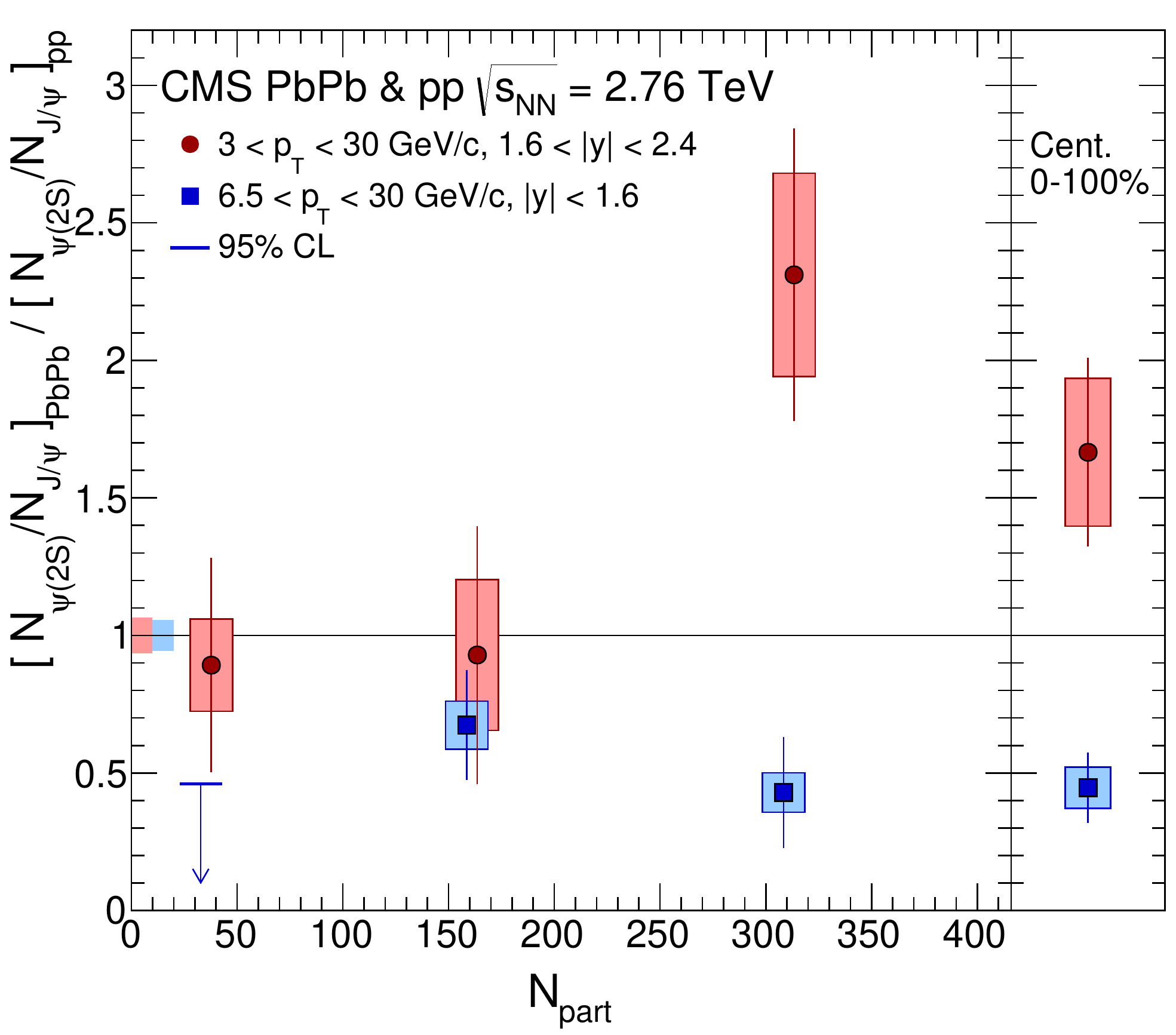} 
\caption{Double ratio of measured yields, \doubleRatioPsi, as a function of 
  centrality, for the mid-rapidity (blue squares) and forward rapidity (red 
  circles, slightly shifted) analysis bins~\cite{Khachatryan:2014bva}. The 
  centrality-integrated results are displayed in the right panel. Statistical 
  (systematic) uncertainties are shown as bars (boxes). The boxes at unity 
  indicate the (global) \pp uncertainties.} 
\label{fig:charm_cms_doubleratio} 
\end{figure}

At the LHC, CMS measured the yields of prompt \jpsi and \psiP in \pb and \pp 
collisions at \snn = 2.76\TeV. The result is presented in 
\fig{fig:charm_cms_doubleratio} as a double ratio, \doubleRatioPsi, as a 
function of event centrality for two kinematic regions: at mid-rapidity, $|y|<1.6$, \psiP are measured with 
$6.5<\pt<30\GeVc$, while at forward rapidity, $1.6<|y|<2.4$, the acceptance 
extends to $3<\pt<30\GeVc$. 
 
A clear difference between the centrality integrated double ratios in the two 
kinematic regions is found. At forward rapidity and low \pt, the double ratio is 
larger than unity, \ie the \psiP to \jpsi ratio is larger in \pb than in \pp. In 
contrast, the \psiP to \jpsi ratio is reduced in \pb compared to the ratio found 
in \pp at mid-rapidity and high \pt. Peripheral and semi-central collisions show 
a double ratio consistent with unity at forward rapidity, whereas the most 
central bin shows an increase of the double ratio. In contrast, the suppression 
of the double ratio at mid-rapidity appears to be independent of centrality. The 
difference between the two kinematic domains is highly unexpected. While the 
mid-rapidity and high-\pt result is in line with the expectation of sequential 
melting, the opposite behaviour is observed at forward rapidity and low \pt. 
While regeneration is not expected to contribute in the investigated \pt ranges, 
it is worth to note that also the statistical hadronisation model predicts a \pt-integrated  
double-ratio of $\approx 0.2$. It remains to be seen which effects 
can explain these results, \eg if regeneration of \psiP can be enhanced relative 
to \jpsi due to the larger binding radius. First attempts have been made to 
explain this observation, arguing that \psiP are regenerated at later stages 
than \jpsi, \ie when a stronger radial flow is present~\cite{Du:2015wha}. On the 
experimental side, it will be important to isolate whether the difference is due 
to the change in rapidity or \pt, and what happens at $\pt=0$. Preliminary 
results from the ALICE Collaboration at forward rapidity ($2.5<y<4$) and 
$\pt>0$ are not precise enough to draw a conclusion~\cite{Arnaldi:2012bg}.


\subsubsection{Bottomonium \raa results}
\label{sec:bottomonium}

With the advent of the LHC, bottomonia have become a new probe of the QGP. While 
their production rate is 200 times smaller than the one of \jpsi, they offer 
several advantages. The three S-wave states \upsa, \upsb, and \upsc have very 
different binding energies and appear at very similar rates in the \mumu decay 
channel. Their relative abundances are $7:2:1$, while the \jpsi to \psiP ratio 
is $50:1$. Hence these three states, which include with the \upsa the strongest 
bound state of all quarkonia, allow one to probe a much wider temperature range 
than previously accessible with charmonia. A further advantage is the absence of 
feed down from heavier-flavour decays, that are a background for high \pt 
charmonium studies. The higher masses also ease theoretical calculations. In the 
context of sequential dissociation, bottomonia may provide another advantage: 
the approximately twenty times smaller beauty production cross section will lead 
to a smaller contribution from regeneration that complicates the picture for 
charmonia. However, the closed to open heavy flavour production ratio for beauty 
is roughly ten times smaller than for charm, which increases the relative 
contribution of recombination to bottomonia and complicates the situation. 
 
Unfortunately, feed down contributions to the \upsa from excited state decays 
that are crucial for a quantitative understanding of a sequential dissociation 
are not very well understood at low \pt. Measurements of feed-down fractions 
exist only for $\pt>6\GeVc$, where about 30\% of \upsa result from decays of 
$\chi_b(nP)$ and \upsbc decays, reaching $\approx50\%$ at higher 
\pt~\cite{Affolder:1999wm,Aaij:2012se,Aad:2011ih,Aaij:2014caa}. 
 
\begin{figure}[t] 
  \centering 
  \includegraphics[width=0.45\textwidth]{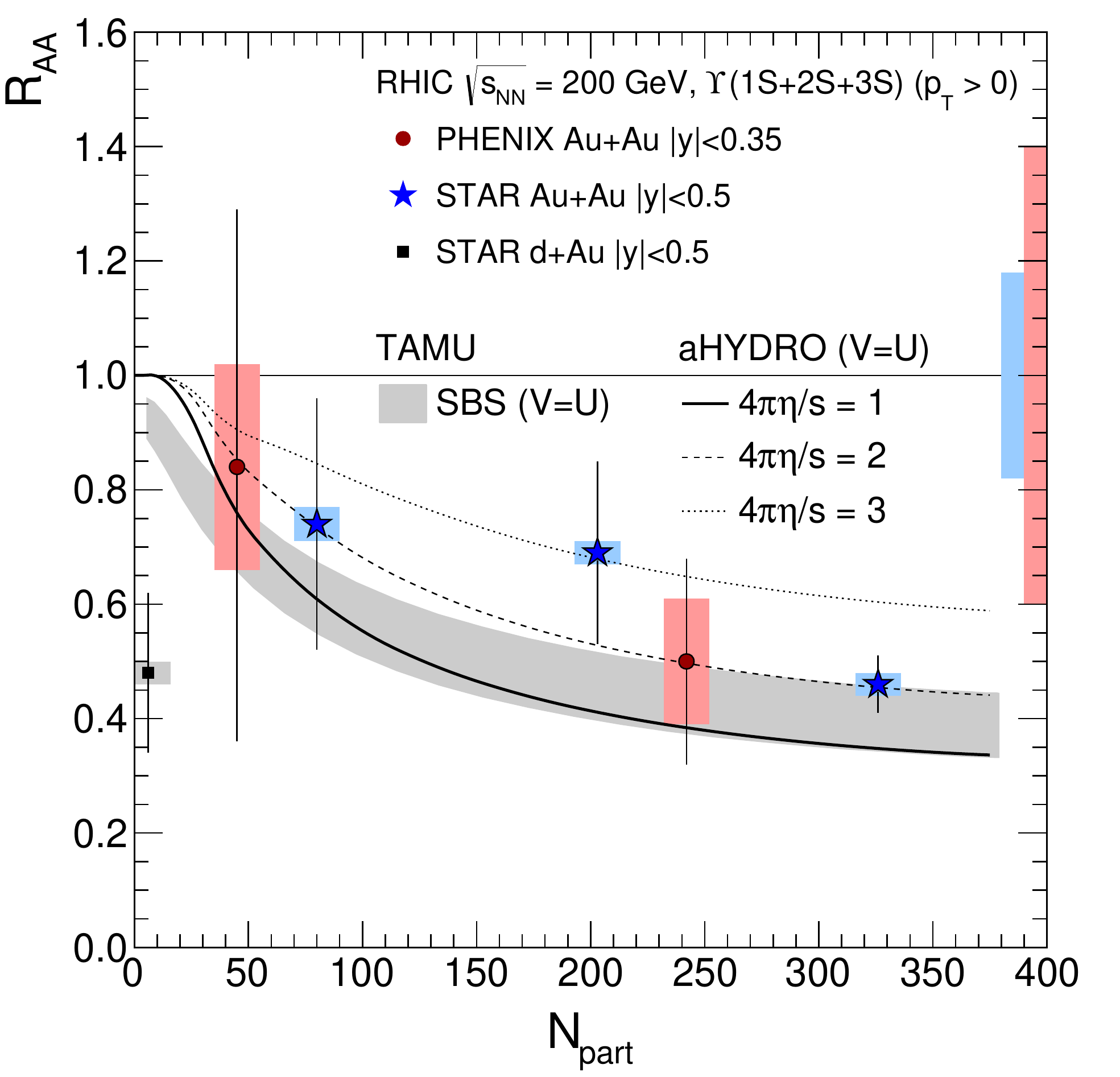} 
  \includegraphics[width=0.45\textwidth]{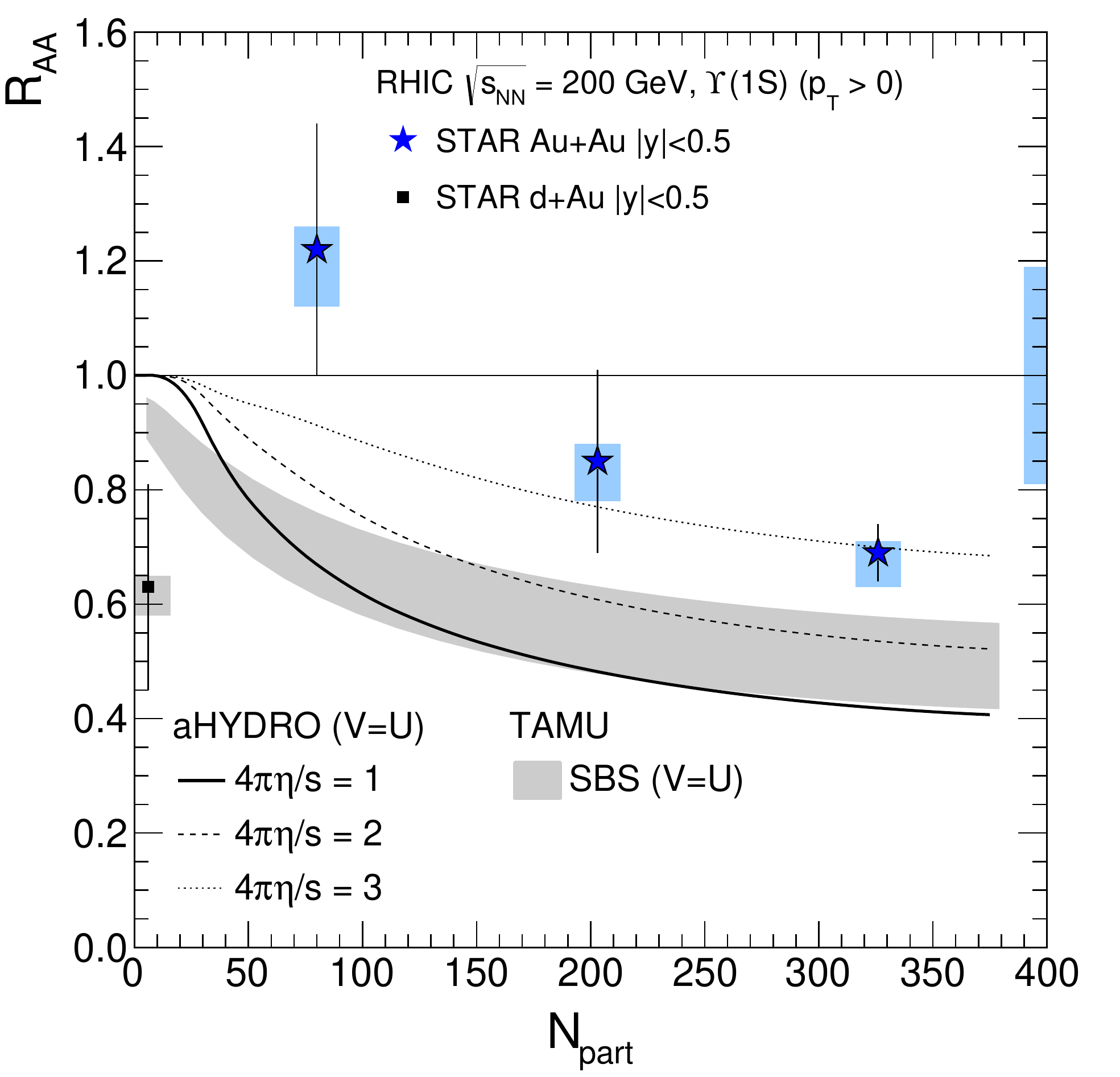} 
  \caption{Left: \raa of \upsabc as a function of centrality measured by PHENIX 
    in $|y|<0.35$~\cite{Adare:2012bv,Adare:2014hje} and STAR in 
    $|y|<0.5$~\cite{Adamczyk:2013poh} compared to TAMU (grey band) and aHYDRO 
    (lines). Right: STAR measurement of the \raa of \upsa with $|y|<0.5$ as a 
    function of centrality compared to the same models.} 
  \label{fig:RHIC_Ups} 
\end{figure}

At RHIC, where the \ups production cross sections are low, a measurement of the 
\ups suppression in \dAu and \AuAu collisions was performed by the PHENIX and 
STAR experiments~\cite{Adare:2012bv,Adare:2014hje,Adamczyk:2013poh}. Integrating 
the yield of the three \ups states, they observe a reduction of the yield in 
central \AuAu collisions, compared to the binary scaled \pp reference as shown 
in the left panel of \fig{fig:RHIC_Ups}. Because of the large statistical 
uncertainties, the experiments cannot yet assess a possible centrality 
dependence in \AuAu. STAR finds in the 10\% most central collisions a nuclear 
modification factor of $\raa = 0.49 \pm 0.13\,\text{(\AuAu stat.)} \pm 
0.07\,\text{(\pp stat.)} \pm 0.02\,\text{(\AuAu syst.)} \pm 0.06\,\text{(pp 
  syst.)}$. Constraining the measurement to the \upsa alone, as shown in the 
right panel of \fig{fig:RHIC_Ups}, only the \raa for the most central \AuAu 
collisions exhibits a significant suppression. Assuming a feed-down contribution 
of $\approx50\%$ this could signal the onset of a suppression of excited states 
in central \AuAu collisions. However, the \raa in most central \AuAu collisions 
is also comparable to the \rdau, so more precise measurements are necessary 
before drawing such a conclusion. 
 
A comparison of TAMU and aHYDRO calculations with the measured \upsabc nuclear 
modification factors shows good agreement within the experimental uncertainties. 
Experimental data cannot yet constrain the $\eta/s$ free parameter of aHYDRO. 
The band of the TAMU curve represents the uncertainty on cold nuclear matter 
effects. These are included by employing nuclear absorption cross sections of 
1.0 and 3.1\,mb, but the data cannot yet constrain their size. The \upsa 
suppression, however, seems to be slightly over-predicted by both models, though 
not beyond the experimental uncertainties, with the data preferring small values 
of $\eta/s$. 
 
\begin{figure}[t] 
  \centering 
  \includegraphics[width=0.45\textwidth]{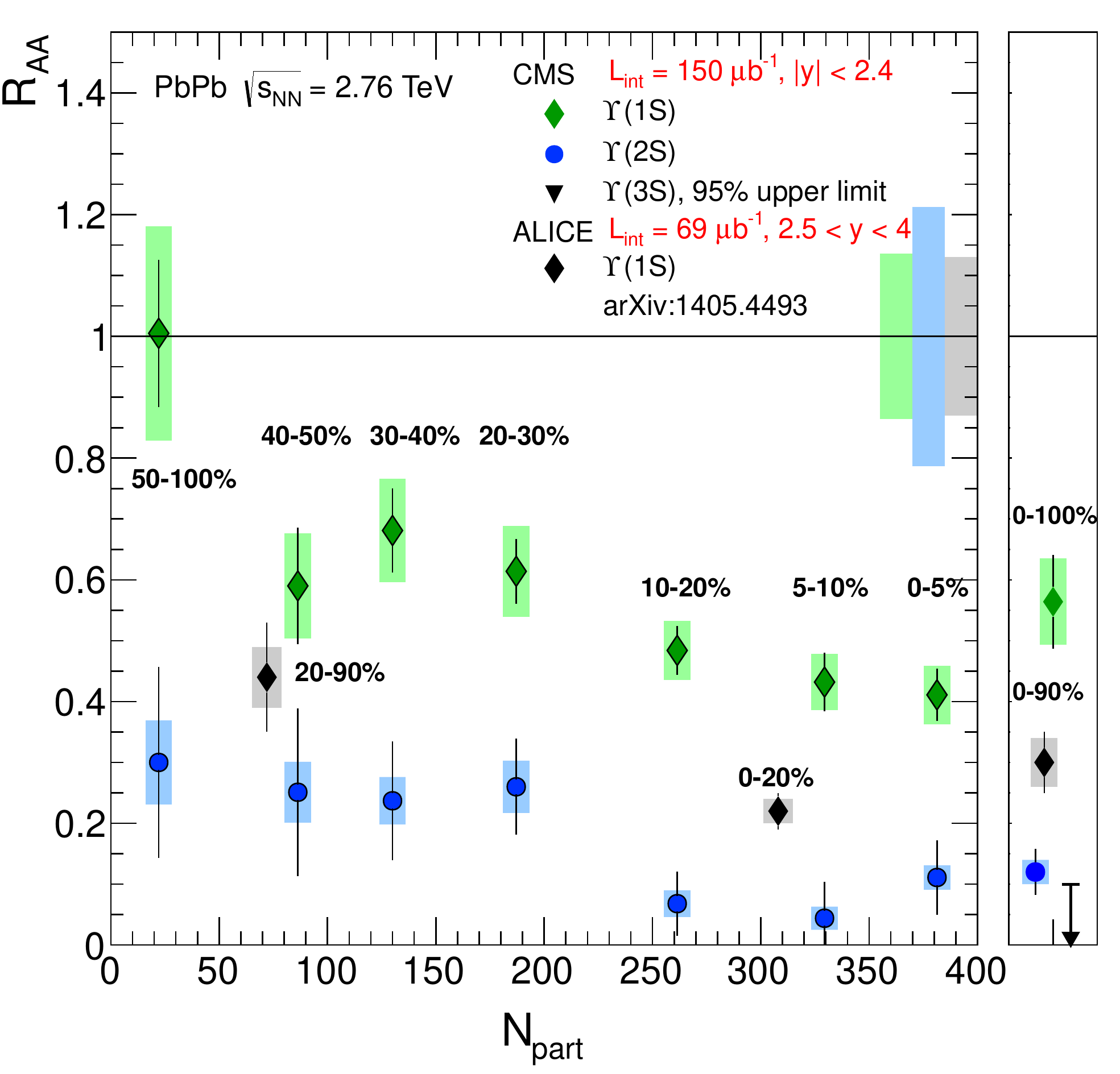} 
  \caption{\upsn \raa as a function of centrality. \upsa \raa is measured in 
    $2.5<|y|<4$ by ALICE~\cite{Abelev:2014nua}. CMS measured the centrality 
    dependence of the \upsa and \upsb \raa at 
    $|y|<2.4$~\cite{Chatrchyan:2011pe,Chatrchyan:2012lxa}. Centrality integrated 
    values are shown in the right panel, including an upper limit at 95\% 
    confidence level of the \upsc \raa by CMS in $|y|<2.4$.} 
  \label{fig:CMS-ALICE_Ups} 
\end{figure}

\begin{figure}[!h] 
  \centering 
  \includegraphics[width=0.44\textwidth]{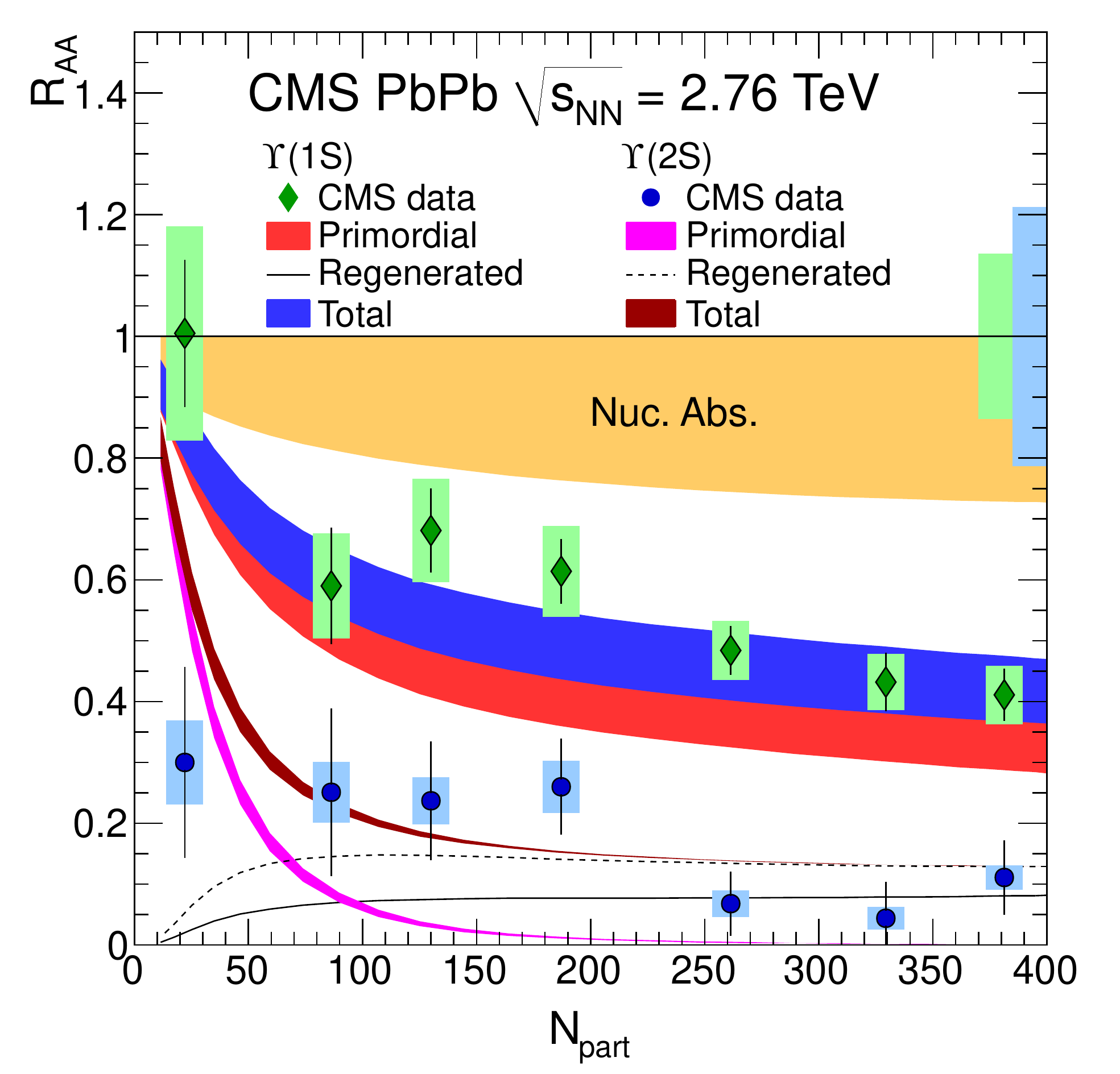} 
  \includegraphics[width=0.44\textwidth]{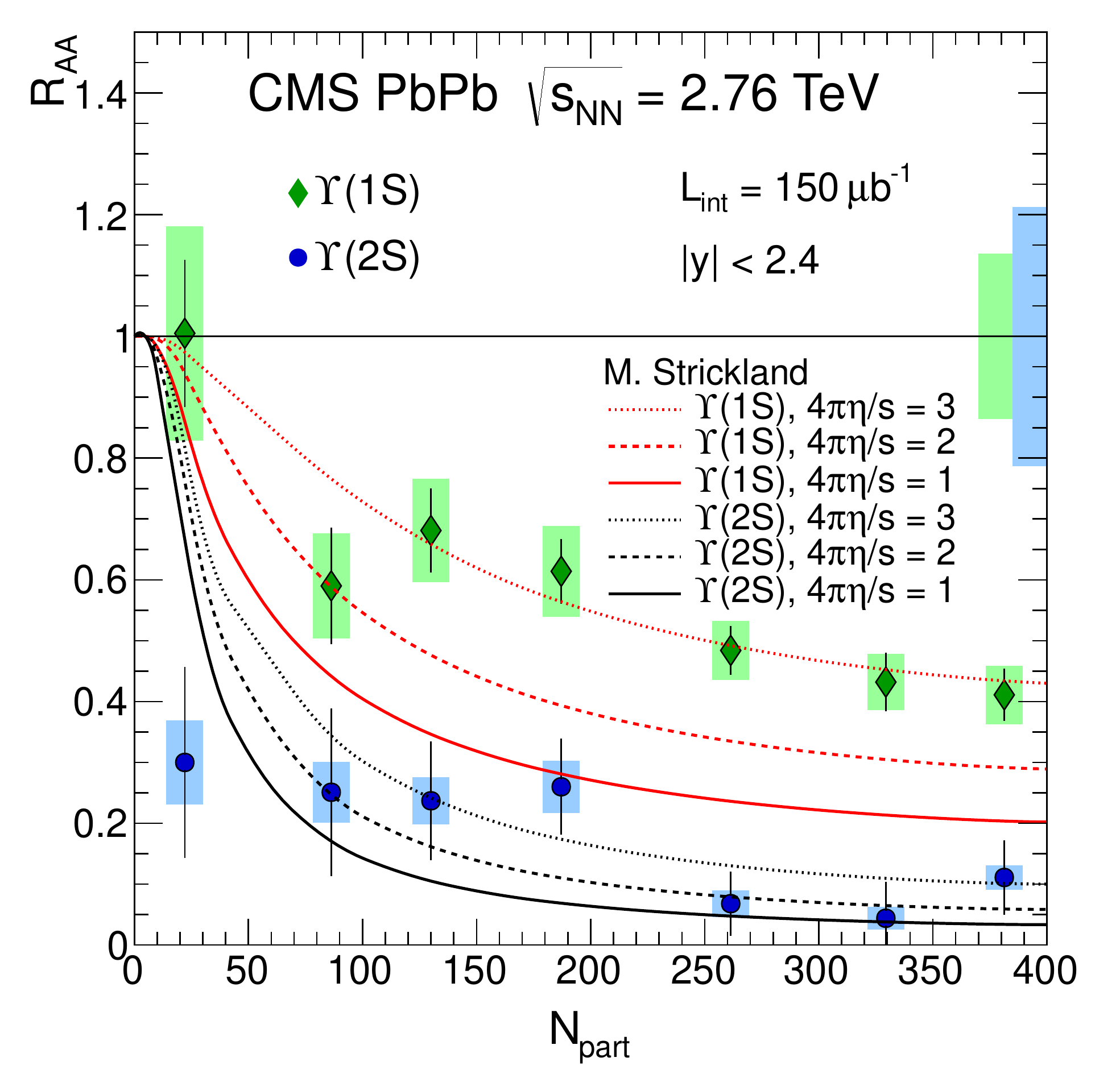} 
  \includegraphics[width=0.44\textwidth]{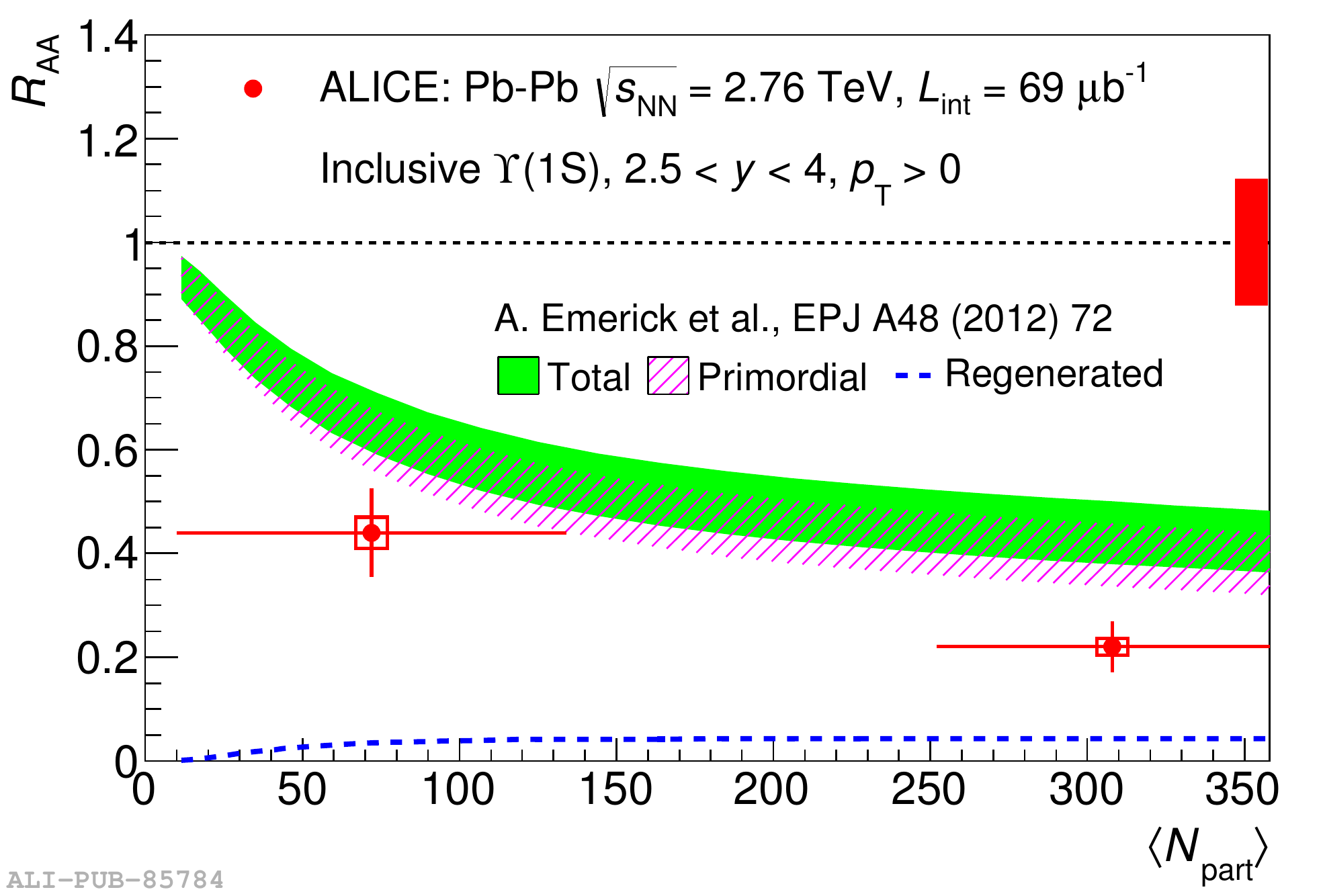} 
  \includegraphics[width=0.44\textwidth]{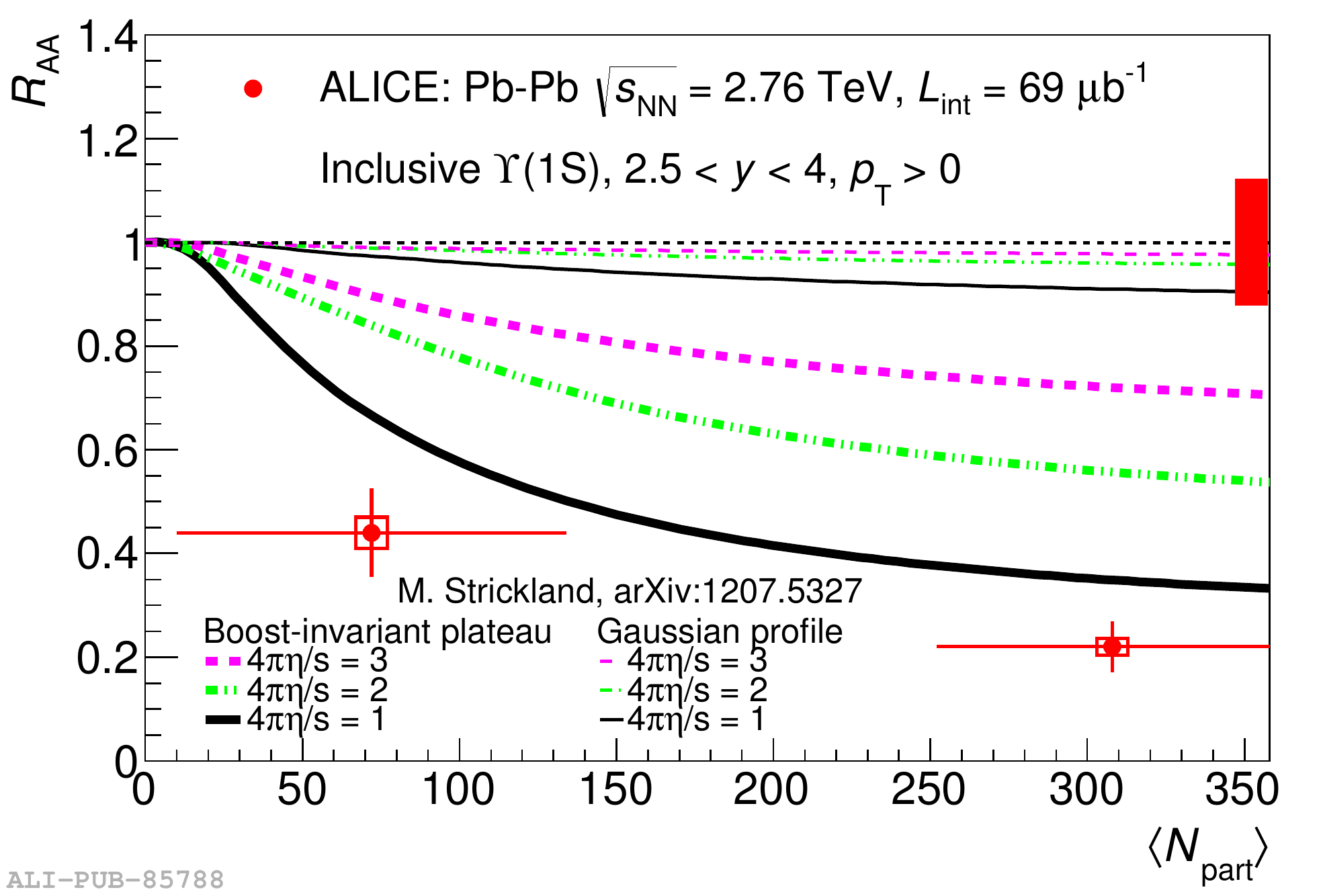} 
  \caption{\upsa\ \raa\ versus centrality at 
    $|y|<2.4$~\cite{Chatrchyan:2011pe,Chatrchyan:2012lxa} (top) and 
    $2.5<y<4$~\cite{Abelev:2014nua} (bottom), compared to TAMU (left) and aHYDRO 
    (right) model calculations discussed in Sections~\ref{sec:transport} 
    and~\ref{sec:non-equilib}.} 
  \label{fig:Upsilon_centrality_theory} 
  \centering 
  \includegraphics[width=0.44\textwidth]{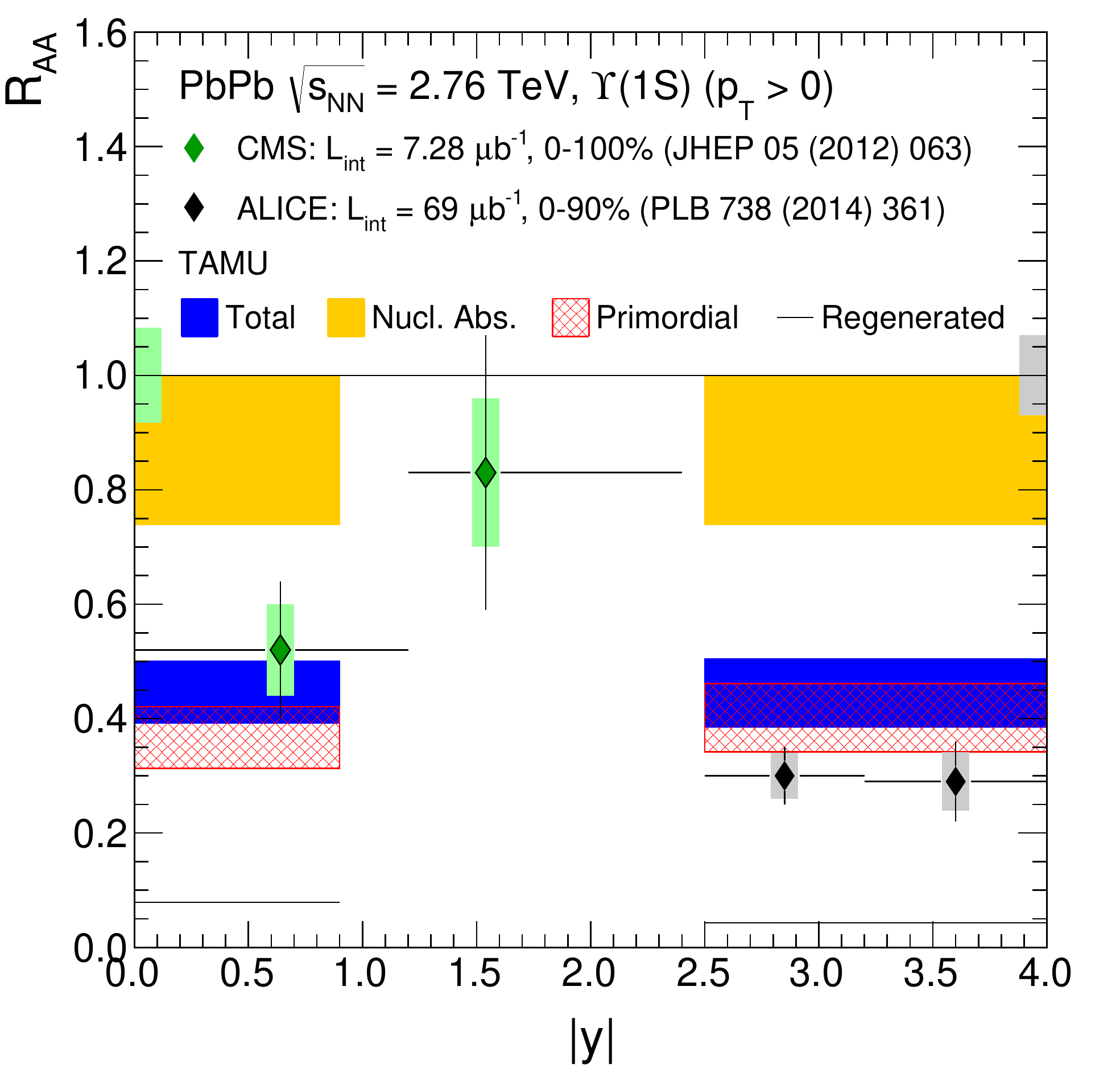} 
  \includegraphics[width=0.44\textwidth]{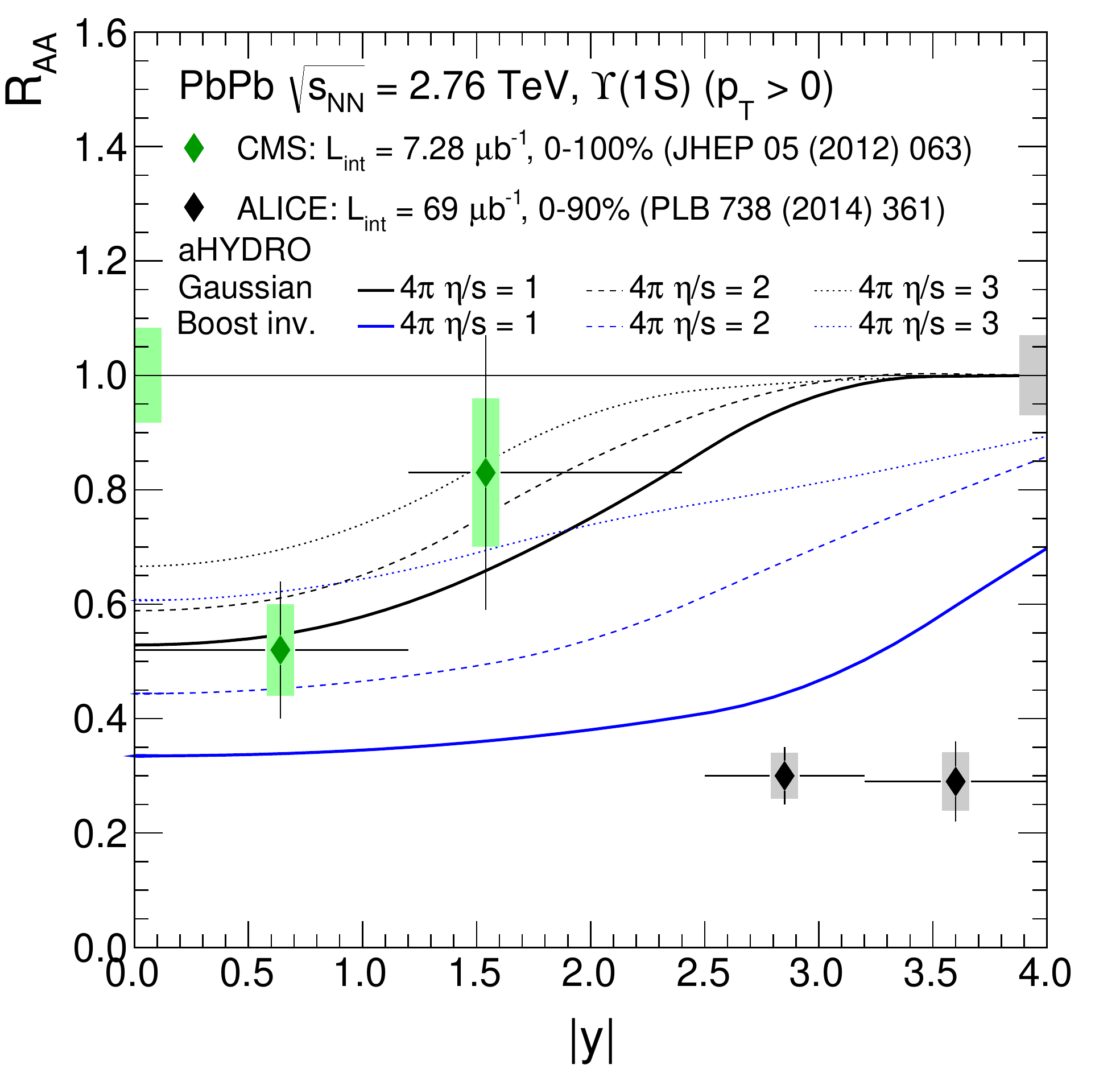} 
  \caption{\upsa\ \raa\ versus rapidity from ALICE~\cite{Abelev:2014nua} and 
    CMS~\cite{Chatrchyan:2011pe,Chatrchyan:2012lxa}. In the bottom row, the \raa 
    is to TAMU (left) and aHYDRO (right) model calculations discussed in 
    Sections~\ref{sec:transport} and~\ref{sec:non-equilib}.} 
  \label{fig:Upsilon_rap_CMS-ALICE} 
\end{figure}

CMS measured the suppression of the first three S-states integrated over all \pt 
and the rapidity range $|y|<2.4$ in \pb collisions at \snn = 
2.76\TeV~\cite{Chatrchyan:2011pe,Chatrchyan:2012lxa}. Following a first 
tantalising indication in 2011 that the excited states are suppressed relative 
to the \upsa, this was confirmed a year later. The centrality integrated \raa 
was measured for all three states, exhibiting a clear ordering with binding 
energy: $\raa(\upsa) = 0.56 \pm 0.08\text{(stat.)} \pm 0.07\text{(syst.)}$, $ 
\raa(\upsb) = 0.12 \pm 0.04\text{(stat.)} \pm 0.02\text{(syst.)}$, and the \upsc 
being so strongly suppressed that only an upper limit of $\raa(\upsc)<0.10$ at 
95\% CL could be quoted. The centrality dependence of the \upsa and \upsb \raa 
are shown in \fig{fig:CMS-ALICE_Ups}. With the \upsb and \upsc essentially 
completely suppressed in central \pb collisions, a more precise understanding of 
the feed down contributions to the \upsa is required to assess whether any 
directly produced \upsa are suppressed in such collisions. Furthermore, the role 
of the $\chib\text{(nP)}$ states in \pb are (and may remain) completely unknown 
so far. 
 
In the top row of \fig{fig:Upsilon_centrality_theory}, the centrality dependence 
of the CMS \upsa and \upsb results are compared to TAMU (left) and aHYDRO 
(right) calculations, described in Sections~\ref{sec:transport} 
and~\ref{sec:non-equilib}, respectively. Both models reproduce the data 
reasonably well, simultaneously describing the \upsa and \upsb suppression over 
the full centrality range. The aHYDRO approach has maybe some slight tension 
describing both states with the same choice for $\eta/s$, though the 
experimental uncertainties are large enough to account for the differences. 
Regarding the TAMU model, it is worth to highlight that it includes a 
non-negligible regeneration contribution. In fact, it is the sole source of 
\upsb in central \pb collisions. It is also interesting to point out that 
regeneration favours the production of \upsb over \upsa, which is opposite to 
the predictions for charmonia. This difference is the result of temperature 
dependent dissociation rates and equilibrium numbers that enter the rate 
equation (\eq{eq:rate}). In contrast to the other states, which all have 
dissociation temperatures in the vicinity of $T_c$, the strong binding energy 
will stop the dissociation of \upsa much earlier, when the equilibrium number is 
still small~\cite{Emerick:2011xu}. Significantly less regeneration of \upsa is 
necessary to reach this equilibrium number. 
 
The production of excited \ups states in \pb collisions has also been reported 
by CMS as fully corrected cross section ratio relative to the \upsa: 
$\sigma(\upsb)/\sigma(\upsa) = 0.09 \pm 0.02\,(\text{stat.}) \pm 
0.02\,(\text{syst.}) \pm 0.01\,(\text{glob.})$ integrated over centrality, \pt, 
and $|y|<2.4$~\cite{Chatrchyan:2013nza}. For the ratio 
$\sigma(\upsc)/\sigma(\upsa)$ an upper limit of 0.04 at 95\% CL has been set. 
These values can be directly compared to theoretical expectations, \eg the 
statistical hadronisation model, which predicts $\sigma(\upsb)/\sigma(\upsa) 
\approx 0.032$~\cite{Andronic:2014sga}. This value is consistent with the 
measured cross section ratio, though quite a bit lower than the central value of 
the measurement. 
 
A comparison of the CMS measurement at mid-rapidity to $\raa(\upsa) = 0.30 \pm 
0.05\text{(stat.)} \pm 0.04\text{(syst.)}$ measured by ALICE at forward rapidity 
($2.5<y<4$)~\cite{Abelev:2014nua}, integrated over \pt and centrality, as well 
as the centrality dependence overlaid in \fig{fig:CMS-ALICE_Ups}, reveal a 
surprising similarity to the \jpsi suppression observed at RHIC: \upsa are more 
suppressed at forward rapidity than at mid-rapidity. At RHIC such rapidity 
dependence was explained with a larger contribution of regeneration at 
mid-rapidity and/or stronger shadowing effects at forward rapidity.  
This similarity is 
also reflected in the centrality integrated rapidity dependence of $\raa(\upsa)$ 
shown in \fig{fig:Upsilon_rap_CMS-ALICE}. However, the large 
statistical uncertainties on the CMS measurement~\cite{Chatrchyan:2012np} that 
is still based on the first \pb and \pp runs at \snn = 2.76\TeV prevents 
conclusions on the \raa in the intermediate rapidity range. 
 
The simultaneous description of ALICE and CMS data provides a real challenge for 
the models so successful in reproducing the mid-rapidity data. As shown in Figures~\ref{fig:Upsilon_centrality_theory} 
and~\ref{fig:Upsilon_rap_CMS-ALICE}, they completely fail to predict the 
rapidity dependence of $\raa(\upsa)$. The aHYDRO model, curves taken from 
Ref.~\cite{Strickland:2012cq}, predicts a disappearance of the suppression at 
forward rapidity and does not get even close to the ALICE data. The TAMU 
transport model approach, including a regeneration component, predicts a rather 
constant rapidity dependence of the suppression though still overshoots the 
forward rapidity data slightly. In both models the \upsa\ suppression is 
dominated by the in-medium dissociation of the higher mass bottomonium states. 
Therefore, a precise measurement of \upsa\ feed-down contributions, as well as 
an accurate estimate of CNM effects in the kinematic ranges probed by ALICE and 
CMS is required in order to make a more stringent comparison with data.

It is interesting to compare the \raa of the three bottomonium states to the 
\raa of the \jpsi and \psiP at high \pt. The charmonium states follow nicely the 
established pattern of the \ups states of a reduced suppression with increasing 
binding energy as predicted by the sequential dissociation picture. If one, 
however, uses the \pt integrated \jpsi \raa, one observes a deviation from this 
pattern that can be explained with a (re)generation contribution. It will be 
interesting to see how low \pt \psiP will fit in. 
 
The picture may be complicated further by the observed multiplicity dependence 
of the $\upsb/\upsa$ and $\upsc/\upsa$ ratios in \pp and \pPb 
collisions~\cite{Chatrchyan:2013nza} that  
is discussed in Sections~\ref{sec:pp:HadCorrelations} and~\ref{sec:CNM:Onia}. 
It is unclear whether the dependence is caused by a suppression of the excited 
states by surrounding particles or by the multiplicity being biased by the 
presence of the \ups states.


\subsection{Alternative references for quarkonium production in nucleus--nucleus collisions}
\label{sec:other_ref}

\subsubsection{Proton--nucleus collisions}
\label{sec:pa_ref}

As discussed in \sect{Cold nuclear matter effects}, proton-nucleus data can 
provide information on CNM effects on quarkonium 
production. Since these mechanisms are present also in \AAcoll collisions, their 
precise evaluation is mandatory to correctly quantify the hot matter effects. However, the extrapolation of CNM effects 
evaluated in \pa to \AAcoll collisions is model dependent and it has to rely on 
assumptions as those discussed in detail in \sect{CNM_pAtoAA}. 
 
The ALICE Collaboration investigated the role of CNM effects on the \jpsi 
\raa in \pb collisions, extrapolating the \rpa results obtained in \pPb 
collisions at \snn = 5.02\TeV~\cite{Abelev:2013yxa}. Although the 
forward-rapidity ALICE \pA data were collected at a higher \snn energy and 
cover a slightly different centre of mass rapidity range with respect to \PbPb 
collisions ($2.03<y<3.53$ and $-4.46<y<-2.96$ in \pPb and $2.5<y<4$ in \PbPb), 
the Bjorken-$x$ regions probed by the \jpsi production process in the colliding 
nuclei are rather similar, differing by less than $\approx10\%$. The $x$ values 
covered in \PbPb collisions are $2\times10^{-5}<x<9\times10^{-5}$ and 
$1\times10^{-2}<x<6\times10^{-2}$, for Pb nuclei moving away from or towards the 
ALICE muon spectrometer, in which the \jpsi at forward rapidity are detected. In 
\pPb collisions the corresponding figures are $2\times10^{-5}<x<8\times10^{-5}$ 
and $1\times10^{-2}<x<5\times10^{-2}$ for the Pb nucleus going away or towards 
the ALICE muon spectrometer. Under the assumptions that shadowing is the main 
nuclear mechanism and that its influence on the two nuclei in \PbPb collisions 
can be factorised, cold nuclear matter effects are then evaluated as the 
product of the \jpsi \rpa computed at forward and backward rapidities, \ie 
$\rpa(y) \times \rpa(-y)$. The \rpa product, in the ALICE forward rapidity 
region, is 0.75$\pm$0.10 (stat) $\pm$0.12 (syst) when integrated over \pt. With 
\raa= 0.57$\pm$0.01 (stat) $\pm$0.09 (syst), this is a hint that the observed 
\jpsi suppression in \PbPb cannot be ascribed to shadowing effects alone. 
Similar conclusions, even if with larger uncertainties, can be obtained from 
ALICE results at mid-rapidity. This observation can be strengthened by comparing 
the \pt dependence of the \jpsi \raa to the one of the CNM effects evaluated as 
$\rpa(y) \times \rpa(-y)$~\cite{Adam:2015iga}. In this case, an opposite transverse momentum 
dependence is observed for the extrapolated shadowing, increasing from low to 
high \pt, and the \jpsi \raa pattern, showing a decrease towards high \pt, with 
a hint of an enhancement at low \pt. In particular, at high \pt, the observed 
\raa suppression is much larger than the shadowing extrapolation. Moreover, 
coherent energy loss effects are also expected to weaken at large 
\pt~\cite{Arleo:2013zua}, unlike the trend of the data. This clearly points to 
the existence of strong hot matter effects~\cite{Adam:2015iga}. 


\subsubsection{Open heavy flavour}
\label{sec:hf_ref}

\begin{figure}[t] 
  \centering 
  \includegraphics[width=0.45\textwidth]{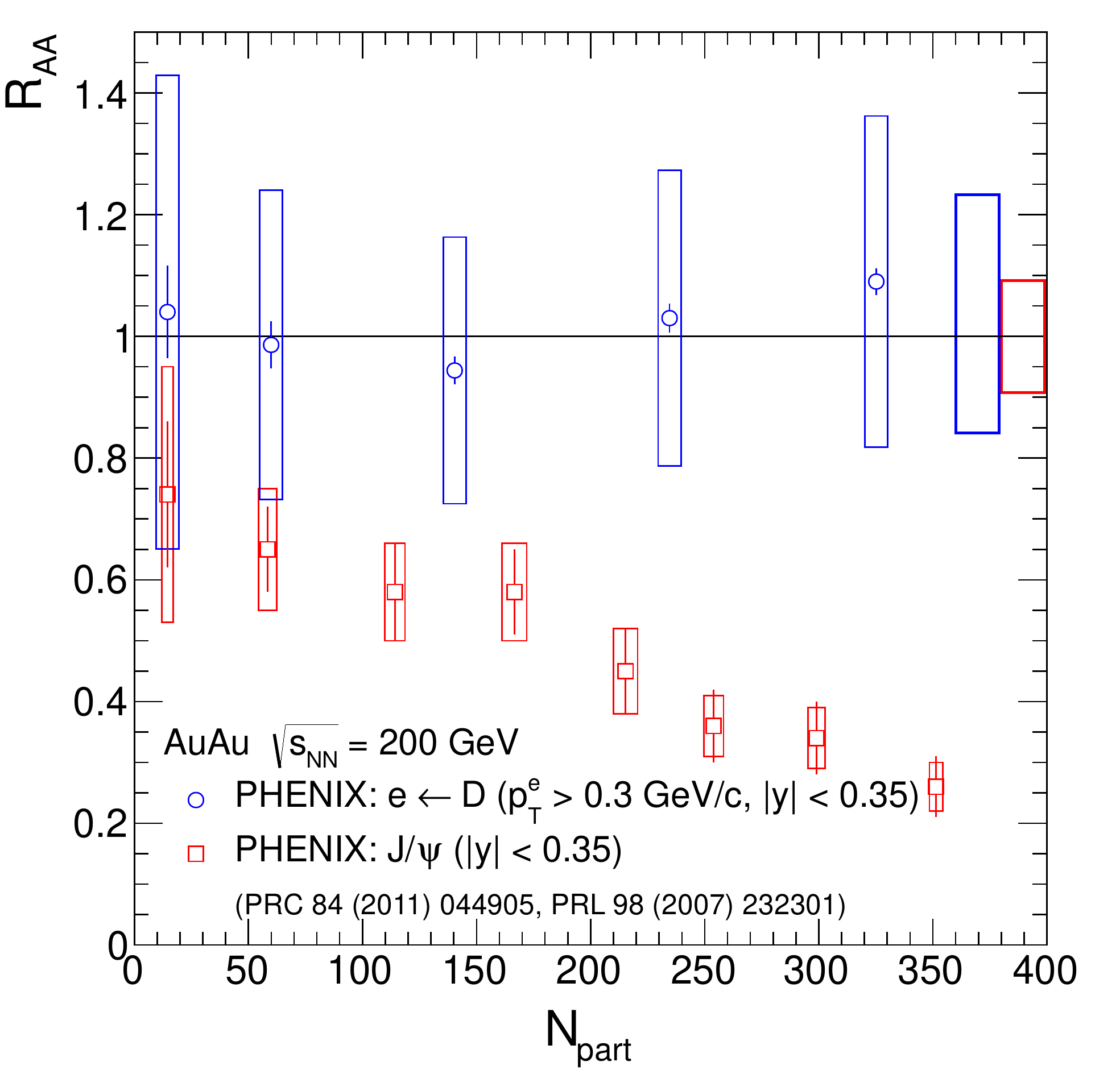} 
  \includegraphics[width=0.45\textwidth]{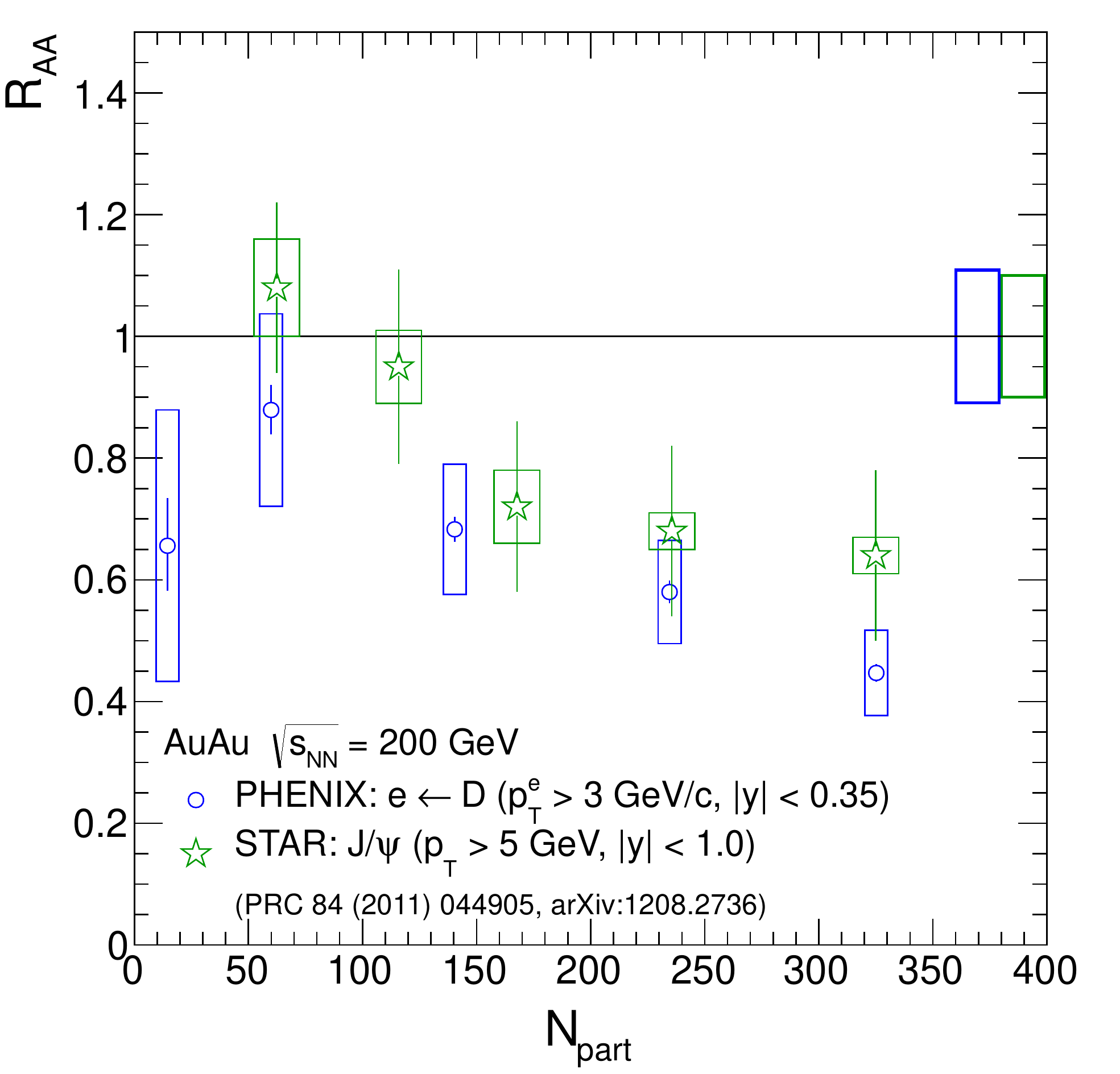}\\ 
  \hspace{0.45\textwidth} 
  \includegraphics[width=0.45\textwidth]{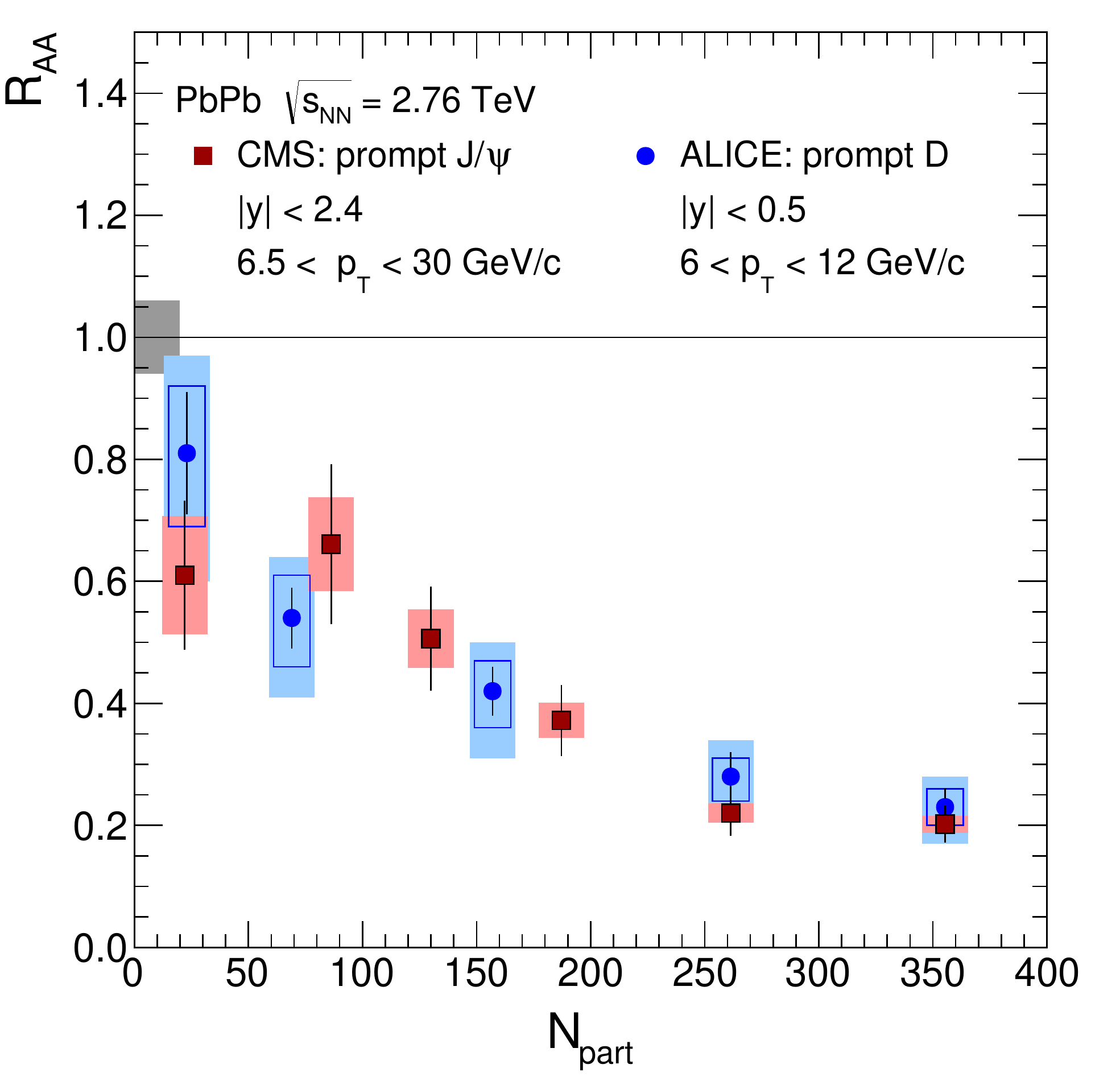} 
  \caption{Comparison of nuclear modification factors for open and closed charm 
    at RHIC (top) and the LHC 
    (bottom)~\cite{Adare:2010de,Adare:2006ns,Adamczyk:2012ey,Chatrchyan:2012np,ALICE:2012ab}. 
    Transverse momentum integrated results are shown on the left, while high \pt 
    \raa are compared on the right. Due to the lack of low \pt open charm data 
    at the LHC, the bottom left panel is missing.} 
  \label{fig:open_closed_HF} 
\end{figure}

To study the effect of a deconfined medium on quarkonium production, we first 
recall the underlying dynamics, using the \jpsi for illustration. The production 
process in elementary hadronic collisions begins with the formation of a \ccbar 
pair; this pair can then either lead to open charm production (about 90\%) or 
subsequently bind to form a charmonium state (about 10\% for all charmonia). 
Since quarkonium 
production is to be used as a tool to study the medium produced in nuclear 
collisions, the primary concern is not if such collisions produce more or fewer 
\ccbar pairs than proton-proton collisions, but rather if the presence of the 
medium modifies the fraction of produced \ccbar pairs going into charmonium 
formation. In other words, the crucial quantity is the amount of charmonium 
production relative to that of open charm.  
Hence the relevant observable is the 
fraction of charmonia to open charm, or more generally, that of quarkonia to the 
relevant open heavy flavour production~\cite{Satz:1993pb,Satz:2013ama}. In this 
quantity, if measured over the entire phase space, down to $\pt=0$, the effects 
of possible initial state nuclear modifications cancel out, so that whatever 
changes it shows relative to the \pp pattern is due to final state effects. Here 
it should be noted that, since the distribution of the different open charm 
channels is in good approximation energy-independent, the measurement of a 
single such channel is sufficient ---it gives, up to a constant, the total open 
charm cross section~\cite{Satz:2013ama}.

A direct comparison of measured open and closed heavy-flavour cross sections has
not been performed yet at RHIC or the LHC. However, one can compare the measured
nuclear modification factors of D mesons (or heavy-flavour decay electrons as
their proxy) and \jpsi. At RHIC, the open charm cross section has been measured
in \pp and \AuAu via non-photonic single electrons from semileptonic charm
decays~\cite{Adare:2010de} as well as with fully reconstructed D mesons via
hadronic decays~\cite{Adamczyk:2014uip}. As shown in the top left of
\fig{fig:open_closed_HF}, the resulting \raa shows no deviation from binary
scaling, though the uncertainties are sizeable. Hence, one can conclude that the
modification of the \jpsi \raa at RHIC~\cite{Adare:2006ns} is a true final state
effect and not just a reduction of charm production by initial state effects. As
evident from the large uncertainties, the measurement of the total charm cross
section is extremely challenging. At the LHC this has not been achieved yet,
preventing such a comparison in the bottom left of \fig{fig:open_closed_HF}.
Instead one can try to make a comparison at high \pt where both, open and closed
charm, have been measured~\cite{Chatrchyan:2012np,ALICE:2012ab}. This then opens
the question which \pt intervals are appropriate for such a comparison. A
comparison of D and \jpsi in the same \pt range will not access the same charm
quarks and/or gluons. This is an issue to be addressed on theoretical grounds.
The comparison is anyway performed, as shown in the bottom right panel of
\fig{fig:open_closed_HF}. The \raa of high-\pt D and \jpsi show a surprising
similar centrality dependence. This, however, is nothing new at the LHC. Already
at RHIC the same trend can be observed~\cite{Adare:2010de,Adamczyk:2012ey}, as
shown in the top right of \fig{fig:open_closed_HF}. The suppression of high-\pt
D mesons has been linked to charm quark energy loss inside the QGP. While the
\jpsi itself is a colourless object, its coloured precursor may be subject to
similar energy loss though current models underestimate the \jpsi suppression at
high \pt in \pb collisions at the LHC~\cite{Sharma:2012dy}.

 

\subsection{Summary and outlook}

With the LHC Run~1 a large wealth of quarkonium measurements has 
enriched and complemented the observations from SPS and RHIC experiments. The 
main results and their current interpretation are summarised in the 
following. 
\begin{itemize} 
\item In the charmonium sector, the \jpsi \raa at high transverse momentum shows a 
  clear suppression. This suppression is stronger than the one observed at RHIC 
  energies, as expected in a sequential melting scenario. 
\item An opposite behaviour was in the low-\pt region, where the 
  \jpsi \raa measured at LHC is larger than the one obtained at lower energies. 
  This observation can be interpreted as an evidence for a new production mechanism 
  setting in at high energies, based on the (re)combination of $c$ and $\bar{c}$ 
  quarks either during the collision history or at the hadronisation. The 
  measurement of a positive \vtwo, for low-\pt \jpsi, is considered a 
  further confirmation of the important role played by this additional 
  contribution. 
\item Theoretical models assuming a fraction of \jpsi produced by 
  (re)combination of the order of 50\% at low \pt and then vanishing at high 
  \pt, provide a fair description of the experimental data. On the contrary, 
  calculations including only shadowing effects cannot account for the observed 
  suppression. Note, however, that coherent energy loss effects in cold nuclear 
  matter are able to reproduce the \jpsi suppression, yet the agreement may be 
  of accidental origin~\cite{Arleo:2014oha}. 
\item 
  For 
  the first time, the \raa for \upsa, \upsb, and \upsc was measured. 
  Results indicate a clear ordering with binding energy, as expected in the 
  sequential melting scenario. 
\end{itemize} 
Even if a qualitative understanding of the quarkonium behaviour at LHC energies 
is nowadays rather well assessed, there are still several aspects which require 
to be furthered: 
\begin{itemize} 
\item In order to quantify the hot matter effects on quarkonium production, a 
  precise knowledge of the cold nuclear matter effects is required. Accurate 
  quarkonium measurements in \pA collisions are, therefore, mandatory to refine 
  the interpretation of the \AAcoll results. 
\item Theoretical calculations are, as of today, still affected by large 
  uncertainties, mainly due to the uncertainties on the cold nuclear matter 
  effects and, for models including a (re)generation component, also on the 
  uncertainties on the \ccbar production cross section. The comparison of the 
  measured \jpsi \raa and theory predictions will benefit from the measurement 
  of the latter down to zero \pt. 
\item Intriguing results have been obtained on the \psiP in the LHC Run~1. Given 
  the observed dependence on rapidity and transverse momentum, the 
  interpretation of the \psiP behaviour will clearly gain from a more 
  differential study feasible with larger data sample. 
\item Similarly, also bottomonia will benefit from multi-differential studies 
  to assess the kinematic dependence of all the \ups states. 
\item Finally, the availability of charmonium and bottomonium results spanning 
  almost three orders of magnitude in \snn and covering very different 
  kinematic regions represents clearly a challenge for all theoretical models, 
  which should now move towards a consistent description of quarkonium data. 
\end{itemize} 
 
The incoming LHC Run~2 data are expected to shed more light, moving 
from a qualitative understanding of the quarkonium fate in a hot medium towards 
a more quantitative one.


\section{Quarkonium photoproduction in nucleus-nucleus collisions}
\label{sec:UPC}

In 2011, the LHC produced collisions of lead ions at a centre-of-mass energy per nucleon pair  $\sqrt{s_{\rm NN}} = 2.76$~TeV. These collisions have been
used to perform different measurements of charmonium  photonuclear production\footnote{The integrated luminosity of the 2010 Pb--Pb run, ten times smaller than in 2011, was not sufficient to perform the measurements described below, although it was enough to measure the coherent production of $\rho^0$ mesons~\cite{Adam:2015gsa}.}.  All but one of the studies described in this section have been carried out using ultra-peripheral collisions (UPC). These are  interactions  where the impact parameter exceeds the sum of the radii of the colliding nuclei. In such collisions, the cross section for hadronic processes is strongly suppressed, while the cross section for  electromagnetic interactions remains large. The  analysis not related to UPC has investigated the photoproduction of $\jpsi$ overlapped with a standard hadronic \PbPb collision. 

Two types of photonuclear production of charmonium have been studied: coherent and incoherent. In the first case, the incoming quasi-real photon interacts coherently with the whole nucleus to produce the charmonium.
The coherence condition, both in the emission of the photon and in the interaction with the nuclear target, constrains the transverse momentum of the produced vector meson to be of the order of the inverse of the nucleus diameter, which translates into approximately 60 MeV/$c$. In the incoherent case, the quasi-real photon couples only to one nucleon, and thus the transverse momentum of the produced vector meson is constrained by the size of the nucleon, which translates into approximately 300 MeV/$c$. In a fraction of the coherent interactions and in all incoherent processes, one or a few neutrons are produced at the  rapidity of the incoming beams. The experimental signature of these processes is therefore a vector meson with fairly small transverse momentum, possibly one or a few neutrons detected at zero degrees, and nothing else in the detector.

In this section, we review these measurements and discuss the models proposed to describe them. Previous reviews addressing these subjects can be found in  \cite{Baur:1998sr,Baur:2001jj,Bertulani:2005ru,Baltz:2007kq}. This section is organised as follows.
First, in Subsection \ref{UPCsec:photons} we discuss the origin and characteristics of the photon flux at the LHC. Subsection \ref{UPCsec:exp} describes previous results from RHIC and  the existing measurements from LHC. Subsection \ref{UPCsec:theory} presents the current theoretical models and the main differences among them. Subsection \ref{UPCsec:comp} discusses
how the models compare to the experimental results. We conclude in Subsection \ref{UPCsec:outlook} with a brief summary of the lessons learnt and with an outlook of what could be possible with the data from the LHC \RunTwo.

\subsection{The flux of photons from lead ions at the LHC}
\label{UPCsec:photons}

The lead beams in the LHC are an intense source of photons, because the electromagnetic field of charged particles accelerated to ultra-relativistic velocities can be seen as a flux of quasi-real photons, according to a proposal made by Fermi \cite{Fermi:1924tc,Fermi:1925fq}, and later refined by Weizs\"acker \cite{vonWeizsacker:1934sx} and Williams \cite{Williams:1934ad}. In this section we discuss the emission of photons from one nucleus. Next section discusses the cross section taking into account the contribution from both nuclei participating in the collision.

The photons are emitted by the nucleus coherently and thus their virtuality is restricted to be of the order of the inverse of the nucleus diameter, which for  lead  implies an upper limit for the virtuality around 30 MeV/$c$; i.e., the photons can be considered as quasi-real.
The intensity of the flux depends on the square of the electric charge of the incoming particle, so it is large for the lead nuclei at the LHC. In  the semi-classical description (see for example \cite{Baur:2001jj}) the photon flux per unit area is given by 

\begin{equation}
\frac{\dd^3n(k,\vec{b})}{\dd k\dd^2\vec{b}} = \frac{\alpha_{\rm em} Z^2}{\pi^2 kb^2}x^2\left[K^2_1(x)+\frac{1}{\gamma^2}K^2_0(x)\right],
\label{UPCeq:FluxPerArea}
\end{equation}
where $\alpha_{\rm em}$ is the fine structure constant, $k$ is the photon energy in the frame where the
photon emitter has Lorentz factor gamma, $Z$ is the electric charge of the lead nucleus, $K_{0}$ and $K_1$ are modified Bessel functions, $\vec{b}$ is the impact-parameter vector with $b$ its magnitude and $x=kb/\gamma$.

The measurements of ultra-peripheral collisions described in the next sections were obtained requiring the absence of a hadronic collision between the incoming nuclei. This requirement is implemented into the computation of the photon flux in two different ways. The simpler option is to integrate \eq{UPCeq:FluxPerArea}
starting from a minimum impact parameter $b_{\rm min}$ given by the sum of the radii of the incoming nuclei. 
In this case, known as the hard-sphere approximation, the flux of quasi-real photons is given by

\begin{equation}
\frac{\dd n(k)}{\dd k} = \frac{2\alpha_{\rm em} Z^2}{\pi k}\left[ \xi K_0(\xi)K_1(\xi)-\frac{\xi^2}{2}\left(K^2_1(\xi)-K^2_0(\xi)\right)\right],
\label{UPCeq:FluxHS}
\end{equation}
with $\xi=kb_{\rm min}/\gamma$. 

Another option is to convolute the flux per unit area  with the probability of no hadronic interaction, which is obtained using the nuclear overlap function and the total nucleon-nucleon interaction cross section and averaging the flux over the target nucleus
(for further details see \cite{Klein:1999qj}) . The resulting integral can be calculated numerically to obtain  the photon flux $n(k)$.

All the measurements described below are elastic in the sense
that the measurement of the charmonium fixes completely the kinematics of the process.
In this case the energy of the photon can be expressed in terms of the mass $M$ of the charmonium and its rapidity $y$ as

\begin{equation}
k = \frac{M}{2}\exp{(y)},
\end{equation}
and thus, the photon flux can be written as:

\begin{equation}
n(y,M)\equiv k\frac{\dd n(k)}{\dd k}.
\end{equation}

\begin{figure}[t]
\begin{center}$
\begin{array}{c}
\includegraphics[width=0.47\textwidth]{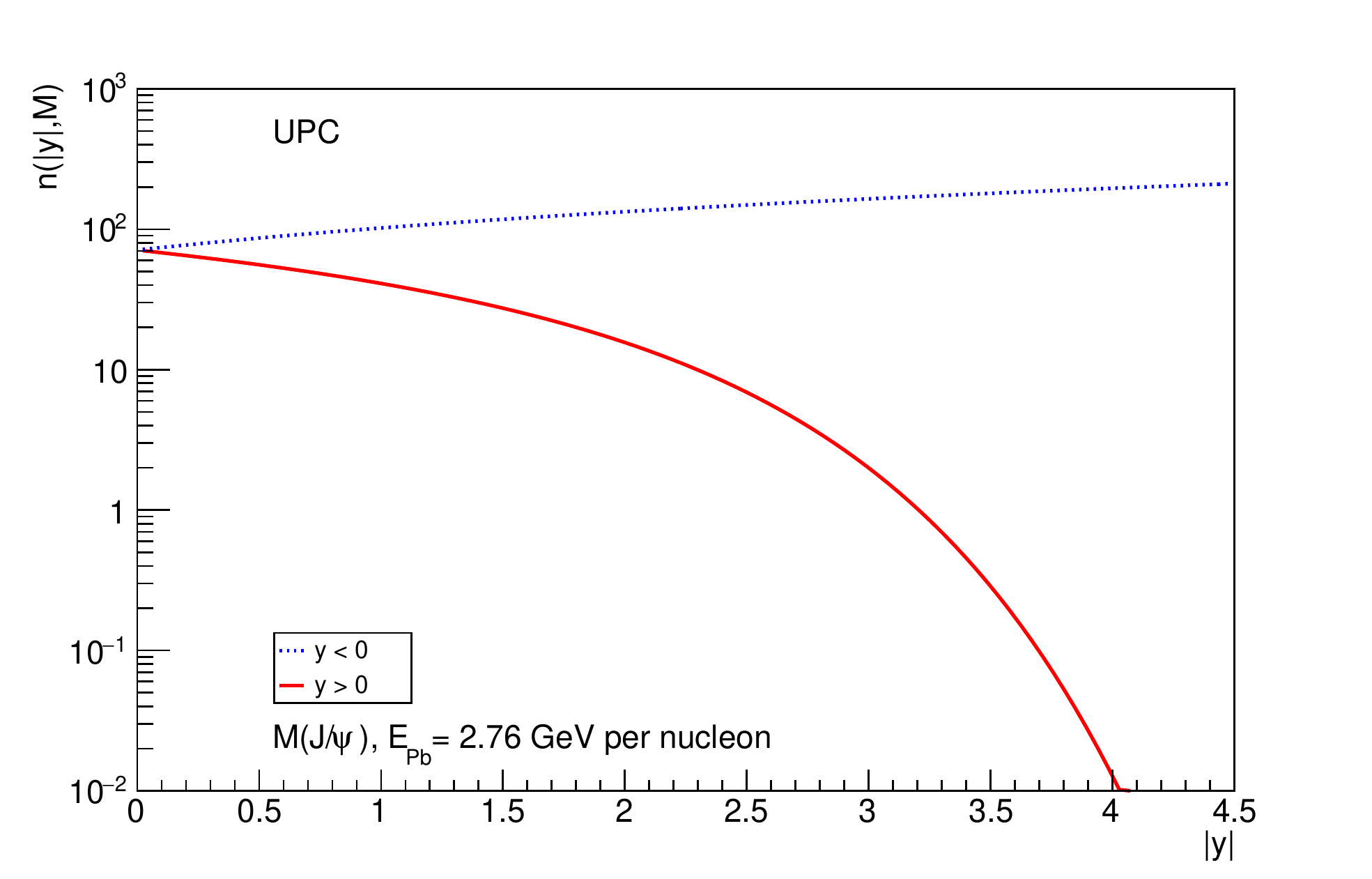} 
\includegraphics[width=0.47\textwidth]{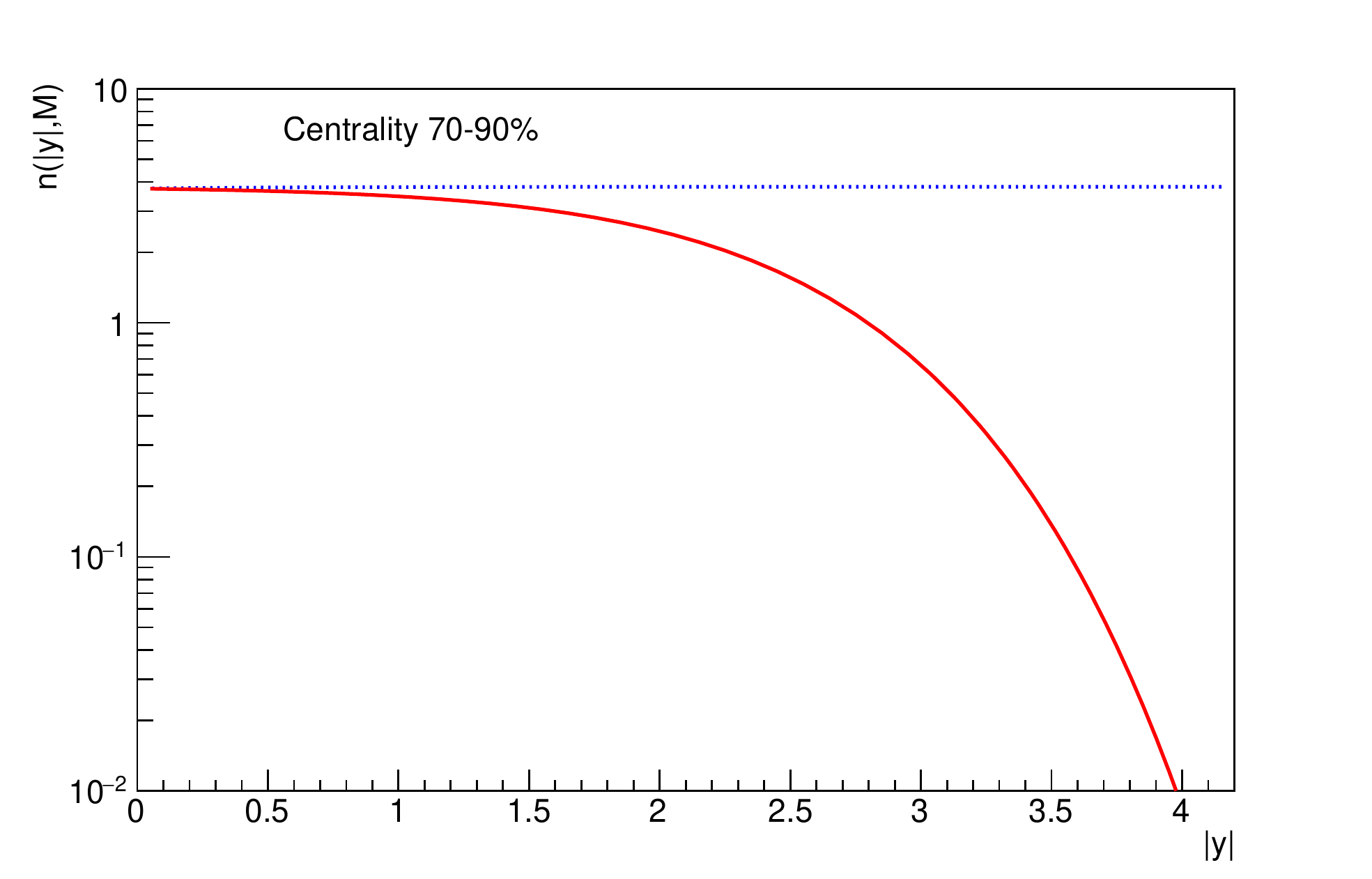} 
\end{array} $
\end{center}
   \caption{Flux of photons emitted by one nucleus for positive and negative values of the rapidity $y$ for the case of  $\jpsi$ coherent photoproduction and an energy of the lead-ion beam of 2.76~GeV per nucleon. Positive rapidities correspond to the direction of the lead-ion. The left panel shows the UPC case given by \eq{UPCeq:FluxHS}, while the right panel shows the integration of \eq{UPCeq:FluxPerArea} for the impact parameters corresponding to the 70--90 \% centrality class in hadronic \PbPb collisions.
  }
\label{UPCfig:FluxHS}
\end{figure}

Figure~\ref{UPCfig:FluxHS} shows the flux of quasi-real photons with the energy required to produce a $\jpsi$ at rapidity $y$. 
At large negative $y$ the centre-of-mass energy of the photon-lead system is not large enough to produce the vector meson. For the energies of the lead beams during \RunOne and the case of a $\jpsi$ this happens at $y\approx -6.8$; see e.g. \eq{eq:W2my}. The left panel shows the UPC case, computed with \eq{UPCeq:FluxHS}: as the rapidity increases the flux decreases. At large rapidities the fast decrease of the flux is related to the behaviour of the Bessel functions. The right panel shows the  integration of \eq{UPCeq:FluxPerArea} for the impact parameters corresponding to the 70--90 \% centrality class in hadronic \PbPb collisions according to \cite{Abelev:2013qoq}. This case is discussed in Section \ref{sec:lowptexcess}.

\subsection{Measurements of photonuclear production of charmonium during the \RunOne at the LHC}
\label{UPCsec:exp}

In both coherent and incoherent photonuclear production of charmonium
the target is not broken by the interaction and in this sense the processes may be considered elastic. Therefore, the measurement of the produced charmonium  completely fixes the kinematics.

\begin{figure}[t]
\begin{center}$
\begin{array}{c}
\includegraphics[width=0.47\textwidth]{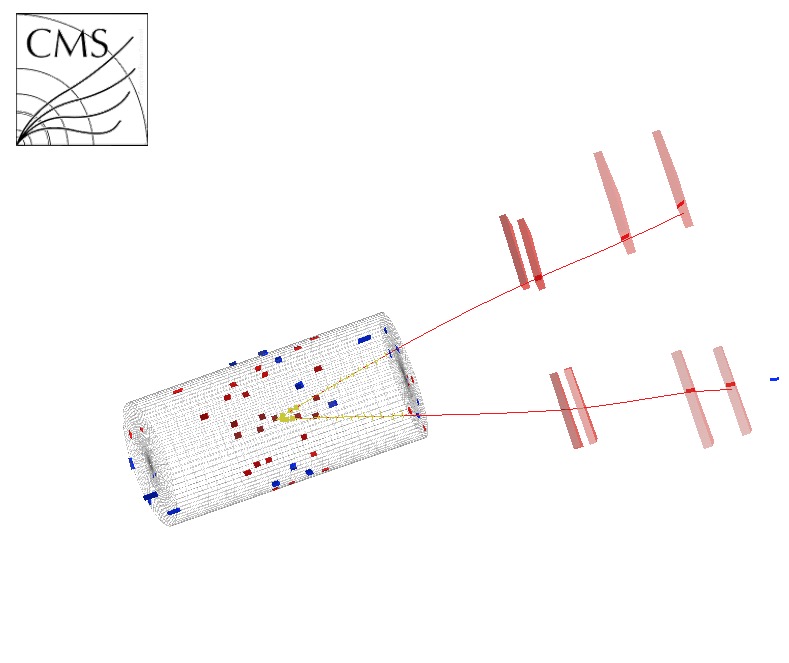} 
\hspace{2mm}
\includegraphics[width=0.47\textwidth]{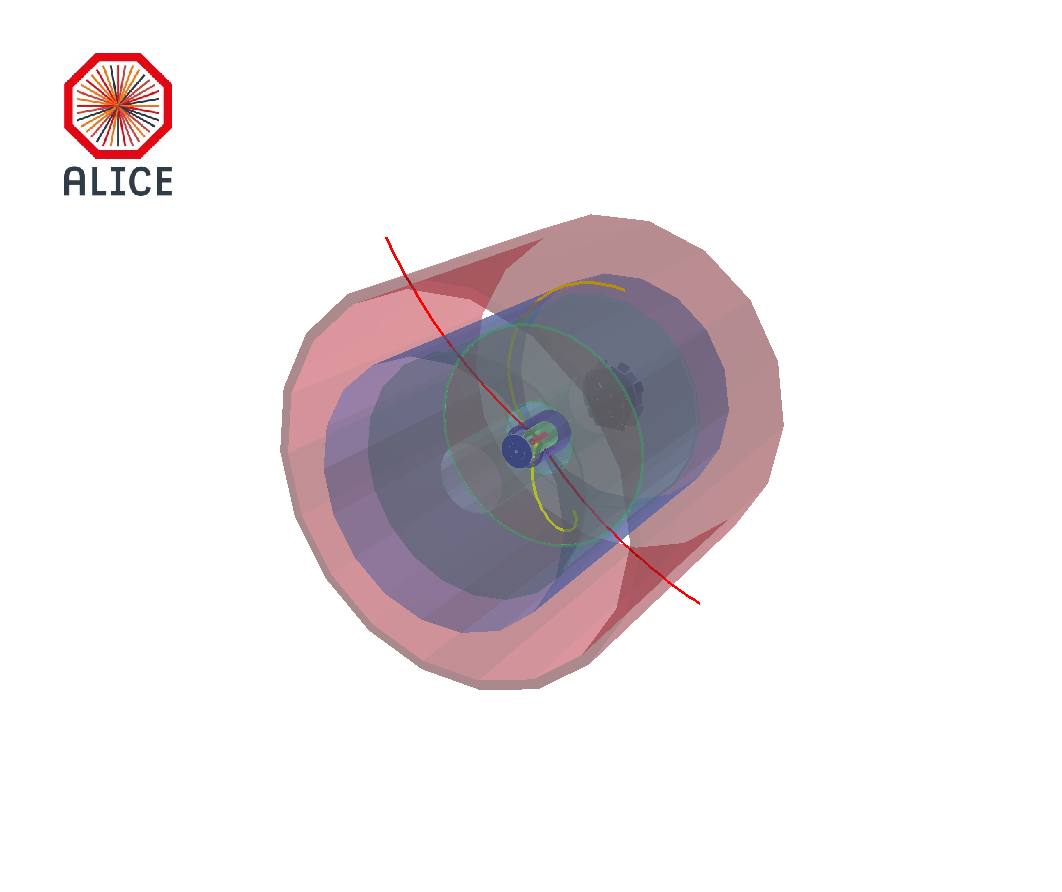} 
\end{array} $
\end{center}
   \caption{Event displays of coherent photonuclear production of $\jpsi \to \mu^+\mu^-$ (left) and $\psiP \to \jpsi \pi^+\pi^- \to  \mu^+\mu^-\pi^+\pi^-$ (right) in UPC as measured with the CMS and ALICE detectors respectively.}
\label{UPCfig:ED}
\end{figure}

As mentioned above, the experimental signature for these processes in UPC consists then in the decay products of a charmonium with fairly small transverse momentum  and in some cases one or a few neutrons at zero degrees. No other event activity is measured in the detector. Figure~\ref{UPCfig:ED} shows two event displays of the coherent photonuclear production of $\jpsi$ and $\psiP$ in UPC as measured with the CMS and ALICE detectors.

The  cross section for these photonuclear processes in \PbPb collisions, with the charmonium measured at rapidity $y$, has two contributions:
\begin{equation}
\frac{\dd\sigma_{\rm PbPb}(y)}{\dd y}= n(y,M)\sigma_{\gamma \rm Pb}(y) + n(-y,M)\sigma_{\gamma \rm Pb}(-y),
\label{UPCeq:SigPbPb}
\end{equation}
where the first term corresponds to one of the incoming lead nucleus acting as the source of the photon and the second term corresponds to the other incoming nucleus acting as the source of the photon.
When the charmonium is measured at mid-rapidities, $y=0$, both terms are equal and can be summed. On the other hand, when the charmonium is measured at rapidities around 3, the flux at positive $y$ is strongly suppressed and  the term at negative $y$ dominates.  Note that for the case of photonuclear production overlapped with a hadronic collision, both fluxes  contribute even at the forward rapidities measured at the LHC. This is illustrated in \fig{UPCfig:FluxHS}.

As mentioned before, the measurement of the charmonium fixes the kinematics. The centre of mass energy of the $\gamma$-Pb system is given by
\begin{equation}
W^2_{\gamma\rm{Pb}} = 2k\sqrt{s_{\rm NN}} = M\exp{(y)}\sqrt{s_{\rm NN}} ,
\label{eq:W2my}
\end{equation}
where in these expressions, both $k$ and $y$ are
evaluated in the nucleus-nucleus centre-of-mass frame. In a leading order pQCD approach $W_{\gamma\rm{Pb}}$ is related to  $x$--Bjorken by
\begin{equation}
x = \frac{M^2}{W^2_{\gamma\rm{Pb}} }.
\end{equation}
According to this prescription, a measurement of  charmonium photonuclear production at large rapidities in UPC samples mainly the low $W_{\gamma\rm{Pb}}$, alternatively large $x$, contribution to \eq{UPCeq:SigPbPb}.

\subsubsection{Photonuclear production of $\jpsi$ at RHIC}

The first measurement of photonuclear production of charmonium in UPC of relativistic heavy ions was performed by the PHENIX Collaboration using \AuAu collisions  at $\sqrt{s_{\rm NN}} = 200$ GeV \cite{Afanasiev:2009hy}.
 The events were triggered by tagging the production of neutrons at zero degrees and the \jpsi were measured at mid-rapidity using the decay to an electron-positron pair. PHENIX found 9.9 $\pm$ 4.1 (stat) $\pm$ 1 (syst) $\jpsi$ candidates. The smallness of the sample did not allow to separate the coherent and incoherent contributions. Their measurement corresponded to $W_{\gamma\rm{Au}}\approx24$ GeV ($x\approx 1.5\cdot 10^{-2}$). The  cross section for \AuAu UPC at mid-rapidity was measured to be 76 $\pm$ 31 (stat) $\pm$ 15 (syst) $\mu$b, which agreed, within the errors, with the theoretical models available at that time~\cite{Klein:1999qj,Baltz:2002pp,Strikman:2005ze,Abreu:2007kv}. Although the large experimental errors precluded setting strong constraints on the models, this study was very important as a proof of principle.

\subsubsection{Coherent production of $\jpsi$ in \PbPb UPC at the LHC}

The coherent photonuclear production of $\jpsi$ has been measured in three different rapidity (equivalently $W_{\gamma\rm{Pb}}$) ranges at the LHC. ALICE has measured it at mid- \cite{Abbas:2013oua}
and forward rapidity \cite{Abelev:2012ba}, while CMS has recently released preliminary results at semi-forward rapidities \cite{CMS:2014ies}. Table~\ref{UPCtab:xs} summarizes  these measurements.

 The ALICE detector \cite{Aamodt:2008zz} 
measures charmonium either in the central barrel using a combination of silicon trackers (ITS), a time projection chamber (TPC) and a time of flight system (TOF); or in the forward part where a muon spectrometer is installed.
In addition to requiring  the decay products of the charmonium to be either in the central barrel or in the muon spectrometer,  the exclusivity condition is realised vetoing activity in a set of two scintillator arrays (VZERO) which cover 4 units of rapidity in the forward/backward region, while the absence of neutrons at zero degrees or the measurements of one or few of them is performed with zero-degree calorimeters (ZDC) located 116 m away and on both sides of the  interaction point.

The first measurement of coherent production of $\jpsi$ in \PbPb UPC was performed by ALICE using the muon spectrometer \cite{Abelev:2012ba}.
 The trigger required a muon above threshold (1 GeV/$c$ of transverse momentum) and no activity in the opposite side of the detector. The coherent contribution was obtained selecting candidates with transverse momentum less than 0.3 GeV/$c$. The cross section in \PbPb UPC was measured to be 1.00 $\pm$ 0.18(stat) $^{+0.24}_{-0.26}$ (syst) mb. 
As this measurement is performed at large rapidities, the dominant contribution to the cross section (about 95\%) originates from the low energy part of the flux (see \fig{UPCfig:FluxHS}) with the corresponding average energy in the centre-of-mass system of the photon and the target being $W_{\gamma\rm{Pb}}\approx20$ GeV ($x\approx 2.2\cdot 10^{-2}$). 
 The second measurement was performed at mid-rapidity using the central barrel detectors \cite{Abbas:2013oua}. The trigger required hits in ITS and TOF (in TOF with a back-to-back topology) and absence of activity in VZERO. In this case, using the PID capabilities of the ALICE TPC,
 two decay channels have been used: $\mu^+\mu^-$ and $e^+e^-$. The transverse momentum distribution of the $\jpsi$ candidates was used to extract the coherent contribution. The measured \PbPb UPC coherent cross section was 2.38 $\pm$ $^{0.34}_{0.24}$ (stat+syst) mb and corresponded to $W_{\gamma\rm{Pb}}\approx 92$ GeV ($x\approx 10^{-3}$). For this sample, the fraction of coherent events with no activity in the ZDC was measured to be 0.70 $\pm$ 0.05 (stat).

The central barrel of the CMS detector \cite{Chatrchyan:2008aa} contains 
a silicon pixel and strip tracker, a lead tungstate crystal electromagnetic calorimeter and a brass/scintillator hadron calorimeter; all of them within a superconducting solenoid of 6 m internal diameter, providing a magnetic field of 3.8 T.  Muons are measured within pseudorapidity $|\eta|< 2.4$ by gas-ionization detectors embedded in the steel return yoke outside the solenoid. The UPC trigger used by CMS requires ($i$) the presence of at least one muon candidate with a minimal transverse-momentum threshold, ($ii$) at least one track in the pixel detector, ($iii$) rejection of events with activity in the scintillator counters covering the pseudorapidity range between 3.9 and 4.4 on both sides of the  interaction point and ($iv$) energy deposit consistent with at least one neutron in either of the ZDCs. This last requirement is similar to what was done by PHENIX and triggers only on a fraction of the cross section. To obtain the total coherent cross section, models of neutron emission in coherent production were used~\cite{Klein:1999qj,Guzey:2013jaa}. Note that CMS result is not corrected for feed-down from $\psiP$ and that
 the contribution of both terms in \eq{UPCeq:SigPbPb} is important, so that it is not possible  to assign a unique value of $W_{\gamma\rm{Pb}}$ to this measurement. The preliminary cross section can be found in~\cite{CMS:2014ies}. It agrees with expectations of models adjusted to describe ALICE data~\cite{Klein:1999qj,Guzey:2013jaa,Adeluyi:2012ph}.

 \subsubsection{Coherent production of $\psiP$ in \PbPb UPC at the LHC}
The ALICE Collaboration carried out a preliminary measurement of the coherent production of $\psiP$ in \PbPb UPC~\cite{Broz:2014wka}  at mid-rapidity using the same trigger and detectors as for the $\jpsi$ case. The $\psiP$ was identified in the following decay channels: to $l^+l^-$ and to $\jpsi\pi^+\pi^-$, with $\jpsi\to l^+l^-$, where $l=e,\mu$. The right panel of \fig{UPCfig:ED} shows an event display of a coherently produced $\psiP \to \jpsi \pi^+ \pi^-$ candidate. 
A preliminary measurement of the ratio of the cross sections for $\jpsi$ and $\psiP$ coherent photonuclear production
was carried out as well~\cite{Broz:2014wka}. 
This measurement is significantly  higher (about a factor of two for the central value) than the ratios 
0.166 $\pm$ 0.007(stat) $\pm$ 0.008(syst) $\pm$ 0.007(BR),  0.14 $\pm$ 0.05 and 0.19 $\pm$ 0.04
 measured by H1 \cite{Adloff:2002re}, CDF \cite{Aaltonen:2009kg}  and  LHCb \cite{Aaij:2013jxj} respectively.

 \begin{table}
 \caption{Summary of published measurements of  photonuclear production of charmonium in \PbPb  UPC at $\sqrt{s_{\rm NN}} = 2.76$\TeV at the LHC (the preliminary results on $\jpsi$ from CMS~\cite{CMS:2014ies} and \psiP from ALICE~\cite{Broz:2014wka} are not included).}
  \label{UPCtab:xs}
\centering
\begin{tabular}{ccccc}
 \hline
Experiment &  Vector meson & d$\sigma/$d$y$ [mb] & Rapidity range & Ref.\\
 \hline
ALICE & Coherent $\jpsi$  & 1.00 $\pm$ 0.18(stat) $^{+0.24}_{-0.26}$ (syst)& $-3.6<y<-2.6$ &  \cite{Abelev:2012ba} \\
ALICE & Coherent $\jpsi$  & 2.38 $^{+0.34}_{-0.24}$ (stat+syst) & $|y|<0.9$ &   \cite{Abbas:2013oua} \\
\hline
ALICE & Incoherent $\jpsi$  & 0.98 $^{+0.19}_{-0.17}$ (stat+sys) & $|y|<0.9$ &  \cite{Abbas:2013oua} \\
\hline 

 \end{tabular}
 \end{table}

\subsubsection{Incoherent production of \jpsi in \PbPb UPC at the LHC}

ALICE has also measured the incoherent production of $\jpsi$ in \PbPb UPC   at mid-rapidity~\cite{Abbas:2013oua} using the same trigger and detectors as for the coherent case. The incoherent contribution was obtained from the distribution of transverse momentum. The centre of mass energy in the $\gamma$-Pb system is the same as for the coherent case. The measured cross section is 0.98 $^{+0.19}_{-0.17}$ (stat+syst) mb.

\subsubsection{Coherent photonuclear production of $\jpsi$ in coincidence with a hadronic \PbPb collision at the LHC}
\label{sec:lowptexcess}

When studying the inclusive distribution of transverse momentum of $\jpsi$ in hadronic \PbPb collisions at large rapidities (in the range 2.5 to 4.0), a significant excess of $\jpsi$ candidates was found for transverse momentum smaller than 0.3 GeV/$c$ for the centrality bin 70--90\% \cite{Lardeux:2013dla} . One possible explanation of this observation is the coherent photonuclear production of the $\jpsi$ in coincidence with a hadronic interaction. Although the possibility of such a process has been discussed in the past \cite{Joakim2010}, currently there is no theoretical calculation available for this process. Such a calculation is a challenge for theorists. Note that an excess is also observed, with reduced significance, in the centrality bin 50--70\% and that there is a framework in place to extract the photonuclear coherent cross section in these cases \cite{LaureQM2014}.

\subsection{Models for photonuclear production of charmonium}
\label{UPCsec:theory}

The following models will be discussed in this section:
\begin{itemize}
\item[] {\bf AB-AN}: Model by Adeluyi and Bertulani \cite{Adeluyi:2012ph} and Adeluyi and Nguyen \cite{Adeluyi:2013tuu}.
\item[] {\bf CSS}: Model by Cisek, Sch\"afer and Szczurek \cite{Cisek:2012yt}.
\item[] {\bf KN}: Model by Klein and Nystrand implemented in the STARLIGHT Monte Carlo program  
\cite{Klein:1999qj,Baltz:2002pp,Klein:2003vd}.
\item[] {\bf LM}: Model by Lappi and Mantysaari \cite{Lappi:2010dd,Lappi:2013am}.
\item[] {\bf GM-GDGM}: Model by Goncalves and Machado \cite{Goncalves:2011vf} and by Gay-Ducati, Griep and Machado \cite{Ducati:2013bya}.
\item[] {\bf RSZ}: Model by Rebyakova,  Strikman, and Zhalov \cite{Rebyakova:2011vf}.
\end{itemize}

All models start from \eq{UPCeq:SigPbPb} which has two ingredients: the photon flux and the photonuclear cross section. The first difference among the models is that some of them (CSS, LM, GM-GDGM) use the hard-sphere approximation of the photon flux; i.e., \eq{UPCeq:FluxHS}, and other models (AB-AN, KN, RSZ-GZ) integrate the convolution of  \eq{UPCeq:FluxPerArea} with the probability of no hadronic interaction. 

Regarding the photonuclear cross section the models contain the following ingredients: ($i$)  an assumption on the nuclear distribution in the transverse plane, ($ii$) an implicit or explicit  prescription for the wave function of the vector meson and finally ($iii$)
all models fix some of the parameters using data on exclusive photoproduction of charmonium off the proton and thus have to include a prescription to link the photoproduction off protons with the photonuclear interaction. In this context the models can be grouped in three different classes: models based on the generalised vector dominance model (KN), on LO pQCD (AB-AN, RSZ) and on  the colour dipole model (CSS, LM, GM-GDGM).

\subsubsection{Models based on  vector dominance}

The only model in this class is KN. There are three main ingredients in this model: ($i$) the vector dominance model (VDM) relates both the $\gamma \mathrm{+Pb}\to\mathrm{Pb+V}$ and the $\gamma \mathrm{+p}\to\mathrm{p+V}$ processes to $\mathrm{Pb+V}\to\mathrm{Pb+V}$ and $\mathrm{p+V}\to\mathrm{p+V}$ respectively (Here $V$ represents a vector meson.); ($ii$) the optical theorem relates these last processes to the total cross section; ($iii$) a classical Glauber model relates the total production cross section off a proton, to that off a nucleus.

In more detail:
\begin{equation}
\sigma_{\gamma \rm Pb}(y) \equiv\sigma(\gamma{\rm + Pb}\to{\rm V+Pb}) =
\left. \frac{\dd\sigma(\gamma{\rm + Pb}\to{\rm V+Pb}) }{\dd t} \right|_{t=0}
\int^\infty_{t_{min}} \dd t |F(t)|^2,
\label{UPCeq:Fwd}
\end{equation}
where  $F(t)$ is the nuclear form factor and $t$ the momentum transferred to the nucleus. 
Using VDM and the optical theorem yields
\begin{equation}
\left. \frac{\dd\sigma(\gamma{\rm + Pb}\to{\rm V+Pb}) }{\dd t} \right|_{t=0}
= \frac{\alpha \sigma^2_{\rm TOT}({\rm V+Pb})}{4f^2_V},
\end{equation}
where $f_V$ is the vector meson photon coupling. A classical Glauber model produces
\begin{equation}
\sigma_{\rm TOT}({\rm V+Pb}) \approx \sigma_{\rm inel}({\rm V+Pb})
=\int \dd^2\vec{b}\left(
1-\exp\left[-\sigma_{\rm TOT}({\rm V+p}) T_{\rm Pb}(\vec{b}) \right]
\right),
\end{equation}
where $T_{\rm Pb}$ is the nuclear thickness function and $\sigma_{\rm TOT}({\rm p+V})$ is obtained from  the optical theorem, now applied at the nucleon level
\begin{equation}
\sigma^2_{\rm TOT}({\rm V+p}) = 16\pi
\left. \frac{\dd\sigma({\rm V+p}\to{\rm V+p}) }{\dd t} \right|_{t=0}.
\end{equation}
Using VDM leads to
\begin{equation}
\left. \frac{\dd\sigma({\rm V+p}\to{\rm V+p}) }{\dd t} \right|_{t=0} = \frac{f^2_V}{4\pi\alpha}
\left. \frac{\dd\sigma(\gamma{\rm + p}\to{\rm V+p}) }{\dd t} \right|_{t=0},
\end{equation}
where  the elementary cross section
\begin{equation}
\left. \frac{\dd\sigma(\gamma{\rm + p}\to{\rm V+p}) }{\dd t} \right|_{t=0}
= b_V\left( XW^\epsilon_{\gamma\rm p} + YW^{-\eta}_{\gamma\rm p}\right)
\end{equation}
is fitted to experimental data to obtain the values for the 
$X$, $Y$, $\epsilon$, $\eta$ and $b_V$  parameters.
 
 \subsubsection{Models based on LO pQCD}

These models start from \eq{UPCeq:Fwd} and use the LO pQCD calculation 
\cite{Ryskin:1992ui,Brodsky:1994kf} for the forward cross section
\begin{equation}
\left. \frac{\dd\sigma(\gamma{\rm + Pb}\to{\rm V+Pb}) }{\dd t} \right|_{t=0}=
\frac{16\pi^3\alpha^2_s \Gamma_{ee}}{3\alpha M^5}
\left[ xG_A(x,Q^2)
\right]^2,
\label{UPCeq:LOpQCD}
\end{equation}
where $\Gamma_{ee}$ is the decay width to electrons and $G_A$ is the nuclear gluon density distribution at a scale $Q^2$, which for the models described below was chosen to be $Q^2 = M^2/4$, although other options are possible and may describe better the experimental data \cite{Guzey:2013qza}. It is important to note that this equation contains implicitly a model for the wave function of the vector meson, but in the final result the only trace of it is the presence of $\Gamma_{ee}$. 

The AB-AN model modifies \eq{UPCeq:LOpQCD} by adding a normalisation parameter to the right side, which should take into account effects missing in the approximation. This factor is then fitted to reproduce HERA data using the same type of equation applied to the $\gamma \mathrm{ + p}\to\mathrm{p}+\jpsi$ case. Nuclear effects are modelled as $G_A(x,Q^2) = g_p(x,Q^2) R^A_g(x,Q^2)$ where $g_p$ is the gluon distribution in the proton and $R^A_g$ is the nuclear modification factor of the gluon distribution.
MSTW08 \cite{Martin:2009iq} is used for the gluon distribution in the proton, while several different choices are made for $R^A_g$ to estimate nuclear effects: EPS08 \cite{Eskola:2008ca},
 EPS09 \cite{Eskola:2009uj}, HKN07 \cite{Hirai:2004wq} and  $R^A_g = 1$ to model the absence of nuclear effects.

The RSZ model computes  $R^A_g$ in the leading twist approach to nuclear shadowing 
\cite{Frankfurt:2011cs}. The
main ingredients are the factorisation theorem for hard diffraction  and the theory of inelastic shadowing by Gribov. The evolution  is done using DGLAP equations. The experimental input to fix the parameters of the model is given by inclusive diffractive parton distribution functions of nucleons as measured at HERA.
For the gluon distribution in the proton the LO distribution from \cite{Martin:2007sb} was used.

\subsubsection{Models based on the colour dipole approach}

\begin{figure}[t]
\begin{center}$
\begin{array}{c}
\includegraphics[width=0.45\textwidth]{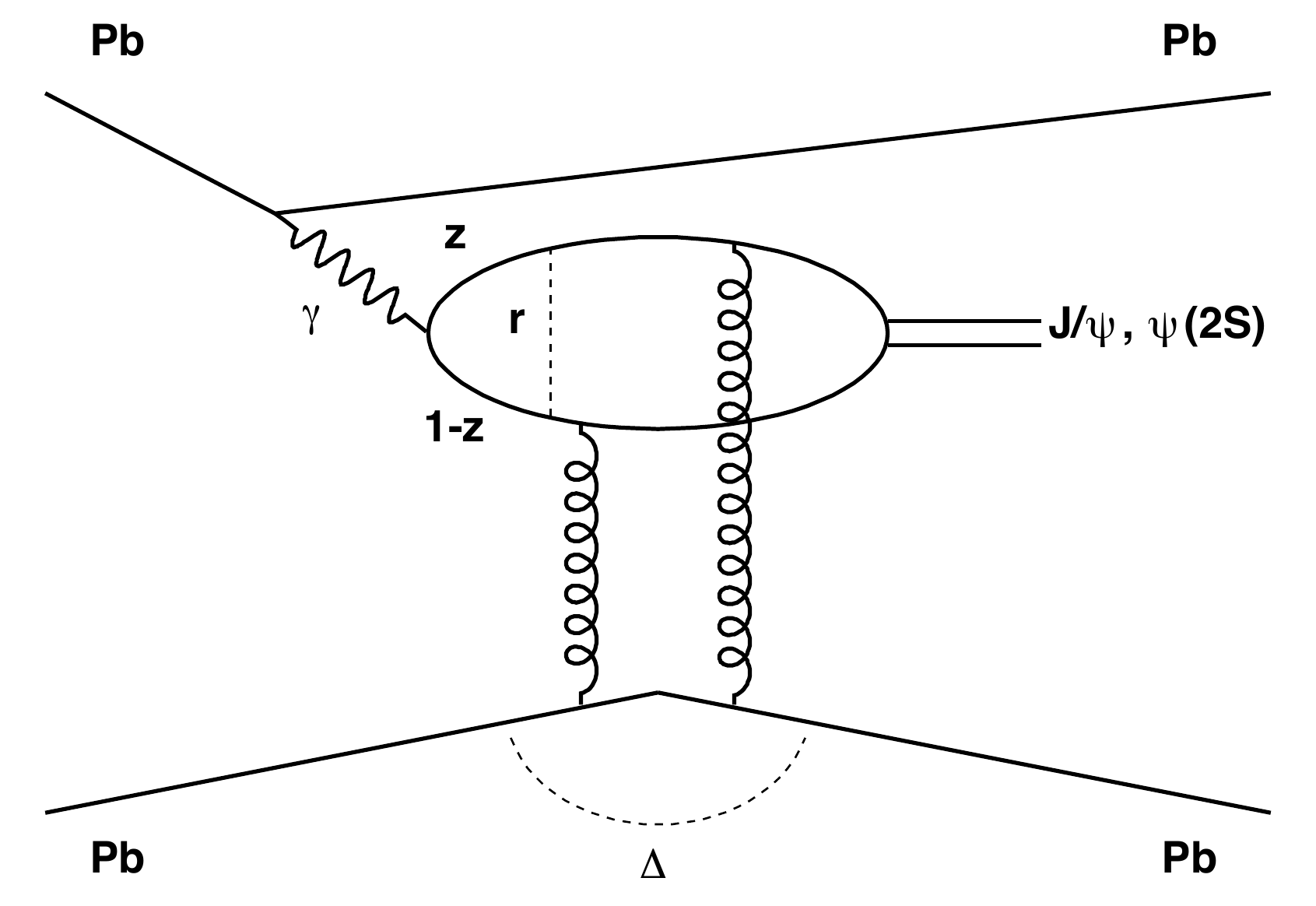} 
\end{array} $
\end{center}
   \caption{Lowest order diagram for the photoproduction of charmonium within the colour dipole model.}
\label{UPCfig:CDM}
\end{figure}	

The basic idea of this formalism is illustrated in \fig{UPCfig:CDM}: long before the interaction, the photon splits into a  quark-antiquark pair, which forms a colour dipole. Then, this dipole interacts with the target and after another long time the dipole creates a vector meson. The cross section in this formalism is given by
\begin{equation}
\frac{\dd\sigma(\gamma{\rm+Pb}\to\jpsi + {\rm Pb}) }{\dd t} =
\frac{R^2_g(1+\beta^2)}{16\pi}\left|
A(x,Q^2,\vec{\Delta})\right|^2,
\end{equation}
where the so called skewdness correction $R^2_g$ compensate for the fact that only one value of $x$ is used, even though the two gluons participating in the interaction have different $x$ \cite{Shuvaev:1999ce}, while $(1+\beta^2)$ is the correction that takes into account the contribution from the real part of the amplitude \cite{Gribov:1968ie}. The imaginary part of the scattering amplitude is given by
\begin{equation}
A(x,Q^2,\vec{\Delta})= \int \dd z \dd^2\vec{r}\dd^2\vec{b} e^{-i(\vec{b}-(1-z)\vec{r})\cdot\vec{\Delta}}
\left[\Psi^*_{\jpsi}\Psi\right] 2\left[
1-\exp\left\{
-\frac{1}{2}\sigma_{\rm dip}T_{\rm Pb}(b)
\right\}
\right],
\end{equation}

where the integration variable $\vec{r}$ represents the distance between the quark and the antiquark in the plane transverse to the collision, $z$ quantifies the fraction of the photon momentum carried by the quark and $b$ is the distance between the centres of the  target and the dipole; $\vec{\Delta}$ is the transverse momentum transferred to the nucleus; the virtuality of the incoming photon is denoted by $Q^2$ and for the case of photoproduction discussed here is zero; $\Psi$ describes the splitting of the photon into the dipole and $\Psi_{\jpsi}$ is the wave function of the $\jpsi$;  the term $i(1-z)\vec{r}\cdot\vec{\Delta}$ in the exponential is a third correction to take into account non-forward contributions to the wave function $\Psi_{\jpsi}$, which is modelled for the forward case \cite{Bartels:2003yj}; and finally $\sigma_{\rm dip}$ is the universal cross section for the interaction of a colour dipole with a nuclear target.  
The models differ in the functional form of $\Psi_{\jpsi}$, in the corrections that were considered and in the formulation of the universal dipole cross section. 

In the case of LM  the non-forward correction to the wave function was not considered. LM use two different prescriptions for the wave function:  the Gauss-LC \cite{Kowalski:2003hm} and the boosted Gaussian \cite{Nemchik:1994fp,Nemchik:1996cw}. $\sigma_{\rm dip}$ is written in terms of the cross section of a dipole and a proton, $\sigma^p_{\rm dip}$; assuming a Gaussian profile in impact parameter for the proton, $\exp(-b^2/(2B_p))$: 
\begin{equation}
\frac{1}{2}\sigma_{\rm dip} = 2\pi B_p A N(r,x),
\end{equation}
where $N(r,x)$ is the dipole target amplitude. LM use two different models for $N(r,x)$: the IIM model \cite{Iancu:2003ge} which is a parametrisation of the expected behaviour of the solution to the BK equation \cite{Balitsky:1995ub,Kovchegov:1999yj,Kovchegov:1999ua} which includes a non-linear term for the evolution of  $N(r,x)$; and the IPsat model \cite{Kowalski:2003hm,Kowalski:2006hc} which uses DGLAP equations to evolve an eikonalised gluon distribution.

The GM-GDGM model uses the boosted Gaussian prescription for the wave function. The dipole cross section is given by $\sigma_{\rm dip} = R^A_g(x,Q^2)\sigma^{\rm p}_{\rm dip}$, where $\sigma^{\rm p}_{\rm dip}$ is given according to the IIM model and the leading twist approximation is used for $R^A_g(x,Q^2)$.
The CSS model is similar to the GM model, but uses the unintegrated gluon distribution of the nucleus including multiple scattering corrections. It also takes into account higher order fluctuations of the incoming photon.  The Gaussian form of the wave function is used.

\subsection{Photonuclear production of charmonium: comparing models to measurements}
\label{UPCsec:comp}

\begin{figure}[t]
\begin{center}$
\begin{array}{c}
\includegraphics[width=0.47\textwidth]{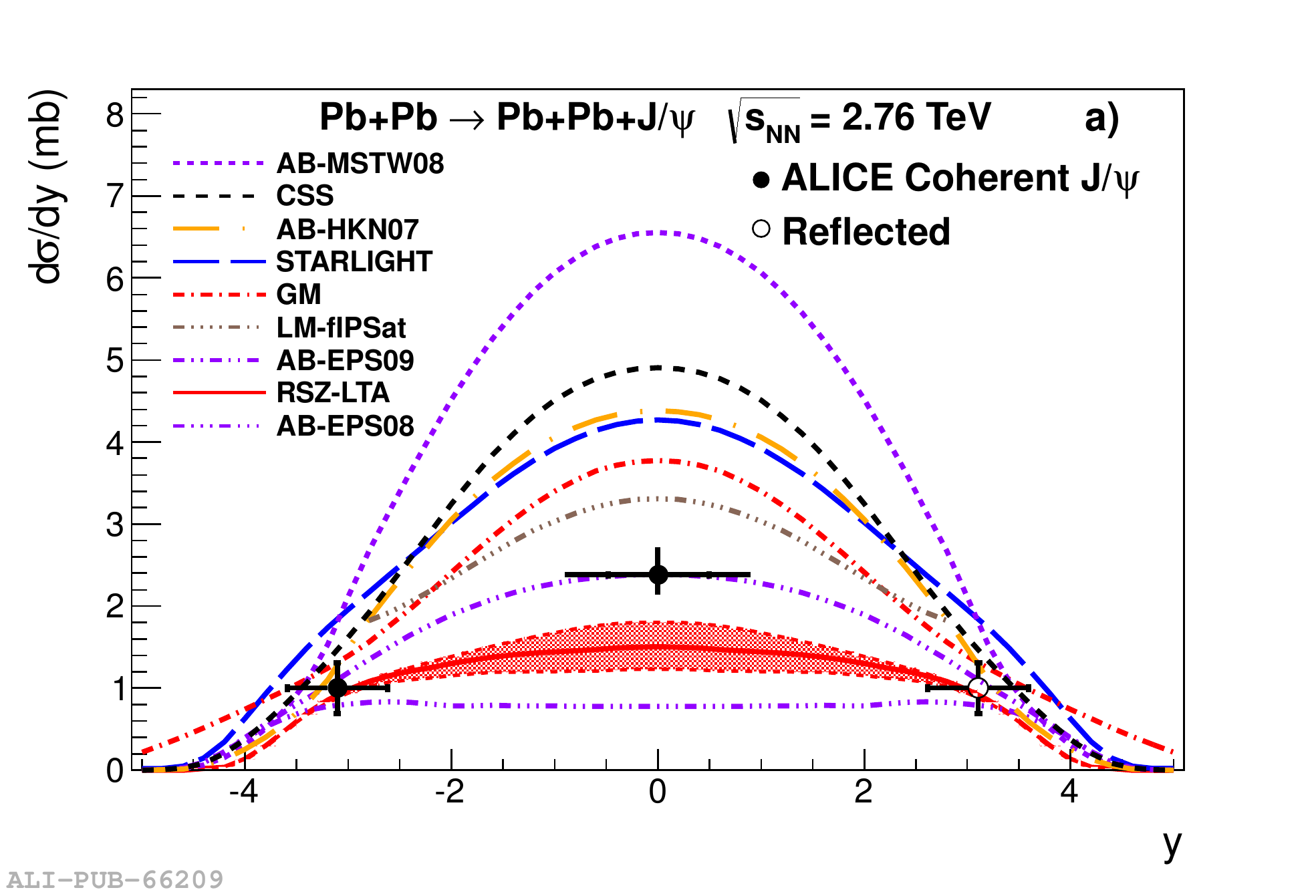} 
\includegraphics[width=0.47\textwidth]{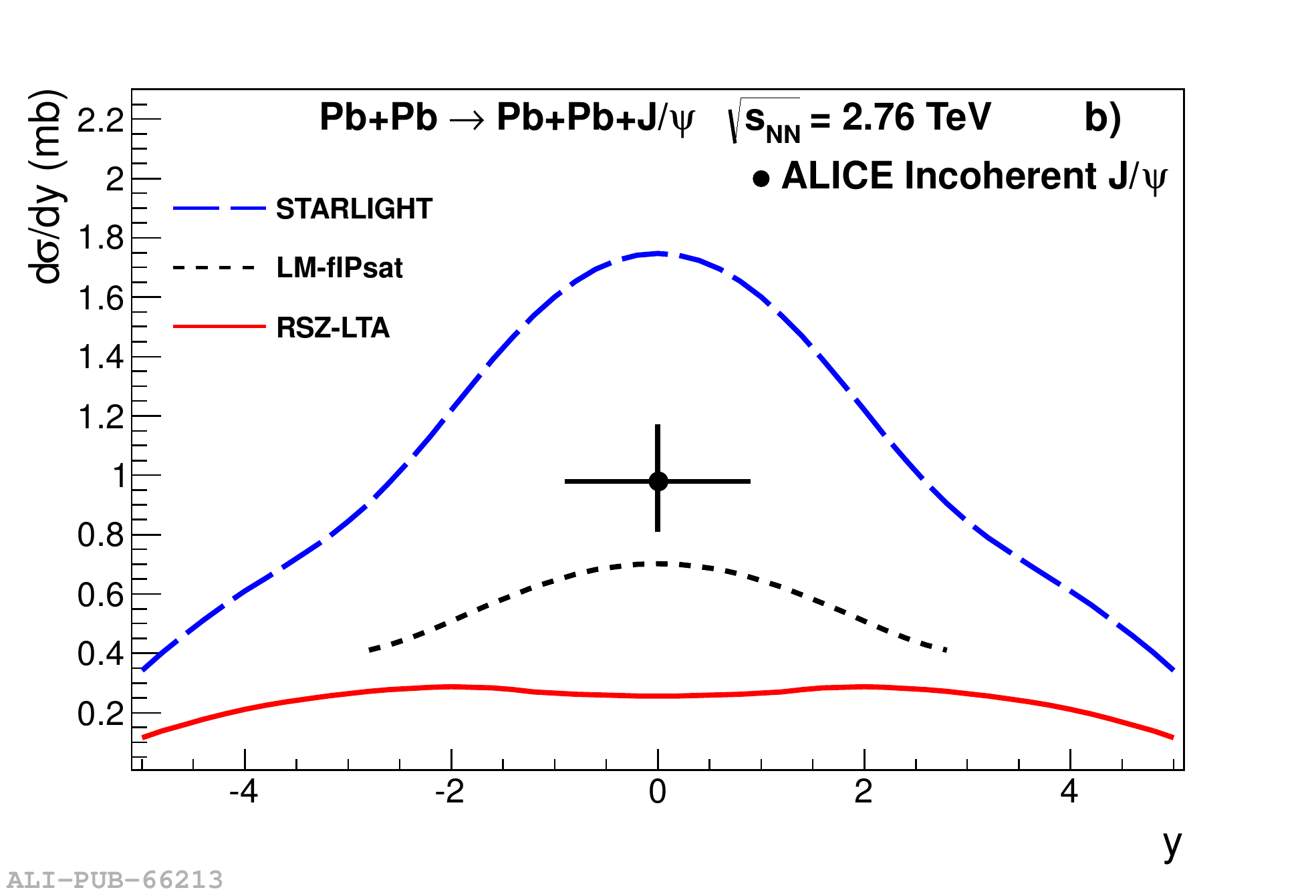} 
\end{array} $
\end{center}
   \caption{Cross section for coherent (left) and incoherent (right) photonuclear production of $\jpsi$ as measured by ALICE~\cite{Abbas:2013oua} compared to theoretical predictions.}
\label{UPCfig:ALICEresults}
\end{figure}	

The comparison of the  cross sections measured with ALICE and the predictions of the different models is shown in \fig{UPCfig:ALICEresults}. Several comments are in order. The spread among the different predictions is quite large, so data impose strong constraints into the different ingredients of the models, in particular  at mid-rapidity, which corresponds to the smallest $x$--Bjorken in a pQCD interpretation. The left panel of the figure shows the measurement for \jpsi coherent production. 
The AB-MSTW08 curve, corresponding to the absence of nuclear effect in that model, and the AB-EPS08 model, corresponding to strong shadowing, are both  disfavoured. 
 The other models are closer to the data and there are several natural refinements that can be done. For example, after the publication of ALICE data the model labelled RSZ-LTA was revisited and a study of the variations of the model with different scales for the coupling constant was performed \cite{Guzey:2013qza}. It was found that using data, one could constrain within this model the value of this scale such that data are correctly described. Similar improvements could be made to other models.
The preliminary data from CMS (not shown in the Figure) is also well described by this updated model and by AB-EPS09, which corresponds to mild shadowing.

The situation with respect to the measurement of incoherent photonuclear production of $\jpsi$ is not that clear. There are less models and less data. The LM--fIPsat model slightly underestimates the data, while the prediction of the same model is above the data for the coherent case.  The modification of the RSZ-LTA model from  \cite{Guzey:2013qza} has not such a large effect in the incoherent case, so that this model is still below the measurement.

The prediction of the cross section for coherent photonuclear production of $\psiP$ and its comparison to data is even more difficult. As mentioned above, all models fixed the predictions such that they reproduce data from HERA on photoproduction off protons. But HERA data for $\psiP$ is less abundant and less precise than for the case of $\jpsi$. One of the consequences is that the STARLIGHT and AB models when all nuclear effects are switched off, have the same prediction for $\jpsi$ but a different prediction for $\psiP$. 
It also happens that the wave function of the $\psiP$ is more complex than that of the $\jpsi$~\cite{Nemchik:2000de} and not all models, see for example~\cite{Adeluyi:2013tuu}, take into account the full complexity of the wave function. 
Finally, the preliminary results from ALICE for the coherent production of $\psiP$ appeared after the publication of the $\jpsi$ measurements, so that some models were already updated to improve the description of ALICE data. Taking into account these caveats  the general conclusion seems to be that models with strong shadowing or without any nuclear effects are disfavoured, while models incorporating a mild form of shadowing are, within the current large experimental uncertainties, relatively close to data.

A somehow surprising result was the ratio of the coherent cross section of $\psiP$ to that of $\jpsi$  at mid-rapidity. The measured value is around two times larger than what has been measured in the photoproduction off protons, while most models expected these ratios to be similar~\cite{Broz:2014wka}. Note  that the experimental uncertainties are still quite large, so this discrepancy is only a bit larger than two sigmas. All models, except AB, are quite close to the ratio measured at HERA, and thus far from that measured by ALICE \cite{Broz:2014wka}. The AB-EPS09 model is closer to the measured ratio, but one has to say that the wave function of the $\psiP$ was not included in its full complexity, and it is not clear what would happen if it were included. In this case, new data and improved models are much needed.

Another area to watch is the measurement of coherent production of charmonium in peripheral or even semi-central \PbPb collisions. As mentioned above, this possibility was discussed some time ago \cite{Joakim2010}, but no full calculation of the process has been performed. 
The possibility of a measurement at different centralities, which will be possible during \RunTwo, will allow  to test different model implementations. For example, if the coherent source were the spectator nucleons, then the distribution of transverse momentum would depend on centrality. 

\subsection{Summary and outlook}
\label{UPCsec:outlook}

The LHC  Collaborations have demonstrated with data from \RunOne that measurements of photonuclear production of charmonium are possible and that these measurements provide valuable information to constrain models and can contribute to a better  understanding of shadowing. 
Present data favour the
existence of a shadowing as predicted by EPS09 in the x range $10^{-3}$--$10^{-2}$.
Furthermore, these data have given a glimpse of two remarkable results: ($i$) the ratio of $\psiP$ to $\jpsi$ cross section for coherent photoproduction seems to be  sensitive to nuclear effects; and ($ii$) it seems to be possible to measure coherent production of charmonium overlapped with hadronic collisions. During \RunTwo, new data with large statistics will be collected and a definitive answer to these two questions may be given. The new data will also allow to explore the dependence of the $\jpsi$ coherent cross section on transverse momentum with great detail and  potentially the measurement of  $\ups$ production.
The  already existing measurements and the forthcoming ones represent  important  milestones in the path going from HERA towards a future dedicated electron-ion facility \cite{Deshpande:2005wd,Dainton:2006wd}.

\newpage
\section{Upgrade programmes and planned experiments}
\label{sec:upgrade}

\subsection{Introduction} 

As seen in the previous \sects, and also summarised in the next one,
alongside the great progress in understanding the physics of heavy quarks
in proton--proton and heavy-ion collisions, a lot of questions emerged too.
Those, as well as the quest for a quantitative description of the hot deconfined 
quark-gluon matter, call for upgrades in existing experiments and also for new ones, 
in which the potential of heavy quarks in answering those questions is fully exploited.
Below, we discuss the ongoing efforts and the possibilities for new experiments.

\subsection{Collider experiments}
\subsubsection{The LHC upgrade programme}

The  LHC roadmap foresees three long shut-downs (LS) of the machine in order to perform major upgrades.
The objective of \LSOne, recently completed, was the preparation of the machine for 6.5--7\TeV operation in 2015 \cite{Bordry:2013} reaching (close to) the design energy and the nominal peak luminosity of $\lumi_{\pp}=10^{34}~{\rm cm}^{-2}{\rm s}^{-1}$ or even higher. 
For heavy ions, interaction rates of 10-15 kHz are expected in \pb in \RunTwo.
The goal of the heavy-ion programme for \RunTwo is to collect about a factor
10 more statistics for \pb collisions with respect to \RunOne, while the new energy of
\snn $\simeq$ 5\TeV will push the frontier of high-energy-density quark-gluon matter.

\LSTwo, scheduled for 2018--2019, will be mainly devoted to a major upgrade of the injectors as well as interventions performed on the LHC itself aiming at increasing its instantaneous luminosity in \pp and heavy-ion running modes to  $\lumi_{\pp}=3\cdot 10^{34}~{\rm cm}^{-2}{\rm s}^{-1}$  and  $\lumi_{\rm PbPb}=6\cdot10^{27}~{\rm cm}^{-2}{\rm s}^{-1}$, respectively.
After \LSTwo, the interaction rate in \pb collisions is foreseen to be 50~kHz.
After \LSThree, envisaged in 2023--2024, the LHC peak luminosity is expected to reach  $\lumi_{\pp}=$5--7$\cdot10^{34}~{\rm cm}^{-2}{\rm s}^{-1}$ levelled down from higher luminosities~\cite{Rossi:2012} (HL-LHC).
The LHC heavy-ion programme is currently planned to extend to Run~4 (2026--2028).

The four major LHC experiments, ALICE, ATLAS, CMS and LHCb, have rich detector upgrade programmes to fully exploit  the accelerator upgrades. Three different phases, corresponding to the three LHC long shut-downs towards the HL-LHC, are planned.
Heavy-flavour and quarkonium physics, both in \pp and \pb collisions, either drive or strongly benefit from the upgrade programmes.

ALICE is the dedicated heavy-ion experiment, whose  strengths, unique at the LHC, are measurements at low \pt and involving a wide range of identified hadrons, and access to forward rapidities ($y\sim 4$) in muon decay channels. This implies primarily minimum-bias (or centrality-selected) collisions, as trigger selectivity for low-\pt observables is obviously weak.

The ATLAS and CMS experiments are general-purpose experiments designed primarily for the investigations of \pp collisions, the Higgs boson discovery \cite{Chatrchyan201230,Aad20121} being the major achievement of \RunOne for this physics programme. 
Both detectors have demonstrated very good performance in heavy-ion collisions too, including measurements devoted to heavy-quarks, as seen in the previous \sects.
The upgrade programmes of both ATLAS and CMS detectors will extend the studies of Higgs and of physics beyond the standard model with improved detector performances to match the LHC luminosity increases. This will benefit the ATLAS and CMS heavy-ion physics programme as well which is focused on higher \pt, complementary to the ALICE programme.

The LHCb experiment is the dedicated experiment for the studies of $b$-quark physics in \pp collisions. The measurements performed by LHCb, for both charm and beauty hadrons, in \pp and \pPb collisions are of a large variety and unique quality. The upgrades of the LHCb detector retain the focus on heavy-flavour studies, which will be performed in \pp and \pPb collisions, as well as in a fixed-target configuration.

\paragraph{The ALICE experiment}
The ALICE Collaboration consolidated and completed the installation of current detectors during \LSOne with the aim to accumulate 1~\nbinv of \pb collisions during \RunTwo corresponding to about 10 times the \RunOne integrated luminosity.
In parallel, the ALICE Collaboration pursues a major effort to upgrade the apparatus, in particular to improve the tracking precision and to enable the read-out of all interactions at 50~kHz, with the goal to accumulate 10~\nbinv of \pb collisions after \LSTwo. A low-B field (0.2 T) run to collect 3~\nbinv is also envisaged. 
The implementation of this upgrade programme~\cite{Abelevetal:2014cna,ALICE_MFT_LoI}, foreseen in \LSTwo, includes: 
a new low-material Inner Tracking System \cite{ALICE_ITS_TDR} with a forward rapidity extension (MFT \cite{ALICE_MFT_TDR}) to add vertexing capabilities to the current Muon Spectrometer; 
the replacement of the Time Projection Chamber (TPC) wire chambers with gas electron multiplier (GEM) readout;
a new readout electronics for most of the detectors and an updated trigger system;  a new set of forward trigger detectors~\cite{ALICEUpgrade_TDR} and a new integrated online--offline system.

The new Inner Tracking System~\cite{ALICE_ITS_TDR}, covering mid-rapidity ($|\eta|<1.3$), 
consists of seven concentric layers.
For the forward region ($2.5<\eta<3.6$) muon tracker (MFT), 5 detection planes are envisaged.
Both systems are composed of CMOS Monolithic Active Pixel Sensors with a pixel cell size of about $20\times 30$~$\mu$m. 
At mid-rapidity, the total material budget per layer is 0.3\% and 0.8\% of $X_0$ for the three inner and four outer layers, respectively.
It is 0.6\% of $X_0$ per detection plane at forward rapidity~\cite{ALICE_MFT_TDR}.
These low material budget, high granularity detectors, in conjunction with the reduction of the beam pipe diameter in the centre of the ALICE detector from the present value of 58~mm to 36~mm, leads to a significantly improved measurement of the track impact parameter (distance of closest approach to the primary vertex). 
It reaches 40~$\mu$m at \pt~$\simeq$ 0.5~\GeVc at mid-rapidity and 90~$\mu$m at \pt~$\simeq$ 1~\GeVc at forward rapidity.

Thanks to this improved performance and the high statistics, a large number of new measurements become possible and the existing measurements will be repeated with improved performance.
In the charm sector, at mid-rapidity the \pt coverage will be extended towards zero \pt  and the uncertainties will be significantly reduced for the D and {\ensuremath{\mathrm{D}_{s}} meson nuclear modification factor and \vtwo. 
The \ensuremath{\Lambda_c} baryon reconstruction in its three-prong decay (p, K and $\pi$) will become possible down to \pt~=~2~\GeVc, allowing the measurement of baryon/meson ratio (\ensuremath{\Lambda_c}/D) crucial for the study of thermalisation and hadronisation of charm quarks in the medium.
Two-particle correlation studies with a D meson as ``trigger" particles (see \sect{OHF}) as well as a measurement of D-jet fragmentation function will be performed.
At forward rapidity the separation of open charm and open beauty production cross sections can be performed via the semi-muonic and \jpsi decay channels.

\begin{figure}[t]
\begin{minipage}{8cm}
\begin{center}
\includegraphics[width=0.99\textwidth]{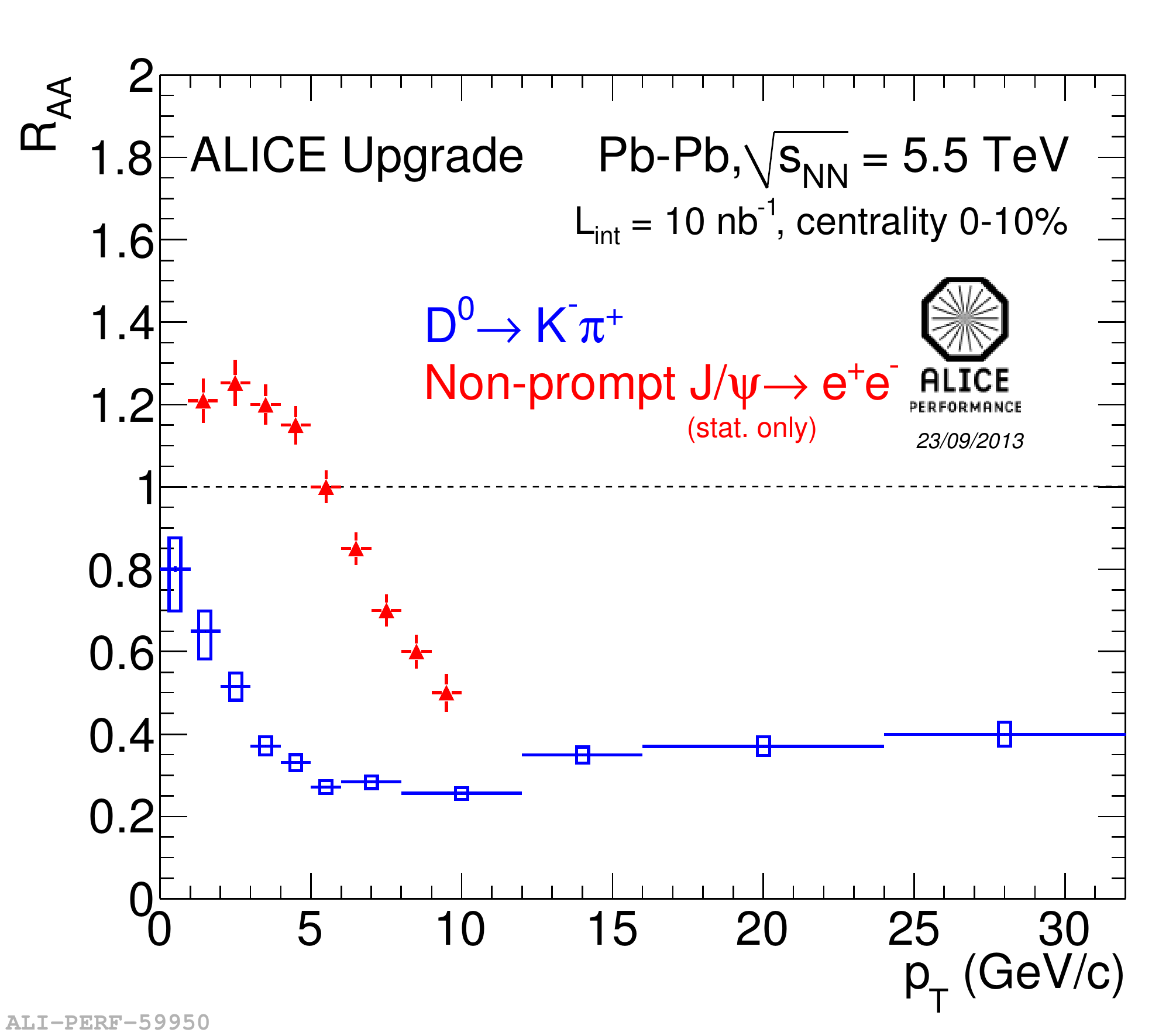}
\end{center}
\end{minipage}
\hfill
\begin{minipage}{8cm}
\begin{center}
\includegraphics[width=0.90\textwidth]{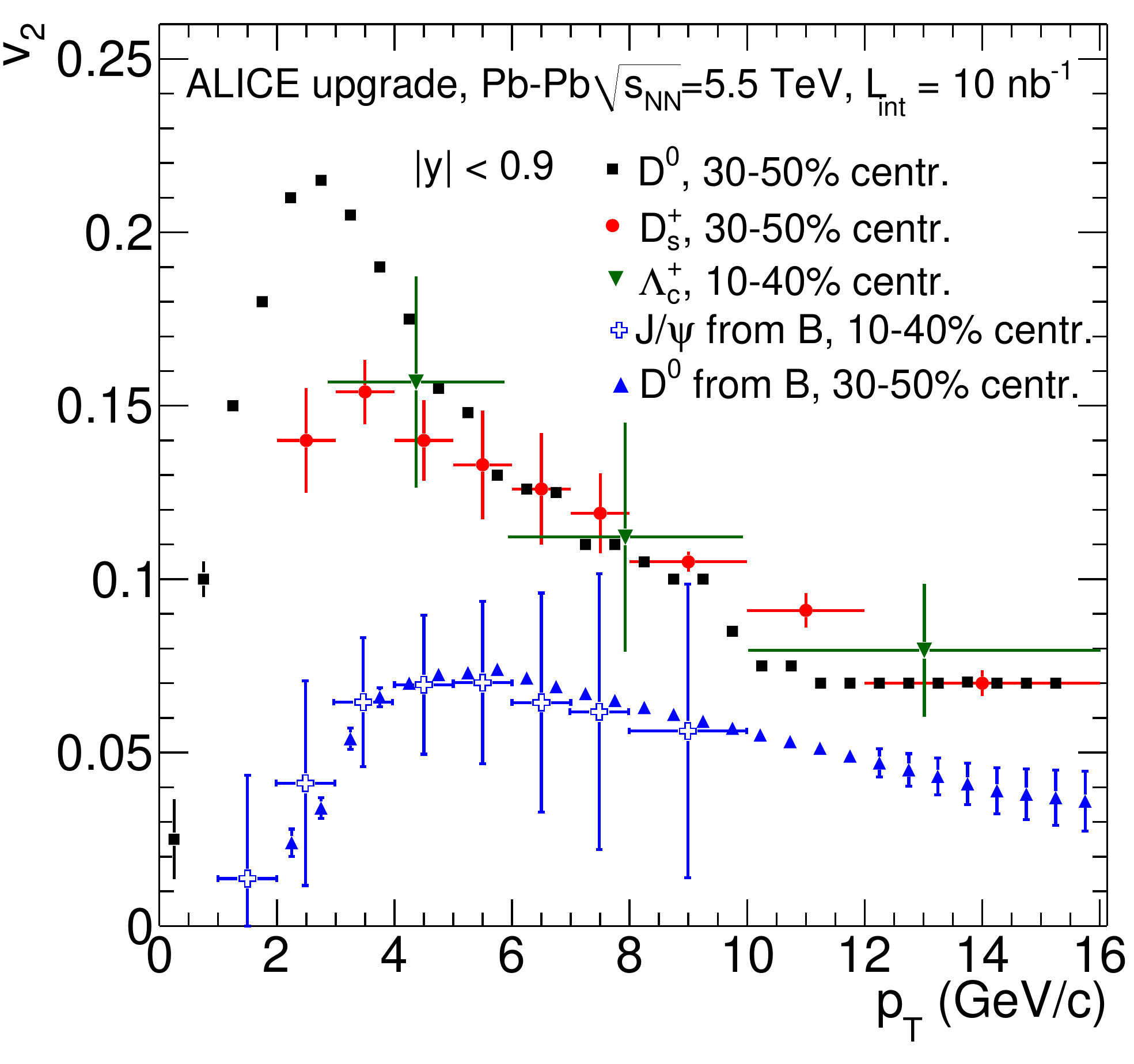}
\end{center}
\end{minipage}
\caption{Estimated performance of open heavy flavour nuclear modification factor
(left) and elliptic flow (right) with the ALICE upgrade.
The \pt dependence of the two observables is assumed based on current measurements and model predictions~\protect\cite{ALICE_ITS_TDR}.}
\label{fig:ALICE_Beauty_v2}
\end{figure}

The ALICE detector upgrade opens up the possibility to fully reconstruct B$^+$ meson (B$^+ \rightarrow \overline{\mathrm{D}^0}\pi^+$) down to \pt~=~2~\GeVc and the $\Lambda_b$ baryon ($\Lambda_b\rightarrow \Lambda_c^+\pi^-$) down to \pt~=~7~\GeVc.
The thermalisation of $b$-quarks in the medium will be studied via the measurement of the elliptic flow in the semi-leptonic as well as \jpsi or D meson decay channels, both at mid- and forward rapidities.
\fig{fig:ALICE_Beauty_v2} gives an example of the expected performance of open heavy flavour nuclear modification factor and elliptic flow measurements~\cite{ALICE_ITS_TDR}.

\begin{figure}[h]
\begin{minipage}{8cm}
\begin{center}
\includegraphics[width=\textwidth]{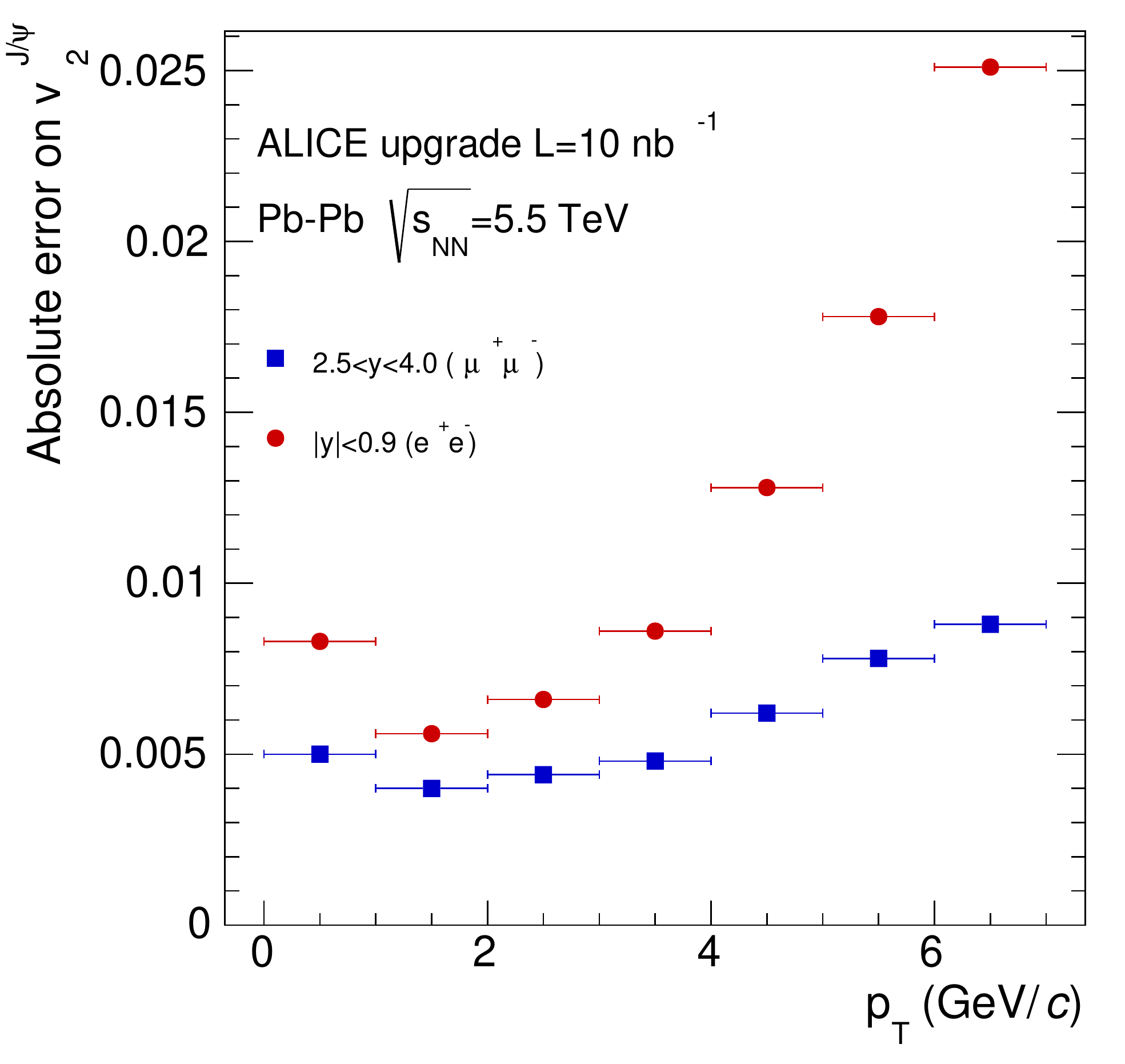}
\end{center}
\end{minipage}
\hfill
\begin{minipage}{8cm}
\begin{center}
\includegraphics[width=\textwidth]{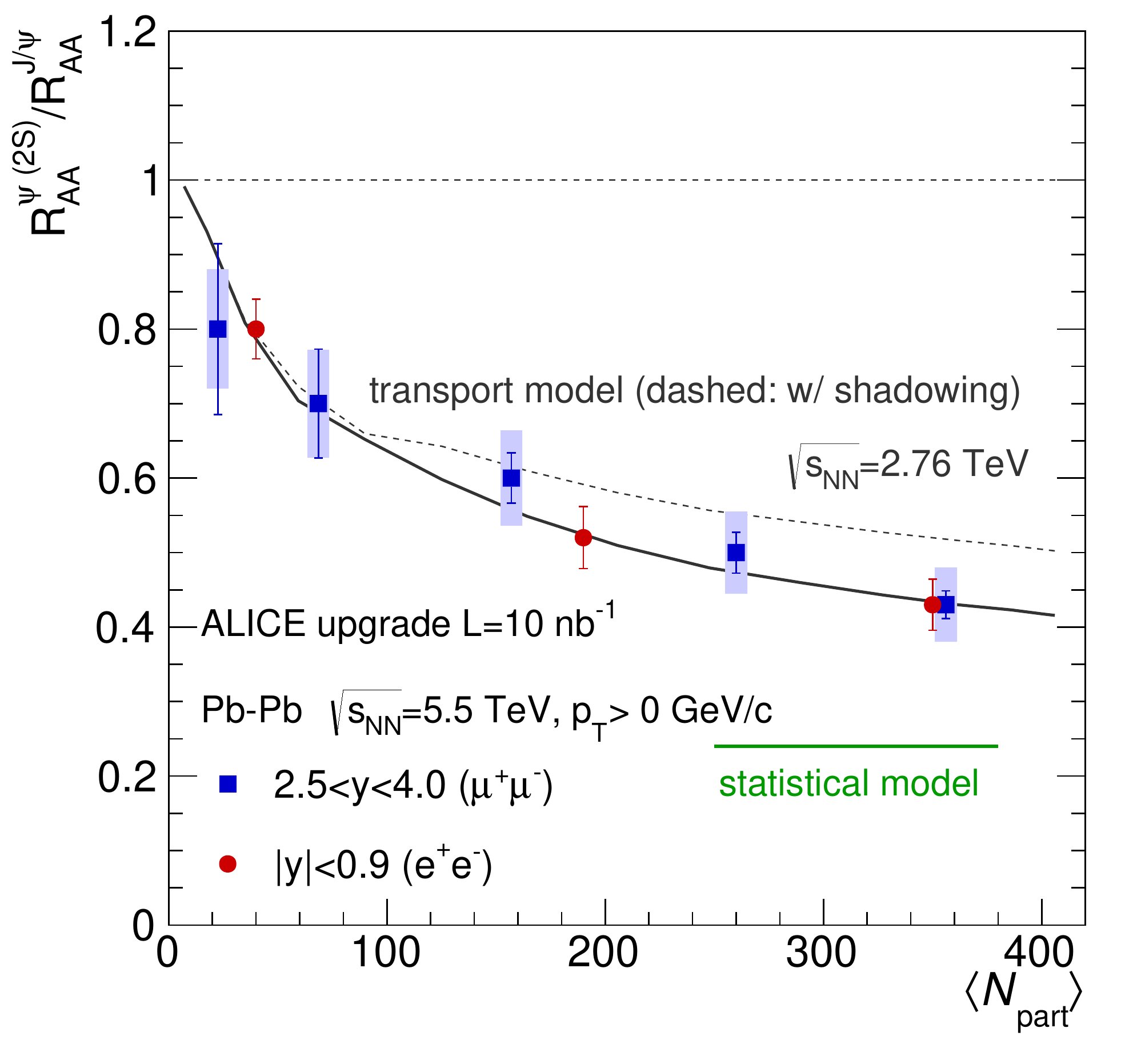}
\end{center}
\end{minipage}
\caption{Left: estimated statistical uncertainties of \vtwo measurement of \jpsi with the ALICE upgrade~\cite{Abelevetal:2014cna}. Right: estimated performance of the measurement of ratio of the nuclear modification factor of \psiP and \jpsi~\protect\cite{ALICE_MFT_LoI} compared to two charmonium production model calculations~\protect\cite{Andronic:2009sv,Zhao:2011cv}.}
\label{fig:ALICE_Jpsi_v2}
\end{figure}

The ALICE detector upgrade will lead to a significant improvement of the (prompt) \jpsi measurement, both at mid- and  forward rapidities. The expected performance of the \vtwo measurement is illustrated in \fig{fig:ALICE_Jpsi_v2} (left). At forward rapidity, the measurement of \jpsi polarization in \pb collisions will become possible, as well as precision measurements of the production of \ups states.

The measurement of the \psiP meson, combined with the \jpsi measurement, offers an important tool to discriminate between different charmonium production models.
The \psiP measurement is a challenge in heavy-ion experiments, in particular at low \pt. 
At forward rapidity, the addition of the MFT will allow a precise measurement of \psiP down to zero \pt even in the most central \pb collisions.
At mid-rapidity the measurement remains very challenging, but a significant result is expected with the full statistics of the ALICE data after the upgrade (\fig{fig:ALICE_Jpsi_v2} right).

Comparable detection performances at mid and forward rapidities will place the ALICE experiment in the position of studying the heavy flavour QGP probes as a function of rapidity.
This will help imposing tighter experimental constraints to theoretical models.

The physics programme for ultra peripheral \AAcoll collisions (UPC, see \sect{sec:UPC}) in ALICE for \RunThree and Run~4 will
have the advantage of increased luminosity by a factor of 10 (100) with respect to \RunTwo (\RunOne).
The large increase in statistics could as well allow the detail study of processes with small cross section
like the coherent production of \ups or the production of \etac in $\gamma\gamma$ collisions.

\paragraph{The ATLAS experiment}
During \LSOne, ATLAS achieved the installation of the Insertable B-Layer (IBL)~\cite{ATLAS:TDR:19}, which is an additional fourth pixel layer, placed closer to the beam pipe at an average radius of 33~mm.
It will add redundancy to the inner tracking system, 
leading to improved tracking robustness.
This new layer provides improved pointing resolution of the inner tracker for \pt as low as 1~\GeVc and a pseudo-proper decay length resolution improved by about 30\% compared to \RunOne, leading to an improved $b$-quark tagging performance~\cite{ATLAS:Btagging}.
The inner tracker will be completely replaced during the \LSThree~\cite{ATLAS:LoI:Phase2} 
in order to cope with the high-luminosity after \LSThree of the LHC. 
This new tracker, composed of silicon pixel and strip layers, will have capabilities equivalent to the current tracker (with the IBL).

 During \LSTwo, ATLAS envisages the installation of new Muon Small Wheels and more selective (``topological") Level-1 trigger criteria ~\cite{ATLAS:LoI:Phase1,ATLAS:TDR:23}, which carry the potential to improve the dimuon acceptance at low \pt. 
The cavern background leads to fake triggers in the forward muon spectrometer, with adverse impact on its physics capability.
In order to be able to handle the high luminosity, it is proposed to replace the first end-cap station by the New Small Wheel (NSW) covering the rapidity range $1.2<|\eta|<2.4$.
The NSW will be integrated into the Level-1 trigger, improving the background rejection.
Two technologies, MicroMegas detectors and small-strip Thin Gap chambers, will be used in order to have both a good position resolution ($<100~\mu$m) and a fast trigger function. 
A longer upgrade plan of the muon end-caps, consisting in the extension of the muon reconstruction coverage to higher rapidities, is under study. 
One of the possible scenarios consists in the addition of a  warm toroid at small angles combined with new muon chambers at high rapidity.
The extension would cover the rapidity range $2.5<|y|<4.0$ and should have a \pt resolution of the order of 15--40\% for \pt ranging from 10 to 100~\GeVc.

The high luminosity of \RunTwo and \RunThree imposes to rise the muon \pt trigger threshold. In order to partially limit this increase, a new Level-1 topological trigger algorithm has been implemented. 
A  \pt threshold of the order of 15~\GeVc at the \jpsi mass which will increase to 30~\GeVc for higher luminosities is foreseen.
The ATLAS quarkonia and heavy flavour programmes will therefore concentrate on the high-\pt range (\pt $\gtrsim$ 30~\GeVc).

\paragraph{The CMS experiment}
Two phases compose the  CMS experiment upgrade programme.
The first one, spread over \LSOne and \LSTwo, involves consolidation of the current detectors.
The upgrade activities in \LSOne and \LSTwo are focused on the inner pixel detector, the hadron calorimeter, the forward muon systems, and the Level-1 trigger~\cite{CMS_L1_Trigger_TDR, CMS_Pixel_TDR, CMS_HCAL_TDR}.
The entire pixel detector will be replaced during the 2016 -- 2017 yearly shut-down.
The new device adds a fourth detection layer for redundancy in tracking and leads to an improved fake-track rejection.
The pointing resolution will be improved thanks to the reduced material budget and by moving the first detection layer closer to the interaction point, which will substantially improve the $b$-quark tagging capability.
The upgrade in the hadron calorimeter Level-1 trigger is motivated by the heavy-ion programme; the significantly-improved selectivity for high-\pt jets opens up 
precision measurements of $b$-tagged jets in Pb--Pb collisions. An illustration of the expected performance is given in \fig{fig:CMS_bJets} for the doubly-tagged $b$-jet asymmetry parameter $A_J$~\cite{CMS:2013gga}.

\begin{figure}[t]
\begin{center}
\includegraphics[width=0.5\textwidth]{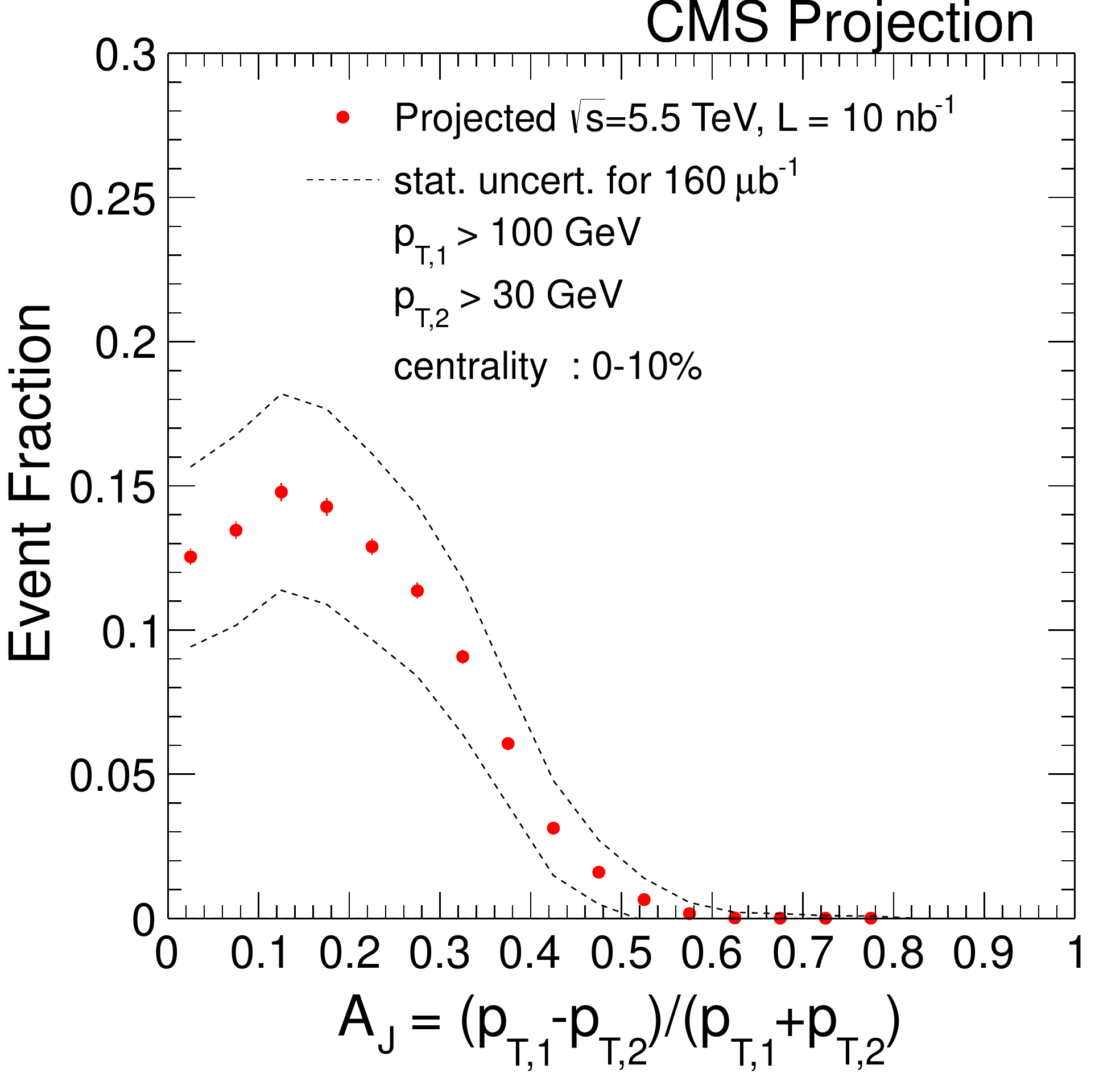}
\end{center}
\caption{Estimated performance of the measurement in CMS of the doubly-tagged $b$-jet asymmetry $A_J$ distribution in $|\eta| < 2$~\cite{CMS:2013gga}.}
\label{fig:CMS_bJets}
\end{figure}

The muon system will be completed during \LSTwo by adding a fourth end-cap layer for $1.2<|\eta |<1.8$ and by improving the read-out granularity at mid-rapidity. 
During \LSOne, Level-1 hardware trigger algorithms was upgraded, resulting in an improved muon trigger selectivity~\cite{CMS:2013gga}. 
The impact on the HI physics programme will be a better non-prompt \jpsi extraction and an improved dimuon mass resolution. 

The second phase of the CMS upgrades~\cite{CMS:Phase2} planned to be completed during \LSThree, leads to the readiness of the experiment for physics at the HL-LHC.
This includes the extension of the inner tracking system to $|\eta|<4$ with triggering capability and additional muon redundancy with a possible extension of the muon system to cover $|\eta|<4$. 
It will result in an improved mass resolution (gain of 1.2--1.5) with an improved quarkonia triggering performance. 
High-statistics measurements of prompt and non-prompt \jpsi, \psiP
and \ups states will become available~\cite{CMS:2013gga}.

With regard to UPC (see \sect{sec:UPC}), thanks to the increase of statistics
expected in \RunThree, and pending detail performance studies, CMS would be able to carry out systematic studies for the various \ups states, as well as detailed studies on UPC dijets produced in photon-nucleus collisions.

\paragraph{The LHCb experiment}
The upgrade programme of the LHCb Collaboration \cite{Bediaga:1443882} has two major facets:
i) the replacement of all front-end electronics\footnote{The replacement of front-end electronics implies, for some of the LHCb detectors, like the silicon trackers, the replacement of the active elements.}, which will enable continuous detector readout at 40~MHz, followed by a full software trigger \cite{CERN-LHCC-2014-016};
ii) detector upgrades designed for operation at a luminosity increased by a factor of 5 compared to current conditions (levelled nominal luminosity will be 2$\cdot10^{33}$~cm$^{-2}{\rm s}^{-1}$ in pp collisions). 
This comprises the replacement of the VELO silicon vertex detector \cite{Collaboration:1624070}, 
new tracker systems before and after the dipole magnet \cite{Collaboration:1647400},
and major upgrades for the systems performing particle identification: the RICH, the calorimeter and the muon system \cite{Collaboration:1624074}.

The upgraded LHCb detector components will be installed during \LSTwo and is envisaged to collect a pp data sample of at least 50~\ensuremath{\text{~fb}^\text{$-$1}}. This will significantly enhance the unique physics capability of LHCb for heavy-flavour measurements also in \pPb collisions.
The focus is on rare observables in connection to physics beyond the standard model, but the high-precision measurements of production cross sections for quarkonia and open charm hadrons is a direct bonus.

\subsubsection{The RHIC programme}
\noindent 
The current plans~\cite{Akiba:2015jwa} envisage measurement at RHIC up to mid-2020's, followed by eRHIC. 
During the 2014 run, thanks to the full implementation of 3D stochastic cooling, RHIC achieved, in \AuAu collisions at \snn~=~200~\GeV, an average stored luminosity of $\lumi=5\cdot10^{27}$~cm$^{-2}{\rm s}^{-1}$ reaching 25 times the design value. The ongoing measurements focus on heavy flavour probes of QGP exploiting the newly-installed silicon vertex detectors in both PHENIX and STAR experiments. These campaigns will extend up to 2016, when the electron cooling of RHIC is expected to enter in operation. The 2015 run modes are scheduled to be \pp, {p--Au}\xspace and possibly  {p--Al}\xspace collisions at \snn=100\GeV.
The RHIC luminosity upgrade plan is to operate the collider in \AuAu collisions at \snn~=~200~\GeV at an average stored luminosity of $\lumi=10^{28}$~cm$^{-2}{\rm s}^{-1}$~\cite{RHIC_Projections}.
The second phase (2018--2019) of the Beam Energy Scan (BES-II), spanning  \snn~=~7 -20~\GeV, opens up the potential of heavy flavour measurements at lowest to-date collider energies.

\paragraph{The sPHENIX project}
The PHENIX Collaboration has a radical upgrade plan consisting on replacing the existing PHENIX central detectors, which have small acceptance and lack hadronic calorimetry, with a compact calorimeter and 1.5~T superconducting magnetic solenoid~\cite{Adare:2015kwa}. 
Full azimuthal calorimeter coverage, both electromagnetic and hadronic, will be available in $|\eta|<1$. 
\begin{figure}[bt]
\begin{minipage}{8cm}
\begin{center}
\includegraphics[width=0.95\textwidth]{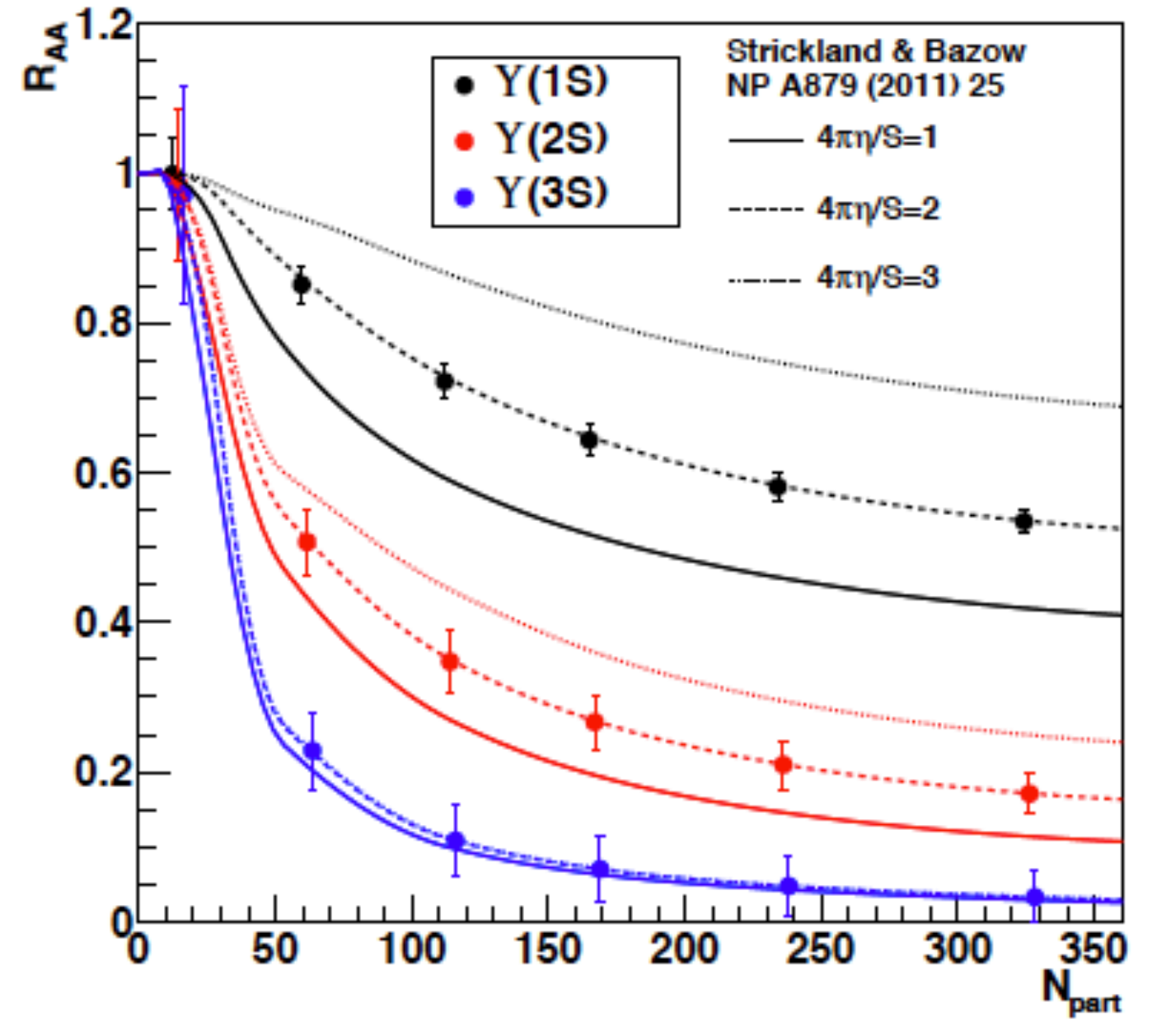}
\end{center}
\end{minipage}
\hfill
\begin{minipage}{8cm}
\begin{center}
\includegraphics[width=0.9\textwidth]{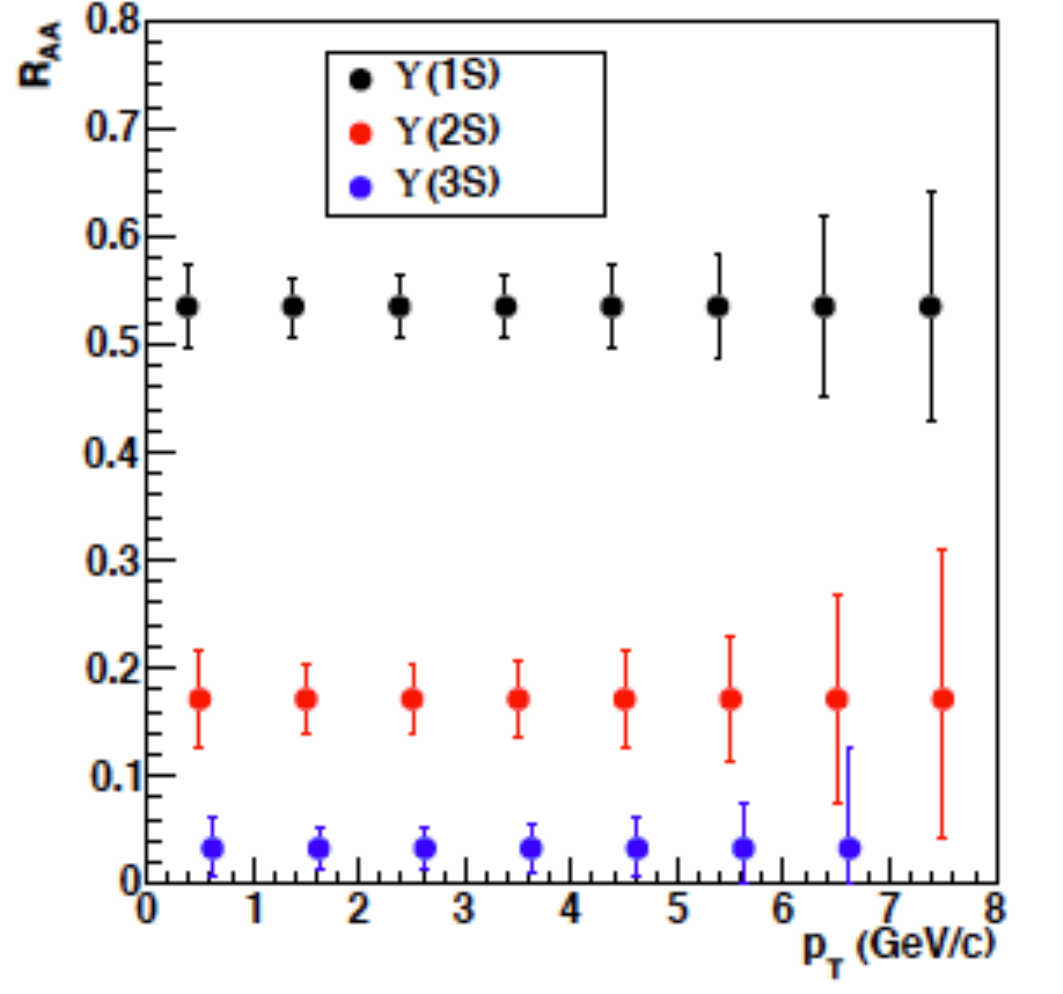}
\end{center}
\end{minipage}
\caption{Estimated performance (quoted are the statistical uncertainties) of the measurement of the \ups  states \raa using sPHENIX in \AuAu collisions at \snn~=~200~\GeV~\protect\cite{Aidala:2012nz}.}
\label{fig:sPHENIX_RAA_UPSILON}
\end{figure}
The sPHENIX detector will be capable of identifying heavy-quark jets and separating \ups states in their dielectron decay channel with a  mass resolution better than 100~{\ensuremath{\mathrm{MeV}/c^2}.
The current PHENIX detectors will be removed during the 2016 shut-down and sPHENIX detectors will be  installed during the 2020 shut-down of RHIC.
The sPHENIX running plan consists of two years of data taking (2021--2022) in \AuAu, \dAu and \pp collisions at \snn~=~200~\GeV.

The inner tracking system will be composed by the currently existing Silicon Vertex Detector (VTX) with three additional silicon layers at larger radii to improve the momentum resolution and the track-finding capability. 
The reconstruction efficiency will be as high as 97\% for \pt$>2$~\GeVc, with a momentum resolution  of the order of 1--1.5\% (depending on \pt).
The pointing resolution will be better than 30~$\mu$m for \pt$>3$~~\GeVc.
The electrons from the \ups decays  are identified using a combination of the electromagnetic calorimeter and the inner hadron calorimeter with a pion rejection power better than 100 at a 95\% electron efficiency.

Exploiting this very good performance, sPHENIX will be able to measure the suppression pattern of the three \ups states.
The high RHIC luminosity and the sPHENIX data acquisition bandwidth (10 kHz) will give to sPHENIX the opportunity  to record 10$^{11}$ \AuAu collisions, leading to an unprecedented precision of \ups measurements, see \fig{fig:sPHENIX_RAA_UPSILON}.

Thanks to the combination of its high-precision inner tracking  and calorimetry systems, sPHENIX will be able to  perform $b$-quark jet tagging by requiring the presence of charged tracks within the jet with a large distance of closest approach to the primary vertex.
Simulations show that for jet \pt~=~20~\GeVc, a $b$-jet purity of 50\% will be reached with an efficiency of the order of 40--50\%~\cite{Aidala:2012nz}, which will allow to extract $b$-jet \raa down to \pt~=~20~\GeVc in central \AuAu collisions. 
Tagging of a $c$-quark jet using the same technique is challenging due to the shorter $c$-hadron lifetime.
Nevertheless, $c$-jet tagging performance is under study by associating fully reconstructed D meson with reconstructed jets in the calorimeter.
Those heavy-quark jet measurements will be an excellent test of in-medium parton energy loss mechanisms and will give some insights on the fragmentation functions of heavy quarks.

\paragraph{STAR experiment}

The recently-installed Heavy Flavour Tracker (HFT) and Muon Telescope Detector (MTD) 
will allow  significantly-improved measurements for heavy flavour observables already 
in the 2016 run \cite{STAR:2015bur}, as illustrated for the $\mathrm{D}^0$ meson elliptic
flow and the \ups \raa in \fig{fig:STAR_perf}.
Among the observables expected to become accessible in this run are $\Lambda_c$ baryon 
and open beauty meson production.

\begin{figure}[bt]
\begin{minipage}{8cm}
\begin{center}
\includegraphics[width=0.97\textwidth]{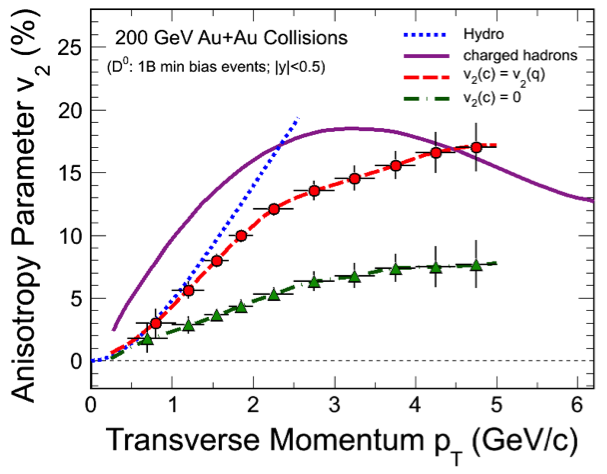}
\end{center}
\end{minipage}
\hfill
\begin{minipage}{8cm}
\begin{center}
\includegraphics[width=0.99\textwidth]{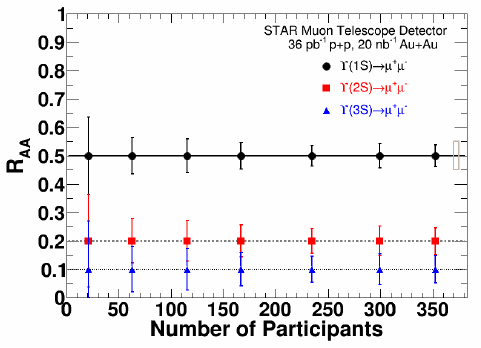}
\end{center}
\end{minipage}
\caption{Estimated performance (quoted are the statistical uncertainties) of the STAR measurement of the $\mathrm{D}^0$ meson elliptic flow (left panel) and the \ups  states \raa (left panel) in \AuAu collisions at \snn~=~200~\GeV~\protect\cite{STAR:2013bur,STAR:2015bur}.}
\label{fig:STAR_perf}
\end{figure}

The next stages of the STAR upgrade programme \cite{STAR_Decadal} include
on a short term (with focus on BES-II, 2018--2019) an upgrade of the
inner part of the TPC readout chambers as well as a new Event Plane
Detector (EPD) covering the rapidity range $1.5<|y|<5$.
The TPC upgrade will improve the tracking and PID performances and extend
the TPC coverage from $|y|<1$ to $|y|<1.7$.
The long term part of the upgrade (foreseen for the 2021--2022 heavy-ion
programme and the  2025+ eRHIC programme) includes upgrades on the HFT for
faster readout and a forward rapidity system with tracking and
calorimetry. The focus will be on measurements relevant for QGP in \AAcoll
collisions, CNM effects in \pA collisions and for the spin programme at
RHIC \cite{Akiba:2015jwa}. All heavy-flavour observables will receive
significant improvements in precision in these measurements.

\subsection{The fixed target experiments} %
Fixed-target  experiments using hadron beams have played a major role in quarkonium physics, starting with
the co-discovery  in 1974 of the \jpsi~\cite{Aubert:1974js} at BNL with a 30 GeV proton beam on a Be target, the discovery of
the $\Upsilon$~\cite{Herb:1977ek} with 400 GeV protons on Cu and Pt targets and the first observation of $h_c$~\cite{Armstrong:1992ae} at Fermilab
with antiprotons on an internal hydrogen jet target. In addition, fixed-target experiments
have revealed, through high-precision quarkonium studies, many novel and mostly unexpected features of quark and gluon
dynamics, which included the anomalous suppression of \jpsi~\cite{Abreu:2000ni}
in \pb collisions at SPS, the strong non-factorising nuclear suppression of \jpsi hadroproduction at high
$x_F$~\cite{Hoyer:1990us} and the large-$x_F$ production of \jpsi pairs~\cite{Badier:1982ae}.

A few fixed-target projects in connection with heavy-flavour and quarkonium are being discussed in our community at the SPS, Fermilab, FAIR and the LHC. These are reported below.

\subsubsection{Low energy projects at SPS, Fermilab and FAIR}
Using a 120~\GeV proton beam extracted from the Fermilab Main Injector, the Fermilab E-906/SeaQuest experiment~\cite{Reimer:2001bt}, 
which is part of a series of fixed target Drell-Yan experiments done at Fermilab, 
aims to examine the modifications to the antiquark structure of the proton from nuclear binding and to better quantify the energy loss of a coloured parton (quark) travelling 
through cold, strongly-interacting matter. In the context of this review, one should stress that their muon spectrometer covering 
dimuon mass from roughly 3 to 7~\ensuremath{\mathrm{GeV}/c^2} also allows one to perform \jpsi and \psiP cross section 
measurements with a good accuracy. 

The COMPASS Collaboration has recently started to look at Drell-Yan measurements using a 190 GeV pion beam~\cite{Quintans:2011zz} with the aim of measuring
single-transverse spin asymmetries and of measuring the quark Sivers functions~\cite{Sivers:1989cc}. Data taken during tests in 2009 have revealed that, with the same set-up, 
they can also measure, with a good accuracy, pion-induced \jpsi (probably also  \psiP) production cross sections at \snn~=~18.8\GeV on nuclear targets.
 
Another experiment at SPS, NA61/SHINE also has plans to move ahead to charm production in the context of heavy-ion physics. Their upgrade relies on the installation 
of a new silicon vertex detector~\cite{Ali:2013yva} which would allow for precise track reconstructions and, in turn, for \Dzero production studies in \PbPb collisions 
at \snn~=~8.6 and 17.1\GeV.

Finally, the FAIR project, presently under construction at GSI Darmstadt, has a nucleus--nucleus collisions 
programme devoted to the study of baryon-dominated matter at high densities~\cite{Friman:2011zz}.
The dedicated Compressed Baryonic Matter (CBM) experiment is designed with the initial priority on the study 
of open charm and charmonium close to production threshold in \AuAu collisions around 25 GeV-per-nucleon beam 
energy (\snn~$\simeq$~7\GeV). The current baseline of the FAIR project envisages collisions with Au beams only 
up to 10~\GeV per nucleon (SIS-100), implying that the charm sector of CBM cannot be covered in nucleus--nucleus collisions. 
Production in proton-nucleus collisions will be studied, while the possible addition of a second ring will bring the accelerator to the initially-designed energy (SIS-300).

\subsubsection{Plans for fixed-target experiments using the LHC beams}

\begin{table}[!ht]
\begin{center}
\caption{Expected yields (assuming no nuclear effects) and luminosities obtained for a 7 (2.76) TeV proton (Pb) beam extracted by means of bent crystal (upper part) and obtained with an internal gas target (lower part).}
\footnotesize
{\begin{tabular}{c  c c c c c c c c c c }
  \hline
  Beam & Flux & Target & \snn & Thickness & $\rho$ & $A$ & $\cal{L}$ & $\int{\cal{L}}$ &${\rm Br}_{\ell\ell} \left.\frac{\dd N_{\jpsi}}{\dd y}\right|_{y=0} $ &${\rm Br}_{\ell\ell} \left.\frac{\dd N_{\Upsilon}}{\dd y}\right|_{y=0} $  \\ 
   & (s$^{-1}$) &  & (GeV) &  (cm)&(g\,cm$^{-3}$) &  & ($\mu$b$^{-1}$s$^{-1}$) & (pb$^{-1}$\,y$^{-1})$ & (y$^{-1}$)& (y$^{-1}$)\\ \hline
  p & $5 \times 10^8$ &Liquid H & 115 & 100 & 0.068 & 1 & 2000 & 20000 & $4.0 \times 10^8$ & $8.0 \times 10^5$  \\
  p & $5 \times 10^8$ &Liquid D & 115 & 100 & 0.16 & 2 & 2400 & 24000 & $9.6 \times 10^8$ & $1.9 \times 10^6$\\
  p & $5 \times 10^8$ &Pb & 115 & 1  & 11.35 & 207 & 16 & 160 & $6.7 \times 10^8$ & $1.3 \times 10^6$ \\ \hline 
  Pb & $2 \times 10^5$ &Liquid H & 72 & 100 & 0.068 & 1 & 0.8 & 0.8 & $3.4 \times 10^6$ & $6.9 \times 10^3$\\
  Pb & $2 \times 10^5$ &Liquid D & 72 & 100 & 0.16 & 2 & 1 & 1 & $8.0 \times 10^6$ & $1.6 \times 10^4$\\
  Pb & $2 \times 10^5$ &Pb & 72 & 1  & 11.35 & 207 & 0.007 & 0.007 & $5.7 \times 10^6$ & $1.1 \times 10^4$ \\ \hline \\ \hline
  Beam & Flux & Target & \snn & Usable gas zone & Pressure &  $A$ & $\cal{L}$ & $\int{\cal{L}}$ &${\rm Br}_{\ell\ell} \left.\frac{\dd N_{\jpsi}}{\dd y}\right|_{y=0} $ &${\rm Br}_{\ell\ell} \left.\frac{\dd N_{\Upsilon}}{\dd y}\right|_{y=0} $  \\
       & (s$^{-1}$)&  &  (GeV) & (cm) & (Bar) &  & ($\mu$b$^{-1}$s$^{-1}$) & (pb$^{-1}$\,y$^{-1})$  & (y$^{-1}$)& (y$^{-1}$) \\ \hline
    p  &  $3 \times 10^{18}$&perfect gas & 115 & 100 & $10^{-9}$ & $A$& 10 & 100 &  $2 \times 10^6 \times A$& $4 \times 10^3\times A$\\ \hline
    Pb &  $5 \times 10^{14}$&perfect gas & 72 & 100 & $10^{-9}$ & $A$ & 0.001 & 0.001 & $4.25 \times 10^3 \times A$& $8.6  \times A$\\ 
  \hline
\end{tabular}}
\end{center}
\label{tab:lumi-after}
\end{table}

Historically, the first proposal to perform fixed-target experiments with the LHC beams  dates back to the early nineties
along with the LHB proposal (see \eg \cite{Costantini:1993rr}) to perform flavour physics studies using the expected $10^{10}$ B mesons produced per year
using an extracted beam with a flux of more than  $10^8$ protons per second obtained with a bent-crystal 
positioned in the halo of the beam. This idea was revived in the mid 2000's~\cite{Uggerhoj:2005xz} 
and it is now being investigated at the LHC along with the smart collimator solution proposed by 
the (L)UA9 Collaboration\footnote{\url{http://home.web.cern.ch/about/experiments/ua9}.}.

More generally, a beam of 7\TeV protons colliding on fixed targets results in a centre-of-mass
energy close to 115\GeV, in a range where few systems have been studied at a limited luminosity.  
With the 2.76\TeV Pb beam, \snn amounts to 72\GeV, approximately
half way between the top \AuAu and \CuCu energy at RHIC and the typical energies studied at the SPS.
As discussed in \cis{Brodsky:2012vg,Lansberg:2012kf,Lansberg:2012wj,Rakotozafindrabe:2012ei}, colliding the LHC proton and heavy-ion
beams on fixed targets offer a remarkably wide range of physics opportunities. 
The fixed-target mode with TeV beams has four critical advantages:
i) very large luminosities, 
ii) an easy access over the full target-rapidity domain,
iii) the target versatility and 
iv) the target polarisation.
This respectively allows for: 
i) decisive statistical precision for many processes,
ii) the first experiment covering the whole negative $x_F$ domain up to $- 1$,
iii) an ambitious spin programme relying on the study of single transverse spin asymmetries and
iv)  a unique opportunity to
study in detail the nuclear matter versus the hot and dense matter formed in heavy-ion collisions, including the formation of the quark-gluon plasma down to the target rapidities.

\paragraph{SMOG -- the first step} A first -- probably decisive -- step towards such a project has been made by the LHCb Collaboration using SMOG, 
 a system designed to perform imaging of the beam profiles towards luminosity determination~\cite{FerroLuzzi:2005em}.
SMOG consists in the injection of a gas (Ne until now) in the VErtex LOcator of LHCb; this also allows to record fixed-target collisions.
During test beams, data have been recorded in {p--Ne} collisions at \snn~=~87~\GeV and {Pb--Ne} at \snn~=~54.4~\GeV. 
The current limited  statistics -- due to the limited gas pressure and the short run durations -- has for now only allowed for strange-hadron reconstruction. 
A handful of \jpsi and charmed mesons might be extracted from these data. In any case, they have illustrated 
that a detector like LHCb has a very good coverage for the fixed-target mode and proved that this system can be used beyond its primary goal and offers new physics opportunities.
LHCb plans in taking more data using SMOG during \RunTwo. The goal is to accumulate about 0.5~\nbinv of  {Pb--Ne} collisions with the aim of studying \jpsi and \Dzero productions.
Let us stress here that, thanks of the boost between the cms and laboratory frame, 
the rapidity shift between them is 4.8 with the 7~\TeV proton beam. Hence,
a detector covering $\eta_{\rm lab} \in [1,5]$ allows for measurements  in essentially the whole backward hemisphere, \ie $y_{\rm cms}\leq 0$ or $x_F \leq 0$.

\paragraph{A Fixed-Target ExpeRiment at the LHC, AFTER@LHC}
With a dedicated set-up and run schedule (see below), \pp and \pA collisions can be studied, during the $10^7$ s LHC proton run,
with luminosities three orders of magnitude larger than at RHIC. \ensuremath{\mathrm{Pb-A}} collisions can be studied, during the $10^6$~s LHC Pb run,  
at a luminosity comparable to that of RHIC and the LHC over the full range of the target-rapidity domain with a large
variety of nuclei. Quarkonium production, open heavy-flavour hadrons
and prompt photons in \pA collisions can thus be investigated~\cite{Brodsky:2012vg,Lansberg:2012kf} with statistics 
previously unheard of (see \tab{tab:lumi-after}) and in the backward region, $x_F < 0$, which is essentially uncharted.
This would certainly complement the studies discussed in \sect{Cold nuclear matter effects}.
In complement to conventional nuclear targets made of Pb, Au, W, Cu, etc., high precision QCD measurements (including some of those discussed in \sect{pp section}) can also obviously be performed in \pp and p--d collisions with hydrogen and deuterium targets.
Finally, looking at ultra-relativistic nucleus--nucleus collisions from the rest frame of one of the colliding nuclei offers the opportunity to study in detail its remnants in collisions where the QGP can be formed. 
Thanks to the use of the recent ultra-granular calorimetry technology,
studies of direct  photons, \chic\ and even \chib\ production  in heavy-ion collisions
-- two measurements not available in any other experimental configuration -- can be envisioned (see~\ci{Arleo:2012dxa} for a similar idea at SPS energies).

To substantiate these claims, we have gathered in \tab{tab:lumi-after} a set of key quantities for two scenarios -- a beam extracted/splitted with a bent crystal with a dense target and a gas target intercepting the full LHC flux --,  
such as the cms energy, the flux through the target, its length, its density/pressure, the instantaneous and yearly luminosities as well as the \jpsi and \ups yields close to $y=0$.
Another possibility, which consists in positioning a 500~$\mu$m thick lead ribbon in the halo of the proton or lead LHC beams, would lead to instantaneous luminosities of 100~mb$^{-1}$s$^{-1}$ and 2.2~mb$^{-1}$s$^{-1}$, respectively~\cite{Kurepin:2011zz}.


\section{Concluding remarks}
\label{sec:summary}

The first Run of the LHC has provided a wealth of measurements in proton--proton and heavy-ion  
collisions for hadrons with open and hidden charm and beauty. The LHC data 
complement the rich experimental programmes at Tevatron, SPS and RHIC,
 extending by factors of about four,  
fourteen and twenty-five the centre-of-mass energies accessible in pp, A--A and p--A collisions, respectively. 
 
The main features of the data are in general understood. However, the current experimental 
precision (statistical) and accuracy (systematic uncertainties) is in most cases
still limited. This, along with the the lack of 
precise enough guidance from theoretical models, still prevents definite 
conclusions on production mechanisms in pp collisions (for quarkonia), their modification in p--A, and extraction of key quantities for the QGP
produced in A--A collisions. 

\vspace{1mm}

In pp collisions, pQCD calculations at NLO or FONLL describe very well the open charm and beauty 
production cross sections
 within, however, rather large theoretical uncertainties, especially for charm 
at low \pt.
At the LHC, this uncertainty also impacts the scaling of the cross sections measured at top pp centre-of-mass energy
to the lower Pb--Pb and p--Pb energies. Therefore, it is crucial that the future LHC programme includes
adequate pp reference runs at the heavy-ion energies.
In the quarkonium sector, there is a large variety of quarkonium production models. 
To date, none describes consistently the available measurements on production cross 
section and polarization. Future data will allow to constrain the models further
and also address the question whether a single production mechanism is responsible for 
the low- and high-\pt quarkonia. Understanding the production process will provide insight on the 
quarkonium formation time, which is an important aspect for the study of medium-induced effects in p--A and A--A collisions. 
For both open and hidden heavy-flavour hadrons, the correlation of production with 
the event multiplicity is an interesting facet that may shed light on production
mechanisms and the general features of proton--proton collisions at high energy.
The connection of these effects with the studies 
in proton--nucleus and nucleus--nucleus collisions is an open and interesting field of theoretical 
and experimental investigation.

\vspace{1mm}

Initially thought as a reference for nucleus--nucleus studies, p--A collisions
provided a host of interesting results of their own. Nuclear medium effects are observed in p--A collisions on open and hidden heavy flavour at both RHIC and LHC, especially for \jpsi production at forward rapidity. None of the individual cold nuclear matter (CNM) effects are able to describe the data in all kinematic regions, suggesting that a mixture of different effects are at work. The approach of nuclear parton distribution functions with shadowing explains the basic features of open and hidden heavy flavour despite large uncertainties at forward rapidity and the coherent energy loss model explains the main characteristics of quarkonium production. 
Theoretical interpretation of quarkonium excited states is still challenging. 
The impact of these CNM studies for the understanding of the nucleus--nucleus data 
in terms of a combination of cold and hot medium effects is yet to be fully understood.
In addition, it is still an open question whether the possible signals of collective behaviour observed 
in high-multiplicity proton-nucleus collisions in the light-flavour sector could manifest also for heavy-flavour production.
This question could become accessible with future higher-statistics proton--nucleus data samples 
at RHIC and LHC.

\vspace{1mm}

The strong electromagnetic field of lead ions circulating in the LHC is an intense source of quasi-real photons, which allows the study of $\gamma\gamma$, $\gamma\rm{p}$ and $\gamma\rm{Pb}$ reactions at unprecedented high energies. 
The coherent and the incoherent photoproduction of \jpsi and \psiP
is a powerful tool to study the gluon distribution in the target hadron and the first data 
from the LHC using Run 1
already set strong constraints to shadowing models.
The statistical precision is one of the main, and in some cases the dominant, sources of uncertainty of the current measurements. The large increase in statistics expected for Run 2 and other future data-taking periods, as well as improvements in the detectors, the trigger and the data acquisition systems, will allow a substantial reduction of the uncertainties. These future measurements will then shed a brighter light on the phenomena of shadowing and the gluon structure of dense sources, like lead ions.

\vspace{1mm}

The measurements of open heavy flavour production in nucleus--nucleus, proton--proton and proton--nucleus collisions 
at RHIC and the LHC allow us to conclude that heavy quarks experience energy loss in the
hot and dense QGP.  
A colour charge dependence in energy loss is not clearly emerging from the data, but it is 
implied by the fair theoretical description of the observed patterns.
A quark mass ordering is suggested by the data (some of them still preliminary, though) 
and the corresponding model comparisons. However, this observation is still limited to a 
restricted momentum and centrality domain.
The important question of thermalisation of heavy quarks appears to be partly answered for 
charm: the positive elliptic flow observed at both RHIC and LHC indicates
that charm quarks take part in the collective expansion of the QGP.  This is consistent with 
thermalisation, but the degree of 
thermalisation is not yet constrained. 
For the beauty sector, thermalisation remains an open issue entirely.
The role of the different in-medium interaction mechanisms, such as radiative, collisional 
energy loss and in-medium hadronisation, is still not completely clarified, although the
comparison of data with theoretical models suggests the relevance of all these effects.

\vspace{1mm}

For the quarkonium families, the LHC data demonstrated the presence of colour screening 
for both charmonium and bottomonium. In case of \jpsi, the LHC data implies the presence
of other production mechanisms, generically called (re)generation. Whether production
takes place throughout the full (or most of the) lifetime of the deconfined state or
rather suddenly at the confinement transition (crossover) can not be disentangled
using the existing measurements.
The $\Upsilon$ production seems to exhibit a sequential pattern, but several assumed
quantities in this interpretation (e.g. the feed-down contributions) make the situation not satisfactory enough.

\vspace{1mm}

The next steps in the study of heavy-flavour hadron production in
heavy-ion collisions will lead to a stage of quantitative understanding of the data, towards 
the extraction of the charm and beauty quarks transport coefficients and the temperature 
history of the deconfined state, including the temperature of the confinement crossover.
An incremental, but nevertheless important, progress is expected with the existing
experimental set-ups at RHIC and the LHC (where in particular the increased collision 
energy enhances the relevance of the data in the next three years).
The ultimate goal can only be achieved with upgraded or new detectors, which will
allow the extension of the set of observables and the precision of the measurements over
a broad range of collision energies.


\vspace{1mm}

This experimental effort needs to be matched on the theory side. Even though the field
of study of extreme deconfined matter with heavy quarks seems to be driven by experiment,
the contribution of theory is of crucial importance. In particular,  
accurate theoretical guidance and modelling are required to interpret the measurements in terms of 
the QGP properties mentioned in 
the previous paragraph.
Ultimately, the quantitative stage 
can only be reached in a close collaboration of experiment and theory.


\section*{Acknowledgements}

The SaporeGravis network was supported by the European Community Research Infrastructures Integrating Activity ``Study of strongly interacting matter" (acronym HadronPhysics3) - Grant Agreement no. 283286 - within the Seventh Framework Programme (FP7) of EU.
The work of T.~Dahms was supported by the DFG cluster of excellence ``Origin and Structure of the Universe". The work of R.~Rapp was supported by the US-NSF grant no. PHY-1306359. 
The work of E.~Ferreiro was supported by the Ministerio de Economia y Competitividad of Spain.
The work of P.B.~Gossiaux was supported by the Region Pays de la Loire through the TOGETHER project.
The work of B.~Kopeliovich was supported by the Fondecyt (Chile) grants 1130543, 1130549, 1100287,and ECOS-Conicyt grant No. C12E04. 
The work of L.~Massacrier was supported 
by the European Community Research Infrastructures Integrating Activity ``Study of strongly interacting matter (acronym HadronPhysics3) - Grant Agreement no. 283286 - within the Seventh Framework Programme (FP7) of EU, and
by the P2IO Excellence Laboratory. 
The work of A.~Mischke was supported by the Vidi grant from the Netherlands Organisation for Scientific Research (project number: 680-47-232) and Projectruimte grants from the Dutch Foundation for Fundamental Research (project numbers: 10PR2884 and 12PR3083). The work of M.~Nahrgang was supported by the Postdoc-Program of the German Academic Exchange Service (DAAD) and the U.S. Department of Energy under grant DE-FG02-05ER41367.
The work of B.~Trzeciak was supported by the European social fund within the framework of realizing the project, Support of inter-sectoral mobility and quality enhancement of research teams at Czech Technical University in Prague, CZ.1.07/2.3.00/30.0034 and by Grant Agency of the Czech Republic, grant No.13-20841S.
The work of I.~Vitev was supported by Los Alamos National Laboratory DOE Office of Science  Contract No. DE-AC52-06NA25396 and the DOE Early Career Program.
The work of R.~Vogt was performed under the auspices of the U.S. Department of Energy by Lawrence Livermore National Laboratory under contract DE-AC52-07NA27344 and supported in part by the JET collaboration, the U.S. Department of Energy, Office of Science, Office of Nuclear Physics (Nuclear Theory).





\bibliographystyle{utphys}
\bibliography{fullBiblio}







\end{document}